\documentclass[11pt]{book}
\pdfoutput=1

\usepackage[utf8]{inputenc}
\usepackage[french,english]{babel}
	
\usepackage[a4paper,left=2cm,right=2cm]{geometry}

\usepackage{fancyhdr,lastpage}

\fancypagestyle{plain}
{
\fancyhf{}
\fancyhead[l]{}
\fancyhead[c]{}
\fancyhead[r]{}
\fancyhead[LE]{\rightmark}
\fancyhead[RO]{\leftmark}

\fancyfoot[c]{--- \textsc{Page} \thepage/\pageref{LastPage} ---}
\fancyfoot[LO,LE]{\textsc{F. Nugier}}
\fancyfoot[RO,RE]{\textsc{UPMC}}
}

\usepackage{xcolor}
	
\usepackage{hyperref}
\hypersetup{
colorlinks=true,
citecolor=green,
linkcolor=blue,
urlcolor=purple
}
 
\usepackage[T1]{fontenc}
\usepackage{ae,aecompl}
\usepackage{amsmath,amsfonts,amssymb,stmaryrd,mathrsfs}
\usepackage{esint}

\usepackage{graphicx,float,caption}

\usepackage[nottoc, notlof, notlot]{tocbibind}

\usepackage{cite}

\usepackage{tikz}
\usetikzlibrary{calc}
\usepackage{pgfplots}

\usepackage[toc]{glossaries}

\usepackage{lipsum}

\usepackage{filecontents}

\usepackage{ifthen}

\usepackage{titlesec}

\usepackage{xspace}

\usepackage[titletoc]{appendix}

\usepackage{multicol}

\renewcommand*{\glossarysection}[2][]{%
  \begin{multicols}{2}[\chapter*{#2}]
  \setlength\glsdescwidth{0.5\linewidth}%
  \glossarymark{\glossarytoctitle}%
}
\renewcommand*{\glossarypostamble}{\end{multicols}}
\makeglossaries

\newacronym{FLRW}{FLRW}{Friedmann-Lema\^{i}tre-Robertson-Walker}
\newacronym{ADM}{ADM}{Arnowitt-Deser-Misner}
\newacronym{CMB}{CMB}{Cosmic Microwave Background}
\newacronym{BAO}{BAO}{Baryonic Acoustic Oscillations}
\newacronym{GCT}{GCT}{General Coordinate Transformation}
\newacronym{GT}{GT}{Gauge Transformation}
\newacronym{RD}{RD}{redshift drift}
\newacronym{dofs}{dofs}{degrees of freedom}
\newacronym{SMC}{SMC}{standard model of cosmology}
\newacronym{DM}{DM}{dark matter}
\newacronym{DE}{DE}{dark energy}
\newacronym{EoM}{EoM}{equation of motion}
\newacronym{IBR}{IBR}{induced backreaction}
\newacronym{RP}{RP}{redshift perturbations}
\newacronym{PG}{PG}{Poisson gauge}
\newacronym{LTB}{LTB}{Lema\^itre-Tolman-Bondi}
\newacronym{RSD}{RSD}{redshift space distortions}
\newacronym{SW}{SW}{Sachs-Wolfe effect}
\newacronym{ISW}{ISW}{integrated Sachs-Wolfe effect}
\newacronym{NG}{NG}{Newtonian gauge}

\newglossaryentry{gauge freedom}{ name = {gauge freedom}, description = {~} }
\newglossaryentry{virial equilibrium} { name = {virial equilibrium}, description = {~} }
\newglossaryentry{standard candels} { name = {standard candels}, description = {~} }
\newglossaryentry{Raychaudhuri's equation} { name = {Raychaudhuri's equation}, description = {~} }
\newglossaryentry{Hamiltonian constraint} { name = {Hamiltonian constraint}, description = {~} }
\newglossaryentry{window function} { name = {window function}, description = {~} }
\newglossaryentry{morphon} { name = {morphon}, description = {~} }
\newglossaryentry{observational coordinates} { name = {observational coordinates}, description = {~} }
\newglossaryentry{geodesic light-cone} { name = {geodesic light-cone}, description = {~} }
\newglossaryentry{backreaction parameter} { name = {backreaction parameter}, description = {~} }
\newglossaryentry{Kullback-Leibler entropy} { name = {Kullback-Leibler entropy}, description = {~} }
\newglossaryentry{commutation rule} { name = {commutation rule}, description = {~} }
\newglossaryentry{integrability condition} { name = {integrability condition}, description = {~} }
\newglossaryentry{flatness problem} { name = {flatness problem}, description = {~} }
\newglossaryentry{horizon problem} { name = {horizon problem}, description = {~} }
\newglossaryentry{transfer function} { name = {transfer function}, description = {~} }
\newglossaryentry{apparent magnitude} { name = {apparent magnitude}, description = {~} }
\newglossaryentry{absolute magnitude} { name = {absolute magnitude}, description = {~} }
\newglossaryentry{distance modulus} { name = {distance modulus}, description = {~} }
\newglossaryentry{deceleration parameter} { name = {deceleration parameter}, description = {~} }
\newglossaryentry{Jeans length} { name = {Jeans length}, description = {~} }
\newglossaryentry{Poisson equation} { name = {Poisson equation}, description = {~} }
\newglossaryentry{gauge invariance} { name = {gauge invariance}, description = {~} }
\newglossaryentry{linear growth function} { name = {linear growth function}, description = {~} }
\newglossaryentry{Jeans mass} { name = {Jeans mass}, description = {~} }
\newglossaryentry{slow-roll parameters} { name = {slow-roll parameters}, description = {~} }
\newglossaryentry{Lie derivative} { name = {Lie derivative}, description = {~} }
\newglossaryentry{cosmological constant problem} { name = {cosmological constant problem}, description = {~} }
\newglossaryentry{expansion tensor} { name = {expansion tensor}, description = {~} }
\newglossaryentry{expansion scalar} { name = {expansion scalar}, description = {~} }
\newglossaryentry{shear tensor} { name = {shear tensor}, description = {~} }
\newglossaryentry{vorticity tensor} { name = {vorticity tensor}, description = {~} }
\newglossaryentry{ADM energy density} { name = {ADM energy density}, description = {~} }
\newglossaryentry{ADM pressure} { name = {ADM pressure}, description = {~} }
\newglossaryentry{Bardeen potentials} { name = {Bardeen potentials}, description = {~} }
\newglossaryentry{conformal geodesic lightcone gauge} { name = {conformal geodesic lightcone gauge}, description = {~} }
\newglossaryentry{conformal double lightcone gauge} { name = {conformal double lightcone gauge}, description = {~} }
\newglossaryentry{Gauss-Weingarten equation} { name = {Gauss-Weingarten equation}, description = {~} }
\newglossaryentry{Gauss-Codazzi equations} { name = {Gauss-Codazzi equations}, description = {~} }
\newglossaryentry{ADM formula} { name = {ADM formula}, description = {~} }
\newglossaryentry{extrinsic curvature} { name = {extrinsic curvature}, description = {~} }
\newglossaryentry{entropy mode} { name = {entropy mode}, description = {~} }
\newglossaryentry{vortical modes} { name = {vortical modes}, description = {~} }
\newglossaryentry{adiabatic modes} { name = {adiabatic modes}, description = {~} }
\newglossaryentry{bias} { name = {bias}, description = {~} }
\newglossaryentry{fitting problem} { name = {fitting problem}, description = {~} }
\newglossaryentry{large scale structure} { name = {large scale structure}, description = {~} }
\newglossaryentry{Hubble diagram} { name = {Hubble diagram}, description = {~} }
\newglossaryentry{coincidence problem} { name = {coincidence problem}, description = {~} }
\newglossaryentry{averaging problem} { name = {averaging problem}, description = {~} }
\newglossaryentry{backreaction problem} { name = {backreaction problem}, description = {~} }
\newglossaryentry{cosmological principle} { name = {cosmological principle}, description = {~} }
\newglossaryentry{luminosity flux} { name = {luminosity flux}, description = {~} }
\newglossaryentry{Chandrasekhar mass} { name = {Chandrasekhar mass}, description = {~} }
\newglossaryentry{light curve} { name = {light curve}, description = {~} }
\newglossaryentry{lapse function} { name = {lapse function}, description = {~} }
\newglossaryentry{shift vector} { name = {shift vector}, description = {~} }
\newglossaryentry{Copernican principle} { name = {Copernican principle}, description = {~} }
\newglossaryentry{Einstein equations} { name = {Einstein equations}, description = {~} }
\newglossaryentry{Meszaros effect} { name = {Meszaros effect}, description = {~} }
\newglossaryentry{stretch factor} { name = {stretch factor}, description = {~} }
\newglossaryentry{color parameter} { name = {color parameter}, description = {~} }
\newglossaryentry{peculiar velocity} { name = {peculiar velocity}, description = {~} }
\newglossaryentry{strong energy condition} { name = {strong energy condition}, description = {~} }
\newglossaryentry{e-fold number} { name = {e-fold number}, description = {~} }
\newglossaryentry{inflaton} { name = {inflaton}, description = {~} }
\newglossaryentry{equation of state} { name = {equation of state}, description = {~} }
\newglossaryentry{speed of sound} { name = {speed of sound}, description = {~} }
\newglossaryentry{decoupling} { name = {decoupling}, description = {~} }
\newglossaryentry{Einstein tensor} { name = {Einstein tensor}, description = {~} }
\newglossaryentry{Ricci tensor} { name = {Ricci tensor}, description = {~} }
\newglossaryentry{stress energy tensor} { name = {stress energy tensor}, description = {~} }
\newglossaryentry{Riemann tensor} { name = {Riemann tensor}, description = {~} }
\newglossaryentry{Christoffel symbols} { name = {Christoffel symbols}, description = {~} }
\newglossaryentry{Hubble horizon} { name = {Hubble horizon}, description = {~} }
\newglossaryentry{scalar shear potential} { name = {scalar shear potential}, description = {~} }
\newglossaryentry{inflationary spectrum} { name = {inflationary spectrum}, description = {~} }
\newglossaryentry{N-body simulations} { name = {N-body simulations}, description = {~} }
\newglossaryentry{density variance} { name = {density variance}, description = {~} }
\newglossaryentry{Etherington law} { name = {Etherington law}, description = {~} }
\newglossaryentry{ensemble-average conditions} { name = {ensemble-average conditions}, description = {~} }
\newglossaryentry{statistical homogeneity condition} { name = {statistical homogeneity condition}, description = {~} }
\newglossaryentry{backreaction} { name = {backreaction}, description = {~} }
\newglossaryentry{spectral coefficients} { name = {spectral coefficients}, description = {~} }
\newglossaryentry{growth factor} { name = {growth factor}, description = {~} }
\newglossaryentry{sound horizon} { name = {sound horizon}, description = {~} }
\newglossaryentry{effective shape parameter} { name = {effective shape parameter}, description = {~} }
\newglossaryentry{Doppler effect} { name = {Doppler effect}, description = {~} }
\newglossaryentry{lensing} { name = {lensing}, description = {~} }
\newglossaryentry{HaloFit model} { name = {HaloFit model}, description = {~} }
\newglossaryentry{lensing dispersion} { name = {lensing dispersion}, description = {~} }
\newglossaryentry{Milne} { name = {Milne}, description = {~} }
\newglossaryentry{angular distance} { name = {angular distance}, description = {~} }
\newglossaryentry{luminosity distance} { name = {luminosity distance}, description = {~} }
\newglossaryentry{Perturbed trajectories} { name = {Perturbed trajectories}, description = {~} }
\newglossaryentry{non-linearity scale} { name = {non-linearity scale}, description = {~} }
\newglossaryentry{curvature of the spectral index} { name = {curvature of the spectral index}, description = {~} }
\newglossaryentry{effective spectral index} { name = {effective spectral index}, description = {~} }
\newglossaryentry{two-halo term} { name = {two-halo term}, description = {~} }
\newglossaryentry{one-halo term} { name = {one-halo term}, description = {~} }
\newglossaryentry{lens-lens coupling} { name = {lens-lens coupling}, description = {~} }
\newglossaryentry{Born approximation} { name = {Born approximation}, description = {~} }
\newglossaryentry{Frame dragging} { name = {Frame dragging}, description = {~} }
\newglossaryentry{Silk damping} { name = {Silk damping}, description = {~} }

\newcommand{\ra}{$\Rightarrow$}
\newcommand{\la}{$\Leftarrow$}

\usepackage{xspace}

\newcommand{\etal}{\emph{et al.}\xspace}

\newcommand{\Gma}[3]{\Gamma^{#1}_{#2#3}}        % Symbole de Christoffel : Exemple d'utilisation : $\Gma abc$ OU $\Gma{a}{b}{c}$
\newcommand{\Rim}[4]{R^{#1}_{#2#3#4}}           % Symbole de Christoffel

\newcommand{\parfrac}[2]{\frac{\partial #1}{\partial #2}}

\newcommand{\tbf}[1]{\textbf{#1}}
\newcommand{\mrm}[1]{\mathrm{#1}}
\newcommand{\rref}[1]{(\ref{#1})}

\newcommand{\lla}{\left\langle}
\newcommand{\rra}{\right\rangle}
\newcommand{\rrao}{\right\rangle_{A_0}}

\let\lbar\l

\newcommand{\tila}{\tilde{a}}

\newcommand{\RRR}{\mathbb{R}\xspace}

\def\beq{\begin{equation}}
\def\eeq{\end{equation}}
\def\bea{\begin{eqnarray}}
\def\eea{\end{eqnarray}}

\def\l {\langle}
\def\re {\rangle}

\def \pa {\partial}
\def \ra {\rightarrow}
\def \ti {\widetilde}
\def \la {\lambda}

\def \La {\Lambda}
\def \Da {\Delta}
\def \da {\delta}

\def \ga {\gamma}
\def \sg {\sigma}
\def \Sg {\Sigma}
\def \da {\delta}
\def \ep {\epsilon}

\def \Om {\Omega}

\def \Mq {{\cal M}_4}

    \def\e{\mbox{e}}

\newcommand{\Ups}{\Upsilon}

\newcommand{\Acal}{\mathcal A}
\newcommand{\Dcal}{\mathcal D}

\newcommand{\Ccal}{\mathcal C}
\newcommand{\Ocal}{\mathcal O}

\newcommand{\Tcal}{\mathcal T}
\newcommand{\Ical}{\mathcal I}
\newcommand{\Pcal}{\mathcal P}
\newcommand{\Mcal}{\mathcal M}

\def \L{\rm L}
\def \NL{\rm NL}
\def \H {\rm H}
\def \Q{\rm Q}
\def \hompc{{\,h\,\rm Mpc^{-1}}}

\usepackage{pifont}

\def\laq{~\raise 0.4ex\hbox{$<$}\kern -0.8em\lower 0.62ex\hbox{$\sim$}~}
\def\gaq{~\raise 0.4ex\hbox{$>$}\kern -0.7em\lower 0.62ex\hbox{$\sim$}~}

\newcommand{\Mpc}{\xspace{\rm Mpc}}

\newcommand{\pc}{\xspace{\rm pc}}
\newcommand{\eV}{\xspace{\rm eV}}

\newcommand{\GeV}{\xspace{\rm GeV}}
\newcommand{\TeV}{\xspace{\rm TeV}}

\newcommand{\dphi}{\dot{\phi}}
\newcommand{\ddphi}{\ddot{\phi}}

\newcommand{\kbf}{\textbf{k}}
\newcommand{\xbf}{\textbf{x}}
\newcommand{\rbf}{\textbf{r}}
\newcommand{\vbf}{\textbf{v}}
\newcommand{\ktbf}{\tilde{\textbf{k}}}
\newcommand{\ebf}{\textbf{e}}
\newcommand{\ubf}{\textbf{u}}

\newcommand{\nbf}{\textbf{n}}
\newcommand{\Tbf}{\textbf{T}}
\newcommand{\qbf}{\textbf{q}}

\newcommand{\rhob}{\overline{\rho}}
\newcommand{\pb}{\overline{p}}
\newcommand{\Phib}{\overline{\Phi}}
\newcommand{\vbbf}{\overline{\textbf{v}}}
\newcommand{\Sb}{\overline{S}}

\newcommand{\Dscr}{\mathscr{D}}
\newcommand{\Lscr}{\mathscr{L}}
\newcommand{\Cscr}{\mathscr{C}}
\newcommand{\Sscr}{\mathscr{S}}

\newcommand{\Drm}{\mathrm{D}}

\newcommand{\Rcal}{\mathcal R}
\newcommand{\Ncal}{\mathcal N}
\newcommand{\Hcal}{\mathcal H}

\newcommand{\Wcal}{\mathcal W}
\newcommand{\Qcal}{\mathcal Q}

\newcommand{\tdev}{\mbox{\Large $\cdot$}}

\newcommand{\cst}{{\rm cst}}

\newcommand{\titheta}{\widetilde{\theta}}

\newcommand{\Prm}{\rm P}

\newenvironment{itemize*}%
{\begin{itemize}%
    \setlength{\itemsep}{5pt}%
    \setlength{\parskip}{5pt}}%
{\end{itemize}}

\begin{document}

% ------------------------- Cover Page ------------------------------ %
\selectlanguage{francais}
\pagestyle{empty}

\setlength{\oddsidemargin}{-0.5cm}
\setlength{\evensidemargin}{-0.5cm}

\begin{tikzpicture}[remember picture,overlay]  
  \node [xshift=-10.3cm,yshift=-2.0cm] at (current page.north east)
    {\includegraphics[width=5cm,height=2.5cm]{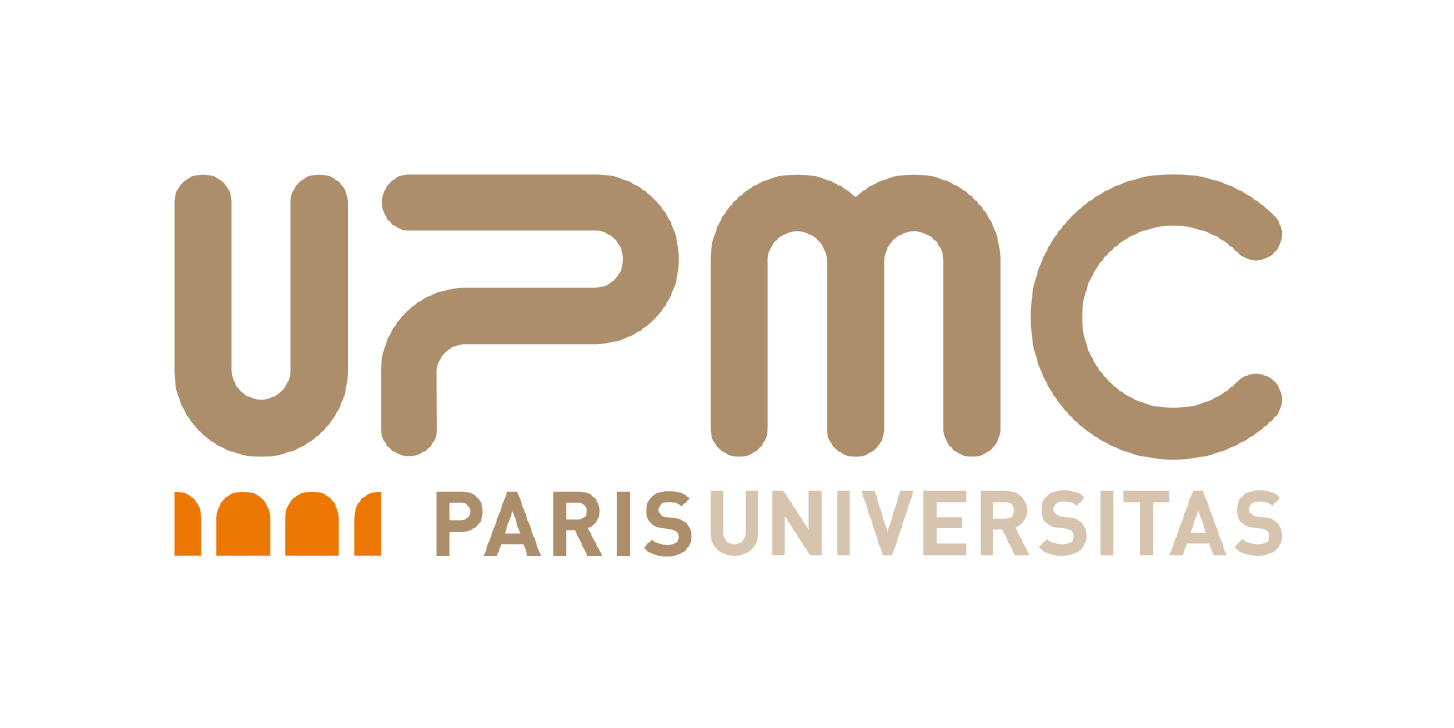}};
\end{tikzpicture}

\vspace{0.5cm}

\begin{center}
\huge{\textsc{Th\`ese de Doctorat \\ de l'Universit\'e Pierre et Marie Curie}}

\bigskip \bigskip
\Large{Sp\'ecialit\'e}

\bigskip

\huge{Physique Th\'eorique}

\bigskip
\Large{(\'{E}cole Doctorale 107)}

\bigskip \bigskip \bigskip

Pr\'esent\'ee par

\smallskip

\Large{\textsc{Fabien Nugier}}

\bigskip

Pour obtenir le grade de

\smallskip

\Large{\textbf{Docteur de l'Universit\'e Pierre et Marie Curie}}

\bigskip \bigskip \bigskip

\underline{Sujet de th\`ese}

\vspace{-0.7cm}
\begin{center} \line(1,0){500} \end{center}

\Huge{
\textsc{\textsc{Moyennes sur le C\^{o}ne de Lumi\`{e}re \\ et Cosmologie de Pr\'{e}cision}}
}

\bigskip

\Large{
\textsc{\textsc{Lightcone Averaging and Precision Cosmology}}
}

\vspace{-1cm}
\begin{center} \line(1,0){500} \end{center}

\smallskip

\Large

Soutenue le

\Large{\textsc{4 Septembre 2013}}

\bigskip
~ devant le jury compos\'{e} de\,:
\bigskip

\vspace{-0.7cm}
\begin{center} \line(1,0){500} \end{center}
\vspace{-0.7cm}
\begin{center}
\begin{tabular}{lcc}
\textsc{Pr. Gabriele Veneziano} & Directeur de Th\`{e}se & Coll\`{e}ge de France \\
\textsc{Pr. Costas Bachas} & Invit\'{e} & \'{E}cole Normale Sup\'{e}rieure \\
\textsc{Pr. Luca Amendola} & Rapporteur & Universit\'{e} de Heidelberg \\
\textsc{Pr. Dominik Schwarz} & Rapporteur & Universit\'{e} de Bielefeld \\
\textsc{Pr. Ruth Durrer} & Examinatrice & Universit\'{e} de Gen\`{e}ve \\
\textsc{Pr. David Langlois} & Examinateur & Universit\'{e} Paris Diderot \\
\textsc{Pr. Jean-Philippe Uzan} & Examinateur & Univ. Pierre et Marie Curie \\
\end{tabular}
\end{center}
\vspace{-0.7cm}
\begin{center} \line(1,0){500} \end{center}

\normalsize

\end{center}

% ------------------------------------------------------- %

\newpage

\pagestyle{empty}

\setlength{\oddsidemargin}{0.2cm}
\setlength{\evensidemargin}{-1.2cm}

% *******************************************

\begin{center}

~

\vspace{1cm}

\large

Cette th\`{e}se a \'{e}t\'{e} pr\'{e}par\'{e}e \\
au sein du LPT ENS

\normalsize

\vspace{5cm}

\begin{tikzpicture}[remember picture,overlay]
\node [xshift=-13.7cm,yshift=-8.0cm] at (current page.north east)
    {\includegraphics[width=3.5cm,height=2cm]{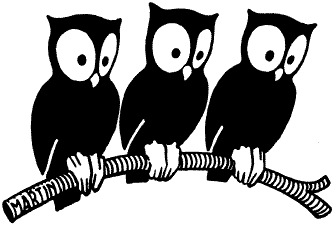}};
\node [xshift=-8.2cm,yshift=-8.0cm] at (current page.north east)
    {\includegraphics[width=2.5cm,height=2.5cm]{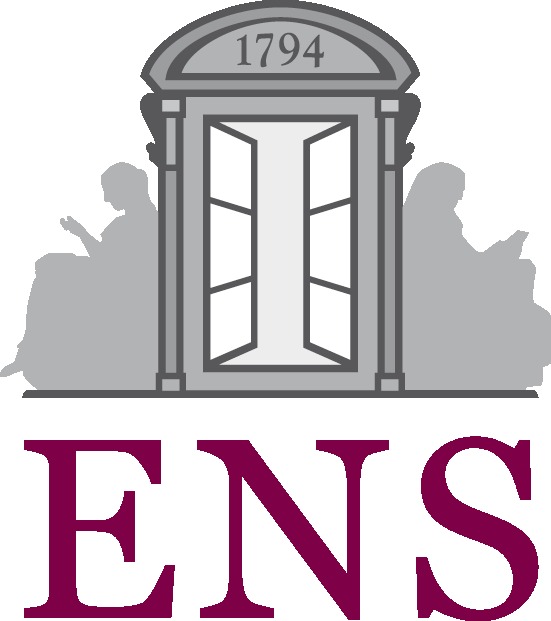}};
\end{tikzpicture}

\Large{Laboratoire de Physique Th\'{e}orique \\
\'{E}cole Normale Sup\'{e}rieure \\
24 rue Lhomond \\
75231 Paris Cedex 05}

\vspace{2cm}

\large

ainsi qu'au \\
Coll\`{e}ge de France

\normalsize

\vspace{4cm}

\begin{tikzpicture}[remember picture,overlay]
\node [xshift=-11.3cm,yshift=-19.5cm] at (current page.north east)
    {\includegraphics[width=5cm,height=3.5cm]{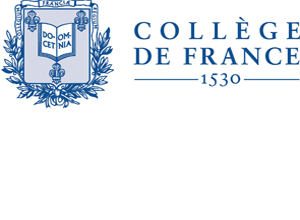}};
\end{tikzpicture}

\Large{
Coll\`{e}ge de France \\
3 rue d'Ulm \\
~ 75005 Paris \!
}

\vspace{2cm}

\large{
Pour me contacter : \\
\texttt{fabien.nugier@lpt.ens.fr} ~~ ou ~~ \texttt{fabien.nugier@etu.upmc.fr} \\
(jusque fin Septembre 2013) \\
\texttt{fabien.nugier@googlemail.com} \\
(ultérieurement, sinon voir autre adresse professionnelle) \\
}

\normalsize

% *******************************************

\end{center}

\setcounter{page}{0}

\newpage

% ****************************************************************

~

\vspace{5cm}

\begin{center} \line(1,0){500} \end{center}

\begin{center}
\Huge{
\textsc{\textsc{Moyennes sur le C\^{o}ne de Lumi\`{e}re \\ et Cosmologie de Pr\'{e}cision}}
}

\bigskip

\Large{
\textsc{\textsc{Lightcone Averaging and Precision Cosmology}}
}
\end{center}

\vspace{-1cm}
\begin{center} \line(1,0){500} \end{center}

\begin{center}
\Large{\textsc{Fabien NUGIER}}
\end{center}

\newpage
\setcounter{page}{0}

~

\vspace{23cm}

\begin{center}
This thesis was written in \LaTeX. \\
A donation for trees plantation was done to compensate for my PhD printings. \\
May these trees grow tall and old.
\end{center}

% ****************************************************************

%\newpage

\pagestyle{empty}

\selectlanguage{english}

% \makeglossaries

\addtocontents{toc}{\protect\thispagestyle{empty}}
\setcounter{tocdepth}{1}
\setcounter{page}{0}

\pagestyle{empty}

\tableofcontents

\pagestyle{empty}

\addtocontents{toc}{~\hfill\textbf{Page}\par}

\newpage

\setcounter{page}{1}
\pagestyle{empty}

% ----- Dedicated to  ----- %
~
\vspace{8cm}

\begin{center}
\Large
\`{A} ma mère, \textsc{Annick}. \\
\`{A} mon père \textsc{Fr\'ed\'eric} \\
et \`{a} ma soeur \textsc{Ana\"is}. \\
\bigskip
Je vous aime.
\normalsize
\end{center}

%\vspace{10cm}
\vspace{12.5cm}

\begin{center}
\emph{Ceux qui vivent, ce sont ceux qui luttent.} \\
Victor Hugo
\end{center}
% ------------------------- %

\newpage

\pagenumbering{roman}

% ----- Acknowledgment ----- %
\setcounter{page}{0}
\setcounter{page}{0}

%********************************************************************************
% ***** To take into account the Acknowledgement session in a proper way *****
%*****************************************************************************

\makeatletter
\newcommand\ackname{Acknowledgements}
\if@titlepage
  \newenvironment{acknowledgements}{%
      \titlepage
      \vspace{-3cm}
      %\null\vfil
      \@beginparpenalty\@lowpenalty
      \begin{center}%
        \bfseries \ackname
        \@endparpenalty\@M
      \end{center}}%
     {\par\vfil\null\endtitlepage}
\else
  \newenvironment{acknowledgements}{%
      \if@twocolumn
	{
        \section*{\abstractname}%
	}
      \else
        \small
        \begin{center}%
          {\bfseries \ackname\vspace{-0.5em}\vspace{\z@}}%
        \end{center}%
        \quotation
      \fi}
      {\if@twocolumn\else\endquotation\fi}
\fi
\makeatother

\makeatletter
\newcommand\absname{Abstract}
\if@titlepage
  \newenvironment{abstract}{%
      \titlepage
      %\null\vfil
      \@beginparpenalty\@lowpenalty
      \begin{center}%
        \bfseries \absname
        \@endparpenalty\@M
     \end{center}}%
     {\par\vfil\null\endtitlepage}
\else
  \newenvironment{abstract}{%
      \if@twocolumn
        \section*{\abstractname}%
      \else
        \small
        \begin{center}%
          {\bfseries \absname\vspace{-.5em}\vspace{\z@}}%
        \end{center}%
        \quotation
      \fi}
      {\if@twocolumn\else\endquotation\fi}
\fi
\makeatother

\hfuzz2pt % Don't bother to report over-full boxes if over-edge is < 2pt 

%********************************************************************************

% Changing the text length
\setlength{\textheight}{27.5cm} % 22.3cm

\begin{acknowledgements}

The first person I would like to thank here is Gabriele Veneziano. I am very grateful to him for accepting me as his student and for proposing me a very interesting subject of research. I have learned a lot from his incredible knowledge and his great human qualities. Thank you ! I would also like to thank Costas Bachas for accepting the codirection of my thesis, and his kindness, and Costas Kounnas for accepting me in the Laboratoire de Physique Th\'eorique de l'E.N.S.

On the funding side, this thesis work has been financed by a Contrat Doctoral with the Universit\'e Pierre et Marie Curie (UPMC). I am very thankful with respect to the Ecole Doctorale 107, the UPMC, the French Ministry of Research for supporting my researches during these three years and I hope they will continue to finance projects in this field. I am also grateful to the Ecole Normale Sup\'{e}rieure (E.N.S.) and the Coll\`{e}ge de France (CdF) which offered me a place to work.

There couldn't be this work without special thanks to my collaborators during this thesis, Giovanni Marozzi for his scientific support during the first two years of my PhD, and the moments of great laugh we had at the Coll\`{e}ge de France, Maurizio Gasperini for his constant kindness and sense of compromise, Ido Ben-Dayan for our discussions and his constant checks in our considerations. I am also thankful to the other people I have met during these years and with whom I have learned about physics. In particular, I would like to thank the different labs I had the occasion to visit : the ICTP in Trieste, the Univ. of Geneva, the ITP of Zurich, the CCPP in New York, the DAS in Princeton, the CPC at UPenn, the Univ. Joseph Fourier in Les Houches, the APC and IAP in Paris. I also want to thank particularly Ed Copeland, and again Gabriele and Maurizio, for the many letters they sent on my behalf for postdoctoral applications. I'm sorry that I bothered you so much with that.

Having worked as a teaching assistant at the UPMC, I would like to thank all the great people I have met in the teaching teams. I have been impressed by their courage and motivation of every day by associating research with lectures, despite the numerous complications. I am particularly grateful to Bruno Sicardy, St\'{e}phane Boucard, Jean Hare, Benoit Semelin, Nicolas Rambaux, Michel Capderou, Philippe Lopez, and all the other teachers that I have met. I also want to thank the students that came into my classes for they were always nice and often very interesting people.

Evolving during four years in the ENS has been an enriching experience for me. I cannot deny that I would have seek for more interaction with people there but this is not so easy. I have nevertheless to sincerely thank several people. I would like to thank the researchers that I had the most interactions with\,: C. Bachas, P. Fayet, J. Troost, A. Kashani-Poor, M.-T. Jaekel, G. Policastro. Outside \textsc{LPT ENS}, I particularly want to thank people in the Cosmology group of the LPNHE (UPMC) for our very interesting discussions\,: J. Guy, P. Astier, C. Balland, D. Hardin and R. Pain. A big thank you also to B\'{e}atrice Fixois, Viviane Sebille, Christelle Cagniart (ENS), and Anna Koczorowski (CdF) for their great administrative support, to Nicole Ribet for her moral help, and their perpetual good spirit.

Lots of people have crossed my world line during these three years, sometimes only for few minutes, sometimes during hours, and I would like to forget none of them. I want to thank my colleagues of the ENS for our good moments\,: Benjamin Assel, No\'{e} Lahaye, Konstantina Kontoudi, Valentina Pozzoli, Alexander Dobrinevski, Swann Piatecki, Sophie Rosay, Bruno Ribstein, Pauline Maury, Evgeni Sobko, Giuseppe Fanizza, Bruno Le Floch, Pawe\lbar\ Lasko\'{s}-Grabowski, Marine Remaud\ldots

Outside ENS I have to thank my long-time friends that I have spent my studies with and who gave me a wonderful support when I was sometimes losing my motivation : Rudy Romain, Noé, Panayotis Akridas-Morel, Pierre-\'{E}lie Larr\'{e}, Nicolas Viens, Benoit Fleury. I like you very much guys ! I hope that our future light-cones will continue to intersect each other as often as possible.

Respect is the primary principle in all human activities which seek for a better world. I have probably missed to thank a lot of people who deserve to be thanked. At the end, even a smile or a nice word from a stranger can bring a new hope in a cloudy day and all our lives are linked in a certain way. I want to thank all the people I have met and spent nice time with. I hope that I have been as kind with you as you have been with me, and I apologize for not putting your names.

Determination is maybe my only gift and I own it from the support I always got from the closest human beings of my life. I want to thank my mother Annick Nugier for her support all along my life and during these three years of hard work, deceptions and excitations, confidence and doubts. There wouldn't be this thesis under the whole Heavens without her help of every day. I also want to thank my father, Frédéric, and my sister, Anaïs, for their support during this writing. You are the beings I love the most on Earth !

\end{acknowledgements}

% Setting back the good length
\setlength{\textheight}{25.5cm} % 22.3cm

\newpage
~
\newpage

% -------------------------- %

\newpage

\pagenumbering{arabic}
\setcounter{page}{1}
%\setcounter{numberofpages}{1}

% Title spacing instruction :
%\titlespacing{command}{left spacing}{before spacing}{after spacing}[right]
%\titlespacing*{\chapter}{0pt}{1\baselineskip}{1.2\baselineskip}

\titlespacing{\chapter}{0pt}{-1.5cm}{0pt}

%  --------------------- Chapter 0 ----------------------   %
% --------------------------------------------------------- %
\vspace{-5cm}
\chapter*{Notations}
\addcontentsline{toc}{chapter}{Notations}
\label{Chap0}
\pagestyle{plain}

For the comfort of the reader, we summarize here the different conventions used in this thesis.

\bigskip

We will adopt the convention of Weinberg \cite{weinberg2008cosmology, Weinberg} and Misner, Thorne \& Wheeler \cite{misner1973gravitation} for the signature of the 4-dimensional metric, namely we will take it to be $(-,+,+,+)$. As a consequence, the line element defined in the 4-dimensional Minkowski spacetime (sometimes denoted by $\Mcal_4$), will be $ds^2 = - (c dt)^2 + \delta_{ij} dx^i dx^j$ and we will take the usual convention $c = 1$. Latin indices will denote the spatial values $1,2,3$ and greek indices will correspond to spacetime indices $0,1,2,3$ (where $0$ will denote the time component). The flat metric will be written as $\eta_{\mu\nu} = \eta^{\mu\nu} =  {\rm diag}(-1,1,1,1)$ and we will adopt the usual convention that repeated indices are summed over. For example, $x_\mu x^\mu$ will denote $\eta_{\mu\nu} x^\mu x^\nu = - (x^0)^2 + (x^1)^2 + (x^2)^2 + (x^3)^2$ in $\Mcal_4$ and $g_{\mu \nu} x^\mu x^\nu$ in a more general spacetime. The angles over an homogeneous sphere will be denoted by $\theta^a$ (with $a=1,2$) or $(\theta, \phi)$.

For 3-dimensional spatial vectors, we will use bold letters instead of putting arrows over them. As an example, we will find $\xbf$ instead of $\vec{x}$ (except for the null vector that we will denote by $\vec{0}$ and the usual nabla operator $\vec{\nabla} \equiv \partial_i$). The scalar products of two such vectors will be denoted by a dot\,: $\xbf \cdot \kbf = x^1 k^1 + x^2 k^2 + x^3 k^3$. Bold letters will also be used to denote tensorial quantities in their general form. The variable dependence $(x)$ will designate $(t, x^i) \equiv (t, \xbf)$.

The covariant derivative with respect to a coordinate $x^\mu$ in a 4-dimensional spacetime will be denoted by $\nabla_\mu$ and sometimes denoted by a semi-colon, e.g. $\nabla_\mu Q \equiv Q_{;\mu}$. We will denote by $\partial_\mu Q \equiv \partial Q / \partial x^\mu \equiv Q_{,\mu}$ the partial $x^\mu$-derivative of a quantity $Q$. We will also encounter the covariant derivation of a quantity with respect to the 3-dimensional spatial metric in a 4-dimensional spacetime. We will use a vertical stroke for a simple space-derivative\,: $\partial_i Q \equiv Q_{|i}$ ; and a double stroke for the covariant derivative\,: ${}^{(3)}\nabla_i Q \equiv Q_{||i}$~. The simple time derivation with respect to $t$, the cosmic time or the synchronous gauge time, will often be denoted by a dot : $\partial_t (\dots) \equiv (\ldots)^{\tdev}$. The partial derivative with respect to the conformal time $\eta$ will be denoted sometimes by a prime : $\partial_\eta (\ldots) \equiv (\ldots)'$.

As it will be defined later, we will use an overline to denote the stochastic average over inhomogeneities : $\overline{(\ldots)}$. The spacetime averages, in their different forms, will be denoted by angle brackets : $\langle \ldots \rangle$ (where the literature, e.g. \cite{peacock1999cosmological}, usually employs that symbol for the stochastic average).

Let us remark that several quantities are described by the same symbol. This is the case for $\phi$ which describes the Bardeen potential ($\Phi$, $\phi^{(2)}$), the luminosity flux ($\Phi$ and $\Phi^{(0),(1),(2)}$), the second angle on the sphere ($\phi$, $\ti \phi$), the inflaton field ($\phi$) and the morphon field ($\Phi_\Dcal$). Note also that $\Theta$ is used for the expansion scalar and for the Heaviside step function. Symbols ($\theta$,$\ti \theta$) are the first angle on the sphere. $\delta$ is used for the matter density contrast and $\delta_{\rm D}$ for the Dirac delta function. No confusion can be made between these different notations and it will be precised only when ambiguous.

Last, we will use sometimes shortcuts of the English language, using ``LHS'' and ``RHS'' for \emph{left} and \emph{right hand side} or ``w.r.t.'' for \emph{with respect to}, or from Latin (like i.e., e.g.). Sometimes also acronyms will be used to lighten the discussions, most of them are summarized in the glossary of this thesis.

% --------------------------------------------------------- %

\setcounter{chapter}{0}

%  --------------------- Chapter 1 ----------------------   %
% --------------------------------------------------------- %
\chapter{Introduction, motivations and outline of this thesis}
\label{Chap1}
\pagestyle{plain}

\section{Cosmology today}
\label{Ch1Sec1}

Cosmology is by far one of the oldest fields of research and at the same time, maybe one of the youngest. Indeed, human beings have always observed the sky to measure time and seasons and wondered about its composition. Nevertheless, most of the breakthroughs in the field have been done thanks to new ways to observe the sky, and only recently we have been able to gather enough knowledge to build up what we now call the \gls{SMC}.

We know from the cosmic microwave background (CMB) that the Universe is almost flat today and that it was created about $13.7 ~ Gyrs$ ago. The precise mechanism behind its creation is not known, but we do know that its temperature was very high and its size very small at this time. We also know that a process of rapid expansion, the so-called inflation mechanism, must have occured to make our Universe so flat and so close to homogeneity on large angular scales. This period was then followed by a phase where photons were tightly coupled to baryonic matter (the \emph{radiation dominated era}) and no matter structure formed. But after some time, expansion cooling down the Universe, neutrons and protons decoupled from light and started to form atoms, and progressively gave birth to stars, heavier elements, galaxies and guys who live in. This is called the \emph{matter dominated era}. We know indeed that matter is nowadays organized in galaxies, regrouped in clusters and superclusters (which size\footnote{We recall that $1 \pc \sim 3.26 ~ly$ (light years) with $1 ~ly = 0.9461~10^{16} {\rm m}$.} is $2-10 ~ \Mpc$) and these last entities are linked by filaments made of galaxies. All these structures are separated by voids which have a size of about $10-80 ~ \Mpc$ and we observe a statistical homogeneity above a scale of about $100-150 ~ \Mpc$ (the value of this scale being still debated). One amazing fact about this structure that we observe around us, the \emph{\gls{large scale structure}}, is that it has evolved from tiny quantum fluctuations in the early stages of the Universe (when it has a size of the order of the Planck length $\sim 10^{-33} {\rm cm}$), which have been stretched out by inflation.

Unfortunately, or fortunately for scientists, this standard model of cosmology still has some dark clouds making us unsatisfied. The biggest clouds in this picture are the true nature of dark matter and dark energy. These two parts of the energy content of the Universe are dark as we cannot observe them directly but only through indirect ways. The first one, \gls{DM}, was discovered by measurements of the velocity of stars in rotating galaxies. Scientists discovered indeed that stars had a much higher velocity than they should have from the measurement of the mass of their galaxy by light emission. It was thus necessary to add about five times more matter than we could infer from light observations to explain these profiles. Dark matter was also needed actually to explain the rapid formation of structures in simulations compared to the structures we observed, and lately became also needed in the measurements of dark energy. Though we still do not know the nature of dark matter, a lot of scientists believe that we may soon enough discover one or more particles (the so-called weakly interacting massive particles, or WIMPS) to explain it.

On the other hand, \gls{DE} is a complete mistery. It was first discovered experimentaly in 1998 by the observation of Type Ia supernovae (SNe Ia) as the fact that the expansion of the Universe seemed to be accelerating. In its simplest version, this exotic type of energy which has a tendency of repulsion under gravity could be simply a term proportional to the metric authorized by Einstein's equations, but we don't understand why its value is so small and comparable to the energy amount of dark+baryonic matter (this is the so-called \gls{coincidence problem}). This type of energy could also be a new particle following a certain potential, a bit like the particle believed to create inflation (the \emph{inflaton}). This scalar field is certainly possible and there has been a very great number of such solutions proposed in the literature (see e.g. the review of \cite{2006IJMPD..15.1753C}). We never directly observed a scalar particle yet, but the discovery at the Large Hadron Collider of a good candidate for the Higgs particle is a strong indication that they may indeed be present in nature. The new game to explain dark energy, when we assume the existence of such a particle, is to lead the expansion of our Universe as it is observed today. We can search for a good potential to explain the evolution of the field (such as in quintessence) or try to couple the field with others, such as matter itself (like with the chameleon field).
Direct modifications of the theory of general relativity are another alternative which, in some cases, can be related to theses particle models.
A lot of involved computations has been done in this direction and especially to learn how we can accommodate these extensions with data.

Theoretical physicists have thus developed a lot of different proposals to explain this wonderful 15-years old discovery of dark energy, but observations have not yet reached enough precision to test these different models and distinguish among them. In science, having a good model to describe a phenomenon is always a good thing. Having two good models is always bad, because it means that we still do not understand something. And when we have tenths of relatively close models to explain this phenomenon, it is natural to start looking for completely new ideas. It is also what happened to dark energy for several reasons. One thing that we should know about dark energy is that it made the expansion of the Universe to accelerate only a few billion years ago ($\sim 5 ~ Gyrs$), and this is close to the time when the large scale structure has formed. Some researchers thus came to the idea that the formation of structure, in other words the content of the Universe, could be responsible for this late time evolution, that is to say with this particular dynamics. In fact this isn't a stupid guess at all, Einstein himself told us that the geometry was no different from its content, that energy influence spacetime. So why matter could not ``mimic'' the role of a dark energy ? This idea was even made more exciting from the emphasize by George Ellis, in the 80's, that looking at an inhomogeneous spacetime and its smoothed out equivalent doesn't lead to the same evolution prediction. Another branch of explanations of dark energy has thus been developed, and it has been regrouped in a maybe not so well adapted term of ``backreaction'', with the idea behind that word being that small inhomogeneities may backreact on spacetime and create a real or an appearant acceleration of the expansion.

%----------------------------------------------------------------------------------------------------

\section{Homogeneous cosmology}
\label{Ch1Sec2}

\subsection{Cosmological principles}
\label{Ch1Sec21}

Cosmology is a particular branch of physics where the experiment -- our whole observable Universe itself -- is not reproducible. It thus imposes to make reasonable but philosophical assumptions to interpret our observations and get reliable predictions. Let us present some of these assumptions.

The usual assumption made in cosmology, and more generally in physics, is the \gls{Copernican principle}. This principle states that the place of experiment, e.g. the solar system, is in no privileged position in the Universe. In the elaboration of cosmological models, one can derive another principle from the Copernican one, which is the \gls{cosmological principle}. In \cite{Buchert:2011yu}, this hypothesis on our position in the Universe, and consequently on the data that we gather, is declined in two versions. The \emph{strong cosmological principle} assumes that the Universe is homogeneous and isotropic at all scales. It is the combination between the Copernican principle and the isotropy assumption which leads to the global homogeneity. This principle is stricto sensu not satisfied in Nature and we have to assume the more realistic version of it, the \emph{weak cosmological principle}, which says that we live in a Universe having an homogeneity scale and which is almost isotropic on distances above that scale. In any case, we should remark that the cosmological principle is more strict than the Copernican one as it implies also assumptions on the non-observable parts of our Universe.

We should notice that, though the Copernican principle has proved to be very useful for all physics experiments so far, it may not be true on cosmological scales. Indeed, one cannot contradict the fact that we are, up to a certain degree, \emph{privileged observers}. What we mean by this degree is that human beings couldn't have been born outside a galaxy, and so in parts of the Universe where the density constrast cannot be random. Moreover, the observations that we have at our disposal to understand our past are all so far light signals on our past lightcone (a 2+1 spacetime hypersurface) and few geological observations giving information on our past timelike world line. These facts must be kept in mind when building up realistic cosmological models and they are important for the discussions of inhomogeneous cosmologies.

\subsection{Evolution equations}
\label{Ch1Sec22}

Let us present the general approach of the standard model of cosmology, based on the cosmological principle.
The most general form of the line element of a free falling observer that observes an isotropic and homogeneous universe around him is given by the \gls{FLRW} line element, which in cosmic time $t$ and spherical coordinates $(\xbf) = (\chi, \theta, \phi)$ reads\,:
\bea
\label{ds2FLRW}
ds^2_{\rm FLRW} &=& - dt^2 + a^2(t)\gamma_{ij}(|\xbf|) dx^i dx^j \nonumber \\
&=& -dt^2 + a^2(t) \left[d\chi^2 + \left( \frac{1}{\sqrt{K}} \sin\left(\sqrt{K} \chi \right) \right)^2 d \Omega^2 \right] ~~, \\
\mbox{ with } ~~~~~ d \Omega^2 &=& d\theta^2 + (\sin\theta)^2 d\phi^2 ~~, \nonumber
\eea
and where $K$ is a constant setting the geometry of spacelike hypersurfaces\,: $K \in \{-1,0,1\}$ in the cases of a \{ open, flat, closed \} Universe\footnote{Indeed, the notation conventionally used in the literature is
\beq
\gamma_{ij}(|\xbf|) dx^i dx^j = d\chi^2 + f_K^2(\chi) d \Omega^2 ~~~\mbox{ with }~~ f_K(\chi) \equiv \left\{ \begin{array}{ll} \sin\left(\sqrt{K}\chi\right) / \sqrt{K} & \mbox{for $K=1$} \\ \chi & \mbox{for $K=0$} \\ \sinh\left(\sqrt{-K}\chi\right) / \sqrt{-K} & \mbox{for $K=-1$} \end{array} \right. ~~, \nonumber
\eeq
but $f_K(\chi)$ is correctly captured by the unique equality $f_K(\chi) = \sin\left(\sqrt{K}\chi\right) / \sqrt{K}$ if we authorize the square root to receive negative arguments, given that $\sqrt{-K} = i \sqrt{K}$, and by considering the complex expression of the sine function.}. We call $a(t)$ the scale factor in cosmic time and $\gamma_{ij}$ is the 3-dimensional metric describing the Universe on a given spatial hypersurface. In this metric, $a(t)$ has the dimension of a distance. We can also use the so-called conformal time $\eta$ such that $d\eta = dt / a(t)$. Defining now a new radial coordinate $r$ such that
\beq
\label{randchi}
r = \sin\left(\sqrt{K}\chi\right) / \sqrt{K} ~~,
\eeq
one can show that the FLRW line element is simply written as
\beq
\label{FLRWlineElement}
d s_{\rm FLRW}^2 = - dt^2 + a(t)^2 \left( \frac{1}{1 - K r^2} dr^2 + r^2 d \Omega^2 \right) ~~,
\eeq
where $a(t)$ can be considered as a dimensionless quantity if $r$ is now expressed in units of a distance (in which case $K$ has the dimension of $r^{-2}$).

Considering the form of the metric presented here, we can compute the \gls{Christoffel symbols} which are at the basis of the connection in general relativity and that have for general definition\,:
\beq
\Gamma_{\sigma\mu\nu} = \frac{1}{2} ( \partial_{\mu} g_{\sigma\nu} + \partial_{\nu} g_{\sigma\mu} - \partial_{\sigma} g_{\mu\nu}) ~~~~~\mbox{and}~~~~~
\Gamma^{\alpha}_{~\mu\nu} = g^{\alpha\sigma} \Gamma_{\sigma\mu\nu} ~~.
\eeq
These symbols are easily computed for this FLRW metric in spherical form and read
\beq
\Gamma^0_{~ij} = H a^2 \gamma_{ij} ~~,~~ \Gamma^i_{~j0} = H \delta^i_j ~~,~~ \Gamma^i_{~jk} = ~ {}^{(3)}\Gamma^i_{~jk} ~~,
\eeq
where we have defined the Hubble parameter $H \equiv \dot{a}/a$ (with the dot denoting the derivative w.r.t. the cosmic time $t$). The $^{(3)}\Gamma^i_{~jk}$ are the Christoffel symbols associated to the spatial metric $\gamma_{ij}$. They define the connection inside the spatial hypersurfaces parametrized by the spherical coordinates $\{ x^i \}$.

The \gls{Riemann tensor} is generally defined as\,:
\beq
R^{\delta}_{~\alpha\beta\gamma} = \partial_{\alpha} \Gamma^{\delta}_{\beta\gamma} - \partial_{\beta} \Gamma^{\delta}_{\alpha\gamma} + \Gamma^{\delta}_{\alpha\epsilon} \Gamma^{\epsilon}_{\beta\gamma} - \Gamma^{\delta}_{\beta\epsilon} \Gamma^{\epsilon}_{\alpha\gamma}
\eeq
and has the following properties\footnote{Other properties of the Riemann tensor are given by the symmetries of its indices and the Bianchi identities\,:
\bea
R_{\alpha\beta\gamma\delta} &=& R_{[\alpha\beta]\gamma\delta} ~\mbox{ and }~ R_{\alpha\beta\gamma\delta} = R_{\alpha\beta[\gamma\delta]} ~~, \nonumber \\
3 R_{\alpha[\beta\gamma\delta]} &=& R_{\alpha\beta\gamma\delta} + R_{\alpha\gamma\delta\beta} + R_{\alpha\delta\beta\gamma} = 0 ~~\mbox{ ($1^{st}$ Bianchi Id.)} ~~, \nonumber \\
3 R_{\alpha\beta[\mu\nu;\tau]} &=& R_{\alpha\beta\mu\nu;\tau} + R_{\alpha\beta\nu\tau;\mu} + R_{\alpha\beta\tau\mu;\nu} = 0 ~~\mbox{ ($2^{nd}$ Bianchi Id.)} ~~, \nonumber
\eea
with the definition of symmetrized and anti-symmetrized tensors\,:
\beq
T_{[\mu_1\mu_2 \ldots \mu_n]} = \frac{1}{n!} \epsilon_{\mu_1\mu_2 ... \mu_n} \sum_{permutations} T_{\mu_1\mu_2 \ldots \mu_n} ~~~~~,~~~~~ T_{(\mu_1\mu_2 \ldots \mu_n)} = \frac{1}{n!} \sum_{permutations} T_{\mu_1\mu_2 \ldots \mu_n} ~~. \nonumber
\eeq
and where $\epsilon_{\mu_1\mu_2 \ldots \mu_n} = \pm 1$ depending on the parity (non-parity) of permutations from $(1,2,\ldots,n)$ to $(\mu_1,\mu_2,\ldots,\mu_n)$.
}\,:
\beq
R_{\alpha\beta\gamma\delta} = -R_{\beta\alpha\gamma\delta} = -R_{\alpha\beta\delta\gamma} = R_{\gamma\delta\alpha\beta} ~~.
\eeq
We then obtain the \gls{Ricci tensor} from the contraction of two indices in the Riemann tensor\,:
\beq
R_{\mu\nu} \equiv R^{\lambda}_{~\mu\lambda\nu} = \partial_{\mu} \Gamma^{\lambda}_{\nu\lambda} - \partial_{\lambda} \Gamma^{\lambda}_{\mu\nu} + \Gamma^{\eta}_{\mu\lambda} \Gamma^{\lambda}_{\nu\eta} - \Gamma^{\lambda}_{\lambda\eta} \Gamma^{\eta}_{\mu\nu} ~~.
\eeq
In the FLRW metric, this leads to the non-zero components
\beq
R_{00} = - 3 \frac{\ddot{a}}{a} ~~,~~ R_{ij} = \left( 2 H^2 + \frac{\ddot{a}}{a} + 2 \frac{K}{a^2} \right) a^2 \gamma_{ij} ~~.
\eeq
The final quantity that we can derive is the 4-dimensional scalar curvature, which is
\beq
R \equiv R^{\mu}_{~\mu} = 6 \left( H^2 + \frac{\ddot{a}}{a} + \frac{K}{a^2} \right) ~~.
\eeq

In this geometry, we then get the \gls{Einstein tensor}\,:
\beq
G_{\mu\nu} \equiv R_{\mu\nu} - \frac{R}{2} g_{\mu\nu} ~~,
\eeq
with the following non-zero components
\beq
\label{NonZeroG}
G_{00} = 3 \left(H^2 + \frac{K}{a^2}\right) ~~,~~ G_{ij} = - \left( H^2 + 2 \frac{\ddot{a}}{a} + \frac{K}{a^2} \right) a^2 \gamma_{ij} ~~.
\eeq
The assumption of isotropy and homogeneity of the Universe, as seen from an observer in geodesic motion, imposes the \gls{stress energy tensor} to have the following components\,:
\beq
T^{00} = \rho(t) ~~,~~ T^{0i} = T_{0i} = 0 ~~,~~ T^{ij} = p(t) a^{-2}(t) \delta_{ij} ~~,
\eeq
where $\rho(t)$ and $p(t)$ are respectively the energy density and pressure of the background fluid of matter.
Expressing this tensor in terms of the velocity of an observer $u^\mu$ which can be taken as $u^\mu = (1,\vec{0})$ in a local coming frame, we obtain\,:
\beq
T^{\mu\nu} = (\rho + p) u^{\mu} u^{\nu} + p \, g^{\mu\nu} ~~,
\eeq
and this is the general expression of the stress energy tensor of a perfect fluid.
Finally, the Einstein equations in the presence of a cosmological constant (see e.g. \cite{MG,PeterUzan}), given by\,:
\beq
\label{EinsteinEqWithLambda}
G_{\mu\nu} + \Lambda g_{\mu\nu} \equiv R_{\mu\nu} - \frac{R}{2} g_{\mu\nu} + \Lambda g_{\mu\nu} = 8 \pi G \, T_{\mu\nu} ~~,
\eeq
take a very simple form given by the $(0,0)$ and $(i,j)$ components, known as the Friedmann equations\,:
\bea
\label{Friedmann1}
H^2(t) &=& \frac{8 \pi G}{3} \rho - \frac{K}{a^2} + \frac{\Lambda}{3} ~~, \\
\label{Friedmann2}
\frac{\ddot{a}}{a}(t) &=& - \frac{4 \pi G}{3} (\rho + 3 p) + \frac{\Lambda}{3} ~~.
\eea
We should add, to these two equations above, the equation of conservation of the fluid given by $\nabla_\mu T^{\mu\nu} = 0$, which reads\,:
\beq
\label{ContinuityEq}
\dot{\rho} + 3 H (\rho + p ) = 0 ~~.
\eeq
We remark that only two out of these three equations are independent.

We can also define a density for the cosmological constant and a pressure in order to have the same dependence as matter on the RHS of these equations. We have
\beq
\rho_{\Lambda} = \frac{\Lambda}{8 \pi G} ~~,~~ p_{\Lambda} = - \frac{\Lambda}{8 \pi G} ~~~\mbox{ imposing that }~~ w_\Lambda \equiv \frac{p_\Lambda}{\rho_\Lambda} = -1 ~~.
\eeq
It is a well-known result that the equation of state $w$ for a cosmological constant is $-1$ and that makes the cosmological constant the most simple explanation for the acceleration of the Universe expansion. Nevertheless, we don't have an explanation for the smallness of this constant, a problem referred to as the \emph{\gls{cosmological constant problem}} in cosmology\footnote{Indeed, the consideration of the energy of the vacuum in quantum field theory predicts a density of energy wich is about 120 orders of magnitude greater that the observed density of dark energy (see \cite{2006IJMPD..15.1753C}).}. It could be that the cosmological constant is smaller than expected, or even (almost) exactly zero, and that the true explanation for the acceleration of the Universe comes from a particle (in most of the models a scalar field). In that case the equation of state $w_{DE}$ of dark energy can differ from $-1$ and even evolve with time.

We can also recast the first Friedmann equation \rref{Friedmann1} by the introduction of the energy density parameters\,:
\beq
\Omega_m = \frac{8 \pi G \, \rho}{3 H^2} ~~~~~,~~~~~ \Omega_\Lambda = \frac{\Lambda}{3 H^2} ~~~~~,~~~~~ \Omega_K = - \frac{K}{H^2 a^2} ~~,
\eeq
where here $\rho$ accounts for both dark and baryonic matter.
In these notations, this equation \rref{Friedmann1} is simply written as\,:
\beq
\label{SumOmega}
\Omega_m + \Omega_\Lambda + \Omega_K ~ \equiv ~ \Omega_{\rm tot} + \Omega_K ~ = ~ 1 ~~.
\eeq
To this equation we could add a radiation contribution $\Omega_r$ which has the same definition as $\Omega_m$ (with $\rho_r$ replacing $\rho$). This contribution is important during the radiation dominated stage of the Universe but is subleading after decoupling. We thus neglect this contribution which is irrelevant for our work.
Let us notice also that for a barotropic fluid with a constant \gls{equation of state} $w = p / \rho$, the Eq. \rref{ContinuityEq} is easily solved and gives the following dependence of the density parameters\,:
\beq
\label{SolutionOmega}
\Omega_m(z) = \Omega_{m0} \left( \frac{a}{a_0} \right)^{-3(1 + w)} \left( \frac{H_0}{H(z)} \right)^2 ~~.
\eeq
For a dust (i.e pressureless) fluid for which $w = 0$, one sees that the density decays as $\sim a^{-3}$. For a cosmological constant, one has a similar relation giving $\Omega_{\Lambda} \sim a^0$.

In conformal time $\eta$, one can similarly show that the Friedmann equations are written as\,:
\bea
\mathcal{H}^2 &=& \frac{8 \pi G}{3} \rho a^2 - K + \frac{\Lambda}{3} a^2 ~~, \\
\mathcal{H}' &=& - \frac{4 \pi G}{3} (\rho + 3p) a^2 + \frac{\Lambda}{3} a^2 ~~, \\
0 &=& \rho' + 3 \mathcal{H} (\rho + p) ~~,
\eea
where we have used the conformal Hubble parameter $\mathcal{H} \equiv a(\eta)' / a(\eta) = a H$. These equations, which we emphasize have only two independent equations, describe the evolution of the density $\rho$ and pressure $p$ of a fluid. In a $\Lambda$CDM model, one then has a specific expression for the Hubble parameter $H$ (or $\Hcal$) and an explicit form for $\rho$ and $p$ in terms of the parameters of the model $\Omega_{m0}$ and $\Omega_{\Lambda 0}$.

%----------------------------------------------------------------------------------------------------

\section{Type Ia supernovae}
\label{Ch1Sec3}

The idea of dark energy, in its minimal version of a cosmological constant, had been proposed by Albert Einstein in order to make the Universe static and was forgotten after the discovery of Hubble that the Universe was expanding \cite{1929PNAS...15..168H}. The acceleration of the expansion of the Universe, or broadly speaking the existence of dark energy (DE), was not considered until 1998 when two different groups of experimentalists studying supernovae (SNe) arrived to the same conclusion that most of the Universe content was of an exotic nature. These groups were the High-z Supernova Search Team \cite{Riess:1998cb} and the Supernova Cosmology Project \cite{Perlmutter:1998np} and they recently received the Nobel Prize in physics (in 2011).

\subsection{Main features}
\label{Ch1Sec31}

% -- Mass of the star & explosion -- *
First of all, one should think about supernovae as really rare events happening only a few times every thousand years per galaxy. They are phenomena emerging after the explosion of a star which is not completely understood. The best explanation we have is that they consist in the result of matter accretion by a white dwarf (a star made of a plasma of unbound nuclei and electrons) in a binary system (i.e. where a star is feeding up the white dwarf) and up to the point where it reaches a mass of $1.4-1.5 ~{\rm M}_{\odot}$, the \gls{Chandrasekhar mass}\footnote{We define the solar mass ${\rm M}_\odot$ by
$ 1 ~ {\rm M}_\odot \equiv \frac{4 \pi^2 \times (1 ~ a.u.)^3}{G \times (1 ~ {\rm year})^2} \simeq 1.98 ~ 10^{30} ~ kg$ where $a.u.$ denotes the astronomical unit (equal to the Sun-Earth distance). The solar luminosity is about $L_\odot \simeq 3.84 ~ 10^{26} ~ W = 3.84 ~ 10^{33} ~ erg ~ s^{-1}$.
}. This second star can also be a white dwarf. A consequence of this nature of supernovae is seen in their chemical composition, as we will see in Fig. \ref{FigSNeSpectra}. What is important in this process is that supernovae are believed to have all roughly the same mass, and that makes them be \emph{\gls{standard candels}}. Indeed, above the Chandrasekhar mass, the pressure from gravitation is stronger than the electron degeneracy pressure inside the star and that leads to a collapse up to a certain radius fixed by the nuclear degeneracy pressure of the star. We then assist to a violent increase in temperature and the conversion of $^{12} C$ and $^{16} O$ into $^{56}Ni$, starting a thermonuclear explosion and the ejection of external cores of the star. The amount of energy emitted by this explosion (and though 99\% of it is formed of neutrinos and 1\% only by light) is so large that a supernova can be more brilliant than its host galaxy, reaching $\sim 10^{10} L_\odot$. As a consequence, supernovea can be seen from really far distances, making them useful for cosmology.

\begin{figure}[ht!]
\centering
\begin{tikzpicture}
  \node [xshift=-7cm,yshift=3cm] at (current page.south east)
    {\includegraphics[width=12cm]{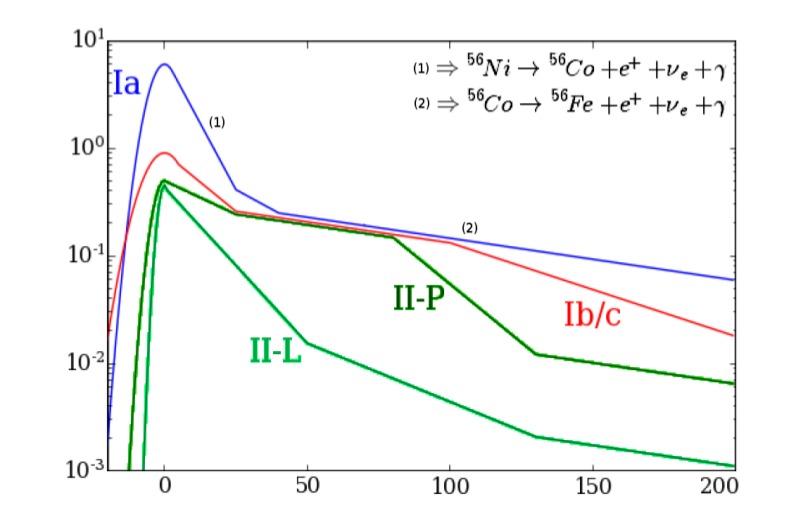}};
\node[anchor=west,rotate=90] at (8.5,1.5) (description) {Luminosity $[10^9 L_{\odot}]$};
\node[anchor=west] at (13.8,-1) (description) {days};
\end{tikzpicture}
\centering
\caption[Idealized light curves for different types of supernov\ae]{Idealized light curves for different types of supernov\ae. We notice that SNe Ia are the brightest and the longest lasting case. Time is given in the rest frame of the supernovae (which for a redshift $z \sim 1$ corresponds to the one half of the observed time on Earth). Figure taken from \cite{fourmanoit:tel-00587450}.}
\label{FigLightCurve}
\end{figure}

\begin{figure}[ht!]
\centering
\includegraphics[width=10.5cm]{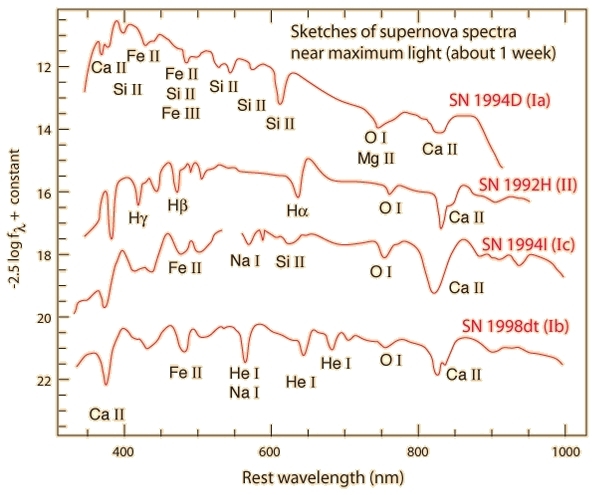}
\centering
\caption[Spectra of different types of supernov\ae]{Spectra of different types of supernov\ae. We can remark the absence of hydrogen and the presence of {\rm Si-II} for the type Ia. Spectra are shown in the rest frame of the supernovae, i.e. corrected from redshift. It is actually such kind of {\rm Si-II} absorption bands which are used \cite{Conley:2011ku} to estimate the redshift of the star. Spectra are usually taken near the light curve maximum for a better accuracy. This figure is taken from \cite{carroll2006introduction}.}
\label{FigSNeSpectra}
\end{figure}

% -- Light curves -- %
Another important point is that SNe must be observed at their peak in luminosity to be standardized and thus to be used. We hence need to detect them at an early stage of their explosive life. Indeed, we observe supernovae in different wavelength bands called U, V, B, R and I (other photometric systems can be used) and we follow their intensity, or their magnitude $m_B$, in the B-band as they are brighter in this band. We see that their intensity follows a growing time period during 15-20 days and is followed by a more or less slow decrease of $\sim 2$ months (corresponding to the radioactive decay of ${}^{56}Ni$ into ${}^{56}Co$, decaying itself into ${}^{56}Fe$). This is what we call the \emph{\gls{light curve}} of a supernova and it has been noticed that the brighter the supernova, the longer the decreasing time. As a consequence, we have classified SNe in different types from their light curves, as can be seen in Fig. \ref{FigLightCurve}. The main two types are I and II, and we split them in subtypes. These different types are actually confirmed by spectroscopy, showing that supernovae also differ significantly in their composition and that is originating from the variety of physical phenomena giving rise to the explosion. The main feature of SNe Ia is that, like white dwarfs, they do not present any trace of hydrogen in their composition. Their second important feature is that they contain silicon (see Fig. \ref{FigSNeSpectra}).

\subsection{Observations and standardization}
\label{Ch1Sec32}

% -- Importance of spectra -- %
Supernovae have been studied during many years now and their frequency of detection has significantly grown during the last decade with the advance of fast computing and automatic detection surveys (cf. the \emph{rolling search} method of SNLS \cite{Conley:2011ku}). The automatic coverage of the sky (or part of it) has greatly enhanced the number of detected SNe. Spectra are also fundamental in the current use of SNe Ia and the reason for that is because they furnish the composition of the star, necessary to confirm its type. They also determine the redshift of the star (by a direct measurement of the wavelengths of absorption bands). Note however that spectroscopy measurements are very costly in terms of observational time, limiting seriously the number of SNe Ia detections. To cure this problem and prepare the next generation of surveys (such as the Large Synoptic Survey, the Dark Energy Survey, or Pan-STAARS), the SNLS and SDSS experiments have tried to include the estimation of the redshift in the fitting process from photometry only \cite{2010arXiv1005.1890P}. Indeed, a supernova which is highly redshifted is more luminous in low frenquency spectral bands. We can thus estimate -- though less accurately but very fast -- their redshift by the comparison of their magnitude in these different bands, i.e. from the color itself instead of spectroscopy.

\begin{figure}[ht!]
\centering
\includegraphics[width=8.4cm]{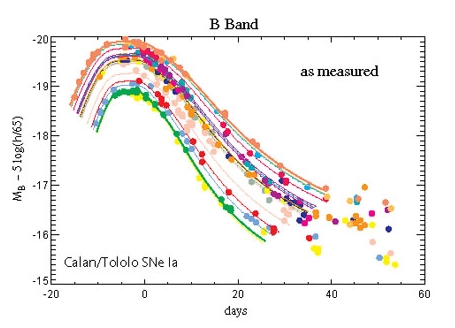}
\includegraphics[width=8.4cm]{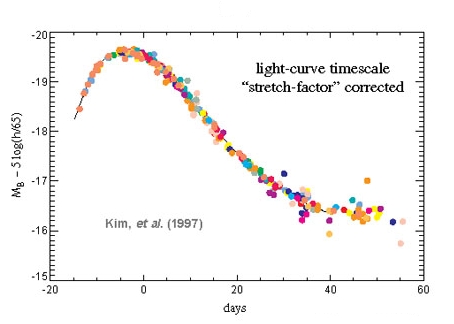}
\centering
\caption[SNe Ia light curves of the Cal\'{a}n-Tololo survey and the stretch correction]{SNe Ia light curves of the Cal\'{a}n-Tololo survey. Left\,: as directly measured in the B-band, replacing them in their supernova rest-frame. Right\,: same light curves after applying the stretch correction. We note the significant reduction in dispersion induced by this correction. Figures borrowed from \cite{Frieman:2008sn}.}
\label{FigStretch}
\end{figure}

The light curve of a SN is a very important source of information and it is used to calibrate the SNe and standardize them. Several methods have been used for their standardization, one can cite particularly the \emph{Multicolor Light Curve Shape} (MLCS) and \emph{Spectral Adaptative Lightcurve Template} (SALT, see e.g. \cite{fourmanoit:tel-00587450,Guy:2007dv}).
From these analyses of the light curves, we can extract two parameters of major importance\,: the stretch factor $s$ and the color parameter $c$. The first one is linked to the lapse in time of the SNe light curve and the second one is a parameter in the SALT fit of the light curves. We can see in Fig. \ref{FigStretch} that the different light curves of SNe Ia can be reshaped by applying to them the simple \gls{stretch factor} to significantly reduce their dispersion. The \gls{color parameter} is more precisely the excess of color between the averaged color of the supernova and its maximal color\,:
\beq
c \equiv (B - V)_{\mbox{max}} - \mbox{Moy}[B - V] ~~.
\eeq
We also observe the magnitude of the supernova in different frequency bands and the total luminosity is thus obtained by reconstruction. For this, the magnitude in the B band is used to correct the measurements from the stretch and color parameters. A relation of the following type is used\,:
\beq
m_B^{\small \mbox{corr} \normalsize} = m_B - \alpha \, (s-1) + \beta \, c ~~,
\eeq
where $\alpha$ and $\beta$ are two parameters evaluated to reduce the dispersion between the light curves of different SNe Ia.
When one considers these corrections, we get a reduction from $\sim 40\%$ to $\sim 15 \%$ on the dispersion of the distance modulus, i.e. directly on the Hubble diagram.

\subsection{Magnitudes}
\label{Ch1Sec33}

Magnitudes are widely used in astronomy. Let us first consider a non-expanding Universe to explain this notion. One defines the \emph{\gls{apparent magnitude}} $m$ of a star at the actual distance $d_L({\rm a.u.})$ (expressed in arbitrary units a.u.) in terms of its \emph{\gls{luminosity flux}} $\Phi$ (using the same distance units) by
\beq
m = -2.5 \log_{10}(\Phi) + {\rm cst} \equiv -2.5 \log_{10} \left(\frac{\Phi({\rm a.u.})}{\Phi_{\rm ref}({\rm a.u.})}\right) ~~.
\eeq
The constant ${\rm cst} \equiv 2.5 \log_{10}\left(\Phi_{\rm ref}({\rm a.u.})\right)$ appearing in this expression makes dimensionless the apparent magnitude. That also means that measuring a luminosity flux is useless if one cannot compare it to a reference source (being another star or a calibration of the apparatus used for the measurement). We also define the \emph{\gls{absolute magnitude}} which is the apparent magnitude for the observed star if we were at a distance of $10 \, \pc$, in that case we denote it by $M$, and we have
\beq
M = -2.5 \log_{10}\left(\frac{\Phi(10\,\pc)}{\Phi_{\rm ref}( \pc)}\right) ~~.
\eeq
Using now the relationship between flux and distance, at a distance expressed in $\rm{a.u.}$, in a homogeneous Universe\,:
\beq
\Phi({\rm a.u.}) = \frac{L}{4 \pi d_L({\rm a.u.})^2} ~~,
\eeq
where $L$ is the intrinsic luminosity of the source (assumed to be constant), we obtain that
\beq
\label{RelmMinusM}
m - M = 5 \log_{10} \left(\frac{d_L({\rm a.u.})}{10\,\pc}\right) ~~.
\eeq
This combination that we call the \emph{\gls{distance modulus}} $\mu \equiv m - M$, is thus explicitly dependent on the unit used for the distance $d_L$. It can be written as\,:
\beq
\mu(\pc) = -5 + 5 \log_{10}(d_L(\pc)) ~~~~~\mbox{or}~~~~~ \mu(\Mpc) = 25 + 5 \log_{10}(d_L(\Mpc)) ~~.
\eeq

Let us now see how this relation is changed in an expanding Universe. In the case of an expanding homogeneous and flat Universe, the luminosity distance can be written in terms of the redshift $z = a(t_0)/a(t) - 1$ and it can be written as
\beq
\label{dLandRz}
d_L(z) = (1 + z) a(t_0) r(z) ~~,
\eeq
where $r(z)$ is the conformal distance from the source to the observer. Chosing a radial lightcone geodesic reaching the observer, the trajectory of light is given by $ds^2 = - dt^2 + a(t)^2 dr^2 = 0$ and we simply have the na\"{i}ve relation\footnote{$t$ being future directed and the system of coordinates centered on the observer, that explains the minus sign here.} $dr = - a(t)^{-1} dt$ that describes the propagation of the photon. This gives
\beq
\label{rInPerturb}
r = - \int_{t_0}^t \frac{dt}{a(t)} = \int_{a(t)}^{a(t_0)} \frac{d a}{a^2 H(a)} = a_0^{-1} \int_{0}^{z} \frac{dz'}{H(z')} ~~,
\eeq
where $t_0$ is ``today'' (and $a_0 \equiv a(t_0)$) and we used $dt / a = da / (a^2 H(a))$ and $dz = - a_0 \, da / a^2 = - (a_0 H / a) ~ dt$. In a general Universe which contains different fluids, all in a pure homogenous state, the function $H(z)$ is not simple (see e.g. Eq. \rref{GeneralH}).

Staying at relatively short observable distances, more precisely in a regime where $z \ll 1$, we can expand $H(z)$ in perturbations of $z$. The Taylor expansion gives 
\beq
H(z) = H_0 + \left( \frac{d H}{dz} \right)_{t= t_0} z + \Ocal(z^2) ~~,
\eeq
where we can use
\beq
\frac{d H}{dz} = - \frac{1}{(1+z) H(z)} \frac{d H(t)}{dt} = - \frac{1}{(1+z) H(z)} \left( \frac{\ddot{a}}{a} - H^2 \right) ~~,
\eeq
which at the observer position (today) is
\beq
\left( \frac{d H}{dz} \right)_{t = t_0} = - H_0 \left( \frac{a \ddot{a}}{\dot{a}^2} - 1 \right)_{t = t_0} \equiv H_0 (1 + q_0) ~~,
\eeq
and where we have introduced the so-called \emph{\gls{deceleration parameter}} $q_0$\,:
\beq
\label{DecelParam}
q_0 \equiv - \left( \frac{a \ddot{a}}{\dot{a}^2} \right)_{t = t_0} = \left( 1 + \frac{\dot{H}}{H^2} \right)_{t = t_0} ~~.
\eeq
We thus obtain that
\beq
H(z) = H_0 \big[ 1 + (1 + q_0) z + \Ocal(z^2) \big]
\eeq
which, after being introduced into Eq. \rref{rInPerturb}, implies the following perturbative expression for the conformal distance $r$
\beq
r(z) = (a_0 H_0)^{-1} \left( z - (1 + q_0)\frac{z^2}{2} + \Ocal(z^3) \right) ~~.
\eeq
This gives at the end, from Eq. \rref{dLandRz}, the perturbative expression of the luminosity distance
\beq
d_L(z) = H_0^{-1} \left[ z + (1 - q_0)\frac{z^2}{2} + \Ocal(z^3) \right] ~~.
\eeq
We notice in passing that a stationary (i.e. constantly evolving) Milne Universe implies $r(z) = (a_0 H_0)^{-1} z$ and thus $d_L^{\, \rm Milne}(z) = z (1 + z/2) H_0^{-1}$ (see Sec. \ref{AppASec6}). We then have $d_L > d_L^{\, \rm Milne}$ if $q_0 > -1$.

We can now come back on the definition of the distance modulus and write a definition which involves the product $H_0 d_L$ in order to have a function of redshift only. This is done by writing\,:
\beq
\mu(z) = \Mcal + 5 \log_{10}(H_0 d_L(z)) ~~~\mbox{ with }~~ \Mcal(\Mpc) \equiv 25 - 5 \log_{10}(H_0) ~~,
\eeq
and where $d_L(z)$ is still in $\Mpc$ and $H_0$ is expressed in $\Mpc^{-1}$. More correctly\,:
\beq
H_0 \equiv h ~ 100 ~ km ~ s^{-1} ~ \Mpc^{-1} = 3.33~10^{-4} ~ h ~ c \, \Mpc^{-1} ~~,
\eeq
where we have used the speed of light $c = 299792,458 ~km ~s^{-1}$ which is taken to be $c=1$ when everything is expressed in $\Mpc$ (like here).
For $h = 0.7$, we get $- 5 \log_{10}(H_0) \simeq 18.16$, leading to
\beq
\mu(\Mpc) = 43.16 + 5 \log_{10}\left(H_0(\Mpc^{-1}) d_L(\Mpc)\right) ~~.
\eeq
We remark that this expression is very well adapted to SNe observations which are sensitive to the product $H_0 d_L$ rather that $d_L$ itself.

\subsection{The Hubble diagram}
\label{Ch1Sec34}

Let us first recall that the luminosity distance $d_L$ of a source at redshift $z$ is related to the angular distance $d_A$ of this source (as seen from the observer) by the \gls{Etherington law} (or reciprocity law) \cite{Et1933}:
\beq
d_L = (1+z)^2 d_A ~~.
\label{EqdL}
\eeq
This relation is valid in any geometry as it relies on the reversibility of light trajectories.
In the particular case of an unperturbed, spatially flat FLRW background, and for a source with redshift $z_s$, the \gls{angular distance} $d_A$ is simply given by
\beq
d_A^{\rm FLRW}(z_s) = a_s \, r_s = a_s(\eta_o-\eta_s) ~~,
\eeq
where $a_s \equiv a(\eta_s)$, while $ \eta_o-\eta_s \equiv \Delta \eta$ denotes the conformal time interval between the emission and observation of the light signal. 

For an unperturbed metric also we have $1+z= a_0/a(t)$, and $d\eta= dt/a=- a_0^{-1} dz/H$, where $H= d(\ln a)/dt$. Hence, using the standard (spatially flat) Friedmann equation for $H$ from Eq. \rref{Friedmann1}, assuming that the homogeneous model has perfect fluid sources with present fractions of the critical density $\Om_{n0}$ and barotropic parameters $w_n = p_n / \rho_n = \cst$ as given by Eq. \rref{SolutionOmega}, i.e. using\,:
\beq
\label{GeneralH}
H^2(z) = H_0^2 \left[ \sum_n \Om_{n0} (1+z)^{3(1+w_n)}\right] ~~,
\eeq
we get the general expression\footnote{
Actually, the most general expression that we can obtain when we consider a non-flat FLRW model with dark matter and a general form of dark energy, i.e. non-necessarily a cosmological constant, is\,:
\beq
d_L(z_s) = \frac{c \, (1 + z_s)}{H_0 \sqrt{|\Omega_{K0}|}} S_K \left( \sqrt{|\Omega_{K0}|} \int_0^{z_s} \frac{dz}{\left[ \Omega_{m0} (1 + z)^3 + \Omega_{K0} (1 + z)^2 + \Omega_{DE,0} \exp \left( 3 \int_0^z \frac{1 + w_{DE}(z')}{1 + z'} dz' \right) \right]^{1/2}}  \right) ~~, \nonumber
\eeq
with the possible addition in the RHS of $\Omega_r \equiv 8 \pi G \, \rho_r / 3 H^2$ the density parameter of radiation (for which $w_r = 1/3$), and where we used $\Omega_{DE} \equiv 8 \pi G \, \rho_{DE} / 3 H^2$. The exponential contribution is obtained by considering the conservation of the dark matter fluid\,: $\rho_{DE} + 3 H (\rho_{DE} + p_{DE}) = 0$ with a general (time dependent) equation of state such that $p_{DE} = w_{DE}(\rho_{DE}) \, \rho_{DE}$. We used $S_K(x) = \sin(x), x, \sinh(x)$ depending on the sign of the curvature $K > 0 ~,~ = 0 ~,~ < 0$~.} in terms of the redshift of a source $z_s$\,:
\bea
d_L^{\rm FLRW}(z_s) &=& (1+z_s) a_0 \int_{\eta_s}^{\eta_o} d\eta = (1+z_s) \int_0^{z_s} {dz\over H(z)} \nonumber \\
&=& {1+z_s\over H_0} \int_0^{z_s} dz \left[ \sum_n \Om_{n0} (1+z)^{3(1+w_n)}\right]^{-1/2} ~~.
\label{standard}
\eea
Expanding in the limit $z_s \ra 0$ we also obtain the expression
\bea
d_L^{\rm FLRW}(z_s) &\simeq& {1\over H_0} \left[z_s + \frac 14 \left(1 - 3 \sum_n  w_n \, \Om_{n0}\right) z_s^2 + O(z_s^3)\right] \nonumber \\
&\equiv& {1\over H_0} \left[z_s + \frac 12 (1 - q_0) z_s^2 + O(z_s^3)\right] ~~,
\label{standard1}
\eea
which shows the well-known sensitivity of the term quadratic in $z_s$ to the composition of the cosmic fluid through the deceleration parameter $q_0$ (see Eq. \rref{DecelParam}).

\begin{figure}[ht!]
\centering
\includegraphics[width=10cm]{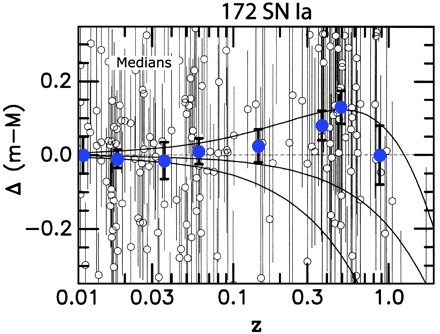}
\centering
\caption[Hubble diagram with binned data]{Hubble diagram with 172 SNe Ia with median values in 8 redshift bins. Solid lines (from bottom to top) correspond to the evaluation from Eq. \rref{dLinLCDM} with $(\Omega_{m0},\Omega_{\Lambda0}) \simeq (1,0)$, $(\Omega_{m0},\Omega_{\Lambda0}) \simeq (0.3,0)$ and $(\Omega_{m0},\Omega_{\Lambda0}) \simeq (0.3,0.7)$. The increase at large redshift in the binned data points shows the existence of a dark energy component. Figure taken from \cite{Kirshner06012004}.}
\label{FigHubbleDiagBin}
\end{figure}

If we assume a flat (i.e. $K=0$) FLRW model with only matter and a cosmological constant, we get the theoretical luminosity distance\,:
\beq
\label{dLinLCDM}
d_L^{\rm \Lambda CDM}(z_s)=  \frac{1+z_s}{H_0} \int_0^{z_s} {\frac{dz'}{\left[\Om_{\Lambda0} + \Om_{m0} (1+z')^{3}\right]^{1/2}}} ~~.
\eeq
This relation is not trivially integrable when its two parameters $\Om_{\Lambda0}$ and $\Om_{m0}$ are non-zero. One thus has to integrate it numerically until we find its best fit with the data.
Measuring magnitudes of SNe Ia with their corresponding redshifts, one can plug the result of Eq. \rref{dLinLCDM} into Eq. \rref{RelmMinusM} and plot the so-called \emph{\gls{Hubble diagram}}, as shown in Fig. \ref{FigHubbleDiagBin} (see \cite{Kirshner06012004} for a short history). From the binned data points, we get that $\Om_{\Lambda0}$ is about 73\% of the total energy content in the Universe and that $\Om_{m0}$ is about 27\% (including baryons)\footnote{See however the new reevaluation of these parameters by the Planck satellite first release \cite{Ade:2013zuv}.}. The densities of radiation and curvature are negligible today. This is such a fit which led to the discovery of dark energy. We can see in Fig. \ref{FigHubbleDiagBin} that the data dispersion is very important. That shows that SNe are very complex objects which are difficult to interpret even after being standardized by a long procedure.

\begin{figure}[ht!]
\centering
\begin{tikzpicture}
  \node [xshift=-7cm,yshift=3cm] at (current page.south east)
    {\includegraphics[width=8cm]{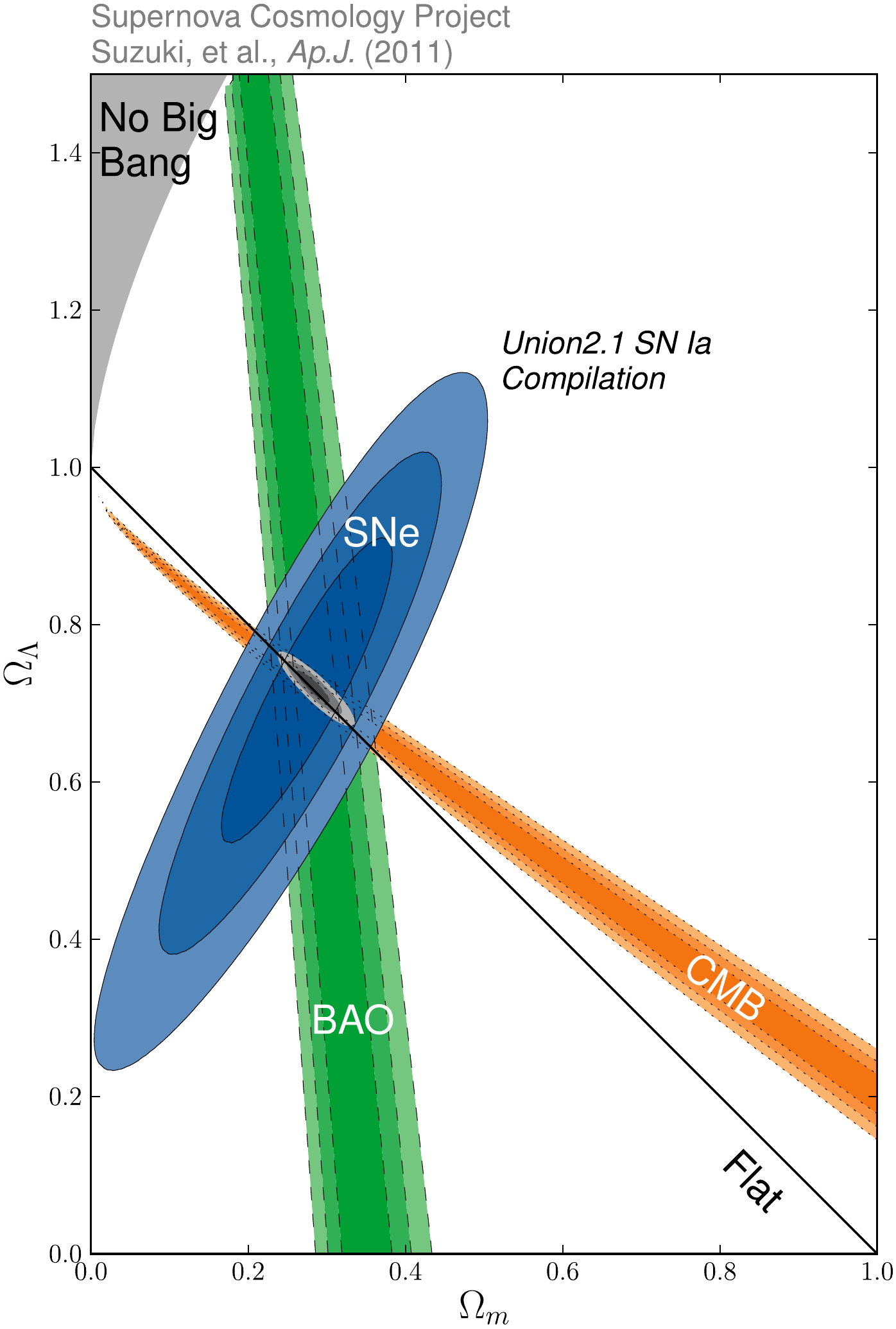}};
\node[anchor=west,rotate=90] at (10.26,3.12) (description) {\scriptsize 0};
\node[anchor=west] at (14.45,-2.8) (description) {\scriptsize 0};
\end{tikzpicture}
\centering
\caption[Cross plot from the Union 2 data set.]{Cross plot from the Union2 data set and other probes. One can notice (within the 68\%, 95\%, and 99.7\% confidence regions) the nice combination of the different observables which are SNe Ia, the CMB and BAOs. The predicted values, with the highest confidence, are $(\Omega_{m0},\Omega_{\Lambda0}) \simeq (0.27,0.73)$. Credit\,: Supernova Cosmology Project (adapted from \href{http://supernova.lbl.gov/Union/}{\texttt{supernova.lbl.gov/Union/}}), see also \cite{Amanullah:2010vv}.}
\label{FigUnion2CrossPlot}
\end{figure}

To improve the estimation of cosmological parameters, we usually use the combination of SNe Ia observations with two other probes which give a precise estimation of the parameters in this homogeneous context. These two other probes are the cosmic microwave background (\gls{CMB}) and the baryonic acoustic oscillations (\gls{BAO}). One can indeed see in Fig. \ref{FigUnion2CrossPlot} the crossing of their different predictions and its great increase in the precision of the estimation of $\Om_{\Lambda0}$ and $\Om_{m0}$. We also remark that these data support a Universe very close to be spatially flat.

\newpage

\section{The large scale structure}
\label{Ch1Sec4}

\subsection{Observations}
\label{Ch1Sec41}

\begin{figure}[ht!]
\centering
\includegraphics[width=10cm]{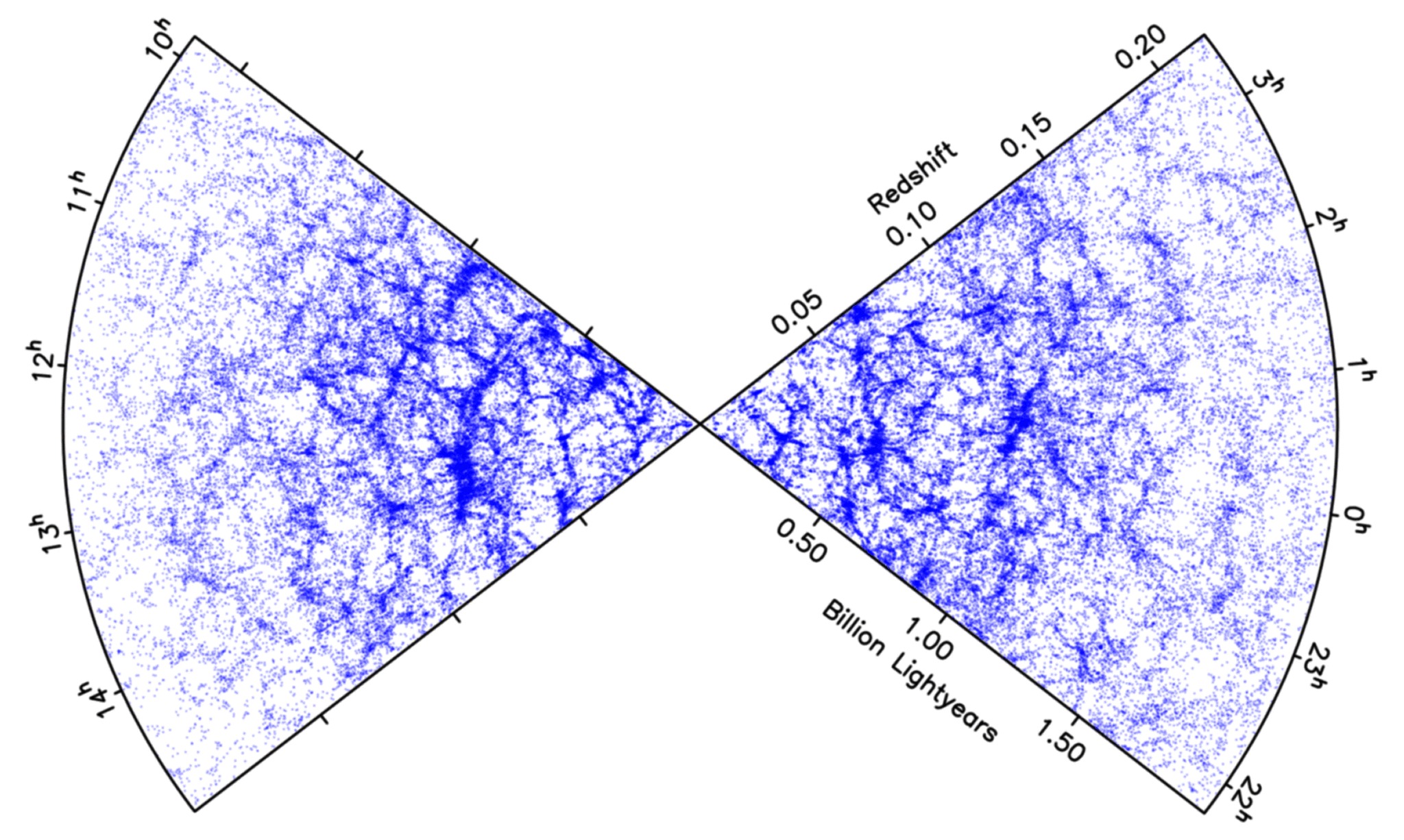}
\centering
\caption[The 2dF Galaxy Redshift Survey (2dFGRS)]{The 2dF Galaxy Redshift Survey (2dFGRS). Two narrow bands on the sky are represented. Data has been stacked in the thin ($\sim 4^{\circ}$) angular direction of the survey to get a two dimensional picture. One can see the homogeneity of structures at large scales. Credit\,: 2dFGRS (adapted from \href{http://www2.aao.gov.au/~TDFgg/}{\texttt{www2.aao.gov.au/$\sim$TDFgg/}}).}
\label{Fig2dF}
\end{figure}

The SDSS (Sloan Digital Sky Survey) and 2dFGRS (Two degree field galaxy redshift survey) are major surveys. SDSS covered about one quarter of the sky and observed millions of objects, with more that a million of galaxies and their distances. On the other hand the 2dFGRS observed a very narrow part of the sky with 250 000 galaxies and their redshift up to $\sim 0.3$. As one can see on Fig. \ref{Fig2dF}, matter is organized in a filamentous structure which looks relatively homogeneous at large scales.

Nevertheless, several complications must be addressed to obtain a real picture of the matter distribution in the Universe.
First, what we observe is light coming from our past lightcone, so we can only probe the galaxy distribution, not the true matter distribution containing dark baryonic objects as well as cold dark matter itself. The link between the galaxy distribution and the true matter distribution is not trivial and struggles with our lack of understanding of the galaxies formation. The density contrast of galaxies $\delta_{\rm gal}$ is related to the density contrast of matter by
\beq
\delta_{\rm gal}(t,\xbf) = b \, \delta (t,\xbf) ~~~~~\mbox{ with}~~~~~ \delta(t,\xbf) = \frac{\rho(t,\xbf) - \rho(t)}{\rho(t)} \equiv \frac{\delta\rho(t,\xbf)}{\rho(t)} ~~,
\eeq
where $\rho(t)$ is the background homogeneous density field and $\rho(t,\xbf)$ the true matter field.
The ``parameter'' $b(t,\xbf)$ is called the \gls{bias} and has been extensively studied in the literature, including its possible spacetime dependence and the complications involved in the non-linear regime.

One other complication is that galaxies are not observed in the real spacetime, namely by two angles on the sky and a spacetime combination as $\eta -r$ on the past lightcone, but are observed in the redshift space, i.e. that radial distances are measured thought the change in light frenquency. As a consequence, velocity and position are entangled in the measurements (like in the so-called \emph{fingers of God} effect).
Nevertheless, several methods have been confronted to describe matter and evaluate its power spectrum. For example, measurements from weak lensing are free from bias effects as they measure the real distribution of matter.

\begin{figure}[ht!]
\centering
\begin{tikzpicture}
  \node [xshift=-7cm,yshift=3cm] at (current page.south east)
    {\includegraphics[width=10cm]{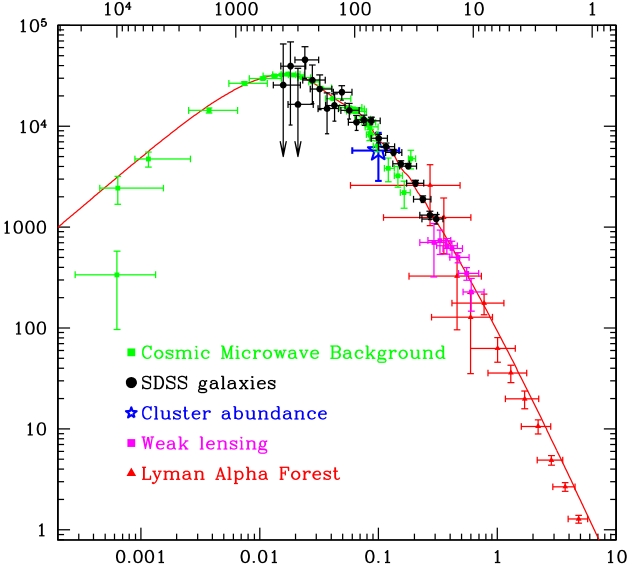}};
\node[anchor=west,rotate=90] at (8.7,-0.5) (description) {\small Current power spectrum $\Delta(k) ~ [(h^{-1} \, \Mpc)^3]$};
\node[anchor=west] at (12.2,7.8) (description) {\small Wavelength $\lambda ~ [h^{-1} \, \Mpc]$};
\node[anchor=west] at (12.2,-1.8) (description) {\small Wavenumber $k ~ [h \, \Mpc^{-1}]$};
\end{tikzpicture}
\centering
\caption[The matter power spectrum as measured from different sources]{The matter power spectrum as measured from different sources. We notice that the linear part of the spectrum at large scales (small $k$) is reconstructed from CMB. Similarly, the linear spectrum shown at small scales (large $k$) is a reconstruction from the non-linear spectrum (as one observes the total, non-linear, spectrum) and is thus model dependent (see Sec. \ref{Ch1Sec43}). Figure adapted from \cite{Tegmark:2003uf}.}
\label{FigExpPS}
\end{figure}
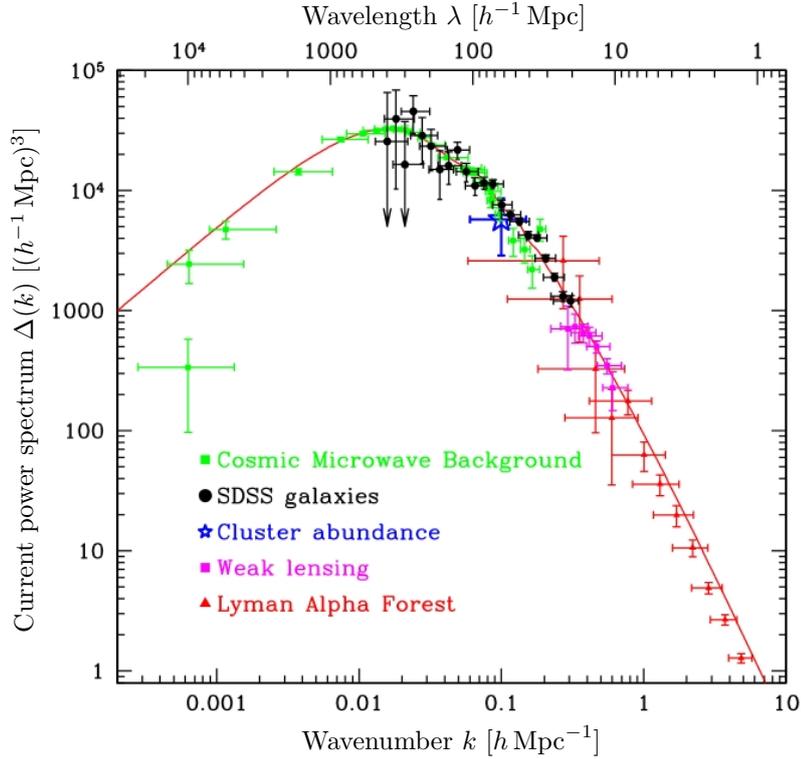

Considering these effects, one can measure the matter density spectrum $\delta(t,\xbf)$ and go in Fourier space. By assuming an ergodic principle, one can take the volume average of the 2-point correlation functions and consider that this measurement is equal to the stochastic average $\overline{\delta_\kbf \delta_\kbf^{\ast}}$. One then obtains the definition of the dimensionless matter power spectrum by the definition\,:
\beq
\Delta^2(t,\kbf) = \frac{k^3}{2 \pi^2} |\delta_\kbf(t)|^2 ~~.
\eeq
From this expression it is possible to plot the experimental power spectrum of Fig. \ref{FigExpPS}. We notice the variety of probes composing this incredible measurement over five orders of magnitude in scales.

\subsection{The linear power spectrum}
\label{Ch1Sec42}

What we will be interested by is the computation of the power spectrum for the gravitational potential in its linear and non-linear regimes. The reason for this is that metric perturbations, that we will study later, are expressed in terms of the gravitational potential and not the matter density. As a consequence, one of the most important equation for us will be the \gls{Poisson equation} which links the two quantities. In synchronous gauge (see Sec. \ref{AppASec33}), this equation is written as\,:
\beq
\nabla^2 \psi(x) = 4 \pi G \rho(t) ~ \delta(x) ~~,
\eeq
where $x=(t,\xbf)$, and $\rho(t)$ is the background matter density. Taking the Fourier transform of this expression brings $k^2 \psi_k \sim H_0^2 \Om_{m0} \, \delta_k$, where $\delta_k$ is the spectral function associated to the matter density contrast $\delta(x)$ and $\psi_k$ the spectral function associated to the gravitational (Bardeen) potential $\psi(x)$ (at first order in perturbations) that we will define in Chapter \ref{Chap4}.

The 2-point correlation function of the gravitational (Bardeen) potential, written as $\overline{\psi(x) \psi(x)}$ with the overbar denoting a statistical (or stochastic) average\footnote{In textbooks, the stochastic average is usually denoted by $\lla \ldots \rra$. In our work this notation denotes sky-average and for that reason we prefer the notation $\overline{(\ldots)}$ to denote stochastic averaging.}, is a dimensionless quantity.
Taking its Fourier transform teaches us that $[|\psi_k|^2] = [k^{-3}]$ (where square brackets denote the dimension of the quantity). Thus, $|\delta_k|^2$ has the same dimension as $|\psi_k|^2$ when we set $c = 1$ (in that case $[k] = [H_0]$).
More precisely, the relation between $|\delta_k|^2$ and $|\psi_k|^2$ is\,:
\beq
|\psi_k|^2 = \left(\frac{3}{2}\right)^2 \frac{\Om_{m}^2(z) \Hcal^4}{k^4} |\delta_k|^2 ~~,
\eeq
where for a CDM flat FLRW Universe we have $\Hcal = a H = 2/\eta$ (with $\eta$ the conformal time, $a(\eta) = C \eta^2$) and hence $\Hcal^4 = \left(\frac{a_0}{a(\eta)}\right)^2 \Hcal_0^4$~.

In a more general context (independent from the Universe model), we have seen in Sec. \ref{Ch1Sec22} that the equation of conservation for a perfect fluid of matter, satisfying $p = w ~ \rho$ with $w = \cst$, gives rise to the solution of Eq. \rref{SolutionOmega}. Using that $\Hcal = H a$, this relation becomes\,:
\beq
\Om_m(z) = \Om_{m0} (1+z) \left( \frac{\Hcal_0}{\Hcal} \right)^2 ~~,
\eeq
and we get, in Fourier space, the Poisson equation at an arbitrary redshift\,:
\beq
\label{PsiDelta}
|\psi_k(z)|^2 = \left(\frac{3}{2}\right)^2 \frac{\Om_{m0}^2 \Hcal_0^4}{k^4} (1+z)^2 |\delta_k(z)|^2 ~~.
\eeq
We notice that there is an easy one-to-one relation between $\eta$ and $z$ in the case of a CDM flat FLRW Universe (where $\Omega_{m0} = 1$). In the case of a $\Lambda$CDM model, the relation $\eta(z)$ is in some way degenerated as it depends on the values of $(\Omega_{m0},\Omega_{\Lambda 0})$ (see Sec. \ref{Ch1Sec34}).

Moreover, the dimensionless power spectrum $\Pcal_\psi(k)$ is related to $|\psi_k|^2$ by its definition\,:
\beq
\Pcal_\psi (k,\eta) \equiv \frac{k^3}{2 \pi^2} |\psi_k(\eta)|^2 ~~,
\eeq
where the time dependence has been written explicitly (but can be expressed in terms of $z$).

The (linear) power spectrum $\Pcal_\psi(k,z)$ is first obtained from the \gls{inflationary spectrum} (i.e. sourced by perturbations at the end of the inflationary era) where its expression is $\left(\frac{3}{5}\right)^2 \Delta_{\cal R}^2$, with\,:
\beq
\Delta_{\cal R}^2=A \left(\frac{k}{k_0}\right)^{n_s-1} ~~\mbox{ and }~~~ A=2.45 ~10^{-9} ~~~~~ , ~~~~~  n_s=0.96 ~~~~~ , ~~~~~ k_0/a_0=0.002 ~{\rm Mpc}^{-1} ~~~.
\eeq
We remark that the value of the spectral index $n_s$ can be related to the \gls{slow-roll parameters} $\epsilon$ and $\eta$ of the single field inflation model (see Sec. \ref{AppASec21}, Eq. \rref{SlowRollParameters}) by
\beq
n_s - 1 = - 6 \epsilon + 2 \eta ~~.
\eeq
We have to add a transfer function $T(k)$ to describe the sub-horizon evolution of the different modes re-entering during the radiation-dominated era (see Sec. \ref{AppASec24} and \ref{AppASec42} for explicit forms). We also add a growth factor $g(z)$ which takes into account the evolution of $\psi_k$ during the dark energy era (assuming here a $\Lambda$CDM Universe). Finally, the expression of the linear power spectrum is\,:
\beq
\label{PsiPIntro}
\Pcal_\psi (k)  = \left(\frac{3}{5}\right)^2 \Delta_{\cal R}^2 T^2(k) \left(\frac{g(z)}{g_{\infty}}\right)^2 ~.
\eeq

\subsection{Short insight in the non-linear regime}
\label{Ch1Sec43}

At small scales, matter is organized in structures and the density contrast can take values which are more important than on cosmological scales. As opposed to the last section, we cannot work here solely in terms of the gravitational potential and we need to go through the matter density spectrum to estimate the non-linear regime. The justification for this change is that the non-linear regime has been studied by \gls{N-body simulations} in which the density is the direct measurable quantity.

So we define what is called the fractional density variance per unit $\ln k$, namely $\Delta^2(k)$, by\,:
\beq
\label{DensityPSIntro}
\Delta^2(k) = \frac{d \sigma^2}{d \ln k} ~~~~~ \mbox{ or equivalently\,:} ~~~~~ \sigma^2 = \int \Delta^2(k) ~ d\ln k
\eeq
with the definition of the \gls{density variance}\,:
\beq
\sigma^2 \equiv \overline{\delta(x)\delta(x)} = \int \frac{d^3k}{(2 \pi)^3} |\delta_{\kbf}|^2 = \int \frac{k^2 dk}{2 \pi^2} |\delta_k|^2 ~~.
\eeq
Here the quantities depend on time (or redshift) and the overbar denotes again the usual stochastic averaging.
These two expressions bring the simple relation between their integrands\,:
\beq
\label{DeltadeltaIntro}
\Delta^2(k,z) = \frac{k^3}{2 \pi^2} |\delta_k(z)|^2 ~~,
\eeq
and confirms that $\Delta^2(k,z)$ is a dimensionless quantity.
Besides this, \cite{Smith:2002dz} defines another dimensionless power spectrum, $P(k)$, which is called the ``initial power spectrum'' (in the sense of structure formation simulations) and is taken to be a simple power law\,:
\beq
P(k) = \Acal~ k^n ~.
\eeq
This spectrum is linked to $\Delta^2(k)$ through the expression (see Eq. (5) in \cite{Smith:2002dz})\,:
\beq
\Delta_{\rm L}^2(k) = \frac{V}{(2 \pi)^3} 4 \pi k^3 P(k) ~~~~~ , ~~~~~\mbox{i.e. }~~~~~ \Delta_{\rm L}^2(k) \sim k^{n+3} ~~~~~,
\eeq
now taken at the linear level and where the so-called $V$ factor is a normalization volume.

Nevertheless, this is not the approach that we will take here. A very first reason for this is that these quantities are only adapted to simulations (e.g. we have no a priori criterion to choose $V$). Furthermore, what we are interested in, here, is the non-linear power spectrum obtained from the linear power spectrum. Both should hence match at large scales. For this reason, the linear power spectrum that we will use is the one from Eq. \rref{PsiPIntro}, and we will now call it $\Pcal_\psi^{\rm L}(k)$. We should also recall that the Poisson equation is valid both in the linear and non-linear regimes, so that we have\,:
\beq
\label{PEIntro}
\Pcal_\psi^{\rm{L,NL}}(k) = \frac{9}{4} \frac{\Om_{m0}^2 \Hcal_0^4}{k^4} (1+z)^2 \Delta_{\rm{L,NL}}^2(k) ~.
\eeq
We will thus obtain the linear density spectrum using the Poisson equation at the linear level\,:
\beq
\label{OurDlIntro}
\Delta_{\rm{L}}^2(k) = \frac{4}{9} \frac{k^4}{\Om_{m0}^2 \Hcal_0^4 (1+z)^2} \Pcal_\psi^{\rm{L}}(k) = \frac{4}{25} \frac{A \Om_{m0}^{-2} \Hcal_0^{-4}}{(1+z)^2} ~ \frac{k^{n_s+3}}{k_0^{n_s-1}} \left(\frac{g(z)}{g_0}\right)^2 T^2(k) ~.
\eeq
We will then use the HaloFit model, as presented in Sec. \ref{AppASec4} and \ref{AppASec5}, to generate from $\Delta_{\L}^2(k)$ the density power spectrum in the non-linear regime, i.e. $\Delta_{\NL}^2(k)$. And as a last step, we will use the Eq. \rref{PEIntro} again, in the non-linear regime, to obtain the desired quantity $\Pcal_\psi^{\rm{NL}}(k)$.

\section{Dealing with the inhomogeneous Universe}
\label{Ch1Sec5}

\subsection{Real nature of the Universe}
\label{Ch1Sec51}

As we have seen, the analysis of the standard model of cosmology takes place within a homogeneous context. The low precision of observations not so far in the past has justified this approach and yet brought amazingly consistent results. This is mainly due to the relative smallness of inhomogeneities as supported by the CMB. Nevertheless, we can notice that SNe Ia observations suffer from a very large dispersion which make their discovery of dark energy not straightforward. On the other hand the true nature of the Universe is inhomogeneous as revealed by the large scale structure below a certain scale. It is thus natural to ask to which extent observations, and in particular SNe Ia, are free from a contamination by the presence of matter (and thus gravitational) inhomogeneities. At the end, these probes are local astrophysical objects in virialized domains of our Universe and light signals that we receive from them propagate over very large distances, encountering possibly a large number of sources of dispersion\footnote{Another problem is the one of curvature as the Universe is mostly filled by voids and thus photons prapagating in it are sensitive to the Weyl curvature instead of the Riemann curvature (see \cite{Clarkson:2011br}). We will not address this here.}. We also know that SNe Ia measurements can be equally well fitted by a LTB model (see e.g. \cite{Celerier:1999hp}) in which we are close to the center of a less dense region of matter. However, experimental constraints on our position inside this underdensity make us unreasonnably close to the center, disproving this possibility. Nevertheless, the question is still there and one can wonder how big is the effect of inhomogeneities and whether or not they could explain (partially maybe) the observation of the acceleration in the expansion. The possibility to explain dark energy with the inhomogeneous Universe is very tempting as it is the most concervative explanation at our disposal, assuming no extra physics, and only reconsidering our interpretation of general relativity. It has also the advantage to solve the coincidence problem, by linking the formation of structures with the appearance of dark energy, and the smallness of the dark energy density.

\subsection{The backreaction problem}
\label{Ch1Sec52}

In relation to the problem of inhomogeneity of our Universe, different questions related to the non-linear nature of general relativity are still unsolved \cite{Clarkson:2011zq}. The question of how to average small-scales, or coarse-graine structures, to describe their large-scale geometry and dynamics, is called the \emph{\gls{averaging problem}} of cosmology. The effect of inhomogeneities on the large-scale dynamics of the Universe is difficult to evaluate. Solving this problem is also solving the \emph{\gls{backreaction problem}} which can be illustrated by the following reasoning. Imagine that we know exactly the metric $g_{\mu\nu}^{(loc)}$ at a (local) subgalactic scale. By the knowledge of a correct smoothing procedure, we could average this metric to get the metric at a larger scale, like the galactic scale\,: $g_{\mu\nu}^{({\rm loc})} \rightarrow g_{\mu\nu}^{({\rm gal})}$. Let us imagine also that this smoothing procedure is applicable to the stress energy tensor so that we have\,: $T_{\mu\nu}^{({\rm loc})} \rightarrow T_{\mu\nu}^{({\rm gal})}$. Assuming that the Einstein equations at the local scale were\,:
\beq
G_{\mu\nu}^{({\rm loc})} = 8 \pi G ~ T_{\mu\nu}^{({\rm loc})} ~~,
\eeq
the non-linear nature of the \gls{Einstein equations}, i.e. the fact that $G \sim \Gamma^2 ~,~ \partial \Gamma \sim (g \partial g)^2 ~,~ \partial (g \partial g)$ (in crude notations), makes that these previous equations will become at galactic scales\,:
\beq
G_{\mu\nu}^{({\rm gal})} = 8 \pi G ~ T_{\mu\nu}^{({\rm gal})} + E_{\mu\nu}^{({\rm gal})} ~~.
\eeq
In this expression, the extra contribution $E_{\mu\nu}^{({\rm gal})} \neq 0$ will emerge from the non-commutation between the averaging procedure of $g_{\mu\nu}^{({\rm loc})}$ and the non-linear aspect of Einstein's equations. By repeating this procedure up to the large scale structure and cosmic scales, we will ideally reach an homogeneous description where the effect of inhomogeneities have been properly taken into account.

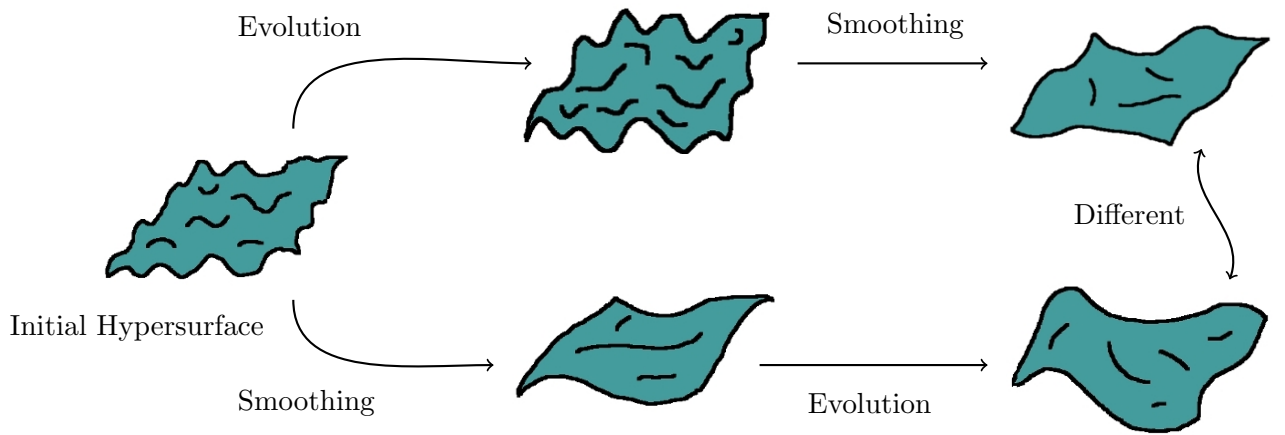
\begin{figure}[ht!]
\centering
\begin{tikzpicture}
\node (label) at (-12cm,0cm){\includegraphics[width=3.5cm]{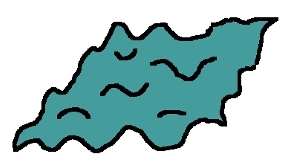}};
\node (label) at (-6.5cm,1.8cm){\includegraphics[width=3.5cm]{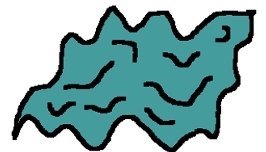}};
\node (label) at (-0cm,1.8cm){\includegraphics[width=3.5cm]{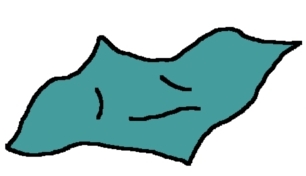}};
\node (label) at (-6.5cm,-1.8cm){\includegraphics[width=3.5cm]{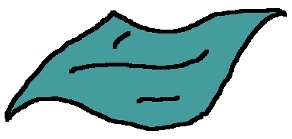}};
\node (label) at (-0cm,-1.8cm){\includegraphics[width=3.5cm]{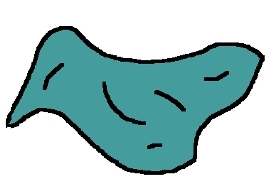}};

\node[anchor=west] at (-15,-1.5) (description) {Initial Hypersurface};

\node[anchor=west] at (-12,+2.5) (description) {Evolution};
\node[anchor=west] at (-12,-2.5) (description) {Smoothing};

\node[anchor=west] at (-4.25,+2.5) (description) {Smoothing};
\node[anchor=west] at (-4.5,-2.5) (description) {Evolution};

\node[anchor=west] at (-1,-0) (description) {Different};

\node[anchor=east] at (-11,1) (text) {};
\node[anchor=west] at (-8,2) (description) {};
\draw (text) edge[out=90,in=180,->,color=black,thick] (description);

\node[anchor=east] at (-11,-1) (text) {};
\node[anchor=west] at (-8.5,-2) (description) {};
\draw (text) edge[out=-90,in=180,->,color=black,thick] (description);

\node[anchor=east] at (-4.5,2) (text) {};
\node[anchor=west] at (-2,2) (description) {};
\draw (text) edge[out=0,in=180,->,color=black,thick] (description);

\node[anchor=east] at (-5,-2) (text) {};
\node[anchor=west] at (-2,-2) (description) {};
\draw (text) edge[out=0,in=180,->,color=black,thick] (description);

\node[anchor=east] at (1,1) (text) {};
\node[anchor=west] at (1,-1) (description) {};
\draw (text) edge[out=-110,in=-290,<->,color=black,thick] (description);

\end{tikzpicture}
\caption[Illustration of the backreaction problem]{Illustration of the backreaction problem\,: we see the non-commutation between the smoothing, or averaging, and time evolution. This is related to the non-linearity of Einstein's equations.}
\label{FigBackreaction}
\end{figure}

These two problems are also related to the \emph{\gls{fitting problem}}, i.e. the question of how to relate observations made from a lumpy environment to an idealized model with a (not really existing) ``backgound'' geometry. We still do not know, nowadays, how to match an inhomogeneous Universe (what we really observe) with an homogeneous FLRW model (used to interpret the data), as illustrated in Fig. \ref{FigBackreaction}.

\newpage

\section{Introduction \& Outline of this thesis}
\label{Ch1Sec6}

% Concordance model & probes
The so-called concordance (or $\Lambda$CDM) model is based on a suitable combination of dark matter, dark energy and baryons for an overall critical density and has become the reference paradigm for the late, i.e. post-equality epoch, evolution of our Universe (see e.g. \cite{Nakamura:2010zzi}). It accounts equally well for the CMB data, the Large Scale Structure and, even more significantly, for the supernovae data in terms of a cosmic acceleration \cite{Riess:1998cb,Perlmutter:1998np}. However, these three tests of the concordance model are not strictly speaking at the same level of theoretical rigor. While the first two have to do, by definition, with the inhomogeneities present in our Universe, the third is based on an ideal homogeneous and isotropic Friedmann-Lema\^itre-Robertson-Walker (FLRW) geometry. It is thus a very amazing fact that these three probes finally give a very consistent picture and this especially as we still lack of a satisfying theoretical explanation for the real nature of dark energy. Establishing the existence of this unknown component and determining its parameters is thus one of the central issues in modern cosmology.

% Need to go for Inhomogeneities
As we said, the analysis of SNe Ia (taken as standard candels) is usually made in the simplified context of a homogeneous and isotropic (FLRW) cosmology. The issue has then been raised about whether inhomogeneities may affect the conclusion of such a theoretically naive assumption. Inhomogeneous models in which we occupy a privileged position in the Universe, for instance, can mimic dark energy (as first pointed out in \cite{Celerier:1999hp}), but look both unrealistic and highly fine-tuned. Nevertheless, we cannot deny that the Hubble diagram is affected by a large dispersion on the data, some SNe Ia being so faint that they look to contradict the reasonable condition $\Omega_{\Lambda 0} \leq 1$. This kind of effect is certainly related to the non-standard nature of some supernovae, but we can still wonder how much is the contamination coming from the inhomogeneous aspect of our Universe.
At the end, cosmology can be summarized as the study of two things\,: the content of the Universe and the dynamics of this content. It is thus a very important question to ask how much structures affect the measurement of distances. It is also clear that, at least for the sake of precision, a better treatment of cosmic acceleration should take inhomogeneities into account, and this at least in the presence of stochastically isotropic and homogeneous perturbations of the kind predicted by inflation. Only when this is done we can establish in a convincing way whether $\Lambda$CDM gives a simultaneous consistent description of the above-mentioned body of cosmological data.

% BR for DE
In additions, it is by now well-known (see e.g. \cite{AppEt1933a}) that averaging solutions of the full inhomogeneous Einstein equations leads, in general, to different results from those obtained by solving the averaged (i.e. homogeneous) Einstein equations. In particular, the averaging procedure does not commute with the non-linear differential operators appearing in the Einstein equations. As a result, the dynamics of the averaged geometry is affected by so-called ``backreaction'' terms, originating from the inhomogeneities present in the geometry and in the various components of the cosmic fluid. Following the discovery of cosmic acceleration on large scales, interest in the possible effects of inhomogeneities for interpreting the data themselves has considerably risen (see \cite{Clarkson:2011zq,Buchert:2011yu,Ellis:2011hk,Buchert:2011sx} for recent reviews).
Indeed, though there is by now general agreement that super-horizon perturbations cannot mimic dark-energy effects \cite{Flanagan:2005dk,Geshnizjani:2005ce,Hirata:2005ei}, the impact of sub-horizon perturbations is by contrast still unsettled \cite{Ishibashi:2005sj,Paranjape:2008jc,Kolb:2011zz,Clarkson:2011uk}, owing to the appearance of ultraviolet divergences while computing their ``backreaction" on certain classes of large-scale averages \cite{Kolb:2011zz,Clarkson:2011uk}. The possibility that these effects may simulate a substantial fraction of dark energy, or that they may at least play some role in the context of near-future precision cosmology, has to be seriously considered.
According to some authors \cite{Buchert:1999mc,Buchert:2002ij,Rasanen:2003fy,Rasanen:2006kp,Barausse:2005nf,Kolb:2004am,Kolb:2005da} present inhomogeneities might explain, by themselves, cosmic acceleration without any need for dark-energy contributions, thereby providing an elegant solution to the well-known ``coincidence problem''. According to others \cite{Flanagan:2005dk, Hirata:2005ei,Ishibashi:2005sj,Paranjape:2008jc,Baumann:2010tm,Green:2010qy} the effect of inhomogeneities is, instead, completely negligible.
The truth may lie somewhere in between, in the sense that a quantitative understanding of inhomogeneities effects could be important in order to put precise constraints on dark-energy parameters, such as the critical fraction of dark-energy density, $\Omega_{\La}$, and the evolution of its effective equation of state, $w_{\La}(z)$.

If inhomogeneities do have an important impact, the actual observations must also correspond in some way to the measurement of averaged quantities.
As a consequence of this, and because of the backreaction issue, much work has been done in these last few years on trying to formulate a suitable ``averaged'' description of inhomogeneous cosmologies. In most of these works, following Buchert's seminal papers \cite{Buchert:1999er,Buchert:2001sa,2000GReGr..32..105B,2001GReGr..33.1381B}, the effective geometry emerging after the smoothing-out of local inhomogeneities has been determined by integrating over three-dimensional spacelike hypersurfaces and computing the ensuing ``backreaction" on the averaged geometry. However, as pointed out long ago \cite{Maartens1,Maartens2}, a phenomenological reconstruction of the spacetime metric and of its dynamic evolution on a cosmological scale -- revealed through the experimental study of the luminosity-distance to redshift relation \cite{Riess:1998cb,Perlmutter:1998np} -- is necessarily based on past light-cone observations (since most of the relevant signals travel with the speed of light). Hence, the averaging procedure should be possibly referred to a null hypersurface coinciding with a past light-cone rather than to some fixed-time spacelike hypersurface. Nonetheless, such a light-cone averaging procedure, whose importance has been repeatedly stressed in the specialized literature (see e.g. \cite{Buchert:2011yu,Maartens1,Maartens2,Marra:2007gc,Li:2008yj,Rasanen:2008be,Rasanen:2009uw,Kolb:2009hn}), has never been implemented in practice.

% My thesis -- Lightcone averaging and GLC metric

In this thesis we introduce a general gauge invariant prescription for averaging scalar quantities on null hypersurfaces, solving thus the long-standing problem of lightcone averaging. We apply it to the past light-cone of a generic observer in the context of an inhomogeneous cosmological metric, and provide the analog of the Buchert-Ehlers commutation rule \cite{Buchert:1995fz} for the derivatives of light-cone averaged quantities.  We also take advantage of the gauge invariance of our formalism by introducing an adapted system of coordinates (defining what we call a ``geodesic light-cone (GLC) frame'', which can be seen as a particular specification of the ``observational coordinates'' introduced in \cite{Maartens1,1985PhR...124..315E}). In these coordinates our averaging prescription greatly simplifies while keeping all the required degrees of freedom for applications to general inhomogeneous metric backgrounds. These GLC coordinates turn out to have other interesting properties for calculations on the past light-cone. In particular, the expression of the redshift $z$ and the luminosity distance $d_L$ are very simple and easy to interpret in this system of coordinates.

% My thesis -- Computation of the luminosity-distance

We then turn our attention to the case of a flat FLRW model filled by scalar perturbations of primordial (inflationary) origin. It is both a realistic model for observations and a simple enough case to authorize an analytical approach. Such perturbations are often conveniently parametrized in the longitudinal (or Newtonian) gauge \cite{Mukhanov:1990me}, we thus derive the transformation of coordinates between the GLC gauge and the Newtonian gauge. As a stochastic average is necessary to obtain a prediction which is independent from a given realization of inhomogeneities and because these latter give an effect from the second order in perturbations, we compute the luminosity distance up to that order. The tranformation of coordinates must then be pushed to second order. This transformation authorizes us then to compute the expression of the distance-redshift relation at second order in scalar perturbations. This very general result\footnote{Following the pioneering work of \cite{Sasaki:1987ad,Kasai:1987ap}, $d_L$ has been already computed to first order in the longitudinal gauge (for a CDM model in \cite{Bonvin:2005ps}, CDM and $\Lambda$CDM in \cite{Pyne:2003bn}), and to second order in synchronous gauge, but only for a dust-dominated Universe, in \cite{Barausse:2005nf}. See also \cite{Umeh:2012pn}.} has an interest on its own (i.e. irrespectively of its subsequent application to light-cone/ensemble averaging) and can possibly find many other applications in precision cosmology. Furthermore, the result presented here for $d_L$ is valid in general, i.e. for any given background model (except if caustics form \cite{Ellis:1998ha}, in which case the area distance is modified).

% My thesis -- Effect of inhomogeneities around FLRW

Applying these two notions that we developed, we study the average of scalar observables around the flat FLRW background and at second order in perturbations. For this we make use of an ensemble average over the inhomogeneities, inserting a realistic power spectrum of stochastic perturbations. The combination of the angular lightcone average and this stochastic average leads to characteristic contributions designated as ``induced backreaction'' terms, arising from a generic correlation between the inhomogeneities present in the variable we want to average (e.g. the luminosity distance) and those appearing in the covariant integration measure\footnote{This integration measure will also induce ``backreaction" terms of the type usually discussed in the literature \cite{Buchert:1999er,Buchert:2001sa}, namely terms that arise from (generalized) commutation rules between differential operators and averaging integrals.}. Applying this developments to scalar quantites such as the luminosity flux or distance of a light source lying on our past light-cone, we compute the effects of a stochastic background of inhomogeneities on the determination of dark-energy parameters in precision cosmology. Unlike the analyses in \cite{Kolb:2011zz,Clarkson:2011uk}, made on spatial hypersurfaces, we find a result always free from ultraviolet divergences and with no significant infrared contributions either. Nevertheless, the induced backreaction terms are not accounting for all the effets engendered by inhomogeneities and a full calculation, including all the second order terms, is made.

Proceeding with the simple model of CDM, we find in particular that the energy flux $\Phi \sim d_L^{-2}$ is practically unaffected by inhomogeneities, while the most commonly used variables (like $d_L$ or the distance modulus $\mu \sim 5 \log_{10} d_L$) receive much larger corrections (much bigger than we could have na\"ively expected). This shows that there are (at least in principle) intrinsic ambiguities in the measurement of the dark-energy parameters, unless the backreaction of stochastic inhomogeneities is properly taken into account. Actually, the advantages of flux averaging for minimizing biases on dark-energy parameters was first pointed out in \cite{Wang:1999bz}, where it was shown how the binning of data in appropriate redshift intervals can reduce the bias due to systematic effects such as weak lensing. It is intriguing that the preferred role played by the flux variable also comes out in this work where we perform a completely different averaging procedure, {\em at fixed redshift}. Our conclusions are not due to a binning of data, but to an application of our covariant spacetime average to different functions of the luminosity distance. It is thus safe to believe that this interesting property of the flux is to be related to the conservation of light along the past lightcone.

Our first conclusions of that analysis, based on the use of a perturbation spectrum valid in the linear regime in a CDM model \cite{Eisenstein:1997ik}, are as follows. On the one hand, such kind of perturbations cannot simulate a substantial fraction of dark energy\,: their contribution to the averaged flux-redshift relation is both too small (at large values of the redshift $z$) and has the wrong $z$-dependence. On the other hand, stochastic fluctuations add a new and relatively important dispersion with respect to the prediction of the homogeneous and isotropic FLRW cosmology. This dispersion is independent of the experimental apparatus, of the observational procedure, of the intrinsic fluctuations in absolute luminosity, and may prevent a determination of the dark-energy parameter $\Omega_{\Lambda}(z)$ down to the percent level -- at least if we are using the luminosity-redshift relation alone. Another important conclusion is that (light-cone averages of) different functions of the same observable get biased in different ways, with the energy flux sticking out as the observable which gets minimally affected by inhomogeneities, irrespectively of the redshift binning utilized. We should recall here that other possible sources of uncertainty, bias and scatter in the Hubble diagram have been studied in many previous papers (see e.g. \cite{Frieman:1996xk,Holz:1998cp,Metcalf:1998ih,Porciani:2000ag,Bolejko:2012uj}).

% My thesis -- Extension to the non-linear regime

The power spectrum used in this first study is valid in the linear perturbative regime. We also extend our treatment by considering a $\Lambda$CDM model, by adding the effect of baryons, and by considering two parametrizations of the HaloFit model \cite{Smith:2002dz,Takahashi:2012em}, describing the density power spectrum in the non-linear regime. It is thus possible, by this choice, to attain a higher level of accuracy by considering the effect from small scale inhomogeneities. We hence use the power spectrum in $\Lambda$CDM to compute the effect of inhomogeneities as in the previous study and restricting ourself to the so-called ``enhanced terms" (i.e the dominant ones in the relevant range of $z$), already identified in the CDM case. We find that the effect of inhomogeneities is enhanced at large redshifts because of the dominant (lensing) term involved. Despite this increase in the size of the effect, inhomogeneities are not strong enough to mimic a dark energy effect. The dispersion, however, is significantly increased and corresponds to a possible statistical deviation of $\sim 10\%$ on $\Omega_{\Lambda0}$. This result underlines the importance of considering properly the effect of inhomogeneities on the dispersion of SNe Ia data. We find, furthermore, that this dispersion is in very good agreement with the Union2 data at small redshifts (where peculiar velocity effects are involved) but too small at large redshifts (where lensing is leading their effect). This constitutes a rigorous proof of the robustness of the standard model of cosmology, but still emphasize the non-negligible aspect of inhomogeneities. Finally, our calculations lead to a precise prediction for the size of the lensing dispersion within the Hubble diagram, a prediction which is in accordance with the recent attempts to detect this signal \cite{Kronborg:2010uj,Jonsson:2010wx} and that should be confirmed in the next few years.

\bigskip

In this \tbf{Chapter} \ref{Chap1} we have presented the basic facts about homogeneous cosmology, stressing the link between the astrophysical object that the SNe Ia are and the inhomogeneous nature of the Universe they live in. We have also presented the equations of evolution of the Universe in the homogeneous case with some of their assumptions. We shall now give the outline of this thesis.

\newpage

% Thesis chapters
% The rest of this thesis is organized as follows.
In \tbf{Chapter} \ref{Chap2} we treat the aspects relative to the averaging of scalars. We first recall the historical approach, à la Buchert, and present the interesting properties that emerge from this effective formalism. We then recall the notion of gauge invariance in averaging and give a more general definition of the average on spatial hypersurfaces. We then present the definition of a gauge invariant light-cone average prescription which has all the required properties to average scalars on null hypersurfaces. This allows us to derive relations analogous to the Buchert-Ehlers commutation rule. We also rigorously show that the average has, as expected, no effect on an homogeneous FLRW model.
In \tbf{Chapter} \ref{Chap3} we introduce a system of coordinates, the ``Geodesic Light-Cone'' (GLC) coordinates, which is very well adapted to calculations on the past light-cone. This system corresponds to a complete gauge fixing of the (recalled) observational coordinates, but it also has characteristic differences with it. In these GLC coordinates, the redshift and the luminosity distance take a very simple form and the null geodesics that photons travel on are parametrized by only one coordinate. We also discuss the link between this system and the synchronous gauge and other of its properties. We finally make use of this GLC coordinates to express our averaging prescriptions in a simple way and consider the example of the redshift drift.
The absence of averaging effets in the homogeneous case leads us to compute, in \tbf{Chapter} \ref{Chap4}, the expression of the luminosity-redshift relation at second order in perturbations around a flat FLRW model. For this, we compute the full second order transformation between the Poisson gauge and the GLC gauge and then make use of the simple expression taken by the distance in this last coordinates. Vector and tensor perturbations are taken into account at second order (as imposed by inflation). The final result, though very involved, encloses terms with a clearly defined physical meaning.
The role of \tbf{Chapter} \ref{Chap5} is to present the combination of the lightcone angular average with a stochastic average over inhomogeneities. We first recall the computation of the luminosity-redshift relation by staying at the first order for a better understanding of the physical terms involved. We then present the stochastic average and the power spectrum describing matter inhomogeneities. The formal combination of this average with the lightcone one gives rise to terms created by the mixing of the averaged scalar $S$ and the measure of integration. These terms are interesting as they autorize an effect of inhomogeneities from the first order of $S$ and they are called induced backreaction (IBR) terms. Using this recall at first order, we are able to compute the effect of inhomogeneities on the variance of $d_L$, an incomplete result, and rigorously derive its dispersion (depending only from first order). We then extend our analysis to the full second order calculation, reorganizing the numerous contributions, and present the calculation of the luminosity flux which turns out to be the quantity which is the least affected by inhomogenenities. We also show that vector and tensor perturbations exactly cancel within our average. We then link the averages of the flux and the distance and address the averaging of the distance modulus. We also present the computation of their respective dispersions and explain how genuine second order contributions are dealt with.
The \tbf{Chapter} \ref{Chap6} presents the numerical computation of the terms previously derived. To that purpose we introduce an explicit form for the transfer function in the power spectrum. This simple expression allows to compute the IBR terms and gives a first estimate of the dispersion on $d_L$ in a CDM model. It also gives an understanding of the dominant effects coming from inhomogeneities in our calculation, turning out to be the peculiar velocities at small redshift and the lensing at large ones. We also explain why our integrals over the wavenumber $k$ of perturbations is finite both in the infrared and ultraviolet regimes. We then proceed to the calculation, still in CDM, of the contributions of the luminosity flux. This derivation gives us an insight on the dominant terms at the quadratic and genuine second orders. It also shows that this latter is always smaller for the redshifts we are interested by. We then move to the study of the realistic $\Lambda$CDM model by the introduction of a growth factor and the use of a more complete transfer function. Using the relations presented in the previous chapter, we compute also the impact of inhomogeneities on the luminosity distance and the distance modulus. Introducing an extra description of the power spectrum in its non-linear regime, through the so-called HaloFit model, we are able to study the enhanced effects from very small scales. Comments are given on the relative size of the effets and are compared with observations of the dispersion of SNe Ia.
We then recall the main results and achievements of this thesis in \tbf{Chapter} \ref{Chap7}. Further details on the ADM formalism, structure formation, gauge invariance and the HaloFit model are given in \tbf{Appendix} \ref{Chap8}.
\\
A French Résumé of this thesis is given in \tbf{Appendix} \ref{Chap10}.

% --------------------------------------------------------- %

%  --------------------- Chapter 2 ----------------------   %
% --------------------------------------------------------- %
\chapter{Gauge-invariant spatial \& lightcone averaging}
\label{Chap2}
\pagestyle{plain}

This chapter is devoted to the mathematical framework of averaging scalar quantities as part of the \gls{averaging problem}. We remark that no easy procedure exists at the moment to average vector and tensor quantities in general relativity. Nevertheless, it is important to stress the existence of a very general formalism, historically developed by R. Zalaletdinov and known as \emph{Macroscopic Gravity}, which addresses the question of averaging and evolving general manifolds\cite{Zalaletdinov:1992cf,Zalaletdinov:1996aj,Coley:2005ei,Paranjape:2009zu}. This work is intimately related to the backreaction issue and deserves a very high consideration. Nevertheless, because of its highly technical aspect and its relative distance with our approach of averaging, it will not be described in this thesis. In a first section, we recall the historical approach of volume averaging over inhomogeneous spacetimes, as mainly developed by T. Buchert \cite{Buchert:2010nt, Buchert:2011yu, Buchert:2011sx, Buchert:1999er} and written in synchronous gauge. We also present the notion of gauge invariant averaging and generalize the volume average to a coordinate-independent formulation. Both sections rely mainly on \cite{Gasperini:2009wp,Gasperini:2009mu}. We then present the formalism of light-cone averaging developed in this thesis and show some of its general properties. We derive in particular the generalized Buchert-Ehlers commutation rules on the light cone. As a final check, and to justify the perturbative calculation of Chapter \ref{Chap4}, we show that the average effects disappear in the FLRW homogeneous case.

\section{Covariant spatial averaging}
\label{Ch2Sec1}

\subsection{Buchert formalism}
\label{Ch2Sec11}

In this formalism, we assume the weak cosmological principle in order to build up effective Friedmann equations inside a simply-connected compact domain $\Dcal(t)$, in a foliated spacetime given by constant-t hypersurfaces. We also assume that the matter content of the Universe is made of a pressureless fluid (dust) and a possible cosmological constant, and we work in synchronous gauge\,:
\beq
\label{SGmetric}
ds^2_{\rm SG} = - dt^2 + g_{ij} dx^i dx^j ~~.
\eeq
The volume $\Dcal$, fixed in these coordinates, conserve the mass $M_\Dcal$ inside it at all times and the average of a scalar quantity $S$ is defined through a Riemannian (i.e. weighted by the metric) volume average\,:
\beq
\label{DefBuchertAv}
\lla S(t,\xbf) \rra_{\Dcal}(t) = \frac{1}{V_\Dcal} \int_\Dcal S(t,\xbf) ~ d\mu_g ~~~\mbox{ with }~~ V_\Dcal = \int_\Dcal d\mu_g ~~~\mbox{ and }~~ d\mu_g \equiv J ~ d^3 \xbf \equiv \sqrt{\det g_{ij}} ~ d^3\xbf ~~.
\eeq
As $g_{ij}$, the measure of integration $J$ is a function of $(t,x^i)$. On the other hand, the volume $\Dcal$ is assumed to keep the same $x^i$ coordinates along its way. In synchronous gauge, the matter geodesics are given by $x^i = \cst$ and the volume is thus comoving with the matter fluid. In that case, assumed in this section, the domain of integration is independent of time (although the measure $J$ is not).

Within synchronous gauge and the assumptions employed here, the Einstein equations can be recast in terms of geometrical quantities such as the \gls{expansion scalar}, the \gls{shear tensor} and the \gls{vorticity tensor}, defined as follows\,:
\beq
\label{DefExpScalarShearVorticity}
\Theta \equiv \nabla_\mu n^\mu ~~~~~,~~~~~ \sigma_{\mu\nu} \equiv h^\alpha_{~\mu} h^\beta_{~\nu} \left( \nabla_{(\alpha} n_{\beta)} - \frac13 h_{\alpha\beta} \Theta \right) ~~~~~,~~~~~ \omega_{\mu\nu} \equiv h^\alpha_{~\mu} h^\beta_{~\nu} \nabla_{[\alpha} n_{\beta]} ~~,
\eeq
and where $n^\mu$ is the normal vector to these spatial hypersurfaces given by $n_\mu = (-1,\vec{0}\,)$ and $n^\mu =(1,\vec{0}\,)$. As we consider a fluid which is at rest in these coordinates, we can also consider $n^\mu$ to be its velocity. The tensor $h_{\mu\nu}$ presented here is the projection tensor on spatial hypersurfaces\,:
\beq
h_{\mu\nu} = g_{\mu\nu} + n_\mu n_\nu ~~~~~,~~~~~ h_{\mu\nu} n^\nu = 0 ~~.
\eeq
The Einstein equations then turn into a system of two constraint and three dynamical equations given by (see Appendix \ref{AppASec13})\,:
\bea
\label{ImportNewConstr1}
\frac12 \Rcal + \frac13 \Theta^2 - \sigma^2 - \omega^2 &=& 8 \pi G \rho + \Lambda ~~, \\
\label{ImportNewConstr2}
\sigma^i_{~j\,||\,i} + \omega^i_{~j\,||\,i} &=& \frac23 \Theta_{|\,j} ~~, \\
\label{ImportNewDyn1}
\dot{\rho} &=& - \Theta \rho ~~, \\
\label{ImportNewDyn2}
(g_{ij})^{\tdev} &=& 2 g_{ik} \sigma^k_{~j} + 2 g_{ik} \omega^k_{~j} + \frac23 \Theta g_{ij} ~~, \\
\label{ImportNewDyn3}
(\sigma^i_{~j} + \omega^i_{~j})^{\tdev} &=& - \Theta (\sigma^i_{~j} + \omega^i_{~j}) - \Rcal^i_{~j} + \frac23 \delta^i_{~j} \left[ \sigma^2 + \omega^2 - \frac{\Theta^2}{3} + 8 \pi G \rho + \Lambda \right] ~~.
\eea
The rates of shear and vorticity appearing in these equations are defined as follows\footnote{Notice the difference in the literature where we sometimes define these rates as
$\sigma^2 \equiv \frac12 \sigma_{ij} \sigma^{ij}$ and $\omega^2 \equiv \frac12 \omega_{ij} \omega^{ij}$. In this case, $\omega^{ij}$ being anti-symmetric, we get the opposite sign for $\omega^2$ in the equations.}\,:
\beq
\sigma^2 \equiv \frac12 \sigma^i_{~j} \sigma^j_{~i} ~~~~~\mbox{and}~~~~~ \omega^2 \equiv \frac12 \omega^i_{~j} \omega^j_{~i} ~~.
\eeq
$\Rcal \equiv \Rcal^i_{~i}$ is the 3-dimensional spatial curvature of constant $t$ hypersurfaces. 
These new forms of the Einstein equations emerge from a more general set of equations in the ADM formalism which is presented in Appendix \ref{AppASec1}. They are obtained by taking the \gls{extrinsic curvature} tensor, describing the embedding of spatial hypersurfaces in the 4-dimensional $\Mcal_4$ spacetime, as the opposite value of the \gls{expansion tensor}\,:
\beq
\label{ImportKijAndThetaij}
K_{ij} = - \Theta_{ij} = - \left( \frac13 h_{ij} \Theta + \sigma_{ij} + \omega_{ij} \right) ~~.
\eeq
It can also be defined in terms of the spatial metric by\,:
\beq
K^i_{~j} = - \frac12 g^{ik} (g_{kj})^{\tdev} ~~.
\eeq
From this framework we also obtain a very important equation, not independent from the system (\ref{ImportNewConstr1}-\ref{ImportNewDyn3}), known as the \gls{Raychaudhuri's equation}\,:
\beq
\label{ImportRauchaudhuriEq}
\dot{\Theta} + \frac{\Theta^2}{3} + 2 \, \sigma^2 + 2 \, \omega^2 + 4 \pi G \rho - \Lambda = 0 ~~.
\eeq
This relation describes the evolution of the geometry of a bundle of geodesic world lines.

Inspired by these forms of the Hamiltonian constraint, Eq. \rref{ImportNewConstr1}, and Raychaudhuri's equation \rref{ImportRauchaudhuriEq} in the ADM formalism, we define a volume scale factor and its associated Hubble parameter as
\beq
a_\Dcal(t) \equiv \left( \frac{V_\Dcal(t)}{V_\Dcal(t_i)} \right)^{1/3} ~~,~~~\mbox{ which implies }~~ \frac{\dot{V}_\Dcal}{V_\Dcal} = 3 \frac{\dot{a}_\Dcal}{a_\Dcal} \equiv 3 H_\Dcal ~~.
\eeq
The relationship between this effective scale factor and the homogeneous scale factor is a bit hard to interpret. Indeed, one can see that $a_\Dcal(t)$ becomes equivalent to $a(t)$ if the matter content is taken to be homogeneous, and this whatever the size and form of the domain $\Dcal$, but on the other hand very different domains can give the same scale factor in an inhomogeneous Universe. One can nevertheless find equations of evolution for $a_\Dcal$ and this is the purpose of the following considerations.

Using the quantity $J(t,\xbf) \equiv \sqrt{\det g_{ij}}$ defined in Eq. \rref{DefBuchertAv}, we obviously have
\beq
\label{RelJdotJ}
(\ln J)^{\tdev} = \frac12 g^{ik} (g_{ki})^{\tdev} = - K ~~~~~\mbox{so that}~~~~~ \dot{J} = - K J = \Theta J ~~.
\eeq
From this last equality we obtain that
\beq
\lla \Theta \rra_\Dcal = \frac{1}{V_\Dcal} \int_\Dcal \frac{\dot{J}}{J} ~ J ~ d^3\xbf = \frac{1}{V_\Dcal} \left( \int_\Dcal J ~ d^3\xbf \right)^{\tdev} = \frac{\dot{V}_\Dcal}{V_\Dcal} ~~.
\eeq
One can see then that the Eq. \rref{ImportNewDyn1} is simply written as
\beq
\label{RelJrho}
(\rho J )^{\tdev} = 0 ~~~~~\mbox{ with the solution }~~~~~ \rho(t,\xbf) = \frac{\rho(t_0,\xbf) J(t_0,\xbf)}{J(t,\xbf)} ~~.
\eeq
This identity implies that the mass contained in the domain $\Dcal$ is, as requested, constant in time\,:
\beq
M_\Dcal = \int_\Dcal \rho J ~ d^3\xbf = {\rm cst} ~~,
\eeq
and so the averaged density, starting with an initial volume $V_{\Dcal_0} \equiv V_\Dcal(t_0)$, becomes
\beq
\lla \rho \rra_\Dcal = \frac{1}{V_\Dcal} \int_\Dcal \rho J ~ d^3\xbf = \frac{M_\Dcal}{V_\Dcal} = \frac{M_\Dcal}{V_{\Dcal_0} a_\Dcal^3} ~~.
\eeq

For all the scalar quantities $S$ encountered, such as $\rho$, we can take the time derivative of its average $\lla S \rra_\Dcal$ and we are lead to the \emph{\gls{commutation rule}}\,:
\beq
\label{NonComRel}
\lla S \rra_\Dcal^{\tdev} - \langle \dot{S} \rangle_\Dcal = \lla S \, \Theta \rra_\Dcal - \lla S \rra_\Dcal \lla \Theta \rra_\Dcal ~~.
\eeq
This relation shows that the time derivation of $S$, in other words its evolution in time, does not commute with its average $\lla \ldots \rra_\Dcal$. This is related to the long-standing problem of backreaction, as presented by Ellis in the 80's (see \cite{AppEt1933a}), and it shows that the spatially averaged quantity $S$ does not follow, on average, the evolution of $S$ when we have a space-dependent expansion scalar (or equivalently an extrinsic curvature).

\subsubsection{Averaging the Einstein equations}
\label{Ch2Sec112}

Averaging \gls{Raychaudhuri's equation} \rref{ImportRauchaudhuriEq} of a pressureless fluid ($p=0$), and using the commutation rule \rref{NonComRel} with $S = \Theta$, leads to
\beq
\lla \Theta \rra_\Dcal^{\tdev} - \frac23 \lla \Theta^2 \rra_\Dcal + \lla \Theta \rra_\Dcal^2 + 2 \lla \sigma^2 \rra_\Dcal + 2 \lla \omega^2 \rra_\Dcal + 4 \pi G \lla \rho \rra_\Dcal - \Lambda = 0 ~~,
\eeq
which is equivalent to
\beq
3 \frac{\ddot{a}_\Dcal(t)}{a_\Dcal(t)} + 4 \pi G \frac{M_\Dcal}{V_{\Dcal_0} a_\Dcal^3} - \Lambda = \Qcal_\Dcal ~~,
\eeq
with the so-called \emph{\gls{backreaction parameter}}\,:
\beq
\Qcal_\Dcal \equiv \frac23 \left[ \lla \Theta^2 \rra_\Dcal - \lla \Theta \rra_\Dcal^2 \right] - 2 \lla \sigma^2 \rra_\Dcal - 2 \lla \omega^2 \rra_\Dcal  ~~.
\eeq
Using now the \gls{Hamiltonian constraint} of Eq. \rref{ImportNewConstr1} and taking its average, one obtains
\beq
\frac12 \lla \Rcal \rra_\Dcal + \frac13 \lla \Theta^2 \rra_\Dcal - 8 \pi G \lla \rho \rra_\Dcal - \Lambda = \lla \sigma^2 \rra_\Dcal + \lla \omega^2 \rra_\Dcal ~~.
\eeq
Removing $ \frac13 \left[ \lla \Theta^2 \rra_\Dcal - \lla \Theta \rra_\Dcal^2 \right]$ on both sides, and using that $\lla \Theta \rra_\Dcal = 3 \frac{\dot{a}_\Dcal}{a_\Dcal}$, one gets
\beq
3 \left(\frac{\dot{a}_\Dcal}{a_\Dcal} \right)^2 + \frac12 \lla \Rcal \rra_\Dcal - 8 \pi G \frac{M_\Dcal}{V_{\Dcal_0} a_\Dcal^3} - \Lambda = - \frac{\Qcal_\Dcal}{2} ~~.
\eeq
We can remark that the \gls{vorticity tensor} plays a similar role as the \gls{shear tensor} in these calculations.

We shall assume now that the considered fluid has no vorticity, so\,: $\omega_{\mu\nu} = 0 ~~\forall~ \mu,\nu$ and $\omega = 0$.
Gathering Raychaudhuri's equation and the Hamiltonian constraint, and allowing both for a constant curvature $K_{\Dcal_i}$ of a FLRW model, one gets the general expressions\,:
\bea
\label{FriedBack1}
3 \frac{\ddot{a}_\Dcal(t)}{a_\Dcal(t)} &=& - 4 \pi G \lla \rho \rra_\Dcal + \Qcal_\Dcal + \Lambda ~~, \\
\label{FriedBack2}
3 H_\Dcal^2 &=& 8 \pi G \lla \rho \rra_\Dcal - \frac{3 K_{\Dcal_i}}{a_\Dcal^2} - \frac12 \left( \Wcal_\Dcal + \Qcal_\Dcal \right) + \Lambda ~~.
\eea
We defined here the deviation of the curvature from the constant curvature by the quantity\,:
\beq
\Wcal_\Dcal = \lla \Rcal \rra_{\Dcal} - 6 ~ \frac{K_{\Dcal_i}}{a_\Dcal^2} ~~.
\eeq

As in the classical Friedmann derivation, these two equations and the conservation equation
\beq
\label{FriedBack3}
\lla \rho \rra_\Dcal^{\tdev} + 3 H_\Dcal \lla \rho \rra_\Dcal = 0
\eeq
only form two independent ones. Differentiating Eq. \rref{FriedBack2} with respect to the cosmic time $t$ and then making use of Eqs. \rref{FriedBack1} and \rref{FriedBack3}, one gets the consistency relation -- or so-called \emph{\gls{integrability condition}} -- involving only geometrical quantities\,:
\beq
\label{FriedBack4}
\dot{\Qcal}_\Dcal + 6 H_\Dcal \Qcal_\Dcal + \dot{\Wcal}_\Dcal + 2 H_\Dcal \Wcal_\Dcal = 0 ~~ \Longleftrightarrow ~~ a_\Dcal^{-6} \left(a_\Dcal^6 \Qcal_\Dcal \right)^{\tdev} + a_\Dcal^{-2} \left( a_\Dcal^2 \Wcal_\Dcal \right)^{\tdev} = 0 ~~.
\eeq
This condition garanties the Friedmann-like form of the above equations, especially for Eq. \rref{FriedBack3}, and gives a dependence between the averaged 3-dimensional curvature $\lla \Rcal \rra_\Dcal$ (an intrinsic curvature invariant) and the backreaction term $\Qcal_\Dcal$ (an \gls{extrinsic curvature} invariant) which involves expansion, shear and vorticity scalars induced by the inhomogeneous aspect of the compact domain $\Dcal$.

One can see from Eq. \rref{FriedBack4} that $\Qcal_\Dcal = 0$ gives $\lla \Rcal \rra_\Dcal \propto a_\Dcal^{-2}$. This case contains the particular one of a FLRW model where there is no structure and the spatial curvature goes like $\Rcal \propto a^{-2}$. One does not encounter any backreaction also in the case where $Q_\Dcal \propto a_\Dcal^{-6}$. On the other hand, when none of these cases is satisfied, the relation between extrinsic and intrinsic curvatures is non-trivial. It is possible (i.e. not scientifically rejected) that small perturbations of the averaged spatial curvature and the averaged expansion rate could grow to lead finally to a global instability of the perturbed FLRW model. Our real Universe tends to be dominated by voids and we can suspect that the averaged curvature tends to be negative. It is believed that this situation, lead from the development of structures, could affect the evolution of the averaged expansion rate and, maybe, explain dark energy.

The backreaction is a profound issue which has also been addressed in \gls{N-body simulations}. The problem of these approaches is that backreaction is also closely related to topological issues. Indeed, when we use a Newtonian description to study structures formation in a spatially flat background, there is no spatial curvature and there are six different orientable space forms out of which we usually take the 3-torus. The backreaction in that case is equivalent to a boundary contribution. Such a study is thus believed to directly kill any possible backreaction mechanism and cannot self-generate an acceleration of the expansion. To be fair, we should also stress that the idea of backreaction could be completely wiped out by mathematics (like a ``\emph{No-backreaction}'' theorem). Indeed, the consideration of general relativity in 2+1 dimensions and the use of the Gauss-Bonnet theorem proves that the volume integral of the spatial curvature in a closed Riemannian 2-space is a topological invariant directly related to the Euler characteristic of the 2-dimensional manifold (see \cite{Buchert:2011yu}). It follows that $\lla \Rcal \rra_\Dcal \propto a_\Dcal^{-2}$ and $Q_\Dcal \propto a_\Dcal^{-4}$ (the right exponent to have no effect in a 2+1 geometry, where the coefficient 6 in Eq. \rref{FriedBack4} is replaced by 4). Thus the backreaction is exactly canceled. Nevertheless, the relation between curvature and topology is more involved in 3+1 gravity, and no theorem has been proved. So backreaction is still an open question in 4-dimensional spacetimes.

\subsubsection{Rewriting of the terms, morphon field}
\label{Ch2Sec113}

One can define the following cosmological-like parameters\,:
\beq
\Omega_m^\Dcal \equiv \frac{8 \pi G}{3} \frac{\lla \rho \rra_\Dcal}{H_\Dcal^2} ~~,~~ \Omega_\Lambda^\Dcal \equiv \frac{\Lambda}{3 H_\Dcal^2} ~~,~~ \Omega_K^\Dcal \equiv - \frac{\lla \Rcal \rra_\Dcal}{6 H_\Dcal^2} ~~,~~ \Omega_\Qcal^\Dcal \equiv - \frac{\Qcal_\Dcal}{6 H_\Dcal^2} ~~,
\eeq
and see that the Hamiltonian constraint (to compare with Eq. \rref{SumOmega}) is simply written as
\beq
\Omega_m^\Dcal + \Omega_\Lambda^\Dcal + \Omega_K^\Dcal + \Omega_\Qcal^\Dcal = 1 ~~.
\eeq
These parameters directly depend on the domain of integration $\Dcal$ and this last equation has to be interpreted with care.
We can also rewrite the averaged Friedmann-like equations derived above in terms of a scalar field which has been called the \gls{morphon}\footnote{This scalar field is indeed sourced by inhomogeneities and thus captures the ``morphological'' signature of structures.}, denoted by $\Phi_\Dcal$. One then defines the effective density and pressure for this field by\,:
\beq
\rho^\Phi_\Dcal \equiv - \frac{1}{16 \pi G} \left( \Qcal_\Dcal + \Wcal_\Dcal \right) ~~,~~ p^\Phi_\Dcal \equiv - \frac{1}{16 \pi G} \left( \Qcal_\Dcal - \frac{\Wcal_\Dcal}{3} \right) ~~.
\eeq
The generalized Friedmann equations then read\,:
\bea
3 \frac{\ddot{a}_\Dcal(t)}{a_\Dcal(t)} &=& - 4 \pi G \left( \lla \rho \rra_\Dcal + \rho^\Phi_\Dcal + 3 p^\Phi_\Dcal \right) + \Lambda ~~, \\
3 H_\Dcal^2 &=& 8 \pi G \lla \rho \rra_\Dcal - \frac{3 K_{\Dcal_i}}{a_\Dcal^2} - \frac12 \left( \Wcal_\Dcal + \Qcal_\Dcal \right) + \Lambda ~~,
\eea
and we have two equations of conservation\,:
\beq
\lla \rho \rra_\Dcal^{\tdev} + 3 H_\Dcal \lla \rho \rra_\Dcal = 0 ~~~~~,~~~~~ \rho^\Phi_\Dcal + 3 H_\Dcal ( \rho^\Phi_\Dcal + p^\Phi_\Dcal ) = 0 ~~.
\eeq
We can easily check that this last condition is equivalent to the integrability condition expressed in Eq. \rref{FriedBack4}. It is an amazing fact to see that the field $\Phi_\Dcal$ defined here has all the properties to effectively describe a cosmological constant.

We can also re-express these quantities by defining directly the field $\Phi_\Dcal$ and its potential $U_\Dcal$\,:
\beq
\rho^\Phi_\Dcal = \frac{e}{2} \dot{\Phi}_\Dcal^2 + U_\Dcal ~~~~~,~~~~~ p^\Phi_\Dcal = \frac{e}{2} \dot{\Phi}_\Dcal^2 - U_\Dcal ~~,
\eeq
where $e=+1$ if $\Phi_\Dcal$ is a standard scalar field and $e=-1$ if it is a phantom scalar field.
One can also show the correspondence between the terms
\beq
\Qcal_\Dcal = - 8 \pi G \left( e \dot{\Phi}_\Dcal^2 - U_\Dcal \right) ~~,~~ \Wcal_\Dcal = - 24 \pi G ~ U_\Dcal ~~,
\eeq
and show that the integrability condition can be recast in a Klein-Gordon equation for $\Phi_\Dcal$\,:
\beq
\ddot{\Phi}_\Dcal + 3 H_\Dcal \dot{\Phi}_\Dcal + e \parfrac{U_\Dcal}{\Phi_\Dcal} \left( \Phi_\Dcal , \lla \rho \rra_\Dcal \right) = 0 ~~.
\eeq
This equation is scale dependent through its dependence in $\Dcal$.
Last, we can define effective kinetic and potential energies of the morphon field by
\beq
E^{\rm kin}_\Dcal = e \, \dot{\Phi}_\Dcal^2 V_\Dcal ~~~~~,~~~~~ E^{\rm pot}_\Dcal = - U_\Dcal V_\Dcal ~~,
\eeq
and we get the simple relation
\beq
2 E^{\rm kin}_\Dcal + E^{\rm pot}_\Dcal = - \frac{\Qcal_\Dcal V_\Dcal}{8 \pi G} ~~.
\eeq
We deduce that an homogeneous Universe, for which $\Qcal_\Dcal = 0$, is equivalent to the \emph{\gls{virial equilibrium} condition} in this effective formalism.

\subsubsection{Kullback-Leibler relative entropy}
\label{Ch2Sec114}

Applying the commutation rule, Eq. \rref{NonComRel}, to the density of matter brings the relation\,:
\beq
\lla \rho \rra_\Dcal^{\tdev} - \lla \dot{\rho} \rra_\Dcal = \lla \rho \, \Theta \rra_\Dcal - \lla \rho \rra_\Dcal \lla \Theta \rra_\Dcal ~~.
\eeq
To describe the relative difference between the inhomogeneous function $\rho(t,x^i)$ inside the domain $\Dcal$ with respect to the homogeneous function $\lla \rho \rra_\Dcal$, one can use the concept of relative information given by the \gls{Kullback-Leibler entropy} \cite{Buchert:2011sx}.
This entropy is defined, for the density function, as\footnote{We should notice that this relative entropy can be related to the Weyl entropy, as recently proved in \cite{Li:2012qh}.}\,:
\beq
\Sscr_\Dcal = \int_\Dcal \rho \ln \left( \frac{\rho}{\lla \rho \rra_\Dcal} \right) ~ J d^3x ~~.
\eeq
One can then show that\,:
\beq
- \frac{\dot{\Sscr}_\Dcal}{V_\Dcal} = - \lla \frac{(\rho J)^{\tdev}}{J} \ln \left( \frac{\rho}{\lla \rho \rra_\Dcal} \right) \rra_\Dcal - \lla \rho \rra_\Dcal \lla \left( \frac{\rho}{\lla \rho \rra_\Dcal} \right)^{\tdev} \rra_\Dcal ~~,
\eeq
and use that $(\rho J)^{\tdev} = 0$, which eliminates the first term. The second one is developed, using Eq. \rref{RelJdotJ}, \rref{RelJrho} and \rref{FriedBack3}, as\,:
\beq
- \frac{\dot{\Sscr}_\Dcal}{V_\Dcal} = - \lla \rho \rra_\Dcal \lla \frac{\dot{\rho}}{\lla \rho \rra_\Dcal} - \rho \frac{\lla \rho \rra_\Dcal^{\tdev}}{\lla \rho \rra_\Dcal^2} \rra_\Dcal = \lla \rho \, \Theta \rra_\Dcal - \lla \rho \rra_\Dcal \lla \Theta \rra_\Dcal ~~.
\eeq
We thus proved that the time variation of the Kullback-Liebler entropy is proportional to the non-commutation terms linking the density of matter and the expansion scalar. One can thus understand that the definition of averaged quantities in cosmology has the first obvious consequence of a loss in information, but secondly has the drawback to introduce a difference in the evolution of averaged quantities with respect to their true non-averaged equivalents.

\subsection{Gauge invariance}
\label{Ch2Sec12}

The average of a quantity naturally depends on the physical hypervolume used to define it. It should not, on the other hand, depend on the chosen coordinates. This fact, called \gls{gauge invariance}, is a general concept which has some subtleties. Here we will solely present the necessary relations to understand the gauge problem that appears when we try to define averaged quantities. More details about the notion of gauge invariance are presented in Appendix \ref{AppASec3} .

Let us consider a spacetime $\Mcal_4$ equipped with a system of coordinates $\{x^\mu \}$ and a Riemannian metric $g_{\mu\nu}$ to describe its geometry. We also consider a scalar function $S(x)$ on these coordinates. Under a \gls{GCT}, where $h$ is a matrix function that we suppose invertible, we have the transformation
\beq
\label{GCTdefin}
x \mapsto \hat{x} = h(x) ~~:~~ S(x) \mapsto \hat{S}(x) ~~~~~ \mbox{such that} ~~~~~ \hat{S}(\hat{x}) = S(x) = S(h^{-1}(\hat{x})) ~~.
\eeq
A \gls{GT} $f$ -- or ``local reparametrization of the fields'', supposed invertible -- on the other hand consists in the evaluation of the field at the same physical point, thus being given by
\beq
\label{GTdefin}
x \mapsto \tilde{x} = f(x) ~~:~~ S(x) \mapsto \tilde{S}(x) = S(f^{-1}(x)) ~~.
\eeq
We can remark that the evaluation of this last equality at $\tilde{x}$ gives $\tilde{S}(\tilde{x}) = S(x)$ which is equivalent to Eq. \rref{GCTdefin} if $f \equiv h$ (and thus $\hat{S} \equiv \tilde{S}$). Under the gauge transformation, the metric is modified in the following way\,:
\beq
g_{\mu\nu}(x) \mapsto \tilde{g}_{\mu\nu}(x) = \left[\parfrac{x^\alpha}{f^\mu} \parfrac{x^\beta}{f^\nu}\right]_{f^{-1}(x)} g_{\alpha\beta}(f^{-1}(x))
\eeq
where the squared brackets contain the product of inverse matrices $(\partial f / \partial x)^{-1}$ in which we finally replace $x$ by $f^{-1}(x)$. The square root of the determinant $g \equiv {\rm det} \, g_{\mu\nu}$ is then transformed as
\beq
\label{GTdetOfg}
\sqrt{-g(x)} \rightarrow \sqrt{-\tilde{g}(x)} = \left| \parfrac{x}{f} \right|_{f^{-1}(x)} \sqrt{-g(f^{-1}(x))} ~~,
\eeq
being thus multiplied by the inverse Jacobian determinant of the gauge transformation $\, x \mapsto f(x)$.

Let us consider now the integration of a scalar quantity $S(x)$ on a spacetime domain $\Dscr \subset \Mcal_4$ spanned by the coordinates $\{ x^\mu \}$. The change of this integration under the gauge transformation $x \mapsto f(x)$, when the domain $\Dscr$ is kept unchanged by this gauge transformation, is given by\,:
\beq
I(S,\Dscr) = \int_{\Dscr(x)} d^4x \sqrt{-g(x)} ~ S(x) ~\longmapsto~ \tilde{I}(\tilde{S},\Dscr) = \int_{\Dscr(x)} d^4x \sqrt{-\tilde{g}(x)} ~ \tilde{S}(x) ~~.
\eeq
We can now proceed to a change of coordinates. Using $\bar{x} = f^{-1}(x)$, we write\,:
\beq
\tilde{I}(\tilde{S},\Dscr) = \int_{\Dscr(f(\bar{x}))} d^4\bar{x} \left|\parfrac{f}{\bar{x}}\right|_{\bar{x}} \sqrt{-\tilde{g}(f(\bar{x}))} ~ \tilde{S}(f(\bar{x})) = \int_{\overline{\Dscr}(\bar{x})} d^4\bar{x} \sqrt{-\tilde{g}(\bar{x})} ~ S(\bar{x}) = I(S,\overline{\Dscr}) ~~,
\eeq
where we used the relations \rref{GTdetOfg} and \rref{GTdefin}, and introduced at the end the new spacetime domain $\overline{\Dscr}$ such that $\overline{\Dscr}(\bar{x}) = \Dscr(x)$. We conclude that the integral $I(S,\Dscr)$ is not gauge invariant if the domain $\Dscr$ is unchanged under the gauge transformation. In other words, the domain of integration must be gauge-dependent and transform in a precise way to keep the integral $I(S,\Dscr)$ gauge invariant.

The transformation of this domain under gauge transformations can easily be found. Indeed, let us assume that the domain $\Dscr$ depends on the coordinates $\{ x^\mu \}$ and let us implement its shape in $\Mcal_4$ by the use of a \gls{window function} $W_{\Dscr}(x)$ in the integral\,:
\beq
\label{DefAvWithWindow}
I(S,\Dscr) = \int_{\Mcal_4} d^4x \sqrt{-g(x)} ~ W_{\Dscr}(x) S(x) ~~.
\eeq
If we assume that this function transforms as
\beq
W_{\Dscr}(x) ~\longmapsto~ \widetilde{W}_{\Dscr}(x) = W_{\Dscr}(f^{-1}(x)) ~~,
\eeq
under the gauge transformation $f$, we then have the new transformation of the integral\,:
\beq
I(S,\Dscr) ~\longmapsto~ \tilde{I}(\tilde{S},\Dscr) = \int_{\Mcal_4} d^4\bar{x} \sqrt{-\tilde{g}(\bar{x})} ~ W_{\Dscr}(\bar{x}) S(\bar{x}) = I(S,\Dscr) ~~,
\eeq
which shows the gauge invariance of the integral as expected. We can understand that if $W_{\Dscr}$ is only made of scalars, such that it is still invariant under $x \mapsto f(x)$, then the gauge invariance is respected (and the GCT invariance is also satisfied). On the other hand, if $W_{\Dscr}$ is not invariant, we say that the domain of integration breaks gauge invariance.

One such example that we can easily visualize is a cylindrical-type domain delimited by two spacelike hypersurfaces (i.e. defined by timelike normal vectors) $A(x) = A_1$ and $A(x) = A_2>A_1$. In properly chosen coordinates, this hypervolume can be restricted by the addition of a timelike hypersurface $B(x)=r_0$ (defined by radial spacelike normal vectors), as we will illustrate in Fig. \ref{FigSpatialAveraging}. $W_{\Dscr}$ is then simply given in terms of Heaviside step functions by\,:
\beq
W_{\Dscr}(x) = \Theta(A(x)-A_1) \Theta(A_2-A(x)) \Theta(r_0-B(x)) ~~,
\eeq
and we have here a domain $\Dscr$ which is gauge invariant if $A$ and $B$ are scalars. Otherwise, the physical domain is not anymore the same after a gauge transformation and breaks gauge invariance of quantities defined on this domain (such as $I(S,\Dscr)$).

\subsection{Gauge invariant generalization of Buchert's spatial average}
\label{Ch2Sec13}

In this section, we wish to present in a more general way the definition of an average on a spatial hypersurface, i.e. the hypersurface orthogonal to a vector $n^\mu$ which is timelike. Indeed, Buchert's definition of the average presented in Sec. \ref{Ch2Sec1} has been written in a given gauge (a synchronous gauge) and we intend to give a definition which does not depend on a gauge choice. We want to derive then the Buchert-Ehlers relation associated to this hypersurface, as it was done in \cite{Gasperini:2009mu}.

\begin{figure}[t]
\centering
\begin{tikzpicture}
\node (label) at (-4cm,0cm){\includegraphics[width=7cm]{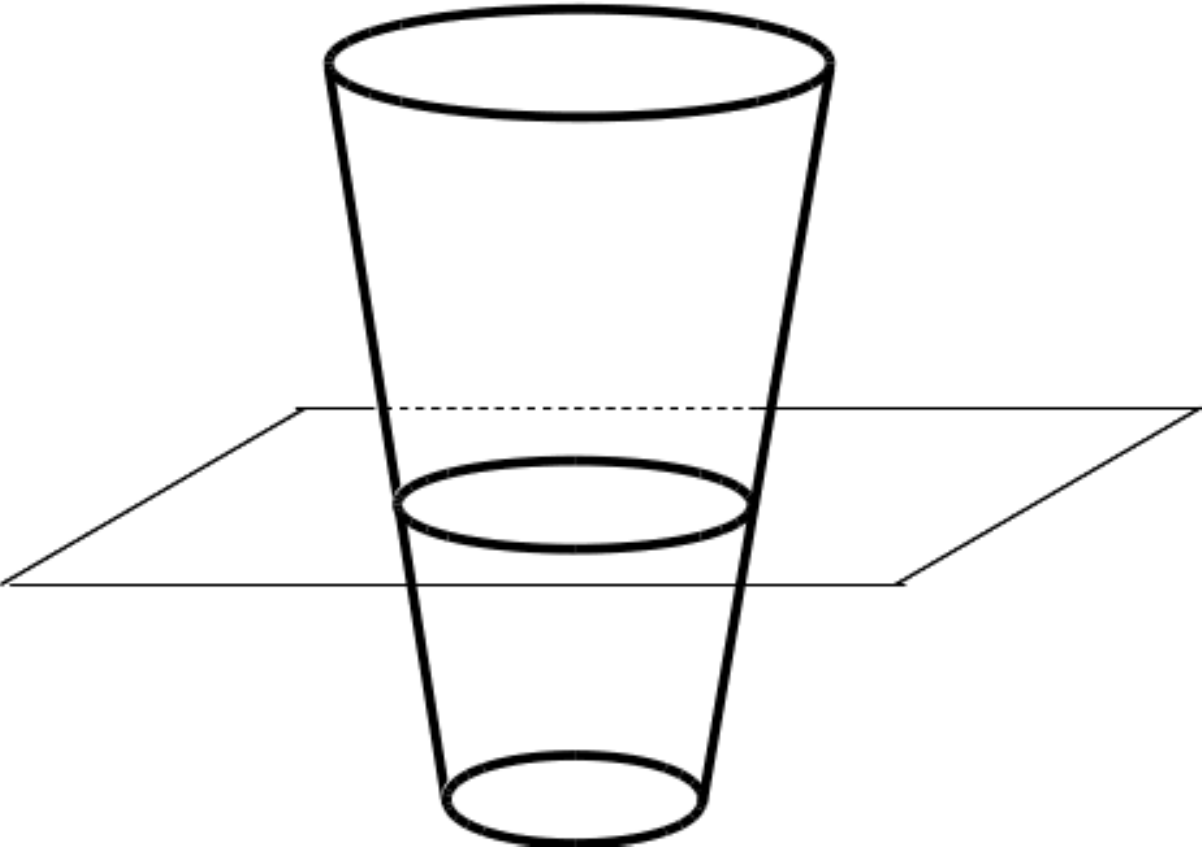}};
      
\node[anchor=west] at (-4.5,1) (description) {\textbf{$r_0$}};

\node[anchor=west] at (-2.5,-0.3) (description) {\textbf{$A_0$}};

\node[anchor=east] at (-2.9,1) (text) {};
\node[anchor=west] at (-1.5,1) (description) {\textbf{$\partial_\mu B$}};
\draw (text) edge[out=0,in=180,->,color=black,thick] (description);

\end{tikzpicture}
\caption[Spatial averaging and introduction of a spatial boundary]{A graphic illustration of the introduction of a new parametrized hypersurface $B(x) = r_0$ to create a boundary inside the constant-$A$ hypersurface. This newly introduced hypersurface has for spacelike gradient $\partial_\mu B$.}
\label{FigSpatialAveraging}
\end{figure}
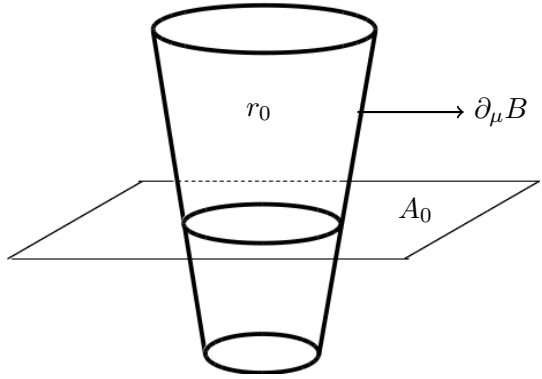

Let us define a spatial hypersurface $\Sigma(A)$ whose points share the same value of a scalar field $A(x)$. This hypersurface can be defined by its (past-directed) normal vector\,:
\beq
n_\mu \equiv - \frac{\partial_\mu A}{\sqrt{-\partial_\mu A \partial^\mu A}}
\eeq
which is a timelike vector as $n_\mu n^\mu = -1$. We choose a finite spatial domain $\Dscr$ inside this hypersurface by adding up a scalar field $B(x)$ which can be defined by its spacelike vector $\propto \partial_\mu B$, see Fig. \ref{FigSpatialAveraging}. We can then write the integration of a scalar $S$, as seen in Eq. \rref{DefAvWithWindow}, by
\beq
\label{IntSpatialAv}
I(S;A_0,r_0) = \int_{\Dscr(x)} d^4x \sqrt{-g(x)} ~ S(x) \equiv \int_{\Mcal_4} d^4x \sqrt{-g(x)} ~ S(x) W_\Dscr(x) ~~,
\eeq
where the window function is
\beq
\label{WindowSpatialAv}
W_{\Dscr}(x) = n^\mu \nabla_\mu \Theta(A(x) - A_0) \Theta(r_0 - B(x)) = \sqrt{-\partial_\mu A \partial^\mu A} ~ \delta_{\rm D}(A(x) - A_0) \Theta(r_0 - B(x)) ~~,
\eeq
and we have used that $n^\mu \nabla_\mu \Theta(A(x) - A_0) = n^\mu \partial_\mu A ~ \delta_{\rm D}(A(x) - A_0)$. As previously shown, this integral is gauge invariant as long as $A$ and $B$ are gauge invariant scalars. We can thus define the gauge invariant average of a scalar $S(x)$ by the volume-weighted ratio\,:
\beq
\lla S \rra_{A_0,r_0} = \frac{I(S;A_0,r_0)}{I(1;A_0,r_0)} ~~.
\eeq

Let us now compute the derivative of this average with respect to the time parameter $A_0$ and obtain a generalization of the so-called Buchert-Ehlers commutation rules \cite{Buchert:1995fz}.

Taking the derivative of $I(S;A_0,r_0)$ with respect to the timelike parameter $A_0$, and using the following relation
\beq
\partial_{A_0} \delta(A(x) - A_0) = - \delta_{\rm D}'(A(x)-A_0) ~~,
\eeq
we get
\beq
\frac{\partial I(S;A_0,r_0)}{\partial A_0} = - \int_{\Mcal_4} d^4x \sqrt{-g(x)} ~ \sqrt{-\partial_\mu A \partial^\mu A} ~ \delta_{\rm D}'(A(x)-A_0) \Theta(r_0 - B(x)) ~ S(x) ~~.
\eeq
We can then use an adapted system of coordinates such as ADM coordinates as it is done in \cite{Gasperini:2009mu}. Nevertheless, a more direct derivation is found as follows. Let us consider the identity\,:
\beq
\label{RelDeltaADeltaPrimeA}
n^\mu \nabla_\mu \delta_{\rm D}(A(x) - A_0) = \sqrt{-\partial_\mu A \partial^\mu A} ~ \delta_{\rm D}'(A(x) - A_0) ~~.
\eeq
From this we get\,:
\bea
\frac{\partial I(S;A_0,r_0)}{\partial A_0} &=&
+ \int_{\Mcal_4} d^4x \sqrt{-g(x)} ~ \delta_{\rm D}(A(x)-A_0) \nabla_\mu \left[ n^\mu ~ \Theta(r_0 - B(x)) ~ S(x) \right] \nonumber \\
&=& \int_{\Mcal_4} d^4x \sqrt{-g(x)} ~ \delta_{\rm D}(A(x)-A_0) \Theta(r_0 - B(x)) ~ \frac{n^\mu \nabla_\mu S(x)}{\sqrt{-\partial_\mu A \partial^\mu A}} \sqrt{-\partial_\mu A \partial^\mu A} \nonumber \\
& & + \int_{\Mcal_4} d^4x \sqrt{-g(x)} ~ \delta_{\rm D}(A(x)-A_0) \Theta(r_0 - B(x)) ~ \frac{\Theta(x) S(x)}{\sqrt{-\partial_\mu A \partial^\mu A}} \sqrt{-\partial_\mu A \partial^\mu A} \nonumber \\
& & - \int_{\Mcal_4} d^4x \sqrt{-g(x)} ~ \delta_{\rm D}(A(x)-A_0) ~ n^\mu \nabla_\mu \Theta(r_0 - B(x)) ~ S(x) ~~,
\eea
where in the first equality we used Eq. \rref{RelDeltaADeltaPrimeA} followed by an integration by parts, with the indentity $\int_{\Mcal_4} d^4x \sqrt{-g(x)} ~ \nabla_\mu(\ldots) = 0$. This is true because the contribution $(\ldots)$ is zero at infinity, either from the fact that the domain of integration is rendered finite by the distributions, or simply because $S$ can be taken to be zero at infinity.
The last integral contains the following function\,:
\beq
n^\mu \nabla_\mu \Theta(r_0 - B(x)) = - n^\mu \nabla_\mu \delta_{\rm D}(r_0 - B(x)) = -2 n^\mu \partial_\mu B \, \Theta(r_0 - B(x)) \delta_{\rm D}(r_0 - B(x)) ~~,
\eeq
where the factor two in the last equality can be understood from the definition of the delta-functions as the zero-width limit of a Gaussian distribution centered on zero.
We understand also that this last integral is related to an integration on the boundary $\partial \Dscr$ of the spatial domain $\Dscr$.

Finally, using the definition of the integral in Eqs. \rref{IntSpatialAv} and \rref{WindowSpatialAv}, we obtain\,:
\bea
\frac{\partial I(S;A_0,r_0)}{\partial A_0} &=& I \left(\frac{\partial_\mu A \partial^\mu S(x)}{\partial_\mu A \partial^\mu A} ~; A_0,r_0 \right) + I \left( \frac{\Theta(x) S(x)}{\sqrt{-\partial_\mu A \partial^\mu A}} ~; A_0,r_0 \right) \nonumber \\
& & ~~~ - 2 I \left( \frac{\partial_\mu A \partial^\mu B}{\partial_\mu A \partial^\mu A} ~ \delta_{\rm D}(r_0 - B(x)) S(x) ~; A_0,r_0 \right) ~~,
\eea
where we have done trivial changes (such as $\partial^\mu A \nabla_\mu S \equiv \nabla^\mu A \nabla_\mu S = \nabla_\mu A \nabla^\mu S = \partial_\mu A \partial^\mu S$ as both $A$ and $S$ are scalars).
Dividing now by the volume $I(1;A_0,r_0)$ and taking the limit $r_0 \rightarrow \infty$, i.e. integrating over the whole spatial hypersurface fixed by $A(x) = A_0$, we have that $\lla \ldots \rra_{A_0,r_0} \rightarrow \lla \ldots \rra_{A_0}$ and we obtain the generalization of the Buchert-Ehlers \gls{commutation rule}\,:
\beq
\label{SpatialBuchertEhlers}
\frac{\partial \lla S \rra_{A_0}}{\partial A_0} = \lla \frac{\partial_\mu A \partial^\mu S(x)}{\partial_\mu A \partial^\mu A} \rra_{A_0} + \lla \frac{\Theta(x) S(x)}{\sqrt{-\partial_\mu A \partial^\mu A}} \rra_{A_0} - \lla \frac{\Theta(x)}{\sqrt{-\partial_\mu A \partial^\mu A}} \rra_{A_0} \lla S(x) \rra_{A_0} ~~.
\eeq

We understand the direct interpretation of this relation by going in the \gls{ADM} coordinates. The spacetime $\Mcal_4$ in ADM is foliated by hypersurfaces of constant time $t$ and the line element is given by\,:
\beq
\label{ADMmetricCh3}
ds_{ADM}^2 = - N^2 dt^2 + g_{ij} (dx^i + N^i dt) (dx^j + N^j dt) ~~,
\eeq
and the normal vectors $n^\mu$ and $n_\mu$ (see Sec. \ref{AppASec11}) are written as
\beq
n_\mu = N (-1,0,0,0) ~~,~~ n^{\mu} = \frac{1}{N} (1, - N^i) ~~.
\eeq
One can then show that the volume \gls{expansion scalar} $\Theta(x)$ in that case is
\beq
\label{ExpScalarADM}
\Theta(x) = \frac{1}{N} \partial_t \ln\left(\sqrt{\det(g_{ij})}\right) ~~,
\eeq
where $g_{ij}$ is the spatial part of the metric $g^{ADM}_{\mu\nu}$ encoding the geometry inside the constant-$t$ hypersurface.
In those coordinates $A$ is homogeneous on $t={\rm cst}$ hypersurfaces. So we can take $A_0 \equiv t$ and we find that the Buchert-Ehlers commutation rule of Eq. \rref{NonComRel} reduces to the simple relation\,:
\beq
\label{SpatialBuchertEhlersInADM}
\frac{\partial \lla S \rra_{A_0}}{\partial A_0} = \lla \frac{\partial S}{\partial A_0} \rra_{A_0} + \lla \frac{\Theta N}{\partial_t A} S \rra_{A_0} - \lla \frac{\Theta N}{\partial_t A} \rra_{A_0} \lla S \rra_{A_0} ~~,
\eeq
where $S = S(A_0,\xbf)$ and we have used $\partial_\mu A \partial^\mu A \equiv g_{ADM}^{\mu\nu} \partial_\mu A \partial_\nu A = g_{ADM}^{tt} (\partial_t A)^2 = - N^{-2} (\partial_t A)^2$. We used here that in ADM coordinates, where $A$ is homogeneous, we can always choose a vanishing shift vector (and so $g^{ADM}_{tt} = - N^2$ and $g^{ADM}_{ti} = 0$) corresponding to a partial gauge fixing.

The interpretation of Eq. \rref{SpatialBuchertEhlersInADM} is similar to the commutation rule derived in Eq. \rref{NonComRel}. Indeed, we see that the time derivative of the average of $S$ is equal to the average of the time derivative of $S$ (first term on the RHS) followed by the addition of two terms which involve the volume expansion $\Theta(x)$ and the scalar $S$ in this schematic form\,: $\lla \Theta S \rra - \lla \Theta \rra \lla S \rra$. If $A(x)$ is taken to be simply $t$ and if we fix the gauge to be synchronous, i.e. $N = 1$, we get exactly the relation presented in Eq. \rref{NonComRel}.

\subsection{Generalization of Friedmann equations in a gauge invariant form}
\label{Ch2Sec14}

Let $u^{\mu}$ be the fluid velocity field and $n^{\mu}$ the field orthonormal to the hypersurface of averaging. This vector $n^{\mu}$ can be seen as the peculiar velocity of the observer if this latter one is not comoving with the fluid (as assumed in Sec. \ref{Ch2Sec11}) but is in general different from it. We define a tilt angle between these two fields by
\begin{equation}
u_{\mu}n^{\mu} = - [1 + \sinh^2(\alpha_T)]^{1/2}= - \cosh(\alpha_T) ~~,
\end{equation}
where the sign is chosen so that we have the normalizations $u_{\mu}u^{\mu} \equiv -1$ and $n_{\mu}n^{\mu} \equiv -1$ (i.e. when $\alpha_T = 0$, we want a timelike vector).
In this section the fluid of matter is assumed to be a perfect fluid of local density $\rho$ and pressure $p$\,:
\begin{equation}
\label{Tmunu}
T_{\mu\nu} = (\rho + p)u_{\mu}u_{\nu} + p \, g_{\mu\nu} ~~.
\end{equation}

From \cite{Gasperini:2009mu} we have the following contractions (\gls{ADM energy density} and \gls{ADM pressure})\,:
\beq
\label{DefinEpsPi}
\epsilon \equiv T_{\mu\nu}n^{\mu}n^{\nu} = \rho + (\rho + p) \sinh^2(\alpha_T) ~~~~~,~~~~~ \pi \equiv \frac{1}{3} T_{\mu\nu}h^{\mu\nu} = p + \frac{1}{3}(\rho + p) \sinh^2(\alpha_T) ~~.
\eeq
Here we have used the projection tensor $h_{\mu\nu}$ orthogonal to the vector $n^\mu$, with the properties\,:
\beq
h_{\mu\nu} = g_{\mu\nu} + n_{\mu} n_{\nu} ~~,~~ h_{\mu\rho} \, h^\rho_\nu = h_{\mu\nu} ~~,~~ h_{\mu\nu} n^\nu = 0 ~~.
\eeq
This system can be inverted and has a unique solution\,:
\begin{equation}
\begin{split}
&\rho = \left[ 1 + \frac{4}{3} \sinh^2(\alpha_T) \right]^{-1} \left[ \left(1 + \frac{1}{3}\sinh^2(\alpha_T)\right) \epsilon - \sinh^2(\alpha_T) \pi \right] ~~, \\
&p = \left[ 1 + \frac{4}{3} \sinh^2(\alpha_T) \right]^{-1} \left[ -\frac{1}{3}\sinh^2(\alpha_T) \epsilon + (1 + \sinh^2(\alpha_T)) \pi \right] ~~,
\end{split}
\end{equation}
that we can insert in the expression of $T_{\mu\nu}$ in Eq. \rref{Tmunu}. We can rewrite it by making use of
\begin{equation}
\label{RelUN}
u_{\mu} h^{\mu\nu} = u^{\nu} - \cosh(\alpha_T) \, n^{\nu} ~~,
\end{equation}
to get an expression depending on $\epsilon$ and $\pi$\,:
\begin{equation}
T_{\mu\nu} = \epsilon \, n_{\mu}n_{\nu} + \pi \, h_{\mu\nu} + \frac{(\epsilon + \pi)}{\left[ 1 + \frac{4}{3}\sinh^2(\alpha_T) \right]} \left[u_{\mu}u_{\nu} - \cosh^2(\alpha_T)n_{\mu}n_{\nu} - \frac{1}{3} \sinh^2(\alpha_T) h_{\mu\nu}\right] ~~.
\end{equation}
Here the third term on the RHS gives zero when contracted with $n^{\mu}n^{\nu}$ or $h^{\mu\nu}$.

We also have the generalized Friedmann's equations for an averaged perfect fluid (i.e. in the general case where $n_{\mu} \neq u_{\mu}$, see \cite{Gasperini:2009mu})\,:
\begin{equation}
\label{F1_original}
\begin{split}
\left( \frac{1}{\tila} \parfrac{\tila}{A_0} \right)^2 = ~ & \frac{8 \pi G}{3} \lla \frac{\epsilon}{[-(\partial A)^2]} \rrao - \frac{1}{6} \lla \frac{\mathcal{R}_S}{[-(\partial A)^2]} \rrao \\
& - \frac{1}{9} \left[ \lla \frac{\Theta^2}{[-(\partial A)^2]} \rrao - \lla \frac{\Theta}{[-(\partial A)^2]^{1/2}} \rrao^2 \right] + \frac{1}{3} \lla \frac{\sigma^2}{[-(\partial A)^2]} \rrao ~~,
\end{split}
\end{equation}
and
\begin{equation}
\label{F2_original}
\begin{split}
- \frac{1}{\tila} \frac{\partial^2 \tila}{\partial {A_0}^2} = ~ & \frac{4 \pi G}{3} \lla \frac{\epsilon + 3 \pi}{[-(\partial A)^2]} \rrao - \frac{1}{3} \lla \frac{\nabla^{\nu}(n^{\mu}\nabla_{\mu}n_{\nu})}{[-(\partial A)^2]} \rrao + \frac{1}{6} \lla \frac{\partial_{\mu}A\partial^{\mu}[(\partial A)^2] \Theta}{[-(\partial A)^2]^{5/2}} \rrao \\
& - \frac{2}{9} \left[ \lla \frac{\Theta^2}{[-(\partial A)^2]} \rrao - \lla \frac{\Theta}{[-(\partial A)^2]^{1/2}} \rrao^2 \right] + \frac{2}{3} \lla \frac{\sigma^2}{[-(\partial A)^2]} \rrao ~~,
\end{split}
\end{equation}
where $\tilde{a}$ is an effective scale factor defined by
\begin{equation}
\frac{1}{\tila} \parfrac{\tila}{A_0} \equiv \frac{1}{3 I(1,A_0)} \parfrac{I(1,A_0)}{A_0} = \frac{1}{3} \lla \frac{\Theta}{[-(\partial A)^2]^{1/2}} \rrao ~~,
\end{equation}
and where in the last equality we took the ADM coordinates and assumed that $B \subset \Sigma_{A_0}$.

In fact one can give the generalization of other quantities that have been derived in the literature. For example, we can define an effective density and an effective pressure to consider the backreaction terms as an effective scalar field. We can also generalize the integrability condition to have a gauge invariant definition of it. Nevertheless, because these expressions are really complicated, and their use very limited, we will not present them in this thesis\footnote{
Let us finally notice that the commutation rule for the energy density, necessary to obtain the integrability condition, is derived in \cite{Gasperini:2009mu}. This derivation makes use of the conservation of the stress-energy tensor for a perfect fluid along the geodesics of $n_{\mu}$, i.e.\,: $n_{\nu}\nabla_{\mu}T^{\mu\nu} = 0$. Developping $T^{\mu\nu}$ from Eq. \rref{Tmunu} in this equation, we get\,:
\beq
u^{\mu}\partial_{\mu}[(\rho + p)u^{\rho}n_{\rho}] + n^{\mu}\partial_{\mu}p + (\rho + p)(\nabla_{\mu}u^{\mu})(u^{\rho}n_{\rho}) + (\rho + p)\left[ n^{\rho}u^{\mu}\nabla_{\mu}u_{\rho} - u^{\mu}\nabla_{\mu}(u^{\rho}n_{\rho}) \right] = 0 ~~, \nonumber
\eeq
where this last term can be written $-(\rho+p)u^{\mu}u^{\rho}\nabla_{\mu}n_{\rho}$.
From the definition of $\Theta^{\mu\nu}$ we have:
\bea
- \Theta^{\mu\nu} u_{\mu} u_{\nu} &=& - \Theta_{\mu\nu} u^{\mu} u^{\nu} = - (\delta_{\mu}^{~\alpha} + n_{\mu}n^{\alpha})(\delta_{\nu}^{~\beta} + n_{\nu}n^{\beta}) (\nabla_{\alpha}n_{\beta}) u^{\mu}u^{\nu} \nonumber \\
&=& - u^{\mu}u^{\nu}(\nabla_{\mu}n_{\nu}) - (u^{\rho}n_{\rho})n^{\mu}u^{\nu}\nabla_{\mu}n_{\nu} ~~, \nonumber
\eea
where we made use of the equality $n^{\mu}n^{\nu}\nabla_{\mu}n_{\nu} = \frac12 n^{\mu}\nabla_{\mu}\left( n^{\nu} n_{\nu} \right) = 0$ (from $n^{\nu} n_{\nu} = -1$). As a consequence, the final equation we obtain is
\beq
\label{eq2dot25ET}
u^{\mu}\partial_{\mu}[(\rho + p)u^{\rho}n_{\rho}] + n^{\mu}\partial_{\mu} p + (\rho + p) [(\nabla_{\mu}u^{\mu})u^{\rho}n_{\rho} - \Theta^{\mu\nu}u_{\mu}u_{\nu} + (u^{\rho}n_{\rho})n^{\mu}u^{\nu}\nabla_{\mu}n_{\nu}] = 0 ~~. \nonumber
\eeq
This last term on the LHS was missing in Eq. (2.25) of \cite{Gasperini:2009mu}. This footnote thus correct this typo.
}.

\section{Light-cone averaging}
\label{Ch2Sec2}

\subsection{Choice of a right scalar}
\label{Ch2Sec21}

Let us first emphasize again the approach given in \cite{Gasperini:2009wp,Gasperini:2009mu} to gauge invariant averaging on a 3-dimensional spacelike hypersurface $\Sg(A)$, embedded in our 4-dimensional spacetime $\Mq$. 
Assuming the hypersurface (or a spacelike foliation) to be defined by an equation involving a scalar field with timelike gradients $A(x)$\,:
\beq
A(x) - A_0 = 0 ~~,
\eeq
the gauge (and hypersurface-parametrization) invariant definition of  the integral of an arbitrary scalar $S(x)$ and of its average on such hypersurface is given by\footnote{In \cite{Gasperini:2009mu} the prescription introduced in \cite{Gasperini:2009wp} is used to give a covariant and gauge invariant generalization of the effective equations presented in \cite{2000GReGr..32..105B,2001GReGr..33.1381B}. Such a generalization
has been recently used  to deal with  the backreaction of quantum fluctuations in an inflationary model \cite{Finelli:2011cw}.}\,:
\beq
\langle S \rangle_{A_0} = \frac{I(S; A_0)}{I(1; A_0)} ~~\mbox{ with }~~ I(S;A_0) =  \int_{\Mq} d^4 x \sqrt{-g(x)}~\delta_{\rm D}(A(x)-A_0)~\sqrt{-\partial_\mu A \partial^\mu A}~ S(x) ~~.
\eeq
Here the spatial hypersurface has no boundary, it goes to infinity. However, as shown in \cite{Gasperini:2009wp,Gasperini:2009mu}, a possible spatial boundary can be added through the following extension of the previous integral\,:
\beq
I(S;A_0; r_0) =  \int_{\Mq} d^4 x \sqrt{-g(x)} \, \delta_{\rm D}(A(x)-A_0) \Theta(r_0- B(x)) \sqrt{-\partial_\mu A \partial^\mu A} ~S(x) ~~,
\label{23}
\eeq
and similarly for the corresponding average ($\Theta$ is the Heaviside step function, and $B$ is a positive function of the coordinates, with spacelike gradient). As already discussed in Sec. \ref{Ch2Sec13} (see also \cite{Gasperini:2009wp,Marozzi:2010qz}), and as illustrated in Fig. \ref{FigSpatialAveraging}, this is still a gauge invariant expression if $B(x)$ transforms as a scalar, while it gives violations of gauge invariance if $B$ is not a scalar and keeps the same form in different coordinate systems. Even in that case, however, gauge invariance violations go to zero  when we choose $r_0$ in such a way that the size of the spatial region goes to infinity \cite{Gasperini:2009wp,Marozzi:2010qz}.

The above procedure unfortunately fails if $A(x) = A_0$ defines a null (light-like) hypersurface, since in that case $\partial_\mu A \partial^\mu A =0$. In order to circumvent this problem let us start with a spacetime integral where the four-dimensional integration region is
bounded by two hypersurfaces, one  spacelike and the other one null (corresponding e.g. to the past light-cone of some observer). Let us choose, in particular, the region inside the past light-cone of the observer bounded in the past by the hypersurface $A(x)=A_0$. Clearly a gauge invariant definition of the integral of a scalar $S(x)$ over such a hypervolume can be written (in a useful notation generalizing the one used above) as 
\beq
\label{Intvol}
I(S;-;A_0,V_0)= \int_{\Mq} d^4x \sqrt{-g}~ \Theta(V_0-V) \Theta(A-A_0) ~ S(x) ~~,
\eeq
where $V(x)$ is a (generalized advanced-time) null scalar satisfying $\partial_{\mu} V \partial^{\mu} V=0$ and $V_0$ specifies the past light cone of a given observer. The LHS symbol ``$-$'' denotes the absence of delta-like window functions. We should remark the analogy between the scalars $V$ and $B$. Indeed, this gauge invariant definition corresponds to the replacement of the timelike hypersurface $B(x) = r_0$ by a null-like hypersurface $V(x) = V_0$.

\subsection{Definition of the averages}
\label{Ch2Sec22}

\begin{figure}[ht!]
\centering
\includegraphics[width=12cm]{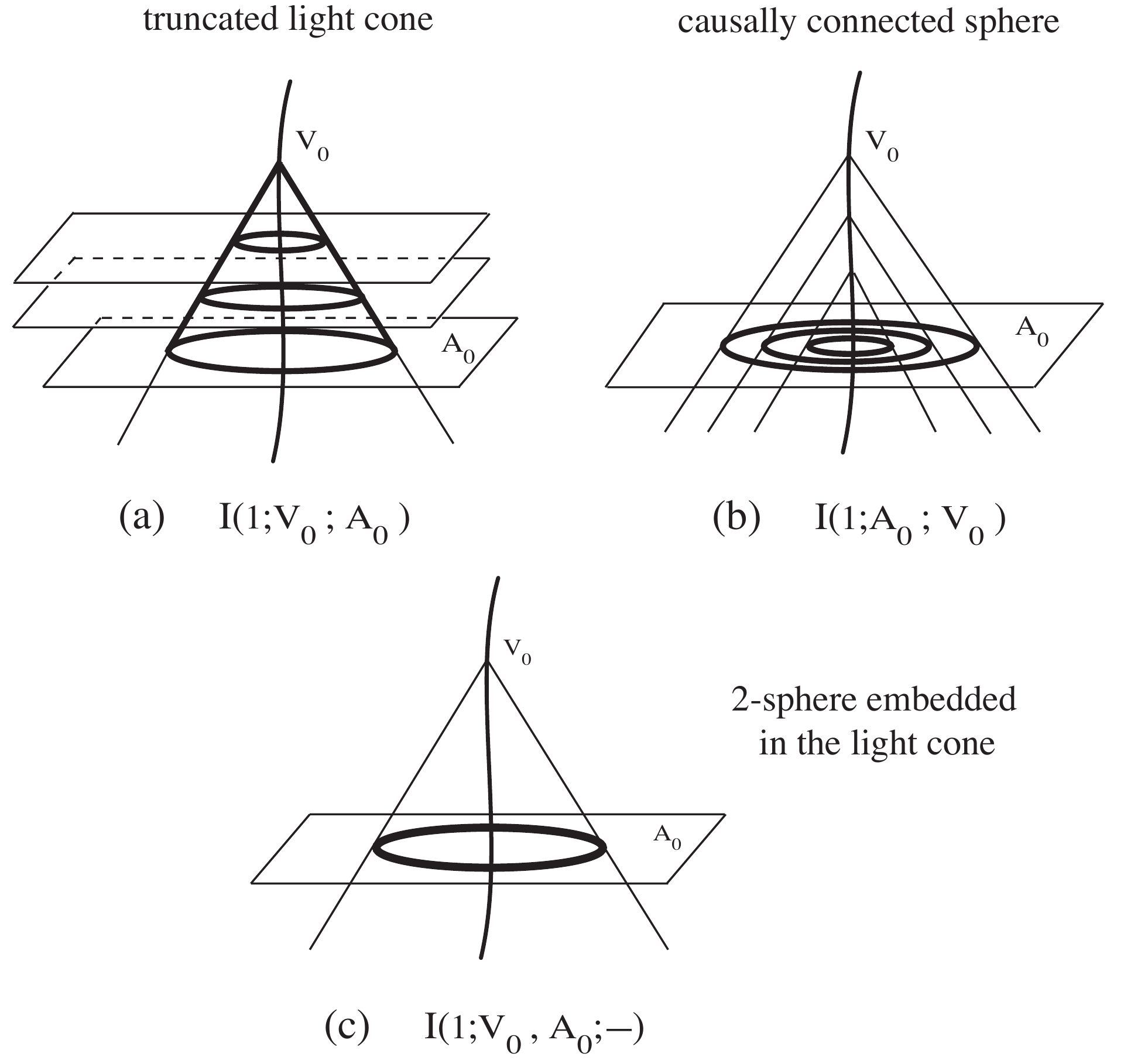}
\caption[Three averaging prescriptions\,: light-cone, spatial hypersurface and embedded 2-sphere]{Three different light-cone averaging prescriptions.   
$(a)$: the average of Eq. (\ref{null3d}) on the light-cone itself starting from a given hypersurface in the past; $(b)$: the average of Eq. (\ref{null3dInv}) on the section of the hypersurface $A(x)=A_0$ which is causally connected with us; $(c)$: the average of Eq. (\ref{null2d}) on a 2-sphere embedded in the light-cone.}
\label{Fig3Averages}
\end{figure}

Starting with this hypervolume integral we can construct gauge invariant hypersurface and surface integrals by applying  to it appropriate differential operators -- or, equivalently,  by applying Gauss's theorem to the volume integral of a covariant divergence. An example of the latter, if we are interested by the variations of the volume averages along the flow lines normal to the reference hypersurface $\Sg(A)$, is obtained by replacing the scalar $S$ with the divergence of the unit normal to $\Sg$,
\beq
n_\mu =- {\pa_\mu A\over \sqrt{-\partial_\nu A \partial^\nu A} } ~~~, ~~~~ n_\mu n^\mu=-1 ~~~,
\label{vel}
\eeq
and leads to the identity\,:
\small
\bea
\int_{\Mq} d^4x \sqrt{-g}~ \Theta(V_0-V)  \Theta(A-A_0)  \nabla^{\mu}  n_{\mu}
&=& -\int_{\Mq} d^4x \sqrt{-g}\, \Theta(V_0-V) \delta_{\rm D}(A-A_0)  
\sqrt{-\partial_\mu A \partial^\mu A} \nonumber \\
&+& \int_{\Mq} d^4x \sqrt{-g} \,\delta_{\rm D}(V_0-V)  \Theta(A-A_0) 
\frac{-\partial_\mu V \partial^\mu A}{\sqrt{-\partial_\nu A \partial^\nu A}} ~~.~~~~~
\label{Gauss1}
\eea
\normalsize
Hence, if we start from Eq. (\ref{Intvol}), and we consider the variation of the average integral by shifting the light-cone 
$V=V_0$ along the flow lines defined by $n_\mu$,  we are led to define the hypersurface integral (with positive measure)\,:
\beq
I(1;V_0;A_0)=\int_{\Mq} d^4x \sqrt{-g} \,\delta_{\rm D}(V_0-V)  \Theta( A-A_0 )  
\frac{|\partial_\mu V \partial^\mu A|}{\sqrt{-\partial_\nu A \partial^\nu A}} ~~.
\label{null3d}
\eeq
Similarly, if we consider the variation of the average integral by shifting the hypersurface $A=A_0$ (along the same 
flow lines defined by $n_\mu$), we are led to another hypersurface integral\,:
\beq
I(1;A_0;V_0)=\int_{\Mq} d^4x \sqrt{-g}\, \Theta(V_0-V)  \delta_{\rm D}( A-A_0 )  
\sqrt{-\partial_\mu A \partial^\mu A} ~~.
\label{null3dInv}
\eeq
In the first case, Eq. (\ref{null3d}), the integration region is on the light-cone itself, and it is spanned by the variation of  $\Sg(A_0)$ along its normal, at fixed light-cone $V_0$ (see Fig. \ref{Fig3Averages}, $(a)$). In the second case of Eq. (\ref{null3dInv}) -- which gives exactly the same integral as in Eq. (\ref{23}) with $V$ replacing $B$ --  the hypersurface $\Sg(A_0)$ is kept fixed, and the integration region describes the causally connected section of $\Sg$ spanned by the variation  of the light-cone hypersurface (see Fig. \ref{Fig3Averages}, $(b)$). 

Further differentiation also leads to the following invariant surface integral
\beq
I(1;V_0,A_0;-)=\int_{\Mq} d^4x \sqrt{-g}~ \delta_{\rm D}(V_0-V) \delta_{\rm D}(A-A_0)  
|\partial_\mu V \partial^\mu A| ~~,
\label{null2d}
\eeq  
with a compact, 2-dimensional integration region defined by the intersection of $\Sg(A_0)$ with the light-cone $V_0$ (Fig. \ref{Fig3Averages}, $(c)$). This integral, as well as the integrals of Eqs. (\ref{null3d}), (\ref{null3dInv}), is not only covariant and gauge invariant but also invariant under separate reparametrizations of the scalar fields $A\rightarrow \tilde{A}(A)$ and $V\rightarrow \tilde{V}(V)$. Eq. (\ref{null2d}), in addition, is a particular case of an invariant integration over an arbitrary codimension-2 hypersurface defined by the conditions $A^{(n)}(x) =  0$, $ n = 1,2$. In general, and in $D$ spacetime dimensions, such an integral can be written as 
\beq
\int_{\cal{M}_D} d^Dx \sqrt{-g} \prod_n \delta_{\rm D}( A^{(n)}(x) ) \sqrt{|\det \bar{g}^{pq}|} ~~\mbox{ with }~~~ \bar{g}^{pq} \equiv  \partial_{\mu} A^{(p)} \partial_{\nu} A^{(q)} g^{\mu \nu} ~~,
\label{v2}
\eeq
(as can be shown by considering the induced metric on the $(D-2)$-hypersurface), and is invariant under the more general  reparametrizations 
$A^{(1)}\rightarrow \tilde{A}^{(1)}\left(A^{(1)},A^{(2)}\right)$ and $A^{(2)}\rightarrow \tilde{A}^{(2)}\left(A^{(1)},A^{(2)}\right)$. It can be easily checked that Eq. (\ref{v2}) reduces to Eq. (\ref{null2d}) if $D=4$ and if $A^{(1)} = A - A_0$ and $A^{(2)} = V- V_0$ are scalar functions with timelike and null gradient, respectively.

In order to make contact with Eqs. (\ref{null3d}), (\ref{null3dInv}),  it may be useful to remark that the integral of Eq. (\ref{null2d}) can also be   obtained starting from the hypervolume integral of Eq. (\ref{Intvol}) by considering the variation of the volume average along the flow lines normal to 
$\Sigma(A)$ for both $\Theta(A)$ and $\Theta(V)$, namely by using the following window function\,:
\bea
&& \!\!\!\!\!\!\!\!\!\! \!\!\!\!\!\!\!\!\!\! 
 -n^{\mu} \nabla_\mu \Theta(A(x)-A_0)n^{\mu} \nabla_\mu \Theta(V_0-V(x))
\nonumber \\
&&   ~~~~~~~~~~~~~~ =
\sqrt{-\partial_\mu A \partial^\mu A} \, \delta_{\rm D}(A(x)-A_0) \frac{-\partial_\mu V 
\partial^\mu A}{\sqrt{-\partial_\mu A \partial^\mu A}} \delta_{\rm D}(V_0-V(x)) ~~. 
\eea

We note, finally, that averages of a scalar $S$ over different (hyper)surfaces are trivially defined, with self explanatory notation, by\,:
\bea
\langle S \rangle_{V_0,A_0} &=& \frac{I(S;V_0,A_0;-)}{I(1;V_0,A_0;-)} ~~,
\label{211} \\  
\langle S \rangle_{V_0}^{A_0} &=& \frac{I(S;V_0;A_0)}{I(1;V_0;A_0)} ~~,
\label{212}\\ 
\langle S \rangle_{A_0}^{V_0} &=& \frac{I(S;A_0;V_0)}{I(1;A_0;V_0)} ~~.
\label{avnull2d}
\label{213}
\eea

\subsection{Compact notations}
\label{Ch2Sec23}

We can give a summary of the normalizations to be used in these three cases, using the general expression for the integral\,:
\beq
\label{GenExprForIntI}
I[S,V_0,A_0]=\int_{\Mq} d^4x \sqrt{-g} ~ \Drm(V_0-V) \Drm(A-A_0) ~ {\mathcal N}(V,A,\partial_\mu) ~ S(x) ~~,
\eeq
where ${\mathcal N}(V,A,\partial_\mu)$ is a normalization, $\Drm(X) = \delta_{\rm D}(X)$ or $\Theta(X)$, and we can even add a fourth case of application of this integral from Eq. \rref{Intvol}, corresponding to the integration over the whole past causal volume and denoting its corresponding average as
\beq
\lla S \rra^{A_0,V_0} = \frac{I(S;-;A_0,V_0)}{I(1;-;A_0,V_0)} ~~.
\eeq
The normalization is choosen in function of the distributions being used, as shown in Table \ref{TabDiffNorm}.

\renewcommand{\arraystretch}{1.5}

\begin{table}[ht!]
\caption[Different normalizations to be used to recover the four average definitions]{\label{TabDiffNorm} Different normalizations to be used in Eq. \rref{GenExprForIntI} to recover the four average definitions.}
\vskip 0.3 cm
\centering
\begin{tabular}{|c|c|c|c|c|}
\hline
$\mbox{Average}_{~\delta_{\rm D}}^{~\Theta}$ & $\lla S \rra_{V_0}^{A_0}$ & $\lla S \rra_{A_0}^{V_0}$ & $\lla S \rra_{V_0,A_0}$ & $\lla S \rra^{V_0,A_0}$ \\
\hline
Integral $I[S,V_0,A_0]$ & $I(S;V_0;A_0)$ & $I(S;A_0;V_0)$ & $I(S;V_0,A_0;-)$ & $I(S;-;A_0,V_0)$ \\
\hline\hline
${\mathcal N}(V,A,\partial_\mu)$ & $\frac{|\partial_\mu V \partial^\mu A|}{\sqrt{-\partial_\nu A \partial^\nu A}}$ & $\sqrt{-\partial_\mu A \partial^\mu A}$ & $|\partial_\mu V \partial^\mu A|$ & $1$ \\
\hline
\end{tabular}
\end{table}

\renewcommand{\arraystretch}{1}

It is interesting also to notice that we can get rid of this normalization $\Ncal$ by defining new distributions that we could call ``gauge invariant distributions''. Indeed, Eq. \rref{GenExprForIntI} is equivalent to
\beq
\label{GenExprForIntInew}
I[S,V_0,A_0]=\int_{\Mq} d^4x \sqrt{-g} ~ \widehat{\Drm}(V_0-V) \widehat{\Drm}(A-A_0) ~ S(x) ~~,
\eeq
where we take
\beq
\widehat{\Drm}(X) ~=~~ \Theta(X) ~~~~~\mbox{ or }~~~~~ \frac{|\partial_\mu X \partial^\mu A|}{\sqrt{-\partial_\nu A \partial^\nu A}} \delta_{\rm D}(X)
\eeq
depending on the spacetime domain we want to integrate over.
We see that this definition is enough to recover the last four cases of integration\,: on spatial hypersurfaces, the past light cone, the embedded 2-sphere and the whole past causal volume.

\section{Buchert-Ehlers commutation rules on the light-cone}
\label{Ch2Sec3}

\subsection{Average on the embedded 2-sphere \& the truncated light-cone}
\label{Ch2Sec31}

For the phenomenological applications of the following sections we are interested, in particular, in the derivatives of the averages defined in Eqs.
(\ref{211}) and (\ref{212}). To this purpose, let us first consider  the derivatives of $I(S;V_0,A_0;-)$ and $I(S;V_0;A_0)$  with respect to  $A_0$ and $V_0$ (quantities that, like the starting integrals themselves, are covariant and gauge invariant).

We will use the identities (here $k_{\mu} \equiv \partial_{\mu} V$)\,:
\beq
\label{RelDeltaThetaPrime}
\delta_{\rm D}'(A-A_0) = \frac{k^{\mu}\partial_{\mu}\delta_{\rm D}(A-A_0)}{k^{\nu}\partial_{\nu}A} 
~~~~~,~~~~~
\delta_{\rm D}(A-A_0) = \frac{k^{\mu}\partial_{\mu}\Theta(A-A_0)}{k^{\nu}\partial_{\nu}A} ~~,
\eeq
\beq
\delta_{\rm D}'(V-V_0) = \frac{\partial^{\mu} A \, \partial_{\mu}\delta_{\rm D}(V-V_0)}{k^{\nu}\partial_{\nu}A}
~~~~~,~~~~~
\delta_{\rm D}(V-V_0) = \frac{\partial^{\mu} A \, \partial_{\mu}\Theta(V-V_0)}{k^{\nu}\partial_{\nu}A} ~~,
\label{RelDeltaThetaPrimeV}
\eeq
and the relation $k^{\mu}\partial_{\mu}\delta_{\rm D}(V-V_0) = k^{\mu}k_{\mu}\delta_{\rm D}'(V-V_0) = 0$. Integrating by parts, we then obtain
\bea
\frac{\partial}{\partial A_0} I(S;V_0,A_0;-) &=& 
I\left(\frac{k \cdot \partial S}{k \cdot \partial A};V_0,A_0;-\right)
+I\left(\frac{\nabla \cdot k}{k \cdot \partial A} S;V_0,A_0;-\right) ~~, \\
\frac{\partial}{\partial A_0} I(S;V_0;A_0) &=&
I\left(\frac{k \cdot \partial S}{k \cdot \partial A};
V_0;A_0\right)
+I\left(\left[\nabla \cdot k -\frac{1}{2} 
\frac{k^\mu \partial_\mu \left((\partial A)^2\right)}{(\partial A)^2} 
\right]\frac{S}{k \cdot \partial A};V_0;A_0\right) ~~, \nonumber
\eea
where, to simplify notations, we have introduced the following definitions\,:
\bea
& k^\mu \partial_\mu S=k \cdot \partial S ~~~~~,~~~~~ k^\mu \partial_\mu A=k \cdot \partial A ~~~~~,~~~~~ \partial_\mu A \partial^\mu S=\partial A \cdot \partial S ~~, \nonumber \\
& \partial_\mu A \partial^\mu A=(\partial A)^2 ~~~~~,~~~~~ \nabla_\mu k^\mu = \nabla \cdot k ~~~~~,~~~~~ \Box=\nabla^\mu \nabla_\mu ~~.
\eea

Using these results we can write a generalized version of the Buchert-Ehlers \gls{commutation rule} \cite{Buchert:1995fz} for light-cone averages. For averages defined over the integration domain of Eq. (\ref{null2d}) we find\,:
\beq
\label{LCeq1}
\frac{\partial}{\partial A_0} \langle S \rangle_{V_0,A_0} = \left\langle \frac{k \cdot \partial S}{k \cdot \partial A}\right\rangle_{V_0,A_0} + \left\langle \frac{\nabla \cdot k}{k \cdot \partial A} S \right\rangle_{V_0,A_0} - \left\langle \frac{\nabla \cdot k}{k \cdot \partial A} \right\rangle_{V_0,A_0}
\langle S \rangle_{V_0,A_0} ~~,
\eeq
while for 3-dimensional averages over the domain of Eq. (\ref{null3d}) we obtain\,:
\bea
\frac{\partial}{\partial A_0} \lla S \rra_{V_0}^{A_0} &=& \lla \frac{k \cdot \partial S}{k \cdot \partial A} \rra_{V_0}^{A_0} + \lla \left[ \nabla \cdot k
- \frac{1}{2}\frac{k^{\mu}\partial_{\mu}\left((\partial A)^2\right)}{(\partial A)^2}\right] \frac{S}{k \cdot \partial A} \rra_{V_0}^{A_0} 
\nonumber \\
& &  - \lla \left[ \nabla \cdot k - \frac{1}{2}\frac{k^{\mu}\partial_{\mu}\left((\partial A)^2\right)}{(\partial A)^2}\right]
\frac{1}{k \cdot \partial A} \rra_{V_0}^{A_0} \lla S \rra_{V_0}^{A_0} ~~.
\label{LCeq2}
\eea

Following a similar procedure we can also evaluate the generalization of the Buchert-Ehlers commutation rule for the derivative of $I(S;V_0,A_0;-)$ and $I(S;V_0;A_0)$  with respect to $V_0$. As we shall see in Sec. \ref{Ch3Sec63}, such derivatives, together with the previous ones, may be useful in evaluating an averaged version of the  ``redshift drift" parameter.
After some calculations we are led to
\bea
\frac{\partial}{\partial V_0} \langle S \rangle_{V_0,A_0} &=& 
\left\langle \frac{\partial A \cdot \partial S}{k \cdot \partial A} \right\rangle_{V_0,A_0} 
-\lla k \cdot \partial S \frac{(\partial A)^2}{(k \cdot \partial A)^2} \rra_{V_0,A_0}
\nonumber \\
\label{LCeq3}
 & & 
+\left\langle
\left[\Box A-\nabla_\mu \left(
k^\mu  
\frac{(\partial A)^2}{k \cdot \partial A}\right)\right] 
\frac{S}{k \cdot \partial A}
\right\rangle_{V_0,A_0}
\nonumber \\ & & 
-\left\langle
\left[\Box A-\nabla_\mu \left(
k^\mu  
\frac{(\partial A)^2}{k \cdot \partial A}\right)\right] 
\frac{1}{k \cdot \partial A}
\right\rangle_{V_0,A_0}
\langle S \rangle_{V_0,A_0} ~~,
\eea
and
\bea
\label{LCeq4}
\frac{\partial}{\partial V_0} \lla S \rra_{V_0}^{A_0} & = & 
\left\langle \frac{\partial A \cdot \partial S}{k \cdot \partial A} \right\rangle_{V_0}^{A_0} 
-\lla k \cdot \partial S \frac{(\partial A)^2}{(k \cdot \partial A)^2} \rra_{V_0}^{A_0}
\nonumber \\ & & 
+\left\langle
\left[\Box A-\partial^\mu A \partial_\mu \ln\left((\partial A)^2\right)
\right] \frac{S}{k \cdot \partial A}
\right\rangle_{V_0}^{A_0}
\nonumber \\
& &
- \left\langle
\left[\nabla \cdot k \frac{(\partial A)^2}{(k \cdot \partial A)^2}+\frac{1}{2} k^\mu \partial_\mu 
\left(\frac{(\partial A)^2}{(k \cdot \partial A)^2}\right)\right] S \right\rangle_{V_0}^{A_0}
\nonumber \\
& &  
- \left\langle
\left[\Box A-\partial^\mu A \partial_\mu \ln\left((\partial A)^2\right)
\right] \frac{1}{k \cdot \partial A}
\right\rangle_{V_0}^{A_0}  \left\langle S \right\rangle_{V_0}^{A_0} 
\nonumber \\
& & 
+ \left\langle
\left[\nabla \cdot k \frac{(\partial A)^2}{(k \cdot \partial A)^2}+\frac{1}{2} k^\mu \partial_\mu 
\left(\frac{(\partial A)^2}{(k \cdot \partial A)^2}\right)\right] \right\rangle_{V_0}^{A_0}
 \left\langle S \right\rangle_{V_0}^{A_0} ~~.
\eea

These expressions will take a simpler form when we consider their averages in the GLC coordinates presented in Chapter \ref{Chap3}. We will show these simplified expressions of the Buchert-Ehlers rules on the lightcone in Sec. \ref{Ch3Sec62}.

\subsection{Average on the causally connected sphere}\
\label{Ch2Sec32}

We have not presented in this last subsection the derivatives of the average $ \lla S \rra_{A_0}^{V_0}$, namely the Buchert-Ehlers relations for the causally connected sphere case. Let us give these general expressions here and we will check in Sec. \ref{Ch3Sec62} that their expressions make us recover the usual average on spatial hypersurfaces.

Starting with the definition of the integral $I(1;A_0;V_0)$ presented in Eq. \rref{null3dInv} we get the derivative with respect to $A_0$ of the average \rref{213}\,:
\bea
\label{LCeq5}
\partial_{A_0} \lla S \rra_{A_0}^{V_0} &=& \lla \frac{\nabla \cdot k}{k \cdot \partial A} S \rra_{A_0}^{V_0} + \lla \frac{k^\mu}{\sqrt{-(\partial A)^2}} \nabla_\mu \left[ \frac{\sqrt{-(\partial A)^2}}{k \cdot \partial A} S \right] \rra_{A_0}^{V_0} \nonumber \\
&& ~~~~~ - \lla S \rra_{A_0}^{V_0} \lla \frac{k^\mu}{\sqrt{-(\partial A)^2}} \nabla_\mu \left[ \frac{\sqrt{-(\partial A)^2}}{k \cdot \partial A} \right] \rra_{A_0}^{V_0} ~~,
\eea
and with respect to the null parameter $V_0$, the derivation gives\,:
\bea
\label{LCeq6}
\partial_{V_0} \lla S \rra_{A_0}^{V_0} &=& \lla \frac{1}{\sqrt{-(\partial A)^2}} \nabla_\mu \left[ \left( \partial^\mu A - k^\mu \frac{(\partial A)^2}{k \cdot \partial A} \right) \frac{\sqrt{-(\partial A)^2}}{k \cdot \partial A} S \right] \rra_{A_0}^{V_0} \nonumber \\
&& ~~~~~ - \lla S \rra_{A_0}^{V_0} \lla \frac{1}{\sqrt{-(\partial A)^2}} \nabla_\mu \left[ \left( \partial^\mu A - k^\mu \frac{(\partial A)^2}{k \cdot \partial A} \right) \frac{\sqrt{-(\partial A)^2}}{k \cdot \partial A} \right] \rra_{A_0}^{V_0} ~~.
\eea

\section{The FLRW Universe}
\label{Ch2Sec4}

In order to prepare the calculation of Chapter \ref{Chap4}, we shall proceed to the consistency check that no backreaction occurs in a general homogeneous FLRW Universe. We will use the general average over spatial hypersurfaces (see Sec. \ref{Ch2Sec13}) but this result can also be verified with lightcone averages.

Taking the \gls{FLRW} metric in cosmic time such as the one defined in Eq. \rref{ds2FLRW}, we can recast it in an ADM-metric form (see Eq. \rref{ADMmetricCh3}) by $N = 1$, $N^i = 0$ and $g_{ij} = a^2(t) \gamma_{ij}(|\xbf|)$. We thus have the expansion scalar from Eq. \rref{ExpScalarADM}\,:
\beq
\Theta(t,\xbf) = \partial_t \ln \left( a^3(t) \sqrt{\det \gamma_{ij}(|\xbf|)} \right) = 3 \frac{\dot{a}(t)}{a(t)} \equiv 3 H(t) ~~.
\eeq
Taking $A(x) \equiv t$ imposes then that the non-commutation terms in Eq. \rref{SpatialBuchertEhlersInADM} cancel out\,:
\beq
\lla \frac{\Theta N}{\partial_t A} S \rra_{A_0} - \lla \frac{\Theta N}{\partial_t A} \rra_{t_0} \lla S \rra_{t_0} = \lla 3 H(t) S \rra_{t_0} - \lla 3 H(t) \rra_{t_0} \lla S \rra_{t_0} = 0 ~~.
\eeq
Thus, no backreaction is present in the homogeneous FLRW model.
In fact, the condition for Eq. \rref{SpatialBuchertEhlersInADM} to exibit a non-commutation form is that $\det(g_{ij})$ must be a function of $(t,\xbf)$ which is non-factorizable as $f(t) \times g(\xbf)$. This is the case of perturbed spacetimes where perturbations of the spatial metric are generally $(t,\xbf)$-dependent. As a consequence, backreaction occurs in many types of inhomogeneous geometries.

Let us now consider the generalization of the Friedmann equations in the context of backreaction, Eq. \rref{F1_original} and \rref{F2_original}, and show that they reduce to the usual Friedmann equation in the homogeneous limit. We know in that case that we can use the ADM formalism with $N^i = 0$ and $N = 1$ and that directly lead to $n_\mu = (-1,\vec{0}\,)$ and $n^\mu = (1,\vec{0}\,)$. We also have that $A(t) \equiv t$ which gives the same result.
In that case we find that the expansion scalar is
\beq
\Theta \equiv \nabla_\mu n^\mu = \partial_\mu n^\mu + \Gamma^{\mu}_{\mu\rho} n^\rho = \Gamma^{\mu}_{\mu 0} = H \delta^i_i = 3 H(t) ~~.
\eeq
We thus find that the effective scale factor reduces to the usual one\,:
\beq
\frac{1}{\tila} \parfrac{\tila}{A_0} \equiv \frac{1}{3} \lla \frac{\Theta}{[-(\partial A)^2]^{1/2}} \rrao = \frac13 \lla 3 H(t) \rra_{t_0} = H(t) \equiv \frac{\partial_t a(t)}{a(t)} ~~.
\eeq

We can also see that the system of Eqs. (\ref{F1_original},\ref{F2_original}) greatly simplifies. Indeed, the facts that $\Theta = 3 H(t)$, $A(t) \equiv t$ and $\lla \ldots \rra = \lla \ldots \rra_{t_0}$ cancel the variance $\lla \Theta^2 \rrao - \lla \Theta \rrao^2$, lead to\,:
\bea
\left( \frac{1}{\tila} \parfrac{\tila}{A_0} \right)^2 &=& \frac{8 \pi G}{3} \lla \epsilon \rrao - \frac{1}{6} \lla \mathcal{R}_S \rrao + \frac{1}{3} \lla \sigma^2 \rrao ~~, \\
- \frac{1}{\tila} \frac{\partial^2 \tila}{\partial {A_0}^2} &=& \frac{4 \pi G}{3} \lla \epsilon + 3 \pi \rrao - \frac{1}{3} \lla \nabla^{\nu}(n^{\mu}\nabla_{\mu}n_{\nu}) \rrao + \frac{1}{6} \lla (\partial_{\mu}A\partial^{\mu}[(\partial A)^2]) \Theta \rrao + \frac{2}{3} \lla \sigma^2 \rrao ~~.~~~~~
\eea
We can remark that $(\partial A)^2 = -1$, so $\partial_{\mu}A\partial^{\mu}[(\partial A)^2] = 0$. In a FLRW model, we have the projection tensor on the spatial hypersurfaces which is
\beq
h_{\mu\nu} = \delta_\mu^i \delta_\nu^j a^2 \gamma_{ij} ~~,~~ h^{\mu\nu} = \delta^\mu_i \delta^\nu_j a^2 \gamma^{ij} ~~,~~ h_{i\mu} h^{\mu j} = \delta_i^j ~~.
\eeq
We thus find that the \gls{shear tensor} defined in Eq. \rref{DefExpScalarShearVorticity} is zero\,:
\bea
\sigma_{\mu\nu} &=& h^\alpha_{~\mu} h^\beta_{~\nu} \left( \nabla_{(\alpha} n_{\beta)} - \frac13 h_{\alpha\beta} \nabla_\gamma n^\gamma \right) = \delta_\mu^\alpha \delta_\nu^\beta \left( - \Gamma^0_{\alpha \beta} - \frac{h_{\alpha\beta}}{3} 3 H(t) \right) \nonumber \\
&=& - \delta_\mu^i \delta_\nu^j \Gamma^0_{ij} - (a^2 \gamma_{ij} \delta_\mu^i \delta_\nu^j) H(t) = 0 ~~,~~ 
\eea
and so the shear scalar $\sigma \equiv \sqrt{\frac12 \sigma_i^{~j} \sigma_j^{~i}}$ is zero. We can also show that the \gls{vorticity tensor} is null\,:
\beq
\omega_{\mu\nu} = h^\alpha_{~\mu} h^\beta_{~\nu} \frac12 \left(\nabla_{\alpha} n_{\beta} - \nabla_{\beta} n_{\alpha} \right) = 0 ~~~\mbox{ as }~~ \nabla_{\alpha} n_{\beta} = \Gamma_{\alpha\beta}^\lambda n_\lambda = - \Gamma^0_{\alpha\beta} = \nabla_{\beta} n_{\alpha} ~~.
\eeq
And we see that $\nabla^{\nu}(n^{\mu}\nabla_{\mu}n_{\nu}) = \nabla^0(n^0\nabla_0 n_0) = 0$ as $\nabla_0 n_0 = \Gamma_{00}^\lambda n_\lambda$ and $\Gamma_{00}^\lambda = 0 ~~ \forall ~ \lambda$.

We thus proved that the system of equations (\ref{F1_original},\ref{F2_original}) is equivalent to
\beq
\left( \frac{1}{\tila} \parfrac{\tila}{A_0} \right)^2 = \frac{8 \pi G}{3} \lla \epsilon \rrao - \frac{1}{6} \lla \mathcal{R}_S \rrao ~~~~~,~~~~~
- \frac{1}{\tila} \frac{\partial^2 \tila}{\partial {A_0}^2} = \frac{4 \pi G}{3} \lla \epsilon + 3 \pi \rrao ~~. \nonumber
\eeq
For a perfect fluid which is homogeneous, we cannot distinguish a particuliar velocity and we can take $u_\mu = (-1,\vec{0}\,)$ and $u^\mu = (1,\vec{0}\,)$. From the definitions of $\epsilon$ and $\pi$ given in Eq. \rref{DefinEpsPi}, we see that the terms going as $(1-(u^\mu n_\mu)^2)$ are zero and we simply get that $\epsilon = \rho$ and $\pi = p$. Also, one has
\bea
& & \sigma^2 = \frac12 \left( \Theta^\mu_{~\nu} \Theta^\nu_{~\mu} - \frac{\Theta^2}{3} \right) = 0 ~~~\Rightarrow~~ \Theta^\mu_{~\nu} \Theta^\nu_{~\mu} = 3 H^2(t) ~~, \\
& & \Rcal_s + \Theta^2 - \Theta^\mu_{~\nu} \Theta^\nu_{~\mu} = 16 \pi G \, \epsilon ~~,
\eea
where the second equation is obtained from Eqs. \rref{Constr1} and \rref{KijAndThetaij} (with $\Lambda = 0$).
We thus have
\beq
\Rcal_s = - 6 H^2 + 16 \pi G \rho = \frac{6 K}{a^2} ~~,
\eeq
as expected for the 3-dimensional spatial curvature in the FLRW model.

We are left with equations which depend only on $\lla \rho \rrao$ and $\lla p \rrao$ on the RHS. These two quantities being only time dependent in the FLRW Universe, we have shown that the generalized Friedmann equations reduce to the Friedmann equations of Eqs. (\ref{Friedmann1},\ref{Friedmann2}) in the FLRW model (where we took $\Lambda = 0$ here).
Finally, we directly see that the averaged conservation equation
\beq
\parfrac{\lla \rho \rrao}{A_0} = - \lla \frac{\Theta ~ p}{[-(\partial A)^2]^{1/2}} \rrao - \lla \rho \rrao \lla \frac{\Theta}{[-(\partial A)^2]^{1/2}} \rrao ~~,
\eeq
simply reduces to the usual conservation equation for a perfect fluid $\dot{\rho} + 3 H (\rho + p) = 0$.

We thus arrived to the conclusion that perturbations are needed to generate a backreaction effect in a FLRW model. In this spirit, we will compute in Chapter 4 the luminosity at second order in perturbations and will exploit this result in the other chapters.

% --------------------------------------------------------- %

%  --------------------- Chapter 3 ----------------------   %
% --------------------------------------------------------- %
\chapter{The GLC metric and its properties}
\label{Chap3}
\pagestyle{plain}

This section is devoted to the geometrical description of a 4-dimensional spacetime in terms of a lightcone foliation. We introduce first an historical system of coordinates called the \emph{observational coordinates} (substantially developed in \cite{Maartens1} and \cite{1985PhR...124..315E}). We then define the \emph{geodesic lightcone} (GLC) \emph{coordinates} that will be a powerful tool in our subsequent calculations. These new coordinates share a large number of properties with the observational coordinates but are different on certain points. They have particularly nice features with, for example, the property to simplify the expressions of the redshift and the luminosity distance, and finally the lightcone averages.

\section{Observational coordinates}
\label{Ch3Sec1}

In the \gls{observational coordinates}, one considers the world line $\Cscr$ of an observer and its past lightcones in a 4-dimensional spacetime. The past lightcones are set by a null coordinate $w \equiv x^0$ and a particular one is given by the condition $w = {\rm cst}$. To fully fix this coordinate, we assume it to be equal to the proper time along the world line $\Cscr$, i.e. $\left. w \right|_{\Cscr} = \left. \tau \right|_{\Cscr}$, and set its value to $w_0$ today. The null geodesic vector field describing the photon path on a lightcone is then given by
\beq
k^\mu = dx^\mu / dv ~~~~~,~~~~~ k_\mu = \partial_\mu w ~~~~~,~~~~~ k_\mu k^\mu = 0 ~~,
\eeq
where $v$ is an affine parameter along this null geodesic (with the condition $v=0$ along $\Cscr$). The consequence of this definition is that $k^\mu$ is a hypersurface-orthogonal ($\nabla_\mu k_\nu = \nabla_\nu k_\mu$) null geodesic ($k^\mu \nabla_\mu k^\nu = 0$) vector field, orthogonal to the constant-$w$ hypersurfaces ($\partial_\mu w \, k^\mu = 0$). We thus have $k_\mu u^\mu = \partial_\mu w \, u^\mu$ and $\left. k_\mu u^\mu \right|_\Cscr = 1$, where $u^\mu$ is the velocity of the observer.

The photon geodesic is a line along a past light cone that keeps the same angles $\theta \equiv x^2$ and $\phi \equiv x^3$ (i.e. $\partial_\mu \theta \, k^\mu = 0 = \partial_\mu \phi \, k^\mu$). They are properly normalized by imposing that they are spherical coordinates along $\Cscr$, so\,:
\beq
\left\{ \lim_{v \rightarrow 0} \left(\frac{ds^2}{v^2}\right) ~~ / ~~ w=\cst , v=\cst \right\} = d\theta^2 + (\sin \theta)^2 d \phi^2 \equiv d^2\Omega ~~.
\eeq

The last coordinate used here, called $y \equiv x^1$, determines the ``distance'' down the null geodesic. Different choices are proposed by \cite{1985PhR...124..315E}\,:
\begin{enumerate}
 \item $y = v$\,: the affine parameter with the conditions $v|_\Cscr = 0$ and $u_\mu k^\mu |_\Cscr = 1$.
 \item $y = R$\,: the angular distance (denoted by $r$ in \cite{1985PhR...124..315E}\footnote{For this section only we will use $R$ to denote the angular distance. After, it will be written as $d_A$.}) from $\Cscr$.
 \item $y = z$\,: the redshift observed from $\Cscr$.
 \item mixed $y$\,: one of the three cases above to set the distance on $w = w_0$ and then we take $y$ to comove with the fluid, i.e. $\partial_\mu y \, u^\mu = 0$. We thus have $y = 0$ on all the world line $\Cscr$.
\end{enumerate}
We can see that the condition $y \rightarrow 0$ is equivalent to $v \rightarrow 0$ or $R \rightarrow 0$.
One should be careful with cases $2.$ and $3.$ as the angular distance and the redshift are not in general monotonic functions of the affine parameter $v$. This is thus the last case, number $4.$, which is chosen in \cite{1985PhR...124..315E}. Its comoving property also makes it easy and unambiguous. Because of the conditions $\partial_\mu w \, k^\mu = 0 = \partial_\mu \theta \, k^\mu = \partial_\mu \phi \, k^\mu$, we have\,:
\beq
k_\mu = \delta_\mu^0 ~~\Rightarrow~~ k^\mu = \frac{d x^\mu}{d v} = \frac{1}{\beta} \delta^\mu_1 ~~~\mbox{ with }~~ \beta = \frac{dy}{dv} > 0 ~~. 
\eeq

The metric describing our geometry in these coordinates $(w,y,\theta,\phi)$ is obtained as follows. From $k_\mu k^\mu = 0 = \partial_\mu w \, \partial_\nu w \, g^{\mu\nu}$ we get that $g^{00} = 0$. The condition $k^\mu = g^{\mu\nu} k_\nu$ imposes $g^{\mu0} = (1 / \beta) \delta^\mu_1$. Introducing new functions for the non-constrained components, we get the general expression of $g^{\mu\nu}$ and we can compute its inverse. We thus have, in $(w,y,\theta,\phi)$-coordinates\,:
\beq
\label{ObsCoordMetric}
g_{\mu\nu} =
\left(
\begin{array}{cccc}
\alpha & \beta & v_2 & v_3 \\
\beta & 0 & 0 & 0 \\
v_2 & 0 & h_{22} & h_{23} \\
v_3 & 0 & h_{23} & h_{33}
\end{array}
\right) ~~~~~,~~~~~
g^{\mu\nu} =
\left(
\begin{array}{cccc}
0 & 1/\beta & 0 & 0 \\
1/\beta & \delta & \sigma_2 & \sigma_3 \\
0 & \sigma_2 & h_{33} / h & -h_{23} / h \\
0 & \sigma_3 & - h_{23} / h & h_{22} / h
\end{array}
\right) ~~,
\eeq
where $h = \det(h_{ab}) \equiv h_{22} h_{33} - (h_{23})^2$ (with $a,b = 2,3$), $\delta \equiv - (\alpha + \beta (v_2 \sigma_2 + v_3 \sigma_3))/\beta^2$, $\sigma_2 = - (v_2 h_{33} - v_3 h_{23}) / \beta h$ and $\sigma_3 = - (v_3 h_{22} - v_2 h_{23}) / \beta h$. One should thus remark that the observational metric depends on seven arbitrary functions which are $\beta$, $\delta$, $\sigma_2$, $\sigma_3$, $h_{22}$, $h_{33}$ and $h_{23}$. The angular distance $R$ can be defined in terms of these functions as\,:
\beq
R^4 (\sin \theta)^2 = h \equiv \det(h_{ab}) ~~.
\eeq

The metric of Eq. \rref{ObsCoordMetric} implies that the hypersurfaces $w = \cst$ are null. Nevertheless, as it stands, this metric does not garantee that these null hypersurfaces are the past lightcones of the geodesic world line $\Cscr$. To set this property, one has to impose some limits near the world line.
When the coordinate $y$ is taken to be $v$ or $R$, these limits are\,:
\beq
\lim_{y \rightarrow 0} \alpha = -1 ~~,~~ \lim_{y \rightarrow 0} \beta = 1 ~~,~~ \lim_{y \rightarrow 0} (v_a / y^2) = 0 ~~,~~ \lim_{y \rightarrow 0} (h_{ab} dx^a dx^b / y^2 ) = d^2 \Omega ~~,
\eeq
where $d^2 \Omega \equiv \sin \theta d \theta d \phi$ is the standard solid angle on the sphere.
When this coordinate $y$ is taken to be $z$ or the last case $4.$ described above, one has a regular relation between $y$ and $v$. The limits are found to be\,:
\beq
\lim_{y \rightarrow 0} \alpha = -1 ~~,~~ \lim_{y \rightarrow 0} \beta = \beta_0(w,x^i) ~~,~~ \lim_{y \rightarrow 0} (v_a) = 0 ~~,~~ \lim_{y \rightarrow 0} (h_{ab} dx^a dx^b / y^2 ) = \beta_0^2(w,x^i) \, d^2 \Omega ~~.
\eeq

These conditions that we described are the necessary conditions to make the null hypersurfaces be the past lightcones of the observer world line $\Cscr$ and they garantee that the angular coordinates are correctly related to the angles as seen from $\Cscr$. More precisely, the direction made by these angles is now based on a parallely propagated tetrad of vectors\footnote{These tetrad vectors $\ebf_\mu$ are defined through the conditions\,:
\beq
\ebf_0 = \ubf ~~,~~ \ubf \cdot \ebf_i = 0 ~~,~~ \ebf_i \cdot \ebf_j = \delta_{ij} ~~, \nonumber
\eeq
and thus satisfy the parallel propagation along $\Cscr$\,:
$\left. \nabla_\ubf \, \ebf_\mu \right|_\Cscr = 0$\,,
with $\ubf$ the velocity of the comoving geodesic observer.
}
along $\Cscr$. We recall also that $w$ is the proper time along $\Cscr$. These properties define the so called \emph{observational coordinates}.

\section{The geodesic light-cone gauge}
\label{Ch3Sec2}

\subsection{Presentation}
\label{Ch3Sec21}

One of the main virtues of using a gauge invariant formalism is the freedom of choosing a gauge particularly adapted to the problem at hand. In the case of spacelike averaging, for instance, a convenient coordinate system corresponds to the case where we average over the constant-time hypersurfaces. 
Indeed, in many applications the averaging hypersurfaces are chosen to be associated to a class of geodesic observers described by the time parameter $t$ of a synchronous gauge (see e.g.  \cite{Marozzi:2010qz}). Similarly, for light-cone averages, it is convenient to identify the null hypersurfaces with those on which a null coordinate takes constant values. It is also useful to have a coordinate to describe the proper time measured by an observer in geodetic motion. For this reason we now turn to a special (adapted) coordinate system $x^\mu= (\tau,w,\tilde{\theta}^a)$, $a=1,2$ \,(with sometimes the change in notations\,: $\tilde{\theta}^1 \equiv \tilde{\theta}$ and $\tilde{\theta}^2 \equiv \tilde{\phi}$). These coordinates define what we have called the ``\gls{geodesic light-cone}'' (GLC) coordinates, or gauge.

\subsection{Mathematical definition}
\label{Ch3Sec22}

The GLC metric depends on six arbitrary functions\,: a scalar $\Ups$, a two-dimensional ``vector" $U^a$ and a symmetric matrix $\gamma_{ab}$ (with inverse $\gamma^{ab}$). Its line element takes the form\,:
\bea
\label{LCmetric}
ds_{GLC}^2 = \Ups^2 dw^2-2\Ups dw d\tau+\gamma_{ab}(d\ti{\theta}^a-U^a dw)(d\ti\theta^b-U^b dw) ~~.
\eea
Clearly $w$ is a null coordinate (i.e. $\pa_\mu w ~ \pa^\mu w=0$), and a past light-cone hypersurface is specified by the condition $w=\cst$. We can also easily check that $\partial_{\mu} \tau$ defines a geodesic flow, i.e. that
\beq
\left( \pa^\nu \tau\right) \nabla_\nu \left( \pa_\mu \tau\right) = 0 ~~,
\eeq
as a consequence of the relation $g_{GLC}^{\tau \tau} = -1$. Thus an observer defined by constant $\tau$ spacelike hypersurfaces is in geodesic motion. 
In matrix form, the metric and its inverse in the $(\tau,w,\titheta^a)$-coordinates are given by\,:
\beq
\label{GLCmetric}
g^{GLC}_{\mu\nu} =
\left(
\begin{array}{ccc}
0 & - \Ups &  \vec{0} \\
-\Ups & \Ups^2 + U^2 & -U_b \\
\vec0^{\,T}  &-U_a^T  & \gamma_{ab} \\
\end{array}
\right) ~~~~~,~~~~~
g_{GLC}^{\mu\nu} =
\left(
\begin{array}{ccc}
-1 & -\Ups^{-1} & -U^b/\Ups \\
-\Ups^{-1} & 0 & \vec{0} \\
-(U^a)^T/\Ups & \vec{0}^{\, T} & \gamma^{ab}
\end{array}
\right) ~~,
\eeq
where $\vec 0=(0,0)$ and $U_b= (U_1, U_2)$, while the $2 \times 2$ matrices $\gamma_{ab}$ and  $\gamma^{ab}$ lower and  raise  the two-dimensional indices.
We also have that\,:
\beq
\sqrt{-g} = \Ups \sqrt{|\gamma|} ~~~~~\mbox{ with }~~~~~ g \equiv \det g^{GLC}_{\mu\nu} ~~~~~\mbox{ and }~~~~~ \gamma \equiv \det\gamma_{ab} ~~.
\eeq
This GLC metric has all the required degrees of freedom for applications to general geometries. It is also really well adapted to computations of quantities related to light signals, as we shall see in Sec. \ref{Ch3Sec3} and Sec. \ref{Ch3Sec4}. We will also see in Sec. \ref{Ch3Sec6} that it greatly simplifies the so-called averages on the light cone.

In the limiting case of a spatially flat \gls{FLRW} geometry, with scale factor $a$, cosmic time $t$, and conformal time parameter $\eta$ such that $d\eta= dt/a$, the transformations to the GLC coordinates and the meaning of the metric components are easily found, from Eq. (\ref{LCmetric}), to be\,:
\bea
\label{FR}
&
\tau=t ~~~~~,~~~~~ w= r+\eta ~~~~~,~~~~~ \titheta^1 = \theta ~~,~~ \titheta^2 = \phi ~~, \nonumber \\
&
\Ups = a(t) ~~~~~,~~~~~ U^a=0 ~~~~~,~~~~~ \gamma_{ab} d \tilde{\theta}^a d\tilde{\theta}^b = a^2(t) r^2 (d \theta^2 +\sin^2 \theta d\phi^2) ~~.
\eea
When we authorize a possible curvature, this limit is simply changed in $w$, which becomes $w = \chi + \eta$ where $\chi$ has been defined in Eq. \rref{ds2FLRW}, and the expression of $\gamma_{ab} d \tilde{\theta}^a d\tilde{\theta}^b$ is unchanged, $r$ being defined by Eq. \rref{randchi}. We notice in passing that $\gamma_{ab} \propto a^2(t)$ in the homogeneous FLRW case (and should not be confused with the metric element $\gamma_{ij}$ of Eq. \rref{ds2FLRW}).
Even though we will be mainly using the GLC gauge for a perturbed FLRW metric in the Newtonian gauge, it is important to stress that the equality between the coordinate $\tau$ and the proper time $t$ of the synchronous gauge holds at the exact, non perturbative level. In fact, it is always possible to choose the GLC coordinates in such a way that $\tau$ and $t$ are identified like in the above FLRW limit. As a consequence we can easily introduce with $\tau$ a family of geodetic reference observers which exactly coincide with the static ones of the synchronous gauge. We shall illustrate this point in Sec. \ref{Ch3Sec51}.

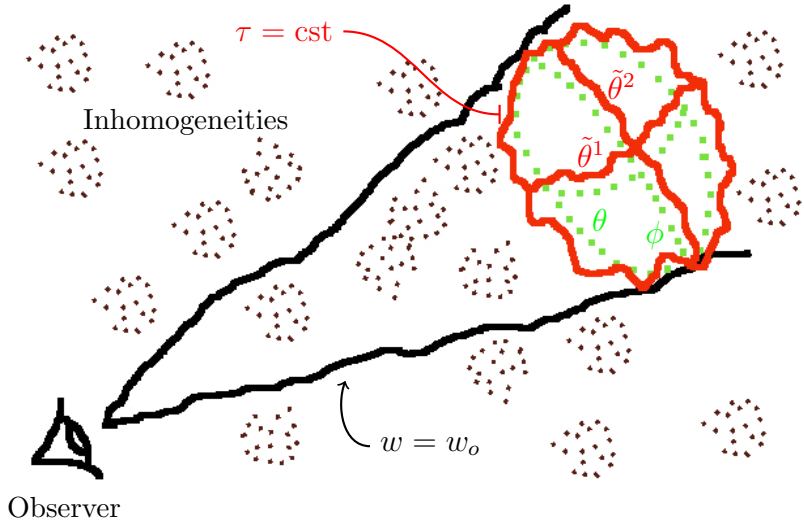
\begin{figure}[t]
\centering
\begin{tikzpicture}
\node (label) at (-5,0) {\includegraphics[width=11cm]{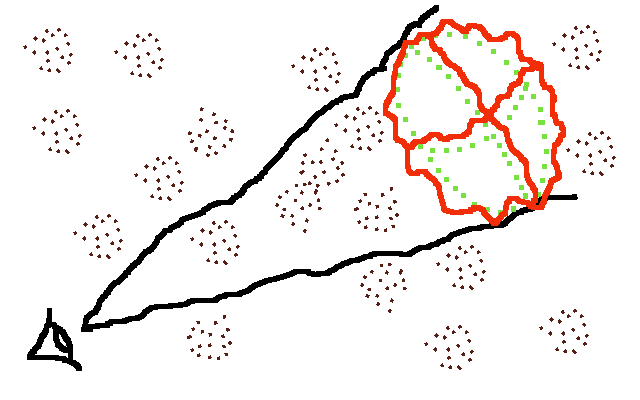}};

\node[anchor=west] at (-9.5,1.8) (description) {Inhomogeneities};
\node[anchor=west] at (-10.5,-3.3) (description) {Observer};

\node[anchor=west] at (-2.8,0.5) (description) {\large \textcolor{green}{$\theta$}};
\node[anchor=west] at (-2.1,0.3) (description) {\large \textcolor{green}{$\phi$}};

\node[anchor=west] at (-3.0,1.4) (description) {\large \textcolor{red}{$\tilde\theta^1$}};
\node[anchor=west] at (-2.6,2.3) (description) {\large \textcolor{red}{$\tilde\theta^2$}};

\node[anchor=east] at (-3.6,1.9) (text) {};
\node[anchor=west] at (-7.5,3.0) (description) {\large \textcolor{red}{$\tau = \cst$}};
\draw (description) edge[out=0,in=180,-|,color=red,thick] (text);

\node[anchor=west] at (-6,-1.5) (text) {};
\node[anchor=east] at (-4,-2.5) (description) {\large $w = w_o$};
\draw (description) edge[out=-180,in=250,->,thick] (text);
\end{tikzpicture}
\caption[Illustration of the GLC coordinates]{Illustration of the GLC coordinates. The observer sees the sky as the superposition of 2-spheres $\Sigma(w_o,\tau)$ embedded in the (lightcone) hypersurface $w = w_o$ at different times $\tau$ (in red). The homogeneous 2-sphere (in green) is parametrized by the angles $\theta^a = (\theta,\phi)$ whereas the presence of inhomogeneities deformes the lightcone and $\Sigma(w_o,\tau)$ is parametrized by the angles $\ti \theta^a = (\ti \theta, \ti \phi)$. Photons travel at constant $(w, \ti \theta^a)$ so that a source keep the same angular coordinates along the way.}
\label{FigGLCMetric}
\end{figure}

To give a more intuitive picture of these coordinates, we present a useful way to visualize the GLC coordinates in Fig. \ref{FigGLCMetric}. The constant $\tau$ and constant $w$ hypersurfaces intersect each other on a 2-dimensional surface $\Sigma(w_o, \tau_s)$ which is topologically a sphere but is deformed with respect to the standard 2-sphere in the general case. Let us notice finally that this 2-sphere can be expressed as a constant redshift hypersurface, by the use of the Eq. \rref{redshift} that we will present later. In that case, it corresponds to the constant $\Ups$ hypersurface on the past light cone.

\subsection{Differences with the observational coordinates}
\label{Ch3Sec23}

When specifying the metric in the GLC gauge, the coordinates $(\tau,w,\ti \theta^a)$ correspond to a complete gauge fixing (modulo residual transformations involving non-generic functions, i.e. not depending on all coordinates) of the so-called observational coordinates, defined e.g in 
\cite{Maartens1,1985PhR...124..315E,Clarkson:2010uz} and recalled in Sec. \ref{Ch3Sec1}. We should underline that descriptions in both systems of coodinates break down when caustics appear on the past light cones (as $w$ becomes multivalued).

Nevertheless, several facts make the GLC metric slightly different from the observational coordinates. We note that these latter ones use the coordinates $(w,y,x^2,x^3)$ whereas the GLC metric uses $(\tau,w,\ti{\theta}^a)$.
The first difference is thus an inversion in the order of the first two coordinates responsible for the apparently exchanged roles of the metrics presented in Eqs. \rref{ObsCoordMetric} and \rref{GLCmetric}. This is an irrelevant difference. The second, much more serious, is that $y$ is a spatial distance or a null distance on the past lightcone. In the GLC metric case, $y$ is replaced by a time coordinate which is $\tau$. This is a significant difference and it is motivated, on the GLC side, by the intention to have a system of coordinates which is both adapted to the observation of the source and the proper time of the observer.
Another difference, less important, is the fact that the past lightcone on which the observations are made is set by the condition $\left. w \right|_{\Cscr} = \left. \tau \right|_{\Cscr}$ (with $\tau$ the proper time of the observer) in the observational coordinates whereas it is set by $w = \int d \tau / a(\tau) \equiv \eta$ in the GLC gauge (where $a(\tau)$ was defined as the homogeneous limit of the $\Ups$ function), i.e. from the conformal time at this local position.
A very important difference as we have seen is also that the observational metric has seven arbitrary functions where the GLC metric only has six. It is in this sense, and forgetting about the other differences mentioned here, that we can say that the GLC gauge is a ``further gauge fixing'' of the observational coordinates.

\section{Redshift in the GLC metric}
\label{Ch3Sec3}

In the GLC coordinates a past light-cone hypersurface is specified by the condition $w=w_o$, and the momentum of a photon traveling on it, being proportional to $k_\mu=\partial_{\mu}w$, is orthogonal both to itself ($k_{\mu}k^{\mu}=0$) and to the 2-surface generated by $\partial_{\theta^a}$ ($\partial_{\mu} {\theta^a} k^{\mu}=0$).
The velocity of a  generic observer defined by a scalar $A$ can be written as\,:
\beq
n_\mu =- {\pa_\mu A\over \sqrt{-\partial_\nu A \partial^\nu A} } ~~~~~\mbox{ with }~~~~~ n_\mu n^\mu = -1 ~~~,
\eeq
and thus its scalar product with $k_\mu$ satisfies\,:
\beq
k_{\mu}n^{\mu} = \frac{1}{\Ups} \frac{\partial_{\tau} A}{\sqrt{-(\partial A)^2}} ~~.
\eeq
For our geodesic observer, $A$ depends only on $\tau$ and we can always set $A=\tau$, so the vector $n_\mu$ becomes $n_\mu = - \partial_\mu \tau = u_\mu$ and the above relation simplifies to\,:
\beq
k_{\mu}u^{\mu} = \Ups^{-1} ~~.
\label{45}
\eeq
We also note for later use that, in these coordinates, the covariant divergence of $k^{\mu}$ takes the simple form\,:
\beq
\nabla_{\mu} k^{\mu} = - \frac{1}{\Ups } \partial_{\tau}\left(\ln \sqrt{|\gamma|} \right) ~~.
\eeq
We also remark that, in GLC coordinates, the null geodesics connecting sources and observer are characterized by the simple tangent vector $k^{\mu} = g_{GLC}^{\mu \nu} \partial_{\nu} w = g_{GLC}^{\mu w} = - \delta^{\mu}_{\tau} \Ups^{-1}$, which means that photons reach the observer travelling at constant $w$ and $\ti \theta^a$ (as it was the case in the observational coordinates). This makes the calculation of the redshift particularly easy.

Let us denote by the subscripts ``o'' and ``s'', respectively, a quantity evaluated at the observer and source spacetime position, and consider a light ray emitted by a static geodetic source lying at the intersection between the past light-cone of a static geodetic observer (defined by the equation $w=w_o$) and the spatial hypersurface $\tau= \tau_s$ where $\tau_s$ is for the moment considered as a constant parameter, the moment of source's emission. Both the observer and the source are in geodetic motion and have a velocity $u_\mu = - \partial_\mu \tau$ in the GLC gauge. They are however static in synchronous gauge. The light ray will be received by the geodetic observer at $\tau=\tau_o>\tau_s$. The redshift $z_s$ associated with this light ray of momentum $k^\mu$ is then given by\,:
\beq
\label{GenRedshift}
(1+z_s) = \frac{(k^{\mu} u_{\mu})_s }{(k^{\mu} u_{\mu})_o} ~~.
\eeq
Note that this object may be regarded as a bi-scalar, as it depends on the ratio of the same scalar quantity calculated in two different spacetime points. Using in particular Eq. \rref{45}, we get\,:
\beq
\label{redshift}
(1+z_s) = \frac{(\partial^{\mu}w \pa_\mu \tau)_s }{(\partial^{\mu}w \pa_\mu \tau)_o} = {\Ups(w_o, \tau_o, \ti \theta^a)\over \Ups(w_o, \tau_s, \ti \theta^a)} \equiv \frac{\Ups_o}{\Ups_s} ~~.
\eeq

We should stress that the redshift, in this GLC gauge as in the FLRW case, is immediately given by an entry of the metric (evaluated at the source and observer positions). This equation will be used in Sec. \ref{Ch4Sec21} in connection with the redshift and luminosity distance.
We shall be interested in averaging the luminosity at fixed redshift also, hence on the two-dimensional surface $\Sigma(w_o, z_s)$ (topologically a sphere) which lies on our past light-cone ($w=w_o$)  and is associated with a fixed redshift ($z = z_s$, see Sec. \ref{Ch3Sec6}). In terms of the $\tau$ coordinate such a surface corresponds to the equation $\tau = \tau_s(w_o, z_s, \tilde{\theta}^a)$. Hereafter $\tau_s$ will denote this (in general angle-dependent) quantity.

\section{Luminosity distance in the GLC gauge}
\label{Ch3Sec4}

In a cosmological context we can define at least three different distances of physical interest (see e.g. \cite{Weinberg})\,: $d_A$, namely the \gls{angular distance} of the source as seen from the observer (in the FLRW limit of Eq. (\ref{FR}) we have $d_A = a(t_s) \, r_s$, where $r_s$ is the comoving radial distance of the source); $d_o$, namely the angular distance of the observer ``as seen from'' the source (in the FLRW limit we have $d_o = a(t_0) \, r_s$); and $d_L$, the so-called \gls{luminosity distance} (as seen from the observer). These three observational distances are always related to each other, independently of the given cosmological model, by redshift factors \cite{AppEt1933a,schneider1999gravitational}\,:
\beq
\label{ReldLdA}
d_L = (1+z) d_o = (1+z)^2 d_A ~~,
\eeq 
as a consequence of the Etherington reciprocity law \cite{Et1933}, valid in any general spacetime.

On the other hand, in a general spacetime, we can also identify (point-by-point on the 2-sphere) the square of the angular distance of the source \cite{Kristian:1965sz}, by\,:
\beq
\label{KS}
d_A^2 \equiv \frac{dS_s}{d^2\Omega_o} ~~,
\eeq
where $d^2\Omega_o$ is the infinitesimal solid angle subtended by the source at the observer position, and $dS_s$ is the area element on the surface orthogonal to both the photon momentum and to the source 4-velocity at the source's position. It is easy to check (see Sec. \ref{Ch3Sec42}) that the surface $dS_s$  is characterized by having constant $w$ and $\tau$, and that the induced 2-metric on it is nothing but $\gamma_{ab}$.

Let us now show, by two different methods, that 
\beq
\label{dAGLC1}
d_A^2 = \frac{d^2 \tilde{\theta} \sqrt{\gamma_s}}{d^2 \tilde{\theta} \sin \tilde{\theta}} = \frac{\sqrt{\gamma_s}}{\sin \tilde{\theta}} ~~,
\eeq
where we will use the fact that photons travel at constant $\tilde{\theta}^a$ in the GLC geometry.
The direct consequence of this identity gives, from Eq. \rref{ReldLdA}, the important non-perturbative expression for $d_L$\,:
\beq
\label{dLNonPert}
d_L = (1 + z)^2 \, \ga_s^{1/4} \left(\sin \tilde{\theta}^1\right)^{-1/2} ~~.
\eeq
Let us note that the averaging surface $\Sigma(w_o, z_s)$, being one of constant $z$, differs from the one of constant $w$ and $\tau$ but -- amusingly -- the same formula holds, locally, for the area element on it, so that Eq. (\ref{dAGLC1}) will also be usable for our light-cone averages defined on $\Sigma(w_o,z_s)$.

This relation of Eq. \rref{dLNonPert} shows that the GLC coordinates are also very well adapted to the computation of the luminosity distance. We can indeed see that this relation involves only the determinant of a GLC metric entry, namely the $2 \times 2$ matrix $\gamma_{ab}$ parametrizing the angular geometry around the source location. This fact will have important consequences on the simplicity of the expression of the averaged flux that will be presented in Sec. \ref{Ch5Sec31}. We now present two methods proving Eq. \rref{dLNonPert}.

\subsection{Proof of Eq. \rref{dLNonPert} -- method 1}
\label{Ch3Sec41}

In our GLC gauge we have $k^\mu=(0, -1/\Upsilon, 0, 0)$ and as already mentioned the 2-sphere embedded in the light-cone is orthogonal to the photon momentum, so that the cross-sectional area element is proportional to $\sqrt{|\gamma|} d \theta^1  d \theta^2$. Therefore, $d_A^2$ can be written as\,:
\beq
d_A^2(w=w_o, \tau_s, \theta^a) = \lim_{\rho \rightarrow 0} ~ \rho^2 \frac{\sqrt{|\gamma_s|}}{\sqrt{|\gamma (\rho)|}} ~~,
\label{d_s2}
\eeq
where $\rho$ is the proper radius of an infinitesimal sphere centered around the observer, $w=w_o$ defines the past light-cone connecting source and observer, and $\tau=\tau_s$ defines the spacelike hypersurface normal to $n^\mu$ at the source position. Eq. (\ref{d_s2}) easily reduces to $d_A = a(t_s) \, r_s$ in the FLRW case of Eq. (\ref{FR}).

Let us now discuss how the well known homogeneous result for $d_L$ is modified when including generic inhomogeneities. We recall, to this purpose, that in a generic metric background the angular distance $d_A$ can be computed by considering the null vector $k^\mu= dx^\mu/ d\la$ tangent to the null ray connecting source and observer (i.e. belonging to the congruence of null geodesics forming the observer's past light-cone). Here $\la$ is an affine parameter along the ray trajectory, chosen in such a way that $\la=0$ at the observer and $\la=\la_s$ at the source position. The expansion $\hat{\Theta}$ of the congruence of null rays is then given by $\hat{\Theta} = \nabla_\mu k^\mu$, and the corresponding angular distance $d_A$ is defined by the differential equation \cite{1985PhR...124..315E,Vanderveld:2007cq}:
\beq
{d \over d \la} \left(\ln d_A\right)= {\hat{\Theta} \over 2}= {1\over 2} \nabla_\mu k^\mu ~~.
\label{210}
\eeq

Consider now a generic metric in the GLC gauge, and the null vector  $k_\mu =\pa_\mu w$ such that $k^\mu = dx^\mu/ d \la= -\da^\mu_\tau \Ups^{-1}$. We have, in this case,
\beq
\nabla_\mu k^\mu= - \Ups^{-1} \pa_\tau \left( \ln \sqrt{\ga}\right)= {d \over d \la} \left(\ln \sqrt{\ga}\right) ~~,
\eeq
where $\ga = \det \ga_{ab}$, and the integration of Eq. (\ref{210}) gives, in general,
\beq
d_A^2(\la)= c \sqrt{\ga(\la)} ~~,
\eeq
where $c$ is independent of $\la$.
In order to fix the constant $c$ we may recall the boundary conditions required to guarantee that the null rays generated by $k^\mu$ belong to the past light cones centered on the world line of our observer. Considering the limit $d_A \ra 0$ (or $\la \ra 0$), and using the already mentioned fact that $\ti \theta^a$ is constant along the null geodesic, such conditions require in particular that \cite{1985PhR...124..315E}
\beq
\lim_{\la \rightarrow 0} \,{\sqrt{\ga}\over d_A^2} = \sin \ti \theta^1 = \sin \theta^1_o ~~,
\eeq
where, in the last equality, we have used $\theta^1_o$ to denote the standard angle on the sphere at the observer position. This clearly fixes $c=1/\sin \tilde{\theta}^1$, and uniquely determines the \gls{angular distance} as
\beq
d_A(\la)= \ga^{1/4}(\la) \left(\sin  \tilde{\theta}^1\right)^{-1/2} ~~.
\label{General_Eq_dA}
\eeq

\subsection{Proof of Eq. \rref{dLNonPert} -- method 2}
\label{Ch3Sec42}

Let us remark that $\gamma_{ab}$ is nothing but the induced metric on the surface $\Sigma(w_o, \tau_s)$ provided this is parametrized in terms of the two ``world-sheet" coordinates $\xi^a \equiv  \tilde{\theta}^a $ and otherwise  given by $w = w_o$, $\tau = \tau_s(w_o, z_s,\tilde{\theta}^a)$. Using the standard definition of an induced metric\,:
\beq
\label{gammaabind}
\gamma_{a b}^{ind} = \frac{\partial x^{\mu}}{\partial \xi^{a}} \frac{\partial x^{\nu}}{\partial \xi^{b}} g^{GLC}_{\mu \nu}(x) ~~,
\eeq
manifestly independent of the spacetime coordinates $x^{\mu}$ being used, we simply find\,:
\beq
\label{gamma=gammaab}
\gamma_{ab}^{ind} = \ga_{ab} ~~,
\eeq
since $w = w_o$ (independently of $\tilde{\theta}^a $) and the only non-zero entry for the metric with a lower index $\tau$ is $g^{GLC}_{\tau w}$.
We can also argue that the area element computed on this surface is orthogonal to the null geodesics and can therefore be identified with $dS_s$ in Eq. (\ref{KS}). Indeed, consider the projection of the photon momentum along the constant-$z_s$ hypersurface (which in this gauge, by Eq. (\ref{redshift}), corresponds to constant $\Ups$ hypersurfaces)\,:
\beq
k_{\mu \|} = k_{\mu} - \frac{k_{\nu} \partial^{\nu} \Ups}{\partial_\sigma \Ups \partial^\sigma \Ups} \partial_{\mu} \Ups ~~,
\eeq
and the particular linear combination 
\beq
n_{\mu}^{(z_s)} =  \alpha ~ \partial_{\mu} w + \beta ~ \partial_{\mu} \Ups ~~,
\eeq
defining the normal to $\Sigma$ lying on the same constant-$z_s$ hypersurface and thus satisfying\,:
\beq
n_{\mu}^{(z_s)}\partial^{\mu} \Ups = 0 ~~~~~ \Rightarrow ~~~~~ (\partial_{\tau} \Ups)~ \alpha = \Ups (\partial_{\mu} \Ups \partial^\mu \Ups)~ \beta ~~.
\eeq
One can easily verify that $n_{\mu}^{(z_s)}$ is exactly parallel to $k_{\mu \|}$. Finally, using their constancy along the null geodesics, we can also identify $\widetilde{\theta}^a$ with  the angular coordinates at the observer's position where, within an infinitesimal region, we can take the metric to be flat. Therefore, as promised\,:
\beq
\label{dAGLC}
d_A^2 = \frac{dS_s}{d^2\Omega_o} = \frac{d^2 \tilde{\theta} \sqrt{\gamma_s}}{d^2 \tilde{\theta} \sin \tilde{\theta}} = \frac{\sqrt{\gamma_s}}{\sin \tilde{\theta}} ~~.
\eeq

\section{Relations of the GLC gauge with other metrics}
\label{Ch3Sec5}

\subsection{Equality between $\tau$ and synchronous gauge (SG) time $t$}
\label{Ch3Sec51}

As we stated in Sec. \ref{Ch3Sec22}, $\tau$ generates a family of geodetic reference observers which exactly coincide with the static ones of the synchronous gauge.
In order to illustrate this point let us consider an arbitrary spacetime metric written in synchronous gauge -- with coordinates $X^{\mu} = (t, X^i)$, $i,j= 1,2,3$ -- where the line element takes the form (already presented in Eq. \rref{SGmetric})\,:
\beq
\label{SGlineElement}
ds_{SG}^2=-dt^2 + g_{ij}dX^i dX^j ~~.
\eeq
Let us impose the condition  $t=\tau$, and check whether we run into any contradiction with the exact metric transformation
\beq
\label{EqBetweenGauges2}
g_{SG}^{\mu\nu}(X)=\frac{\partial X^\mu}{\partial x^\rho} \frac{\partial X^\nu}{\partial x^\sigma} g_{GLC}^{\rho\sigma}(x) ~~.
\eeq
Using Eq. (\ref{GLCmetric}) for the GLC metric, and considering the transformation for the $g_{SG}^{t \mu}$ components, we then obtain the conditions
\beq
g_{SG}^{t\mu}=\{-1,\vec{0}\}=-\left[\partial_{\tau}+\Ups^{-1}(\partial_w+U^a\partial_a)\right]X^{\mu}= u^{\nu}\partial_{\nu}X^{\mu}= \frac{dX^{\mu}}{d\lambda} ~~,
\eeq
where $u_{\mu} = - \partial_{\mu}\tau$ is the four-velocity of the geodesic GLC observer, and $\la$ the affine parameter along the observer world line. So the requirement $\tau=t$ boils down to the statement that along the geodesic flow of the vector field $u^\mu$  the SG coordinates $X^i$ are constant. This clearly defines the coordinate transformation in a non-perturbative way, and also shows that the geodesic observer of the GLC gauge corresponds to a static (and geodesic) observer in synchronous gauge. It follows that the identification $t=\tau$ can always be taken for any spacetime metric, and that this simple connection between GLC and synchronous gauge has validity far beyond the particular FLRW case or  its perturbed generalizations.

\subsection{First-order transformation from synchronous to GLC gauge}
\label{Ch3Sec52}

We will mainly be interested in the following chapters by scalar perturbations of a conformally flat FLRW background and their expression in the Poisson gauge (see Chapter \ref{Chap4}). It is nevertheless interesting to consider, as an example, the first-order transformation between the GLC and the synchronous gauge (SG).

The perturbed metric of a general spacetime can be written in synchronous gauge as described by Eq. \rref{SGlineElement}. When we use cosmic time $t$ and polar coordinates $(r, \theta^a)$ (where $\theta^a=(\theta, \phi)$)), this metric can be parametrized at first order as follows\,:
\beq
g_{SG}^{\mu\nu} = \left(
\begin{array}{ccc}
-1 & 0 & \vec{0} \\
0 & a^{-2} (1 + 2\hat{\Psi}) & a^{-2} S^a \\
\vec{0}^T & a^{-2} (S^a)^T & a^{-2} h^{ab}
\end{array}
\right) ~,
\eeq
where $h^{ab}= \ga_0^{ab} + \ep^{ab}$ (with $a,b =1,2$), $\ga_0^{ab} = r^{-2} {\rm diag} \left( 1, \sin^{-2} \theta \right)$ and where $\hat{\Psi}$, $S^a$, $\ep^{ab}$ are first-order perturbations. These scalar perturbations can be related to the $\ti \psi$ and $\ti E$ perturbations described in Appendix \ref{AppASec33} and thus describe only two scalar degrees of freedom.

Solving perturbatively the inverse transformation of Eq. \rref{EqBetweenGauges2}, i.e. $g_{GLC}^{\mu\nu}(x)=\frac{\partial x^\mu}{\partial X^\rho} \frac{\partial x^\nu}{\partial X^\sigma} g_{SG}^{\rho\sigma}(X)$, for the particular equations $(\mu,\nu) = (\tau,\tau)$, $(w,w)$ and $(w,\ti\theta^a)$, we find that the coordinate transformation from SG to GLC gauge is given to first order by\,:
\bea
\label{SGtoGLC1}
\tau &=& t ~~, \\
\label{SGtoGLC2}
w &=& r + \eta  + \frac12 \int_{\eta_+}^{\eta_-} dx~ \hat{\Psi} (\eta_+, x, \theta^a) ~~, \\
\label{SGtoGLC3}
\ti{\theta}^a &=& \theta^a + \frac{1}{2} \int_{\eta_+}^{\eta_-} dx \left[ S^a (\eta_+, x,\theta^a) + \ga_0^{ab} \pa_b w (\eta_+, x,\theta^a) \right] ~~,
\eea
where the first equality has been proved in the last section and we have used the zeroth order lightcone coordinates $\eta_\pm = \eta \pm r$ with the conformal time $\eta$ defined as usual by $\eta = \int dt/a(t)$.
From Eq. \rref{SGtoGLC2} we get the last term of Eq. \rref{SGtoGLC3}\,:
\beq
\pa_b w = \frac12 \int_{\eta_+}^{\eta_-} dy~ \partial_b \hat{\Psi}(\eta_+,y,\theta^a) ~~.
\eeq

The components $\Ups$, $U^a$ and $\gamma^{ab}$ of the transformed GLC metric $g_{GLC}^{\mu\nu}$ are found, by the use of the $(\mu,\nu) = (\tau,w)$, $(\tau,\ti\theta^a)$ and $(\ti\theta^a,\ti\theta^b)$ equations, to be\,:
\bea
&&
\Ups = a(\eta) \left[ 1  - \frac{1}{2}\hat{\Psi}(\eta_+, \eta_-, \theta^a) +{1\over 2} \hat{\Psi}(\eta_+, \eta_+, \theta^a) - \frac{1}{2}\int_{\eta_+}^{\eta_-} dx \, \partial_+ \hat{\Psi}(\eta_+,x,\theta^a) \right] ~~, \nonumber \\
&& \\
\label{24}
&&
U^a = \frac{1}{2} \left[ S^a(\eta_+, \eta_-, \theta^a) -S^a(\eta_+, \eta_+, \theta^a) + \frac{1}{2} \gamma^{ab}_0(\eta,r,\theta^a) \int_{\eta_+}^{\eta_-} dx \, \pa_b\hat{\Psi}(\eta_+,x,\theta^a) \right. \nonumber \\
& & \left. ~~~~~~~
+ \int_{\eta_+}^{\eta_-} dx \,\partial_+\left(S^a(\eta_+, x, \theta^a) + \frac12 \gamma^{ab}_0(\eta_+, x, \theta^a) \int_{\eta_+}^{x} dy\, \pa_b\hat{\Psi}(\eta_+, y, \theta^a) \right)
\right] ~~, \\
&&
a^2 {\gamma}^{ab} =  h^{ab} +{1\over 2} \gamma_{0}^{ac} \int_{\eta_+}^{\eta_-} dx\, \pa_c \xi^b (\eta_+,\eta_-,\theta^a)+ a \leftrightarrow b ~~,
\eea
where
\beq
\xi^b(\eta_+,\eta_-,\theta^a) = S^b + \frac12
\gamma_{0}^{bc} \int_{\eta_+}^{\eta_-} dx \, \pa_c \hat{\Psi} ~~.
\eeq
From this last equation for $\ga^{ab}$ we can easily compute also the determinant of the $2 \times 2$ matrix $\ga_{ab}$, and we obtain\,:
\beq
{\gamma}^{-1} \equiv \det \ga^{ab}= a^{-4}r^{-4} \sin\theta^{-2} \left[ 1 + \ep^c\,_c+ \gamma_0^c\,_d \int_{\eta_+}^{\eta_-} dx\, \pa_c\xi^d\right] .
\eeq

\subsection{GLC gauge and LTB model}
\label{Ch3Sec53}

Let us show in this section another application of the GLC coordinates defined in Sec. \ref{Ch3Sec2}. We explain how to make a link between this observer adapted system of coordinates and the usual coordinates describing the \gls{LTB} model\footnote{See \cite{Sussman:2011na} for a recent paper about backreaction in LTB model and \cite{Cosmai:2013iga} for a discussion involving LTB and SNe Ia.}. As a first approximation we consider a Universe close to a homogeneous FLRW one and we expand our quantities only up to first order in perturbations around their FLRW limits.

The LTB model, which is a spherically symmetric exact solution of Einstein's equations with a source made of dust, is described by the following metric\,:
\begin{equation}
\label{LTBmetric}
ds^2_{LTB} = - dt^2 + \frac{\partial_r R^2(r,t)}{1+\beta(r)} dr^2 + R(r,t)^2 \left(d \theta^2 + (\sin \theta)^2 d\phi^2 \right)
\end{equation}
where $R(r,t)$ is a function of the cosmic time $t$ and of the radial coordinate $r$, called the \emph{area radius}, and $\beta(r)$ is instead an arbitrary function only of the radial coordinate and is called the \emph{energy function}. Here primes refer to derivation with respect to $r$ and dots with respect to the proper (cosmic) time $t$.

In order to connect the GLC coordinates to a LTB model we restrict ourselves to the simpler case in which the metric of Eq. \rref{LCmetric} has $U^a=0$ (a=1,2). This assumption is justified from the interpretation of $U^a$ as a shift vector made in Sec. \ref{Ch3Sec51} and the spherical symmetry of the LTB model. The GLC metric goes to a homogeneous FLRW Universe if we take $w=\eta+r$, $\tau=t$, $\Upsilon=a(t)$ and $\gamma_{ab}=a^2 \gamma^{(0)}_{ab}$ with $\gamma^{(0)}_{ab} d\theta^a d\theta^b = r^2 (d\theta^2 + (\sin \theta)^2 d\phi^2)$.
In this case, the first order non-trivial perturbations around the FLRW Universe can be written in the following way\,:
\bea
\Ups &=& a(t) (1 + \delta \Ups) ~~, \\
\label{dw}
w &=& (\eta + \chi) + \delta w ~~, \\
\label{dtau}
\tau &=& t + \delta \tau ~~, \\
\gamma_{ab} &=& a(t)^2 (1 + \delta \gamma) \gamma^{(0)}_{ab} ~~,
\eea
where $a(t)$ is the scale factor, $\eta$ the conformal time, $\chi$ the radius defined in Eq. \rref{randchi} by $\chi = \arcsin(\sqrt{K} r) / \sqrt{K}$ and $\gamma^{(0)}_{ab}$ the angular metric defined above.

On the other hand, LTB quantities $g_{tt}$, $R(r,t)$ and $\beta(r)$ have the following limit in the homogeneous FLRW metric\,: $g_{tt} = -1$, $R(r,t)= a \, r$ and $\beta(r)= - K \, r^2$. So they are expanded at first order in perturbation theory as
\bea
g_{tt} = -1 + \delta t ~~,~~ R = a(t) r \, (1 + \delta R(r,t)) ~~,~~ \beta(r) = - K r^2 + \delta \beta(r) ~~,
\eea
with $K$ the curvature of the FLRW background.

Starting from Eqs. \rref{dw} and \rref{dtau} we obtain the following differential relations
\bea
d w &=& \frac{1}{a}(1+a\dot{\delta w}) d t + (\frac{1}{\kappa} + \delta w ')dr ~~, \nonumber \\
d \tau &=& (1+\dot{\delta \tau})dt + \delta \tau' dr ~~,
\eea
where $\kappa \equiv \sqrt{1 - K \, r^2}$.

Using this last equalities, writing down the two metrics in Eqs. \rref{LCmetric} and \rref{LTBmetric} up to first order in the perturbations and comparing the results, we obtain the system of equations that links the perturbations $\delta \Ups$, $\delta w$, $\delta \tau$ and $\delta \gamma$ to the perturbations $\delta R$ and $\delta \beta$.
The system we find is
\bea
-2\dot{\delta \tau} &=& \delta t ~~, \label{Pert1} \\
\delta \Ups + \kappa \delta w' - \frac{\kappa}{a} \delta\tau' &=& \widetilde{\delta R} ~~, \label{Pert2} \\
\delta \Ups + a\dot{\delta w} - \frac{1}{a} \delta\tau' &=& - \frac{\delta t}{2} ~~, \label{Pert3} \\
\delta \gamma &=& 2 \, \delta R \label{Pert4} ~~,
\eea
with $\widetilde{\delta R} = \delta R + r \, \delta R' - \frac{1}{2} \frac{\delta \beta}{1-Kr^2}$.

To this system we could think to add the condition that $w$ is a null-coordinate, that is to say
\beq
g^{\mu\nu}_{LTB}(t,r) \, \partial_{\mu}(\eta + r + \delta w) \, \partial_{\nu}(\eta + r + \delta w) = 0 ~~,
\eeq
which is equivalent to
\beq
\label{wnull}
a\dot{\delta w} - \kappa \delta w' = - \widetilde{\delta R} - \frac{\delta t}{2} ~~. \\
\eeq
Nevertheless, this condition is guaranteed by construction of the GLC metric and this is directly seen from Eqs. \rref{Pert1}, \rref{Pert2} and \rref{Pert3}.

We now restrict our study to the case $\delta t = 0$ and $K=0$ (so $\kappa=1$). From Eq. \rref{Pert1} one obtains $\delta \tau = \delta \tau (r)$ and Eq. \rref{wnull} gives $\delta w$ in terms of $\delta R$ as
\beq
\delta w = -\frac{1}{2}\left[\int^{\eta-r} \widetilde{\delta R}(\tilde{\eta} - \tilde{r}, \eta + r)
\, d(\tilde{\eta} - \tilde{r})\right] + \delta\bar{w}(\eta + r) \equiv
-\frac{1}{2}\left[\int^{\eta_-} \widetilde{\delta R}(\tilde{\eta}_-,\eta_+) \, d\tilde{\eta}_-\right] +
\delta\bar{w}(\eta_+) ~~,
\eeq
where $\eta_+ \equiv \eta + r$, $\eta_- \equiv \eta - r$, and $\delta\bar{w}(\eta_+)$ is an integration function.
This solution determines $\delta \Ups(r,t)$ in terms of $\widetilde{\delta R}$ by using Eq. \rref{Pert2}, after a function is chosen for $\delta \tau (r)$, as\,:
\beq
\delta \Ups = \frac{1}{2} \left[\int^{\eta_-} \partial_{+} \widetilde{\delta R}(\tilde{\eta}_-,\eta_+)
\, d\tilde{\eta}_- + \widetilde{\delta R}(\eta_-,\eta_+)\right] - \partial_{+}\delta\bar{w}(\eta_+) + \frac{1}{a} \delta \tau'
~~.
\eeq
where we made use of the first of the two following relations
\bea
a \partial_t \delta w &=& -\frac{1}{2}\partial_{\eta} \int^{\eta_-} \widetilde{\delta R}(\tilde{\eta}_-,\eta_+)
\, d\tilde{\eta}_- = -\frac{1}{2}\left[\int^{\eta_-} \partial_{+} \widetilde{\delta R}(\tilde{\eta}_-,\eta_+)
\, d\tilde{\eta}_- + \widetilde{\delta R}(\eta_-,\eta_+)\right] + \partial_{+}\delta\bar{w}(\eta_+) ~~, \nonumber \\
\partial_r \delta w &=& -\frac{1}{2}\partial_{r} \int^{\eta_-} \widetilde{\delta R}(\tilde{\eta}_-,\eta_+)
\, d\tilde{\eta}_- = -\frac{1}{2}\left[ \int^{\eta_-} \partial_{+} \widetilde{\delta R}(\tilde{\eta}_-,\eta_+)
\, d\tilde{\eta}_- - \widetilde{\delta R}(\eta_-,\eta_+) \right] + \partial_{+}\delta\bar{w}(\eta_+) ~~. \nonumber
\eea

As a consequence, we have shown that when a perturbed LTB metric is given (i.e. $\delta R(r,t)$ and $\delta\beta(r)$ are given), we can match this metric with a GLC metric up to a function $\delta \tau(r)$ that we can choose at our convenience. In the case where $\delta t \neq 0$, the solution is determined by doing the replacement $\widetilde{\delta R} \rightarrow \widetilde{\delta R} + \frac{\delta t}{2}$ in the integrals. The case $K \neq 0$ is solved by introducing, instead of $r$, the quantity $\chi \equiv \int^r \frac{dr'}{\sqrt{1-Kr'^2}}$ already presented in Eq. \rref{randchi}, so that Eq. \rref{wnull} is written as
\beq
\partial_\eta \delta w - \partial_\chi \delta w = - \widetilde{\delta R} - \frac{\delta t}{2} ~~,
\eeq
and we then use $\eta + \chi$\,, $\eta - \chi$ for the variables $\eta_+$, $\eta_-$ present in the previous solutions.

The results given here are expressed in terms of undetermined integrals and, to be complete, should be associated to initial boundary conditions.

\subsection{Analogy between GLC and ADM metrics}
\label{Ch3Sec54}

In the \gls{ADM} formalism, the Minskowski spacetime $\Mcal_4$ is foliated by constant time hypersurfaces in the following way\,:
\beq
ds_{ADM}^2 = -(N^2 - g_{ij} N^i N^j) dt^2 + 2 g_{ij} dx^i N^j dt + g_{ij} dx^i dx^j
\eeq
with the so-called \emph{\gls{lapse function}} $N$ and \emph{\gls{shift vector}} $N^i$. Physically, the lapse function sets the ``distance'' between two different constant-time hypersurfaces whereas $N^i$ takes care of the change in spatial coordinates $x^i$ when we move from a constant-time hypersurface to another.

\begin{figure}[ht!]
\centering
\begin{tikzpicture}

% The ADM picture
  \node [xshift=-25cm,yshift=2cm] at (current page.south east) {\includegraphics[width=7cm]{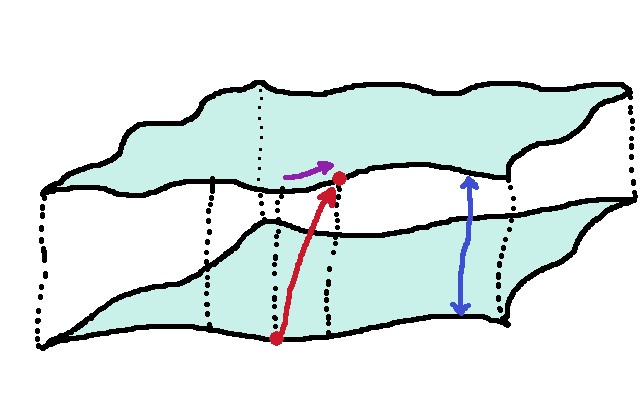}};

\node[anchor=west] at (-3.1,1.3) (description) {$N$};
\node[anchor=west] at (-4.5,2.8) (description) {$N^i$};
\node[anchor=west] at (-6.2,0.2) (description) {$t=t_0$};
\node[anchor=west] at (-6.7,3.1) (description) {$t=t_1$};

% The GLC picture
  \node [xshift=-17cm,yshift=2cm] at (current page.south east) {\includegraphics[width=7cm]{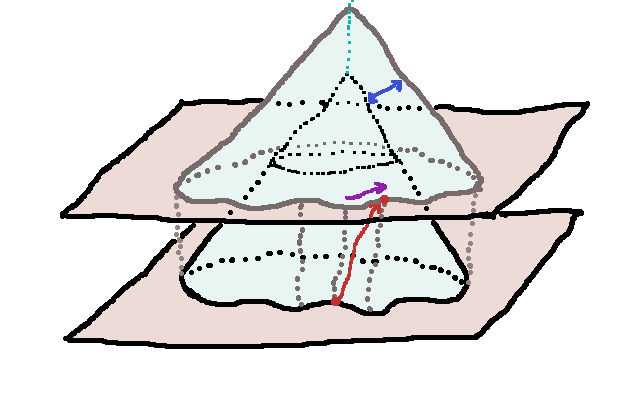}};

\node[anchor=west] at (4.3,3.5) (description) {$\Ups$};
\node[anchor=west] at (3.9,2.2) (description) {$U^a$};
\node[anchor=west] at (0.8,0.1) (description) {$\tau = \tau_0$};
\node[anchor=west] at (0.6,1.4) (description) {$\tau = \tau_1$};

\node[anchor=west,rotate around={45:(1.5,2.5)}] at (0.9,3.7) (description) {\small $w=w_1$};

\node[anchor=east] at (5.2,0.9) (text) {};
\node[anchor=west] at (5.2,0.1) (description) {$\Sigma(w_0,\tau_0)$};
\draw (description) edge[out=180,in=270,|->,thick] (text);

\node[anchor=east] at (5.8,2.3) (text) {};
\node[anchor=west] at (5.5,3.8) (description) {$\Sigma(w_1,\tau_1)$};
\draw (description) edge[out=270,in=0,|->,color=gray,thick] (text);

\end{tikzpicture}
\centering
\caption[Illustration of the comparative forms of the ADM and GLC metrics]{Illustration of the comparative forms of the ADM (left) and GLC (right) metrics. One can see that the GLC metric foliates $\Mcal_4$ by lightcones and spatial hypersurfaces whereas the ADM metric uses only spatial hypersurfaces. $\Ups$ plays the role of a lapse function ($N$ in ADM), $U^a$ the role of the shift vector ($N^i$ in ADM) that takes care of the change in angular coordinates when we go from one 2-sphere $\Sigma(w_0,\tau_0 )$ to another $\Sigma(w_1,\tau_1)$. The angular part of the metric, $\gamma_{ab}$ defines the geometry inside the 2-sphere (as $g_{ij}$ in the 3-dimensional spatial volume in ADM).}
\label{FigADMandGLC}
\end{figure}
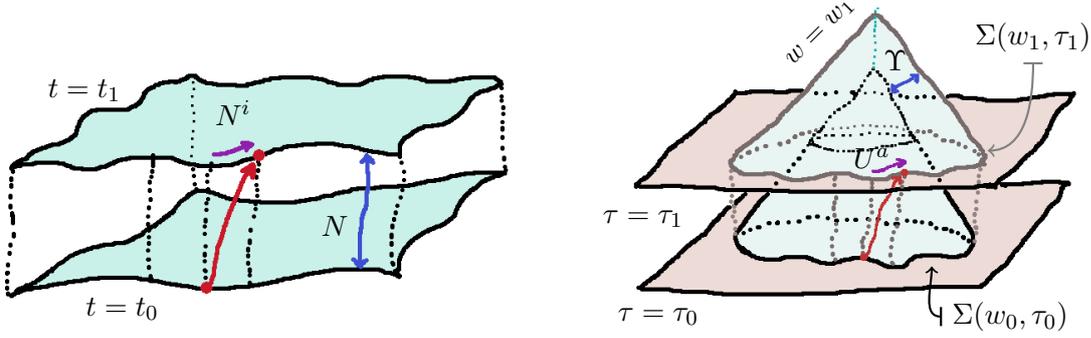

On the other hand, the GLC line element given in Eq. \rref{GLCmetric} is trivially reorganized as follows\,:
\beq
\label{GLCmetricReorg}
ds_{GLC}^2 = (\Ups^2 + \gamma_{ab} U^a U^b) dw^2  - 2 \gamma_{ab} d\tilde\theta^a U^b dw + \gamma_{ab} d\ti{\theta}^a d\ti\theta^b - 2\Ups dw d\tau ~~.
\eeq
Hence we can see that $\Ups$ acts in a way similar to the lapse function by setting the ``distance'' between two lightcones $w=w_0$ and $w=w_1$. We will also see in the study of the redshift drift (see Sec. \ref{Ch3Sec63}) that $\Delta w = \Ups^{-1} \Delta \tau$, which clearly goes in the same line of arguments. On the other hand, $U^a$ plays a role similar to the shift vector, though in a reduced dimensional case, taking care of the change in angular coordinates between two 2-spheres $\Sigma(w_0,\tau_0)$ and $\Sigma(w_1,\tau_1)$ (see Fig. \ref{FigADMandGLC}). One can nevertheless notice a difference between the GLC metric of Eq. \rref{GLCmetricReorg} and the ADM metric\,: it contains an extra piece $- 2\Ups dw d\tau$ related to the fact that we are doing a foliation with both constant-$w$ and constant-$\tau$ coordinates which are intersecting each other. This term shows that $w$ is the coordinate acting like time here whereas $\tau$ looks more nicely comparable with a third spatial component of the ADM metric. We can again remark in this spirit that $\tau$ is playing the role of $y$ in the observational coordinates and, if the foliation of spacetime is here seen as a foliation by null cones, the $\tau = \cst$ hypersurfaces are definitely setting the radial distance along these cones.

\section{Applications to the lightcone average}
\label{Ch3Sec6}

\subsection{Averaging on constant redshift hypersurfaces}
\label{Ch3Sec61}

Let us recall here, in GLC coordinates, that the covariant average of a scalar quantity $S(\tau, w, \ti \theta^a)$ over the compact 2-dimensional surface $\Sg$, defined by the intersection of our past light cone $w=w_o$ with the spacelike hypersurface $\tau= \tau_s$, is simply given (using $\ga \equiv \det \ga_{ab}$) by\,:
\bea
\label{tauaverageexact}
\langle S \re_{w_o, \tau_s} &=& {\int_\Sg d^4 x \sqrt{-g}\, \delta_{\rm D}(w-w_o) \delta_{\rm D}(\tau-\tau_s) \left| \pa_\mu \tau \pa^\mu w\right| S(\tau, w, \ti \theta^a) \over 
\int_\Sg d^4 x \sqrt{-g} \,\delta_{\rm D}(w-w_o) \delta_{\rm D}(\tau-\tau_s) \left| \pa_\mu \tau \pa^\mu w\right|}
\nonumber \\
&=& {\int d^2 \ti \theta \sqrt{ \ga (w_o, \tau_s, \ti \theta^a)} \,S(w_o, \tau_s, \ti \theta^a)\over 
\int d^2 \ti \theta \sqrt{ \ga (w_o, \tau_s, \ti \theta^a)}} ~~.
\eea

To get an average which is still written in terms of the proper time of the geodetic observer, but corresponding to the average over the 2-sphere at constant redshift $z_s$, i.e. $\Sigma(w_o,z_s)$, we can express $\tau_s$ in the last equation in terms of the redshift $z_s$, i.e. $\tau_s = \tau_s (z_s, w_o, \ti \theta^a)$ (or in different words choosing $\tau (z, w_o, \theta^a)$ to be the solution of Eq. \rref{redshift}\,: $\Ups(w_o, \tau , \theta^a) / \Ups_o = 1 / (1 + z_s)$). We arrive at an exact expression for the light-cone average of $S$ as a function of $z_s$\,:
\beq
\label{zaverageexact}
\langle S \re_{w_o, z_s} =  {\int d^2 \ti \theta \sqrt{ \ga (w_o, \tau (z_s, w_o, \ti \theta^a), \ti \theta^b)} \, S(w_o, \tau (z_s, w_o, \ti \theta^a), \ti \theta^b) \over \int d^2 \ti \theta \sqrt{ \ga (w_o,  \tau (z_s, w_o, \ti \theta^a), \ti \theta^b)}} ~~.
\eeq
This average will be of great use in the next chapters as it is a well adapted expression for averaging observational quantities. To emphasize this point, let us present the effect of this change of parametrization of $\Sigma$ for two physical observables related to supernovae\,: the luminosity distance $d_L$ and the redshift $z$ \footnote{Note that their relation has been studied within a gauge invariant approach, for a linearly perturbed FLRW metric, in \cite{Sasaki:1987ad,Kasai:1987ap}.}.

Using the above result of Eq. \rref{tauaverageexact} for $S = d_L$, we obtain, for a general inhomogeneous metric background\,:
\beq
\langle d_L \rangle_{w_o,\tau_s}=\frac{\int d^2\theta  \sqrt{|\gamma(w_o, \tau_s, \theta^a)|}
\left[ \Ups^{2}(w_o, \tau_o, \theta^a)/ \Ups^{2}(w_o, \tau_s, \theta^a)\right]
d_A(w_o, \tau_s, \theta^a)}{\int d^2\theta \sqrt{|\gamma(w_o, \tau_s, \theta^a)|}} ~~,
\label{Int_dL}
\eeq
where $d_A$ is given by Eq. \rref{General_Eq_dA} and we used Eq. \rref{redshift}. Since our coordinates are not pathological near $w=w_o$ and $\tau=\tau_o$ we have $\lim_{\tau\rightarrow \tau_o} \Ups (w=w_o, \tau, \theta^a)=\Ups (w=w_o, \tau_o)\equiv\Ups_o$ independent of $\theta^a$. Hence, in Eq. (\ref{Int_dL}), the factor $\Upsilon(w_o, \tau_o, \theta^a)=\Upsilon_o$ behaves like a constant, leaving a well defined integral of a scalar object over  the 2-sphere.
In our case of interest, namely light-cone averages on surfaces of constant redshift $z= z_s$, one then obtains\,:
\beq
\langle d_L \re_{w_o, z_s} = (1+z_s)^2 {\int d^2 \ti \theta ~ \sqrt{ \ga (w_o, \tau_s (z_s, w_o, \ti \theta^a), \ti \theta^b)} d_A (w_o, \tau_s (z_s, w_o, \ti \theta^a), \ti \theta^b) \over \int d^2 \ti \theta ~  \sqrt{ \ga (w_o, \tau_s (z_s, w_o, \ti \theta^a), \ti \theta^b)}} ~~.
\label{dLaverageexact}
\eeq
This equation, being exact, can be applied in principle to any given highly inhomogeneous cosmology. It can for example be applied to a Lema\^itre-Tolman-Bondi (LTB) model \cite{Lemaitre:1933gd,Tolman:1934za,Bondi:1947av} even in the case of a generic observer shifted away from the symmetry centre of the isotropic geometry.

Let us compare with Eq. \rref{Int_dL} the average of $1 + z$ on the constant $\tau$ surface, which reads\,:
\beq
\langle 1+ z \rangle_{w_o,\tau_s}=\frac{\int d^2\theta \sqrt{|\gamma(w_o, \tau_s, \theta^a)|}~
\left[ \Ups(w_o, \tau_o, \theta^a)/ \Ups(w_o, \tau_s, \theta^a)\right]}
{\int  d^2\theta \sqrt{|\gamma(w_o, \tau_s, \theta^a)|}} ~~.
\label{Int_z}
\eeq
When we move to the constant $z$ average presented in Eq. \rref{zaverageexact}, this relation is trivially satisfied.

The above Eqs. (\ref{Int_dL}) and (\ref{Int_z}) can in principle be used to see how the usual redshift to luminosity-distance relation gets affected by the inhomogeneities of the cosmological geometry. The average of the luminosity distance on constant redshift surfaces, on the other hand, corresponds to an average where the geodesic observers measuring the redshift do not coincide any longer with the ``observers" associated to the flow lines of the reference hypersurface $\Sigma(z)$, chosen to specify the averaging region. In practice, this last expression of Eq. \rref{zaverageexact} will be the one that we will use because of its adaptation to the measured quantities in observational cosmology.

\subsection{Simplified Buchert-Ehlers rules in the GLC gauge}
\label{Ch3Sec62}

In the GLC gauge the averaging integrals introduced in Sec. \ref{Ch2Sec2} greatly simplify, especially in the case where the reference hypersurface $\Sg(A)$ defines a geodesic observer. We will concentrate on the integrals of Eqs. (\ref{null2d}) and (\ref{null3d}) which, setting $V= w$ and $V_0=w_o$, are now given by\,:
\bea
\label{null2d_SLC}
I(S;w_o,A_0;-) &=& \int d^2\theta  dw d\tau \sqrt{|\gamma|}~ \left|\partial_\tau A \right| \delta_{\rm D} (w-w_o) \delta_{\rm D}(A-A_0) S  ~~, \\
\label{null3d_SLC}
I(S;w_o;A_0) &=& \int d^2\theta  dw d\tau \sqrt{|\gamma|}~ \frac{|\partial_\tau A|}{\sqrt{-(\partial A)^2}} \delta_{\rm D}(w-w_o) \Theta( A-A_0 ) S ~~.
\eea
The case of the causally connected spheres of Eq. \rref{null3dInv}, closer to the spatial average definition, will be treated at the end of this section.

For a geodesic reference observer, with $A=\tau$ (and $A_0=\tau_s$), we obtain by the use of the full set of GLC coordinates that the two definitions above reduce to the simple integrals\,:
\bea
\label{null2d_SLC_geodetic}
I(S;w_o,\tau_s;-)&=&\int d^2\theta  \sqrt{|\gamma(w_o, \tau_s, \theta^a)|}~ S(w_o, \tau_s, \theta^a) ~~, \\
\label{null3d_SLC_geodetic}
I(S;w_o;\tau_s)&=&\int d^2\theta d\tau \sqrt{|\gamma(w_o, \tau, \theta^a)|}~ \Theta( \tau-\tau_s ) S(w_o, \tau, \theta^a) ~~.
\eea
This has important consequences and shows one of the many reasons to use the GLC coordinates, as we also saw in Sec. \ref{Ch3Sec61}, but let us now see how the generalized Buchert-Elhers rules introduced in Sec. \ref{Ch2Sec31} simplify in the GLC gauge.

\paragraph{Average on the 2-sphere embedded in the light cone}\

Let us consider separately, as before, derivatives with respect to $A_0$ and $V_0= w_o$. Starting with $A_0$, Eq. \rref{LCeq1} can now be written as
\beq
\partial_{A_0} \lla S \rra_{w_o,A_0} = \lla \frac{\partial_{\tau} S}{\partial_{\tau} A}\rra_{w_o,A_0} + \lla S \frac{\partial_{\tau}\ln\sqrt{|\gamma|}}{\partial_{\tau} A}\rra_{w_o,A_0} - \lla \frac{\partial_{\tau}\ln\sqrt{|\gamma|}}{\partial_{\tau} A}\rra_{w_o,A_0} \lla S \rra_{w_o,A_0} ~~,
\eeq
and for the special case where $\Sigma(A)$ defines a geodesic observer (i.e. $A=\tau$, $A_0=\tau_s$), it reduces to
\beq
\label{413}
\partial_{\tau_s} \lla S \rra_{w_o,\tau_s} = \lla \partial_{\tau} S \rra_{w_o,\tau_s} + \lla S \partial_{\tau}\ln\sqrt{|\gamma|}\rra_{w_o,\tau_s} - \lla \partial_{\tau}\ln\sqrt{|\gamma|} \rra_{w_o,\tau_s} \lla S \rra_{w_o,\tau_s} ~~.
\eeq
Note that, in this last case with $A=\tau$,  the partial derivative with respect to $A_0$ reduces to a standard derivative with respect to the proper (cosmic) time if we consider the limit of a homogeneous FLRW Universe (see Sec. \ref{Ch3Sec21}).

In the geodesic case we have also the following simplification for  the derivative with respect to $w_o$ as given in Eq. (\ref{LCeq3})\,:
\bea
\label{414}
\partial_{w_o} \lla S \rra_{w_o,\tau_s} &=& \lla \partial_{w} S + U^a \partial_a S \rra_{w_o,\tau_s} \\
& & + \lla \left[\partial_{w}\ln\sqrt{|\gamma|} \right] S \rra_{w_o,\tau_s} - \lla \partial_{w}\ln\sqrt{|\gamma|}\rra_{w_o,\tau_s} \lla S \rra_{w_o,\tau_s} \nonumber \\
& & + \lla \left[ \partial_a U^a +U^a \partial_a \ln\sqrt{|\gamma|}\right] S \rra_{w_o,\tau_s} - \lla \partial_a U^a +U^a \partial_a \ln\sqrt{|\gamma|} \rra_{w_o,\tau_s} \lla S \rra_{w_o,\tau_s} ~~. \nonumber
\eea

\paragraph{Average on the truncated light cone}\

For a generic hypersurface we cannot specify $(\pa A)^2$, and Eq. (\ref{LCeq2}) just simplifies as
\beq
\label{415}
\partial_{A_0} \lla S \rra_{w_o}^{A_0} = \lla \frac{\partial_{\tau} S}{\partial_{\tau} A}\rra_{w_o}^{A_0} + \lla \frac{\partial_{\tau}\ln\sqrt{|\gamma|}- \partial_{w}(\partial A \cdot \partial A)}{\partial_{\tau} A} S \rra_{w_o}^{A_0} - \lla \frac{\partial_{\tau}\ln\sqrt{|\gamma|}- \partial_{w}(\partial A \cdot \partial A)}{\partial_{\tau} A}\rra_{w_o}^{A_0} \lla S \rra_{w_o}^{A_0} ~~.
\eeq
\normalsize
However, in the case where $\Sigma(A)$ defines a geodesic observer (with $A=\tau$ and $A_0=\tau_s$) one obtains\,:
\beq
\partial_{\tau_s} \lla S \rra_{w_o}^{\tau_s} = \lla \partial_{\tau} S \rra_{w_o}^{\tau_s} + \lla S \partial_{\tau}\ln\sqrt{|\gamma|} \rra_{w_o}^{\tau_s} - \lla \partial_{\tau}\ln\sqrt{|\gamma|} \rra_{w_o}^{\tau_s} \lla S \rra_{w_o}^{\tau_s} ~~.
\eeq
Comparing with Eq. (\ref{413}) we find, for this last case, the same commutation rule independently of the averaging prescription used.

Finally, for a hypersurface $\Sigma(A)$ associated to a geodesic observer, we  also have the following simplification of Eq. (\ref{LCeq4}):
\bea
\label{416}
\partial_{w_o} \lla S \rra_{w_o}^{\tau_s} &=& \lla \partial_{w} S + U^a \partial_a S \rra_{w_o}^{\tau_s} + \lla \left[ \partial_{w}\ln\sqrt{|\gamma|} \right] S \rra_{w_o}^{\tau_s} - \lla \partial_{w}\ln\sqrt{|\gamma|}\rra_{w_o}^{\tau_s} \lla S \rra_{w_o}^{\tau_s} \nonumber \\
& & + \lla \left[ \partial_a U^a +U^a \partial_a \ln\sqrt{|\gamma|}\right] S \rra_{w_o}^{\tau_s} - \lla \partial_a U^a +U^a \partial_a \ln\sqrt{|\gamma|} \rra_{w_o}^{\tau_s} \lla S \rra_{w_o}^{\tau_s} ~~.
\eea
Hence, we obtain the same commutation rule (see Eq. (\ref{414})) for derivatives with respect to $w_o$.

This is one other nice property of the GLC gauge to make the very different Buchert-Ehlers relations of Eqs. \rref{LCeq1}, \rref{LCeq2} and of \rref{LCeq3}, \rref{LCeq4} take the same form (though keeping their own definitions of the average).

\paragraph{Average on the causally connected sphere}\

We can now consider the effect of the GLC gauge (where $V = w$ and we can choose $A = \tau$) on the integral of Eq. \rref{null3dInv}. In this gauge, this integral simplifies as\,:
\beq
I(1;\tau_s;w_o) = \int d^2 \theta dw \Ups(\tau_s,w,\theta^a) \sqrt{|\gamma(\tau_s,w,\theta^a)|} ~ \Theta (w_o - w) ~~.
\eeq
One can notice by the comparison of this equation with Eq. \rref{null2d_SLC_geodetic} and \rref{null3d_SLC_geodetic} that the full metric determinant $\Ups \sqrt{|\gamma|}$ appears in the integral. This is a normal thing considering the fact that this average is over a 3-dimensional spatial volume whereas the integrals defined in Eq. \rref{null2d_SLC_geodetic} and \rref{null3d_SLC_geodetic} were only including the 2-sphere as part of their spatial integration.
The consideration of the previous equations \rref{LCeq5} and \rref{LCeq6} in the GLC gauge lead to the following simplifications\,:
\bea
\label{417}
\partial_{\tau_s} \lla S \rra_{\tau_s}^{w_o} &=& \lla \partial_{\tau} S \rra_{\tau_s}^{w_o} + \lla S \partial_{\tau}\ln(\Ups \sqrt{|\gamma|}) \rra_{\tau_s}^{w_o} - \lla \partial_{\tau}\ln(\Ups \sqrt{|\gamma|}) \rra_{\tau_s}^{w_o} \lla S \rra_{\tau_s}^{w_o} ~~, \\
\label{418}
\partial_{w_o} \lla S \rra_{\tau_s}^{w_o} &=& \lla \partial_{w} S + U^a \partial_a S \rra_{\tau_s}^{w_o} + \lla \left[ \partial_{w}\ln( \Ups \sqrt{|\gamma|}) \right] S \rra_{\tau_s}^{w_o} - \lla \partial_{w}\ln( \Ups \sqrt{|\gamma|}) \rra_{\tau_s}^{w_o} \lla S \rra_{\tau_s}^{w_o} \nonumber \\
& & + \lla \left[ \partial_a U^a +U^a \partial_a \ln( \Ups \sqrt{|\gamma|}) \right] S \rra_{\tau_s}^{w_o} - \lla \partial_a U^a + U^a \partial_a \ln( \Ups \sqrt{|\gamma|}) \rra_{\tau_s}^{w_o} \lla S \rra_{\tau_s}^{w_o} ~~. ~~~~~ ~~~~~
\eea
We can remark the profound resemblance between these equations and those obtained in the truncated lightcone and the embedded 2-sphere, with the small difference of having replaced $\sqrt{|\gamma|}$ by $\Ups \sqrt{|\gamma|}$. This difference is again due to the fact that the 3-dimensional spacetime determinant of the metric is $\Ups \sqrt{|\gamma|}$. Considering the expression of the volume expansion $\Theta$ given by Eq. \rref{ExpScalarADM} in synchronous gauge, one can directly see that our simplified average of Eq. \rref{417} gives back, after the identification $\tau = t$ valid in this gauge, the Buchert-Ehlers relation generalized in Eq. \rref{SpatialBuchertEhlersInADM}.

\subsection{The redshift drift}
\label{Ch3Sec63}

\begin{figure}[ht!]
\centering
\includegraphics[width=8cm]{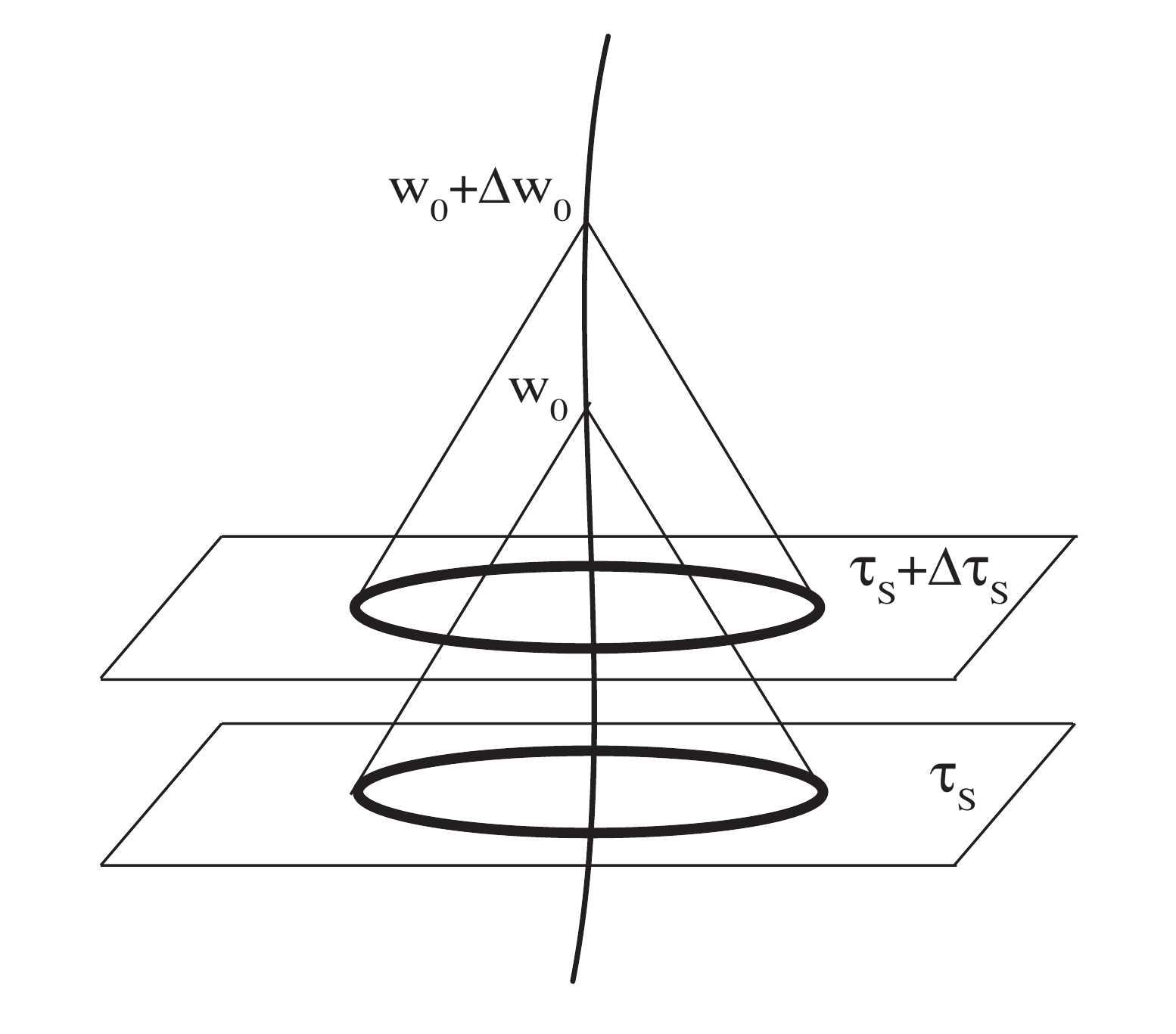}
\caption[A graphic illustration of the redshift drift effect]{A graphic illustration of the redshift drift effect. A possible variation of the cosmological expansion rate is detected by comparing observations performed on two different past light cones.}
\label{fig2}
\end{figure}

The \gls{RD} is defined as the rate of change of the redshift of a given source with respect to the observer's proper time (see \cite{1962ApJ...136..319S,Loeb:1998bu}).
Since both the observer and the source simultaneously evolve in time, the relevant hypersurfaces for the RD effect will consist of two light cones with different bases (see Fig. \ref{fig2}). Assuming that source and observer are in geodesic motion and have negligible peculiar velocities, the RD in a FLRW Universe, and in the proper-time interval $\Da t_0$, can be simply expressed as (see e.g. \cite{Quercellini:2010zr})\,:
\beq
\frac{\Delta z}{\Delta t_0} = (1+z) H_0 - H_s = \frac{\dot{a}_o - \dot{a}_s}{a_s} ~~,
\label{526}
\eeq
where $H_0$ is the present value of the Hubble parameter while $H_s$ is its value at the emission time. Clearly the RD effect is a good indicator of cosmic acceleration as a function of $z$ which {\it does not} use hypothetical standard candles. Nevertheless, in practice it needs observations over several decades.

In order to generalize this expression to a general inhomogeneous Universe (and attempt some averaging of it) let us consider again the geodesic observer of the GLC gauge, with coordinates $x^{\mu} = (\tau, w, \ti\theta^a)$. As shown in Eq. (\ref{redshift}), $z$ is in principle a function of seven independent variables, namely of $w_o$, $\tau_o$, $\ti\theta^a_o$, $\tau_s$ and $\ti\theta^a_s$. Note that, since we are assuming that source and observer are on the same light cone $w=w_o$ at $\tau=\tau_o$, they will be both on the light cone $w = w_o + \Delta w_o$ at $\tau = \tau_o + \Delta \tau_o$, see Fig. \ref{fig2}.
We can also note that $\Ups_o$ is independent of $\theta^a_o$.
As a consequence  we have only five independent contributions to the variation of $1+z$, and we can write, in general\,:
\beq
\Delta(1+z) = \frac{\partial (1+z) }{\partial w_o} \Delta w_o + \frac{\partial (1+z)}{\partial \tau_o} \Delta \tau_o + \frac{\partial (1+z)}{\partial \tau_s} \Delta \tau_s + \frac{\partial (1+z) }{\partial \ti\theta^a_s} \Delta \ti\theta^a_s ~~.
\label{GVZ} 
\eeq

As shown in Eq. (\ref{FR}),  in the homogeneous limit the coordinate $\tau$ goes to the proper (cosmic) time $t$ of the synchronous gauge. So, locally around our geodesic observer, we choose to evaluate the redshift drift $\Da z$ with respect to his/her time parameter $\tau$, and we need, to this purpose, an explicit relation between $\Da \tau_o$ and the variation of the other coordinates involved. For a geodesic observer with $n_\mu= -\pa_\mu \tau$ we have  $\dot{x}^{\mu} \sim  n^\mu=(1/\Upsilon, 1, U^a/\Upsilon)$, so that we can express $\Da x^\mu$ in terms of $\Da w$, at all times, as\,:
\beq
\Delta \tau=\Upsilon \Delta w ~~~~~,~~~~~ \Delta \ti\theta^a=U^a \Delta w ~~.
\label{59}
\eeq
Using Eq. (\ref{59}) we find $\Delta w_o=\Delta \tau_o/\Ups_o=\Delta \tau_s/\Ups_s$, and the final result for the RD can be written in terms of $z$ and its derivatives as\,:
\beq
\frac{ \Delta z}{\Delta \tau_o} = (1+z) \widetilde{H}_0 
+ \frac{1}{\Upsilon_o}{\pa \over \pa w_o} (1+z) 
+ {\pa \over \pa \tau_s} \ln(1+z) 
+ \frac{U^a_s}{\Upsilon_o}{\pa \over \pa \ti\theta^a_s} (1+z) ~~,
\label{Hetother}
\eeq
where\,:
\beq
\widetilde{H}_0\equiv \frac{1}{\Upsilon_o}\frac{\partial \Upsilon_o}{\partial \tau_o} ~~.
\eeq
Let us note again that $\partial_{w_o}$ acts on both metric factors $\Upsilon$ contained in $(1+z)$. This result is valid for a general inhomogeneous metric, and can be compared, as a useful consistency check, with a similar result previously obtained in the particular case of a spherically symmetric geometry \cite{Uzan:2008qp}. If we move from our coordinates to the adapted coordinates used in \cite{Uzan:2008qp} we find that our expression of Eq. (\ref{Hetother}) exactly reduces to the expression for the RD  reported in Eq. (5) of Ref. \cite{Uzan:2008qp}.

The quantities appearing in the above equation can now be averaged over the 2-sphere embedded, at $\tau=\tau_s$, in the light cone $w=w_o$. Using our prescription based on Eq. (\ref{null2d_SLC_geodetic}) we find that both $\Ups_o$ and $\widetilde{H}_0$ can be taken out from the averaging integrals (which are performed over the variables $\ti\theta^a$), and we obtain\,:
\small
\beq
\frac{\lla \Delta z \rra_{w_o, \tau_s}}{\Delta \tau_o} = \lla 1+z \rra_{w_o, \tau_s} \widetilde{H}_0
+ \frac{1}{\Upsilon_o} \lla \partial_{w} (1+z) \rra_{w_o, \tau_s} 
+ \lla \partial_{\tau} \ln(1+z) \rra_{w_o, \tau_s} 
+ \frac{1}{\Ups_o} \lla U^a \partial_a (1+z) \rra_{w_o, \tau_s} ~~.
\eeq
\normalsize
For the derivatives performed with respect to the hypersurface parameters $w$ and $\tau$ we can now apply the Buchert-Ehlers commutation rules in the simplified form of Sec. \ref{Ch3Sec62}. Using in particular Eqs. (\ref{413}) and (\ref{414}) we are lead to\,:
\beq
\frac{\lla \Delta z \rra_{w_o, \tau_s}}{\Delta \tau_o} = \lla 1+z \rra_{w_o, \tau_s} \widetilde{H}_0 + \frac{1}{\Upsilon_o} \partial_{w_o}  
\lla 1+z \rra_{w_o, \tau_s} + \partial_{\tau_s} \lla \ln(1+z) \rra_{w_o, \tau_s} + Q_w(z) + Q_\tau(z) ~~,
\label{533}
\eeq
where
\bea
Q_\tau(z) &=& - \left\langle \ln(1+z) \pa_\tau \ln\sqrt{|\ga|} \right\re_{w_o, \tau_s}
+\left\langle \ln (1+z) \right\re_{w_o, \tau_s} 
\left\langle \pa_\tau \ln\sqrt{|\ga|} \right\re_{w_o, \tau_s} ~~, \\
Q_w(z) &=& \frac{1}{\Upsilon_o} \left\{-\left\l (1+z) \pa_w \ln\sqrt{|\ga|}  \right\re_{w_o, \tau_s} - \left\l (1+z)
\left[ \pa_a U^a+U^a \partial_a \ln\sqrt{|\gamma|} \right]  \right\re_{w_o, \tau_s}
\right.
\nonumber \\
& &\left. +\l 1+z  \re_{w_o, \tau_s} 
\left[\left\l \pa_w \ln\sqrt{|\ga|}  \right\re_{w_o, \tau_s} + \left\langle
 \pa_a U^a+U^a \partial_a \ln\sqrt{|\gamma|} \right\re_{w_o, \tau_s}\right]\right\} ~~.
\eea
These last terms, together with the others of Eq. (\ref{533}) control the difference between the {\em averaged redshift drift} and the {\em drift} for the {\em averaged redshift} and represent a light-cone analog of the backreaction terms due to the geometric inhomogeneities, first computed by Buchert \cite{2000GReGr..32..105B,2001GReGr..33.1381B} in the context of spacelike averages. Unlike Buchert's terms, however, the backreaction in this case is physically controlled by the geometry of a 2-dimensional surface, with metric $\ga_{ab}$. Also, the possible physical meaning of these backreaction terms is in principle strongly model-dependent, and their physical effects are to be explicitly computed for any given model chosen to parametrize deviations from FLRW geometry.

For a homogeneous FLRW metric the averages disappear, and the backreaction terms $Q_\tau$ and $Q_w$ are identically vanishing (see also Sec. \ref{Ch2Sec4}). In that case $\widetilde{H}_0=H_0$ and, using the homogeneous limit of Eq. (\ref{FR}), one also finds $\pa_w(1+z)=0$ and $\pa_\tau \ln(1+z)_s=-H_s$, so that Eq. (\ref{533})  gives back the result of Eq. (\ref{526}). However, for inhomogeneous models (such as the \gls{LTB} Universe \cite{Lemaitre:1933gd,Tolman:1934za,Bondi:1947av}, see in Sec. \ref{Ch3Sec53}), we see that several new contributions appear. These backreaction effects may be relevant for determining the kinematics of the Universe (and the equation of state of its energy components as a function of $z$) from the forthcoming RD experiments (see e.g. \cite{2005Msngr.122...10P}).

% --------------------------------------------------------- %

%  --------------------- Chapter 4 ----------------------   %
% --------------------------------------------------------- %
\chapter{The luminosity-redshift relation at second order}
\label{Chap4}
\pagestyle{plain}

In this section, we present the computation, in the Poisson (or longitudinal) gauge, of the luminosity distance at second order in perturbations. For that we start by doing the transformation at second order between the GLC and the Poisson gauge (PG), expressing the GLC gauge coordinates in terms of the PG coordinates and metric. We then impose the condition that the sources lie on $\Sigma(w_o,z_s)$ and finally invert the second order transformation to express the luminosity distance in terms of the $\ti{\theta}^a$ angles. We conclude with a brief physical description of its enclosed terms. Let us notice for the reader not interested by this technical (but physically very relevant) calculation that the essential of the discussion presented here will be sketched at first order at the beginning of Chapter \ref{Chap5}.

\section{From Poisson to GLC gauge at second order}
\label{Ch4Sec1}

\subsection{The Poisson gauge}
\label{Ch4Sec11}

Let us consider a non-homogeneous spacetime approximated by a spatially flat FLRW Universe plus scalar, vector and tensor perturbations. 
In the so-called \gls{PG} \cite{1995STIN...9622249B,Bombelli:1994zh}, a generalization of the Newtonian (or longitudinal) gauge beyond first order, the corresponding metric takes the following standard form in cartesian coordinates\,:
\beq
ds_{PG}^2 = a^2(\eta) \left( -(1+ 2 \Phi) d\eta^2 + 2 \omega_i \, d\eta d x^i + \left[(1- 2 \Psi)\delta_{ij} + h_{ij}   \right] dx^i dx^j \right) ~~,
\label{PGmetricstandard}
\eeq
where $\Phi$ and $\Psi$ are scalar perturbations, $\omega_i$ is a transverse vector ($\partial^i \omega_i=0$) and $h_{ij}$ is a transverse and traceless
tensor ($\partial^i h_{ij}=0=h^i_{~i}$). This metric depends on six arbitrary functions, hence it is completely gauge fixed. Up to second order the (generalized) Bardeen potentials $\Phi$ and $\Psi$ are defined as follows \cite{Mukhanov:1990me}\,:
\beq
\label{PhiAndPsiInPG}
\Phi \equiv \psi + \frac{1}{2} \phi^{(2)} ~~,~~ \Psi \equiv \psi + \frac{1}{2} \psi^{(2)} ~~,
\eeq
where we have assumed, for simplicity, that there is no (or a negligible) anisotropic stress of the matter distribution, thus setting $\Psi = \Phi = \psi$ at first order. We will consider $\omega_i$ and $h_{ij} $ to be second order quantities as, in inflationary cosmology, first order scalar perturbations are the dominant contributions (see App. \ref{AppASec23}). On the other hand vector and tensor perturbations are automatically generated from scalar perturbations at second order (see e.g. \cite{Matarrese:1997ay,Bartolo:2005kv}).

\subsection{The second order transformation for scalar perturbations}
\label{Ch4Sec12}

We intend to write here the second order transformation between the GLC gauge and the PG. Before carrying on, let us mention that using the GLC approach means that we are taking many physical effects into account already at the level of the metric. In this process, the geodesic equations for the observer and light rays are solved non-perturbatively, and their solutions are expressed in terms of the $\tau,w$ coordinates. The outcome is that physical phenomena that we will describe later -- such as redshift perturbations (RP), redshift space distortions (RSD), Sachs-Wolfe effect (SW), integrated Sachs-Wolfe effect (ISW), peculiar velocities, lensing and others -- are manifestly encoded in the metric, and are derived from a coordinate transformation. This is different from the usual approach, which first takes some perturbed metric, and then solves the geodesic equations order by order to construct physical observables (see, for example, \cite{Bonvin:2005ps,DiDio:2012bu}).

Since by our assumption vector and tensor modes appear only at second order and, as a consequence, will be decoupled from scalar perturbations,  
we can neglect them momentarily and add them at the end. Therefore, considering only scalar perturbations in spherical coordinates $(r, \theta^a)=(r, \theta, \phi)$, the PG defined in Eq. (\ref{PGmetricstandard}) becomes the \gls{NG}, with inverse metric\,:
\beq
g_{PG}^{\mu \nu} = a(\eta)^{-2} {\rm diag}( -1 + 2 \tilde{\Phi}, 1 + 2 \tilde{\Psi}, (1 + 2 \tilde{\Psi}) \gamma_0^{ab} ) ~~,
\eeq
where
\beq
\ga_0^{ab} = {\rm diag} \left( r^{-2}, r^{-2} \sin^{-2} \theta \right) ~~~~~,~~~~~ \tilde{\Phi} = \psi + \frac{1}{2} \phi^{(2)} - 2 \psi^2 ~~~~~,~~~~~ \tilde{\Psi} = \psi + \frac{1}{2} \psi^{(2)} + 2 \psi^2 ~~.
\eeq
We compute the GLC gauge (inverse) metric through\,:
\beq
g_{GLC}^{\rho\sigma}(x)=\frac{\partial x^\rho}{\partial y^\mu} \frac{\partial x^\sigma}{\partial y^\nu} g_{PG}^{\mu\nu}(y) ~~,
\label{EqBetweenGauges}
\eeq
where we indicate with $y^\mu=(\eta, r, \theta^a)$ the PG coordinates and with $x^\nu=(\tau, w,\ti{\theta}^a)$ the GLC ones. Let us also introduce the useful (zeroth-order) light-cone variables with corresponding derivatives\,:
\beq
\eta_\pm= \eta \pm r ~~~\mbox{ with\,: }~~~~~ \pa_\eta = \pa_+ + \pa_- ~~~,~~~~~ \pa_r = \pa_+ - \pa_- ~~~,~~~~~\pa_\pm= {\pa \over \pa \eta_\pm}={1\over 2} \left( \pa_\eta \pm \pa_r \right) ~~.
\eeq

Using these variables we solve the four differential equations obtained from Eq. (\ref{EqBetweenGauges}) for the components $g_{GLC}^{\tau \tau} = -1$, $g_{GLC}^{ww} = 0$, $g_{GLC}^{wa} = 0$, by imposing the boundary conditions that $i)$ the transformation is non singular around $r=0$, and $ii)$ that the 2-dimensional spatial sections $r=$ cst are locally parametrized at the observer position by standard spherical coordinates, $\ti \theta^a(0)= \theta^a=(\theta, \phi)$.

To this purpose we also introduce the following auxiliary quantities\,:
\beq
P(\eta, r, \theta^a) = \int_{\eta_{in}}^\eta d\eta' \frac{a(\eta')}{a(\eta)} \psi(\eta',r,\theta^a)
\,\,\,\,\,\,\,\,,\,\,\,\,\,\,\, Q(\eta_+, \eta_-, \theta^a) = \int_{\eta_+}^{\eta_-} dx~ \hat{\psi}(\eta_+,x,\theta^a) ~~,
\label{PQ}
\eeq
where, hereafter, we use a hat to denote a quantity expressed in terms of $(\eta_+,\eta_-,\theta^a)$ variables, for instance $\hat{\psi}(\eta_+,\eta_-,\theta^a) \equiv \psi(\eta,r,\theta^a)$.
The sought for transformation can then be written, to second order in perturbation theory and with self-explanatory notations, 
as follows\,:
\bea
\label{tau2order}
\tau &=& \tau^{(0)}+\tau^{(1)}+\tau^{(2)} \\
&\equiv& \int_{\eta_{in}}^\eta d\eta' a(\eta') + a(\eta) P(\eta, r, \theta^a) 
+ \int_{\eta_{in}}^\eta d\eta' \frac{a(\eta')}{2} \left[ \phi^{(2)} - \psi^2 + ( \partial_r P )^2 + \gamma_0^{ab} ~ \partial_a P ~ \partial_b P \right] (\eta', r, \theta^a)~, \nonumber \\
\label{w2order}
w &=& w^{(0)}+w^{(1)}+w^{(2)} \\
&\equiv& \eta_+ + Q(\eta_+, \eta_-, \theta^a) + \frac{1}{4} \int_{\eta_+}^{\eta_-} dx~ \left[ \hat{\psi}^{(2)} + \hat{\phi}^{(2)} + 4 \hat{\psi} \partial_+ Q + \hat{\gamma}_0^{ab} ~ \partial_a Q ~ \partial_b Q \right] (\eta_+, x, \theta^a) ~~, \nonumber \\
\label{thetatilde2orderShort}
\tilde{\theta}^a &=& \tilde{\theta}^{a (0)}+\tilde{\theta}^{a (1)}+\tilde{\theta}^{a (2)} \\
&\equiv& \theta^a + \frac12 \int_{\eta_+}^{\eta_-} dx~ \left[ \hat{\gamma}_0^{ab} \partial_b Q \right] (\eta_+,x,\theta^a) 
+ \int_{\eta_+}^{\eta_-} dx~ \left[ \hat{\gamma}_0^{ac} \zeta_c + \hat{\psi} ~ \xi^a + \lambda^a \right] (\eta_+,x,\theta^a) ~~, \nonumber
\eea
where $\eta_{in}$ represents an early enough time when the perturbation (or better the integrand) was negligible. In other words, all the relevant integrals (i.e. for all scales of interest) are insensitive to the actual value of $\eta_{in}$. Furthermore, we have used the following shorthand notations\,:
\bea
\label{ExprChi}
\zeta_c(\eta_+,x,\theta^a) &\equiv& \frac12 \partial_c w^{(2)} (\eta_+,x,\theta^a) \\
&=& \frac18 \int_{\eta_+}^x du~ \partial_c \left[ \hat{\psi}^{(2)} + \hat{\phi}^{(2)} + 4 \hat{\psi} ~ \partial_+ Q + \hat{\gamma}_0^{ef} ~ \partial_e Q ~ \partial_f Q \right] (\eta_+,u,\theta^a) ~~, \nonumber
\\
\label{ExprXi}
\xi^a(\eta_+,x,\theta^a) &\equiv& \partial_+ \tilde{\theta}^{a(1)}(\eta_+,x,\theta^a) +2\partial_x\tilde{\theta}^{a(1)}(\eta_+,x,\theta^a) \\
&=& \partial_+ \left( \frac12 \int_{\eta_+}^x du~ [\hat{\gamma}_0^{ac} \partial_c Q] (\eta_+,u,\theta^a) \right) + [\gamma_0^{ac} \partial_c Q] (\eta_+,x,\theta^a) ~~, \nonumber
\\
\label{ExprLamb}
\lambda^a(\eta_+,x,\theta^a) &\equiv& \partial_x \tilde{\theta}^{d(1)} (\eta_+,x,\theta^a)\left(\partial_d \tilde{\theta}^{a(1)} (\eta_+,x,\theta^a)-\delta_d^a  \partial_+ Q (\eta_+,x,\theta^a)\right) \\
&=& \frac14 [ \hat{\gamma}_0^{dc} ~ \partial_c Q ] (\eta_+,x,\theta^a) \left( \int_{\eta_+}^x du ~ \partial_d \left[ \hat{\gamma}_0^{ae} \partial_e Q \right] (\eta_+,u,\theta^a) \right) - \frac12 \left[ \partial_+ Q ~ \hat{\gamma}_0^{ab} \partial_b Q \right] (\eta_+,x,\theta^a) ~~. \nonumber
\eea
Let us now compute the various non-trivial entries of the GLC metric. Let us use on one side that $\Ups^{-1} = - \partial_{\mu} w ~\partial_{\nu} \tau ~g_{PG}^{\mu \nu}$ and take the following expansion\,:
\beq
\Ups^{-1} = \frac{1}{a(\eta)} \left( 1 + \epsilon^{(1)} + \epsilon^{(2)} \right) ~~.
\label{InvUps}
\eeq
In terms of quantities implicitly defined in Eqs. \rref{tau2order}-(\ref{w2order}) we find\,:
\bea
\epsilon^{(1)} &=& \partial_+ Q - \partial_r P ~~,
\label{eps1} \\
\epsilon^{(2)} &=& \partial_{\eta}w^{(2)} + \frac{1}{a}(\partial_\eta - \partial_r) \tau^{(2)} - \psi \partial_{\eta}Q - \phi^{(2)} + 2\psi^2 - \partial_r P \partial_r Q - 2 \psi \partial_r P - \gamma^{ab}_0 \partial_a P \partial_b Q ~~,~~~~~
\label{eps2}
\eea
where
\bea
\partial_+ Q (\eta_+,\eta_-,\theta^a) &=& - \hat{\psi}(\eta_+,\eta_+,\theta^a) + \int_{\eta_+}^{\eta_-} dx \, \partial_+ \hat{\psi}(\eta_+,x,\theta^a) ~~, \\
\partial_r P(\eta,r,\theta^a) &=& \int_{\eta_{in}}^\eta d\eta' \frac{a(\eta')}{a(\eta)} \partial_r \psi(\eta',r,\theta^a) ~~,
\eea
and $\hat{\psi}(\eta_+,\eta_+, \theta^a)$ denotes (at any $\eta$) the value of the perturbation potential evaluated at the tip of the light-cone connecting the origin ($r=0$) to the point $y^\mu=(\eta, r, \theta^a)$ (namely, $\hat{\psi}(\eta_+,\eta_+, \theta^a) = \psi(\eta+r, 0, \theta^a)$ with $r = 0$).
The full explicit expression for $\epsilon^{(2)}$ can be then written as follows\,:
\bea
\epsilon^{(2)} &=&  \frac{1}{4} \left( \psi^{(2)} - \phi^{(2)} \right) + \frac{\psi^2}{2}
+ \frac12 (\partial_r P)^2  - (\psi + \partial_+ Q) \, \partial_r P \nonumber \\
&+& \frac14 \gamma_0^{ab}\left( 2\partial_a P \, \partial_b P + \partial_a Q \, \partial_b Q  - 4  \partial_a Q \, \partial_b P \right) \nonumber \\
&+& \frac14~ \partial_+ \int_{\eta_+}^{\eta_-} dx~  \left[ \hat{\psi}^{(2)} + \hat{\phi}^{(2)} + 4 \hat{\psi} ~ \partial_+ Q + \hat{\gamma}_0^{ab} \partial_a Q \, \partial_b Q \right] (\eta_+,x,\theta^a) \nonumber \\
&-& \frac12 \int_{\eta_{in}}^\eta d\eta'~ \frac{a(\eta')}{a(\eta)} \partial_r \left[ \phi^{(2)} - \psi^2 + \left( \partial_r P \right)^2 + \gamma_0^{ab} \partial_a P \, \partial_b P \right] (\eta',r,\theta^a) ~~,
\label{eps2Full} 
\eea
where the variables $(\eta,r, \theta^a)$ have been omitted for the sake of conciseness.
The computation of the GLC functions $U^a$ gives\,:
\bea
U^a &=&-\left\{-\partial_{\eta}\tilde{\theta}^{a (1)}+\frac{1}{a}\gamma_0^{ab}\partial_b \tau^{(1)}-\partial_{\eta}\tilde{\theta}^{a (2)} +\frac{1}{a}\gamma^{ab}_0\partial_b \tau^{(2)} + \frac{1}{a} \partial_r \tau^{(1)} \partial_r \tilde{\theta}^{a(1)} \right. \cr
& +& \left. \psi \left(\partial_\eta \tilde{\theta}^{a (1)}+\frac{2}{a}\gamma^{ab}_0 \partial_b \tau^{(1)} \right)+\frac{1}{a}\gamma_0^{cd}\partial_c \tau^{(1)} \partial_d \tilde{\theta}^{a (1)}-\epsilon^{(1)} \left(-\partial_{\eta} \tilde{\theta}^{a (1)}+\frac{1}{a}\gamma^{ab}_0 \partial_b \tau^{(1)}\right)\right\} ~~,~~~~~
\label{Ua}
\eea
where $\tau^{(1),(2)}$ and $\tilde \theta^{a (1),(2)}$ are implicitly defined in the coordinate transformations of Eqs. (\ref{tau2order}) and (\ref{thetatilde2orderShort}). The vector $U^a$ is a measure of anisotropy of spacetime in GLC coordinates. Substituting in Eq. \rref{Ua} the explicit values of $\tau^{(1),(2)}$ and $\tilde{\theta}^{a (1),(2)}$, we obtain the explicit expression for $U^a$\,:
\bea
U^a &=& \Bigg\{ \gamma_0^{ab} \left( \frac12 \partial_b Q - \partial_b P \right) + \partial_+ \left( \frac12 \int_{\eta_+}^{\eta_-} dx ~ \left[ \hat{\gamma}_0^{ab} ~ \partial_b Q \right] (\eta_+,x,\theta^a) \right) \Bigg\} \nonumber \\
&+& \Bigg\{ - \left( \psi + \partial_+ Q \right) \partial_+ \left( \frac12  \int_{\eta_+}^{\eta_-} dx ~ \left[ \hat{\gamma}_0^{ab} ~ \partial_b Q \right] (\eta_+,x,\theta^a) \right) + \left( - \psi + 2 \partial_r P - \partial_+ Q \right) \frac12 \gamma_0^{ab} ~ \partial_b Q \nonumber \\
& & ~~~~~ - 2 \psi \gamma_0^{ab} ~ \partial_b P - \frac12 \gamma_0^{cd} ~ \partial_c P \int_{\eta_+}^{\eta_-} dx~ \partial_d \left[ \hat{\gamma}_0^{ab} ~ \partial_b Q \right] (\eta_+,x,\theta^a) + ( \partial_+ Q - \partial_r P ) \gamma_0^{ab} \partial_b P \nonumber \\
& & ~~~~~ - \frac12 \gamma_0^{ab} \int_{\eta_{in}}^{\eta} d\eta' \frac{a(\eta')}{a(\eta)} \partial_b \left[ \phi^{(2)} - \psi^2 + (\partial_r P)^2 + \gamma_0^{cd} \partial_c P \partial_d P \right](\eta',r,\theta^a) \nonumber \\
& & ~~~~~ + (\partial_+ + \partial_-) \int_{\eta_+}^{\eta_-} dx~ \left[ \hat{\gamma}_0^{ac} \zeta_c + \hat{\psi} ~ \xi^a + \lambda^a \right] (\eta_+,x,\theta^a) \Bigg\} ~~.
\eea
Finally, starting from $\gamma^{a b} = \frac{\partial \tilde{\theta}^a}{\partial y^\mu} \frac{\partial \tilde{\theta}^b}{\partial y^\nu} g_{PG}^{\mu\nu}(y)$, we find\,:
\bea
a(\eta)^2 \gamma^{ab} &=& \gamma_0^{ab} \left(1 +  2 \psi\right) + \left[\gamma_0^{a c} \partial_c \tilde{\theta}^{b (1)}+ (a\leftrightarrow b) \right] \nonumber \\
&+& \gamma_0^{ab} \left(\psi^{(2)} + 4 \psi^2 \right)-\partial_\eta \tilde{\theta}^{a (1)}\partial_\eta \tilde{\theta}^{b (1)}+\partial_r \tilde{\theta}^{a (1)}\partial_r \tilde{\theta}^{b (1)} \nonumber \\
&+& 2 \psi \left[\gamma_0^{a c} \partial_c \tilde{\theta}^{b (1)}+ (a\leftrightarrow b) \right]+\gamma_0^{c d} \partial_c \tilde{\theta}^{a (1)}
 \partial_d \tilde{\theta}^{b (1)}+\left[\gamma_0^{a c} \partial_c \tilde{\theta}^{b (2)}+ (a\leftrightarrow b) \right] ~~.
\label{gammaab1}
\eea
More explicitly, in terms of the quantities defined in Eqs. (\ref{PQ}) and (\ref{ExprChi})-(\ref{ExprLamb})\,:
\bea
a(\eta)^2 \gamma^{ab} &=& \gamma_0^{ab} \left(1 +  2 \psi\right) + \frac12 \bigg\{\gamma_0^{ad} \int_{\eta_+}^{\eta_-} dx~ \partial_d \left[ \hat{\gamma}_0^{bc} \partial_c Q \right] (\eta_+,x,\theta^a) + (a\leftrightarrow b) \bigg\} \nonumber \\
&+& \left(\psi^{(2)} + 4 \psi^2 \right) \gamma_0^{ab} - \bigg\{ \gamma_0^{ac} \partial_c Q ~ \partial_+ \left(\frac12 \int_{\eta_+}^{\eta_-} dx~  \left[ \hat{\gamma}_0^{bd} ~ \partial_d Q \right] (\eta_+,x,\theta^a) \right) + (a\leftrightarrow b) \bigg\} \nonumber \\
&+& \psi \bigg\{ \gamma_0^{ad} \int_{\eta_+}^{\eta_-} dx~ \partial_d \left[ \hat{\gamma}_0^{bc} \partial_c Q \right] (\eta_+,x,\theta^a) + ( a \leftrightarrow b ) \bigg\} \nonumber \\
&+& \frac{1}{4} \gamma_0^{cd} \bigg(\int_{\eta_+}^{\eta_-} dx~ \partial_c \left[ \hat{\gamma}_0^{ae} \partial_e Q \right] (\eta_+,x,\theta^a) \bigg) \bigg( \int_{\eta_+}^{\eta_-} d\bar{x}~ \partial_d \left[ \hat{\gamma}_0^{bf} \partial_f Q \right] (\eta_+,\bar{x},\theta^a) \bigg) \nonumber \\
&+& \bigg\{ \gamma_0^{ac} \int_{\eta_+}^{\eta_-} dx~ \partial_c \left[ \hat{\gamma}_0^{bd} ~ \zeta_d + \hat{\psi} ~ \xi^b + \lambda^b \right] (\eta_+,x,\theta^a) + ( a \leftrightarrow b ) \bigg\} ~~.
\label{gammaab2}
\eea

\subsection{The second order transformation for vector and tensor perturbations}
\label{Ch4Sec13}

As already mentioned we can add the contributions of the tensor and vector perturbations by considering them separately. 
Using spherical coordinates $\{x^i\} = (r, \theta^a)\equiv(r, \theta, \phi)$, the tensor and vector part of the PG metric defined in Eq. \rref{PGmetricstandard} can be rewritten as\,:
\beq
\label{NewPGmetric}
ds_{PG}^2 = a^2(\eta) \left[ -d\eta^2 + 2 v_i d\eta d x^i + [(\gamma_0)_{ij} + \chi_{ij}] d x^i d x^j \right] ~~,
\eeq
corresponding to\,:
\beq
\label{PGgmunu}
g_{PG}^{\mu\nu}(\eta,r,\theta^a) = a^{-2}(\eta)
\left(
\begin{array}{cc}
-1 & v^i \\
v^j & \gamma_0^{ij} - \chi^{ij} \\
\end{array}
\right) ~~,
\eeq
and where $\gamma_0^{ij} = {\rm diag}(1, r^{-2}, r^{-2} (\sin\theta)^{-2} )$ is the (inverse) flat 3-metric. Here $v^i$ and $\chi^{ij}$  are the vector and 
tensor perturbations in spherical coordinates equivalent to the more standard definition $\omega^i$ and $h^{ij}$ used in cartesian coordinates of Eq. \rref{PGmetricstandard}. They satisfy $\nabla_i v^i=\nabla_i \chi^{i j}=0$ and $(\gamma_0)_{ij} \chi^{ij}=0$ with $\nabla_i$ the flat 3-dimensional covariant derivative in spherical coordinates.

Proceeding as for the scalar part of the metric, we first note that $\tau$ is not affected by vector and tensor perturbations\,:
\beq
\label{tauVT}
\tau = \int_{\eta_{in}}^\eta d\eta' a(\eta') ~~, 
\eeq
while the light-cone coordinate $w$ is\,:
\beq
\label{wVT}
w = \eta_+  + Q^{(\alpha)}(\eta_+,\eta_-,\theta^a) ~~,
\eeq
and where
\beq
\label{Qalpha}
Q^{(\alpha)}(\eta_+,\eta_-,\theta^a) = \int_{\eta^+}^{\eta^-} dx ~ \hat{\alpha}^r (\eta^+,x,\theta^a)~~~~~ \mbox{ with } ~~~~~ \alpha^r \equiv \frac{v^r}{2} - \frac{\chi^{rr}}{4} ~~.
\eeq
As before, hats mean that $(\eta_+,\eta_-,\theta^a)$-coordinates are used.
Finally, we find\,:
\bea
\label{thetaVT}
\tilde{\theta}^a &=& \tilde{\theta}^{a (0)}+\tilde {\theta}^{a (2)} \\
&=& \theta^a + \frac{1}{2} \int_{\eta_+}^{\eta-} dx \left( [\hat{v}^a - \hat{\chi}^{ra}](\eta_+,x,\theta^a) + \hat{\gamma}^{ab}_0(\eta_+,x,\theta^a) \int_{\eta_+}^x dy~ \partial_b \hat{\alpha}^r (\eta_+,y,\theta^a) \right) ~~. \nonumber
\eea
Following the same steps as for scalar perturbations, we then compute the non-trivial entries of the GLC metric\,:
\small
\bea
a(\eta) \Ups^{-1} &=& 1 - \frac{1}{2} v^r - \frac{1}{4} \chi^{rr} + \partial_+  \int_{\eta_+}^{\eta_-} dx~  \hat{\alpha}(\eta_+,x,\theta^a) ~~,
\label{UpsVT} \\
a(\eta)^2 \gamma^{ab} &=& \gamma_0^{ab} - \chi^{ab} \\
& +& \left[ \frac{\gamma_0^{ac}}{2} \int_{\eta_+}^{\eta_-} dx~ \partial_c \left( [\hat{v}^b - \hat{\chi}^{rb}](\eta_+,x,\theta^a) + \hat{\gamma}_0^{bd} (\eta_+,x,\theta^a) \int_{\eta_+}^{x} dy~ \partial_d \hat{\alpha}^r (\eta_+,y,\theta^a) \right) + (a \leftrightarrow b) \right] ~~, \label{gammaVT} \nonumber \\
U^a &=& - v^a + \partial_{\eta}\tilde {\theta}^{a (2)} = -v^a+ \partial_{-}\tilde {\theta}^{a (2)} + \partial_{+}\tilde {\theta}^{a (2)}
= \frac{1}{2} \left( - v^a - \chi^{ra} + \gamma^{ab}_0 \int_{\eta_+}^{\eta_-} dx ~ \partial_b \hat{\alpha}^r (\eta_+,x,\theta^a) \right) \nonumber \\
&+& \frac{1}{2} \partial_+ \left( \int_{\eta_+}^{\eta-} dx~ \left[ [\hat{v}^a - \hat{\chi}^{ra}](\eta_+,x,\theta^a) + \hat{\gamma}^{ab}_0(\eta_+,x,\theta^a) \int_{\eta_+}^x dy~ \partial_b \hat{\alpha}^r (\eta_+,y,\theta^a) \right] \right) ~~,
\eea
\normalsize
where, again, the variables $(\eta,r,\theta^a)$ have been omitted for the sake of  conciseness. Of course, these vector and tensor corrections to the FLRW metric have to be added to the scalar ones of the previous subsection. We will see in Sec. \ref{Ch5Sec36} that these modes play a negligible role in our formalism.

\section{Detailed expression for $d_L( z, \theta^a)$}
\label{Ch4Sec2}

\subsection{The scalar contribution}
\label{Ch4Sec21}

We now apply the above coordinate transformations to find the final expression of the luminosity distance in terms of perturbations in the PG, of the observed redshift, and of the observer's angular coordinates beginning, once more, with the scalar contribution.
From Eq. \rref{dLNonPert} we have\,:
\beq
d_L=(1+z_s)^2~\gamma^{1/4}~(\sin \tilde{\theta})^{-1/2} ~~.
\label{ExactdL}
\eeq
For a source emitting light at time $\eta_s$ and radial distance $r_s$ we obtain from Eq. (\ref{gammaab1})\,:
\bea
\gamma^{-1} \equiv \det \gamma^{ab} &=& 
(a_s^2r_s^2\sin\theta)^{-2} \left\{1 + 4 \psi_s + 2 \partial_a \tilde{\theta}^{a (1)} +2 \psi_s^{(2)}
+12 \psi_s^2+ 2 \partial_a \tilde{\theta}^{a (2)}
-4 \gamma_{0 a b} \partial_+ \tilde{\theta}^{a (1)}\partial_- \tilde{\theta}^{b (1)}
\right. \nonumber \\
& & ~~~~~ ~~~~~ ~~~~ ~~~~ ~~~ \left.
+ ~ 8 \psi_s \partial_a \tilde{\theta}^{a (1)}+2 \partial_a \tilde{\theta}^{a (1)}
\partial_b \tilde{\theta}^{b (1)}
-\partial_a \tilde{\theta}^{b (1)}\partial_b \tilde{\theta}^{a (1)}
\right\} ~~.
\label{det}
\eea
We also need the expression for $\sin \tilde{\theta}$ up to second order in perturbations. This is easily given as\,:
\beq
\sin \tilde{\theta}= \sin \theta \left[1+\cot \theta \left(  \tilde{\theta}^{ (1)}
+\tilde{\theta}^{ (2)}\right)-\frac{1}{2}\left(\tilde{\theta}^{ (1)}\right)^2\right] ~~.
\label{sinThetaTilde}
\eeq
Using Eqs. (\ref{det}) and (\ref{sinThetaTilde}), the Eq. (\ref{ExactdL}) yields\,:
\bea
d_L &=& (1+z_s)^2 (a_s r_s) \left\{1-\psi_s - J_2 - \frac{1}{2}\psi_s^{(2)} - \frac{1}{2} \psi_s^2 - K_2 + \psi_s J_2 + \frac{1}{2} (J_2)^2 + \frac{1}{4 \sin^2 \theta}\left(\tilde{\theta}^{ (1)}\right)^2 \right. \nonumber \\
& & ~~~~~ ~~~~~ ~~~~~ ~~~~ ~~~~ \left. + ~(\gamma_0)_{a b} \partial_+ \tilde{\theta}^{a (1)}\partial_- \tilde{\theta}^{b (1)}
+ \frac{1}{4}\partial_a \tilde{\theta}^{b (1)}\partial_b \tilde{\theta}^{a (1)}
\right\} ~~,
\label{d_L}
\eea
where\,:
\beq
J_2 = \frac{1}{2}\left[\cot \theta ~ \tilde{\theta}^{(1)}+\partial_a \tilde{\theta}^{a (1)}\right] \equiv \frac12 \nabla_a \tilde{\theta}^{a (1)}
\,\,\,\,\,\,\,\,\,,\,\,\,\,\,\,\,\,\,
K_2 = \frac{1}{2}\left[\cot \theta ~ \tilde{\theta}^{(2)}+\partial_a \tilde{\theta}^{a (2)}\right] \equiv \frac12 \nabla_a \tilde{\theta}^{a (2)}
~~.
\eeq
All the above quantities are evaluated at the source (apart for $\psi_s$ we neglect the suffix $s$ for simplicity).

At this point, for the explicit expression of the luminosity distance $d_L$ at constant redshift, we need the  first and second-order expansion of the factor $a_sr_s\equiv a(\eta_s) r_s$ appearing in Eq. (\ref{d_L}). To this purpose, we start from the explicit expression for the redshift parameter $z_s$ (see Eq. (\ref{redshift})), considered as a constant parameter localizing, together with the $w = w_o$ condition, the source on $\Sigma(w_o, z_s)$. We then look for approximate solutions for $\eta_s= \eta_s(z_s, \theta^a)$ and $r_s= r_s(z_s, \theta^a)$. 

Let us first define the zero-order solution ${\eta}_s^{(0)}$  through the (exact) relation\,:
\beq
\label{OnePlusZHom}
\frac{a({\eta}_s^{(0)})}{a_o} = \frac{1}{1+z_s} ~~,
\eeq
where $a_o \equiv a(\eta_o)$.
Inserting now the result of Eq. (\ref{InvUps}) into  Eq. (\ref{redshift}) and expanding $a(\eta_s)$ and $\Ups^{-1}$ with respect to the background solutions $\eta_s^{(0)}$ and $r_s^{(0)}$ (where we define $\eta_s=\eta_s^{(0)} + \eta_s^{(1)} + \eta_s^{(2)}$ and
$r_s=r_s^{(0)} + r_s^{(1)} + r_s^{(2)}$), we obtain\,:
\bea
\label{OnePlusZInom}
\frac{1}{1 + z_s} &=& \frac{a(\eta_s^{(0)})}{a(\eta_o)} \Bigg\{1 \frac{}{} + \left[ {\mathcal H}_s \eta_s^{(1)} + \epsilon_o^{(1)} - \epsilon_s^{(1)} \right] 
+ \bigg[ {\mathcal H}_s \eta_s^{(2)} + ({\mathcal H}_s' + {\mathcal H}_s^2)\frac{(\eta_s^{(1)})^2}{2} + \epsilon_o^{(2)} - \epsilon_s^{(2)} - \epsilon_s^{(1\rightarrow 2)}
\nonumber \\
& & 
\,\,\,\,\,\,\,\,\,\,\,\,\,\,\,\,\,\,\,\,\,\,\,\,\,\,+ (\epsilon_s^{(1)})^2 - \epsilon_o^{(1)} \epsilon_s^{(1)} 
+ {\mathcal H}_s \eta_s^{(1)} (\epsilon_o^{(1)} - \epsilon_s^{(1)}) \bigg] \Bigg\} ~~, 
\label{1over1plusz}
\eea
where ${\mathcal H}_s = a'(\eta_s^{(0)}) / a(\eta_s^{(0)})$ and
\beq
\epsilon_o^{(1)}=\epsilon^{(1)}(\eta_o, 0, \theta^a) ~;~
\epsilon_s^{(1)}=\epsilon^{(1)}(\eta_s^{(0)}, r_s^{(0)}, \theta^a) ~;~
\epsilon_o^{(2)}=\epsilon^{(2)}(\eta_o, 0, \theta^a) ~;~
\epsilon_s^{(2)}=\epsilon^{(2)}(\eta_s^{(0)}, r_s^{(0)}, \theta^a) ~~,
\label{epsis}
\eeq
\beq
\epsilon_s^{(1\rightarrow 2)} = \left[\partial_\eta \epsilon^{(1)}\right](\eta_s^{(0)}, r_s^{(0)}, \theta^a) ~ \eta_s^{(1)} + \left[\partial_r \epsilon^{(1)}\right](\eta_s^{(0)}, r_s^{(0)}, \theta^a) ~ r_s^{(1)} ~~.
\label{eps12}
\eeq

Similarly, in order to compute $ r_s^{(1)}$ and $ r_s^{(2)}$, we need to expand the $w = w_o$ constraint by writing:
\beq
\label{ws1,2}
w_o = \left\{ \eta_s^{(0)} + r_s^{(0)} \right\} + \left\{ \eta_s^{(1)} + r_s^{(1)} + w_s^{(1)} \right\} + \left\{ \eta_s^{(2)} + r_s^{(2)} + 
w_s^{(2)} + w_s^{(1 \rightarrow 2)} \right\}
\eeq
and where
\beq
\label{ws12a}
w_s^{(1)} = w^{(1)}(\eta_s^{(0)}, r_s^{(0)}, \theta^a) \,\,\,\,\,,\,\,\,\,\, w_s^{(2)} = w^{(2)}(\eta_s^{(0)}, r_s^{(0)}, \theta^a) ~~,
\eeq
\beq
\label{ws12}
w_s^{(1 \rightarrow 2)} = \left[\partial_\eta w^{(1)}\right](\eta_s^{(0)}, r_s^{(0)}, \theta^a) ~ \eta_s^{(1)} + \left[\partial_r w^{(1)}\right] (\eta_s^{(0)}, r_s^{(0)}, \theta^a) ~ r_s^{(1)} ~~.
\eeq
The additional terms $\epsilon_s^{(1 \rightarrow 2)}$, $w_s^{(1 \rightarrow 2)}$ appearing in the above equations stand for the second order contributions coming from Taylor expanding $\epsilon_s^{(1)}$, $w_s^{(1)}$,  around the background source 
position~\footnote{There is no equivalent contribution at the observer position as $\eta_o$ and $r_o = 0$ are fixed quantities, with no perturbative corrections.}. 
More precisely,
they come from the fact that, at first order, quantities that are already first order  are integrated along the \emph{unperturbed} line of sight, while, at second order, first order terms have to be integrated along the \textit{perturbed} line of sight.

From Eq. (\ref{eps1}) we obtain\,:
\beq
\epsilon_s^{(1)} - \epsilon_o^{(1)} = J 
\eeq
with
\beq
J ~~ \equiv ~ ([\partial_+ Q]_s - [\partial_+ Q]_o) - ([\partial_r P]_s - [\partial_r P]_o) ~~,
\eeq
and where, for example, the term $[ \partial_r P ]_s$ denotes the expression of $\partial_r P$ with $\eta$ and $r$ replaced by $\eta_s^{(0)}$ and $r_s^{(0)}$, we also remark that $[\partial_+ Q]_o = - \psi_o$. 
Then, from Eqs. (\ref{1over1plusz}) and (\ref{ws1,2}), recalling Eq. (\ref{w2order}), we compute\,:
\bea
\eta_s^{(1)}=\frac{J}{\Hcal_s} ~~~~~,~~~~~ r_s^{(0)} = \eta_o - \eta_s^{(0)} \equiv \Delta \eta ~~~~~,~~~~~ r_s^{(1)} = - Q_s - \frac{J}{{\mathcal H}_s} ~~~.
\eea
We wish to note, already at this point, that the $\epsilon$ terms correspond to \gls{RP}. The first order term, $\epsilon^{(1)}$, gives rise to the \gls{Doppler effect} due to the peculiar velocities $\partial_rP$, and the \gls{SW} and \gls{ISW} are combined together in $[\partial_+ Q]_s - [\partial_+ Q]_o$.
Let us, in fact, recall that $\partial_r P$ can be rewritten as\,:
\beq
\label{doppler1Def}
\partial_r P =  \vbf \cdot \nbf ~~, 
\eeq
where $\nbf$ is the unit tangent vector along the null geodesic connecting source and observer, and where 
\beq
\vbf = -\int_{\eta_{in}}^{\eta}d\eta'\frac{a(\eta')}{a(\eta)} \vec{\nabla} \psi(\eta', r, \theta^a)
\eeq
is the \emph{\gls{peculiar velocity}} associated to a geodesic trajectory perturbed up to first order in the PG. 

Let us now move to the second order quantities appearing in Eqs. (\ref{epsis}), (\ref{eps12}) and (\ref{ws12a}), (\ref{ws12}).
We simply have, from Eq. (\ref{eps2Full}),
\bea
\label{eps2Smoins0}
\epsilon_s^{(2)} - \epsilon_o^{(2)} &=& - \frac{1}{4} \left( \phi_s^{(2)} - \phi_o^{(2)} \right) + \frac{1}{4} \left( \psi_s^{(2)} - \psi_o^{(2)} \right) + \frac12 \left( \psi_s^2 - \psi_o^2 \right) \nonumber \\
& & + \frac12 ([\partial_r P]_s)^2 - \frac12 ([\partial_r P]_o)^2 - \left( \psi_s + [\partial_+ Q]_s \right) \, [\partial_r P]_s
\nonumber \\
& &
+ \frac14 (\gamma_0^{ab})_s \left( 2 \partial_a P_s \, \partial_b P_s + \partial_a Q_s \, \partial_b Q_s - 4 \partial_a Q_s \, \partial_b P_s \right) - \frac12 \lim_{r\rightarrow 0} \left[\gamma_{0}^{ab} \partial_a P \, \partial_b P \right] \nonumber \\
& & 
- \frac12 \int_{\eta_{in}}^{\eta_s^{(0)}} d\eta' \frac{a(\eta')}{a(\eta_s^{(0)})} \partial_r \left[ \phi^{(2)} - \psi^2 + (\partial_r P)^2 + \gamma_0^{ab} \partial_a P \, \partial_b P \right] (\eta', \Delta \eta, \theta^a) \nonumber \\
& & 
+ \frac12 \int_{\eta_{in}}^{\eta_o} d\eta' \frac{a(\eta')}{a(\eta_o)}  \partial_r \left[ \phi^{(2)} - \psi^2 + (\partial_r P)^2 + \gamma_0^{ab} \partial_a P \, \partial_b P \right] (\eta', 0, \theta^a) \nonumber \\
& & 
+ \frac14 \int_{\eta_s^{(0)+}}^{\eta_s^{(0)-}} dx~ \partial_+ \left[ \hat{\phi}^{(2)} + \hat{\psi}^{(2)} + 4 \hat{\psi} ~ \partial_+ Q + \hat{\gamma}_{0}^{ab} \, \partial_a Q \, \partial_b Q \right] (\eta_s^{(0)+}, x, \theta^a) ~~, 
\eea
while
\beq
\label{eps2oneTotwo}
\epsilon_s^{(1 \rightarrow 2)} = Q_s \left\{ - [\partial_+^2 Q]_s + [\partial_+ \hat{\psi}]_s + [\partial_r^2 P]_s \right\} + \frac{J}{{\mathcal H}_s} \left\{ [\partial_\eta \psi]_s + {\mathcal H}_s [\partial_r P]_s + [\partial_r^2 P]_s \right\} ~~.
\eeq
We then have\,:
\bea
w_s^{(1 \rightarrow 2)} &=& \frac{2}{{\mathcal H}_s} \psi_s J + Q_s \left( \psi_s - [\partial_+ Q]_s \right) ~~.
\eea
Using Eqs. (\ref{OnePlusZInom}) and (\ref{ws1,2}), and recalling again Eq. (\ref{w2order}), we can now calculate $\eta_s$ and $r_s$ to second order, obtaining\,:
\bea
\eta_s^{(2)}&=&-\frac{1}{\Hcal_s}\left\{({\mathcal H}_s' + {\mathcal H}_s^2)\frac{(\eta_s^{(1)})^2}{2} + \epsilon_o^{(2)} - \epsilon_s^{(2)} - \epsilon_s^{(1\rightarrow 2)} 
+ (\epsilon_s^{(1)})^2 - \epsilon_o^{(1)} \epsilon_s^{(1)} 
+ {\mathcal H}_s \eta_s^{(1)} (\epsilon_o^{(1)} - \epsilon_s^{(1)})\right\}\nonumber \\
&=&-\frac{1}{\Hcal_s}\left\{\frac{{\mathcal H}_s' + {\mathcal H}_s^2}{\Hcal_s^2}\frac{J^2}{2} + \epsilon_o^{(2)} - \epsilon_s^{(2)} - \epsilon_s^{(1\rightarrow 2)} 
+ \epsilon_s^{(1)}J 
- J^2\right\} ~~,
\eea 
and
\bea
r_s^{(2)}&=&-\left( \eta_s^{(2)} + w^{(2)}_s + w_s^{(1 \rightarrow 2)} \right) \nonumber \\
 &=& - \frac{1}{{\mathcal H}_s} \left( \epsilon_s^{(2)} - \epsilon_o^{(2)} + \epsilon_s^{(1 \rightarrow 2)} \right) - \frac{J}{{\mathcal H}_s} \left( 2 \psi_s + \psi_o + [\partial_r P]_o \right) + \frac{{\mathcal H}_s^2 + {\mathcal H}_s'}{2{\mathcal H}_s^3} J^2 + \left( - \psi_s + [ \partial_+ Q ]_s \right) Q_s \nonumber \\
& & - \frac14 \int_{\eta_s^{(0)+}}^{\eta_s^{(0)-}} dx~ \left[ \hat{\phi}^{(2)} + \hat{\psi}^{(2)} + 4 \hat{\psi} ~ \partial_+ Q + \hat{\gamma}_0^{ab} ~ \partial_a Q ~ \partial_b Q \right] (\eta_s^{(0)+},x,\theta^a) ~~.
\eea

In the second order terms we have the expected couplings between first order terms as well as the (also expected) genuine second order \gls{SW} and \gls{ISW} effects such as $(\psi_s^{(2)} - \psi_o^{(2)})$ and $\int dx ~ \partial_{+}\hat{\psi}^{(2)}$. However, at second order, new effects come into play\,: most notably the \emph{tangential peculiar velocity} $\partial_aP$, the \emph{tangential variation of the photon path} $\partial_aQ$, and a \gls{RSD} due to the \emph{peculiar acceleration} $\partial^2P$. The somewhat surprising appearance of tangential derivatives in $\eta_s^{(2)}$ and $r_s^{(2)}$ is simply a reflection of working on a fixed-$z$ surface. As a consequence, redshift perturbations (\gls{RP}) originating from those of $\tau$, Eq. (\ref{tau2order}), feed back on $\eta_{s}, r_s$ and eventually on $d_L(z)$.

To conclude, combining these results, we obtain\,:
\bea
\frac{a(\eta_s) r_s}{a(\eta_s^{(0)}) \Delta\eta} &=&
1 + \left\{ \Xi_s J - \frac{Q_s}{\Delta \eta} \right\} +
\Bigg\{ \Xi_s \left( \epsilon_s^{(2)} - \epsilon_o^{(2)} + \epsilon_s^{(1 \rightarrow 2)} \right) - \frac{1}{{\mathcal H}_s \Delta \eta} \left( 1 - \frac{{\mathcal H}_s'}{{\mathcal H}_s^2} \right) \frac{J^2}{2} - \frac{2}{{\mathcal H}_s \Delta \eta} \psi_s J \nonumber \\
& & ~ + \Xi_s \left( \psi_o + [\partial_r P]_o \right) J + \left( - \psi_o - \psi_s + [\partial_r P]_s - [\partial_r P]_o \right) \frac{Q_s}{\Delta \eta} \nonumber \\
& & ~ - \frac{1}{4 \Delta\eta} \int_{\eta_s^{(0)+}}^{\eta_s^{(0)-}} dx~ \left[ \hat{\phi}^{(2)} + \hat{\psi}^{(2)} + 4 \hat{\psi} ~ \partial_+ Q + \hat{\gamma}_{0}^{ab} ~ \partial_a Q ~ \partial_b Q \right] (\eta_s^{(0)+},x,\theta^a) \Bigg\} ~~, 
\label{asrs}
\eea
where (using $\Delta \eta \equiv \eta_o - \eta_s^{(0)}$) we defined\,:
\beq
\Xi_s \equiv 1 - \frac{1}{\Hcal_s \Delta \eta} ~~.
\eeq

Let us also note that in Eq. (\ref{d_L}) there are two other first order terms that have to be Taylor expanded up to second order around the background solution connected to the observed redshift $z_s$, i.e. $\psi_s$ and $J_2$.
We find\,:
\beq
\psi_s =  \psi_s^{(1)}  + \psi_s^{(1 \rightarrow 2)} = \psi(\eta_s^{(0)}, \Delta \eta, \theta^a)+\frac{J}{{\mathcal H}_s} 
\left[ \partial_\eta \psi - \partial_r \psi \right] (\eta_s^{(0)}, \Delta \eta, \theta^a) - Q_s [\partial_r \psi](\eta_s^{(0)}, \Delta \eta, \theta^a) ~~,
\label{ExpPsis}
\eeq
\bea
\label{ExpJ2}
J_2 &=& J_2^{(1)} +  J_2^{(1 \rightarrow 2)} \\
&=& \frac{1}{\Delta\eta} \int_{\eta_s^{(0)}}^{\eta_o} d \eta' \,\frac {\eta' - \eta_s^{(0)}}{\eta_o - \eta'} \Delta_2 \psi(\eta', \eta_o-\eta', \theta^a) - \left( \frac{J}{{\mathcal H}_s} + \frac{Q_s}{2} \right) \frac{1}{\Delta\eta^2} \int_{\eta_s^{(0)}}^{\eta_o} d \eta' \Delta_2 \psi(\eta', \eta_o-\eta', \theta^a) \nonumber \\
& & - Q_s ~ \partial_+ \left(\int_{\eta_s^{(0)+}}^{\eta_s^{(0)-}} dx \frac{1}{\big(\eta_s^{(0)+}-x\big)^2} \int_{\eta_s^{(0)+}}^x d y ~ \Delta_2 \hat{\psi}(\eta_s^{(0)+}, y, \theta^a) \right) ~~,  \nonumber
\eea
where we have used the 2-dimensional Laplacian  $\Delta_2 \equiv \partial^2_{\theta} + \cot \theta\, \partial_{\theta} + (\sin \theta)^{-2} \partial^2_{\phi}$ ~.

Collecting all the results obtained up to now, and inserting them in Eq. (\ref{d_L}), we  write our final result on the effect of scalar perturbations in the following concise form\,:
\beq
\frac{d_L(z_s, \theta^a)}{(1+z_s)a_o \Delta \eta}
= {d_L(z_s, \theta^a)\over d_L^{\rm FLRW}(z_s)} =  1 + \delta_S^{(1)}(z_s, \theta^a) + \delta_S^{(2)}(z_s, \theta^a) ~~,
\eeq
where\,:
\bea
\delta_S^{(1)}(z_s, \theta^a) &=& \Xi_s J - \frac{Q_s}{\Delta \eta} - \psi_s^{(1)} - J_2^{(1)} ~~, \nonumber \\
\delta_S^{(2)}(z_s, \theta^a)   &=& - \left(\Xi_s J - \frac{Q_s}{\Delta \eta} \right) \left( \psi_s^{(1)}  + J_2^{(1)} \right)  - \psi_s^{(1 \rightarrow 2)} - J_2^{(1 \rightarrow 2)} + X^{(2)} + Y^{(2)} ~~.
\label{finaldL} 
\eea 
Here $\psi_s^{(1)}$, $\psi_s^{(1 \rightarrow 2)}$, $J_2^{(1)}$ and $J_2^{(1 \rightarrow 2)}$ are implicitly defined in Eqs. (\ref{ExpPsis}), (\ref{ExpJ2}) and $X^{(2)}$, $Y^{(2)}$ are the second order terms appearing in Eq. (\ref{d_L}) and (\ref{asrs}), namely\,:
\small
\bea
\label{X2Y2}
X^{(2)} &=& - \frac{1}{2}\psi_s^{(2)} - \frac{1}{2} \psi_s^2 - K_2 + \psi_s J_2 + \frac{1}{2} (J_2)^2 \\
& & + \frac{1}{4 \sin^2 \theta}\left(\tilde{\theta}^{ (1)}\right)^2 + (\gamma_0)_{a b} \partial_+ \tilde{\theta}^{a (1)}\partial_- \tilde{\theta}^{b (1)}
+\frac{1}{4}\partial_a \tilde{\theta}^{b (1)}\partial_b \tilde{\theta}^{a (1)} ~~, \nonumber \\
Y^{(2)} &=& \Xi_s \left( \epsilon_s^{(2)} - \epsilon_o^{(2)} + \epsilon_s^{(1 \rightarrow 2)} \right) - \frac{2}{{\mathcal H}_s \Delta \eta} \psi_s J  + \Xi_s \left( \psi_o + [\partial_r P]_o \right) J  + \left( [\partial_r P]_s - \psi_o - \psi_s - [\partial_r P]_o \right) \frac{Q_s}{\Delta \eta}\nonumber \\
&  - &  \frac{1}{{\mathcal H}_s \Delta \eta} \left( 1 - \frac{{\mathcal H}_s'}{{\mathcal H}_s^2} \right) \frac{J^2}{2}
~ - \frac{1}{4 \Delta\eta} \int_{\eta_s^{(0)+}}^{\eta_s^{(0)-}} dx~ \left[ \hat{\phi}^{(2)} + \hat{\psi}^{(2)} + 4 \hat{\psi} ~ \partial_+ Q + \hat{\gamma}_{0}^{ab} ~ \partial_a Q ~ \partial_b Q \right] (\eta_s^{(0)+},x,\theta^a) ~~. \nonumber
\eea 
\normalsize

On the other hand, as already stressed, photons reach the observer traveling at constant $\tilde{\theta}^a$. Therefore, the observer's angles are given by the $\tilde{\theta}^a$ which coincide with ${\theta}^a$ at the observer position but not at the source, hence $d_L$ should be written in terms of $\tilde{\theta}^a$ rather  than of ${\theta}^a$. As a consequence let us consider the inverse form of Eq. (\ref{thetatilde2orderShort})\,:
\small
\bea
{\theta}^a &=& {\theta}^{a (0)}+{\theta}^{a (1)}+{\theta}^{a (2)}
= \tilde{\theta}^a - \frac12 \int_{\eta_+}^{\eta_-} dx~ \hat{\gamma}_0^{ab}(\eta_+,x,\tilde{\theta}^a) \int_{\eta_+}^x dy~ \partial_b \hat{\psi}(\eta_+,y,\tilde{\theta}^a)\nonumber \\
&+&\frac{1}{4}\left[\int_{\eta_+}^{\eta_-} dx~ \hat{\gamma}_0^{cb}(\eta_+,x,\tilde{\theta}^a) \int_{\eta_+}^x dy~ \partial_b \hat{\psi}(\eta_+,y,\tilde{\theta}^a)\right]
\partial_c \left[\int_{\eta_+}^{\eta_-} dx~ \hat{\gamma}_0^{ad}(\eta_+,x,\tilde{\theta}^a) \int_{\eta_+}^x dy~ \partial_d \hat{\psi}(\eta_+,y,\tilde{\theta}^a)\right]
\nonumber \\
&-&\int_{\eta_+}^{\eta_-} dx~ \left[ \hat{\gamma}_0^{ac} \zeta_c + \hat{\psi} ~ \xi^a + \lambda^a \right] (\eta_+,x,\tilde{\theta}^a) ~~.
\label{thetatilde1orderShort_Inverted}
\eea
\normalsize
The luminosity distance $\bar{d}_L(z_s, \tilde{\theta}^a)$ will then be given by Taylor expanding $d_L(z_s, {\theta}^a)$ around $\tilde{\theta}^a$ (we use a bar to denote that the luminosity distance is now expressed in terms of $\tilde{\theta}^a$). Using Eq. (\ref{thetatilde1orderShort_Inverted}) we obtain\,:
\beq
\frac{\bar{d}_L(z_s, \tilde{\theta}^a)}{(1+z_s)a_0 \Delta \eta}
= {\bar{d}_L(z_s, \tilde{\theta}^a)\over d_L^{\rm FLRW}(z_s)} = 1 + \bar{\delta}_S^{(1)}(z_s, \tilde{\theta}^a) + \bar{\delta}_S^{(2)}(z_s, \tilde{\theta}^a) ~~, \nonumber 
\eeq
\beq
\mbox{ with } ~~~
\bar{\delta}_S^{(1)}(z_s, \tilde{\theta}^a) = \delta_S^{(1)}(z_s, \tilde{\theta}^a) ~~~~~,~~~~~
\bar{\delta}_S^{(2)}(z_s, \tilde{\theta}^a) = \delta_S^{(2)}(z_s, \tilde{\theta}^a) + \partial_b \left[ \delta_S^{(1)}(z_s, \tilde{\theta}^a)\right] \theta^{b (1)} ~~.
\label{finaldbarL}
\eeq

Let us note the presence of two particular terms which appear in the Eqs. \rref{X2Y2}, \rref{finaldbarL} and are well-known in the literature (see e.g. \cite{Dodelson:2005zj}). The first term, present in \rref{X2Y2}, is called the \gls{lens-lens coupling} and stands as (using Eq. \rref{thetatilde2orderShort})\,:
\beq
\frac{1}{4}\partial_a \tilde{\theta}^{b (1)}\partial_b \tilde{\theta}^{a (1)} = \frac{1}{16} \partial_a \left( \int_{\eta_s^{(0)+}}^{\eta_s^{(0)-}} dx ~ \left[ \hat{\gamma}_0^{bc} ~ \partial_c Q \right] (\eta_s^{(0)+},x,\tilde{\theta}^a) \right) \partial_b \left( \int_{\eta_s^{(0)+}}^{\eta_s^{(0)-}} d\bar{x} ~ \left[ \hat{\gamma}_0^{ad} ~ \partial_d Q \right] (\eta_s^{(0)+},\bar{x},\tilde{\theta}^a) \right) .
\eeq
The second, revealed in Eq. \rref{finaldbarL}, describes the first deviations from the line of sight and is named the ``first correction to the \gls{Born approximation}''. It is given by the term\,:
\beq
\partial_b \left[ \delta_S^{(1)}(z_s, \tilde{\theta}^a)\right] \theta^{b (1)} ~ \supset ~ \frac12 \partial_a J_2^{(1)} \left( \int_{\eta_s^{(0)+}}^{\eta_s^{(0)-}} dx ~ \left[ \hat{\gamma}_0^{ab} ~ \partial_b Q \right] (\eta_s^{(0)+},x,\tilde{\theta}^a) \right) ~~.
\eeq

Equations (\ref{finaldL}), (\ref{finaldbarL}), supplemented with  the vector and  tensor contribution discussed in the next subsection, are our main result about the \gls{luminosity distance}. More explicit expressions, where terms with different physical meaning are collected separately, can also be written down. Indeed, the second order corrections of $\bar{\delta}_S^{(2)}(z_s, \tilde{\theta}^a)$ appearing in Eq. (\ref{finaldbarL}) can be conveniently grouped as follows\,:
\beq
\bar{\delta}_S^{(2)}(z_s, \tilde{\theta}^a) =  \bar{\delta}_{path}^{(2)} +  \bar{\delta}_{pos}^{(2)} +  \bar{\delta}_{mixed}^{(2)}
\label{deltabardl2}
\eeq
where $\bar{\delta}_{path}^{(2)}$ is for the terms concerning the photon path, $\bar{\delta}_{pos}^{(2)}$ for the terms generated by the source and observer positions, and $\bar{\delta}_{mixed}^{(2)}$ is a mixing of both effects. Their explicit expressions are\,:
\vspace{-0.5cm}
\small
\bea
\bar{\delta}_{path}^{(2)} &=& \Xi_s \Bigg\{ - \frac{1}{4} \left( \phi_s^{(2)} - \phi_o^{(2)} \right) + \frac{1}{4} \left( \psi_s^{(2)} - \psi_o^{(2)} \right) + \frac{1}{2} \left( \psi_s - \psi_o \right)^2 - \psi_o J_2^{(1)} \nonumber \\
&+& ( \psi_o - \psi_s - J_2^{(1)} ) [\partial_+ Q]_s  + \frac14 (\gamma_{0}^{ab})_s \partial_a Q_s \partial_b Q_s + Q_s \left( - [\partial_+^2 Q]_s + [\partial_+ \psi]_s \right) + \frac{1}{{\mathcal H}_s} ( \psi_o + [\partial_+ Q]_s ) 
[\partial_\eta \psi]_s \nonumber \\
&-& \frac12 \int_{\eta_{in}}^{\eta_s^{(0)}} d\eta' \frac{a(\eta')}{a(\eta_s^{(0)})} \partial_r \left[ \phi^{(2)} - \psi^2 \right] (\eta', \Delta \eta, \tilde{\theta}^a) + \frac12 \int_{\eta_{in}}^{\eta_o} d\eta' \frac{a(\eta')}{a(\eta_o)} \partial_r \left[ \phi^{(2)} - \psi^2 \right] (\eta', 0, \tilde{\theta}^a) \nonumber \\
&+& \frac14 \int_{\eta_s^{(0)+}}^{\eta_s^{(0)-}} dx~ \partial_+ \left[ \hat{\phi}^{(2)} + \hat{\psi}^{(2)} + 4 \hat{\psi} ~ \partial_+ Q + \hat{\gamma}_{0}^{ab} ~ \partial_a Q ~ \partial_b Q \right] (\eta_s^{(0)+},x,\tilde{\theta}^a) \nonumber \\
&-& \frac12 \, \partial_a (\psi_o + \partial_+ Q_s) \left( \int_{\eta_s^{(0)+}}^{\eta_s^{(0)-}} dx ~ \left[ \hat{\gamma}_0^{ab} ~ \partial_b Q \right] (\eta_s^{(0)+},x,\tilde{\theta}^a) \right)
\Bigg\} \nonumber \\
&-& \frac{1}{2}\psi_s^{(2)} - \frac{1}{2} \psi_s^2 - K_2 + \psi_s J_2^{(1)} +\frac{1}{2}(J_2^{(1)})^2 + ( J_2^{(1)} - \psi_o ) \frac{Q_s}{\Delta \eta} - \frac{1}{\Hcal_s \Delta\eta} \left( 1 - \frac{\Hcal_s'}{\Hcal_s^2} \right) \frac12 ( \psi_o + [\partial_+ Q]_s )^2 \nonumber \\
&-& \frac{2}{\Hcal_s \Delta \eta} \psi_s ( \psi_o + [\partial_+ Q]_s ) + \frac12 \partial_a \left( \psi_s + J_2^{(1)} + \frac{Q_s}{\Delta \eta} \right)  \left( \int_{\eta_s^{(0)+}}^{\eta_s^{(0)-}} dx ~ \left[ \hat{\gamma}_0^{ab} ~ \partial_b Q \right] (\eta_s^{(0)+},x,\tilde{\theta}^a) \right) \nonumber \\
&+& \frac14 \partial_a Q_s ~ \partial_+ \left( \int_{\eta_s^{(0)+}}^{\eta_s^{(0)-}} dx ~ \left[ \hat{\gamma}_0^{ab} ~ \partial_b Q \right] (\eta_s^{(0)+},x,\tilde{\theta}^a) \right) \nonumber \\
&+& \frac{1}{16} \partial_a \left( \int_{\eta_s^{(0)+}}^{\eta_s^{(0)-}} dx ~ \left[ \hat{\gamma}_0^{bc} ~ \partial_c Q \right] (\eta_s^{(0)+},x,\tilde{\theta}^a) \right) \partial_b \left( \int_{\eta_s^{(0)+}}^{\eta_s^{(0)-}} d\bar{x} ~ \left[ \hat{\gamma}_0^{ad} ~ \partial_d Q \right] (\eta_s^{(0)+},\bar{x},\tilde{\theta}^a) \right) \nonumber \\
&-& \frac{1}{4 \Delta \eta} \int_{\eta_s^{(0)+}}^{\eta_s^{(0)-}} dx~ \left[ \hat{\phi}^{(2)} + \hat{\psi}^{(2)} + 4 \hat{\psi} ~ \partial_+ Q + \hat{\gamma}_{0}^{ab} ~ \partial_a Q ~ \partial_b Q \right] (\eta_s^{(0)+},x,\tilde{\theta}^a) \nonumber \\
&+& \frac{1}{\Hcal_s} ( \psi_o + [\partial_+ Q]_s ) \left\{ - [\partial_\eta \psi]_s + [\partial_r \psi]_s + \frac{1}{\Delta \eta^2} \int_{\eta_s^{(0)}}^{\eta_o} d\eta' \Delta_2 \psi (\eta', \eta_o - \eta', \tilde{\theta}^a) \right\} \nonumber \\
&+& Q_s \Bigg\{ [\partial_r \psi]_s + \partial_+ \left(\int_{\eta_s^{(0)+}}^{\eta_s^{(0)-}} d x \frac{1}{\big(\eta_s^{(0)+}-x\big)^2} \int_{\eta_s^{(0)+}}^x d y \Delta_2 \hat{\psi}(\eta_s^{(0)+}, y, \tilde{\theta}^a) \right) \nonumber \\
&+& \frac{1}{2 \Delta\eta^2} \int_{\eta_s^{(0)}}^{\eta_o} d \eta' \Delta_2 \psi(\eta', \eta_o-\eta', \tilde{\theta}^a) \Bigg\}
+ \frac{1}{16 \sin^2 \tilde{\theta}} \left( \int_{\eta_s^{(0)+}}^{\eta_s^{(0)-}} dx ~ \left[ \hat{\gamma}_0^{1b} ~ \partial_b Q \right] (\eta_s^{(0)+},x,\tilde{\theta}^a) \right)^2  ~~,
\eea

\bea
\bar{\delta}_{pos}^{(2)} &=& \frac{\Xi_s}{2} \Bigg\{ ([\partial_r P]_s - [\partial_r P]_o)^2 + (\gamma_0^{ab})_s \partial_a P_s \, \partial_b P_s - \lim_{r\rightarrow 0} \left[\gamma_0^{ab} \partial_a P \, \partial_b P \right] \nonumber \\
&-& \frac{2}{\Hcal_s} \left( [ \partial_r P ]_s - [ \partial_r P ]_o \right) \left( \Hcal_s [\partial_r P]_s + [\partial_r^2 P]_s \right) \nonumber \\
&-& \int_{\eta_{in}}^{\eta_s^{(0)}} d\eta' \frac{a(\eta')}{a(\eta_s^{(0)})} \partial_r \left[ (\partial_r P)^2 + \gamma_0^{ab} \partial_a P \, \partial_b P \right] (\eta',\Delta\eta,\tilde{\theta}^a) \\
&+& \int_{\eta_{in}}^{\eta_o} d\eta' \frac{a(\eta')}{a(\eta_o)} \partial_r \left[ (\partial_r P)^2 + \gamma_0^{ab} \partial_a P \, \partial_b P \right] (\eta',0,\tilde{\theta}^a) \Bigg\} - \frac{1}{2 \Hcal_s \Delta\eta} \left( 1 - \frac{\Hcal_s'}{\Hcal_s^2} \right) \left( [ \partial_r P ]_s - [ \partial_r P ]_o \right)^2  ~~, \nonumber
\eea

\bea
\bar{\delta}_{mixed}^{(2)} &=& \Xi_s \Bigg\{ \left( 2 \psi_o - \psi_s + \partial_+ Q_s - \frac{Q_s}{\Delta \eta} \right) [\partial_r P]_o  - ( [\partial_r P]_s - [\partial_r P]_o ) \left( \frac{1}{\Hcal_s} [\partial_\eta \psi]_s - J_2^{(1)} \right) \nonumber \\
&-& (\gamma_0^{ab})_s \partial_a Q_s \partial_b P_s + \frac{1}{\Hcal_s} (\psi_o + [\partial_+ Q]_s) [\partial_r^2 P]_s + Q_s \, [\partial_r^2 P]_s \nonumber \\
&+& \frac12 \, \partial_a ( [\partial_r P]_s - [\partial_r P]_o ) \left( \int_{\eta_s^{(0)+}}^{\eta_s^{(0)-}} dx ~ \left[ \hat{\gamma}_0^{ab} ~ \partial_b Q \right] (\eta_s^{(0)+},x,\tilde{\theta}^a) \right)
\Bigg\} \nonumber \\
&+& \frac{1}{\Delta \eta} ( [\partial_r P]_s - [\partial_r P]_o ) \Bigg\{ \frac{1}{\Hcal_s} \left( 1 - \frac{\Hcal_s'}{\Hcal_s^2} \right) (\psi_o + [\partial_+ Q]_s) + \frac{2}{\Hcal_s} \psi_s + Q_s \Bigg\} \nonumber \\
&+& \frac{1}{\Hcal_s} ( [\partial_r P]_s - [\partial_r P]_o ) \left\{ [\partial_\eta \psi]_s - [\partial_r \psi]_s - \frac{1}{\Delta \eta^2} \int_{\eta_s^{(0)}}^{\eta_o} d\eta' \Delta_2 \psi (\eta', \eta_o - \eta', \tilde{\theta}^a) \right\}  ~~.
\eea
\normalsize

\subsection{The Vector and Tensor contribution}
\label{Ch4Sec22}

Following the procedure just presented for scalar perturbations we start from the general expression for $d_L$ 
considering now just vector and tensor perturbations~\footnote{Note that an expression for the contribution of vectors and tensors to $d_L$ has been derived \cite{DiDio:2012bu}. We can check that our results agree with this independent calculation except for a possible velocity contribution at the observer position.}.
We obtain:
\beq
d_L = (1 + z_s)^2 (a_s r_s) \bigg\{ 1  + \frac{1}{4} [(\gamma_0)_{ab} \chi^{ab}](\eta_s,r_s,\theta^a) - J_2^{(\alpha)} - \frac{1}{4} \int_{\eta_s^+}^{\eta_s^-} dx ~ \nabla_a \left[\hat{v}^a - \hat{\chi}^{ra}\right](\eta_s^+,x,\theta^a) \bigg\} ~~,
\eeq
where, in terms of the quantity $Q^{(\alpha)}$ defined in Eq. (\ref{Qalpha}), we have\,:
\bea
J_2^{(\alpha)} & \equiv & \int_{\eta_s^+}^{\eta_s^-} dx ~ \frac{1}{\big(\eta_s^{(0)+}-x\big)^2} \Delta_2 Q^{(\alpha)} (\eta_s^{(0)+},x,\theta^a) ~~,
\eea
and where $a_s r_s$ is a quantity that still needs to be expanded with respect to the observed redshift.
In order to do that, we first write the analogue of Eq. (\ref{OnePlusZInom})\,:
\bea
\label{1+zVT}
\frac{1}{1+z_s} = \frac{a(\eta_s)}{a_o} \left\{ 1 + \left[\frac{v^r}{2} + \frac{\chi^{rr}}{4}\right](\eta_s, r_s, \theta^a) - \left[\frac{v^r}{2} + \frac{\chi^{rr}}{4}\right](\eta_o, 0, \theta^a) - J^{(\alpha)} \right\} ~~,
\eea
with
\beq
J^{(\alpha)} \equiv \int_{\eta_s^+}^{\eta_s^-} dx~ \partial_+ \hat{\alpha}^r (\eta_s^+,x,\theta^a) = \hat{\alpha}_o^r + \partial_+ Q_s^{(\alpha)} ~~.
\eeq
Expanding  $\eta_s$ as $\eta_s = \eta_s^{(0)} + \eta_s^{(2)}$ and imposing $(1 + z_s) a(\eta_s^{(0)}) = a_o$, we get
\beq
\eta_s = \eta_s^{(0)} + \frac{1}{\Hcal_s} \left\{ \left[\frac{v^r}{2} + \frac{\chi^{rr}}{4}\right](\eta_o, 0, \theta^a) - \left[\frac{v^r}{2} + \frac{\chi^{rr}}{4}\right](\eta_s^{(0)}, r_s^{(0)}, \theta^a) + J^{(\alpha)} \right\} ~.
\eeq
Using also the transformation of $w$, Eq. (\ref{wVT}), we get the expression of $r_s$\,:
\beq
r_s = \Delta \eta+ \frac{1}{\Hcal_s}\left\{ \left[\frac{v^r}{2} + \frac{\chi^{rr}}{4}\right](\eta_s^{(0)}, r_s^{(0)}, \theta^a) - \left[\frac{v^r}{2} + \frac{\chi^{rr}}{4}\right](\eta_o, 0, \theta^a) - J^{(\alpha)} \right\} - Q^{(\alpha)} ~~,
\eeq
and  finally reach the conclusion that the luminosity distance at linear order in vector and tensor perturbations (regarded themselves as second order quantities) is
\bea
\frac{d_L^{(V,T)}}{d_L^{\rm FLRW}} &=& 1 + \delta^{(2)}_{V,T} \\
&=& 1 - \frac{Q_s^{(\alpha)}}{\Delta \eta}
+ \Xi_s\left\{\left(\frac{v^r_o}{2} + \frac{\chi^{rr}_o}{4}\right) - \left(\frac{v^r_s}{2} + \frac{\chi^{rr}_s}{4}\right)
+ J^{(\alpha)}\right\} + \frac{1}{4} [ (\gamma_0)_{ab} \chi^{ab} ](\eta_s^{(0)},r_s^{(0)},\theta^a) \nonumber \\
& & ~ - \int_{\eta_s^{(0)+}}^{\eta_s^{(0)-}} dx~ \Bigg\{\frac{1}{4} \nabla_a [\hat{v}^a -\hat{\chi}^{ra}](\eta_s^{(0)+},x,\theta^a) + \frac{1}{\big(\eta_s^{(0)+}-x\big)^2}\Delta_2 Q^{(\alpha)} (\eta_s^{(0)+},x,\theta^a) \Bigg\} ~~. \nonumber
\eea
Using  the transversality and trace-free conditions on the perturbations\,:
\bea
\nabla_a v^a = -\left(\partial_r+\frac{2}{r}\right)v^r ~~,~~  \nabla_a \chi^{ra} = -\left(\partial_r+\frac{3}{r}\right)\chi^{rr} ~~,~~ 
(\gamma_0)_{ab} \chi^{ab} = - \chi^{rr} ~~,
\eea
we finally get an expression that depends only on $v^r$ and $\chi^{rr}$\,:
\bea
\label{dLVTfinal}
\frac{d_L^{(V,T)}}{d_L^{\rm FLRW}} &=&  1 - \frac{Q_s^{(\alpha)}}{\Delta \eta}
+ \Xi_s\left\{\left(\frac{v^r_o}{2} + \frac{\chi^{rr}_o}{4}\right) - \left(\frac{v^r_s}{2} + \frac{\chi^{rr}_s}{4}\right)
+ J^{(\alpha)}\right\}  - \frac{1}{4} \chi^{rr}_s \nonumber \\
& & ~ + \frac{1}{4} \int_{\eta_s^{(0)+}}^{\eta_s^{(0)-}} dx~ \left[\left(\partial_r+\frac{2}{r}\right)v^r- \left(\partial_r+\frac{3}{r}\right)\chi^{rr} - \frac{1}{r^2} \Delta_2 Q^{(\alpha)}\right](\eta_s^{(0)+},x,\theta^a) ~~,
\eea
where one should interpret $r$ as $r= (\eta_s^{(0)+}-x) / 2$ inside the last integral.

We note, once more, the nature of the terms appearing in Eq. (\ref{dLVTfinal}). In the first line we see a SW term as well as an average (or integrated) SW effect for the vector/tensor perturbation. The second line involves frame-dragging and a ``magnification" term (lensing) for vectors/tensors proportional to the Laplacian of the perturbation on the 2-sphere.
Our final expression for $d_L$ is thus\,:
\beq
\label{dLwithSVT}
{\bar{d}_L(z_s, \tilde{\theta}^a)\over d_L^{\rm FLRW}(z_s)} = 1 +  \bar{\delta}_S^{(1)}(z_s, \tilde{\theta}^a)+
 \bar{\delta}_S^{(2)}(z_s, \tilde{\theta}^a) + \bar{\delta}^{(2)}_{V,T}(z_s, \tilde{\theta}^a) ~~,
\eeq
where we directly replaced $\theta^a$ with $\tilde{\theta}^a$ in $\delta^{(2)}_{V,T}$ to get $\bar{\delta}^{(2)}_{V,T}$ since this is already a second-order quantity.

\subsection{Physical interpretation of terms in $\bar{d}_L(z, \ti{\theta}^a)$}
\label{Ch4Sec23}

We have obtained a ``local" expression for $\bar{d}_L(z, \ti{\theta}^a)$, expression that can find a large number of possible applications in cosmology. The different terms appearing in $\bar \delta_S^{(1)}$, $\bar \delta_S^{(2)}$ and $\bar \delta_{V,T}^{(2)}$, in Eq. \rref{dLwithSVT}, can be roughly classified as follows\,:

\begin{itemize*}
\renewcommand{\labelitemi}{$\bullet$}
\item Redshift Perturbations (\gls{RP}).
These are the $\epsilon_s^{(2)} - \epsilon_o^{(2)}$ terms as well as the analogous vector/tensor $v^r ,~ \chi^{rr}$ terms appearing in Eq. (\ref{1+zVT}).
As mentioned in the text, at first order they include \gls{Doppler effect} of peculiar velocities $([\partial_rP]_s - [\partial_rP]_o) $, SW and ISW 
(in agreement with the results obtained in \cite{Bonvin:2005ps}). At second order additional effects such as the tangential peculiar velocity $\partial_aP$, the tangential variation of the photon path $\partial_aQ$, and \gls{RSD} also appear. We will see that peculiar velocities are the dominant contribution at low redshift $z\lesssim0.2$, when we average generic functions of the luminosity distance $d_L(z)$ (see Chapter \ref{Chap6}).
\item \gls{Perturbed trajectories}. At second order, first order integrated quantities are evaluated along the perturbed geodesics giving rise to $(1\ra 2)$ terms. 
\item \gls{SW} and \gls{ISW} effects coming from the evaluation of the area distance. Once again, in our notation, they are simply combined as
 $([\partial_+Q]_s - [\partial_+Q]_o)$. There is also an equivalent effect in the vector/tensor contribution.
\item The \gls{lensing} effect. These are the magnification $J_2$, $K_2$ terms as well as the shear in $\partial_a \tilde{\theta}^{b(1)}\partial_b \tilde{\theta}^{a(1)}$. They are the most important contributions at high redshifts $z \gtrsim 0.5$, when we average generic functions of the luminosity distance $d_L(z)$.
\item \gls{Frame dragging}, in the vector contribution (described at the end of the last section).
\end{itemize*}
\vspace{0.5cm}

We have also identified the terms that have been computed in the literature, such as the lens-lens coupling and the Born correction. The presence of these expected terms shows the validity of our calculation and clearly supports its usefulness in precision cosmology. This calculation indeed accounts for all the possible effects at second order in perturbations from scalar perturbations and in a realistic way (but effectively first order) for vector and tensor perturbations.

\bigskip \bigskip

We will start the next chapter, in Sec. \ref{Ch5Sec1}, by recalling the transformation that we presented here but keeping only the first order contributions. It will then be easier to interpret the terms that we have presented and give more details about their physical meaning.

% --------------------------------------------------------- %

%  --------------------- New Chapter 5 ----------------------   %
% --------------------------------------------------------- %
\chapter{Combining light-cone and stochastic averaging}
\label{Chap5}
\pagestyle{plain}

In this section, we describe how the light-cone average presented in Chapter \ref{Chap2} can be combined with the stochastic average over inhomogeneities. This procedure gives us a rigorous way to consider an observable (such as the luminosity flux or distance) and average it over the whole sky (to get an isotropic prediction) and over inhomogeneities (to properly consider the non-homogeneous aspect of structures). We have seen in Chapter \ref{Chap4} that the full computation of the luminosity distance up to second order in perturbations, around the flat FLRW Universe, was leading to a complicated expression. We can ask ourself to which extent we are able to understand the effect of inhomogeneities if we only consider the second order coming from the quadratic terms of first order. We will see that most of the physics can already be understood from this linear order because of the reduced number of dominant terms. We then address the more complicated task of a full calculation, giving a complete result that will be used in the next chapter to give numerical predictions.

\section{Recalling the first order equations}
\label{Ch5Sec1}

The consideration of the transformation of coordinates, Eq. \rref{EqBetweenGauges}, between the geodesic light-cone gauge and the Newtonian gauge (\gls{NG}) (we forget about vectors and tensors here), assuming the simplication $\Phi = \Psi = \psi$ at first order and the appropriate boundary conditions, leads to the expression for $(\tau, w,\ti{\theta}^a)$-coordinates (see Eqs. \rref{tau2order} to \rref{thetatilde2orderShort})\,:
\bea
\label{tauwtheta}
\tau &=& \int_{\eta_{in}}^\eta d\eta' a(\eta')\left[1 + \psi(\eta', r, \theta^a)\right] ~~, \\
w &=& \eta_+ + \int_{\eta_+}^{\eta_-} dx\, \hat{\psi}(\eta_+, x, \theta^a) ~~,
\label{tauwtheta1}\\
\ti{\theta}^a &=& \theta^a + \frac{1}{2} \int_{\eta_+}^{\eta_-} dx \, \hat{\gamma}^{ab}_0 
(\eta_+, x,\theta^a) \int_{\eta_+}^x dy\, \pa_b \hat{\psi}(\eta_+, y, \theta^a) ~~,
\label{theta}
\eea
where $\hat{\psi}(\eta_+,\eta_-,\theta^a)\equiv \psi(\eta,r,\theta^a)$, $\hat \gamma^{ab}(\eta_+,\eta_-,\theta^a) \equiv \gamma^{ab}(\eta,r,\theta^a)$
and $\eta_{in}$ is, as already said, an early time when the integrand of $\tau$ was negligible.
Similarly, Eq. \rref{EqBetweenGauges} determines the non-trivial entries of the GLC metric (see Eqs. \rref{InvUps} to \rref{gammaab2})\,:
\scriptsize
\bea
\Ups &=& a(\eta) \left[ 1 + \hat{\psi}(\eta_+,\eta_+, \theta^a) - \int_{\eta_+}^{\eta_-} dx \,\partial_+ \hat{\psi}(\eta_+, x, \theta^a)\right]
+ \int_{\eta_{in}}^\eta d\eta' a(\eta')\partial_r \psi(\eta', r, \theta^a) ~~,
\label{Ups} \\
U^a &=& \frac{1}{2}  \hat{\gamma}^{ab}_0 \int_{\eta_+}^{\eta_-} dx\, \pa_b \hat{\psi}(\eta_+, x, \theta^a) - \frac{1}{a(\eta)}  \gamma^{ab}_0 \int_{\eta_{in}}^\eta d\eta' a(\eta') \, \pa_b \psi(\eta', r, \theta^a)
\nonumber \\
&& +\frac{1}{2} \int_{\eta_+}^{\eta_-} dx~ \partial_+ \left[\hat{\gamma}^{ab}_0(\eta_+,x,\theta^a) \int_{\eta_+}^x dy~ \pa_b\hat{\psi}(\eta_+, y, \theta^a) \right] -\frac{1}{2} \lim_{x\rightarrow \eta_+} \left[\hat{\gamma}^{ab}_0(\eta_+,x,\theta^a) \int_{\eta_+}^x dy~ \pa_b\hat{\psi}(\eta_+, y, \theta^a) \right] ~~, \\
\gamma^{ab} &=& \frac{1}{a(\eta)^2}\left\{
\left[1+2 \psi(\eta, r, \theta^a)\right]\gamma_0^{ab} 
+{1\over 2}\left[  \hat{\gamma}_0^{ac}  \int_{\eta_+}^{\eta_-} dx\,
\partial_c \left(
\hat{\gamma}_0^{bd}(\eta_+, x,\theta^a) \int_{\eta_+}^x dy\, \pa_d\hat{\psi}(\eta_+, y, \theta^a)\right)
+ a \leftrightarrow b \right]\right\} ~~.
\label{gammaab}
\eea
\normalsize

From Eq. \rref{redshift} one finds that the redshift of a source in the past light-cone of the observer is\,:
\beq
\label{OnePlusZ1}
1+z_s = \frac{a(\eta_o) }{a({\eta}_s)} \Big[1  + J(z_s, \theta^a) \Big],
\eeq
where we defined in \cite{P2}\,:
\bea
\label{IplusIr}
J =  I_+ - I_r ~~~~~ \mbox{with\,:} ~~~~~
I_+ &\equiv& ([\partial_+ Q]_s - [\partial_+ Q]_o) = \int_{\eta_s^+}^{\eta_s^-} dx \,\partial_+ \hat{\psi}(\eta_s^+,x,\theta^a) \nonumber \\
&=& \psi_s - \psi_o - 2 \int_{\eta_s}^{\eta_o} d\eta' \, \partial_r \psi (\eta', \eta_o-\eta', \theta^a) ~~,
\eea
\vspace{-0.5cm}
\beq
\mbox{and\,:} ~~~~~ I_r \equiv ([\partial_r P]_s - [\partial_r P]_o) = \int_{\eta_{in}}^{\eta_s} d\eta' \frac{a(\eta')}{a(\eta_s)} \partial_r \psi(\eta',\eta_o-\eta_s,\theta^a) - \int_{\eta_{in}}^{\eta_o} d\eta' \frac{a(\eta')}{a(\eta_o)} \partial_r \psi(\eta',0,\theta^a) ~~.
\label{Ir}
\eeq
and where the integrals $P$ and $Q$ have been defined in Eq. \rref{PQ}.
We have defined $\eta_s^\pm = \eta_s \pm r_s$, $\psi_s = \psi(\eta_s, \eta_o - \eta_s, \theta^a)$, $\psi_o = \psi(\eta_o, 0, \theta^a)$ and the zeroth-order light-cone condition $\eta_s^+ = \eta_o^+ = \eta_o$. We also used the change of coordinates\,:
\beq
\label{ChgtVarxeta}
x = 2 \eta - \eta_o ~~~~~\mbox{ with\,: }~~~~~ \left\{ \begin{array}{l} x = \eta_s^- ~\Leftrightarrow~ \eta = \eta_s \\ x = \eta_s^+ \equiv \eta_o ~\Leftrightarrow~ \eta = \eta_o \end{array} \right. ~~,
\eeq
and the general relation\,:
\beq
\label{deta}
\frac{d}{d\eta} \equiv \partial_\eta + n^i \nabla_i = \partial_\eta - \partial_r ~~,
\eeq
where $n^i$ are the components of $\nbf$, the unit tangent vector along the null geodesic connecting source and observer (and pointing from the last to the first).
It should be stressed, however, that while the integrals appearing in $I_r$ are evaluated at constant $r$ (namely along timelike geodesics), the integrals in $I_+$ are evaluated at fixed $\eta_+$ (i.e.  along null geodesics on the observer's past light-cone). 

The contribution associated to $I_r$ can also be rewritten as
\beq
\label{doppler1}
I_r = (\vbf_s  - \vbf_o ) \cdot \nbf ~~~~~ \mbox{ with } ~~~~~ \vbf_{s,o}=-\int_{\eta_{in}}^{\eta_{s,o}}d\eta'\frac{a(\eta')}{a(\eta_{s,0})}\vec{\nabla} \psi(\eta', r, \theta^a) ~~,
\eeq
the ``peculiar velocities" of source and observer associated to a geodesic configuration perturbed up to first order in the Newtonian gauge. 
In a realistic situation one should add to Eq. \rref{OnePlusZ1} similar terms taking into account a possible intrinsic (non-perturbative) motion of source and observer, unless our theoretical predictions are compared with data already corrected for these latter Doppler contributions. We suppose here that these terms are zero and thus state that the source and the observer are static w.r.t. each other (apart from the peculiar velocity coming from the gravitational perturbations). The expression for the redshift, Eq. \rref{OnePlusZ1}, is in full agreement with the expression obtained in Eq. (38) of \cite{Bonvin:2005ps} (where $z_s$ is computed for the particular case of a CDM-dominated model).

To obtain the angular distance, given by Eq. \rref{General_Eq_dA}, of a source emitting light at time $\eta_s$ and radial distance $r_s$, one computes the determinant $\gamma$ from Eq. (\ref{gammaab}), to first order\footnote{Note that $\gamma^{-1} = (\gamma_{11} \gamma_{22})^{-1}$ at first order and, for these diagonal matrix elements, $\pa_b$ commutes with $\gamma_0^{ab}$.}, and we get\,:
\beq
\gamma^{-1}(\la_s) \equiv \det \gamma^{ab}(\la_s) = (a_s^2r_s^2\sin\theta)^{-2} \Big[1 + 4 \psi_s + 4 \ti{J}_2(z_s, \theta^a) \Big] ~~,
\label{detOnlyOrder1}
\eeq
where (using a simple integration by parts)\,:
\bea
\ti{J}_2 &=& \frac{1}{4} \int_{\eta_s^+}^{\eta_s^-} dx\, \hat{\gamma}_0^{ab}(\eta_s^+,x,\theta^a) \int_{\eta_s^+}^x dy\, \pa_a \pa_b\hat{\psi}(\eta_s^+,y,\theta^a) \nonumber \\
&=& {\frac{1}{\eta_o-\eta_s} \int_{\eta_s}^{\eta_o} d \eta' \frac {\eta' - \eta_s}{\eta_o - \eta'} \Big[ \partial^2_{\theta} + (\sin\theta)^{-2}\partial^2_{\phi}\Big] \psi(\eta', \eta_o-\eta', \theta^a)} ~~.
\eea
We then get\,:
\beq
d_A(\la_s) = a_s r_s \Big[ 1 - \psi_s - \ti{J}_2 \Big] \left(\frac{\sin \ti \theta^1}{\sin \theta}\right)^{-1/2} = a_s r_s \left[ 1 - \psi_s - J_2 (z_s, \theta^a) \right] ~~,
\label{d_s2OnlyOrder1}
\eeq
where in the second equality we developed the term $\ti\theta^1$ from Eq. \rref{theta} to define $J_2$ as\,:
\bea
J_2 &=& \frac{1}{\eta_o - \eta_s} \int_{\eta_s}^{\eta_o} d \eta \,\frac {\eta - \eta_s}{\eta_o - \eta} \Big[ \partial^2_{\theta} + \cot \theta\, \partial_{\theta} + (\sin \theta)^{-2} \partial^2_{\phi}\Big] \psi(\eta, \eta_o - \eta, \theta^a) \nonumber \\
&\equiv& \frac{1}{\eta_o - \eta_s} \int_{\eta_s}^{\eta_o} d \eta \, \frac {\eta - \eta_s}{\eta_o - \eta} \, \Delta_2 \psi (\eta, \eta_o - \eta, \theta^a) ~~,
\label{J2_Equation}
\eea
and where $\Delta_2$ is the two-dimensional Laplacian operator on the unit two-sphere.

We finally obtain the full explicit expression of $d_L$ by the first-order expansion of $a_sr_s\equiv a(\eta_s) r_s$ appearing in Eq. \rref{d_s2OnlyOrder1}. To this purpose we start from Eq. (\ref{OnePlusZ1}), considering $z_s$  as a constant parameter localising the given light source on the past light-cone ($w=w_o$) of our observer, and we look for approximate solutions for $\eta_s= \eta_s(z_s, \theta^a)$. Expanding Eq. \rref{OnePlusZ1} with respect to the parameter $\delta \eta = {\eta}_s - {\eta}_s^{(0)}$, we find\,:
\beq
\label{OnePlusZ2}
\frac{1}{1+z_s} = \frac{a({\eta}_s^{(0)})}{a_0}  [1 + {\mathcal H}_s ~\delta \eta - J(z_s, \theta^a) ] \equiv
\frac{1}{1+z_s} [1 + {\mathcal H}_s \, \delta \eta - J(z_s, \theta^a)] ~~, \\
\eeq
where ${\mathcal H}_s = d (\ln a({\eta}_s^{(0)}))/d\eta_s^{(0)}$, so that ${\mathcal H}_s\, \delta \eta =  J(z_s, \theta^a)$. On the other hand, applying Eq. \rref{tauwtheta1} to the light-cone $w = w_o$ at the source position gives 
\beq
\label{Relw0pert}
w_o = \eta_s^+ + Q_s = \eta_o ~~,
\eeq
where we use $\Delta \eta = \eta_o- {\eta}_s^{(0)}$ and $Q_s$ is the average value of $\psi$ along the (unperturbed) null geodesic connecting source and observer\,:
\beq
Q_s = \int_{\eta^s_+}^{\eta^s_-} dx \, \hat\psi(\eta^s_+, x, \theta^a) = - 2 \int_{\eta_s}^{\eta_o} d \eta' \, \psi (\eta', \eta_o-\eta', \theta^a)  ~~.
\label{Psi_average_Equation}
\eeq
Eq. \rref{Relw0pert} thus brings the value of the radial coordinate $r_s(z_s, \theta^a)$ for the given redshift $z_s$\,:
\bea
{r}_s(z_s, \theta^a) = w_o - {\eta}_s^{(0)}(z_s) - \delta\eta - Q_s = \Delta \eta \left[ 1 - \frac{J(z_s, \theta^a)}{{\mathcal H}_s \Delta \eta } - \frac{Q_s}{\Delta \eta} \right] ~~,
\eea
and we finally obtain the value of $a_sr_s$ on the constant-$z_s$ 2-surface\,:
\beq
[{a}_s {r}_s](z_s, \theta^a) = a({\eta}_s^{(0)}) \Delta \eta \left[ 1 - \frac{Q_s}{\Delta\eta} + \Xi_s J(z_s, \theta^a) \right] ~~.
\label{asrsOnlyOrder1}
\eeq
The angular distance of Eq. (\ref{d_s2OnlyOrder1}), for a source at redshift $z_s$, can now be written as\footnote{Let us note that in the first order terms we can always safely identify $\eta_s$ with its zeroth-order value $\eta_s^{(0)}$.}
\bea
d_A(z_s, \theta^a) = a({\eta}_s^{(0)}) \Delta \eta \left[1 - \frac{Q_s}{\Delta\eta} + \Xi_s J - \psi({\eta}_s,\eta_o-{\eta}_s, \theta^a) - J_2\right] ~~, 
\eea
and the expression of $d_L$ in our perturbed background is finally given, using $\psi_s \equiv \psi({\eta}_s,\eta_o-{\eta}_s,\theta^a)$, by\,:
\beq
\label{finaldLOrder1}
\frac{d_L(z_s, \theta^a)}{(1+z_s)a_0 \Delta \eta} \equiv {d_L(z_s, \theta^a)\over d_L^{\rm FLRW}(z_s)} = 1 - \psi_s - \frac{Q_s}{\Delta\eta} + \Xi_s J - J_2 ~~.
\eeq

If we apply this general result to the particular case of a CDM-dominated  Universe we find almost full agreement with the result for the luminosity distance at constant redshift computed in \cite{Bonvin:2005ps}. After several manipulations, in fact, it turns out that our $d_L$ is equivalent to the one found in \cite{Bonvin:2005ps}, modulo a term which can be written as $\vbf_o \cdot \nbf$ (in the notations of Eq. (\ref{doppler1})). Such a term gives a subleading contribution to the backreaction and can be neglected with no impact on our final results.

It should be noted, however, that by expanding the above expression in the small $z_s$ limit, and comparing the result with the analogous expansion of the homogeneous distance in Eq. (\ref{standard1}), we could easily introduce  a redefined value of $H_0$, say $H_0^{\rm \,ren}$, such that its inverse corresponds to the coefficient of the linear term in $z_s$ for the perturbed relation. With such a ``renormalized'' Hubble parameter we have that $H_0^{\rm \,ren} d_L$ tends smoothly to $H_0 d_L^{\rm FLRW}$ for $z_s \ra 0$, and we recover full agreement with \cite{Bonvin:2005ps} provided the same renormalization is applied there. Such a renormalization of $H_0$ is also suggested by (and closely related to) the first-order computation of the scalar expansion factor $\nabla_\mu u^\mu$ for the flow world lines $u_\mu = - \pa_\mu \tau$ of a local geodesic observer. 

We could thus choose, for the next considerations, to take the average of $H_0^{\rm \,ren} d_L(z_s, \theta^a)$ instead of $d_L(z_s, \theta^a)$, computing the associated backreaction, and evaluating the corrections to the standard homogeneous relation given by $H_0 d_L^{\rm FLRW}(z_s)$. We have performed that exercise, but we have found that the contribution of the renormalization terms give large $z_s$-independent contributions to the variance. Furthermore, renormalizing $H_0$ at $z_s=0$ could be physically incorrect since, at very small-$z_s$, the backreaction is dominated  by short-scale inhomogeneities which are deeply inside the non-linear regime (where even the concept of a Hubble flow becomes inappropriate). We could instead try to renormalize $H_0$ at some small but finite $z_s$, e.g. at a redshift corresponding to the closest used supernovae (say $z_s \sim 0.015$, see e.g. \cite{Amanullah:2010vv}), but then the results (although much better behaved) would depend on the choice of the particular ``renormalization point". Thus, it seems better to consider just the full ``unrenormalized" expression of Eq. (\ref{finaldLOrder1}) in a limited region of $z_s$ where one can trust the approximations  made, and use that expression for a phenomenological parametrization of the backreaction effects which could possibly include a redefinition of $H_0$.

\section{Combining lightcone and ensemble averages}
\label{Ch5Sec2}

\subsection{Definition of the stochastic average}
\label{Ch5Sec21}

The simplest way to implement the ensemble average of our stochastic background of (first order) scalar perturbations $\psi$ is to consider their Fourier decomposition in the form\,:
\beq
\psi(\eta, \xbf) = \frac{1}{(2 \pi)^{3/2}} \int d^3 k \, \e^{i \kbf \cdot \xbf} \, \psi_k(\eta) E(\kbf) ~~,
\label{PsiFourier}
\eeq
where, assuming that the fluctuations are statistically homogeneous, $E$ is a unit random variable satisfying the \emph{\gls{statistical homogeneity condition}}\,:
\beq
E^*(\kbf) = E(-\kbf) ~~,
\eeq
as well as  the \emph{\gls{ensemble-average conditions}}\,:
\beq
\label{PropRandomVar}
\overline{E(\kbf)} = 0 ~~~~~ \forall ~ \kbf ~~~~~,~~~~~ \overline{E(\kbf_1) E(\kbf_2)} = \delta_{\rm D}(\kbf_1 + \kbf_2) ~~.
\eeq
These relations actually claim that two modes of different wavelengths are independent. We should keep in mind that this is true in a general FLRW model only when flatness is assumed. That is indeed the case which is taken all the way here and such a decomposition is hence justified.

In order to proceed with some qualitative and quantitative considerations it is important now to specify some of the properties of the power spectrum of $\psi$, denoted $\Pcal_{\psi}(k)$, and defined in general by\,:
\beq
\label{RecallLinPS}
\Pcal_\psi (k, \eta) \equiv \frac{k^3}{2 \pi^2} |\psi_k(\eta)|^2 ~~.
\eeq
Limiting ourselves to sub-horizon perturbations, and considering the standard CDM model, we can simply obtain $\psi_k$ by applying an appropriate, time-independent \gls{transfer function} to the primordial (inflationary) spectral distribution (see e.g. \cite{DurCos}). The power spectrum of the Bardeen potential is then given by
\beq
\label{PsiP}
\Pcal_\psi (k) = \left(\frac{3}{5}\right)^2 \Delta_{\cal R}^2 T^2(k) ~~~~~,~~~~~ \Delta_{\cal R}^2=A \left(\frac{k}{k_0}\right)^{n_s-1} ~~,
\eeq
where $T(k)$ is a time-dependent transfer function which takes into account the sub-horizon evolution of the modes re-entering during the radiation-dominated era, and  $\Delta_{\cal R}^2$ is the primordial power spectrum of curvature perturbations, amplified by inflation, outside the horizon. The typical parameters of such a spectrum, namely the amplitude $A$,  the spectral index $n_s$ and the scale $k_0$, are determined by the results of WMAP observations \cite{Komatsu:2010fb}. In our computations we will use, in particular, the following approximate values\,:
\beq
 A=2.45 \times 10^{-9} ~~~~~,~~~~~  n_s=0.96 ~~~~~,~~~~~ k_0/a_0=0.002 \,{\rm Mpc}^{-1} ~~.
\eeq
Finally, we will have to assume an explicit form of the transfer function $T(k)$. That will be done in Chapter \ref{Chap6} where different cases will be considered from \cite{Eisenstein:1997ik}, in CDM and $\Lambda$CDM models, with and without baryons.

In the next sections we will discuss the corresponding backreaction effects on $\overline{\langle d_L \rangle}$, together with its dispersion, for the scalar perturbation spectrum described by Eq. (\ref{PsiP}) and evaluate their spectral coefficients in order to prepare their numerical estimation in Chapter \ref{Chap6}.

\subsection{Combination with the lightcone average\,: induced backreaction terms}
\label{Ch5Sec22}

Let us consider that the inhomogeneous deviations from the standard FLRW quantities are sourced by a stochastic background of primordial perturbations and, according to Eq. \rref{PropRandomVar}, satisfy\,:
\beq
\overline \psi = 0 ~~~~~,~~~~~ \overline{\psi^2} \neq 0 ~~.
\eeq
Hence, if we limit ourselves to a first-order computation of a scalar quantity $S$, we immediately obtain $\overline{S}= S^{\rm FLRW}$. Non-trivial effects can only be obtained from quadratic and higher-order perturbative corrections, or from the spectrum of the two-point correlation function $\overline{S(z, \theta^a) S(z', \theta^{\prime a})}$ (as discussed in detail in \cite{Bonvin:2005ps} for $S = d_L$). 
 
In this chapter we will consider the ensemble average not of $S$ but of $\langle S \rangle$, where the angular brackets refer to the light-cone angular average defined in Eq. \rref{zaverageexact}. We should notice that several studies, such as \cite{Li:2007ny,Clarkson:2009hr,Clarkson:2011uk}, considered the combination of the ensemble average with averages over spacelike hypersurfaces. However, its combination with lightcone averages has never been addressed before this work. We will see in this section that the light-cone average automatically induces quadratic (and higher-order) backreaction terms, due to the inhomogeneities present both in $d_L$ and in the covariant integration measure\footnote{This is also the case for spatial averaging. Nevertheless, the integration measure is modified and the generated terms are directly affected by this geometrical difference (see Sec. \ref{Ch5Sec24}).}, and since the ensemble average of such terms is non-vanishing we obtain in general that $\overline{\langle S \rangle} \neq S^{\rm FLRW}$.
We will now present general considerations on how to combine lightcone and ensemble averaging, and how to isolate those terms in $\overline{\langle S \rangle}$ that we may genuinely call ``\gls{backreaction}" effects, i.e. effects on averaged quantities due to inhomogeneities. We shall also discuss how to estimate the variance around mean values due to such fluctuations\footnote{The importance of the cosmic variance for a precise measurement of the cosmological parameters, taking into account backreaction effects from averaging on domains embedded in a spatial hypersurface (according to \cite{Buchert:1999er,Buchert:2001sa}), has been recently pointed out also in \cite{Wiegand:2011je}.}.

Let us consider a typical  average over the compact surface $\Sg$ (topologically equivalent to a 2-sphere) embedded on the past light-cone $w=w_o$ at constant $z_s$. We simply denote this average by\,:
\beq
\label{AvS}
\langle S  \rangle_\Sg =  \frac{\int_\Sg d^2 \mu \,S}{ \int_\Sg d^2 \mu} ~~,
\eeq
where $d^2 \mu$ is the appropriate measure provided by our gauge invariant prescription (see Eq. \rref{zaverageexact}) and $S$ is a (possibly non local) scalar observable.
We can conveniently extract, from both $d^{\,2} \mu$ and from  $S$, a zeroth-order homogeneous contribution by defining\,:
\bea
\label{exp}
d^2 \mu =  (d^2 \mu)^{(0)}(1 + \mu) ~~~~~,~~~~~  S = S^{(0)} (1 + \sigma) ~~,
\eea
and use the possibility of rescaling both integrals in Eq. (\ref{AvS}) by the same constant, in order to normalize $\int (d^2 \mu)^{(0)} = 1$.
We then easily get:
\bea
\label{AvS1}
\left\langle \frac{S}{S^{(0)}}  \right\rangle =  \frac{\int (d^2 \mu)^{(0)}(1+ \mu)(1+\sigma) }{\int (d^2 \mu)^{(0)}(1+ \mu)} = 
{\langle (1+ \mu)(1+\sigma)\rangle_0  \over 1+ \langle \mu \rangle_0} ~~,
\eea
where we have dropped, for simplicity, the subscript of the averaging region $\Sg$, and where we have defined by $\langle \dots \rangle_0$ averages with respect to the unperturbed measure $(d^2 \mu)^{(0)}$. For simplicity, we shall even drop the subscript $0$ hereafter.

Let us now perform the ensemble average, paying attention to the fact that ensemble averages do {\it not} factorize, i.e. $\overline{AB} \ne \overline{A} ~\overline{B}$. This simple combination leads to\,:
\bea
\label{EnsAv}
\overline{\langle S/S^{(0)} \rangle} = 1 + \overline{(\langle \sigma\rangle+ \langle \mu \sigma\rangle)  (1+ \langle \mu \rangle)^{-1}} ~~.
\eea
This last equation is supposedly exact but, as such, pretty useless. It becomes an interesting equation, though, if we can expand the quantities $\mu$ and $\sigma$ in perturbative series\,:
\bea
\mu = \sum_i \mu_i ~~~~~,~~~~~~ \sigma = \sum_i \sigma_i ~~,
\eea
and we further assume that the first order quantities $\mu_1$ and $\sigma_1$ have vanishing ensemble averages (as it is the case for typical cosmological perturbations coming from inflation).
In that case we can easily expand the result and obtain, for instance\,:
\beq
\label{EnsAvExp}
\overline{\langle S/S^{(0)}  \rangle}  = 1 +  \overline{\langle \sigma_2 \rangle} + \rm{IBR}_2 + \overline{\langle \sigma_3 \rangle} + \rm{IBR}_3 + \dots 
\eeq
where
\bea
\label{ibr2}
\rm{IBR}_2 &=&  \overline{\langle \mu_1 \sigma_1\rangle} - \overline{\langle \mu_1\rangle  \langle \sigma_1\rangle} ~~, \\
\rm{IBR}_3 &=&  \overline{\langle \mu_2 \sigma_1\rangle} - \overline{\langle \mu_2\rangle  \langle \sigma_1\rangle} +
\overline{\langle \mu_1 \sigma_2\rangle} - \overline{\langle \mu_1\rangle  \langle \sigma_2\rangle} - \overline{\langle \mu_1\rangle  \langle \mu_1 \sigma_1\rangle} + \overline{\langle \mu_1\rangle \langle \mu_1\rangle \langle \sigma_1\rangle} ~~,
\eea
and where we have used again the non-factorization property, i.e. $ \overline{\langle \mu_1\rangle  \langle \sigma_1\rangle} \neq \overline{\langle \mu_1\rangle}~  \overline{\langle \sg_1\rangle}$, and so on. We see that the result contains, to a given order, both terms that depend on expanding $S$ to that order (but {\it not} on the precise averaging prescription), and \gls{IBR} terms  that  depend on correlations between the fluctuations of $S$ and those in the measure. These latter terms only depend on lower-order perturbations of $S$ and the measure separately. Note also that, whenever the fluctuations of $S$ and $d^2 \mu$ are uncorrelated, all IBR effects drop out.

Let us now discuss the issue of the variance, i.e. of how broad is the distribution of values for $S/S^{(0)} $ around its mean value $\overline{\langle S/S^{(0)} \rangle}$. This dispersion is due to both the fluctuation on the averaging surface and to those due to ensemble fluctuations. Let us thus define\,:
\beq
\label{vargen}
{\rm Var}[S/S^{(0)}] \equiv \overline{\lla \left(S/S^{(0)} - \overline{\langle S/S^{(0)}\rangle} \right)^2 \rra}  = \overline{\lla (S/S^{(0)})^2 \rra}  - \left(\overline{\lla S/S^{(0)} \rra}\right)^2 ~~.
\eeq
Inserting the definitions of Eq. (\ref{exp}), we get, after a little algebra\,:
\beq
\label{vargenexp}
{\rm Var}[S/S^{(0)}] =  \overline{\langle \sigma^2 (1 + \mu) \rangle(\langle 1 + \mu \rangle)^{-1}}  - \left(\overline{\langle \sigma (1 + \mu) \rangle (1+ \langle  \mu \rangle)^{-1}} \right)^2 ~~.
\eeq
If we now make the same assumptions as before on expanding $\mu$ and $\sigma$, we find that the second term
in Eq. (\ref{vargenexp}) is at least of fourth order.  
Therefore, for the leading term in the variance we find the amazingly simple result (see also \cite{Wiegand:2011je})\,:
\beq
\label{vargen2nd}
{\rm Var}[S/S^{(0)}] = \overline{\langle \sigma_1^2  \rangle} ~~.
\eeq
As in the case of IBR$_2$ we only need to know the first-order perturbation, but this time the effect is completely independent of the averaging measure. As we shall see, for $S = d_L$ the dispersion (which is actually the square root of the variance) turns out to be larger than the averaging corrections due to IBR$_2$ (see Chapter \ref{Chap6}).

We may note, at this point, that we could have also considered the dispersion of the angular average $\langle S/S^{(0)}\rangle$ due to the stochastic fluctuations, namely\,:
\beq
\label{vargen1}
{\rm Var'}[S/S^{(0)}] \equiv \overline{\left(\langle S/S^{(0)}\rangle- \overline{\langle S/S^{(0)}\rangle} \right)^2} = \overline{\left(\lla S/S^{(0)}\rra\right)^2}  - \left(\overline{\lla S/S^{(0)} \rra}\right)^2 ~~,
\eeq
which, after calculations similar to the ones above, gives
\beq 
\label{varp}
{\rm Var'}[S/S^{(0)}] = \overline{\langle \sigma_1 \rangle^2} ~~.
\eeq
Such a quantity is much smaller than the previous one, as can be inferred from $S = d_L$, indicating that the main reason for the dispersion lies in the angular scatter of the data rather than in their stochastic distribution due to the ensemble (see e.g. Sec. \ref{Ch6Sec1}). One can indeed understand that for an early enough time we are integrating inhomogeneities over a very large sphere $\Sigma(w_o,z_s)$ and thus the main source of dispersion, when we consider different realizations of the inhomogeneous universe, comes from the combination of lightcone and ensemble average acting on $S$ (involving the fluctuation of $S$) rather than the ensemble average of $\lla S \rra$ alone (which averages the different realizations of $\Sigma(w_o,z_s)$).

\subsection{Application to $S = d_L$ up to the quadratic first order terms}
\label{Ch5Sec23}

This section will be devoted to a systematic evaluation of the various terms contributing to the induced backreaction IBR$_2$ defined in the previous section, as well as to the variance associated to the (angle and ensemble)-averaged value of $d_L(z_s, \theta^a)$.

Indeed, to describe the real impact of inhomogeneities on the observational data we have to consider the scalar $S$ corresponding to the true observed quantity. In the case of the supernovae data \cite{Riess:1998cb,Perlmutter:1998np} this quantity is the received flux of radiation, which is proportional to $d_L^{-2}$. On the other hand, we can think to consider the average of $d_L$ instead of the average of the flux. One can thus take (among others) two different ways to compute the distance, with the following expressions\,:
\beq
\label{InterestofIBR2sigma1and2}
(\overline{\langle d_L^{-2} \rangle})^{-1/2} = d_L^{\rm FLRW} \left( 1 + \overline{\langle \sigma_2 \rangle} + \rm{IBR}_2 - \frac{3}{2}\overline{\langle \sigma_1^2 \rangle} \right) ~~~~~,~~~~~ \overline{\langle d_L \rangle} = d_L^{\rm FLRW} \left( 1 + \overline{\langle \sigma_2 \rangle} + \rm{IBR}_2 \right) ~~.
\eeq
As we can notice, the difference between the two quantities only appears at second order. The genuine second-order contribution $\overline{\langle \sigma_2 \rangle}$ cannot be determined from the first order transformation that we use here, but is only attainable through the full second order transformation that we will treat in the next section, Sec. \ref{Ch5Sec3}. We remark also the interesting fact that the evaluation of $\mu_2$, the second order contribution of the measure, is not needed for the full second order computation. We shall thus present the calculation of only $\rm{IBR}_2$ and $\overline{\langle \sigma_1^2 \rangle}$ here, i.e. the effect that comes from the above interplay of $\mu_1$ and $\sg_1$ or $\sg_1$ with itself, and consider the full case later. One last remark is that the distance modulus, rather than being computed as a function of $(\overline{\langle d_L^{-2} \rangle})^{-1/2}$ or $\overline{\langle d_L \rangle}$ should be directly averaged to test its sentivity to inhomogeneities (see Sec. \ref{Ch5Sec33}).

In such a case, from Eq. (\ref{finaldLOrder1}) we immediately find\,:
\beq 
\label{sigma1}
\sigma_1 = A_1 + A_2 + A_3 + A_4 + A_5 ~~,
\eeq
where we have defined\,:
\beq
\label{defA}
A_1 = -\psi_s ~~,~~  A_2 =  - \frac{Q_s}{\Delta \eta} ~~,~~  A_3 = \left(1 - \frac{1}{{\mathcal H}_s\Delta \eta}\right) I_+ ~~,~~ A_4 = - \left(1 - \frac{1}{{\mathcal H}_s\Delta \eta}\right) I_r ~~,~~ A_5= - J_2 ~~.
\eeq
On the other hand, in our particular case, the general measure $d^2\mu$ is given by Eq. \rref{zaverageexact}\,:
\beq
d^2\mu=d^2 \ti{\theta} \sqrt{\gamma(w_o,\tau(z_s,w_o,\tilde{\theta}^a),\tilde{\theta}^a)} ~~.
\eeq
We need to transform the integral in $d^2 \ti{\theta}$ (over the ``$2$-sphere" $\Sg$) to the standard polar coordinates  $d^2 \theta= d\theta d\phi$ of the Newtonian gauge. To first order, it is easy to check (by taking into account the Jacobian determinant of the transformation $\theta^a \ra \ti \theta^a$, see Eq. (\ref{theta})) that\,:
\beq
\int d^2 \ti{\theta} \sqrt{\gamma} = \int d\phi d\theta \sin \theta \left([a_s r_s](z_s, \theta^a)\right)^2 (1- 2 \psi_s) ~~.
\eeq
The normalized unperturbed measure is then given by $(d^2 \mu)^{(0)}= d^2 \Om/ 4 \pi$, where $d^2\Om= d \phi d\theta \sin \theta$.
Using now the expansion of $a_s r_s$ presented in Eq. \rref{asrsOnlyOrder1}, we obtain\,:
\beq
\label{mu1}
\mu_1 = - 2 \psi_s - 2 \frac{Q_s}{\Delta \eta} + 2 \left(1 - \frac{1}{{\mathcal H}_s\Delta \eta}\right) J(z_s, \theta^a) = 2\,( A_1 + A_2 + A_3 + A_4) ~~.
\eeq

The second-order induced backreaction IBR$_2$, and the variance of $d_L/d_L^{\rm FLRW}$, can now be obtained by inserting the results of Eqs. (\ref{sigma1}), (\ref{mu1}) into Eqs. (\ref{ibr2}) and (\ref{vargen2nd}), respectively. According to these expressions, all corrections we need to compute are bilinear terms in the potential $\psi$, always occurring in the combination $\overline{\langle A_i A_j \rangle}$ or $\overline{\langle A_i \rangle \langle A_j \rangle} $, where  the quantities $A_i$, $A_j$ are defined in Eq. (\ref{defA}).

Let us illustrate a typical computation with the simplest contribution associated to $A_1= -\psi_s$. Following previous notations, we will denote with simple angular brackets the integration over the two-surface $\Sg$ embedded on the light-cone (at $w=w_o$, $z=z_s$). The measure of integration is the unperturbed normalized one, i.e. $d^2 \Om/ 4 \pi$. We then obtain\,:
\bea
\label{TrivialExample1}
\overline {\lla \psi_s \psi_s \rra} &=& \int \frac{d^3 k ~d^3 k'}{(2 \pi)^{3}} \overline{E(\kbf) E(\kbf')} \int \frac{d^2 \Omega}{4 \pi} \left[ \psi_k(\eta_s^{(0)}) e^{i r \kbf \cdot \hat{\xbf}}\right]_{r=\eta_o - \eta_s^{(0)}} \left[ \psi_{k'}(\eta_s^{(0)}) e^{i r \kbf' \cdot \hat{\xbf}} \right]_{r=\eta_o - \eta_s^{(0)}} \nonumber \\
&=& \int \frac{d^3 k}{(2 \pi)^{3}} ~ |\psi_k(\eta_s^{(0)})|^2 \int_{-1}^{1} \frac{d(\cos\theta)}{2} \left[ e^{i k \Delta \eta \cos\theta} \right] \left[ e^{-i k \Delta \eta \cos\theta} \right] \nonumber \\
&=& \int \frac{d^3 k}{(2 \pi)^{3}}  ~ |\psi_k(\eta_s^{(0)})|^2 = \int_0^{\infty} \frac{d k}{k} ~ \Pcal_{\psi}(k,\eta_s^{(0)}) ~~,
\\
\label{TrivialExample1a}
\overline {\lla \psi_s \rra \lla \psi_s \rra} &=& \int \frac{d^3 k ~d^3 k'}{(2 \pi)^{3}} \overline{E(\kbf) E(\kbf')} \left[ \int \frac{d^2 \Omega}{4 \pi} \psi_k(\eta_s^{(0)}) e^{i r \kbf \cdot \hat{\xbf}} \right]_{r=\eta_o - \eta_s^{(0)}}   \left[ \int \frac{d^2 \Omega'}{4 \pi} \psi_{k'}(\eta_s^{(0)}) e^{i r \kbf' \cdot \hat{\xbf}'} \right]_{r=\eta_o - \eta_s^{(0)}} \nonumber \\
&=& \int \frac{d^3 k}{(2 \pi)^{3}} ~ |\psi_k(\eta_s^{(0)})|^2 \left[ \int_{-1}^{1} \frac{d(\cos\theta)}{2} e^{i k \Delta \eta \cos\theta} \right] \left[ \int_{-1}^{1} \frac{d(\cos\theta')}{2} e^{-i k \Delta \eta \cos\theta'} \right] \nonumber \\
&=& \int \frac{d^3 k}{(2 \pi)^{3}}  ~ |\psi_k(\eta_s^{(0)})|^2 \left( \frac{\sin(k \Delta\eta)}{k \Delta \eta} \right)^2 = \int_0^{\infty} \frac{d k}{k}  ~ \Pcal_{\psi}(k,\eta_s^{(0)}) \left( \frac{\sin(k \Delta\eta)}{k \Delta \eta} \right)^2 ~~,
\eea
where in the second line of both equations we made use of isotropy (i.e. $\psi_k$ only depends on $k = |\kbf|$), and defined $\theta$ and $\theta'$ as the angles between $\kbf$ and $\xbf \equiv r \hat{\xbf}$ and between $\kbf'$ and $\xbf' \equiv r \hat{\xbf}'$. We also made use of the (so-called dimensionless) power spectrum defined in Eq. \rref{RecallLinPS}. More complicated examples, that contain almost all the subtleties of these computations, are presented in Sec. \ref{Ch5Sec24}.

In general we have many contributions like the above ones, appearing in both the induced backreaction and the variance, and generated by all the $A_i$ terms of Eq. (\ref{defA}). Here, for a start, we only consider the particularly simple case of a CDM-dominated background geometry, with a time-independent spectral distribution of sub-horizon scalar perturbations, $\pa_\eta \psi_k=0$. In such a case all the relevant contributions can be parameterized in the form\,:
\bea
\overline{\langle A_iA_j \rangle} &=& \int_0^{\infty} \frac{d k}{k} ~ \Pcal_{\psi}(k) \, {\mathcal C}_{ij}(k,\eta_o,\eta_s^{(0)}) ~~,
\label{aiaj}
\\  
\overline{\langle A_i \rangle \langle A_j \rangle} &=& \int_0^{\infty} \frac{d k}{k} ~ \Pcal_{\psi}(k) \, {\mathcal C}_{i}(k,\eta_o,\eta_s^{(0)})~ {\mathcal C}_{j} (k,\eta_o,\eta_s^{(0)}) ~~,
\label{ai}
\eea
where ${\mathcal C}_{ij}$ and ${\cal C}_i$ are constant \gls{spectral coefficients}. This form is valid for any given model of time-independent scalar perturbation spectrum. With such a parametrization, the leading-order induced backreaction, called $\rm{IBR}_2$ in the expansion of Eq. \rref{EnsAvExp}, of $\overline{\langle d_L  \rangle}/ d_L^{\rm FLRW}$, can be written as\,:
\bea
{\rm IBR}_2 &=& 2 \sum_{i=1}^4 \sum_{j=1}^5  \Big[ \overline{\langle A_iA_j \rangle} - \overline{\langle A_i \rangle \langle  A_j \rangle} \Big] \nonumber \\
&=& \int_0^{\infty} \frac{d k}{k}  ~ \Pcal_{\psi}(k) \times 2 \sum_{i=1}^4 \sum_{j=1}^5  \Big[ {\mathcal C}_{ij}(k,\eta_o,\eta_s^{(0)}) - {\mathcal C}_{i}(k,\eta_o,\eta_s^{(0)})~ {\mathcal C}_{j} (k,\eta_o,\eta_s^{(0)}) \Big] ~~.
\label{BR2backreaction}
\eea
The dispersion, from Eq. \rref{vargen2nd}, takes instead the form\,:
\bea
\label{BRVarTh}
\left({\rm Var}\left[\frac{d_L}{d_L^{\rm FLRW}}\right]\right)^{1/2} &=& \sqrt{\overline{\lla  \sigma_1^2 \rra}} ~~ = ~~ \left[ \sum_{i=1}^5 \sum_{j=1}^5 \overline{\langle A_iA_j \rangle}\right]^{1/2} \nonumber \\
&=& \left[\int_0^{\infty} \frac{d k}{k} ~ \Pcal_{\psi}(k) \sum_{i=1}^5 \sum_{j=1}^5  {\mathcal C}_{ij}(k,\eta_o,\eta_s^{(0)})\right]^{1/2} ~~.
\eea

The analytical values of all the required coefficients ${\mathcal C}_{ij}$, ${\cal C}_i$ are given in Table \ref{Tab1OfP2} for ${\mathcal C}_{ij}(k,\eta_o,\eta_s^{(0)})$, and in Table \ref{Tab2OfP2} for ${\mathcal C}_{i}(k,\eta_o,\eta_s^{(0)})$. In Table \ref{Tab1OfP2} we also show the small-$k$ limit ($k \Delta \eta \ll 1$) of ${\mathcal C}_{ij}$ and of the products ${\mathcal C}_{i} \, {\mathcal C}_{j}$. For notational convenience we have introduced in the tables the dimensionless variable $l= k \Da \eta$, and we have used the definition of the sine integral\,:
\beq
\label{DefSinInt}
{\rm SinInt}(l)= \int_0^l dx \, \frac{\sin x}{x} ~~.
\eeq
It should be noted that the computation of some of the terms requires the explicit expression of the scale factor $a(\eta)$. It is the case of the terms involving the integral $P(\eta,r,\ti{\theta}^a)$ (see Eq. \rref{PQ}). In those cases we have assumed a dust-dominated CDM scale factor with $a(\eta) = a(\eta_o)(\eta/\eta_o)^2$, and we have defined\,:
\beq
f_{o,s} \equiv \int_{\eta_{in}}^{\eta_{o,s}} d\eta \frac{a(\eta)}{a(\eta_{o,s})} = \frac{\eta_{o,s}^3 - \eta_{in}^3}{3\eta_{o,s}^2} ~ \simeq \frac13 \eta_{o,s} ~~,
\eeq
where we recall that $\eta_{in}$ satisfies, by definition, $\eta_{in} \ll \eta_{o,s}$. 
Using such a CDM-dominated model also imposes the following (zeroth-order) relation between $z_s$ and $\Delta \eta$\,:
\beq
\label{DefDeltaEta}
\Delta \eta = \frac{2}{a_o H_0} \left[1-\frac{1}{\sqrt{1+z_s}}\right] ~~.
\eeq

We can notice on this table that the difference $\overline{\langle A_iA_j \rangle} - \overline{\langle A_i \rangle \langle  A_j \rangle}$ is giving spectral coefficients which are all at least of order $(k \Delta \eta)^2 = l^2$. This states that the terms $\overline{\langle A_i \rangle \langle A_j \rangle}$, though negligible in general with respect to $\overline{\langle A_iA_j \rangle}$ guarantee the good behaviour of the terms in the limit $z_s \rightarrow 0$. Indeed, one can see from Eq. \rref{DefDeltaEta} that at small redshifts one has $\Delta \eta \simeq z_s / a_o H_0$, thus each of these contributions, for a fixed mode $k$, goes to zero in the limit $k \Delta \eta \sim z_s \rightarrow 0$. Though being clear, this nice limit property will not appear in the plots of Chapter \ref{Chap6} and the reason is that the turning point of this limit happens at really small redshifts ($\lesssim 10^{-4}$ depending on the modes). Looking at the $l^2$ dependence of the terms, one can easily see in Table \ref{Tab1OfP2} that the dominant contributions are $\overline{\langle A_4 A_4 \rangle}$ and $\overline{\langle A_5 A_5 \rangle}$ (as computed in Sec. \ref{Ch5Sec24}). These contributions are the \gls{Doppler effect} (peculiar velocities) and \gls{lensing} from inhomogeneities and will play a major role in the rest of this document.

\bigskip \bigskip

\newpage

~

\bigskip

\renewcommand{\arraystretch}{1.7}

\begin{table}[ht!]
\centering
\caption[The spectral coefficients ${\mathcal C}_{ij}(k,\eta_o,\eta_s^{(0)})$ for the $\overline{\lla A_i A_j \rra}$ terms.]{\label{Tab1OfP2} The spectral coefficients ${\mathcal C}_{ij}(k,\eta_o,\eta_s^{(0)})$ for the  $\overline{\lla A_i A_j \rra}$ terms defined by Eq. (\ref{aiaj}). We also give the $k \Delta \eta \ll 1$ limit (up to  leading order in $k \Delta \eta$) of ${\mathcal C}_{ij}$ and of the products ${\mathcal C}_{i} \, {\mathcal C}_{j}$ for the coefficients defined in Table \ref{Tab2OfP2}. We have used the definition $\Xi_s \equiv \left(1 - \frac{1}{{\mathcal H}_s\Delta \eta}\right)$.}
% \vskip{0.3cm}
\small
\begin{tabular}{|c|c|c|c|}
\hline
$\overline{\lla A_i A_j \rra}$ & ${\mathcal C}_{ij}(k,\eta_o,\eta_s^{(0)})$ & ${\mathcal C}_{ij}$ , $k \Delta \eta \ll 1$ & ${\mathcal C}_i~{\mathcal C}_j$ , $k \Delta \eta \ll 1$\\
\hline\hline
$\overline{\lla A_1 A_1 \rra}$ & 1 & 1 & $1-\frac{l^2}{3}$ \\ \hline
$\overline{\lla A_1 A_2 \rra}$ & $-\frac{2}{l} {\rm SinInt}(l)$ & $-2+\frac{l^2}{9}$ & $-2 +\frac{4}{9} l^2$\\ \hline
$\overline{\lla A_1 A_3 \rra}$ & $\Xi_s \left[1 - \frac{\sin(l)}{l} \right]$ & $\Xi_s \frac{l^2}{6}$ & $- \Xi_s \frac{l^2}{6}$ \\ \hline
$\overline{\lla A_1 A_4 \rra}$ & $\Xi_s \frac{f_0}{\Delta \eta}[\cos l - \frac{\sin(l)}{l}]$ & 
$-\frac{f_0}{\Delta \eta} \Xi_s \frac{l^2}{3}$ & 
$-\frac{f_s}{\Delta \eta} \Xi_s \frac{l^2}{3}$ \\ \hline
$\overline{\lla A_1 A_5 \rra}$ & $- 2 \left[1 - \frac{\sin(l)}{l} \right]$ & $-\frac{l^2}{3}$ & 0 \\ \hline
$\overline{\lla A_2 A_2 \rra}$ & $\frac{8}{l^2} \left[ -1 + \cos l + l {\rm SinInt}(l) \right]$ & 
$4-\frac{l^2}{9}$ & $4-\frac{4}{9} l^2$ \\ \hline
$\overline{\lla A_2 A_3 \rra}$ & 0 & 0 & 
$\Xi_s \frac{l^2}{3}$\\ \hline
$\overline{\lla A_2 A_4 \rra}$ & $2 \Xi_s  \frac{f_0 + f_s}{\Delta \eta}[1-\frac{\sin(l)}{l}]$ & 
$\frac{f_0+f_s}{\Delta \eta} \Xi_s \frac{l^2}{3}$ & 
$\frac{f_s}{\Delta \eta} \Xi_s \frac{2}{3}l^2$ \\ \hline
$\overline{\lla A_2 A_5 \rra}$ & $\frac{2}{3l^2} \left[-4 + (4 + l^2)\cos l + l \sin l + l^3 {\rm SinInt}(l) \right]$ & 
$\frac{l^2}{3}$ & 
0 \\ \hline
$\overline{\lla A_3 A_3 \rra}$ & $2 \Xi_s^2 \left[ 1 - \frac{\sin(l)}{l} \right]$ & 
$\Xi_s^2 \frac{l^2}{3} $ & $\Xi_s^2 \frac{l^4}{36}$ \\ \hline
$\overline{\lla A_3 A_4 \rra}$ & $\Xi_s^2  \frac{f_0 - f_s}{\Delta \eta} \left[ \cos l - \frac{\sin(l)}{l} \right]$ & 
$-\frac{f_0-f_s}{\Delta \eta} \Xi_s^2 \frac{l^2}{3} $ & 
$\frac{f_s}{\Delta \eta} \Xi_s^2 \frac{l^4}{18}$ \\ 
\hline
$\overline{\lla A_3 A_5 \rra}$ & $-2 \Xi_s \left[1 - \frac{\sin(l)}{l} \right]$ & 
$- \Xi_s \frac{l^2}{3}$ & 
0 \\ \hline
$\overline{\lla A_4 A_4 \rra}$ & $\Xi_s^2  \left[ \frac{f_0^2 + f_s^2}{\Delta \eta^2} \frac{l^2}{3} - \frac{2 f_0 f_s}{\Delta \eta^2} \left( 2 \cos l + (-2 + l^2) \frac{\sin l}{l} \right) \right]$ & 
$\left(\frac{f_0-f_s}{\Delta \eta}\right)^2 \Xi_s^2 \frac{l^2}{3}$ & 
$\left(\frac{f_s}{\Delta \eta}\right)^2 \Xi_s^2 \frac{l^4}{9}$ \\ \hline
$\overline{\lla A_4 A_5 \rra}$ & $\Xi_s \left[\frac{f_0 + 3 f_s}{\Delta \eta} \cos l + \frac{f_0 - f_s}{\Delta \eta} \frac{\sin l}{l} + \frac{(f_0 + f_s) (-2 + l {\rm SinInt}(l))}{\Delta \eta} \right]$ & 
$\frac{f_0-f_s}{\Delta \eta} \Xi_s \frac{l^2}{3}$ & 
0 \\ \hline
$\overline{\lla A_5 A_5 \rra}$ & $\frac{1}{15 l^2} \left[-24 + 20 l^2 + (24 - 2 l^2 + l^4) \cos(l) \right.$ & 
$\frac{l^2}{3}$ & 
0 \\
 & $\left. + l (-6 + l^2) \sin(l) + l^5 {\rm SinInt}(l) \right]$ &  &  \\ \hline
\end{tabular}
\normalsize
\end{table}

\renewcommand{\arraystretch}{1}

~

\renewcommand{\arraystretch}{1.5}

\begin{table}[ht!]
\centering
\caption[The spectral coefficients ${\mathcal C}_i$ for the  $\overline{\lla A_i \rra \lla A_j \rra}$ terms]{\label{Tab2OfP2} The spectral coefficients ${\mathcal C}_i$ for the  $\overline{\lla A_i \rra \lla A_j \rra}$ terms defined by Eq. \rref{ai}. The signs \textcolor{blue}{$(-1)\times$} indicated for $A_1$ and $A_2$ are emphasized as they where presented with the wrong sign in \cite{P2} (though having no consequences on the plots).}
\vskip 0.3 cm
\begin{tabular}{|c|c|}
\hline
$A_i$ & ${\mathcal C}_i(k,\eta_o,\eta_s^{(0)})$ \\
\hline\hline
$A_1$ & \textcolor{blue}{$(-1)\times$} $\frac{\sin l}{l}$ \\ \hline
$A_2$ & \textcolor{blue}{$(-1)\times$} $-\frac{2}{l} {\rm SinInt}(l)$ \\ \hline
$A_3$ & $-\left(1  -  \frac{1}{{\mathcal H}_s\Delta \eta}\right) \left(1 - \frac{\sin l}{l}\right)$ \\ \hline
$A_4$ & $ \left(1  -  \frac{1}{{\mathcal H}_s\Delta \eta}\right) \frac{f_s}{\Delta \eta} \left( \cos l - \frac{\sin l}{l} \right)$ \\ \hline
$A_5$ & 0 \\ \hline
\end{tabular}
\end{table}

\renewcommand{\arraystretch}{1}

~

\newpage

\subsection{Other examples of backreaction integrals}
\label{Ch5Sec24}

Here we present a detailed computation of some of the terms contributing to the induced backreaction and to the dispersion associated to the light-cone average of $d_L(z_s, \theta^a)$. Such computations also explain the corresponding spectral coefficients appearing in Table \ref{Tab1OfP2} and Table \ref{Tab2OfP2}. In slightly different notations and, more importantly, the consideration of all the necessary terms up to second order, it will also define the spectral coefficients presented in Sec. \ref{Ch5Sec34}.

To give an example slightly more involved than $\overline{\lla \psi_s \psi_s \rra} \equiv \overline{\lla A_1 A_1 \rra}$ or $\overline{\lla \psi_s \rra \lla \psi_s \rra} \equiv \overline{\lla A_1 \rra \lla A_1 \rra}$ of Eqs. \rref{TrivialExample1} and \rref{TrivialExample1a}, we can consider the average of a term like $\psi_s {Q_s}/{\Delta \eta} \equiv A_1 A_2$. Again we consider here the CDM case where the power spectrum has no time dependence. We get\,:
\small
\bea
\overline{\lla A_1 A_2 \rra} &=& - \frac{2}{\Delta \eta} \int \frac{d^3 k d^3 k'}{(2 \pi)^3}\overline{E(\kbf) E(\kbf')} \int \frac{d^2 \Omega}{4 \pi} \left[ \psi_{k}(\eta_s^{(0)})  e^{i \kbf \cdot \hat{\xbf} \Delta\eta} \int_{\eta_s^{(0)}}^{\eta_o} d\eta\, \psi_{k'}( \eta)  e^{i \kbf' \cdot \hat{\xbf} (\eta_o - \eta)} \right] \nonumber \\
&=& - 2 \int_{0}^{\infty} \frac{d k}{k} \Pcal_{\psi} (k) \frac{ {\rm SinInt}(k \Delta \eta)}{k \Delta \eta} = \overline{\lla \psi_s \frac{Q_s}{\Delta \eta} \rra} ~~,
\label{TrivialExample2}
\eea
\bea
\overline{\lla A_1 \rra \lla A_2 \rra} &=& - \frac{2}{\Delta \eta} \int \frac{d^3 k d^3 k'}{(2 \pi)^3} \overline{E(\kbf) E(\kbf')} \left[ \int \frac{d^2 \Omega}{4 \pi} \psi_{k}(\eta_s^{(0)}) e^{i \kbf \cdot \hat{\xbf} \Delta\eta} \right] \left[ \int \frac{d^2 \Omega'}{4 \pi} \int_{\eta_s^{(0)}}^{\eta_o} d\eta \,\psi_{k'}( \eta) e^{i \kbf' \cdot \hat{x'} (\eta_o - \eta)} \right] \nonumber \\
&=& - 2  \int_{0}^{\infty} \frac{d k}{k} \Pcal_{\psi}(k)  ~ \frac{\sin(k \Delta\eta)}{k \Delta \eta} \frac{ {\rm SinInt}(k \Delta \eta)}{k \Delta \eta} = \overline{\lla \psi_s \rra \lla \frac{Q_s}{\Delta \eta} \rra} ~~,
\label{TrivialExample2a}
\eea
\normalsize
which indeed correspond to the results presented in Tables \ref{Tab1OfP2} and \ref{Tab2OfP2} for $\overline{\lla A_1 A_2 \rra}$ and $\overline{\lla A_1 \rra \lla A_2 \rra}$. As one can see, the angular average is making the results completely different in the two cases. We remark that the presence of the ${\rm SinInt}$ function (see Eq. \rref{DefSinInt}) is a direct consequence of the integration over time in $Q_s$. Consequently, the non-local nature of the backreaction terms is reflected in the form of the corresponding spectral coefficients.

Let us now present $\overline {\lla A_2 A_5 \rra} \propto \overline{\lla Q_s J_2 / \Delta \eta \rra}$ and $\overline {\lla A_2 \rra \lla A_5 \rra}$. These terms are interesting as they correspond to the mixing between an integrated perturbation $\psi$ and the lensing term. Using Eqs. (\ref{J2_Equation}), (\ref{Psi_average_Equation}), \rref{defA}, and working in the hypothesis of time-independent $\psi_k$, we obtain\,:
\small
\bea
\overline {\lla A_2 A_5 \rra} &=& \int \frac{d^3 k ~d^3 k'}{(2 \pi)^{3}} \overline{E(\kbf) E(\kbf')} \int \frac{d^2 \Omega}{4 \pi} \nonumber \\
& & ~~~~~ ~~~~~ ~~~~~ \times \left[ \frac{2}{\Delta\eta} \int_{\eta_s}^{\eta_o} d\eta' ~ \psi_k \, e^{i (\eta_o-\eta') \kbf \cdot \hat{\xbf}} \right] \left[ - \int_{\eta_s}^{\eta_o} \frac{d\eta''}{\Delta \eta} \frac{\eta'' - \eta_s}{\eta_o - \eta''} ~ \Delta_2 \left(\psi_{k'} \, e^{i (\eta_o-\eta'')  \kbf' \cdot \hat{\xbf}} \right) \right] \nonumber \\
&=& \int \frac{d^3 k}{(2 \pi)^{3}} ~ |\psi_k|^2 \int_{-1}^{1} \frac{d(\cos\theta)}{2} \left[ \frac{2}{\Delta\eta} \int_{\eta_s}^{\eta_o} d\eta' ~ e^{i k (\eta_o - \eta') \cos\theta} \right] \nonumber \\
& & \times \left[ \frac{1}{\Delta \eta} \int_{\eta_s}^{\eta_o} d\eta'' \frac{\eta'' - \eta_s}{\eta_o - \eta''} \left( k^2 (\eta_o - \eta'')^2 \sin^2 \theta - 2 i k (\eta_o - \eta'') \cos\theta \right) e^{- i k (\eta_o - \eta'') \cos\theta} \right] \nonumber \\
&=&  \int_0^{\infty} \frac{d k}{k} ~ \Pcal_{\psi}(k) \nonumber \\
& & ~~~~~ ~~~~~ \times \frac{2}{3 (k \Delta \eta)^2} 
\left[-4 + (4 + (k \Delta \eta)^2)\cos(k \Delta \eta) + k \Delta \eta \sin(k \Delta \eta) + (k \Delta \eta)^3 {\rm SinInt}(k \Delta \eta) \right] ~~, ~~~~~ ~~~~~
\label{lA2A5r}
\eea
\bea
\overline {\lla A_2 \rra \lla A_5 \rra} &=& \int \frac{d^3 k ~d^3 k'}{(2 \pi)^{3}} \overline{E(\kbf) E(\kbf')} \left[ \int \frac{d^2 \Omega}{4 \pi} \frac{2}{\Delta\eta} \int_{\eta_s}^{\eta_o} d\eta' ~ \psi_k \, e^{i (\eta0 - \eta') \kbf \cdot \hat{\xbf}} \right] 
\nonumber \\
& & ~~~~~ ~~~~~~ ~~~~~ \times
\left[ - \int \frac{d^2 \Omega'}{4 \pi} \int_{\eta_s}^{\eta_o} \frac{d\eta''}{\Delta \eta} \frac{\eta'' - \eta_s}{\eta_o - \eta''} ~ \Delta_2 \left(\psi_{k'} \, e^{i (\eta_o - \eta'')  \kbf' \cdot \hat{\xbf}} \right) \right] \nonumber \\
&=& \int \frac{d^3 k}{(2 \pi)^{3}} ~ |\psi_k|^2 \left[ \int_{-1}^{1} \frac{d(\cos\theta)}{2} \frac{2}{\Delta\eta} \int_{\eta_s}^{\eta_o} d\eta' ~ e^{i k (\eta_o - \eta') \cos\theta} \right] \nonumber \\
& & \times\left[ \int_{-1}^{1} \frac{d(\cos\theta')}{2} \frac{1}{\Delta \eta} \int_{\eta_s}^{\eta_o} d\eta'' \frac{\eta'' - \eta_s}{\eta_o - \eta''} \left( k^2 (\eta_o - \eta'')^2 \sin^2 \theta' - 2 i k (\eta_o - \eta'') \cos\theta' \right) e^{- i k (\eta_o - \eta'') \cos\theta'} \right] \nonumber \\
&=& \int_0^{\infty} \frac{d k}{k} ~ \Pcal_{\psi}(k) \left[ \frac{2}{k\Delta \eta} {\rm SinInt}(k\Delta \eta) \right] \times 0 ~ = ~ 0 ~~.
\label{lA2rlA5r}
\eea
\normalsize
In the second equalities of Eqs. (\ref{lA2A5r}) and (\ref{lA2rlA5r}) we have applied Eq. \rref{PropRandomVar} to remove the integration over $\kbf'$, and we have used the isotropy of the scalar product $\kbf \cdot \hat{\xbf}$.

We can see that the terms presented until now are not dominant for a general power spectrum as they are, at most, going like $k \Delta \eta$. We consider now the two leading contributions given by $\overline{\lla A_4 A_4 \rra}$ and $\overline{\lla A_5 A_5 \rra}$ (and, for completeness, the associated terms $\overline{\lla A_4 \rra \lla A_4 \rra}$ and $\overline{\lla A_5 \rra \lla A_5 \rra}$). We will show in Sec. \ref{Ch6Sec1} that these are the two leading contributions to IBR$_2$ and to the dispersion and they correspond respectively to the ``Doppler squared'' and ``lensing squared terms''. Starting from Eq. \rref{Ir}, and using the same hypotheses as in the previous calculations, we have\,:
\bea
I_r &=& \int_{\eta_{in}}^{\eta_s} d\eta' \frac{a(\eta')}{a(\eta_s)} \pa_r \psi(\eta',\eta_o-\eta_s,\theta^a)- \int_{\eta_{in}}^{\eta_o} d\eta' \frac{a(\eta')}{a(\eta_o)} \pa_r \psi(\eta', 0, \theta^a) \nonumber \\
&=& \int \frac{d^3 k}{(2 \pi)^{3/2}} \left[ \int_{\eta_{in}}^{\eta_s} d\eta' \frac{\eta'^2}{\eta_s^2} (i k \cos\theta) \psi_k \, e^{i \Delta \eta \, \kbf \cdot \hat{\xbf}} - \int_{\eta_{in}}^{\eta_o} d\eta' \frac{\eta'^2}{\eta_o^2} (i k \cos\theta) \psi_k \right] \nonumber \\
&=& \int \frac{d^3 k}{(2 \pi)^{3/2}} \psi_k \, \left( f_s e^{i k \Delta\eta \cos\theta} - f_0 \right) (i k \cos\theta) ~~.
\eea
Then, according to the definitions of Eq. (\ref{defA})\, we have $\overline {\lla A_4 A_4 \rra} = \Xi_s^2 \, \overline {\lla I_r I_r \rra}$ where\,:
\small
\bea
\overline {\lla I_r I_r \rra} &=& \int \frac{d^3 k}{(2 \pi)^{3}} ~ |\psi_k|^2 \int_{-1}^{1} \frac{d(\cos\theta)}{2} \left[ \left( f_s e^{i k \Delta\eta \cos\theta} - f_0 \right) (i k \cos\theta) \right] \left[ \left( f_s e^{-i k \Delta\eta \cos\theta} - f_0 \right) (- i k \cos\theta) \right] \nonumber \\
&=& \int_0^{\infty} \frac{d k}{k} ~ \frac{\Pcal_{\psi}(k)}{3 \Delta\eta^2} \left[ (f_0^2 + f_s^2) ( k \Delta\eta)^2 - 12 f_0 f_s \cos(k \Delta\eta) - 6 f_0 f_s ( -2 + (k \Delta\eta)^2)\frac{\sin(k \Delta\eta)}{k \Delta\eta} \right] ~~, ~~~~~
\eea
\normalsize
and where we have used the following exact integral result\,:
\beq
\int_{-1}^{1} \frac{d(\cos\theta)}{2} (\cos\theta)^2 e^{\pm i k \Delta\eta \cos\theta} = \frac{2 k \Delta\eta \cos(k \Delta\eta) + (-2 + (k \Delta\eta)^2) \sin(k \Delta\eta)}{(k \Delta\eta)^3} ~~.
\eeq

In the same way we obtain $\overline {\lla A_4 \rra \lla A_4 \rra} = \Xi_s^2 \, \overline {\lla I_r \rra \lla I_r \rra}$ where\,:
\bea
\overline {\lla I_r \rra \lla I_r \rra} &=& \int \frac{d^3 k}{(2 \pi)^{3}} ~ |\psi_k|^2 \left[ \int_{-1}^{1} \frac{d(\cos\theta)}{2} \left( f_s e^{i k \Delta\eta \cos\theta} - f_0 \right) (i k \cos\theta) \right] 
\\
&& \times \left[ \int_{-1}^{1} \frac{d(\cos\theta')}{2} \left( f_s e^{- i k \Delta\eta \cos\theta'} - f_0 \right) (- i k \cos\theta') \right] \nonumber \\
&=& \int_0^{\infty} \frac{d k}{k} ~ \Pcal_{\psi}(k) \left[ \frac{f_s}{\Delta \eta} \left( \cos(k \Delta\eta) - \frac{\sin(k \Delta\eta)}{k \Delta\eta} \right) \right]^2 ~~,
\eea
and where we have used the integral
\beq
\int_{-1}^{1} \frac{d(\cos\theta)}{2} ~ \cos\theta e^{\pm i k \Delta\eta \cos\theta} = \pm i \left(\frac{- k \Delta\eta \cos(k \Delta\eta) + \sin(k \Delta\eta)}{(k \Delta\eta)^2}\right) \,.
\eeq

Following a similar procedure, Eq. (\ref{J2_Equation}) leads us to the lensing contribution\,:
\bea
\overline {\lla A_5 A_5 \rra} &=& \int \frac{d^3 k}{(2 \pi)^{3}} ~ |\psi_k|^2 \int_{-1}^{1} \frac{d(\cos\theta)}{2} \nonumber \\
& & \times \left[ \int_{\eta_s}^{\eta_o} \frac{d\eta'}{\Delta \eta} \frac{\eta' - \eta_s}{\eta_o - \eta'} \left( k^2 (\eta_o - \eta')^2 \sin^2 \theta + 2 i k (\eta_o - \eta') \cos\theta \right) e^{i k (\eta_o - \eta') \cos\theta} \right] \nonumber \\
& & \times \left[ \int_{\eta_s}^{\eta_o} \frac{d\eta''}{\Delta \eta} \frac{\eta'' - \eta_s}{\eta_o - \eta''} \left( k^2 (\eta_o - \eta'')^2 \sin^2 \theta - 2 i k (\eta_o - \eta'') \cos\theta \right) e^{- i k (\eta_o - \eta'') \cos\theta} \right] \nonumber \\
&=& \int_0^{\infty} \frac{d k}{k} ~ \Pcal_{\psi}(k) \frac{1}{15 (k \Delta\eta)^2} \left[-24 + 20 (k \Delta\eta)^2 + (24 - 2 (k \Delta\eta)^2 + (k \Delta\eta)^4) \cos(k \Delta\eta) \right. \nonumber \\
& & ~~~~~ ~~~~~ ~~~~~ ~~~~~ \left. - 6 k \Delta\eta \sin(k \Delta\eta) + (k \Delta\eta)^3 \sin(k \Delta\eta) + (k \Delta\eta)^5 {\rm SinInt}(k \Delta\eta) \right] ~~.
\eea
Note that, for the same reason for which  $\overline {\lla A_2 \rra \lla A_5 \rra} = 0$, i.e. for the fact that the integrand of $A_5$ gives exactly zero when averaged over the angles, we finally also obtain $\overline {\lla A_5 \rra \lla A_5 \rra} = 0$.

\section{Averaging the luminosity flux at second order}
\label{Ch5Sec3}

In this section we treat the full computation of the light-cone and stochastic average of the luminosity flux $\overline{\langle \Phi \rangle} \sim \overline{\langle d_L^{-2}\rangle}$ and the luminosity distance $\overline{\langle d_L \rangle}$, including the contributions from the genuine second order. The calculations will thus rely on the full transformation of coordinates as presented in Chapter \ref{Chap4}. We will further consider all types of perturbations, of scalar, vector and tensor kind. Because the number of terms is even more important than with the consideration of the terms generated by first-order only, namely $\rm{IBR}_2$ and $\overline{\langle \sigma_1^2 \rangle}$, we will proceed to a different regroupment of the terms.

\subsection{Exact expression for $\langle d_L^{-2}\rangle$ in the geodesic light-cone gauge}
\label{Ch5Sec31}

The \gls{luminosity flux}, $\Phi \sim d_L^{-2} = (1+z_s)^{-4} d_A^{-2}$, appears as an important and extremely simple observable to average over the 2-sphere $\Sigma(w_o, z_s)$ embedded in the light-cone. In fact, we have\,:
\beq 
\label{Phitheory}
\langle d_L^{-2} \rangle(w_o,z_s) = (1+z_s)^{-4}\frac{\int dS \frac{d^2\Omega_o}{dS}}{\int dS} = (1+z_s)^{-4}\frac{\int d^2\Omega_o}{\int dS} = (1+z_s)^{-4}\frac{4\pi}{\Acal(w_o, z_s) } ~~,
\eeq
where
\beq
\label{AreaDeformed2sphere}
\Acal(w_o, z_s) = \int_{\Sigma(w_o,z_s)} d^2 \tilde{\theta}^a \sqrt{\gamma (w_o, \tau_s(w_o,z_s,\ti{\theta}^a), \ti{\theta}^a)} ~~
\eeq
is the proper area of $\Sigma(w_o,z_s)$ computed with the metric $\gamma_{ab}$, and expressed in terms of internal coordinates $(w_o,z_s)$ parametrizing this deformed 2-sphere. One could decide to parametrize this integral by two other angular coordinates $\xi^a$. One would then have the same expression\,:
\beq
\label{Nambu}
\Acal(w_o,z_s) = \int_{\Sigma(w_o,z_s)} d^2 \xi \sqrt{\gamma (w_o, \tau_s(w_o,z_s,\ti{\xi}^a), \ti{\xi}^a)} ~~~~~\mbox{with}~~~~~
\gamma_{ab}(\xi) = \frac{\partial y^\mu}{\partial \xi^a} \frac{\partial y^\nu}{\partial \xi^b} g_{\mu\nu}^{PG}(y) = \gamma_{ab} ~~,
\eeq
and where $\{y^{\mu}\}$ are, for instance, the Poisson gauge coordinates such that $y^\mu = y^\mu(w_o,\tau_s(w_o,z_s,\tilde\xi^a),\tilde\xi^b)$.
In fact, one can notice that $\gamma_{ab}(\xi)$ is the metric induced by the coordinates $\{y^\mu\}$ on $\Sigma(w_o,z_s)$, so we can write $\gamma_{ab}(\xi) \equiv \gamma_{ab}^{ind}(\xi)$. From this remark and the consideration of Eq. \rref{Nambu}, we can see very amusingly that $\Acal$ is similar to the reparametrization-invariant Nambu-Goto action in string theory.

The Eq. \rref{Phitheory} holds non-perturbatively for any spacetime geometry and is the starting point for the computation of the average flux.
It can also be written in an elegant form in which the flux of an inhomogeneous Universe is compared to that of a FLRW one\,:
\bea
\label{avfluxgen}
\langle \Phi/\Phi^{\rm FLRW} \rangle = \frac{4\pi \left( d_A^{\rm FLRW}\right)^2}{\Acal} ~~.
\eea
This expression directly shows that a deviation by the presence of inhomogeneities from the standard FLRW result is related to a deformation of the 2-sphere $\Sigma(w_o,z_s)$ because of these inhomogeneities.

Coming back to the notations introduced in Chapter \ref{Chap4}, we can in particular recast Eq. \rref{Phitheory} into\,:
\beq
\label{dLminus2}
\langle d_L^{-2} \rangle(w_o, z_s) = (1+z_s)^{-4} \left[ \int \frac{d^2 \tilde{\theta}^a}{4 \pi} \sqrt{\gamma(w_o, \tau_s(w_o, z_s,\tilde{\theta}^a), \tilde{\theta}^a)} \right]^{-1} \equiv \left(d_L^{\rm FLRW}\right)^{-2} I_\phi^{-1} (w_o, z_s)~~,
\eeq
where we have defined 
\beq
\label{Integral}
I_\phi(w_o, z_s) \equiv \frac{{\cal A}(w_o, z_s)}{4 \pi \left[a(\eta_s^{(0)}) \Delta \eta \right]^2} ~~,
\eeq
and where $d_L^{\rm FLRW}= (1+z_s)^2 a(\eta_s^{(0)}) \Delta \eta$ is the luminosity distance for the unperturbed FLRW geometry, with scale factor $a(\eta)$. Here $\eta$ is the conformal time coordinate, $\Da \eta = \eta_o-\eta_s^{(0)}$, and we have denoted with $\eta_s^{(0)}$ the background solution of the equation for the source's conformal time $\eta_s= \eta_s(z_s, \ti \theta^a)$ (see Sec. \ref{Ch4Sec21}). Note again that, according to the above equation,  the interpretation of $I_\phi(w_o, z_s)$ is straightforward\,: it is simply the ratio of the area of the 2-sphere at redshift $z_s$ on the past light-cone (deformed by inhomogeneities), over the area of the corresponding homogeneous 2-sphere.

\subsection{Second-order expression for $\langle d_L^{-2}\rangle$ in the Poisson gauge}
\label{Ch5Sec32}

Let us consider a spacetime geometry that can be approximated by a spatially flat FLRW metric distorted by the presence of scalar, vector and tensor perturbations, in the Poisson gauge and as used in Chapter \ref{Chap4}.
It is important to stress that for this thesis we can safely restrict our discussion to pure scalar perturbations. In fact, it is true that at second order different perturbations get mixed\,: vector and tensor perturbations, $\omega_i$ and $h_{ij}$, are automatically generated from scalar perturbations (see e.g. \cite{Matarrese:1997ay,Bartolo:2005kv}), while second-order scalar perturbations are generated from first-order vector and tensor perturbations. However, a single vector or tensor perturbation does not contribute to our angular averages on $\Sigma(w_o,z_s)$, as it will be proved in Sec. \ref{Ch5Sec36}. Furthermore, we treat $\omega_i$ and $h_{ij}$ as second order quantities, i.e. we assume the first-order perturbed metric to be dominated by scalar contributions (as perturbations generated by a phase of standard slow-roll inflation, see e.g. \cite{DurCos,MG}).
 
We have established a connection between the second-order perturbative expression of the luminosity distance $d_L$ and of the integrand of $I_\phi$ (controlling $\langle d_L^{-2}\rangle$), both written in terms of the PG perturbations and the observer's angles $\tilde{\theta}^a$ (remember that photons travel at constant $\tilde{\theta}^a$). In particular, recalling Eq. \rref{dLwithSVT} where the contribution $\bar{\delta}^{(2)}_{V,T}$ has been forgotten (see Sec. \ref{Ch5Sec36}), we have\,:
\beq
\frac{{d}_L(z_s, \tilde{\theta}^a)}{(1+z_s)a_o \Delta \eta} = {{d}_L(z_s, \tilde{\theta}^a)\over d_L^{\rm FLRW}(z_s)} = 1 + \bar{\delta}_S^{(1)}(z_s, \tilde{\theta}^a) + \bar{\delta}_S^{(2)}(z_s, \tilde{\theta}^a) ~~, 
\label{215}
\eeq
\vspace{-0.3cm}
\beq
\label{Iphi}
I_\phi(z_s)=\int \frac{d^2 \tilde{\theta}^a}{4 \pi} \sin \tilde{\theta} \left(1+{\cal I}_1+{\cal I}_{1,1}+{\cal I}_2\right) ~~,
\eeq
and we can show that\footnote{As already mentioned, the terms appearing on the RHS of Eq. \rref{215} should be corrected, in principle, for the peculiar velocity of the observer. However, such a correction has no effect on the integral $I_{\phi}$ itself, since it can be compensated by a Lorentz transformation of the angular variables.}\,:
\beq
\label{Ideltarel}
{\cal I}_1 = 2 \bar{\delta}_S^{(1)} + (\rm{t.~d.})^{(1)} ~~~~~,~~~~~ {\cal I}_{1,1} + {\cal I}_2 = 2 \bar{\delta}_S^{(2)} + ( \bar{\delta}_S^{(1)})^2 +  (\rm{t.~d.})^{(2)} ~~.
\eeq
where the terms $(\rm{t.~d.})^{(1),(2)}$ denote total derivative terms w.r.t. the $\tilde{\theta}^a$ angles,  giving vanishing contributions either by periodicity in $\ti{\phi}$ or by the vanishing of the integrand at $\ti{\theta} = 0, \pi$. We only recall here, for later use, the first-order total derivative\,:
\beq
({\rm t.~d.})^{(1)}= 2\, J_2^{(1)} ~~~~~,~~~~~ J_2^{(1)} = \frac{1}{\Delta\eta} \int_{\eta_s^{(0)}}^{\eta_o} d \eta \,\frac {\eta - \eta_s^{(0)}}{\eta_o - \eta} \Delta_2 \psi(\eta, \eta_o-\eta, \ti \theta^a) ~~,
\label{216}
\eeq
where $\Delta_2= \pa^2_{\ti\theta} + \cot \ti\theta \pa_{\ti\theta} + \sin^{-2} \ti\theta \pa^2_{\ti\phi}$\,. It is also convenient to rewrite the PG metric using spherical coordinates (but still considering that photons travel at constant $\tilde{\theta}^a$), and use the following quantities\,:
\bea
&
P(\eta, r, \ti\theta^a) = \int_{\eta_{in}}^\eta d\eta' \frac{a(\eta')}{a(\eta)} \psi(\eta',r,\ti\theta^a) ~~~~~,~~~~~ Q(\eta_+, \eta_-, \ti\theta^a) = \int_{\eta_+}^{\eta_-} dx~ \hat{\psi}(\eta_+,x,\ti\theta^a) ~~,~~  \nonumber \\
&
\Xi_s =  1 - \frac{1}{\Hcal_s \Delta \eta} ~~~~~,~~~~~ J  = \left([\partial_+ Q]_s - [\partial_+ Q]_o\right) - \left([\partial_r P]_s - [\partial_r P]_o\right) ~~,
\eea
which were defined in Chapter \ref{Chap4}. We recall that the radial gradient of $P$ is related to the \gls{Doppler effect} (due to peculiar velocities of source and observer), while the gradient of $Q$ with respect to $\pa_+$ represents the Sachs-Wolfe (\gls{SW}) and the integrated Sachs-Wolfe (\gls{ISW}) effects. The last term $J$ corresponds to a combination of the three mentioned effects.
 
The results obtained in Eqs. \rref{Iphi} and \rref{Ideltarel} can be reported in the following form\,:
\beq
\label{DecomIinTterms}
{\cal I}_{1} = \sum_{i = 1}^{3} \Tcal_i^{(1)} ~~~~~,~~~~~ {\cal I}_{1,1} = \sum_{i = 1}^{23} \Tcal_i^{(1,1)} ~~~~~,~~~~~ {\cal I}_{2} = \sum_{i = 1}^{7} \Tcal_i^{(2)} ~~~~~,
\eeq
where ${\cal I}_1,~ {\cal I}_{1,1},~ {\cal I}_2 $ are, respectively, the first-order, quadratic first-order, and genuine second-order  contributions of our stochastic fluctuations, and where these sums are made of the terms\,:
\beq
\label{DefI1}
\Tcal_1^{(1)} = - 2 \psi (\eta_s^{(0)}, r_s^{(0)}, \ti\theta^a)
~~~~~;~~~~~
\Tcal_2^{(1)} = 2 \Xi_s J
~~~~~;~~~~~
\Tcal_3^{(1)} = - \frac{2}{\Delta \eta} Q_s
~~;
\eeq
\bea
\label{DefI11part1}
& &
\Tcal_1^{(1,1)} = \Xi_s \left[ \psi_s^2 - \psi_o^2 \right]
~~;~~
\Tcal_2^{(1,1)} = \Xi_s \left( ([\partial_r P]_s)^2 - ([\partial_r P]_o)^2 \right)
~~;~~
\Tcal_3^{(1,1)} = - 2 \Xi_s \left(\psi_s + [\partial_+ Q]_s\right) [\partial_r P]_s
~~;~~ \nonumber \\
& &
\Tcal_4^{(1,1)} = \frac12 \Xi_s (\gamma_0^{ab})_s \left( 2 \partial_a P_s \partial_b P_s + \partial_a Q_s \partial_b Q_s - 4 \partial_a Q_s \partial_b P_s \right)
~~;~~
\Tcal_5^{(1,1)} = - \Xi_s \lim_{r\rightarrow 0} \left[\gamma_0^{ab} \partial_a P \partial_b P \right] 
~~;~~ \nonumber \\
& &
\Tcal_6^{(1,1)} = 2 \Xi_s Q_s \left( 2 \partial_r \psi_o + 2 \partial_\eta \psi_o - \partial_r \psi_s + 2 \int_{\eta_s^{(0)}}^{\eta_o}d \eta' \partial_r^2
\psi\left(\eta',\eta_o-\eta',\ti\theta^a\right) + [ \partial_r^2 P ]_s \right)
~~;~~ \nonumber \\
& &
\Tcal_7^{(1,1)} = 2 \Xi_s \frac{J}{{\mathcal H}_s} \left( [\partial_\eta \psi]_s + {\mathcal H}_s [\partial_r P]_s \right)
~~;~~
\Tcal_8^{(1,1)} = 2 \Xi_s \frac{J}{{\mathcal H}_s} [\partial_r^2 P]_s
~~;~~ \nonumber \\
& &
\Tcal_9^{(1,1)} = - \Xi_s \int_{\eta_{in}}^{\eta_s^{(0)}} d\eta' \frac{a(\eta')}{a(\eta_s^{(0)})} \partial_r \left[ - \psi^2 + (\partial_r P)^2 + \gamma_0^{ab} \partial_a P \partial_b P \right] (\eta', \Delta \eta, \tilde{\theta}^a)
~~;~~ \nonumber \\
& &
\Tcal_{10}^{(1,1)} = \Xi_s \int_{\eta_{in}}^{\eta_o} d\eta' \frac{a(\eta')}{a(\eta_o)}  \partial_r \left[ - \psi^2 + (\partial_r P)^2 + \gamma_0^{ab} \partial_a P \partial_b P \right] (\eta', 0, \tilde{\theta}^a)
~~;~~ \nonumber \\
& &
\Tcal_{11}^{(1,1)} = 2 \Xi_s \int_{\eta_s^{(0)+}}^{\eta_s^{(0)-}} dx~ \partial_+ \left[ \hat{\psi} ~ \partial_+ Q + \frac14 \hat{\gamma}_{0}^{ab} \partial_a Q \partial_b Q \right] (\eta_s^{(0)+}, x, \tilde{\theta}^a)
~~;~~
\nonumber
\eea
\bea
\label{DefI11part2}
& &
\Tcal_{12}^{(1,1)} = \left[ \Xi_s^2 - \frac{1}{{\mathcal H}_s \Delta \eta} \left( 1 - \frac{{\mathcal H}_s'}{{\mathcal H}_s^2} \right) \right] J^2
~~;~~
\Tcal_{13}^{(1,1)} = - 4 \psi_s J
~~;~~
\Tcal_{14}^{(1,1)} = 2 \Xi_s \left\{ \psi_o + [\partial_r P]_o - \frac{Q_s}{\Delta \eta} \right\} J
~~;~~ \nonumber \\
& &
\Tcal_{15}^{(1,1)} = - 2 (J - 2 \psi_s) \frac{Q_s}{\Delta \eta}
~~;~~
\Tcal_{16}^{(1,1)} = - 2 \left( \psi_s - \partial_+ Q_s \right) \frac{Q_s}{\Delta \eta}
~~;~~
\Tcal_{17}^{(1,1)} = \left(\frac{Q_s}{\Delta \eta}\right)^2
~~;~~ \nonumber \\
& &
\Tcal_{18}^{(1,1)} = \frac{1}{\Hcal_s} (\gamma_0^{ab})_s \partial_a Q_s \partial_b J
~~;~~
\Tcal_{19}^{(1,1)} = \frac12 (\gamma_0^{ab})_s \partial_a Q_s \partial_b Q_s
~~;~~
\Tcal_{20}^{(1,1)} = 2 \frac{J}{{\mathcal H}_s} (-[\partial_\eta \psi]_s + [\partial_r \psi]_s)
~~;~~ \nonumber \\
& &
\Tcal_{21}^{(1,1)} = 2 Q_s [\partial_r \psi]_s
~~;~~
\Tcal_{22}^{(1,1)} = - \frac{2}{\Delta\eta} \int_{\eta_s^{(0)+}}^{\eta_s^{(0)-}} dx~ \left[ \hat{\psi} ~ \partial_+ Q + \frac14 \hat{\gamma}_{0}^{ab} ~ \partial_a Q ~ \partial_b Q \right] (\eta_s^{(0)+},x,\tilde{\theta}^a)
~~;~~ \nonumber \\
& &
\Tcal_{23}^{(1,1)} =  \frac{1}{8\sin \tilde{\theta}} \frac{\partial}{\partial \tilde{\theta}} \left\{ \cos \tilde{\theta} ~ \left( \int_{\eta_s^{(0)+}}^{\eta_s^{(0)-}} dx~ [\hat{\gamma}_{0}^{1b} ~ \partial_b Q](\eta_s^{(0)+},x,\tilde{\theta}^a) \right)^2 \right\}~;
\eea
\bea
\label{DefI2}
& &
\Tcal_1^{(2)} = - \frac{1}{2} \Xi_s \left( \phi_s^{(2)} - \phi_o^{(2)} \right)
~~;~~
\Tcal_2^{(2)} = \frac{1}{2} \Xi_s \left( \psi_s^{(2)} - \psi_o^{(2)} \right)
~~;~~ \nonumber \\
& &
\Tcal_3^{(2)} = - \Xi_s \int_{\eta_{in}}^{\eta_s^{(0)}} d\eta' \frac{a(\eta')}{a(\eta_s^{(0)})} [\partial_r \phi^{(2)}] (\eta', \Delta \eta, \tilde{\theta}^a)
~~;~~ \nonumber \\
& &
\Tcal_4^{(2)} = \Xi_s \int_{\eta_{in}}^{\eta_o} d\eta' \frac{a(\eta')}{a(\eta_o)} [\partial_r \phi^{(2)}](\eta', 0, \tilde{\theta}^a)
~~;~~ \nonumber \\
& &
\Tcal_5^{(2)} = \frac12 \Xi_s \int_{\eta_s^{(0)+}}^{\eta_s^{(0)-}} dx~ \partial_+ \left[ \hat{\phi}^{(2)} + \hat{\psi}^{(2)} \right](\eta_s^{(0)+}, x, \tilde{\theta}^a)
~~;~~ \nonumber \\
& &
\Tcal_6^{(2)} = - \psi_s^{(2)}
~~;~~
\Tcal_7^{(2)} = - \frac{2}{\Delta\eta} \int_{\eta_s^{(0)+}}^{\eta_s^{(0)-}} dx~ \left[ \frac{\hat{\phi}^{(2)} + \hat{\psi}^{(2)}}{4} \right] (\eta_s^{(0)+},x,\tilde{\theta}^a)
~~.
\label{Tcal2}
\eea
In the above equations $\ga_0^{ab}= {\rm diag} (r^{-2}, r^{-2} \sin^{-2} \ti \theta)$, and for $\Tcal_6^{(1,1)}$ we used the following identity\,:
\beq
\label{GenevaIdentity}
- [\partial_+^2 Q]_s + [\partial_+ \hat{\psi}]_s = 2 \partial_r \psi_o + 2 \partial_\eta \psi_o - \partial_r \psi_s + 2 \int_{\eta_s^{(0)}}^{\eta_o}d \eta' \partial_r^2 \psi\left(\eta',\eta_o-\eta', \ti\theta^a\right) ~~.
\eeq
This relation can be proved by first noticing that\,:
\bea
\label{GenevaIdentityIntermediary}
- [\partial_+^2 Q]_s + [\partial_+ \hat{\psi}]_s &=& \bigg\{ \partial_+ \hat{\psi}(\eta_+,\eta_-,\ti{\theta}^a) + \partial_+ \hat{\psi}(\eta_+,\eta_+,\ti{\theta}^a) + \lim_{x \rightarrow \eta_+} \left[ \partial_+ \hat{\psi}(\eta_+,x,\ti{\theta}^a)\right] \nonumber \\
& & ~~~~~ ~~~~~ - \int_{\eta_+}^{\eta_-} dx \, \partial_+^2 \hat{\psi}(\eta_+,x,\ti{\theta}^a) \bigg\} (\eta_s^{(0)+},\eta_s^{(0)-,\ti{\theta}^a}) ~~.
\eea
We then convert the integral term in an integration over the time $\eta$ and convert the derivatives by the use of Eqs. \rref{ChgtVarxeta} and \rref{deta}. We can hence perform the integrations involving a total derivative and we are left with integrands at observer and source positions and integrals involving partial derivatives of $\psi$ with respect to $r$. The last subtlety to consider is at the tip of the observer light cone. Indeed, the relation $\partial_+ \equiv (\partial_\eta + \partial_r) / 2$ leads to\,:
\beq
\lim_{x \rightarrow \eta_s^{(0)+}}\left[ \partial_+ \hat{\psi}(\eta_+,x,\ti{\theta}^a)\right]_s = \lim_{\eta \rightarrow \eta_o}\left[ \frac12 (\partial_\eta + \partial_r) \psi(\eta,\eta_o - \eta, \ti{\theta}^a) \right] = \frac12 \left( [ \partial_\eta \psi]_o + [ \partial_r \psi]_o \right) ~~,
\eeq
where the subscript ``s'' denotes $\eta_+ \rightarrow \eta_s^{(0)+}$, but on the other hand we have to take for the second term in the RHS of Eq. \rref{GenevaIdentityIntermediary}\,:
\beq
[\partial_+ \hat{\psi}(\eta_+,\eta_+,\ti{\theta}^a)]_s = [ \partial_\eta \psi ]_o ~~.
\eeq
This last result is well understood by seeing that this derivative is a derivative at the tip of the light cone, thus acting in all angular directions, giving thus a double derivation in the $\eta$-direction and a cancelation in $r$-direction. After a few simplifications we recover Eq. \rref{GenevaIdentity}.

Let us point out, finally, that the last term in Eq. (\ref{DefI11part2}) corresponds to a total derivative, and thus to a boundary contribution, that superficially looks non vanishing. This term is probably the result of a naive treatment of the angular coordinate transformation, which becomes singular near the poles of the 2-sphere. This contribution, indeed, has the same form as the irrelevant one coming from an overall $SO(3)$ rotation. In any case, we will see that its spectral coefficient is zero (Sec. \ref{Ch5Sec34}).

Let us conclude this section about the flux by giving the explicit expressions for $\Ical_1$, $\Ical_{1,1}$ and $\Ical_2$. Gathering the terms from Eqs. \rref{DecomIinTterms} and \rref{DefI1}, we get the explicit expression of ${\cal I}_1$ which is\,:
\bea
{\cal I}_1 &=& 2 \left( \Xi_s J - \frac{1}{\Delta \eta} Q_s - \psi (\eta_s^{(0)}, r_s^{(0)}, \theta^a) \right) ~~.
\label{I1}
\eea
This expression can be compared with the one of $\bar{\delta}_S^{(1)}$ given in Eq. (\ref{finaldL}). Apart from an obvious factor of two, the expression for ${\cal I}_1$ lacks the $J_2^{(1)}$ (lensing) term which is precisely a typical one that vanishes upon angular integration.

Still dropping irrelevant total derivatives, the two second order terms appearing in Eq. (\ref{Iphi}) take the following explicit form\,:
\small
\bea
{\cal I}_{1,1} &=& 2 \Xi_s ~ \Bigg\{ ~~~ \frac12 \left[ \psi_s^2 - \psi_o^2 \right] + \frac12 ([\partial_r P]_s)^2 - \frac12 ([\partial_r P]_o)^2 - \left(\psi_s + [\partial_+ Q]_s\right) \, [\partial_r P]_s \nonumber \\
~~~~~ ~~~~~ &+& \frac14 (\gamma_0^{ab})_s \left( 2 \partial_a P_s \, \partial_b P_s + \partial_a Q_s \, \partial_b Q_s - 4 \partial_a Q_s \, \partial_b P_s \right) - \frac12 \lim_{r\rightarrow 0} \left[\gamma_0^{ab} \partial_a P \, \partial_b P \right] \nonumber \\
~~~~~ ~~~~~ &+& Q_s \left( - [\partial_+^2 Q]_s + [\partial_+ \hat{\psi}]_s+ [\partial_r^2 P]_s \right) + \frac{J}{{\mathcal H}_s} \left( [\partial_\eta \psi]_s + {\mathcal H}_s [\partial_r P]_s + [\partial_r^2 P]_s \right) \nonumber \\
~~~~~ ~~~~~& -& \frac12 \int_{\eta_{in}}^{\eta_s^{(0)}} d\eta' \frac{a(\eta')}{a(\eta_s^{(0)})} \partial_r \left[ - \psi^2 + (\partial_r P)^2 + \gamma_0^{ab} \partial_a P \, \partial_b P \right] (\eta', \Delta \eta, \tilde{\theta}^a) \nonumber \\
~~~~~ ~~~~~ &+& \frac12 \int_{\eta_{in}}^{\eta_o} d\eta' \frac{a(\eta')}{a(\eta_o)}  \partial_r \left[ - \psi^2 + (\partial_r P)^2 + \gamma_0^{ab} \partial_a P \, \partial_b P \right] (\eta', 0, \tilde{\theta}^a) \nonumber \\
~~~~~ ~~~~~ &+& \int_{\eta_s^{(0)+}}^{\eta_s^{(0)-}} dx~ \partial_+ \left[ \hat{\psi} ~ \partial_+ Q + \frac14 \hat{\gamma}_{0}^{ab} \, \partial_a Q \, \partial_b Q \right] (\eta_s^{(0)+}, x, \tilde{\theta}^a) ~~ \Bigg\} \nonumber \\
&+& \left[ \Xi_s^2 - \frac{1}{{\mathcal H}_s \Delta \eta} \left( 1 - \frac{{\mathcal H}_s'}{{\mathcal H}_s^2} \right) \right] J^2 - 4 \psi_s J + 2 \Xi_s \left( \psi_o - \frac{Q_s}{\Delta \eta} + [\partial_r P]_o \right) J + \left(\frac{Q_s}{\Delta \eta}\right)^2 \nonumber \\
&+& 2 \left( \psi_s - \psi_o + [\partial_r P]_s - [\partial_r P]_o \right) \frac{Q_s}{\Delta \eta} + (\gamma_0^{ab})_s \partial_a Q_s \partial_b \left( \frac{Q_s}{2} + \frac{J}{\Hcal_s} \right) - 2 \frac{J}{{\mathcal H}_s} [\partial_\eta \psi]_s \nonumber \\
&+& 2 \left(  \frac{J}{{\mathcal H}_s}+Q_s\right) [\partial_r \psi]_s - \frac{2}{\Delta\eta} \int_{\eta_s^{(0)+}}^{\eta_s^{(0)-}} dx~ \left[ \hat{\psi} ~ \partial_+ Q + \frac14 \hat{\gamma}_{0}^{ab} ~ \partial_a Q ~ \partial_b Q \right] (\eta_s^{(0)+},x,\tilde{\theta}^a)
\nonumber \\
&+&\frac{1}{8} \frac{1}{\sin \tilde{\theta}} \frac{\partial}{\partial \tilde{\theta}} \left\{ \cos \tilde{\theta} ~ \left( \int_{\eta_s^{(0)+}}^{\eta_s^{(0)-}} dx~ [\hat{\gamma}_{0}^{1b} ~ \partial_b Q](\eta_s^{(0)+},x,\tilde{\theta}^a) \right)^2 \right\} ~~,
\label{I11}
\eea

\bea
{\cal I}_2 &=& 2 \Xi_s ~ \Bigg\{ - \frac{1}{4} \left( \phi_s^{(2)} - \phi_o^{(2)} \right) + \frac{1}{4} \left( \psi_s^{(2)} - \psi_o^{(2)} \right) - \frac{1}{2} \int_{\eta_{in}}^{\eta_s^{(0)}} d\eta' \frac{a(\eta')}{a(\eta_s^{(0)})} [\partial_r \phi^{(2)}] (\eta', r_s^{(0)}, \tilde{\theta}^a) \nonumber \\
&+& \frac{1}{2} \int_{\eta_{in}}^{\eta_o} d\eta' \frac{a(\eta')}{a(\eta_o)} [\partial_r \phi^{(2)}](\eta', 0, \tilde{\theta}^a) + \frac14  \int_{\eta_s^{(0)+}}^{\eta_s^{(0)-}} dx~ \partial_+ \left[ \hat{\phi}^{(2)} + \hat{\psi}^{(2)} \right](\eta_s^{(0)+}, x, \tilde{\theta}^a) \Bigg\} \nonumber \\
 & -& \psi_s^{(2)} - \frac{2}{\Delta\eta} \int_{\eta_s^{(0)+}}^{\eta_s^{(0)-}} dx~ \left[ \frac{\hat{\phi}^{(2)} + \hat{\psi}^{(2)}}{4} \right] (\eta_s^{(0)+},x,\tilde{\theta}^a) ~~,
\label{I2}
\eea
\normalsize
where it is important to stress that all these quantities have their angular dependence expressed in terms of $\tilde{\theta}^a$.
One can explicitly check (through a long but straightforward calculation) that Eq. (\ref{Ideltarel}) is indeed satisfied.

To conclude, using the results of Eqs. (\ref{I1}-\ref{I2}) in Eq. (\ref{Iphi}) and considering the well-known ensemble average (see e.g. \cite{Li:2007ny,Clarkson:2009hr,Clarkson:2011uk, Marozzi:2012ib}) of $\langle d_L^{-2} \rangle (w_o,z_s)$ for a stochastic spectrum of inhomogeneities, we obtained the full second order contributions of the averaged flux.

\subsection{Combining light-cone and ensemble average for other observables}
\label{Ch5Sec33}

Now that we have presented the procedure to combine ensemble average with lightcone average for the quadratic first order terms (Sec. \ref{Ch5Sec2}) and presented also the full calculation of the luminosity flux (Sec. \ref{Ch5Sec31} and \ref{Ch5Sec32}), we can address the complete task of computing the stochastic average of the luminosity distance $d_L$ and the distance modulus $\mu ~\sim 5 \log_{10} d_L$. We will also recall the general necessary facts on the flux $\Phi \sim d_L^{-2}$.

Non-trivial effects on the {\em  ensemble} average of $d_L$, or of a generic function of it, can only originate either from quadratic and higher-order perturbative corrections, or from the spectrum of correlation functions such as $\overline{d_L(z, \ti\theta^a) d_L(z', \ti\theta^{\prime a})}$ (see \cite{Bonvin:2005ps}). Here, and as before, we deal with the {\em ensemble} average of $\langle d_L \rangle$ instead of $d_L$.
As already stressed in Sec. \ref{Ch5Sec22}, given the covariant (light-cone) average of a perturbed (inhomogeneous) observable $S$, the average of a generic function of this observable  differs, in general, from the function of its average, i.e. $\overline{\langle F(S) \rangle} \not= F(\overline{\langle S\rangle})$ (as a consequence of the nonlinearity of the averaging process).
Expanding the observable  to second order as $S=S_0+S_1+S_2$, one finds\,:
\beq
\overline{\langle F(S) \rangle} = F(S_0)+ F'(S_0) \overline{\langle S_1+S_2  \rangle}+F''(S_0) \overline{\langle S_1^2/2 \rangle} ~~,
\label{3}
\eeq
where in general $\overline{\langle S_1\rangle}\neq 0$ as a consequence of the so-called ``induced backreaction" (\gls{IBR}) terms, arising from the coupling between the inhomogeneity fluctuations of $S$ and those of the integration measure (see Sec. \ref{Ch5Sec22}). The overall correction to $\overline{\langle F(S) \rangle}$ thus depends not only on the intrinsic inhomogeneity of the observable $S$, but also on the covariance properties of the adopted averaging procedure. The Eq. (\ref{3}) implies, in our case, that different functions of the luminosity distance (or of the flux) may be differently affected by the process of averaging out the inhomogeneities, and  may require different ``subtraction" procedures for an unbiased determination of the relevant observable quantities. 

Let us consider, in particular, the lightcone averaged luminosity flux $\lla \Phi \rra \sim \lla d_L^{-2} \rra$ computed in Sec. \ref{Ch5Sec31}. Performing the stochastic average of Eq. (\ref{dLminus2}) (and using Eq. (\ref{Iphi})) we have\,:
\beq
\overline{\langle d_L^{-2} \rangle}(z) = (d_L^{\rm FLRW})^{-2} \, \overline{(I_\Phi(z))^{-1}} \equiv (d_L^{\rm FLRW})^{-2} \left[1 + f_\Phi(z) \right] ~~,
\label{7}
\eeq
where\,:
\beq
f_\Phi(z) \equiv \overline{\lla \Ical_1 \rra^2} - \overline{\lla \Ical_{1,1} + \Ical_2 \rra} ~~.
\label{7a}
\eeq
We can now apply the general result of Eq. (\ref{3}) to the flux variable, by setting $S= \Phi$ and consider two important functions of the flux\,: $F(\Phi)= \Phi^{-1/2} \sim d_L$ and  $F(\Phi)= -2.5 \log_{10}\Phi + {\rm cst} \sim \mu$ (the distance modulus). For the \gls{luminosity distance} we can introduce a fractional correction $f_d$, in analogy with Eq. (\ref{7}), such that\,:
\beq
\overline{\langle d_L \rangle}(z)=d_L^{\rm FLRW} \left[1+f_d(z)\right] ~~.
\label{2}
\eeq
Then, by using the general expression of Eq. (\ref{3}), we find the important relation\,:
\beq
f_d = -{1\over 2}f_\Phi+{3\over 8}\,\overline{\langle \left(\Phi_1/\Phi_0\right)^2 \rangle} ~~,
\label{13}
\eeq
where, in terms of the quantities defined in Eq. (\ref{215}), we have
\beq
\overline{\langle \left(\Phi_1/\Phi_0\right)^2 \rangle} = 4 \, \overline{\langle (\bar{\delta}_S^{(1)})^2 \rangle} ~~,
\eeq
and where $f_\Phi$ is defined by Eq. (\ref{7a}). For the distance modulus we obtain, instead, 
\beq
\overline{\langle \mu \rangle}- \mu^{\rm FLRW}= - 1.25 (\log_{10}e)
\Big[ 2f_\Phi-\overline{\langle \left(\Phi_1/\Phi_0\right)^2 \rangle}
\Big] ~~.
\label{14}
\eeq

We can also consider, for any given averaged variable $\langle S \rangle$, the associated dispersion $\sg_S$ controlling how broad is the distribution of a perturbed observable $S$ around its mean value $\overline{\langle S  \rangle}$. This dispersion is due to both the geometric fluctuations of the averaging surface and to the statistical {\em ensemble} fluctuations, and is defined in general by (see Eq. \rref{vargen})\,:
\begin{equation}
\sigma_S \equiv \sqrt{\overline{\lla \left(S - \overline{\langle S \rangle} \right)^2 \rra}} = \sqrt{ \overline{\lla S^2 \rra} - \left(\overline{\lla S \rra}\right)^2} ~~.
\end{equation}
The dispersion associated with the flux and the distance are thus given by\,:
\bea
& \sigma_\Phi = \sqrt{ \overline{\langle \left(\Phi/\Phi_0\right)^2 \rangle}- \left( \overline{\langle \Phi/\Phi_0 \rangle}\right)^2} = \sqrt{\overline{\langle \left(\Phi_1/\Phi_0\right)^2 \rangle}} ~~, \\
& \sigma_{d_L} = \sqrt{ \overline{\lla \left( d_L / d_L^{\rm FLRW} \right)^2 \rra} - \left( \overline{\lla d_L / d_L^{\rm FLRW} \rra}\right)^2} = \frac12 \sqrt{\overline{\langle \left(\Phi_1/\Phi_0\right)^2 \rangle}} ~~,
\label{16}
\eea
while for the distance modulus we find\,:
\beq
\sigma_\mu=\sqrt{ \overline{\langle \mu^2 \rangle}- \left( \overline{\langle \mu \rangle}\right)^2}=  2.5 (\log_{10} e) \sqrt{\overline{\langle \left(\Phi_1/\Phi_0\right)^2 \rangle}} ~~.
\label{15}
\eeq
The above results will be applied to the case of a realistic background of cosmological perturbations of inflationary origin in the following chapter.

Let us conclude this section by recalling the convenient spectral parametrization, slightly different from Eqs. \rref{aiaj} and \rref{ai}, to be used for the various terms contributing to the fractional corrections of our observables. All the relevant contributions to the averaged functions of the luminosity redshift relation, at second order, can be parameterized in the form\,:
\begin{eqnarray}
\overline{\langle X \rangle} &=& \int_0^{\infty} \frac{d k}{k} ~ {\cal P}_{\psi}(k, \eta_o) \, \Ccal_X(k,\eta_o,\eta_s^{(0)}) ~~,  \\
\overline{\langle X' \rangle \langle Y' \rangle} &=& \int_0^{\infty} \frac{d k}{k} ~ {\cal P}_{\psi}(k, \eta_o) \, \Ccal_{X'}(k,\eta_o,\eta_s^{(0)}) \, \Ccal_{Y'}(k,\eta_o,\eta_s^{(0)}) ~~,
\end{eqnarray}
valid for any given model of perturbation spectrum. Here, $X'$ and $Y'$ ($X$) are first (second) order  generic functions of $(\eta,r,\ti \theta^a)$, and the $\Ccal$ are the associated spectral coefficients. In the particularly simple case of a CDM-dominated  geometry the spectral distribution of sub-horizon scalar  perturbations is time-independent ($\pa_\eta \psi_k=0$) and, as we shall see later, all the spectral coefficients $\Ccal$ can be calculated analytically. In the more general case where the power spectrum is a function of time, like for a $\Lambda$CDM model, the above equations are still satisfied but the spectral coefficients also depend on the power spectrum (usually through time integrals of it, convolved with other functions).
We should also stress that when performing numerical calculations the above integration limits will be replaced by appropriate cut-off values determined by the physical range of validity of the considered spectrum.

\bigskip

The next three sections will present the detailed expression of the spectral coefficients for the flux, considerations relative to the genuine second order terms and the proof that vector and tensor perturbations average out under the lightcone average. The reader not interested by these technical considerations can skip this part and follow his/her reading from Chapter \ref{Chap6}.

\newpage

\subsection{Computation of the $\Ccal(\Tcal^{(1,1)}_i)$ spectral coefficients of $\overline{ \lla \Ical_{1,1} \rra }$}
\label{Ch5Sec34}

We give here the result for the $\Ccal(\Tcal^{(1,1)}_i)$ spectral coefficients of $\overline{ \lla \Ical_{1,1} \rra }$ computed in the CDM case, using the notations $l=k \Delta \eta$ and ${\rm Sinc} (l)= \sin (l) /l$, and enclosing in a box the leading contributions.

\small
\bea
\Ccal(\Tcal_{1}^{(1,1)}) &=& 0 ~~~~~,~~~~~ 
\Ccal(\Tcal_{2}^{(1,1)}) = \boxed{\Xi_s \frac{f_s^2 - f_o^2}{\Delta\eta^2} \frac{l^2}{3}} ~~~~~,~~~~~
\Ccal(\Tcal_{3}^{(1,1)}) = 0 ~~, \nonumber \\
\Ccal(\Tcal_{4}^{(1,1)}) &=& \boxed{2 \Xi_s \frac{f_s^2}{\Delta\eta^2} \frac{l^2}{3}} + \frac{4 \Xi_s}{3 l^2} \left\{ 2 - 3l^2 + (l^2 -2) \cos(l) + l \sin(l) + l^3 {\rm SinInt}(l) \right\} \nonumber \\
& & ~~~~~ ~~~~~ ~~~~~ ~~~~~ + \, 4 \Xi_s \frac{f_s}{\Delta \eta} \left( \cos l-2+{\rm Sinc}(l)+l {\rm SinInt}(l) \right) ~~, \nonumber \\
\Ccal(\Tcal_{5}^{(1,1)}) &=& \boxed{- 2 \Xi_s \frac{f_o^2}{\Delta\eta^2} \frac{l^2}{3}} ~~~~~,~~~~~
\Ccal(\Tcal_{6}^{(1,1)}) = 4 \Xi_s [ 1 - {\rm Sinc}(l) ] - 4 \Xi_s \frac{f_s}{\Delta\eta} \left( \cos(l) - {\rm Sinc}(l) \right) ~~, \nonumber \\
\Ccal(\Tcal_{7}^{(1,1)}) &=& - \frac{2 \Xi_s}{3} \left\{ \boxed{\frac{f_s^2}{\Delta\eta^2} l^2} - 3 \frac{f_s}{\Delta \eta}( \cos(l) - {\rm Sinc}(l) ) - \frac{f_s}{\Delta \eta} \frac{2}{\Hcal_0 \Delta\eta} [ 2 \cos(l) + ( l^2 -2)  {\rm Sinc}(l) ] \right\} ~~, \nonumber \\
\Ccal(\Tcal_{8}^{(1,1)}) &=& \frac{2 \Xi_s}{\Hcal_s} \frac{f_s}{3 \Hcal_0 \Delta\eta^3} \left\{ \boxed{\Hcal_0 \Delta\eta ~ l^2} + 2 [- 3 \Hcal_0 \Delta\eta + (l^2 - 6) ] \cos(l) - 3 ( 2 + \Hcal_0 \Delta\eta ) ( -2 + l^2 ) {\rm Sinc}(l) \right\} ~~, \nonumber \\
\Ccal(\Tcal_{9}^{(1,1)}) &=& 0 ~~,~~
\Ccal(\Tcal_{10}^{(1,1)}) = 0 ~~, \nonumber \\
\Ccal(\Tcal_{11}^{(1,1)}) &=&  8 \Xi_s \left\{\frac{1}{2}- \frac{1}{3 l^2}+\left[\frac{1}{3 l^2}-\frac{1}{6}\right] \cos(l)-\frac{1}{6}{\rm Sinc}(l)-\frac{l}{6} {\rm SinInt}(l)\right\} ~~, \nonumber \\
\Ccal(\Tcal_{12}^{(1,1)}) &=& \left[ \Xi_s^2 - \frac{1}{\Hcal_s \Delta\eta} \left( 1 - \frac{\Hcal_s'}{\Hcal_s^2} \right) \right] \Bigg\{ 2 ~ [1 - {\rm Sinc}(l)] + 2 \frac{f_o - f_s}{\Delta\eta} [cos(l) - {\rm Sinc}(l)] \nonumber \\
& & ~~~~~~ ~~~~~ ~~~~~ ~~~~~ ~~~~~ + ~ \boxed{\frac{f_o^2 + f_s^2}{\Delta\eta^2} \frac{l^2}{3}} - \frac{2 f_o f_s}{\Delta\eta^2} [ 2 \cos(l) + ( l^2 -2) {\rm Sinc}(l) ] \Bigg\} ~~, \nonumber \\
\Ccal(\Tcal_{13}^{(1,1)}) &=& 4 \left\{ 1 - {\rm Sinc}(l) + \frac{f_o}{\Delta\eta}\left[cos(l) - {\rm Sinc}(l)\right] \right\} ~~, \nonumber \\
\Ccal(\Tcal_{14}^{(1,1)}) &=& 2\Xi_s \Bigg\{ \frac{f_o-f_s}{\Delta \eta}(\cos l- {\rm Sinc}(l)) \nonumber \\
& & ~~~~~ ~~~~~ + \, \boxed{\frac{f_o^2}{3\Delta \eta^2}l^2} - \frac{f_of_s}{\Delta \eta^2}\left(2 \cos(l) + ( l^2 -2){\rm Sinc}(l)\right) +\left(1+2\frac{f_o+f_s}{\Delta \eta}\right)\left(1-{\rm Sinc}(l)\right) \Bigg\} ~~, \nonumber \\
\Ccal(\Tcal_{15}^{(1,1)}) &=& 4 ~ \Xi_s \frac{f_o + f_s}{\Delta \eta} \left\{ 1 - {\rm Sinc}(l) \right\} - \frac{8}{l} {\rm SinInt}(l) ~~, \nonumber \\
\Ccal(\Tcal_{16}^{(1,1)}) &=& \frac{8}{l}{\rm SinInt}(l) ~~,~~
\Ccal(\Tcal_{17}^{(1,1)}) = \frac{8}{l^2} \left\{ -1 + \cos(l) + l ~ {\rm SinInt}(l) \right\} ~~, \nonumber \\
\Ccal(\Tcal_{18}^{(1,1)}) &=& - \frac{2}{\Hcal_s \Delta \eta} \Bigg\{ 2 - \cos(l) - {\rm Sinc}(l) - l ~ {\rm SinInt}(l) \nonumber \\
& & ~~~~~ ~~~~~ ~~~~~ + \frac{2 f_s}{\Delta \eta} \left( 1 + \boxed{\frac{l^2}{3}} - \cos(l) - l ~ {\rm SinInt}(l) \right) + \frac{2 f_o}{\Delta \eta} \left[ 1 - {\rm Sinc}(l) \right] \Bigg\} ~~, \nonumber \\
\Ccal(\Tcal_{19}^{(1,1)}) &=& \frac{4}{3 l^2} \left\{ 2 - 3 l^2 + ( l^2 -2 ) \cos(l) + l \sin(l) + l^3 {\rm SinInt}(l) \right\} ~~, \nonumber \\
\Ccal(\Tcal_{20}^{(1,1)}) &=& -\frac{2}{\Hcal_s \Delta \eta} \left\{ - \cos (l) + {\rm Sinc}(l) + \boxed{\frac{f_s}{3 \Delta \eta} l^2} - \frac{f_o}{\Delta\eta} [ 2 \cos(l) + ( l^2 - 2) {\rm Sinc}(l) ] \right\} ~~, \nonumber \\
\Ccal(\Tcal_{21}^{(1,1)}) &=& 4 ~ [ 1 - {\rm Sinc}(l) ] ~~, \nonumber \\
\Ccal(\Tcal_{22}^{(1,1)}) &=& -4 + \frac{4}{3 l^2} \left[ -2(2+3l^2) + (4 + l^2) \cos(l) + l \sin(l) + l (6 + l^2) {\rm SinInt}(l) \right] ~~, \nonumber \\ 
\Ccal(\Tcal_{23}^{(1,1)}) &=& 0 ~~.
\eea
\normalsize

\subsection{Dynamical evolution of scalar perturbations up to second-order}
\label{Ch5Sec35}

For a full computation of the fractional correction $f_\Phi$, what we need is the combined angular and ensemble averages of the three basic quantities, $\Ical_{1}$, $\Ical_{1,1}$ and $\Ical_{2}$. We must evaluate, in particular, the spectral coefficients $\Ccal(\Tcal_i^{(1)})$, $\Ccal(\Tcal_i^{(1,1)})$ and $\Ccal(\Tcal_i^{(2)})$, related to the terms of Eqs. (\ref{DefI1}-\ref{DefI2}) in terms of the first and  second-order Bardeen potentials. For this we need to know the dynamical evolution of the scalar fluctuations, at first order for the computation of  $ \Ical_{1}$, $ \Ical_{1,1}$ and at second order for $\Ical_{2}$.

Let us consider, first of all, a general  model with cosmological constant and dust sources. For the evolution of the scalar degrees of freedom in the Poisson gauge we will follow the analysis performed, up to second order, in \cite{Bartolo:2005kv}. In a general $\Lambda$CDM model the linear scalar perturbation obeys the evolution equation\,:
\beq
\psi''+3\Hcal \psi'+\left(2 \Hcal'+\Hcal^2\right) \psi=0 ~~.
\eeq
Considering only the growing mode solution we can set
\beq
\psi(\eta,\vec x)=\frac{g(\eta)}{g(\eta_o)} \psi_o(\vec x) ~~,
\label{SolutionGeneralPsi}
\eeq
where $g(\eta)$ is the so-called ``growth-suppression factor", or -- more precisely -- the least decaying mode solution, and $\psi_o$ is the present value of the gravitational potential. This ``\gls{growth factor}'' can be expressed analytically in terms of elliptic functions \cite{Eisenstein:1997ij} (see also \cite{PeterUzan}), and it is well approximated by a simple function of the critical-density parameters of non-relativistic matter $\Om_m$ and cosmological constant $\Om_\La$\,:
\beq
g(z) = \frac{5\,\Omega_m(z)/2}{\Omega_m(z)^{4/7}-\Omega_{\Lambda}(z)+(1+\Omega_m(z)/2)(1+\Omega_{\Lambda}(z)/70)} g_{\infty} ~~.
\label{gformula}
\eeq
Here $g_{\infty}$ represents the value of $g(\eta)$ at early enough times when the cosmological constant was negligible, and  is fixed by the condition $g(\eta_o) = g(z=0) = 1$.

The second-order potentials obey a similar evolution equation containing, however, an appropriate source term.
Their final expression in terms of  $\psi_o$ has been given in \cite{Bartolo:2005kv} and reads\,:
\begin{eqnarray}
\label{PSI}
\psi^{(2)}(\eta)&=&\left( 
B_1(\eta)-2g(\eta)g_{\infty} -\frac{10}{3}(a_{\rm nl}-1)g(\eta)g_{\infty}
\right)\psi_o^2 
+\left( B_2(\eta) - \frac{4}{3}g(\eta)g_{\infty}\right) \Ocal^{ij} \partial_j \psi_o \partial _i \psi_o
\nonumber \\ & &  
+ B_3(\eta)\, \Ocal_3^{ij} \partial_j \psi_o \partial _i \psi_o +B_4(\eta)\,  \Ocal_4^{ij} \partial_j \psi_o \partial _i \psi_o  ~~,\\
\label{PHI}
\phi^{(2)}(\eta)&=&\left( B_1(\eta)+4g^2(\eta) - 2g(\eta)g_{\infty} - \frac{10}{3}(a_{\rm nl}-1)g(\eta)g_{\infty}
\right)\psi_o^2 \nonumber \\
& & +\Bigg( B_2(\eta)+\frac{4}{3} g^2(\eta) \left( e(\eta)+\frac{3}{2} \right)
-\frac{4}{3}g(\eta)g_{\infty} \Bigg) \Ocal^{ij} \partial_j \psi_o \partial _i \psi_o \nonumber \\
& & + B_3(\eta)\, \Ocal_3^{ij} \partial_j \psi_o \partial _i \psi_o +B_4(\eta)\,  \Ocal_4^{ij} \partial_j \psi_o \partial _i \psi_o ~~,
\end{eqnarray}
where
\beq
\label{Ocals}
\Ocal^{ij} = \nabla^{-2} \left( \delta^{ij}-3 \frac{\partial^i \partial^j}{\nabla^2} \right) ~~~~~,~~~~~ \Ocal_3^{ij} = \frac{\partial^i \partial^j}{\nabla^2} ~~~~~,~~~~~ \Ocal_4^{ij} =\delta^{ij} ~~.
\eeq
We have introduced the functions $B_A(\eta)={\mathcal H}_o^{-2} \left[l(\eta_o)+3 \Omega_{m}(\eta_o)/2 \right]^{-1} \tilde{B}_A(\eta)$, with $A=1,2,3,4$, and the following definitions\,:
\begin{eqnarray}
\label{B1}
& \tilde{B}_1(\eta) \!= \int_{\eta_{m}}^\eta d\tilde{\eta} \,{\mathcal H}^2(\tilde{\eta}) 
(l(\tilde{\eta})-1)^2 C(\eta,\tilde{\eta}) ~~, \\
\label{B2}
& \tilde{B}_2(\eta) \!=\! 2\int_{\eta_{m}}^\eta d\tilde{\eta} \, {\mathcal H}^2(\tilde{\eta}) 
\Big[2 (l(\tilde{\eta})-1)^2-3 +3 \Omega_m(\tilde{\eta}) \Big] C(\eta,\tilde{\eta}) ~~, \\
& \tilde{B}_3(\eta) \!=\! \frac{4}{3} \int_{\eta_{m}}^\eta d\tilde{\eta} \left(e(\tilde{\eta})
+\frac{3}{2} \right) C(\eta,\tilde{\eta}) ~~,~~
\tilde{B}_4(\eta) \!=\! - \int_{\eta_{m}}^\eta d\tilde{\eta} \,C(\eta,\tilde{\eta}) ~~,
\label{B3tildeB4tilde}
\end{eqnarray}
and with 
\begin{equation}
C(\eta,\tilde{\eta})= g^2(\tilde{\eta}) \frac{a(\tilde{\eta})}{a(\eta_o)} 
\Big[ g(\eta){\mathcal H}(\tilde{\eta})-g(\tilde{\eta}) 
\frac{a^2(\tilde{\eta})}{a^2(\eta)} {\mathcal H}(\eta) \Big] ~~~ , ~~~ e(\eta)=\frac{l^2(\eta)}{\Omega_m(\eta)} ~~~,~~~ l(\eta)=1+\frac{g'}{\Hcal g} ~~.
\label{labelC}
\end{equation}
Here $\eta_m$ denotes the time when full matter domination starts \cite{Bartolo:2005kv}. Its precise value is irrelevant since the region of integration around $\eta_m$ is strongly suppressed. Finally, $a_{{\rm nl}}$ is the so-called non-gaussianity parameter (see \cite{Bartolo:2005kv}), which approaches unity in the standard inflationary scenario.

For further use let us now evaluate the ensemble (and angular/light-cone) average of the different operators defined in Eq. (\ref{Ocals}), when applied to $\partial_i \psi_o \partial_j \psi_o$. Considering first the ensemble average of $\Ocal^{ij} \partial_i \psi_o \partial_j \psi_o$, and Fourier-expanding $\psi_o$, we get (see \cite{Marozzi:2011zb})\,:
\beq
\overline{\Ocal^{ij} \partial_i \psi_o \partial_j \psi_o} = 
\int \frac{d^3q ~ d^3 k}{(2 \pi)^3} \delta^{(3)}(\qbf) e^{i r \qbf \cdot \hat{\xbf}} ~ \psi_{|\kbf|} \psi^*_{|\kbf-\qbf|} \left[ -\frac{2 (\kbf\cdot\qbf) + |\kbf|^2}{ |\qbf|^2} + 3 \frac{(\kbf\cdot\qbf)^2}{|\qbf|^4} \right] ~~.
\eeq
From this point, and up to the end of this section, we will neglect all suffixes ``$o$" present in terms inside the integrals.
By using the Taylor expansion of $\psi^*_{|\kbf-\qbf|}$ around $\qbf = \vec{0}$ we have\,:
\beq
\psi^*_{|\kbf-\qbf|} \simeq \psi^*_{k} - \frac{\kbf\cdot\qbf}{k} \partial_k \psi^*_{k} + \frac12 \left\{ \left( \frac{q^2}{k} - \frac{(\kbf\cdot\qbf)^2}{k^3} \right) \partial_k \psi^*_{k} + \frac{(\kbf\cdot\qbf)^2}{k^2} \partial_k^2 \psi^*_{k} \right\} + \Ocal(q^3) ~~,
\eeq
where $k \equiv |\kbf|$, $q \equiv |\qbf|$, and where the latter terms have been obtained by using the Hessian matrix\,:
\beq
H^{ij} = \partial_{k_i} \partial_{k_j} \psi^*_k = \partial_{k_i} \left[ \frac{k^j}{k} \partial_{k} \psi^*_k \right] ~~.
\eeq
Combining the  last two results, writing the integral over $k$ as $\int_0^\infty 2 \pi k^2 dk \int_{-1}^{+1} d\cos\alpha \, (...)$, where  $\kbf\cdot\qbf = k q \cos\alpha$, and integrating over $\cos\alpha$, we obtain\,:
\beq
\overline{\Ocal^{ij} \partial_i \psi_o \partial_j \psi_o} = \int \frac{d^3q}{(2 \pi)^3} \delta^{(3)}(\qbf) e^{i r \qbf \cdot \hat{\xbf}} \int_0^\infty 2 \pi k^2 dk ~ \left[ \frac{16 k}{15} ~ \psi_k \partial_k \psi^*_k + \frac{4 k^2}{15} ~ \psi_k \partial_k^2 \psi^*_k \right] ~~.
\label{Op}
\eeq
Note that the integrand's dependence on the angle $\theta$ between $\xbf$ and $\qbf$ only arises from the exponential term $\exp(i r \qbf \cdot \hat{\xbf} )$, which disappears in the presence of $\delta^{(3)}(\qbf)$. As a consequence, the angular average has no impact on this particular term and we get\,:
\beq
\overline{\lla \Ocal^{ij} \partial_i \psi_o \partial_j \psi_o \rra} = 
\overline{\Ocal^{ij} \partial_i \psi_o \partial_j \psi_o} = \int_0^\infty dk \frac{k^2}{2 \pi^2} ~ \left[ \frac{8k}{15} ~ \psi_k \partial_k \psi^*_k + \frac{2 k^2}{15} ~ \psi_k \partial_k^2 \psi^*_k \right] ~~.
\label{Final_overline_O}
\eeq
Let us note also that, quite generally, $\partial_k \psi_k \sim \psi_k/k$, and thus the above term is of the same order as  
$\overline{\lla \psi_o \psi_o \rra}$.

By repeating exactly the same procedure for the $\Ocal_3^{ij} \partial_i \psi_o \partial_j \psi_o$ term, we find\,:
\beq
\overline{\Ocal_3^{ij} \partial_i \psi_o \partial_j \psi_o} = \int \frac{d^3q}{(2 \pi)^3} \delta^{(3)}(\qbf) e^{i r \qbf \cdot \hat{\xbf} } \int_0^\infty 4 \pi k^2 dk ~ \left[ \frac{k^2}{3} 
|\psi_k|^2 + \frac{k q^2}{5} ~ \psi_k \partial_k \psi^*_k + \frac{k^2 q^2}{20} ~ \psi_k \partial_k^2 \psi^*_k \right] \,,
\label{O3p}
\eeq
where the  last  two contributions  are vanishing  because of the $q$-integration. We are thus left with the following simple result (insensitive to the angular average, as before)\,:
\beq
\overline{\lla \Ocal_3^{ij} \partial_i \psi_o \partial_j \psi_o \rra} = 
\overline{\Ocal_3^{ij} \partial_i \psi_o \partial_j \psi_o}
=\int_0^\infty \frac{dk}{k} ~ \left[ \frac{k^2}{3} {\cal P}_\psi(k,\eta_o)\right] ~~.
\label{Final_overline_O3}
\eeq
Finally, the last term $\Ocal_4^{ij} \partial_i \psi_o \partial_j \psi_o=|\vec{\nabla} \psi_o|^2$ is trivial and gives 
\beq
\overline{\lla \Ocal_4^{ij} \partial_i \psi_o \partial_j \psi_0 \rra} = 
\overline{\Ocal_4^{ij} \partial_i \psi_o \partial_j \psi_o}
=\int_0^\infty \frac{dk}{k} ~ \left[ k^2 {\cal P}_\psi(k,\eta_o)\right] ~~.
\label{Final_overline_O4}
\eeq

\newpage

\subsection{Second-order vector and tensor perturbations}
\label{Ch5Sec36}

We have stressed in Chapter \ref{Chap4} that vector and tensor perturbations automatically appear, at second order, as sourced by the squared first-order perturbation terms. Hence, vector and tensor perturbations  must be included in a consistent second-order computation of the luminosity distance, even if their contributions are negligible at first order (as expected, in particular, for a background of super-horizon perturbations generated by a phase of slow-roll inflation). 

Working in the Poisson gauge, and moving to spherical coordinates $x^i=(r, \theta, \phi)$, the PG metric of Eq. (\ref{NewPGmetric}), implying Eq. \rref{PGgmunu}, is\,:
\beq
ds_{PG}^2 = a^2 \left[ -d\eta^2 + 2 v_i \, d\eta d x^i \right] + a^2\left[(\gamma_0)_{ij} + \chi_{ij}\right] d x^i d x^j ~~,
\eeq
where $\ga_0^{ij} = {\rm diag}(1, r^{-2}, r^{-2} \sin^{-2} \theta)$, and where $v^i$ and $\chi^{ij}$ are the vector and tensor perturbations written in spherical polar coordinates. They satisfy the conditions $\nabla_i v^i=0= \nabla_i \chi^{ij}$ and $\ga_0^{ij} \chi_{ij}=0$, where $\nabla_i$ is the covariant gradient of the 3-dimensional Euclidean space in spherical coordinates.

Following the same procedure as in the scalar case we can evaluate the vector and tensor contributions to the coordinate transformation connecting Poisson and GLC gauge, and then express the perturbed GLC metric, up to second order, including the vector and tensor variables $v^i$ and $\chi^{ij}$. Such a detailed computation has already been performed in Sec. \ref{Ch4Sec13}. We recall here the vector and tensor contributions to the coordinate transformation between $\theta$ and $\ti \theta$\,:
\bea
\label{TrThetaRecall}
\tilde{\theta}^a &=& \tilde{\theta}^{a (0)}+\tilde {\theta}^{a (2)} \\
&=& \theta^a + \frac{1}{2} \int_{\eta_+}^{\eta-} dx \left( \hat{v}^a - \hat{\chi}^{ra} + \hat{\gamma}^{ab}_0 \int_{\eta_+}^x dy~ \partial_b \hat{\alpha}^r (\eta_+,y,\theta^a) \right) (\eta_+,x,\theta^a) ~~, \nonumber
\eea
and to the $2 \times 2$ matrix $\ga^{ab}$ appearing in the GLC metric\,:
\bea
\label{TrGammaRecall}
a(\eta)^2 \gamma^{ab} &=& \gamma_0^{ab} - \chi^{ab} + \\
& & + \left[ \frac{\gamma_0^{ac}}{2} \int_{\eta_+}^{\eta_-} dx~ \partial_c \left( \hat{v}^b - \hat{\chi}^{rb} + \hat{\gamma}_0^{bd} \int_{\eta_+}^{x} dy~ \partial_d \hat{\alpha}^r (\eta_+,y,\theta^a) \right) (\eta_+,x,\theta^a) + (a \leftrightarrow b) \right] ~~, \nonumber
\eea
where $\alpha^r \equiv ({v^r}/{2}) - ({\chi^{rr}}/{4})$. Both results are needed, in fact, for the computation of the averaged flux as presented in Eq. \rref{dLminus2}.

If we take into account $v^i$ and $\chi^{ij}$ and compute $d_L$, we find that the right-hand side of Eq. \rref{215} has to be modified by the addition of $\da^{(2)}_{V,T} (z_s, \ti \theta^a)$ as presented in Eq. \rref{dLwithSVT}, representing the effect of the vector and tensor part of the perturbed geometry. In fact it turns out that no modification is induced on the corresponding equation for $I_\phi(z_s)$ controlling the light-cone average of $\lla d_L^{-2} \rra$, so that Eq. \rref{Iphi} holds even in the presence of vector and tensor perturbations.
The sought average, in fact, is proportional to the proper area of the deformed 2-sphere $\Sg(w_o, z_s)$, and is given by $I_\phi \sim \int d^2 \ti\theta \sqrt{\ga}$ (see Eq. (\ref{dLminus2})). Considering the vector and tensor contributions to $\ga^{-1}= \det \ga^{ab}$ (obtained from Eq. (\ref{TrGammaRecall})), and computing from Eq. (\ref{TrThetaRecall}) the Jacobian determinant $| \pa \ti \theta/\pa \theta |$, we can  express $I_\phi$ as an angular integral over the 2-sphere with unperturbed measure $d^2 \Om= \sin \theta d \theta d \phi$. In that case many terms cancel among each other, and we end up with the result\,:
\beq
I_\phi(w_o,z_s)-1 = {1\over 4 \pi} \int_0^\pi  \sin \theta d \theta \int_0^{2\pi} d\phi\, f(\eta, r, \theta, \phi),
\label{A3}
\eeq
where the integrand $f(\eta, r, \theta, \phi)$ is a simple expression proportional to the components of the vector and tensor perturbations. 

The above angular integrals are all identically vanishing, as we can check by expanding the perturbations in Fourier modes $v^i_\kbf$ and $\chi^{ij}_\kbf$. For each mode $\kbf$ we can choose, without loss of generality, the $x^3$ axis of our coordinate system aligned along the direction of $\kbf$. Considering, for instance, tensor perturbations, we can then write the most general perturbed line-element, in Cartesian coordinates (omitting, for simplicity, the Fourier index), as follows\,:
\beq
h_{ij} dx^i dx^j = h_{+}(\eta, x^3) (dx^1dx^1 - dx^2dx^2 ) +  2 h_{\times}(\eta,x^3) dx^1dx^2  ~~.
\eeq
We have called $h_{+}$ and $h_{\times}$, as usual, the two independent polarization modes. After transforming to spherical coordinates, using the standard definitions 
\beq
x ^1 = r \sin \theta \cos \phi ~~,~~ x^2 = r \sin \theta \sin \phi ~~,~~ x^3 = r \cos \theta ~~,
\eeq
we easily obtain\,:
\bea
\chi^{rr} = \sin^2 \theta \left( h_+ \cos 2 \phi +h_\times \sin 2 \phi \right) ~~&,&~~
\chi^{r\theta} = {\sin 2 \theta\over 2 r } \left( h_+\cos 2 \phi+h_\times \sin 2 \phi \right) ~~,
\\
\chi^{r\phi} = {1\over  r }\left( -h_+\sin 2 \phi + h_\times \cos 2 \phi \right) ~~&,&~~
\chi^{\theta\theta} = { \cos^2 \theta\over r^2}  \left( h_+ \cos 2 \phi +h_\times \sin 2 \phi \right) ~~,
\nonumber \\
\chi^{\theta\phi} = {\cos \theta \over  r^2 \sin \theta} \left( -h_+\sin 2 \phi + h_\times \cos 2 \phi \right) ~~&,&~~
\chi^{\phi\phi} = -{1\over r^2\sin^2 \theta} \left( h_+ \cos 2 \phi +h_\times \sin 2 \phi \right) ~~.
\nonumber
\eea
Since $h_{+} = h_+(\eta, r \cos \theta)$ and $h_{\times} = h_\times(\eta, r \cos \theta)$, all perturbation components depend on $\phi$ only through  $\cos2 \phi$ or $\sin 2 \phi$, so that their contribution averages to zero when inserted into Eq. \rref{A3}. The same is true for the case of vector perturbations, with the only difference that the $\phi$ dependence of $v^i$, in spherical coordinates, goes through $\cos \phi$ or $\sin \phi$ (corresponding to waves of helicity one instead of helicity two as in the tensor case). Also the vector contribution thus averages to zero when inserted into Eq. \rref{A3}.

% --------------------------------------------------------- %

%  --------------------- New Chapter 6 ----------------------   %
% --------------------------------------------------------- %
\chapter{Effect of inhomogeneities on cosmological observables}
\label{Chap6}
\pagestyle{plain}

We now present the final results of this thesis. We first address the incomplete stage of quadratic first order terms. We see that it allows a rough estimation of the size of the effect of inhomogeneities, giving a rigorous prediction for the dispersion but leaving open the question of the average itself. We also explain why the computed effects are infrared and ultraviolet convergent, showing the robustness of our approach. We then turn our attention to the full calculation and use the power spectrum in its non-linear regime, and with a proper transfer function. We finally compute the full effect of inhomogeneities from second order perturbations, explaining the approximations made in the process to circumvent the problems of numerical calculations.

\section{Induced backreaction in CDM and IR \& UV convergence}
\label{Ch6Sec1}

\subsection{The linear power spectrum of the CDM case}
\label{Ch6Sec11}

In order to perform the numerical computations that we intend to do, we need to insert the explicit form of the power spectrum. Limiting ourselves to sub-horizon perturbations in a CDM model, we can obtain this spectrum by applying an appropriate, time-independent, transfer function to the primordial (inflationary) spectral distribution (see e.g. \cite{DurCos}). The power spectrum is then given by Eq. \rref{PsiPIntro}\,:
\beq
\label{PsiPch6}
{\cal P}_\psi (k) = \left(\frac{3}{5}\right)^2 \Delta_{\cal R}^2 T^2(k) ~~~~~,~~~~~ \Delta_{\cal R}^2=A \left(\frac{k}{k_0}\right)^{n_s-1} ~~,
\eeq
where $T(k)$ is the  transfer function  which takes into account the sub-horizon evolution of the modes re-entering during the radiation-dominated era, and  $\Delta_{\cal R}^2$ is the primordial power spectrum of  curvature perturbations outside the horizon. The typical parameters of such a spectrum, namely the amplitude $A$,  the spectral index $n_s$ and the scale $k_0$, are determined by the results of recent WMAP observations  \cite{Komatsu:2010fb}. In our computations we will use, in particular, the following approximate values\,:
\beq
A=2.41 \times 10^{-9} ~~~~~,~~~~~  n_s=0.96 ~~~~~,~~~~~ k_0 = 0.002 \,{\rm Mpc}^{-1} ~~.
\eeq
Finally, since our main purpose here is to present an illustrative example, it will be enough for our needs  to approximate $T(k)$ by the effective shape of the \gls{transfer function} for density perturbations without  baryons, namely $T(k)=T_0(k)$, where \cite{Eisenstein:1997ik}:
\beq
\label{EH97}
T_0(q) = \frac{L_0}{L_0+q^2C_0(q)}  ~~,~~ L_0(q) = \ln(2e+1.8q) ~~,~~ C_0(q) = 14.2+\frac{731}{1+62.5q} ~~,~~ q = \frac{k}{13.41 k_{\rm eq}} ~~,
\eeq
and where $k_{\rm eq}\simeq 0.07 \, \Omega_{m0} ~ h^2 \Mpc^{-1}$ is the scale corresponding to matter-radiation equality, with $h\equiv H_0/(100 \,{\rm km \,s}^{-1} {\rm Mpc}^{-1})$.
We can easily check that the above transfer function goes to 1 for $k \ll k_{\rm eq}$,  while it falls like  $k^{-2}(\log k)$ for $k \gg k_{\rm eq}$. 
For the numerical estimates we will use $h=0.7$ and we will set $a_o =1$, $\Omega_m =1$. In that case we obtain  
$k_{\rm eq}\simeq 0.036 \,{\rm Mpc}^{-1}$ (see \cite{Eisenstein:1997ik}), and we can more precisely define the asymptotic regimes of our transfer function as $T_0 \simeq 1$ for $k \laq 10^{-3}\, {\rm Mpc}^{-1}$, and $T_0 \sim k^{-2}(\log k)$ for $k \gaq 2.5\, {\rm Mpc}^{-1}$. This second scale is already deep inside the so-called non-linear regime, roughly corresponding to $k \gaq 0.1\, h \, {\rm Mpc}^{-1}$ (see e.g. \cite{Smith:2007sb}). Further information on the transfer function is given in the Appendix sections \ref{AppASec24} and \ref{AppASec42}.

\subsection{Dominant contributions in induced backreaction terms and convergence}
\label{Ch6Sec12}
 
Let us start by comparing the behaviour of the power spectrum with the behaviour of the coefficients in Tables \ref{Tab1OfP2} and \ref{Tab2OfP2}. As shown in Table \ref{Tab1OfP2}, all the coefficients ${\mathcal C}_{ij}$ and ${\mathcal C}_{i} {\mathcal C}_{j}$ will go at most as ${\mathcal O}(1)$ while their combination ${\mathcal C}_{ij}-{\mathcal C}_{i} {\mathcal C}_{j}$ will vanish at least as ${\mathcal O}(k^2 \Delta \eta^2)$ in the infrared (IR) limit $k \Delta \eta \ll 1$. On the other hand the power spectrum is almost constant in this limit ($\Pcal_\psi (k) \sim k^{n_s - 1} \simeq k^{-0.04}$). As a consequence, the infrared part of the terms in the induced backreaction IBR$_2$ and in the dispersion $\overline{\lla \sigma_1^2 \rra}$ will give a subleading contribution, and we can safely fix our infrared cut-off to be $k= H_0$ -- i.e. we can limit ourselves to all the sub-horizon modes -- without affecting the final result.

\begin{figure}[ht!]
\centering
\includegraphics[width=8cm]{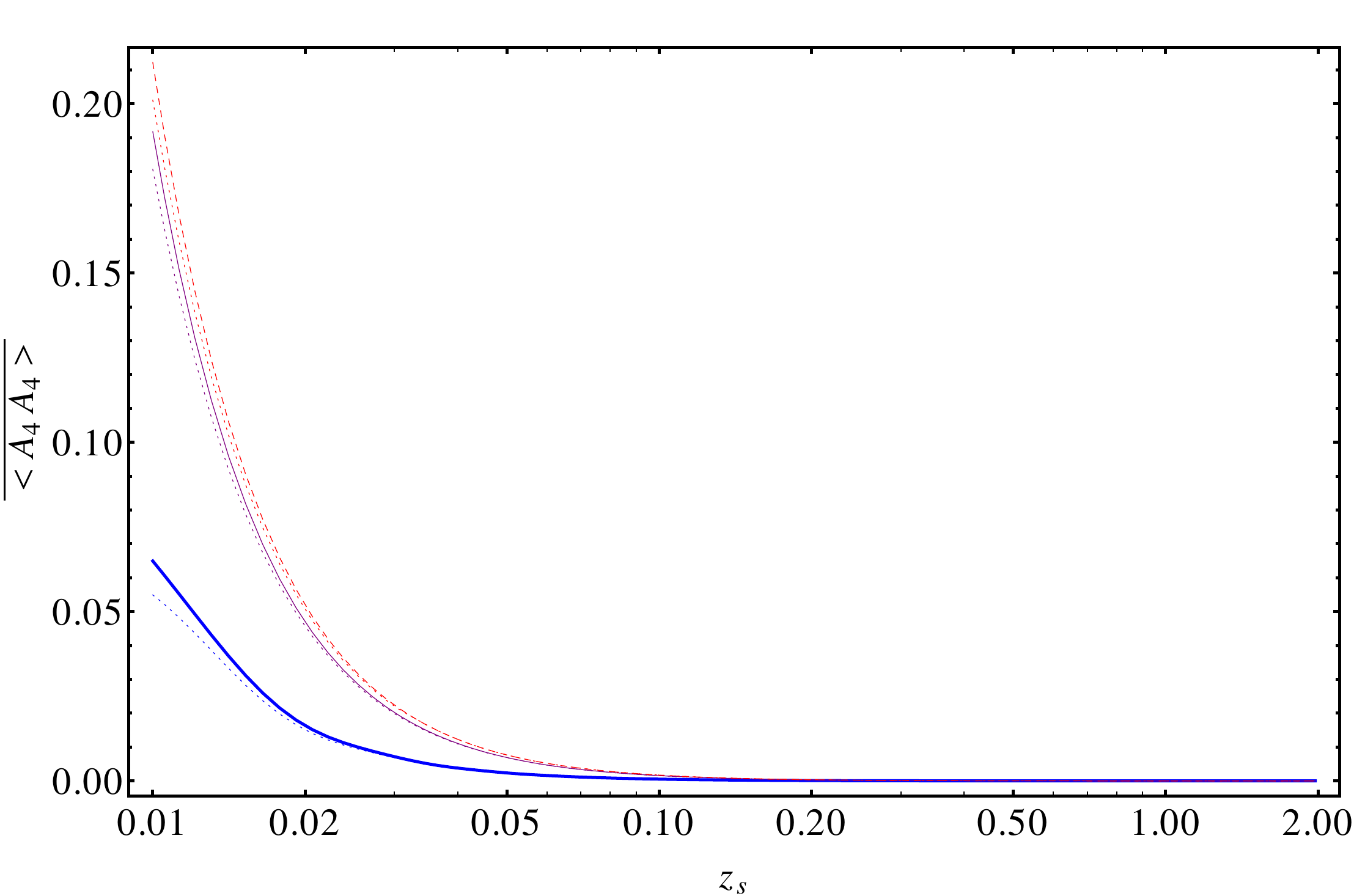}
\includegraphics[width=8cm]{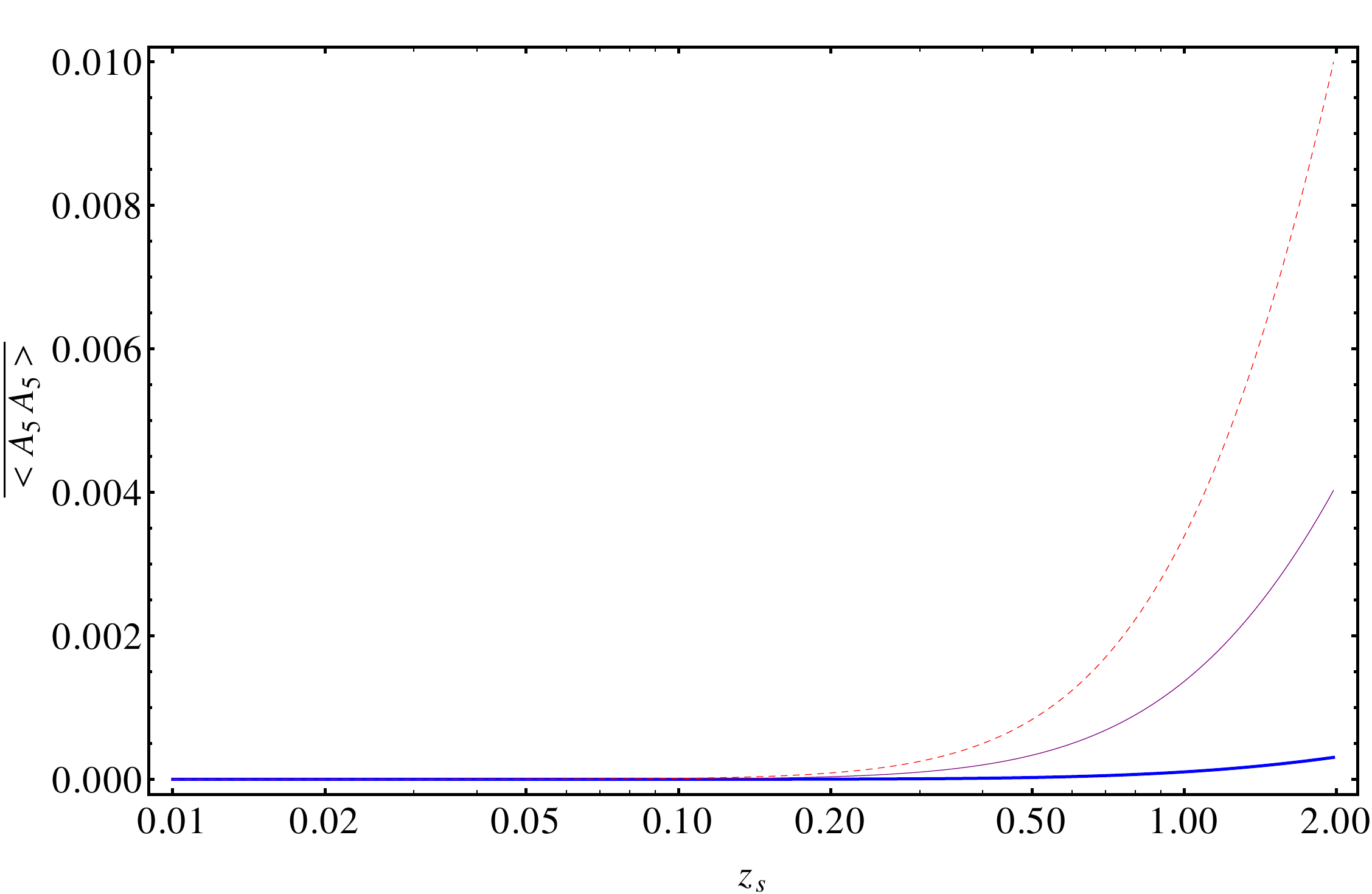}
\centering
\caption[Doppler contribution $\overline{\lla A_4 A_4 \rra}$ with non-baryonic transfer function and different cut-offs]{The result of the numerical integration of Eq. (\ref{aiaj}) for $\overline{\lla A_4 A_4 \rra}$ (\emph{Left}) and $\overline{\lla A_5 A_5 \rra}$ (\emph{Right}) is plotted as a function of $z_s$ for three different values of the UV cut-off: $k= 0.1 \,{\rm Mpc}^{-1}$ (thick blue line), $k=1 \,{\rm Mpc}^{-1}$ (thin purple line), $k=+\infty$ (dashed red line). \emph{Left:} Also shown (by the corresponding dotted curves) is the plot of the IBR$_2$ contribution $ \overline{\langle A_4A_4 \rangle}  -\overline{\langle A_4 \rangle \langle  A_4 \rangle}$.}
\label{Fig1and2ofP2}
\end{figure}

Furthermore, as can be seen in Tables \ref{Tab1OfP2} and \ref{Tab2OfP2}, all the coefficients ${\mathcal C}_{ij}$ and ${\mathcal C}_{i} {\mathcal C}_{j}$ will go at most as ${\mathcal O}(k^3 \Delta \eta^3)$ in the ultraviolet (UV) limit $k \Delta \eta \gg 1$. As a consequence, all the integrals involved in the induced backreaction and in the dispersion will be UV-convergent and their main contribution will come from the range $1/\Delta \eta \ll k \lesssim 2.5 \, {\rm Mpc}^{-1}$. In this range, some of the backreaction coefficients grow as positive power of $k$ while the transfer function is not yet decreasing as 
$k^{-2} (\log k)$. In particular, there are only two leading contributions corresponding to the integrals controlled by the coefficients\,: ${\mathcal C}_{44}$ and ${\mathcal C}_{55}$. This is the reason why the dispersion of the angular average of $d_L$, Eq. (\ref{varp}), is smaller than the one of $d_L$ itself, Eq. (\ref{vargen2nd}). For such coefficients we have in fact the following behaviour at $k \Delta \eta \gg 1$:
\bea
{\mathcal C}_{44} &\simeq & \left(1 - \frac{1}{{\mathcal H}_s\Delta \eta}\right)^2 \frac{(f_0^2 + f_s^2)}{\Delta \eta^2} \frac{k^2 \Delta \eta^2}{3} ~~,
\\
{\mathcal C}_{55} &\simeq & \frac{k^3 \Delta \eta^3}{15} {\rm SinInt}(k \Delta \eta) ~~.
\eea
From the  above expressions it is easy to understand that $\overline{\lla A_4 A_4 \rra}$ will give the largest contribution for $z_s\ll 1$ while $\overline{\lla A_5 A_5 \rra}$ will give the largest contribution for $z_s\gg 1$ (see Fig. \ref{Fig1and2ofP2}).

It is also clear that both the induced backreaction IBR$_2$ and the dispersion depend in principle on the UV cut-off $k_{UV}$ eventually used to evaluate the integrals. On the other hand, when $k_{UV}$ is taken inside the regime where  the spectrum goes like $k^{-4} (\log k)^2$, the dependence on the cut-off is not too strong. In particular the leading contribution to IBR$_2$, given by $2 \overline{\lla A_4 A_4 \rra}$, depends very weakly on the particular value of $k_{UV}$.  The leading contribution to the dispersion of Eq. (\ref{BRVarTh}), controlled by $\overline{\lla A_5 A_5 \rra}$, has a somewhat stronger dependence on $k_{UV}$ (because of the extra power of $k$  in the integrand of $\overline{\lla A_5 A_5 \rra}$). The numerical integrations of $\overline{\lla A_4 A_4 \rra}$ and $\overline{\lla A_5 A_5 \rra}$ are presented in Fig. \ref{Fig1and2ofP2}, where we illustrate the magnitude of the backreaction effects as a function of the redshift (in the range of values relevant to supernovae observations), and as a function of its dependence on  the  cut-off (ranging from $k_{UV}=0.1 \,{\rm Mpc}^{-1}$ to $k_{UV}=+\infty$). We should emphasize that the two abovementioned contributions have a clear and distinct physical meaning in the Newtonian gauge. Going back to their explicit expressions it appears that $\overline{\lla A_4 A_4 \rra}$ represents the \gls{Doppler effect} while $\overline{\lla A_5 A_5 \rra}$ is associated with \gls{lensing}, both in qualitative agreement with previous claims in the literature \cite{Vanderveld:2007cq,Bonvin:2005ps}. Note that the latter term can only appear in the variance, because it contributes to $\overline{\lla \sigma_1^2 \rra}$, while $\overline{\lla A_4 A_4 \rra}$ appears also in IBR$_2$ since $A_4$ is  present in both $\mu_1$ and $\sigma_1$. In the left panel of Fig. \ref{Fig1and2ofP2} we have also plotted the corrected leading contribution of IBR$_2$, namely $ \overline{\langle A_4A_4 \rangle} - \overline{\langle A_4 \rangle \langle A_4 \rangle}$ (see Eq. (\ref{BR2backreaction}))\,: we can see that the difference from the behavior of the $ \overline{\langle A_4A_4 \rangle}$ term is very small, with no qualitative effect on our discussion.

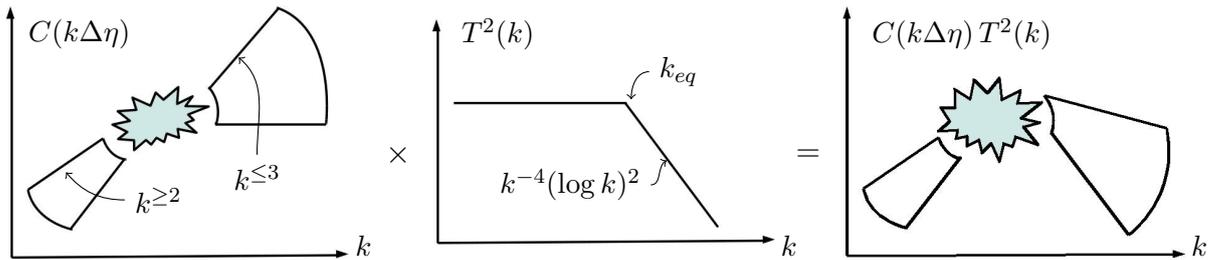
\begin{figure}[ht!]
\centering
\begin{tikzpicture}
  \node [xshift=-21.5cm,yshift=3.cm] at (current page.south east)
    {\includegraphics[width=4.5cm]{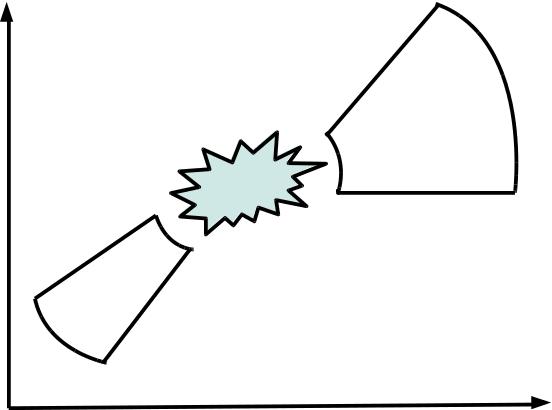}};      
  \node [xshift=-15.9cm,yshift=3.cm] at (current page.south east)
    {\includegraphics[width=4.5cm]{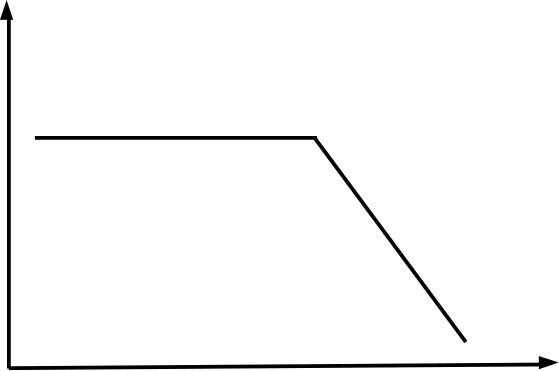}};
  \node [xshift=-10.5cm,yshift=3.cm] at (current page.south east)
    {\includegraphics[width=4.5cm]{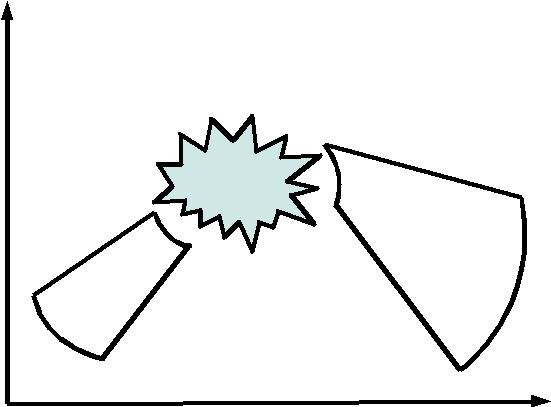}};

\node[anchor=west] at (2.1,2.7) (description) {\large $\times$};
\node[anchor=west] at (7.5,2.7) (description) {\large $=$};

\node[anchor=west] at (-2.6,4.3) (description) {$C(k \Delta \eta)$};
\node[anchor=west] at (1.7,1.5) (description) {$k$};

\node[anchor=east] at (-1.95,2.6) (text) {};
\node[anchor=west] at (-1.2,2.1) (description) {$k^{\geq 2}$};
\draw (description) edge[out=180,in=310,->] (text);

\node[anchor=east] at (0.3,4.1) (text) {};
\node[anchor=west] at (0.1,2.4) (description) {$k^{\leq 3}$};
\draw (description) edge[out=90,in=320,->] (text);

\node[anchor=west] at (3.1,4.3) (description) {$T^2(k)$};
\node[anchor=west] at (7.3,1.5) (description) {$k$};

\node[anchor=east] at (6.2,2.7) (text) {};
\node[anchor=west] at (3.6,2.3) (description) {$k^{-4} (\log k)^2 $};
\draw (description) edge[out=0,in=210,->] (text);

\node[anchor=east] at (5.6,3.3) (text) {};
\node[anchor=west] at (5.7,3.8) (description) {$k_{eq}$};
\draw (description) edge[out=180,in=90,->] (text);

\node[anchor=west] at (8.5,4.3) (description) {$C(k \Delta \eta) \, T^2(k)$};
\node[anchor=west] at (12.7,1.5) (description) {$k$};

\end{tikzpicture}
\centering
\caption[Illustration of the absence of IR and UV divergences]{Illustration of the absence of IR and UV divergences. One can see that the functions of $k$ which appear from the spectral coefficients of our terms, when multiplied by the transfer function, give rise to an integrand giving a finite contribution both in IR ($k \rightarrow 0$) and in UV ($k \rightarrow \infty$) regimes.}
\label{FigSchemeNoDiv}
\end{figure}

It is important to stress that, although all considered backreaction contributions are (at any redshift) UV-finite, they can induce relatively big effects at the distance scales relevant for supernovae observations, provided the given spectrum is extrapolated to sufficiently large values of $k$. We also stress that the decoupling of the small-distance (high-$k$) scales from the considered large-scale backreaction is due to the efficient suppression of the linear perturbation modes inside the horizon, an effect well described by the transfer function of Eq. (\ref{EH97}). The overall behaviour of the integrated terms, both in the IR and UV regimes, and the crucial role of the transfer function to make them finite is illustrated in Fig. \ref{FigSchemeNoDiv}.

Let us now sum up all contributions to the induced backreaction IBR$_2$ of Eq. (\ref{BR2backreaction}), and to the dispersion of Eq. (\ref{BRVarTh}), and compare the results for $\overline{\langle d_L \re} \pm d_L^{CDM}\sqrt{\overline{\langle \sg_1^2 \re}}$ with the homogeneous luminosity-distance of a pure CDM model and of a successful  $\La$CDM model. We include into $\overline{\langle d_L \re}$ only the IBR$_2$ contribution, namely we will set $\overline{\langle d_L \re}= d_L^{CDM}(1+ {\rm IBR}_2)$. It is thus an incomplete calculation but a full computation will be presented in the next section. The extra contributions arising from second-order perturbations of $d_L$, like the term called $\overline{\langle \sg_2 \re}$ in Eq. (\ref{EnsAvExp}), are a priori expected to be of the same order of $\overline{\lla \sigma_1^2 \rra}$ and we could even believe that (modulo cancellations) our computation may estimate a reliable ``lower limit" on the strength of the possible corrections to the luminosity-redshift relation. The reality, as we will see in the next section, is more complexe and we will find that the answer is highly dependent of the observable which is considered.

\begin{figure}[ht!]
\centering
\includegraphics[width=8cm]{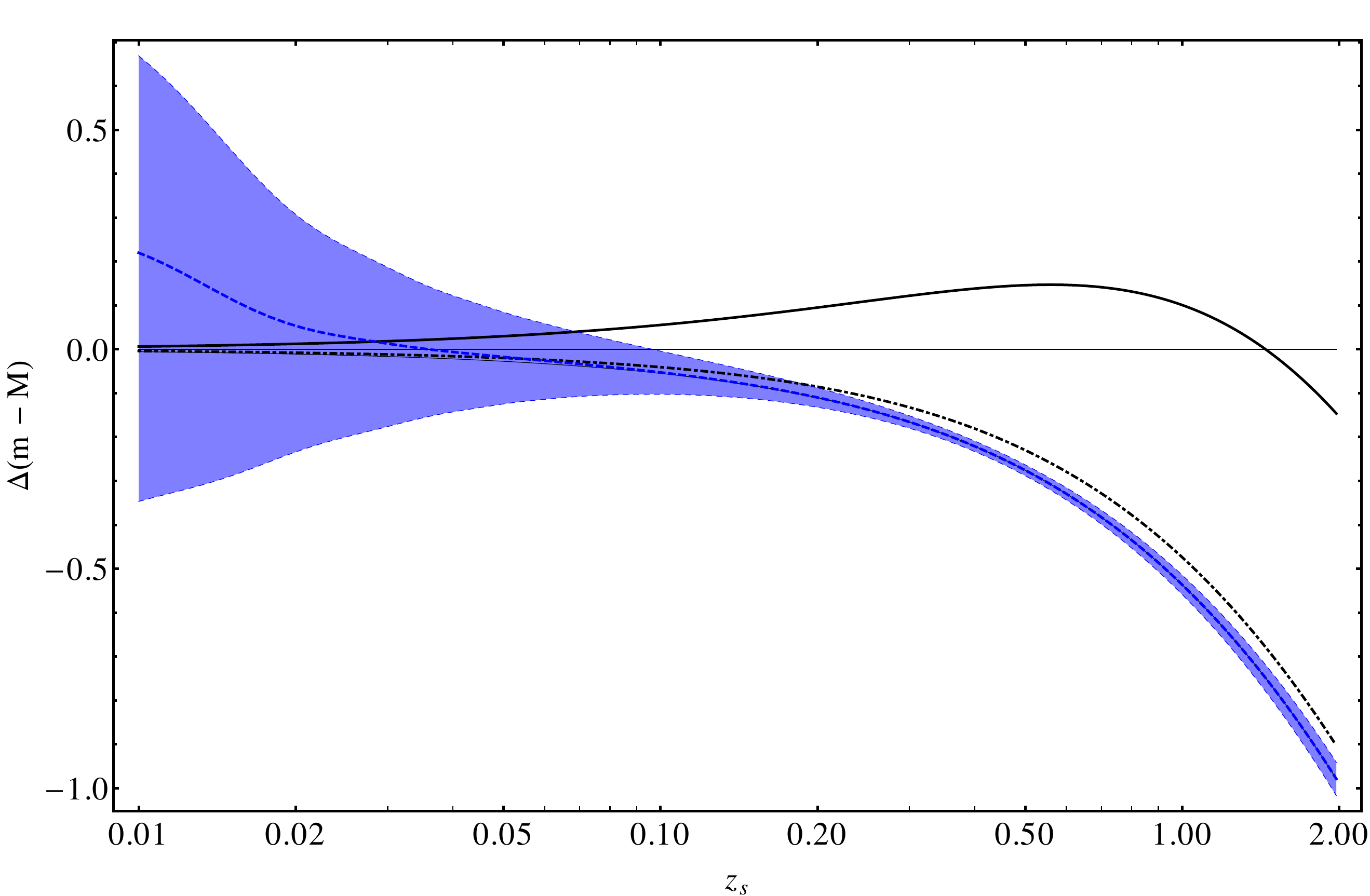}
\includegraphics[width=8cm]{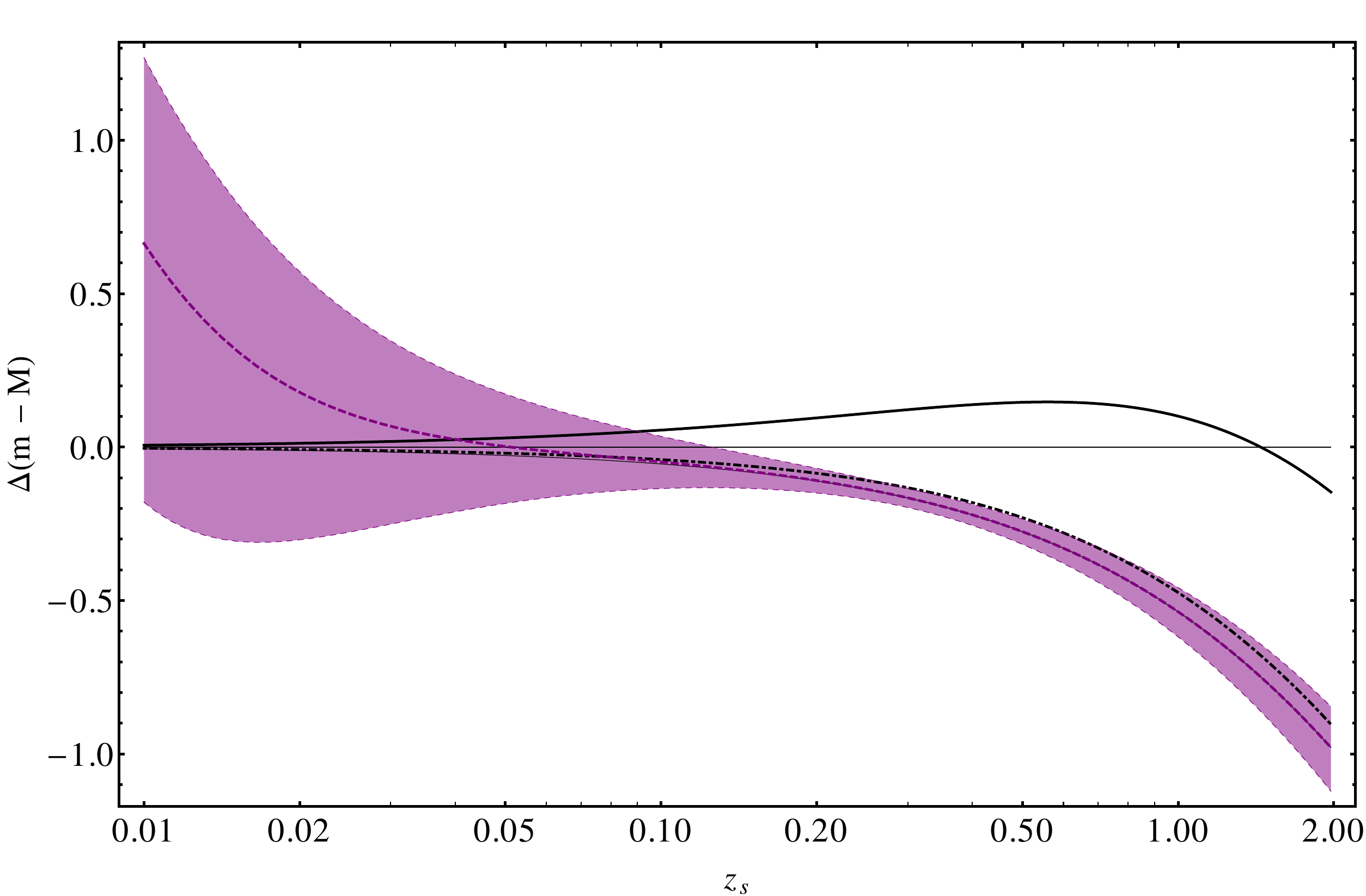}
\centering
\caption[Distance-modulus and dispersion for an inhomogeneous CDM model, $k_{UV}=0.1 \,{\rm Mpc}^{-1}$]{The distance-modulus difference $\mu(\overline{\lla d_L \rra}) - \mu^{M}$ of Eq. (\ref{modulus}) plotted for pure homogeneous models\,: CDM (thin line), $\Lambda$CDM with $\Om_\Lambda=0.73$ (thick) and $\Om_\Lambda=0.1$ (dashed-dot thick)\,; and for a CDM model including the contribution of IBR$_2$ (dashed line) plus/minus the dispersion (coloured region). \emph{Left\,:} We used for backreaction integrals the cut-off $k=0.1 \,{\rm Mpc}^{-1}$. \emph{Right\,:} same with $k=1 \,{\rm Mpc}^{-1}$.}
\label{Fig3and4ofP2}
\end{figure}

The comparison between the homogeneous and inhomogeneous (averaged) values of $d_L$ can be conveniently illustrated by plotting the so-called distance modulus $\mu \equiv m-M = 5 \log_{10}d_L$ or, even better, by plotting the difference between the distance modulus of the considered model and that of a flat, linearly expanding \gls{Milne}-type geometry, used as a reference value (see e.g. \cite{MG} and Appendix \ref{AppASec6}). In such a case we can plot, in particular, the following quantity\,:
\beq
\Delta(m-M) \equiv \mu(\overline{\lla d_L \rra}) - \mu^M = 5 \log_{10} \left[ \overline{\langle d_L \re} \right] - 5 \log_{10}\left[\frac{(2+z_s)z_s}{2 H_0}\right] ~~. \label{modulus}
\eeq
The results are illustrated in Fig. \ref{Fig3and4ofP2} for the cases of cut-off $k_{UV}=0.1 \,{\rm Mpc}^{-1}$ and $k_{UV}=1 \,{\rm Mpc}^{-1}$. The averaged luminosity-redshift relation of our inhomogeneous model is compared, in particular, with that of a pure CDM model and with that of a $\La$CDM model with $\Om_\La=0.1$ and with $\Om_\La= 0.73$. We have also explicitly shown the expected dispersion around the averaged result, by plotting the curves corresponding to $\overline{\langle d_L \re} \pm d_L^{CDM}\sqrt{\overline{\langle \sg_1^2 \re}}$ (bounding the coloured areas appearing in the figures).

Let us stress again that the choice of the cut-off may affect (even if not dramatically) the final result when the values of $k_{UV}$ are varying in the range $(0.1 - 1) \,{\rm Mpc}^{-1}$, while the precise choice becomes less important at higher values of $k_{UV}$. We should recall, however, that the inhomogeneous model adopted here is fully under control only in the linear perturbative regime, and that the spectrum cannot be extrapolated at scales higher than about $k\sim1 \,{\rm Mpc}^{-1}$ without taking into account the complicated effects of its non-linear dynamical evolution. 
The approximate coincidence between the above limiting value of $k_{UV}$ and the scale marking the beginning of the non-linear perturbative regime is not only a particular consequence of the transfer function adopted here (see Eq. (\ref{EH97})), but also an avoidable property of realistic non-linear perturbations spectra (see e.g. \cite{Smith:2002dz}).

\subsection{Conclusions from quadratic first order}
\label{Ch6Sec13}

As clearly shown by Fig. \ref{Fig3and4ofP2}, the corrections induced by IBR$_2$ on the luminosity distance of a homogeneous CDM model, 
even taking into account the expected dispersion of values around $ \overline{\langle d_L \re}$, cannot be used to successfully simulate realistic dark-energy effects. In the figures we have also plotted, for illustrative purposes,  an example of a standard $\La$CDM model with $\Om_\La=0.1$, which seems to be compatible (within the allowed region defined by the dispersion) with the prediction of our simple CDM+IBR$_2$ model, at least for sufficiently high values of the cut-off scale (see \emph{Right} figure of Fig. \ref{Fig3and4ofP2}).  However, we would be wrong to conclude that our backreaction can mimic a fraction of dark energy of the order of $\Om_\La \sim 0.1$, because a similar conclusion might be reliably reached only by averaging inhomogeneities on a background which already includes a significant amount of dark energy from the beginning. This situation will be realized in Sec. \ref{Ch6Sec3} where we will perform a calculation in a $\Lambda$CDM model.

Several arguments relative to this incomplete evaluation suggest the need of the full calculation that we are going to present in the next sections. First, as already stressed, a consistent second-order computation of the backreaction should include the contribution of $\overline{\langle \sg_2 \re}$. Nevertheless, we recall that from Eq. \rref{InterestofIBR2sigma1and2}, we can believe that $\overline{\lla \sigma_2 \rra}$ could contain, among others, contributions of the type $\overline{\lla A_5 A_5 \rra}$ already computed in this section. The behaviour of this term, in the asymptotic regime $k\Delta \eta \gg 1$, is very different from the behaviour of  terms like $\overline{\lla A_4 A_4 \rra}$ which give the leading contribution to IBR$_2$. The contribution of $\overline{\lla A_5 A_5 \rra}$, in particular, grows at large redshifts, as illustrated in Fig. \ref{Fig1and2ofP2}. 
Using our results for $\overline{\langle A_5 A_5 \rangle}$, and barring an unlikely cancellation, we could suggest that a full computation of  $\overline{\langle \sg_2 \re}$ may strongly enhance the overall backreaction effects at large $z_s$, with respect to the effects due to IBR$_2$ discussed previously. We will see that it is the case, $\overline{\lla \sigma_2 \rra}$ does contain this lensing contribution, but a cancellation with the $\overline{ \lla \sigma_1^2 \rra}$ term does happen, miraculously, as a nice property of the luminosity flux.

Secondly, we expect to see an enhancement of the effect of inhomogeneities by the use of a reliable estimate of contributions from the non-linear regime\footnote{The importance of such a non-linear regime was also underlined in \cite{Clarkson:2011br}, following a different approach to describe the impact of inhomogeneities on the supernovae observations.}. Also, the different behaviour of the backreaction contributions, at small and large $z_s$, can represent an important signature to distinguish the effects due to the averaged inhomogeneities from the more conventional dynamical effects of homogeneous dark energy sources. This is another justification for an accurate and robust second order calculation.

Let us note however that although a reliable estimate of the full backreaction on the averaged luminosity distance requires a full second-order calculation, some suitable linear combinations of averages of different powers of $ d_L$ only depend on the first-order quantity $\sigma_1$ (defined by the expansion of $d_L$). As an example, one can show that the following equality holds at second order\,:
\beq
\overline{\lla \left(d_L/d_L^{\rm FLRW}\right)^{\alpha} \rra}  - \alpha  \overline{\lla d_L/d_L^{\rm FLRW} \rra} = 1-\alpha + \frac{\alpha (\alpha -1)}{2} \, \overline{\langle \sigma_1^2 \rangle} ~~~~~ \forall \, \alpha \in \RRR ~~.
\eeq
This quantity can be plotted for a given inhomogeneous model, and compared with  its (deterministic) value in a $\Lambda$CDM model, for different values of  $\alpha$. The result is  that the two models disagree for a generic $\alpha$, bringing to the conclusion that realistic  inhomogeneities added to CDM lead to a model that can be distinguished, in principle,  from the conventional $\Lambda$CDM scenario. In practice, however, we only have a single quantity measured by SNe Ia experiments (basically the received flux of radiation), and one cannot  exclude that the two models happen to give the same result for that particular observable.

\section{Full calculation for the CDM model}
\label{Ch6Sec2}

In this section we evaluate the averages of the terms $\Ical_{1}$, $\Ical_{1,1}$ and $\Ical_{2}$ for the particular case of the standard CDM model, as more than a simple illustrative example. The results we will obtain shall confirm that the dominant terms are those already obtained for IBR$_2$ and $\overline{\lla \sigma_1^2 \rra}$ but within a complete study. The described dominant terms will be selected also for the discussion of the phenomenologically more relevant case, the $\Lambda$CDM model, presented in the forthcoming sections.

\subsection{The quadratic first-order contributions $\overline{\lla \Ical_1 \rra^2}$ and $\overline{\lla \Ical_{1,1} \rra}$}
\label{Ch6Sec21}

Let us now explicitly calculate the spectral coefficients for the first two  backreaction  terms (induced by the averaging)  contributing to Eq. \rref{7a}. The genuine second-order term $\overline{\lla \Ical_2 \rra}$ will be discussed in the next subsection. We start considering the first term $\overline{\lla \Ical_1 \rra^2}$. From Eq. (\ref{DefI1}) we obtain\,:
\beq
\overline{\lla \Ical_1 \rra^2} = \sum_{i=1}^{3} \sum_{j=1}^{3} \overline{\lla {\Tcal^{(1)}_i} \rra \lla {\Tcal^{(1)}_j} \rra} 
= \int_0^{\infty} \frac{d k}{k} ~ {\cal P}_{\psi}(k, \eta_o) ~ \sum_{i=1}^3 \sum_{j=1}^3  \Ccal_{\Tcal^{(1)}_i}(k,\eta_o,\eta_s)~ 
\Ccal_{\Tcal^{(1)}_j}(k,\eta_o,\eta_s)
\label{Eq_with_spectralI_1^2}
\eeq
(notice that, from now on, the background solution $\eta_s^{(0)}$ will be simply denoted by $\eta_s$). 
For a dust-dominated phase the spectral distribution of sub-horizon scalar perturbations is time independent, $\partial_\eta \psi_k=0$, and the scale factor $a(\eta)$ can be written as $a(\eta) = a(\eta_o)(\eta/\eta_o)^2$. We can also define
\beq
f_{o,s} \equiv \int_{\eta_{in}}^{\eta_{o,s}} d\eta \frac{a(\eta)}{a(\eta_{o,s})} = \frac{\eta_{o,s}^3 - \eta_{in}^3}{3\eta_{o,s}^2} ~ \simeq \frac13 \eta_{o,s}
\label{fCDM}
\eeq
(recall that $\eta_{in}$ satisfies, by definition, $\eta_{in} \ll \eta_{o,s}$). All the spectral coefficients of Eq. (\ref{Eq_with_spectralI_1^2}) can then be easily calculated, and the result is reported in Table \ref{Tab1}. This table is equivalent, within different notations, to the already presented Tab. \ref{Tab2OfP2}. In a similar way, the second contribution to Eq. (\ref{7a}) can be expressed as\,:
\beq
\label{IntkI11}
\overline{\lla \Ical_{1,1} \rra} = \int_0^{\infty} \frac{d k}{k} ~ {\cal P}_{\psi}(k, \eta_o) ~ \sum_{i=1}^{23} \Ccal_{\Tcal_i^{(1,1)}} (k,\eta_o,\eta_s) ~~,
\eeq
and the explicit form of the spectral coefficients $\Ccal(\Tcal_i^{(1,1)})$ for CDM has been presented in Sec. \ref{Ch5Sec34}.

\renewcommand{\arraystretch}{1.7}

\begin{table}[t]
\centering
\caption[]{\label{Tab1} The spectral coefficients $\Ccal(\Tcal^{(1)}_i)$ for the  $\overline{\lla {\Tcal^{(1)}_i} \rra \lla {\Tcal^{(1)}_j} \rra}$ terms.}
\vskip 0.2 cm
\begin{tabular}{|c|c|}
\hline
$\Tcal^{(1)}_i$ & $ \Ccal_{\Tcal^{(1)}_i} (k,\eta_o,\eta_s)$ \\
\hline\hline
$\Tcal^{(1)}_1$ & $-2 \frac{\sin k \Da \eta}{k \Da \eta}$ \\ \hline
$\Tcal^{(1)}_2$ & $-2 \left(1 - \frac{1}{{\mathcal H}_s\Delta \eta}\right) \left(1 - \frac{\sin k \Da \eta}{k \Da \eta}\right)+2 \left(1 - \frac{1}{{\mathcal H}_s\Delta \eta}\right) \frac{f_s}{\Delta \eta}  \left(\cos k \Da \eta - \frac{\sin k \Da \eta}{k \Da \eta}\right) $ \\ \hline
$\Tcal^{(1)}_3$ & $ \frac{4}{k \Da \eta} {\rm SinInt}(k \Da \eta)$ \\ \hline
\end{tabular}
\end{table}

\renewcommand{\arraystretch}{1}

Considering all contributions generated by $\Ical_1$ and $\Ical_{1,1}$, we find that the dominant ones are all contained in $\overline{\lla \Ical_{1,1} \rra}$, not $\overline{\lla \Ical_1 \rra^2}$, and are characterized by spectral coefficients proportional to $k^2$. Such dominant terms have been  emphasized, in Sec. \ref{Ch5Sec34}, by enclosing them in a rectangular box. They correspond, in particular, to the terms $\Tcal_i^{(1,1)}$ with $\{i=2,4,5,7,8,12,14,18,20\}$. Including only such dominant contributions we find that, to leading order, 
\bea
\label{SumT11lead}
\left[ \sum_{i=1}^{23}\Ccal_{\Tcal_i^{(1,1)}}\right]_{\mbox{Lead}} &=& \aleph_s \frac{f_s^2 + f_o^2}{3} k^2 + \frac{2 (\Xi_s - 3)}{\Hcal_s} \frac{f_s}{3} k^2 + \Xi_s \frac{f_s^2}{3} k^2 - \Xi_s \frac{f_o^2}{3} k^2 \nonumber \\
&=& f_o^2 k^2 \left(\frac{\aleph_s}{3} - \frac{\Xi_s}{3} \right) + f_s^2 k^2 \left(\frac{\aleph_s}{3} + \frac{4 \Xi_s}{3} - 3 \right)\nonumber \\
&=& - \frac{k^2}{\Hcal_0^2} ~ \ti f_{1,1}(z) ~~,
\eea
where we have defined 
\beq
\aleph_s ~\equiv~ \Xi_s^2 - \frac{1}{\Hcal_s \Delta\eta}\left(1 - \frac{\Hcal_s'}{\Hcal_s^2}\right) ~~,
\eeq
and we have used the relation $\Hcal_s \simeq 2 / (3 f_s)$ (valid in CDM). Also, we have included into the function $\ti f_{1,1}(z)$ all the $z$-dependence of these leading contributions. After some simple algebra we  find\,:
\begin{equation}
\ti f_{1,1}(z) = \frac{10-12 \sqrt{1+z}+5 z \left(2+\sqrt{1+z}\right)}{27\, (1+z) \left(-1+\sqrt{1+z}\right)^2} ~~.
\label{f_11}
\end{equation}
This function (and thus the corresponding backreaction) flips sign around  $z^{\ast}=0.205$ (as illustrated in Fig. \ref{Figf11f2}, also visible in Fig. \ref{f1}).
It should be noted, finally, that some of the genuine second-order terms contained into $\Ical_2$ are also associated to spectral coefficients proportional to $k^2$. However, as we shall see in the next subsection, the corresponding contributions to Eq. \rref{7a} turn out to be roughly an order of magnitude smaller than the above ones, this because of approximate cancellations.

\subsection{The genuine second-order contribution $\overline{\lla \Ical_2 \rra}$}
\label{Ch6Sec22}

In order to complete the calculation of $f_\Phi(z)$ we still have to consider the genuine second-order backreaction term  $\overline{\lla \Ical_2 \rra}$. In particular, we  must evaluate the spectral coefficients $\Ccal(\Tcal_i^{(2)})$, corresponding to the various contributions defined  in Eq. \rref{DefI2}, in terms of the first-order Bardeen potential. By using the results of Sec. \ref{Ch5Sec35}, and starting from Eqs. (\ref{PSI}) and (\ref{PHI}), we can easily see that the only possible  $k^2$-enhanced contributions arising from the coefficients  $\Ccal(\Tcal_i^{(2)})$ (which  only contain functions of $\phi^{(2)}$ and $\psi^{(2)}$) should correspond to the terms $B_3(\eta) \nabla^{-2} \partial_i\partial^j(\partial^i \psi_o \partial_j \psi_o )$ and $B_4(\eta) \partial^i \psi_o \partial _i\psi_o$. Therefore, we should obtain  $\phi^{(2)}\simeq \psi^{(2)}$ at leading order.

Let us discuss and estimate all possible contributions  for the CDM model we are considering in this section.  In this simple case  
$l(\eta)=1$ and $B_1(\eta)=B_2(\eta)=0$ (see Eqs. \rref{B1}, \rref{B2}). Furthermore, restricting our attention to the standard inflationary scenario, we can set $a_{nl}=1$. Eqs. \rref{PSI} and \rref{PHI} thus reduce to\,:
\bea
\psi^{(2)} &=&  -2\psi_o^2 - \frac{4}{3} ~ \Ocal^{ij} \partial_i \psi_o \partial_j \psi_o + B_3(\eta) ~ \Ocal_3^{ij} \partial_i \psi_o \partial_j \psi_o
+ B_4(\eta) ~ \Ocal_4^{ij} \partial_i \psi_o \partial_j \psi_o ~~,
\label{412}
\\
\phi^{(2)} &=& 2\psi_o^2 + 2 ~ \Ocal^{ij} \partial_i \psi_o \partial_j \psi_o + B_3(\eta) ~ \Ocal_3^{ij} \partial_i \psi_o \partial_j \psi_o
+ B_4(\eta) ~ \Ocal_4^{ij} \partial_i \psi_o \partial_j \psi_o ~~.
\label{412a}
\eea
Finally, we can use (as before) the  scale factor  $a(\eta) = a(\eta_o) \left({\eta}/{\eta_o}\right)^2$, and obtain (according to Eq. (\ref{B3tildeB4tilde}))\,:
\beq
B_3(\eta)=\frac{20}{21}\frac{1}{ {\cal H}^2}  ~~~~,~~~~  B_4(\eta)=-\frac{2}{7}\frac{1}{{\cal H}^2} ~~.
\eeq

We are now in the position of evaluating the genuine second-order terms, by exploiting the results given in  Eqs. (\ref{Final_overline_O}), (\ref{Final_overline_O3}) and (\ref{Final_overline_O4}). Following the classification of Eq. (\ref{Tcal2}) we can see that the first two terms $\Ccal(\Tcal_1^{(2)}) $ and $\Ccal(\Tcal_2^{(2)})$ exactly cancel for the CDM case (while they cancel only at leading order for the case of a $\Lambda$CDM model). For CDM we have, in particular, 
\beq
\Ccal(\Tcal_1^{(2)})= -\Ccal(\Tcal_2^{(2)})= - \frac{\Xi_s}{252} (\eta_s^2 - \eta_o^2) k^2 ~~.
\eeq

Another interesting simplification concerns the terms $\Tcal_{3}^{(2)}$, $\Tcal_{4}^{(2)}$ and $\Tcal_{5}^{(2)}$, namely those terms for which the integrand contains $\partial_r\psi^{(2)}$ or $\pa_r \phi^{(2)}$. From our previous results, in particular from Eqs. (\ref{Final_overline_O}), (\ref{Final_overline_O3}) and (\ref{Final_overline_O4}), it is easy to see that the $\psi^{(2)}$ and $\phi^{(2)}$ contributions are unchanged when one averages over the 2-sphere (i.e. over $\theta$ by isotropy), since the $\theta$-dependence is removed by the presence of $\delta^{(3)}(\qbf)$. On the other hand, the presence of the $r$-derivative brings a further factor $|\qbf| \cos \theta$, and the $q$-integration gives zero, even before performing the angular average. It follows that, for a general model,
\beq
\Ccal(\Tcal_{3}^{(2)}) = 0 ~~~~~,~~~~~ \Ccal(\Tcal_{4}^{(2)}) = 0 ~~.
\eeq
The contribution of $\Tcal_{5}^{(2)}$, on the contrary, is non-vanishing because of the presence of the partial derivative $\partial_\eta$, acting on the $B_A(\eta)$ coefficients. For the CDM model we find, in particular,
\beq
\Ccal(\Tcal_{5}^{(2)}) = \frac{\Xi_s}{126} (\eta_s^2 - \eta_o^2) k^2 ~~.
\eeq
Finally, for the last two terms $\Tcal_{6}^{(2)}$ and $\Tcal_{7}^{(2)}$ we obtain\,:
\bea
\Ccal(\Tcal_{6}^{(2)}) &=& -2 - \frac{k^2 \eta_s^2}{126} + \frac{32 k}{45}
\frac{\psi_k \partial_k \psi_k^\ast}{|\psi_k|^2} + \frac{8 k^2}{45}
\frac{\psi_k \partial_k^2 \psi_k^\ast}{|\psi_k|^2} ~~, \nonumber \\
\Ccal(\Tcal_{7}^{(2)}) &=& \frac{\eta_o^3 - \eta_s^3}{189 \Delta\eta}k^2 +
\frac{16 k}{45} \frac{\psi_k \partial_k \psi_k^\ast}{|\psi_k|^2} + \frac{4
k^2}{45} \frac{\psi_k \partial_k^2 \psi_k^\ast}{|\psi_k|^2} ~~.
\eea

\begin{figure}[ht!]
\centering
\includegraphics[width=12cm]{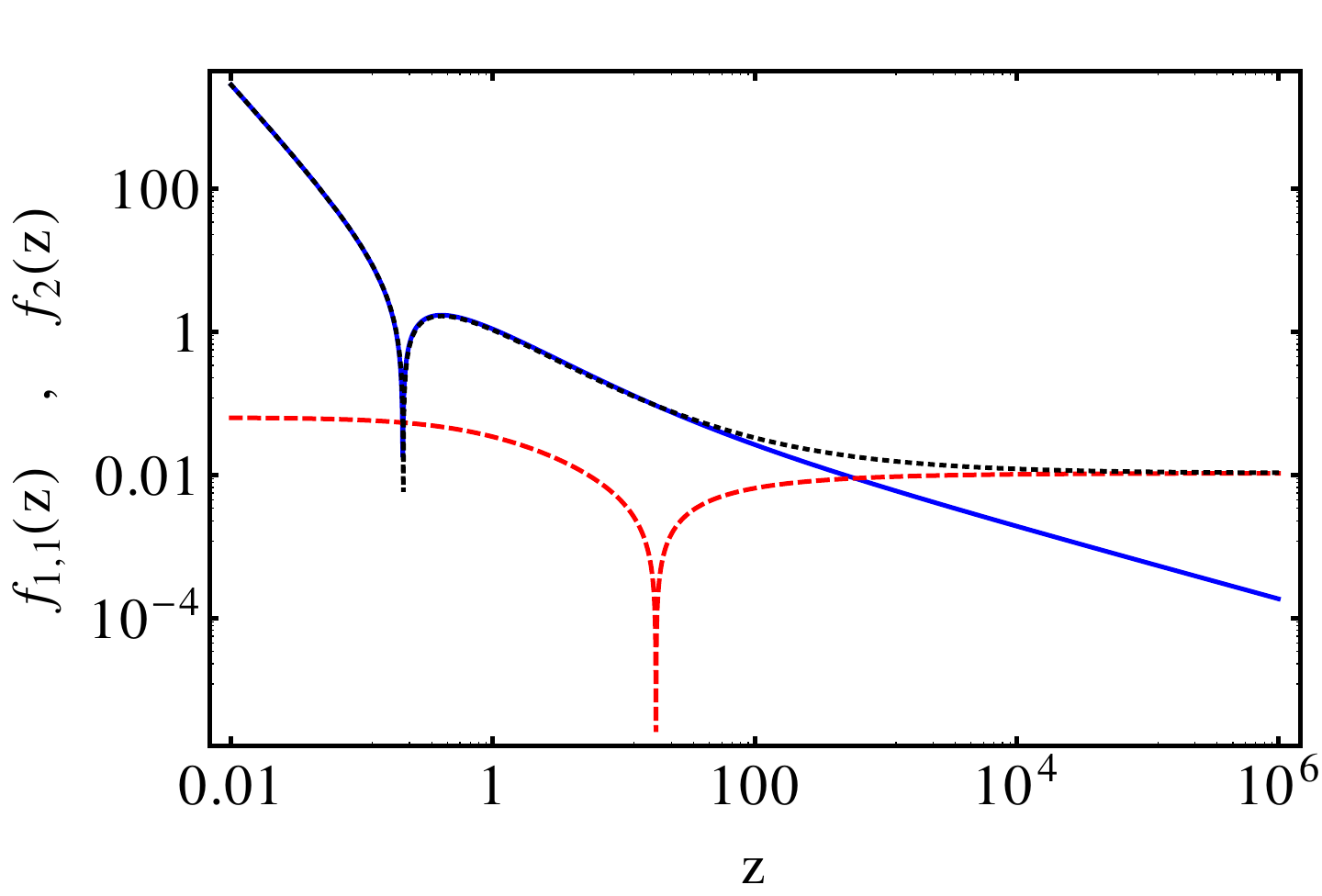}
\centering
\caption[Plot of the redshift functions $\widetilde{f}_{1,1}(z)$ and $\widetilde{f}_2(z)$.]{Plot of the redshift functions $\widetilde{f}_{1,1}(z)$ and $\widetilde{f}_2(z)$ for a very wide range of redshifts. One notice the different roles played by $\widetilde{f}_{1,1}(z)$ (solid blue) and $\widetilde{f}_2(z)$ (dashed red). The sum $\widetilde{f}_{1,1}(z) + \widetilde{f}_2(z)$ (dotted black) is also presented for the understanding of Eq. \rref{9}.}
\label{Figf11f2}
\end{figure}

The sum of all contributions  then leads to\,:
\beq
\label{GenSecOrder}
\sum_{i = 1}^{7} \Ccal(\Tcal_i^{(2)}) = - 2 + {1\over 7}\left[ \frac{\Xi_s}{2}
(f_s^2 - f_o^2) - \frac{f_s^2}{2} - \frac{f_s^3 - f_o^3}{ \Delta \eta}
\right] k^2 + \frac{16 k}{15} \frac{\psi_k \partial_k
\psi_k^\ast}{|\psi_k|^2} + \frac{4 k^2}{15} \frac{\psi_k \partial_k^2
\psi_k^\ast}{|\psi_k|^2} ~~.
\eeq
All the above spectral coefficients are now to be numerically evaluated by using the power spectrum of the CDM model. We can easily check, however, that the leading $k^2$-contributions of these coefficients are given by\,:
\beq
\left[ \sum_{i = 1}^{7} \Ccal(\Tcal_i^{(2)}) \right]_{\mbox{Lead}} = {1\over 7}\left[ \frac{\Xi_s}{2} (f_s^2 - f_o^2) - \frac{f_s^2}{2} - \frac{f_s^3 - f_o^3}{\Delta \eta} \right] k^2 = - \frac{k^2}{{\cal H}_o^2} \, \ti{f}_2(z) ~~,
\eeq
where\,:
\beq
\ti{f}_2(z)= -\frac{1}{189} \frac{2-2 \sqrt{1+z}+z \left(9-2 \sqrt{1+z}\right)}{(1+z)(\sqrt{1+z}-1)} ~~.
\label{f_2}
\eeq
Such second-order contributions turn out, for redshifts $z \sim 0.1 - 2$, to be about one or two orders of magnitude smaller than the leading contributions of the squared first-order terms of Sec. \ref{Ch6Sec21}, as can be easily checked by comparing the plots of $\widetilde{f}_{1,1}$ and $\widetilde{f}_{2}$ in Fig. \ref{Figf11f2}. We can also remark that it flips sign at a value of the redshift equal to $z^{\ast} = 17.59$, a value which is much too large for impacting the observed data. Nevertheless, we can see from this plot that the genuine second order is constant at large redshift, making it more important than the quadratic first order in the early Universe. This answers the question of how large is the genuine second order term $\overline{\lla \sigma_2 \rra}$ with respect to the induced backreaction terms IBR$_2$.

Considering the sum of these two contributions that will appear in the next section, we can deduce that there is no blowing up of the effect from inhomogeneities on the flux when we go to arbitrary early times (within the classical description employed here).

\subsection{Full numerical results for the CDM model}
\label{Ch6Sec23}

We made use here of the transfer function presented in Sec. \ref{Ch6Sec11} to represent the CDM model. Following the results of the previous subsections, especially Eq. \rref{7a}, we have found\,:
\beq
f_\Phi(z) = \int_0^{\infty} \frac{d k}{k} \, {\cal P}_\psi(k, z=0) \Big[f_{1,1}(k,z) +  f_{2}(k,z) \Big] ~~,
\label{9}
\eeq
where $f_{1,1}$ and  $f_{2}$ are complicated -- but known for the CDM case -- analytic functions of their arguments. 
However, as already stressed, the leading contributions in the range of $z$ relevant to dark-energy phenomenology are sourced by terms of the type $f(k, z) \sim (k/ {\cal H}_o)^2 \ti f(z)$. In that range of $z$ we can thus write, to a very good accuracy\,:
\beq
f_\Phi(z) \simeq \Big[\ti f_{1,1}(z) +  \ti f_{2}(z) \Big] \int_0^{\infty} \frac{d k}{k} \,\left( \frac{k}{{\cal H}_o}\right)^2{\cal P}_\psi(k, z=0) ~~, 
\label{10}
\eeq 
where $\ti f_{1,1}(z)$ and $\ti f_{2}(z)$ are given, respectively, by Eqs. (\ref{f_11}) and (\ref{f_2}). 

\begin{figure}[ht!]
\centering
\includegraphics[width=12cm]{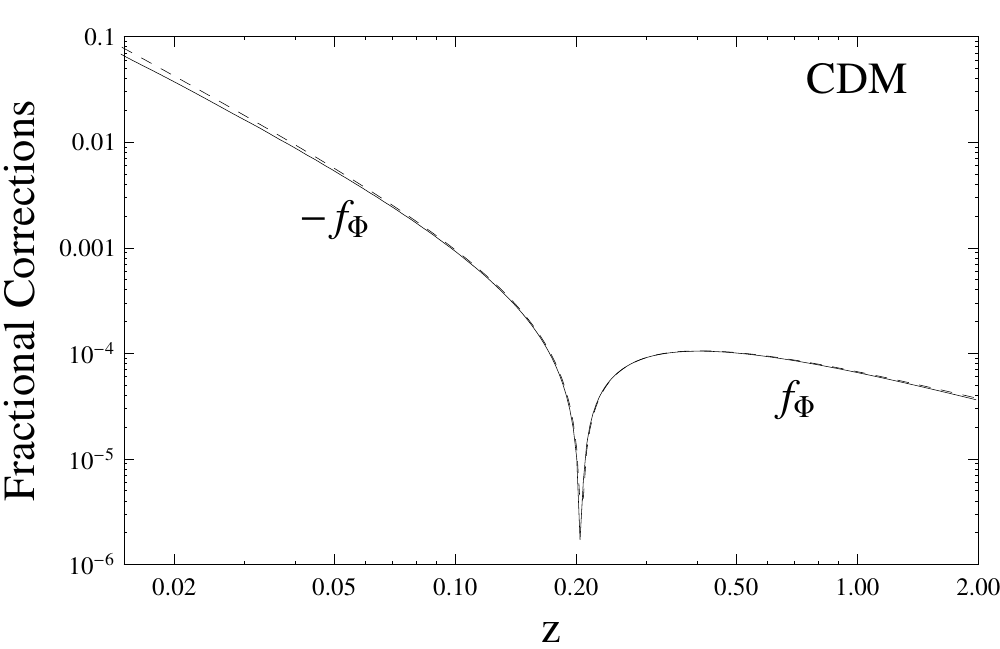}
\centering
\caption[Flux correction $f_\Phi$ in CDM, with non-baryonic $T(k)$ in $\Pcal_{\L}(k)$, with $k_{UV}=1 {\rm Mpc}^{-1}$]{The fractional correction $f_\Phi$ of Eq. (\ref{9}) (solid curve), compared with the same quantity given to leading order by Eq. (\ref{10}) (dashed curve), in the context of an inhomogeneous CDM model. We have used for the spectrum the one defined in Eq. (\ref{PsiPch6}). The plotted curves refer, as an illustrative example, to an UV cutoff $k_{UV}=1 \, {\rm Mpc}^{-1}$.}
\label{f1}
\end{figure}

To proceed, we need to insert  a power spectrum as well as infrared (IR) and ultraviolet (UV) cutoffs in Eq. (\ref{9}). The former can be identified with the present horizon $\Hcal_o^{-1}$. However, considering the used spectra, larger scales give a completely negligible contribution. 
On the other hand, in spite of the fact that our expressions converge in the UV for any reasonable power spectrum, some mild sensitivity to the actual UV cutoff will be shown to occur in certain observables.
The absolute value (and sign) of $f_\Phi(z)$ for the CDM model, obtained from both Eqs. (\ref{9}) and (\ref{10}), are illustrated in Fig. \ref{f1}. We can explicitly check  the accuracy of the leading order terms of Eq. (\ref{10}). The figure also confirms that the backreaction of a realistic  spectrum of stochastic perturbations induces  negligible corrections to the averaged flux at large $z$ (the larger corrections at small $z$, due to ``Doppler terms", have been already discussed also in Sec. \ref{Ch6Sec12}). In addition, it shows that such corrections have the wrong $z$-dependence (in particular, they change sign at some $z$) for simulating even a tiny dark-energy component.

~

\section{The $\Lambda$CDM model\,: power spectrum in the linear regime}
\label{Ch6Sec3}

We will now extend the procedure of the previous section to the case of the so-called ``concordance" cosmological model, using first  a  power spectrum computed in the linear  regime, and then in Sec. \ref{Ch6Sec4}, adding the effects of non-linearities following the parametrizations proposed in \cite{Smith:2002dz} and \cite{Takahashi:2012em}. In both cases, we will restrict our attention only to the $k^2$-enhanced terms already identified in the CDM model for the light-cone average of the flux variable, as well as to the $k^3$-enhanced terms which, as we will see, will appear in the variance, or in the averages of other functions of $d_L(z)$. Hereafter in all numerical computations we will use, in particular, the following numerical values\,:
\beq
\Omega_{\Lambda 0} = 0.73 ~~~~~,~~~~~ \Omega_{m0}=0.27 ~~~~~,~~~~~ \Omega_{b0}=0.046 ~~~~~,~~~~~ h=0.7 ~~. \nonumber
\eeq

\subsection{Second-order corrections to the averaged luminosity flux}
\label{Ch6Sec31}

The scalar power spectrum of the $\Lambda$CDM model is, in general, time-dependent. Considering for the moment only the linear regime, this spectrum can be written, starting from Eq. (\ref{SolutionGeneralPsi}), as\,:
\beq
\label{PsiPpar}
\Pcal_\psi (k,\eta) = \left[\frac{g(\eta)}{g(\eta_o)}\right]^2 \Pcal_\psi (k,\eta=\eta_o) ~~,
\eeq
and in this case we can easily extend the results previously obtained for the CDM model, concerning the leading ($k^2$-enhanced) contributions to the averaged-flux integral $I_\phi$. Using the general definitions of the $\Tcal_i^{(1,1)}$ terms, we already noticed, in Eq. \rref{SumT11lead}, that such enhanced contributions arise from the following particular (sub)-terms of Eq. (\ref{DefI11part2})\,:
\beq
\label{DefI11_Leading}
\Tcal_{2,L}^{(1,1)} = \Xi_s \left( ([\partial_r P]_s)^2 - ([\partial_r P]_o)^2 \right) ~~,
\nonumber 
\eeq
\beq
\Tcal_{4,L}^{(1,1)} = \frac12 \Xi_s (\gamma_0^{ab})_s \left( 2 \partial_a P_s \partial_b P_s \right)
~~~~~,~~~~~
\Tcal_{5,L}^{(1,1)} = - \Xi_s \lim_{r\rightarrow 0} \left[\gamma_0^{ab} \partial_a P  \partial_b P \right] 
~~, \nonumber 
\eeq
\beq
\Tcal_{7,L}^{(1,1)} = -2 \Xi_s ([\partial_r P]_s)^2
~~~~~,~~~~~
\Tcal_{8,L}^{(1,1)} = 2 \Xi_s \frac{1}{{\mathcal H}_s} \left(\psi_s-2\int_{\eta_s}^{\eta_o} d\eta' \partial_r \psi(\eta',\eta_o-\eta', 
\tilde{\theta}^a)\right) [\partial_r^2 P]_s
~~, \nonumber 
\eeq
\beq
\Tcal_{12,L}^{(1,1)} = \left[ \Xi_s^2 - \frac{1}{{\mathcal H}_s \Delta \eta} \left( 1 - \frac{{\mathcal H}_s'}{{\mathcal H}_s^2} \right) \right] 
\left( ([\partial_r P]_s)^2 + ([\partial_r P]_o)^2 \right)
~~~~~,~~~~~
\Tcal_{14,L}^{(1,1)} = 2 \Xi_s \left([\partial_r P]_o\right)^2
~~, \nonumber 
\eeq
\beq
\Tcal_{18,L}^{(1,1)} = - \frac{1}{\Hcal_s} (\gamma_0^{ab})_s \partial_a Q_s \partial_b [\partial_r P]_s 
~~~~~,~~~~~
\Tcal_{20,L}^{(1,1)} = -2 \frac{1}{{\mathcal H}_s} [\partial_r P]_s  [\partial_r \psi]_s ~~,
\eeq
where we have added the suffix ``$L$" to stress that we are reporting here only the leading terms. 

In order to calculate the corresponding spectral coefficients in $\Lambda$CDM, we note that their time dependence can be factorized with respect to 
the $k$-dependence whenever the time variable does not appear in the exponential factor $\exp(i \, \kbf \cdot \xbf)$ present in our integrals (see for instance the examples of Sec. \ref{Ch5Sec24}). This is indeed the case for the terms $\Tcal_{2,L}^{(1,1)}$, $\Tcal_{4,L}^{(1,1)}$, $\Tcal_{5,L}^{(1,1)}$, $\Tcal_{7,L}^{(1,1)}$, $\Tcal_{12,L}^{(1,1)}$, $\Tcal_{14,L}^{(1,1)}$ and $\Tcal_{20,L}^{(1,1)}$ which involve the integral $P(\eta,r,\ti\theta^a)$. In that case the previous CDM results for the leading  spectral coefficients can be simply generalized to the $\Lambda$CDM case through the following procedure\,: $(i)$ by inserting a factor $g(\eta_s)/g(\eta_o)$ whenever $\psi_s$ is present in the initial term, and $(ii)$ by replacing the $f_{o,s}$ factors (see Eq. (\ref{fCDM})), arising from the presence of $P_{o,s}$ terms, by\,:
\beq
\ti{f}_{o,s} \equiv \int_{\eta_{in}}^{\eta_{o,s}} d\eta \frac{a(\eta)}{a(\eta_{o,s})}  \frac{g(\eta)}{g(\eta_{o})} ~~.
\label{ftildeLambdaCDM}
\eeq

For  the remaining two terms $\Tcal_{8,L}^{(1,1)}$ and $\Tcal_{18,L}^{(1,1)}$, the integrals are performed along the path $r=\eta_o-\eta$, and 
the time dependence cannot be fully factorized. Therefore, the evaluation of the double integrals over $\eta$ and $k$ is much more involved than in the previous cases. However, a good approximation of the exact result can be obtained by replacing  $\psi(\eta',\eta_o-\eta', \tilde{\theta}^a)$, appearing in the integrands of $\Tcal_{8,L}^{(1,1)}$ and $\Tcal_{18,L}^{(1,1)}$, with $\psi(\eta_s,\eta_o-\eta', \tilde{\theta}^a)$. This is possible since the leading contributions to the time integral arise from a range of values of $\eta$ approaching $\eta_s$. By adopting such an approximation we can follow the same procedure as before, and we obtain\,:
\beq
\Ccal(\Tcal_{8,L}^{(1,1)}) = \frac{2}{3} \Xi_s \frac{\ti{f}_s}{\Hcal_s} k^2 ~~~~~,~~~~~
\Ccal(\Tcal_{18,L}^{(1,1)}) = - \frac{4}{3} \frac{\ti{f}_s}{\Hcal_s} k^2 ~~.
\eeq
This is formally the same result as in the CDM case (see Sec. \ref{Ch5Sec34} for the leading terms of $\Tcal_{8,L}^{(1,1)}$ and $\Tcal_{18,L}^{(1,1)}$), with the only difference that ${f}_s$ is replaced by $\ti{f}_s$.

Let us now move to the evaluation of the leading contributions present in the genuine second-order part $\Ical_2$. The final results of Sec. \ref{Ch6Sec22}, for the particular case of a CDM model, can be easily generalized to the $\Lambda$CDM case starting from the observation that the leading contributions can only arise  from terms containing the operators  ${\cal O}_3^{ij}$ and ${\cal O}_4^{ij}$ in Eqs. (\ref{PSI}), (\ref{PHI}) and \rref{Ocals}. 
As a consequence, the first two terms $\Tcal_1^{(2)}$ and $\Tcal_2^{(2)}$ will give a subleading overall contribution, while the general result that the terms $\Tcal_{3}^{(2)}$ and $\Tcal_{4}^{(2)}$ give identically zero still holds. The remaining leading contributions can be easily obtained, using the results in Eqs. (\ref{Final_overline_O3}) and (\ref{Final_overline_O4}), as follows\,:
\begin{eqnarray}
\Ccal(\Tcal_{5,L}^{(2)}) &=& -\Xi_s\left[ \frac{1}{3}\left(B_3(\eta_o)-B_3(\eta_s)\right)+\left(B_4(\eta_o)-B_4(\eta_s)\right)\right] k^2 ~~,
\\ 
\Ccal(\Tcal_{6,L}^{(2)}) &=& -\left( \frac{1}{3} B_3(\eta_s) + B_4(\eta_s)\right) k^2 ~~,
\\ 
\Ccal(\Tcal_{7,L}^{(2)}) &=& \frac{2}{\Delta\eta} \int_{\eta_s}^{\eta_o} d\eta' \left( \frac{1}{3} B_3(\eta') + B_4(\eta')\right) k^2 ~~.
\end{eqnarray}
These leading contributions can be now evaluated using Eqs. (\ref{B3tildeB4tilde}) and (\ref{labelC}) and moving to redshift space, where we can write\,:
\bea
\Hcal(z) &=& \frac{\Hcal_0}{1+z} \left[\Omega_{m0}(1+z)^3+\Omega_{\Lambda0}\right]^{1/2} ~~, \\
\Omega_m(z) &=& \frac{\Omega_{m0}(1+z)^3}{\Omega_{m0}(1+z)^3+\Omega_{\Lambda0}} ~~, \\
\Omega_{\Lambda}(z) &=& \frac{\Omega_{\Lambda0}}{\Omega_{m0}(1+z)^3+\Omega_{\Lambda0}} ~~,
\eea
where the suffix ``$0$" appended to $\Om_m$ and $\Om_\La$ denotes the present value of those fractions of critical density. We note that, as in CDM, the leading genuine second-order contributions are about one to two orders of magnitude smaller than the leading squared first-order contributions, evaluated above.

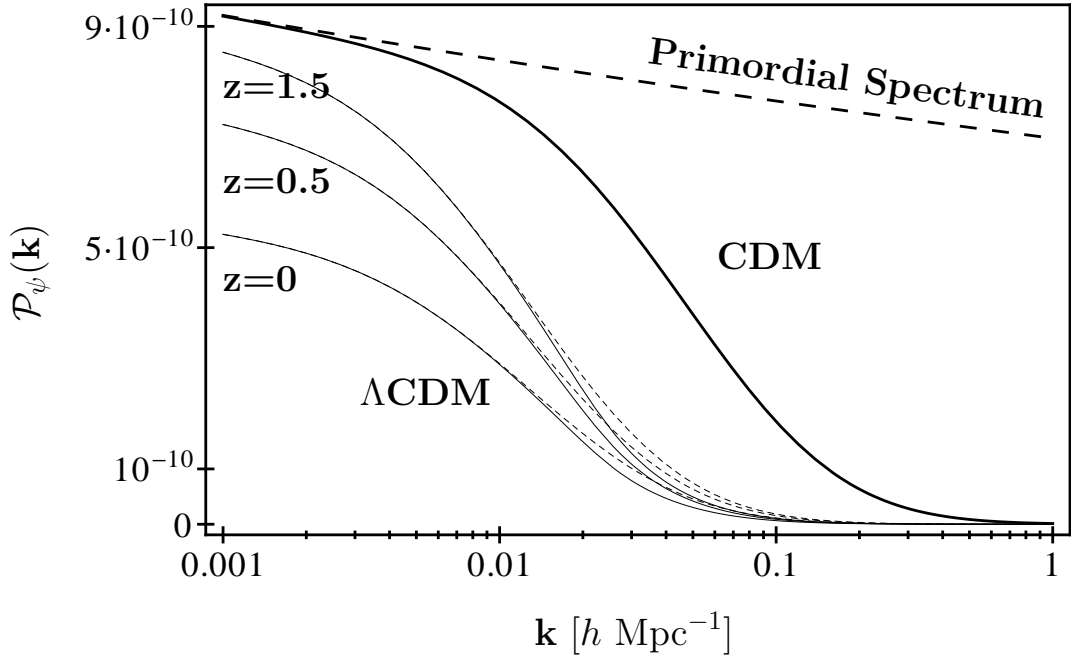
\begin{figure}[t]
\centering
\begin{tikzpicture}
\node (label) at (0,0){\includegraphics[width=13cm]{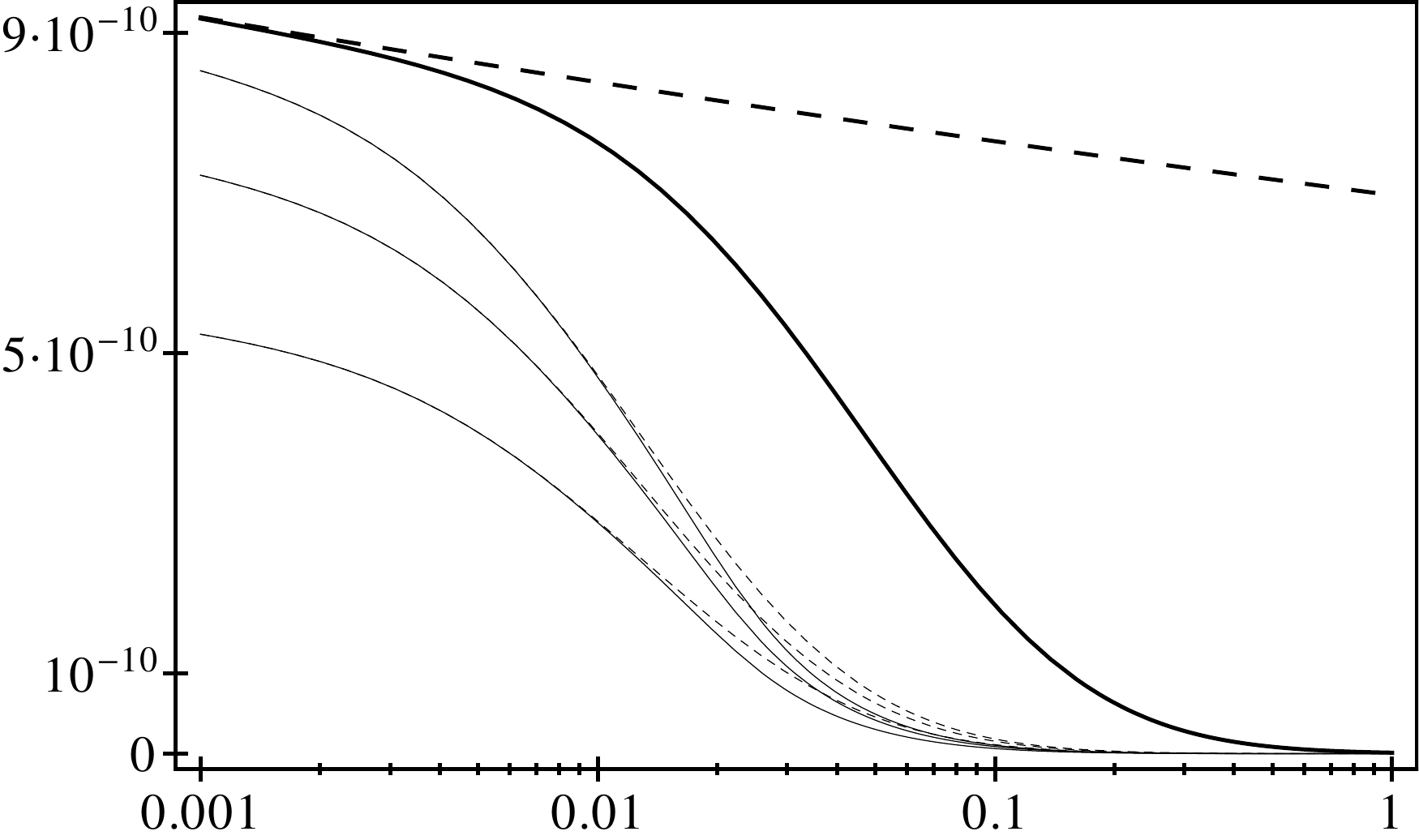}};
      
\node[anchor=west] at (-3,-1.3) (description) {\Large \textbf{$\Lambda$CDM}};
\node[anchor=west] at (1.7,0.5) (description) {\Large \textbf{CDM}};
\node[anchor=west] at (-4.8,0.2) (description) {\Large \textbf{z=0}};
\node[anchor=west] at (-4.8,1.5) (description) {\Large \textbf{z=0.5}};
\node[anchor=west] at (-4.8,2.75) (description) {\Large \textbf{z=1.5}};

\node[anchor=west,rotate=-8] at (0.8,3.2) (description) {\Large \textbf{Primordial Spectrum}};

\node[anchor=west,rotate=90] at (-7.2,-0.5) (description) {\Large $\Pcal_{\psi}(\textbf{k})$};

\node[anchor=west] at (-0.7,-4.5) (description) {\Large $\textbf{k} ~[h ~{\rm Mpc}^{-1}]$};
\end{tikzpicture}
\caption[Linear power spectra $\Pcal_{\L}(k)$ in CDM and $\La$CDM, with baryonic and non-baryonic $T(k)$]{A comparison of the primordial inflationary spectrum (long-dashed curve) with the spectrum of the CDM model neglecting baryons (thick solid curve) and of a $\Lambda$CDM model (thin solid curves), at various values of $z$. The dotted curves for the  $\La$CDM case describe the spectrum obtained by neglecting the baryon contribution (hence without taking into account the Silk damping effect).}
\label{Fig2}
\end{figure}

We need now to insert the explicit form of the power spectrum. Considering the general solution of Eq. (\ref{SolutionGeneralPsi}) we can also re-express the $z$-dependence of the power spectrum, in the linear regime, as follows\,:
\beq
\label{PsiPz}
\Pcal_\psi (k,z) = \left(\frac{3}{5}\right)^2 \Delta_{\cal R}^2 T^2(k) \left(\frac{g(z)}{g_{\infty}}\right)^2 ~~,
\eeq
where the previous CDM result of Eq. (\ref{PsiPch6}), based on the transfer function $T(k)$ given in \cite{Eisenstein:1997ik}, is modified by the presence of the  factor $g(z)/g_\infty$ originating from the time dependence of the gravitational perturbations. Another modification with respect to the CDM result, implicitly contained into the transfer function $T(k)$, concerns the different numerical value of the equality scale $k_{\rm eq}$ (which now turns out to be lower because of the lower value of $\Omega_{m0}$). The effects of such modifications are illustrated in Fig. \ref{Fig2} by comparing the $\Lambda$CDM  spectrum of Eq. (\ref{PsiPz}), at different values of $z$, with two $z$-independent spectra\,: the primordial spectrum of scalar perturbations $(3/5)^2  \Delta_{\cal R}^2$, and the ``transferred" spectrum of the CDM model, introduced in Sec. \ref{Ch6Sec11}. Notice that, as expected, the $\Lambda$CDM and CDM spectra tend to coincide at large enough values of $z$. In the $\La$CDM case the solid curves are obtained with a transfer function which takes into account the presence of baryonic  matter, while the dashed curves correspond to the transfer function $T(k)=T_0(k)$ of Eq. \rref{EH97} (without baryons). Here, however, we are always using a spectrum  evaluated in the linear regime.

As illustrated in Fig. \ref{Fig2}, the effect of neglecting the baryonic fraction of $\Omega_{m0}$ (and thus the associated \gls{Silk damping} effect) may lead to an overestimation up to 40\%  of the corresponding transfer function for scalar perturbations, in the range $k \gaq 0.01 ~ h \,{\rm Mpc}^{-1}$ (see  \cite{Eisenstein:1997ik}).  In order to take into account this baryonic contribution we have to replace the value of $q$ used in the previous section, i.e. $q  = k/(13.41 \, k_{\rm eq})$, with the more accurate value given by \cite{Eisenstein:1997ik}\,:
\bea
q&=&\frac{k \times \Mpc}{h \Gamma} ~~,\\
\Gamma&=&\Omega_{m0}h\left(\alpha_{\Gamma}+\frac{1-\alpha_{\Gamma}}{1+(0.43 k s)^4}\right) ~~,\\
\alpha_{\Gamma}&=&1-0.328\ln(431 \Omega_{m0}h^2)\frac{\Omega_{b0}}{\Omega_{m0}}+0.38 \ln(22.3 \Omega_{m0} h^2)\left(\frac{\Omega_{b0}}{\Omega_{m0}}\right)^2 ~~,\\
s&=&\frac{44.5 \ln(9.83/\Omega_{m0}h^2)}{\sqrt{1+10(\Omega_{b0}h^2)^{3/4}}} \,{\rm Mpc} ~~.
\eea
Here $\Omega_{b0}$ is the baryon density parameter, $s$ is the \gls{sound horizon} and $\Gamma$ is the $k$-dependent \gls{effective shape parameter}.

We have compared the above \gls{transfer function} to the one which includes baryon acoustic oscillations (\gls{BAO}) \cite{Eisenstein:1997ik}, and to a transfer function calculated numerically by using the so-called ``code for anisotropies in the microwave background" (CAMB) \cite{Lewis:1999bs}. 
We have checked, in particular,  that the above simple form of transfer function is accurate to within a few percent compared to the one calculated numerically by CAMB, for all scales of interest. In addition, the effect of including BAO only produces oscillations of the spectrum around the above value. Since we are considering here integrals over a large range of $k$, the presence of BAO has a negligible effect on our final results, and will be neglected in the rest of the paper. 

We are now in the position of computing the fractional corrections to the averaged flux variable in a perturbed  $\Lambda$CDM geometry. As discussed before, there are complicated average integrals which can be performed only by using some approximations. Once this is done, the remaining integration over $k$ can be done numerically,  exactly as in the case of the CDM model.
In a $\Lambda$CDM context we may generally expect smaller corrections to the averaged flux, due to the fact that the perturbation spectrum $\Pcal_\psi$ is  suppressed by the presence of $g(z)$. In addition (and as already stressed) the transfer function  \cite{Eisenstein:1997ik} turns out to be suppressed, at large $k$, because of a smaller value of the parameter  $k_{\rm eq}$ (see Eq. (\ref{EH97})). These expectations are fully confirmed by an explicit numerical computation of $|f_\Phi|$, which we have performed with and without the inclusion of the baryon contributions into the transfer function $T(k)$. 

\begin{figure}[t]
\centering
\includegraphics[width=8.3cm]{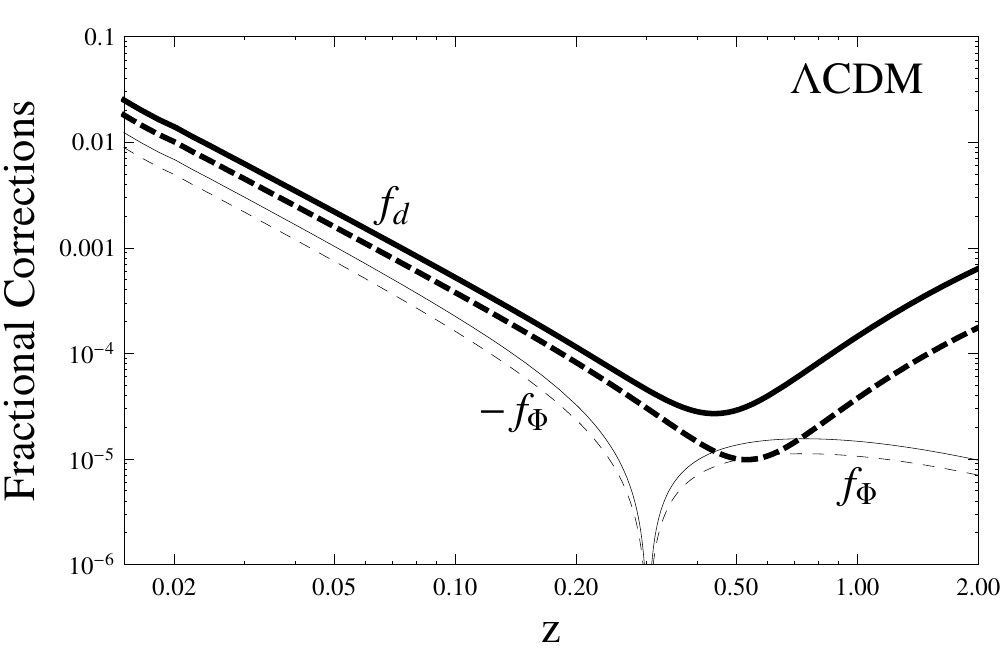}
\includegraphics[width=8.3cm]{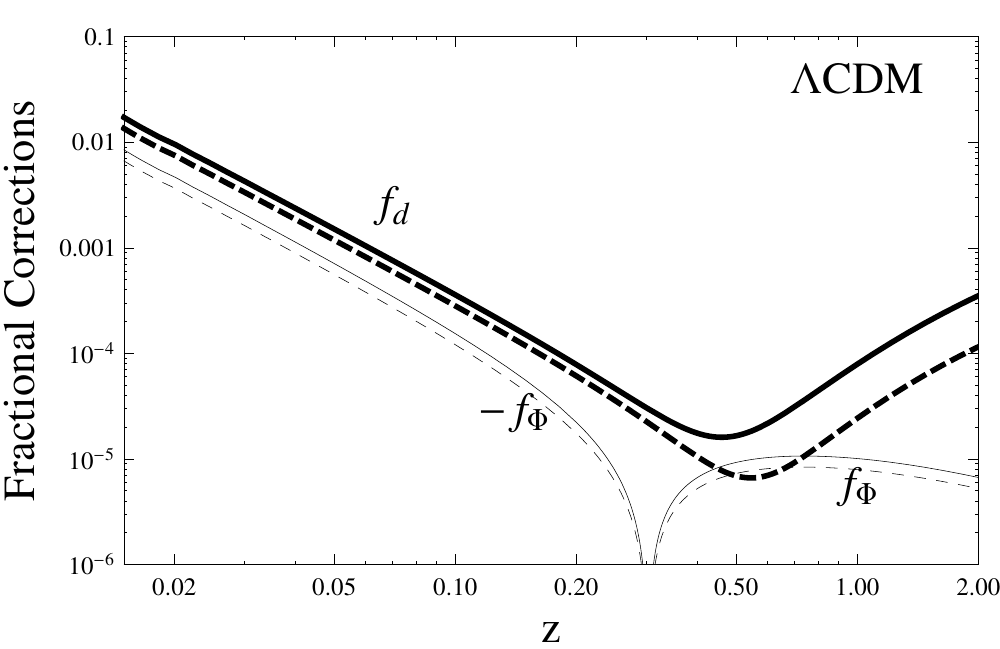}
\centering
\caption[Flux $f_\Phi$ and distance $f_d$ corrections in $\La$CDM, using $\Pcal_{\rm L}(k)$ with/without baryon $T(k)$]{The fractional correction to the flux $f_\Phi$ of Eq. (\ref{7a}) (thin curves) is plotted together with the fractional correction to the luminosity distance $f_d$  of Eq. (\ref{13}) (thick curves), for a $\La$CDM model with $\Om_{\La 0}=0.73$. We have used two different cutoff values\,: $k_{UV}=0.1 \, {\rm Mpc}^{-1}$ (dashed curves) and $k_{UV}=1 \, {\rm Mpc}^{-1}$ (solid curves). \emph{Left}\,: results obtained with a linear spectrum without baryon contributions. \emph{Right}\,: effects of including baryons (we have used, in particular, $\Omega_{b0}=0.046$).}
\label{Fig3}
\end{figure}

The results of such a computation are illustrated in Fig. \ref{Fig3}, and a comparison with Fig. \ref{f1} clearly shows that $|f_\Phi|$ is smaller in the  $\Lambda$CDM case than in the CDM case, and further (slightly) depressed when we take into account the presence of a small  fraction of baryonic matter (see the curves presented in the right panel). In any case, the small values of $|f_\Phi|$ at relatively large $z$, for a realistic $\Lambda$CDM scenario,  lead us to conclude that the averaged flux is a particularly appropriate quantity for extracting from the observational data the ``true" cosmological parameters. As we will discuss now, the situation is somewhat different for other functions of $d_L$. 

\subsection{Second-order corrections to other observables and dispersions}
\label{Ch6Sec32}

Let us now consider other observables, beyond the flux, to see  how the impact of the inhomogeneities may change. We will treat, in particular, the two important examples extensively presented in Sec. \ref{Ch5Sec33}, namely the luminosity distance $d_L$ and the distance modulus $\mu$. The fractional corrections to their averages (see Eqs. (\ref{2}--\ref{14})) are qualitatively different from those of the averaged flux (represented by $f_\Phi$), because of the presence of  extra  contributions, unavoidable for any non-linear function of the flux and  proportional to the square of the first order fluctuation $(\Phi_1/\Phi_0)^2$. 

For a better understanding of such contributions let us start with the results obtained in Sec. \ref{Ch5Sec32}, concerning the second-order perturbative expansion of the luminosity distance, $d_L = d_{L}^{(0)} + d_{L}^{(1)} + d_{L}^{(2)}$ (see in particular Eqs. (\ref{215}) and (\ref{Ideltarel})). Using those results we obtain\,:
\beq
\frac{\Phi_1}{\Phi_0}=-2 \frac{d_{L}^{(1)}}{d_{L}^{(0)}} = - {\cal I}_1 + ({\rm t.d.})^{(1)} = 
2\left(-\Xi_s J+\frac{Q_s}{\Delta \eta}+\psi_s+J_2^{(1)}\right) ~~.
\label{OBcorrection}
\eeq
In order to determine the leading corrections we first notice that, by applying the procedure of Sec. \ref{Ch6Sec21}, and computing the averages for the CDM case, we obtain, to leading order,
\beq
\overline{\left\langle \left(\frac{\Phi_1}{\Phi_0}\right)^2 \right\rangle}_{ L}=4\left\{\overline{\left\langle \left(J_2^{(1)}\right)^2 \right\rangle}+
\Xi_s^2 \left[\overline{\langle  ([\partial_r P]_s)^2  \rangle}+\overline{\langle  ([\partial_r P]_o)^2  \rangle}\right]\right\} ~~.
\label{615}
\eeq
Working in the context of a $\Lambda$CDM model, considering only these leading terms, and limiting ourselves to the linear regime, we find that the terms multiplying $\Xi_s^2 $ on the right hand side of the above equation can be calculated without approximations (as seen in the previous subsection). Also, we find that their contribution is controlled by the spectral factor $k^2 {\cal P}_\psi(k, \eta_o)$ and they correspond to the \gls{peculiar velocity} terms. The first term, on the contrary, is due to the so-called ``\gls{lensing} effect", dominates at large $z$ (as already shown in Sec. \ref{Ch6Sec12} for a CDM model), and has leading spectral contributions of the type $k^3 {\cal P}_\psi(k, \eta_o)$. For such term, however, the integrals over time cannot be factorized with respect to the $k$ integrals, and we must use the approximation already introduced in the previous subsection (namely, we have to replace in the integrands $\psi(\eta',\eta_o-\eta', \tilde{\theta}^a)$  with $\psi(\eta_s,\eta_o-\eta', \tilde{\theta}^a)$). The full result for the spectral coefficient can then be finally written as follows\,:
\beq
\Ccal(\left(\Phi_1/\Phi_0\right)^2_L)\simeq \frac{4}{3}\Xi_s^2 \left(\ti f_s^2 + \ti f_o^2\right)k^2 +
\frac{4}{15} \left[\frac{g(\eta_s)}{g(\eta_o)}\right]^2 \Da \eta^3 k^3\,{\rm SinInt}(k \Da \eta) ~~.
\eeq

Because of the new term (due to lensing) affecting the average of $\mu$ and $d_L$ -- but not of the flux $\Phi$ -- we may expect larger fractional corrections for these variables, as well as for other functions of $\Phi$, at higher redshifts. This is indeed confirmed by the plots presented in Fig. \ref{Fig3}, reporting the results of an explicit numerical integration and comparing, in particular,  the value of $f_d$ with the absolute value of $f_\Phi$. We obtain $|f_\Phi| \ll f_d$, at large values of $z$ where the lensing term dominates, both in the presence and in the absence of the baryon contribution to the total energy density. It should be stressed, however, that also the new $k^3$-enhanced contributions are free from IR and UV divergences, at least for the class of models we are considering.

\begin{figure}[t]
\centering
\includegraphics[width=8.4cm]{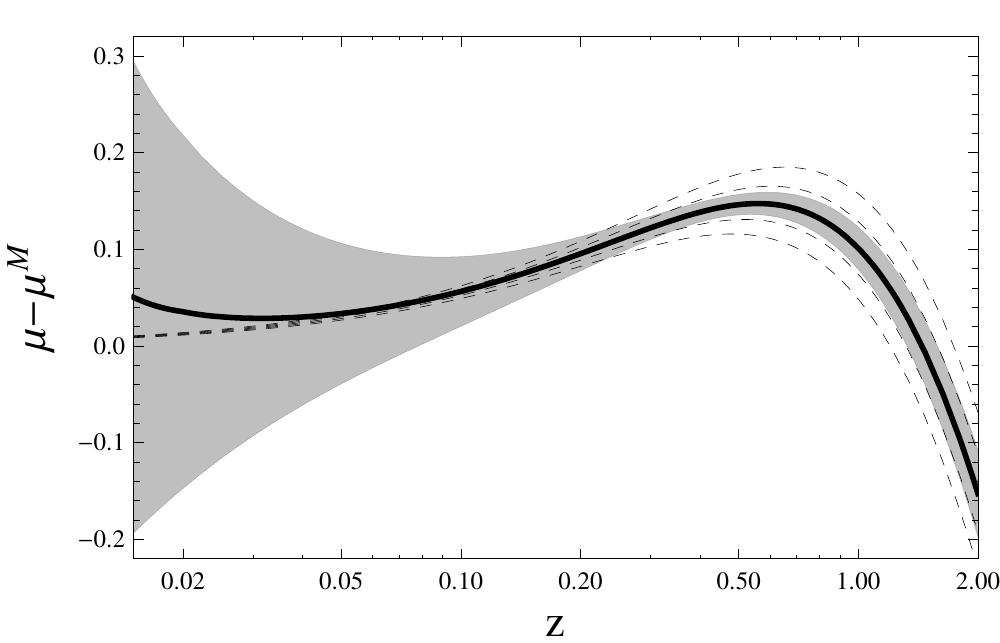}
\includegraphics[width=8.4cm]{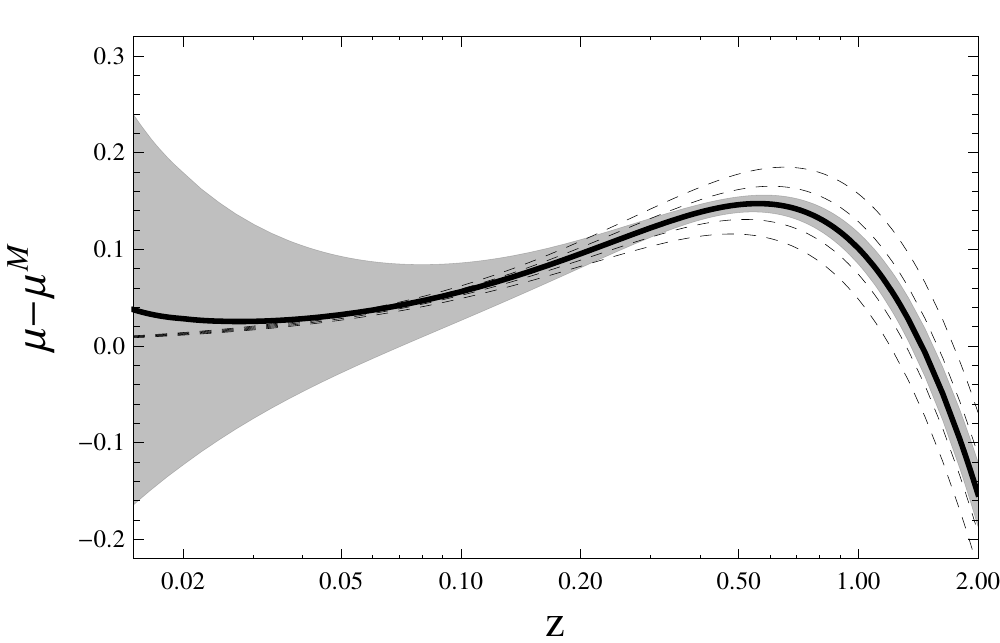}
\centering
\caption[Average $\overline{\langle \mu \rangle} -\mu^M$ and dispersion, in CDM with $\Pcal_{\L}(k)$ up to $k_{UV} = 1 \, h ~\Mpc^{-1}$]{The averaged distance modulus $\overline{\langle \mu \rangle} -\mu^M$ of Eq. (\ref{14}) (thick solid curve), and its dispersion of Eq. (\ref{15}) (shaded region) are computed for $\Om_{\La 0}=0.73$ and compared with the homogeneous value  for the unperturbed $\La$CDM models with, from bottom to top,  
$ \Om_{\La 0}= 0.69$, $0.71$, $0.73$, $0.75$, $0.77$ (dashed curves). We have used  $k_{UV}=1\,\rm{Mpc}^{-1}$. The \emph{left} panel shows the results obtained with a linear spectrum without baryon contributions. The \emph{right} panel illustrates the effects of including baryons, with $\Omega_{b0}=0.046$.}
\label{Fig4}
\end{figure}

Let us now discuss to what extent the enhanced corrections due to the square of the first-order flux fluctuation can affect the determination of the dark-energy parameters, if quantities other than the flux are used to fit the observational data. To this purpose we may consider the much used (average of the) distance modulus given in Eq. (\ref{14}), referring it to a homogeneous \gls{Milne} model with $\mu^M= 5 \log_{10} [(2+z)z/(2H_0)]$.  
Considering Eqs. (\ref{14}) and (\ref{15}), where the averaged value of the distance modulus and its dispersion are given in function
of $f_\Phi$ and of $\overline{\left\langle \left({\Phi_1}/{\Phi_0}\right)^2 \right\rangle}$, we have investigated the magnitude of the effect in this case. 
In particular, in Fig. \ref{Fig4} we have compared the averaged value $\overline{\langle \mu \rangle} -\mu^M$ with the corresponding expression for  homogeneous $\La$CDM models with different values of $\Om_{\La 0}$. We have also illustrated the expected dispersion around the averaged result, represented by the dispersion previously reported in Eq. (\ref{15}) (and already computed in CDM).

We have found that the  given inhomogeneities, on the average, may affect the determination of $\Om_{\La 0}$ obtained from the measure of the distance modulus, at large $z$, only at the third decimal figure (at least if the spectral contributions are computed in the linear  regime). As we can see from Fig. \ref{Fig4}, the curves for $\overline{\langle \mu \rangle}$ and for the corresponding unperturbed value $\mu^{\rm FLRW}$ (with the same $\Om_{\La 0}$) practically coincide at large enough $z$. It should be stressed, also, that the dispersion on the distance modulus computed from Eq. (\ref{15}) reaches,  at large redshift, a value which is comparable with a change of about 2\% in the dark energy parameter $\Omega_{\Lambda 0}$ (cf. Sec. \ref{Ch6Sec12}, considering however that baryonic effects have now been added in the transfer function). We shall see in the next section that this effect is enhanced by the use of  non-linear power spectra. 

\section{The $\Lambda$CDM model\,: power spectrum in the non-linear regime}
\label{Ch6Sec4}

The linear spectra considered so far are sufficiently accurate only up to scales of $\sim 0.1 \, h \, {\rm Mpc}^{-1}$. 
If we want to better study the effect of shorter-scale inhomogeneities on our light-cone averages we need to go beyond such linear approximation, taking into consideration the non-linear  evolution of the gravitational perturbations. This can be done by using the so-called ``HaloFit'' model, which is known to reproduce quite accurately the results of cosmological $N$-body simulations. In particular, the \gls{HaloFit model} of \cite{Smith:2002dz} and its recent upgrade of  \cite{Takahashi:2012em} provide an accurate fitting formula for the power spectrum up to a wavenumber $k \simeq 30 \, h \,{\rm Mpc}^{-1}$. 

In our previous section, Sec. \ref{Ch6Sec3}, the analysis was limited to $k<1 \,{\rm Mpc}^{-1}$, but with the use of the non-linear power spectra we can now extend our analysis up to the maximum scale of the mentioned HaloFit models. The main difficulty with such an extension is that the time (i.e. $z$) dependence of the spectrum becomes more involved than in the linear case, since different scales no longer evolve independently. Hence, the need for introducing  approximations in performing the integrals becomes even more essential in this case. 

Let us start by recalling a few details of the HaloFit model. A more complete presentation is given in Appendix \ref{AppASec5}. The fractional \gls{density variance} per unit $\ln k$ is represented by the variable $\Delta^2(k)$,  defined by \cite{Smith:2002dz}\,:
\beq
\sigma^2 \equiv \overline{\delta(x)\delta(x)} = \int \frac{d^3k}{(2 \pi)^3} |\delta_{\vec{k}}|^2 = \int \Delta^2(k) ~ d\ln k ~~,
\eeq
which implies
\beq
\label{DeltadeltaResults}
\Delta^2(k) = \frac{k^3}{2 \pi^2} |\delta_k|^2 ~~,
\eeq
and where $\delta(x)= \delta \rho(x)/\rho(t)$ is the fractional density perturbation of the gravitational sources, and $\da_k$ is the associated Fourier component. 
On the other hand, the power spectrum of scalar perturbations, $\Pcal_\psi(k,z)$, is related to $\Delta^2(k,z)$ by the Poisson equation, holding at both the linear (L) and non-linear (NL) level \cite{PeterUzan}\,:
\beq
\label{PE}
\Pcal_\psi^{\rm{L,NL}}(k,z) = \frac{9}{4} \frac{\Om_{m0}^2 \Hcal_0^4}{k^4} (1+z)^2 \Delta_{\rm{L,NL}}^2(k,z) ~~.
\eeq
The linear part of the spectrum is used to introduce the normalization equation, defining the non-linearity length scale $k_\sigma^{-1}(z)$, as follows\,:
\beq
\sigma^2(k_\sigma^{-1}) \equiv \int \Delta^2_{\rm{L}}(k,z) \, \exp(-(k/k_\sigma)^2)\;d\ln k \equiv 1 ~~.
\eeq
It is then obvious that the \gls{non-linearity scale} $k_{\sigma}^{-1}$ is redshift-dependent. Since the non-linear power spectra obtained from the models \cite{Smith:2002dz} and \cite{Takahashi:2012em} are using such a scale, they will be characterized by an implicit $z$-dependence which is more complicated than the one following from the usual growth factor of Eq. (\ref{gformula}), and which cannot be factorized. 

Besides $k_{\sigma}$, the linear spectrum also determines two additional  parameters, important for the construction of the HaloFit model\,:
the \gls{effective spectral index} $n_{\rm eff}$ and the parameter $C$, controlling  the \gls{curvature of the spectral index} at the scale $k_{\sigma}^{-1}$. They are defined by\,:
\beq
3+n_{\rm eff} \equiv -\left.{d \ln \sigma^2(R) 
\over d \ln R}\right|_{R=k_\sigma^{-1}} ~~~~~,~~~~~
C \equiv - \left.{d^2 \ln \sigma^2(R) \over d \ln R^2}
\right|_{R=k_\sigma^{-1}} ~~.
\eeq
Once the values of $k_\sigma$, $n_{\rm eff}$ and $C$ are determined, they can be inserted into a given HaloFit model \cite{Smith:2002dz,Takahashi:2012em}  to produce the non-linear power spectrum $\Delta_{\rm{NL}}^2(k,z)$. One finally goes back to $\Pcal_\psi^{\rm{NL}}(k,z)$ using again the Poisson equation (\ref{PE}), as already explained in Sec. \ref{Ch1Sec43}.

\begin{figure}[t]
\centering
\begin{tikzpicture}
\node (label) at (0,0){\includegraphics[width=14cm]{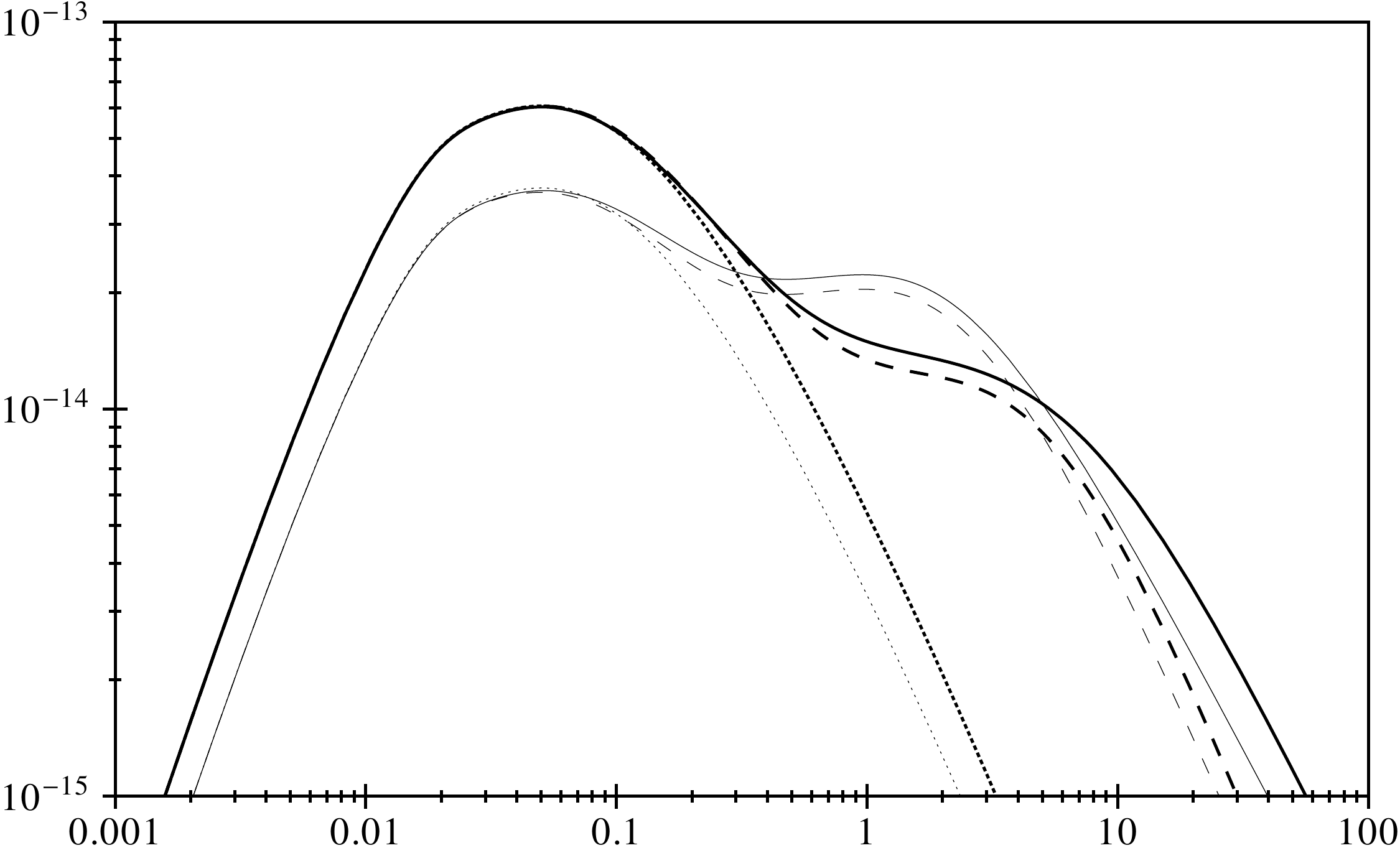}};

\node[anchor=west] at (-5.5,1) (description) {\Large $\textbf{z=1.5}$};
\node[anchor=west] at (-4,-2) (description) {\Large $\textbf{z=0}$};

\node[anchor=west] at (3.8,0.8) (description) {\Large $\textbf{k}^2 ~ \Pcal_{\psi}^{\NL}(\textbf{k})$};
\node[anchor=west] at (-0.5,-2.5) (description) {\Large $\textbf{k}^2 ~ \Pcal_{\psi}^{\L}(\textbf{k})$};

\node[anchor=west,rotate=90] at (-7.5,-2.5) (description) {\Large $\textbf{k}^2 ~ \Pcal_{\psi}^{\L,\NL}(\textbf{k}) ~~[h^2 ~ {\rm Mpc}^{-2} ]$};

\node[anchor=west] at (-1,-4.7) (description) {\Large $\textbf{k} ~[h ~{\rm Mpc}^{-1}]$};
\end{tikzpicture}
\caption[Linear and non-linear spectra $k^2 ~ \Pcal_\psi^{\rm{L,NL}}$ within the HaloFit model, for $z=0 ~,~ 1.5$]{The linear spectrum $\Pcal_\psi^{\rm{L}}$ (dotted curves) and the non-linear spectrum $\Pcal_\psi^{\rm{NL}}$ for the HaloFit model of \cite{Smith:2002dz} (dashed curves) and of \cite{Takahashi:2012em} (solid curves). In all three cases the spectrum is multiplied by $k^2$ (for graphical convenience) and is given for $z=0$ (thin curves) and $z=1.5$ (thick curves). We have included  baryons with $\Om_{b0}=0.046$.}
\label{Fig5}
\end{figure}
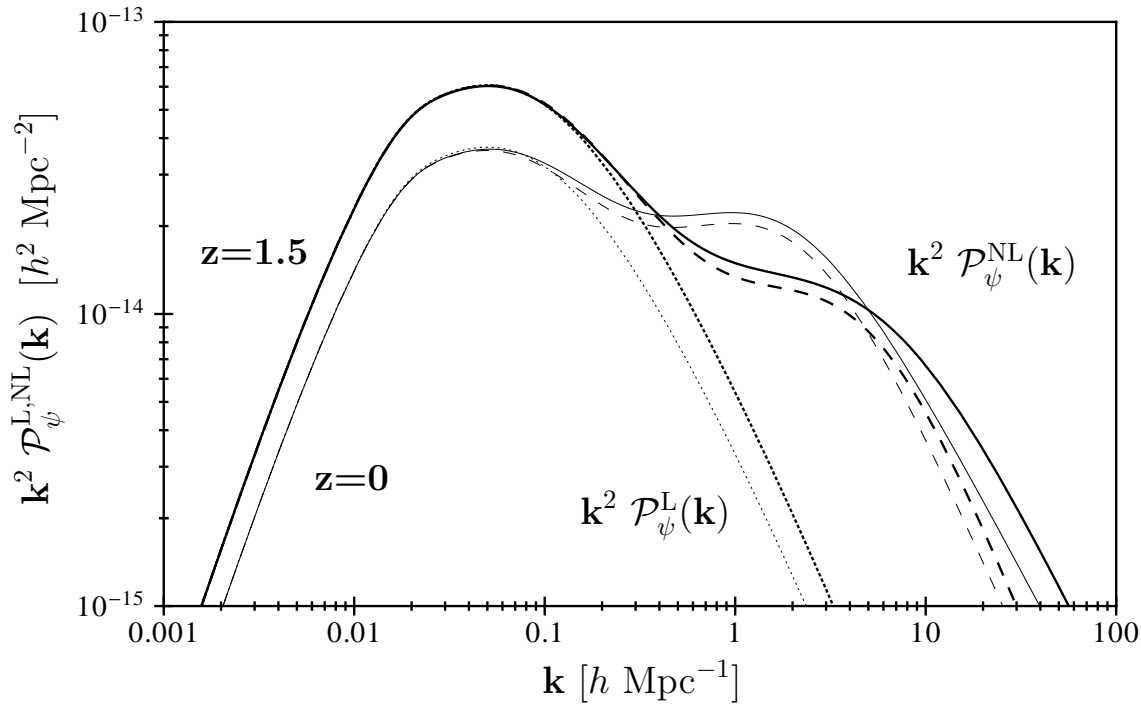

The non-linear spectra obtained in this way, and based on the previous linear spectrum with baryon contributions, are illustrated in Fig. \ref{Fig5} for the two HaloFit models \cite{Smith:2002dz} and \cite{Takahashi:2012em}. The results are compared with the spectrum of the linear regime, for different values of the redshift ($z=0$ and $z=1.5$). We can see that the spectra intersect each other in the non-linear regime ($k\gaq 0.1 \, h \, {\rm Mpc}^{-1}$), as a result of the intricate redshift dependence. We  can also observe that the two HaloFit models lead to the same result at low values of $k$ (including baryons, but without BAO).

\subsection{Numerical results and comparison with the linear regime}
\label{Ch6Sec41}

In order to evaluate our  integrals using the non-linear power spectra we need  to face the additional (already mentioned) problem that the non-linear spectra -- unlike the linear ones -- cannot be factorized  as a function of $k$ times a function of $z$. In that case the full two-dimensional integration is highly non-trivial, and we have thus exploited a further approximation. In the presence of time-integrals of the mode functions (like those appearing, for instance, in the leading terms of $\overline{\langle \Ical_{1,1} \rangle}$), we have parametrized the non-linear power spectrum in a factorized form as follows\,:
\beq
\Pcal_\psi^{\rm{NL}}(k, z) = \frac{g^2(z)}{g^2(z^*)} \Pcal_\psi^{\rm{NL}}(k, z^*) ~~,
\label{76}
\eeq
and we have chosen $z^*=z_s/2$ to try to minimize the error. 

Let us briefly comment on the validity of the approximation we have introduced. Our analysis being focussed on the redshift range $0.015 \leq z \leq 2$,  the above approximation is most inadequate only when we consider $z_s=2$ (i.e. $z^*=1$), and our mode functions are evaluated inside the integrals at the values of $z$ most distant from $z^*$ (namely, $z=0.015$ and $z=2$). In that case we are led to underestimate the spectrum by about a $40\%$ factor for $z=0.015$, and to overestimate it by about a $80\%$ factor for $z=2$. However, this only occurs in two  narrow bands of $k$ centered around the values $k=1 \, h \,{\rm Mpc}^{-1}$  and $k=2 \, h \,{\rm Mpc}^{-1}$, while, outside these bands, our approximation is good. Since these errors are limited to only  a part of the region of integration, both in $z$ and in $k$, we can estimate an overall accuracy at the $10\%$ level, at least,  for the results given by the adopted approximation. We have also checked the sensitivity of the numerical results to changes in $z^*$, such as $z^*=z_s$, and checked that our final results are only weakly dependent (typically at the $1\%$ level) on the  choice of $z^*$.

\begin{figure}[ht!]
\centering
\includegraphics[width=13cm]{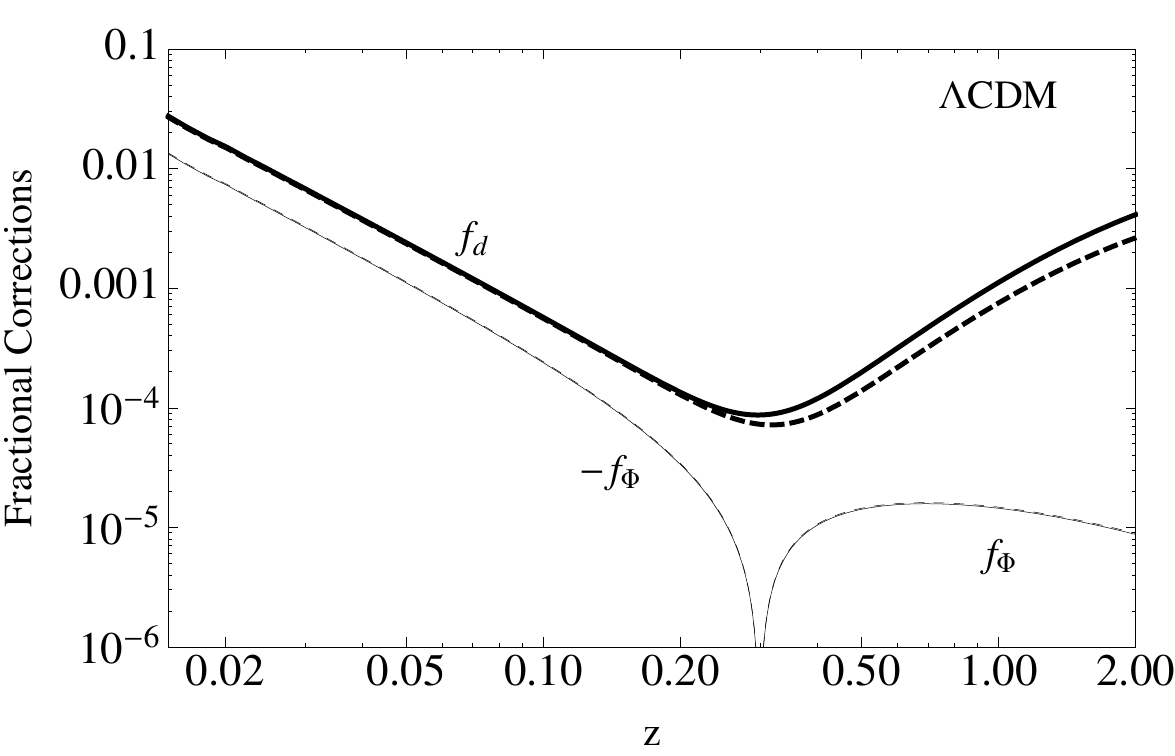}
\centering
\caption[Flux $f_\Phi$ and distance $f_d$ corrections in $\La$CDM, using $\Pcal_{\NL}(k)$ up to $k_{UV}=10 ~,~ 30 \, h \, {\rm Mpc}^{-1}$]{The fractional correction to the flux ($f_\Phi$, thin curves) and to the luminosity distance ($f_d$, thick curves), for a perturbed $\La$CDM model with $\Omega_{\Lambda 0}=0.73$. Unlike in Fig. \ref{Fig3}, we have taken into account the non-linear contributions to the power spectrum given by the HaloFit model of \cite{Takahashi:2012em} (including baryons), and we have used the following cutoff values\,: $k_{UV}=10 \, h \,{\rm Mpc}^{-1}$ (dashed curves) and $k_{UV}=30 \, h \,{\rm Mpc}^{-1}$ (solid  curves).}
\label{Fig6}
\end{figure}

\begin{figure}[ht!]
\centering
\includegraphics[width=8.4cm]{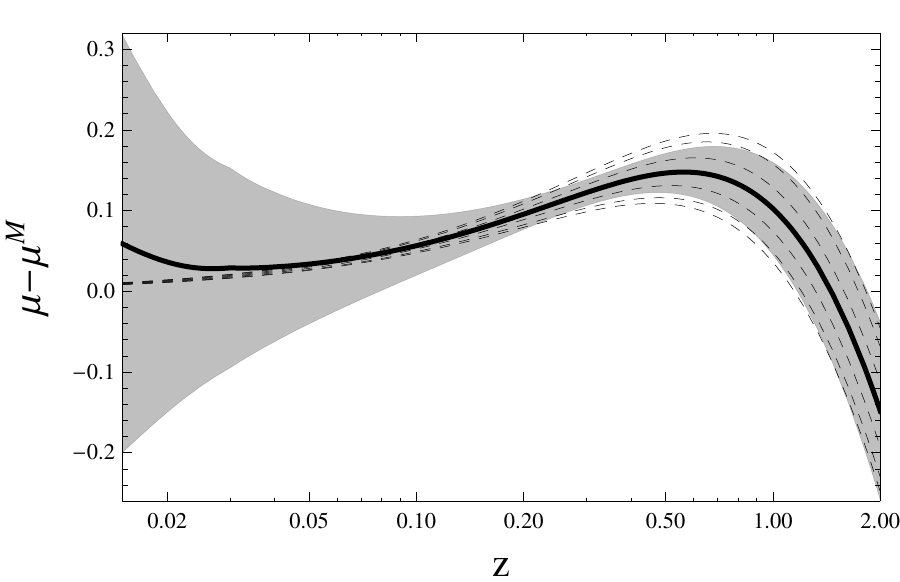}
\includegraphics[width=8.4cm]{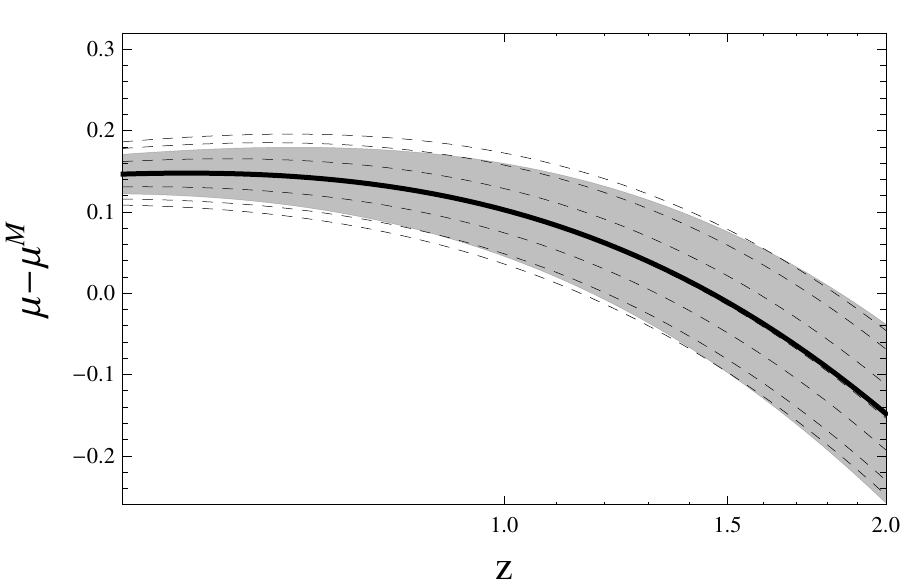}
\centering
\caption[Average $\overline{\langle \mu \rangle} -\mu^M$ and dispersion, in $\La$CDM with $\Pcal_{\NL}(k)$ up to $k_{UV} = 30 \, h ~\Mpc^{-1}$]{The averaged distance modulus $\overline{\langle \mu \rangle} -\mu^M$ of Eq. (\ref{14}) (thick solid curve), and its dispersion of Eq. (\ref{15}) (shaded region), for a perturbed $\La$CDM model with $\Omega_{\Lambda 0}=0.73$. Unlike Fig. \ref{Fig4}, we have taken into account the non-linear contributions to the power spectrum given by the HaloFit model of \cite{Takahashi:2012em} (including baryons), and used the cut-off $k_{UV}=30 \, h ~\Mpc^{-1}$. The averaged results are compared with the homogeneous values of $\mu$ predicted by  unperturbed $\La$CDM models with  (from bottom to top) $\Omega_{\Lambda 0}= 0.68$, $0.69$, $0.71$, $0.73$, $0.75$, $0.77$, $0.78$ (dashed curves). The right panel simply provides a zoom of the same curves, plotted in the smaller redshift range $0.5 \leq z \leq 2$.}
\label{Fig7}
\end{figure}

Once we have established the range of validity of the parametrization (\ref{76}), we proceed in evaluating  the $z$ (or $\eta$) integrals as we did in Sec. \ref{Ch6Sec3}. The final results of this procedure are illustrated by the curves plotted in Figs. \ref{Fig6} and \ref{Fig7}, computed with the non-linear power spectrum following from  the HaloFit model of \cite{Takahashi:2012em} and including baryon contributions\footnote{The two HaloFit models quoted above give similar results, and we have chosen to present here, for simplicity, only those obtained with the parametrization of \cite{Takahashi:2012em}.}. As illustrated for instance in Fig. \ref{Fig6}, the fractional correction to the \gls{luminosity distance} $d_L$ turns out to be of order of a few parts in $10^{-3}$ around $z=2$, and smaller in the rest of the intermediate redshift range relevant for cosmic acceleration. On the other hand, in the same redshift range, the fractional correction to the flux is about two orders of magnitude smaller.

Comparing with the results obtained using the linear spectrum  (see Figs. \ref{Fig3}, \ref{Fig4}), we can see that taking into account the non-linearity distortions (and using higher cut-off values) enhances the backreaction effects on the considered functions of the luminosity distance, but not enough to reach a (currently) observable level. We see in Fig. \ref{Fig7}, especially on the right panel, that the average is very weakly affected by inhomogeneities. On the other hand, the dispersion, already large in the linear case, is further enhanced when non-linearities are included. In particular, the dispersion on $\mu$ due to inhomogeneities is of order $10\%$ around $z=2$, which implies that the predictions of homogeneous models with $\Omega_{\Lambda 0}$ ranging from $0.68$ to $0.78$ lie inside one standard deviation with respect to the averaged predictions of a perturbed (inhomogeneous) model with $\Omega_{\Lambda 0}=0.73$.

\subsection{Comparing theory and observations via the intrinsic dispersion of data}
\label{Ch6Sec42}

\begin{figure}[t]
\centering
\includegraphics[width=12cm]{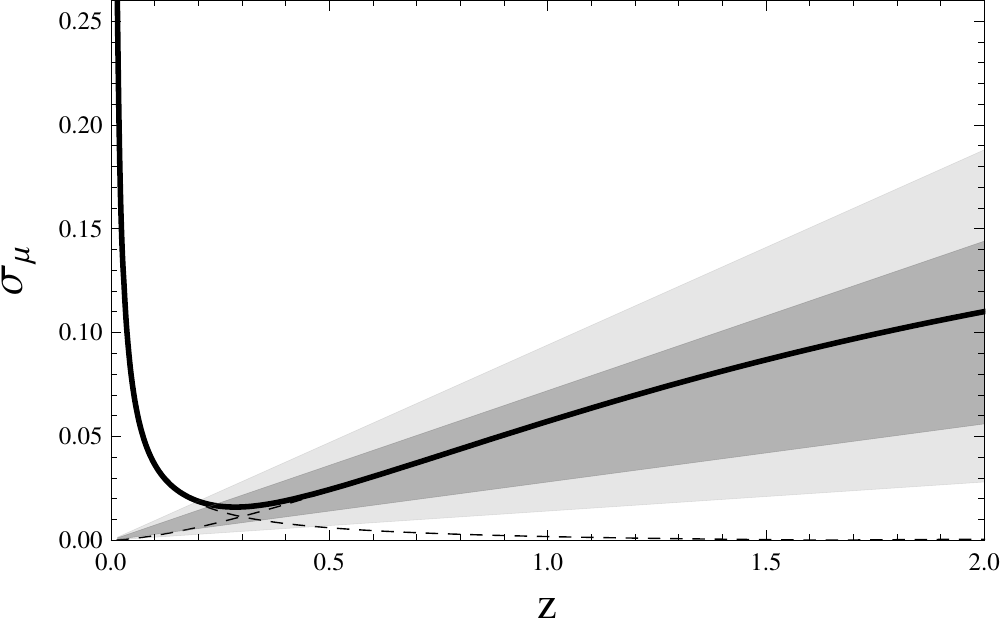}
\centering
\caption[The $z$-dependence of the total dispersion $\sg_\mu$ compared with experimental estimates]{The $z$-dependence of the total dispersion $\sg_\mu$ is illustrated by the thick solid curve, and it is separated into its ``Doppler" part (dashed curve dominant at low $z$)  and ``lensing" part (dashed curve dominant at large $z$). The slope of the dispersion in the lensing-dominated regime is compared with the experimental estimates of  Kronborg \etal \cite{Kronborg:2010uj} (dark shaded area), and  of J\"onsson \etal \cite{Jonsson:2010wx} (light shaded area).}
\label{Fig8}
\end{figure}

Let us now consider in more detail our prediction for the dispersion $\sigma_{\mu}$ induced by the presence of the inflationary perturbation background, and compare it with the intrinsic dispersion of the distance modulus inferred from SNe Ia data. Our results for  the dispersion are already implicitly contained in Fig. \ref{Fig7}  but, for the sake of clarity,  we have  separately plotted our value of $\sg_\mu$  in Fig. \ref{Fig8}, where the thick solid curve represents the value of $\sg_\mu$ obtained from Eq. (\ref{15}), and plotted as a function of $z$.  We can see from the figure that $\sigma_{\mu}$ has a characteristic $z$-dependence, with a minimal value of about $0.016$ reached around $z = 0.285$. Also,  the total dispersion of Eq. (\ref{15}) nicely interpolates between the leading Doppler contribution obtained from the two last terms of Eq. (\ref{615}) (represented  by the dashed curve approaching zero at large $z$), and the leading lensing contribution obtained from  the first term of Eq. (\ref{615}) (represented  by the dashed curve approaching zero at small $z$). 

The total variance $\sigma_{\mu}^{\rm obs}$ associated with the observational data,  on the other hand, can be decomposed in general  as follows (see e.g. \cite{March:2011xa,Conley:2011ku,Vishwakarma:2010nc})\,:
\beq
(\sigma_{\mu}^{\rm obs})^2 =  (\sigma_{\mu}^{\rm fit})^2 + (\sigma_{\mu}^{z})^2  + (\sigma_{\mu}^{\rm int})^2 ~~.
\eeq
Here $\sigma_{\mu}^{\rm fit}$ is  the statistical uncertainty due, for instance, to the method adopted for fitting the light curve (e.g. the so-called SALT-II method \cite{Guy:2007dv}), but also to the uncertainty in the modeling of the supernova process. The term $\sigma_{\mu}^{z}$ represents instead the uncertainty in redshift due to the peculiar velocity of the supernova  as well as to the precision of spectroscopic measurements. 
Finally, $\sigma_{\mu}^{\rm int}$ is an unknown phenomenological quantity, needed to account for the remaining dispersion of the data with respect to the chosen homogeneous model. This part of the dispersion can be subsequently redefined whenever we are able to estimate some of the possible contributions it contains. The contribution we are mainly interested in here is the one originating from the \gls{lensing} effect, which is dominant at large redshift. We can thus write, at large $z$\,:
\beq
(\sigma_{\mu}^{\rm int})^2 = (\widehat{\sigma_{\mu}^{\rm int}})^2 + (\sigma_{\mu}^{\rm lens})^2  ~~,
\eeq
where  $\widehat{\sigma_{\mu}^{\rm int}}$ is the remaining source of intrinsic dispersion.

Given the typical precision of current data \cite{Kowalski:2008ez,Guy:2007dv}, a reasonable fit of the Hubble diagram does not seem to require a strong $z$-dependence of the parameter $\sigma_{\mu}^{\rm int}$\,: for instance, a nearby sample gives $\sigma^{\rm int} = 0.15 \pm 0.02$, to be compared with the value $\sigma^{\rm int} = 0.12 \pm 0.02$ obtained for distant supernovae. On the other hand, as illustrated in Fig. \ref{Fig8}, the results of our computations at $z \gaq 0.3$ is very well captured by a linear behaviour which can be roughly fitted by $\sigma_{\mu}^{\rm lens}(z) = 0.056 \, z$. We should also note that this contribution stays below $0.12$ up to $z \sim 2$, which makes it perfectly compatible with observations so far performed.

It is remarkable that the (above mentioned) simple linear fit of our curve for $\sg_\mu(z)$ at $z \gaq 0.3$ turns out to be very close to the experimental estimate reported in \cite{Kronborg:2010uj}, namely\,:
\beq
\left( \sigma_{\mu}^{\rm lens} \right)_{\mbox{Kronborg}} = (0.05 \pm 0.022) \, z ~~.
\eeq
Also, such a fit is well compatible with the results of \cite{Jonsson:2010wx}\,:
\beq
\left( \sigma_{\mu}^{\rm lens} \right)_{\mbox{J\"onsson}} = (0.055~^{+\,0.039}_{-\,0.041}) \, z ~~.
\eeq
These two observational estimates of the \gls{lensing dispersion}, with the relative error bands, are illustrated by the shaded areas of Fig. \ref{Fig8}. Our result is also in relatively good agreement with the simulations carried out in \cite{Holz:2004xx} which predicted an effect of $0.088 \,z$. Other authors \cite{Williams:2004jr, Menard:2004sm, Karpenka:2012ys}, however, have found no indication of this $z$-dependence of $\sg_\mu$, and such a signal certainly waits for a better observational evidence (see \cite{Amendola:2013twa,Quartin:2013moa} for a recent statistical approach and its applications).

It is likely that future improvements in the accuracy of SNe Ia data will detect (or disprove) this effect at a higher confidence level. In this respect, our results  for $\sg_\mu(z)$ stand out as a challenging prediction, which, we believe, could represent a further significative test of the concordance model. We finally remark, incidentally, that our prediction for the Doppler-related dispersion at small $z$ is also consistent with previous findings \cite{Hui:2005nm} on the so-called Poissonian \gls{peculiar velocity} contribution to $\sigma_{\mu}$. We have checked that our result for the Doppler contribution (see Fig. \ref{Fig8}) is well fitted by an inverse power law\,:
\beq
\sg_\mu^{\rm Doppler}(z) \sim 0.00323 \, z^{-1} ~~,
\eeq
in very good agreement, shape-wise, with the corresponding result in \cite{Hui:2005nm}. The fact that our prediction is about a factor 1.7 larger than the one of \cite{Hui:2005nm} is due, presumably, to the use of somewhat different power spectra (linear or non-linear, with or without baryons).

\subsection{Comparing theory and observations of the Union 2 data set}
\label{Ch6Sec43}

\begin{figure}[t]
\centering
\includegraphics[width=12cm]{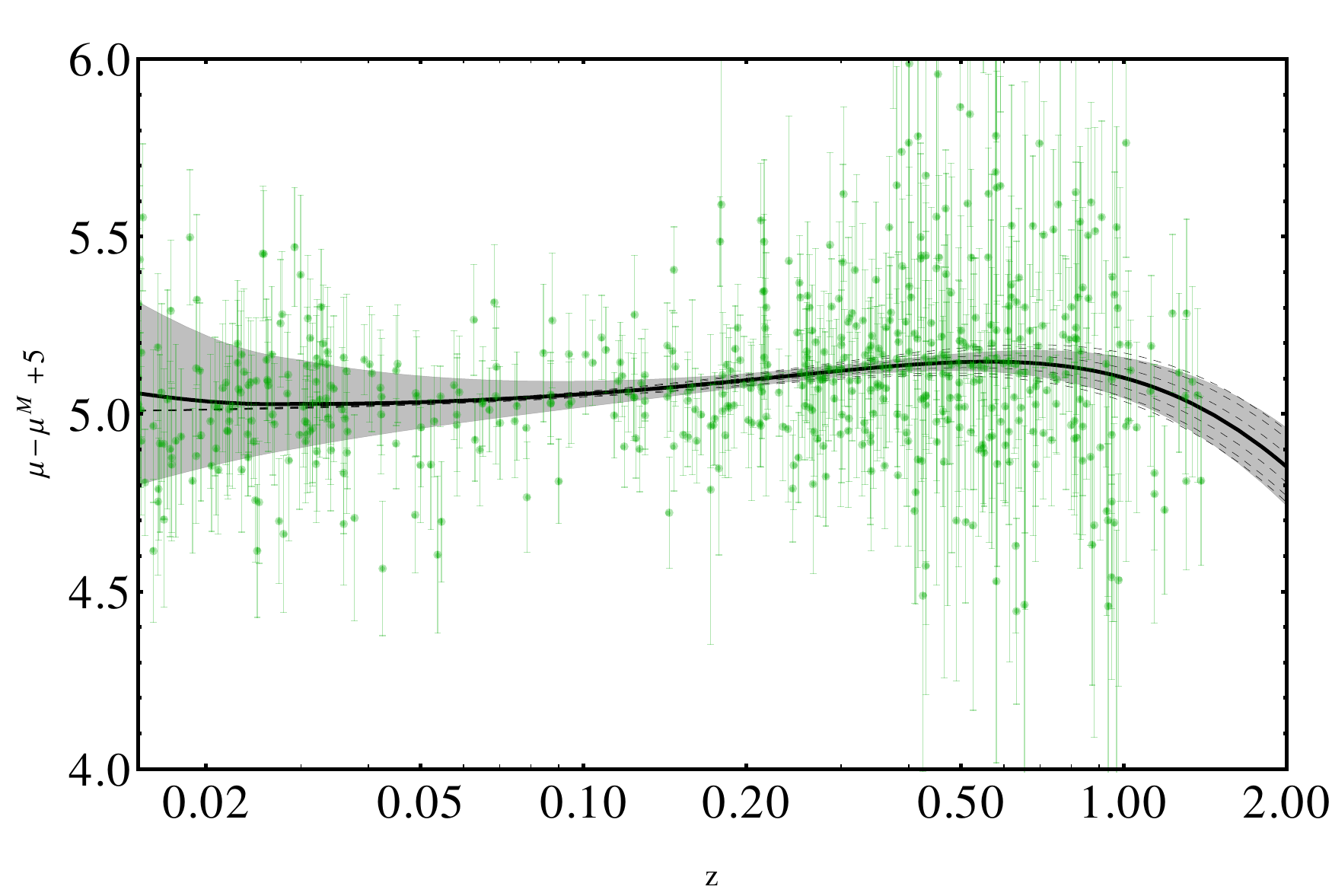}
\centering
\caption[Comparison of the Union 2 data set with our theoretical prediction]{Comparison of the Union 2 data set with the theoretical prediction obtained in section \ref{Ch6Sec42} by the use of the non-linear power spectrum and a cut-off $k_{UV} = 30 ~h \, {\rm Mpc}^{-1}$. We have introduced an arbitrary shift, by a value of 5, to avoid data points with $\overline{\lla \mu \rra} - \mu^M$ reaching negative values.}
\label{ThAndData}
\end{figure}

Taking the data of the Union 2 compilation\footnote{We warmly thank Luigi Tedesco for giving us access to these observational data.}, we can see the relative size of our theoretical prediction on the Hubble diagram, and especially compare our prediction on the dispersion $\sigma_\mu$ with the data points. This comparison has been done and is presented in Fig. \ref{ThAndData}.

As we can see on this plot by eye, the dispersion on the data is very important compared to the theoretical dispersion that we have computed in this thesis. This difference appears especially clearly for large redshift supernovae. On the other hand, the dispersion we predict for nearby SNe Ia is in pretty good agreement with the data. This seems to indicate that the main source of dispersion in SNe Ia at small redshifts is mainly due to their peculiar velocity. On the other hand, we observe a much serious contamination of data at large redshifts which cannot be explained only by weak lensing (and subleading physical effects that we have pointed out in our calculation of the distance at second order in perturbations). This clearly means that a lot of systematic effects are still to be understood at these distances. The main source of this dispersion could be related to the supernova process itself, but a lot of other contamination effects may be involved. For example, absorption of light by some matter clouds can reduce the luminosity of SNe Ia. Another effect, not taken into account in our study, is the case of strong lensing, where a perturbative treatment breaks down.

We can nevertheless be glad to observe the relative smallness of the effect that we computed with respect to the dispersion of data. Indeed, if data would have turned out to be less dispersed than our theoretical variance, that would have meant that they could not have been used alone for studying dark energy and the whole detected shift in data to upper magnitudes would have to be considered as a statistical effect. The only hope that we would have had then would have been to remove lensing and peculiar velocity contaminations for each observed SN Ia, to disantangle the statistics from the real physical effects. As we said, it is quite fortunate that inhomogeneities do not play such a big role, that is why we discovered dark energy so easily (in practice with a lot of efforts, but from a very simple theoretical framework). Nevertheless, we should stress that the effect is not negligible and has been recently noticed in other independent studies (see \cite{Li:2007ny,Li:2008yj,Amendola:2013twa,Fleury:2013uqa,Fleury:2013sna,Marra:2013rba}).

We have thus proved, from our complete second order calculations and the simplifying assumptions described in the last sections, that inhomogeneities were playing a reasonably small effect on the Hubble diagram. Nevertheless, and it is important to be stressed, their effect is much bigger than it could have been thought and it depends strongly on the averaged observable. Also, the study of the induced dispersion in SNe Ia data by the inhomogeneous nature of our Universe seems to give an effect which is small enough to strengthen our confidence in the standard model of cosmology, but big enough to expect a more direct detection of its influence in the next decade. From this perspective, it looks that inhomogeneous models and their influence on observations have a major role to play in the future.

% --------------------------------------------------------- %

%  --------------------- Chapter 7 ----------------------   %
% --------------------------------------------------------- %
\chapter{Summary and conclusions}
\label{Chap7}
\pagestyle{plain}

This thesis has addressed the general problem of lightcone averaging and its implications in cosmology in order to improve the understanding of the Universe in its real inhomogeneous nature.

We have first presented (in \tbf{Chapter} \ref{Chap1}) the basic facts about the standard model of cosmology. In particular, we gave the equations used to interpret the data within the concordance $\Lambda$CDM model (the Friedmann equations). We also described one of the main experimental probes of the standard model, namely the type Ia supernov\ae. These astrophysical objects are indeed at the basis of the discovery of dark energy and their conclusions are the strongest model depend ones (compared to CMB or baryonic acoustic oscillations). We also recalled the basic facts about the large scale structure in the Universe.

In \tbf{Chapter} \ref{Chap2} we gave the definition of the average of scalar quantities over null hypersurfaces. This definition, first inspired by Buchert's spatial averaging formalism \cite{Buchert:1999er,Buchert:2001sa}, shares common properties with the spatial average defined in a gauge invariant way in \cite{Gasperini:2009wp,Gasperini:2009mu}. In particular, this definition is gauge invariant, invariant under general coordinate transformations and under reparametrizations of the hypersurfaces composing it. Following this definition, we derived the equivalents of the Buchert-Ehlers commutation rule which stands at the basis of the backreaction problem. This lightcone average, whose importance has been emphasized since a long time, has never been defined in the literature before. 

We then presented (in \tbf{Chapter} \ref{Chap3}) a system of coordinates that simplifies greatly the description of physical quantities on our past light cone. We called these coordinates the ``Geodesic Light-Cone'' coordinates (GLC) and they turned out to be very useful for a much wider range of applications than the average. These coordinates are very close to another set known in the literature as the observational coordinates \cite{1985PhR...124..315E,Maartens1}. Our gauge in fact corresponds to a further gauge fixing of these latter one in the sense that it only has six arbitrary functions instead of seven in the observational coordinates. It also has other differences, the main one being that the GLC coordinates use the proper time of the observer $\tau$ where the observational ones use a distance $y$ inside the past lightcone (either null or spatial). The study of this gauge is likely to bring new interesting results for all applications of light signals in cosmology (see e.g. \cite{Fanizza:2013doa}). The GLC coordinates constitute a system where photons propagate on a constant lightcone $w = w_o$ and fixed angles $\ti \theta^a$. This nice property which consist in redefining the angles such as to follow the photon path from a source to the observer, as we saw, makes the expression of the redshift $z$ and the luminosity distance $d_L$ in this gauge to be very simple (and directly expressed in terms of the GLC metric components). We also discussed, as simple examples of applications, the link at first order in perturbations between these coordinates and both the synchronous gauge and a LTB model.
These GLC coordinates were finally used, and are very well adapted, to the gauge invariant average on the past lightcone as they simplify greatly its expression. They also simplify the generalized Buchert-Ehlers commutation rules. We applied its definition on scalars over hypersurfaces standing at the intersection between the lightcone $w=w_o$ and the spatial $\tau = \cst$ hypersurfaces but then turned to a better adapted definition for observations, i.e. in terms of $z = \cst$ hypersurfaces. A first, formal, application was then made by studying the average of the redshift drift.

Backreaction, or in more common terms the effect of inhomogeneities, vanishes in an homogeneous flat FLRW spacetime and for this reason we computed in \tbf{Chapter} \ref{Chap4} the perturbed luminosity-redshift relation. As the inhomogeneities are assumed to be stochastic, their average disappears at first order when we take their ensemble average. The perturbative calculation of the distance was thus pushed to second order and expressed in terms of the metric elements of the Poisson gauge, as we know the power spectrum of inhomogeneities in that gauge (and not in the GLC). The final expression for the distance is rich in physical terms, including peculiar velocities, SW and ISW effects, lensing, redshift distortions and a very large number of combined effects at second order. This result is very important for cosmology and could be used in a large set of problems. One should remark also that the GLC metric takes an important role in this calculation which is a proof of its interesting properties. Vector and tensor perturbations have not been neglected either, though considered to be of second order (because of their inflationary origin).

We then addressed, in \tbf{Chapter} \ref{Chap5}, the question of the relative size of the influence from inhomogeneities on the luminosity distance, the flux and the distance modulus. For that we combined the lightcone average with the ensemble average over the stochastic inhomogeneities. In a first preliminary step, we did an incomplete calculation which was based only on (quadratic) first order but enough to reveal the presence of terms that we called the ``induced backreaction'' terms. These terms are interesting as they show the direct multiplication, within the average, between the expression of the considered scalar and the measure of integration itself. It was also enough to obtain the exact expression of the dispersion of the distance. We gave explicit examples of computation of these terms where an important intermediary result consists of the evaluation of their respective spectral coefficients. We then extended the perturbative expressions of our averaged observables and their variance to the full second order. We found that the correction to the average of the luminosity distance differs from the one of the flux by the (involving flux) contribution $\overline{\lla (\Phi_1 / \Phi_0)^2 \rra}$. This correction is proportional to the average of the first order distance correction. It includes, in particular, terms which involve peculiar velocities and lensing. We then proved that vector and tensor perturbations exactly cancel under the angular (lightcone) average, which render them useless for our study. We also properly dealt with the genuine second order terms by relating them to the first order (using \cite{Bartolo:2005kv}).

Taking these theoretical results, we proceeded in \tbf{Chapter} \ref{Chap6} to their numerical estimation by assuming an explicit form for the transfer function of the (linear) power spectrum describing inhomogeneities (and sourced by a primordial inflationary spectrum). We first considered, as a warm up exercise, the simple case of a CDM universe with a simple transfer function involving no baryonic fraction. We applied this case to the incomplete calculation of induced backreaction terms and the variance $\overline{\lla \sigma_1^2 \rra}$. This case already showed that two effects dominate over the others in the dispersion. The first one happens at small redshifts and is related to the peculiar velocity of sources (SNe Ia) because of gravitational inhomogeneities, the second one is a lensing effect at large redshifts. This case has also proved that our integrals over the momentum of perturbations is free from both infrared (IR) and ultraviolet (UV) divergences. This result is important and is to be contrasted with the fact that spatial averages lead, in general, to UV divergences. This absence of divergence is intrinsically due to our lightcone average and its gauge invariance. Another conclusion which appeared is that the average of the distance is merely unaffected by inhomogeneities, except at small redshifts where the peculiar velocity terms are also present. However, from this incomplete calculation of quadratic contributions, one could not draw a definitive conclusion on the average.

We then presented the numerical result of the full computation of the luminosity flux, still in the CDM case. We noticed from this quantity that the flux is not affected at large redshifts and its relative correction stays at a rough order of $\sim 10^{-5}$. This revealed that the luminosity flux is a special quantity for which, when averaged over the lightcone, the lensing term in the genuine second order cancels the lensing contribution from the quadratic first order (i.e. in $\overline{\lla \sigma_1^2 \rra}$). This fact is related to the conservation of the intensity of light over the past lightcone. Nevertheless, this flux calculation also showed that inhomogeneities do not create the same effect on the flux as that of a cosmological constant\,: the correction is both too small and has the wrong sign at large redshifts. The effect on the distance, on the other hand, is always positive but has a characteristic shape due to the combined effects of peculiar velocity and lensing. The correction to the distance in that case can reach a value of $\sim 10^{-2}$ at $z=2$. The consideration of the distance modulus has shown that the averaged effect is negligible (except as before at small redshifts) but the dispersion implies an equivalent error in the estimate of $\Omega_{\Lambda0}$ of $\sim 1-2 \, \%$.

We then addressed the more realistic $\Lambda$CDM model and added the effect of baryons (Silk damping) in the transfer function \cite{Eisenstein:1997ij}. These two modifications have a direct impact on the linear power spectrum, reducing its values, and decrease the corrections of the flux and the distance by about one order of magnitude. We finally considered the non-linear regime of the power spectrum through the HaloFit model \cite{Smith:2002dz,Takahashi:2012em}. This had for consequence to strengthen the effect of inhomogeneities, especially on the distance because of its sensitivity to lensing (reaching $\sim 10^{-3}$ at $z=2$). The distance modulus then showed a shift in its average which stays negligible but a dispersion which corresponds to a possible deviation of $\sim 10\%$ in the value of $\Omega_{\Lambda0}$. This result is important and shows that inhomogeneities cannot be neglected, especially in the dispersion of observational quantities. We finally compared our theoretical prediction of the dispersion (based on a realistic power spectrum) with the Union2 data set and the recent estimates of velocity and lensing contributions to the intrinsic dispersion of supernov\ae\ data. Our results are found to be in amazingly good agreement with the data and should be confirmed in the next few years by surveys of high cosmological precision.

\bigskip

Our main conclusions, reached with the help of a convenient choice of coordinates defining the GLC gauge and by the introduction of a gauge-invariant prescription to average on the light cone, can be summarized as follows. We found that the combination of the lightcone average with a properly defined ensemble average describing stochastic inhomogeneities around a flat FLRW spacetime, applied to different observables ($d_L$, $\Phi \propto d_L^{-2}$ and $\mu \sim 5 \log_{10} d_L$), could not mimic a dark energy component. This result can give us trust in the robustness of the standard model of cosmology. On the other hand, averaged inhomogeneities have a much bigger impact than it was naively expected. Their influence is mainly caracterised by peculiar velocities and lensing and they affect the distance more than the flux (which is actually the least affected quantity at large $z$), by a factor $\sim 100$ at $z = 2$. This clearly selects the flux as the ideal quantity to be measured but also shows that other observables are much more sensitive to inhomogeneities. Finally we have compared our predictions to the intrinsic dispersion of SNe Ia, showing that part of this dispersion could be explained and that future experiments should be able to detect (and remove) this contaminating signal. As a consequence, this work clearly underlines that though inhomogeneities are very unlikely to explain dark energy (at least within this framework), we can strongly suspect that neglecting their effect will prevent a precise determination of the dark energy parameters in the next decade and could even lead to a misinterpretation of data, for example by the claim of a possible but in fact inhomogeneity-related spacetime dependence of dark energy.

% --------------------------------------------------------- %

\begin{appendices}
\appendixpage
\noappendicestocpagenum
\addappheadtotoc

%  --------------------- Chapter 8 ----------------------  %
% --------------------------------------------------------- %
\chapter{Complementary topics}
\label{Chap8}
\pagestyle{plain}

We present in these appendices the detailed study of some particular topics which were too long, or not fully necessary, to be explained within the chapters of this thesis. We start by recalling technical relations of the ADM formalism and use them in a restrictive case. We then provide the basic elements relative to the linear theory of structure formation, including some recall about inflation, fluid equations and gravitational instability. The section which follows presents the main features of cosmological perturbations and their gauge invariant issues. We then present in two consecutive sections the study of the power spectrum in both its linear and non-linear regimes (through the HaloFit model). Finally we give a short derivation of the distance modulus of the Milne universe.

\section{ADM formalism (Arnowitt-Deser-Misner)}
\label{AppASec1}

We present here some of the basic equations related to the splitting of spacetime in spatial hypersurfaces, i.e. the so-called \gls{ADM} formalism. We will mainly follow the explanations of \cite{misner1973gravitation} and will make use of this formalism in Sec. \ref{AppASec13}.

\subsection{The ADM metric}
\label{AppASec11}

Let us try to build up a 4-dimensional geometry from 3-dimensional hypersurfaces in the Minkowski spacetime $\Mcal_4$ (described by coordinates $(\tau,X^i)$). One starts with a lower hypersurface at time $t$ for which we know the inside geometry given by the 3-metric $g_{ij}$ and the elementary distance is given by $g_{ij}(t,x^i) dx^i dx^j$. Putting an upper spatial hypersurface describing our geometry at time $t + dt$, one has a large variety of choices to fix the shape of it. One can first set this new hypersurface at a ``distance above'' the lower one. This is done by setting that the proper time, i.e. the time orthogonal to the spatial 3-hypersurface $\Sigma$, is equal to $\tau = N(t,x^i) dt$ where $N$ is called the \emph{\gls{lapse function}}. Having done this first necessary step does not fix the geometry inside this upper 3-hypersurface and one needs to describe how spatial coordinates $x^i$ change when we go from the lower to the upper hypersurfaces. This is given by introducing a 3-vector called the \emph{\gls{shift vector}} and such that $x^i_{upper}(x^k) = x^i - N^i(t,x^k) dt$. We thus obtain the line element in the upper geometry, and more generally in all hypersurface built up according to this framework\,:
\bea
\label{ADMmetric}
ds_{ADM}^2 = g^{ADM}_{\mu\nu} dx^\mu dx^\nu &=& - N^2 {dt}^2 + g_{ij} (dx^i + N^i dt)(dx^j + N^j dt) \\
&=& -\left( \begin{array}{c}\mbox{proper time from lower} \\ \mbox{to upper 3-geometry}\end{array} \right)^2 + \left( \begin{array}{c}\mbox{proper distance }\\ \mbox{in base 3-geometry}\end{array} \right)^2 ~~.
\eea
One should understand also that $t$ is the time inside the 3-geometry (defining this hypersurface). As a consequence, the upper hypersurface inherits from the time $t + dt$ inside its hypersurface, but the proper distance used in $\Mcal_4$ to set the ``time distance'' between the two is $d\tau = N dt$.

We thus have the metric form for $g_{\mu\nu}^{ADM}$, and its inverse $g^{\mu\nu}_{ADM}$, in $(t,x^i)$ coordinates\,:
\beq
g_{\mu\nu}^{ADM} = \left( \begin{array}{cc} -( N^2 - g_{ij}N^i N^j) & N_i \\ N_j & g_{ij} \end{array} \right) ~~,~~ g^{\mu\nu}_{ADM} = \left( \begin{array}{cccc} - \frac{1}{N^2} & \frac{N^i}{N^2} \\ \frac{N^j}{N^2} & g^{ij} - \frac{N^i N^j}{N^2} \end{array} \right) ~~,
\eeq
where $g_{ij} = g_{ji}$ and we have used that $N_i = g_{ij} N^j$ and $g^{ij}$ is the inverse metric of $g_{ij}$ in the 3-hypersurface. The unit timelike covariant vector $n_\mu$ normal to the constant-t hypersurface and its contravariant counterpart $n^\mu$ are given by
\beq
n_{\mu} = (-N, \vec{0}) ~~~\mbox{ and }~~ n^\mu = (1/N , - N^i/N) ~~~\mbox{ such that }~~ n_\mu n^\mu = -1 ~~.
\eeq

\subsection{The extrinsic curvature}
\label{AppASec12}

Let us illustrate now the role played by the extrinsic curvature tensor $K_{ij}$ that we will encounter in the next section.
For that we consider an elementary spatial displacement on the hypersurface given by the ADM foliation\,:
\beq
d \ell = \ebf_i \, dx^i ~~.
\eeq
In this equation, $\ebf_i$ are the three basis tangent vectors ($\ebf_i \equiv \partial_i$) dual to the three coordinate 1-forms $dx^i$. A tangent vector field $\Tbf$ is then defined simply by
\beq
\Tbf = \ebf_i \, T^i ~~~~~,~~~~~ \Tbf \cdot \ebf_i = T^j (\ebf_i \cdot \ebf_j) = T^j g_{ij} = T_i ~~.
\eeq
The covariant derivative of $\Tbf$ in the i-th direction is then given by the expression\,:
\beq
{}^{(4)}\nabla_{\ebf_i} \Tbf \equiv {}^{(4)}\nabla_i \Tbf = {}^{(4)}\nabla_i ( \ebf_j T^j ) = \ebf_j \parfrac{T^j}{x^i} + {}^{(4)}\Gamma^\mu_{ji} \, \ebf_\mu \, T^j ~~,
\eeq
where the last term has been obtained by using the identity\,: ${}^{(4)}\nabla_i \ebf_j = {}^{(4)}\Gamma^\mu_{ji} \, \ebf_\mu$~.
The covariant derivative of $T_i$ in the $j$-th direction, intrinsic to the 3-dimensional spatial geometry, is given by
\beq
T_{i\,||\,j} = \ebf_j {}^{(3)}\nabla_i \Tbf = \parfrac{T_i}{x^j} - {}^{(3)}\Gamma_{kji} T^k ~~,
\eeq
where the 3-dimensional connection can be expressed in terms of the three basis vectors and the covariant 3-derivative\,: ${}^{(3)}\Gamma_{kji} \equiv \ebf_k {}^{(3)}\nabla_{i} \ebf_j$.

Let us consider the spatial hypersurface $\Sigma$ on which we have a normal (timelike) vector $\nbf$. When we consider this vector at a spatial position $\ell$ and we do an infinitesimal transportation of this vector, parallely to itself, to another point $\ell + d \ell ~ \in ~ \Sigma$, the vector we obtain is not anymore orthogonal to $\Sigma$. The substraction of this transported vector to the normal vector at this new location, called $d \nbf$, is related to the \gls{extrinsic curvature} by
\beq
d \nbf = - \textbf{K}(d \ell) ~~\mbox{(as an operator)}~~ ~~ \Rightarrow ~~ {}^{(4)}\nabla_i \nbf = - \textbf{K}(\ebf_i) = - K_{~i}^j \ebf_j ~~\mbox{(with components)} ~~.
\eeq
One can easily show that this tensor, or more precisely $K_{ij} = K_{~i}^k \, g_{kj} = K_{~i}^k (\ebf_k \cdot \ebf_j)$, is symmetric, i.e. $K_{ij} = K_{ji}$.
We understand by this reasoning that the extrinsic curvature of the 3-dimensional hypersurface $\Sigma$ has a meaning only when it is embedded in a 4-dimensional geometry. We remark also that it is defined in such a way that it is negative when the vectors $\nbf$ at two neighbouring positions tend to point to the same future event, i.e. when the spatial hypersurfaces tend to contract (and not to expand, in which case the extrinsic curvature is positive).

The transportation of the unit vectors $\ebf_i$ can be written in terms of the extrinsic curvature and the connection inside the hypersurface, by the so-called \gls{Gauss-Weingarten equation}\,:
\begin{equation}
^{(4)}\nabla_i \tbf{e}_j = K_{ij} \frac{\tbf{n}}{\tbf{n}\cdot\tbf{n}} + \, ^{(3)}\Gma{h}{j}{i} \, \tbf{e}_h ~~.
\end{equation}
The parallel transportation of a vector $\Tbf$ inside the 4-geometry (i.e. parallel to the geometry of the spacetime) can now be obtained directly by the expression\,:
\beq
{}^{(4)}\nabla_i \Tbf = T^j_{~||\,i} \, \ebf_j + K_{ij} T^j \frac{\nbf}{\nbf\cdot\nbf} ~~.
\eeq
Using the covariant ADM representation of $\nbf$, where $n_\mu = (N, \vec{0})$, we can show that $K_{ij}$ can be expressed in terms of the shift vector and the metric of the spatial hypersurface, by the \gls{ADM formula}\,:
\begin{equation}
K_{ij} = \frac{1}{2N}\left[ N_{i\,||\,j} + N_{j\,||\,i} - \parfrac{g_{ij}}{t} \right] ~~.
\end{equation}

By the use of a self-dual basis, with the vector basis being $(\ebf_0 , \ebf_i) \equiv (\tbf{n}, \ebf_i) = (N^{-1}(\partial_t - N^m \partial_m) ~,~ \partial_i )$ and the associated basis of 1-forms 
$(\tbf{w}^0 , \tbf{w}^i) = ( N \mathrm{d}t ~,~ \mathrm{d}x^i + N^i \mathrm{d}t ) = ( (\tbf{n}\cdot\tbf{n}) \tbf{n} ~,~ \mathrm{d}x^i + N^i \mathrm{d}t )$, one can compute the terms ${}^{(4)}\nabla_{\tbf{e}_j} {}^{(4)}\nabla_{\tbf{e}_k} \tbf{e}_i$ and ${}^{(4)}\nabla_{\tbf{e}_k} {}^{(4)}\nabla_{\tbf{e}_j} \tbf{e}_i$.
Using now the action of the 4-dimensional Riemann curvature as an operator, we have by definition\,:
\beq
\Rcal (\ebf_i, \ebf_j ) \ebf_k = {}^{(4)}\nabla_i {}^{(4)}\nabla_j \ebf_k - {}^{(4)}\nabla_j {}^{(4)}\nabla_i \ebf_k - {}^{(4)}\nabla_{[\ebf_i,\ebf_j]} \ebf_k ~~,
\eeq
where the last term is zero as $[\tbf{e}_i,\tbf{e}_j] = 0 ~~ \forall ~ i,j$.
We then obtain the expression of the 4-dimensional Riemann tensor\,:
\begin{equation}
\mathcal{R}(\tbf{e}_j,\tbf{e}_k)\tbf{e}_i = \left( K_{ik\,||\,j} - K_{ij\,||\,k} \right) \frac{\tbf{n}}{\tbf{n}\cdot\tbf{n}} + \left[ (\tbf{n}\cdot\tbf{n})^{-1} (K_{jk}K^m_i - K_{ij}K^m_k) + \Rim m i j k \right]\tbf{e}_m
\end{equation}
From this expression, one obtain the time and space components of the curvature tensor, the \gls{Gauss-Codazzi equations}, by\,:
\bea
{^{(4)}\Rim 0 i j k} &=& \frac{1}{\tbf{n} \cdot \tbf{n}} {^{(4)}R_{0 i j k}} ~ = ~ - \frac{1}{\tbf{n}\cdot\tbf{n}} \left( K_{ij\,||\,k} - K_{ik\,||\,j} \right) ~~, \\
{^{(4)}\Rim m i j k} &=& {^{(3)}\Rim m i j k} + \frac{1}{\tbf{n}\cdot\tbf{n}} \left( K_{ij}K^m_k - K_{ik}K^m_{~j} \right) ~~.
\eea

One can then directly deduce the following components of the Einstein tensor\,:
\bea
\label{EinsteinTensorInADM1}
-G^0_{~0} &=& \frac{1}{2} R - \frac{1}{2} (\tbf{n}\cdot\tbf{n})^{-1}\left[(\mrm{Tr} \, \tbf{K})^2 - \mrm{Tr}(\tbf{K}^2)\right] ~~, \\
\label{EinsteinTensorInADM2}
G^0_{~i} &=& - \frac{1}{\tbf{n}\cdot\tbf{n}}\left[K^m_{i\,||\,m} - (\mrm{Tr} \, \tbf{K})_{||\,i}\right] ~~,
\eea
where we have\,:
\beq
\label{TracesOfK}
\mrm{Tr} (\tbf{K}) = g^{ij} K_{ij} = g_{ij} K^{ij} = K^j_{~j} \equiv K ~~~~~,~~~~~ \mrm{Tr} (\tbf{K}^2) = K_j^{~m}K_m^{~j} ~~,
\eeq
and where the trace of $K_{ij}$ denoted by $K$ should not be confused with the similar symbol denoting the intrinsic (and not the extrinsic) curvature in Friedmann equations (such as Eq. \rref{NonZeroG}).

The component $G^0_{~0}$ is a measure of curvature independent from how curved one cuts a spacelike slice, it determines how mass (or energy) fixes the curvature. $G^0_{~i}$ determines the transfer of momentum by the geometry to the matter content. On the other hand, $G^i_{~j}$ is hard to evaluate (except in the particular case of \emph{Gaussian normal coordinates}) and the dynamics of the 6 components that it contains is evaluated by an Hamiltonian formalism.

\subsection{Einstein's equations in ADM}
\label{AppASec13}

One can address the problem of the dynamics of the Universe inside the ADM formalism and consider the coordinates $(t,x^i)$ in a fixed gauge. Choosing a synchonous form for this gauge, i.e. taking $N=1$ and $N^i=0$, the line element is given by\,:
\beq
ds^2 = - dt^2 + g_{ij} dx^i dx^j ~~.
\eeq
One considers a fluid with density $\rho$, possibly pressure $p$ and a velocity $u^\mu$.
We can also define a projection tensor $h_{\mu\nu}$ into the hypersuface orthogonal to $n_\mu = (-1,\vec{0}\,)$ and satisfying\,:
\beq
h_{\mu\nu} = g_{\mu\nu} + n_\mu n_\nu ~~~,~~ h_{\mu\nu} n^\nu = 0 ~~.
\eeq
One can see then that\,:
\beq
h_{00} = 0 ~~,~~ h_{0i} = 0 ~~,~~ h_{ij} = g_{ij} ~~.
\eeq
In Sec. \ref{Ch2Sec11} we do the further assumption that $u_\mu$ is also orthogonal to the spatial hypersurfaces given by the foliation (static fluid), so we can also interpret $h_{\mu\nu}$ as the projection tensor on the hypersurfaces orthogonal to $u_\mu$. Here this vector $u_\mu$ is kept general but we define a density $\epsilon$ and a pressure $\pi$ (the \gls{ADM energy density} and \gls{ADM pressure}) which are given by the projections\,: $\epsilon = T_{\mu\nu} n^\mu n^\nu$ and $\pi = \frac13 T_{\mu\nu} h^{\mu\nu}$ (see Eq. \rref{DefinEpsPi}). These quantities will thus coincide with $\rho$ and $p$ in the limit $u_\mu \rightarrow n_\mu$ taken in Sec. \ref{Ch2Sec11} (see also the discussion about the ``tilt angle'' in Sec. \ref{Ch2Sec14}).
We can also define the so-call extrinsic curvature tensor (or ``second fundamental form'' of the constant-t hypersurface)\,:
\beq
K_{ij} \equiv - h^\alpha_i h^\beta_j \nabla_\beta n_{\alpha} ~~.
\eeq

One has from the considerations of Eq. \rref{EinsteinTensorInADM1}, \rref{EinsteinTensorInADM2} and \rref{TracesOfK} that the \gls{Einstein equations} (see Eq. \rref{EinsteinEqWithLambda}) and the continuity equation reduce to the constraint equations
\bea
\label{Constr1}
\frac12 \left( \Rcal + K^2 - K^i_{~j} K^j_{~i} \right) &=& 8 \pi G (\epsilon + 3 \pi) + \Lambda ~~, \\
\label{Constr2}
K^i_{~j\,||\,i} - K_{|\,j} &=& 0 ~~,
\eea
and the evolution equations
\bea
\label{Dyn1}
& \dot{\rho} = K (\epsilon + \pi) ~~, \\
\label{Dyn2}
& (g_{ij})^{\tdev} = - 2 ~ g_{ik} K^k_{~j} ~~, \\
\label{Dyn3}
& (K^i_{~j})^{\tdev} = K K^i_{~j} + \Rcal^i_{~j} - \left(4 \pi G (\epsilon + 3 \pi) + \Lambda \right) \delta^i_{~j} ~~,
\eea
where $\Rcal \equiv \Rcal^i_{~i}$ and $K \equiv K^i_{~i}$ are the traces of respectively the spatial Ricci tensor $\Rcal_{ij}$ and the extrinsic curvature $K_{ij}$.

Considering the covariant derivative of the vector $\nbf$, normal to the spatial hypersurfaces, one has in general the result\,:
\beq
n_{\mu;\nu}  = \Theta_{\mu\nu} - a_{\mu}n_{\nu} ~~\mbox{ with }~~ a_{\mu} = n_{\mu;\nu} n^{\nu} ~. \nonumber
\eeq
In the RHS of this expression, the first term is the so-called \gls{expansion tensor} and the second term involves the peculiar acceleration of the world lines defined by $\nbf$. When we consider that this vector $\nbf$ describes a fluid or an observer in geodetic motion, one then has $a_{\mu} = n_{\mu;\nu} n^{\nu} = 0$. This is the case we will consider here. The quantity $n_{\mu;\nu}$ is thus equal to the expansion tensor, defined as follows\,:
\beq
\Theta_{\mu\nu} = \frac13 h_{\mu\nu} \Theta + \sigma_{\mu\nu} + \omega_{\mu\nu} ~~,
\eeq
where the three RHS terms are the \gls{expansion scalar}, the \gls{shear tensor} and the \gls{vorticity tensor}, defined as follows\,:
\beq
\Theta \equiv \nabla_\mu n^\mu ~~~~~,~~~~~ \sigma_{\mu\nu} \equiv h^\alpha_\mu h^\beta_\nu \left( \nabla_{(\alpha} n_{\beta)} - \frac13 h_{\alpha\beta} \Theta \right) ~~~~~,~~~~~ \omega_{\mu\nu} \equiv h^\alpha_\mu h^\beta_\nu \nabla_{[\alpha} n_{\beta]} ~~.
\eeq
We could assume from now that the vorticity tensor is null, but we will keep this tensor in the calculations to show the similar roles played by the shear and the vorticity tensors in these calculations.

The shear tensor is symmetric and traceless as
\beq
\sigma^\mu_{~\mu} = h^{\alpha\beta} \nabla_\alpha n_\beta - \frac13 h^\alpha_{~\alpha} \Theta = \Theta - \frac13 (3\Theta) = 0 ~~,
\eeq
where we have used the symmetry of $h^{\alpha\beta}$, $h^\alpha_{~\alpha} = 3$ and $n^\beta \nabla_\alpha n_\beta = 0$. It is also the case for the vorticity tensor which is anti-symmetric, and traceless as
\beq
\omega^\mu_{~\mu} = h^{\alpha\beta} \nabla_{[\alpha} n_{\beta]} = 0 ~~,
\eeq
because $h^{\alpha\beta}$ is symmetric in $\alpha \leftrightarrow \beta$ whereas $\nabla_{[\alpha} n_{\beta]}$ is antisymmetric.
One can obtain the vorticity vector from its tensor\,:
\beq
\omega^\mu = \frac12 n_\alpha \epsilon^{\alpha \mu \beta \gamma} \omega_{\beta \gamma} ~~.
\eeq

In the ADM formalism, one can deal with spatial quantities only and the extrinsic curvature is directly linked to the spatial components of the expansion tensor $\Theta_{\mu\nu}$ by
\beq
\label{KijAndThetaij}
K_{ij} = - \Theta_{ij} = - \left( \frac13 h_{ij} \Theta + \sigma_{ij} + \omega_{ij} \right) ~~.
\eeq
Given that $\sigma^0_{~0} = 0 = h^0_{~0}$ we have that the 3-dimensional trace of $\sigma_{\mu\nu}$ and $\omega_{\mu\nu}$ are also zero, leading to the result\,:
\beq
K = - \Theta ~~.
\eeq
We will also use, in the following, the definitions
\bea
\sigma^2 &\equiv & \frac12 \sigma^i_{~j} \sigma^j_{~i} = \frac14 \left( \nabla_\alpha n^\beta \nabla_\beta n^\alpha + \nabla_\alpha n^\beta \nabla^\alpha n_\beta \right) - \frac{\Theta^2}{3} ~~, \\
\omega^2 &\equiv & \frac12 \omega^i_{~j} \omega^j_{~i} = \frac14 \left( \nabla_\alpha n^\beta \nabla_\beta n^\alpha - \nabla_\alpha n^\beta \nabla^\alpha n_\beta \right) ~~.
\eea
The quantity $\sigma$ is called the rate of shear and represents how much a geodesic bundle is stretched out. The quantity $\omega$ similarly represents the rate of rotation of the bundle. Let us notice also that the literature employs sometimes the definition $\omega^2 \equiv \frac12 \omega_{ij} \omega^{ij}$. As $\omega_{ij}$ is an anti-symmetric tensor, this leads to a sign difference for the terms in $\omega^2$ with respect to our notations here\footnote{This last definition of the literature is useful to understand that the rate of shear is a quantity which gravitates, tending to gather the geodesic bundle, whereas the vorticity anti-gravitates. In our case, keeping the symmetry between $\omega^2$ and $\sigma^2$, which play the same role, is more useful.}.

Taking the trace of Eq. \rref{Dyn3} and making use of the expression of $K_{ij}$ given in Eq. \rref{KijAndThetaij}, we obtain the so-called \emph{\gls{Raychaudhuri's equation}}\,:
\beq
\label{RauchaudhuriEq}
\dot{\Theta} + \frac{\Theta^2}{3} + 2 \, \sigma^2 + 2 \, \omega^2 + 4 \pi G (\epsilon + 3 \pi) - \Lambda = 0 ~~.
\eeq
To get this expression we used the Eq. \rref{Constr1} which is the so-called \emph{\gls{Hamiltonian constraint}}\,:
\beq
\label{HamiltonianConstraint}
\frac12 \Rcal + \frac13 \Theta^2 - \sigma^2 - \omega^2 = 8 \pi G (\epsilon + 3 \pi) + \Lambda ~~.
\eeq
We should notice that the Eq. \rref{RauchaudhuriEq} and the continuity Eq. \rref{Dyn1} are analog in the Newtonian regime. The Hamiltonian constraint \rref{HamiltonianConstraint}, however, has no Newtonian equivalent (and is at the source of the backreaction problem, as seen in Sec. \ref{Ch2Sec1}).
Considering also the Eq. \rref{Dyn3}, we can write the general result\,:
\beq
(\sigma^i_{~j} + \omega^i_{~j})^{\tdev} + \Theta(\sigma^i_{~j} + \omega^i_{~j}) = - \left( \Rcal^i_{~j} - \frac13 \delta^i_{~j} \Rcal \right) ~~,
\eeq
in which we can replace $\Rcal$ by the use of Eq. \rref{Constr1}.

The two constraint equations -- Eq. \rref{Constr1}, \rref{Constr2} -- and the three dynamical equations -- Eq. \rref{Dyn1}, \rref{Dyn2} and \rref{Dyn3} -- are finally rewritten by the use of Eq. \rref{KijAndThetaij} as\,:
\bea
\label{NewConstr1}
\frac12 \Rcal + \frac13 \Theta^2 - \sigma^2 - \omega^2 &=& 8 \pi G (\epsilon + 3 \pi) + \Lambda ~~, \\
\label{NewConstr2}
\sigma^i_{~j\,||\,i} + \omega^i_{~j\,||\,i} &=& \frac23 \Theta_{|\,j} ~~, \\
\label{NewDyn1}
\dot{\rho} &=& - \Theta (\epsilon + \pi) ~~, \\
\label{NewDyn2}
(g_{ij})^{\tdev} &=& 2 g_{ik} \sigma^k_{~j} + 2 g_{ik} \omega^k_{~j} + \frac23 \Theta g_{ij} ~~, \\
\label{NewDyn3}
(\sigma^i_{~j} + \omega^i_{~j})^{\tdev} &=& - \Theta (\sigma^i_{~j} + \omega^i_{~j}) - \Rcal^i_{~j} + \frac23 \delta^i_{~j} \left[ \sigma^2 + \omega^2 - \frac{\Theta^2}{3} + 8 \pi G (\epsilon + 3 \pi) + \Lambda \right] ~~.
\eea
We recall that $||i$ denotes the covariant derivative with respect to the 3-dimensional spatial metric $g_{ij}$ and that $|i$ denotes the simple derivative $\partial_i$. For $\Theta$, which is a scalar, we have $\Theta_{||i} = \Theta_{|i}$.

\newpage

\section{Linear theory of structure formation}
\label{AppASec2}

We will give in this section the necessary elements to understand the basic principles about structure formation. For that we will mainly follow the review of \cite{Knobel:2012wa}. These notions are important to better understand the considerations relative to the transfer function and the power spectrum. 

\subsection{Some words about inflation}
\label{AppASec21}

\subsubsection{General facts}
\label{AppASec211}

The first seeds of matter fluctuations that we observe today in the \gls{large scale structure} are believed to emerge from quantum fluctuations in the primordial states of the Universe. This very important fact is at the basis of the construction of the matter power spectrum. It is thus very important, in the context of this thesis, to recall some of the basic facts about inflation.

The first thing to notice is that inflation is not a precise model, but more a scenario created to solve several problems, opening the possibility of different precise realizations of this scenario. From CMB measurements, it appears that the cosmological curvature parameter today, $\Omega_{K0}$, is very close to zero and our Universe is very close to be flat. In other words, it appears that we live in a Universe which has a curvature radius much bigger than its apparent radius. If we look back in the past, this flatness is even more puzzling. Indeed, the curvature parameter can be written as
\beq
\Omega_K(t) = - K \left(\frac{d_{\H}(t)}{R_0}\right)^2 ~~\mbox{ with }~~ d_{\H}(t) \equiv \frac{c}{H a} = \frac{c}{\dot{a}} ~~,
\eeq
where $K= \{ ~ -1,0,1 ~ \}$ depending on the real nature of our Universe. $d_{\H}(t)$ is the ``Hubble length'' (roughly the observable radius of our Universe), or more commonly the \emph{\gls{Hubble horizon}}, and $R_0$ is the curvature radius of our Universe. If we forget about dark energy for a while, as it appeared only recently in the history of the Universe, we realize that going further in the past is going through the eras of matter and radiation domination. In these periods, the Hubble length $d_{\H}(t)$ is increasing with $t$ (going from past to present), because $\dot{a}(t)$ is decreasing (the expansion is decelerating). We thus realize that our Universe needed to be very close to flatness in its earlier times. Why is that so ? This question is called the \emph{\gls{flatness problem}}.

The second problem that inflation tackles is the following one. From the expression of $d_{\H}$, which corresponds to the causal horizon between two points, and its value at the time of \gls{decoupling} (happening for $z_{\rm dec} \sim 1100$), the angular parts of the sky that have been able to interact in their past should not exceed $1^\circ$. Nevertheless, the relative differences in temperature over the whole sky observed in the CMB are less than $\sim 10^{-5}$. This homogeneity in temperatures cannot be explained \emph{a priori} and this is called the \emph{\gls{horizon problem}}.

The concordance model does not give any answer to that questions and this is why we need an inflation mechanism to satisfy this observable facts. It also gives clues of why we do not observe any topological defects (such as cosmic strings) and gives the initial seeds for the formation of structures.

Inflation is a stage of the Universe where the expansion is accelerating, $\ddot{a} (t) > 0$, thus having $\dot{d}_{{\rm H}}(t) < 0$ leading to $\partial_t |\Omega_K(t)| < 0 $. As a consequence, inflation reduces the amount of curvature of the Universe and, if it last long enough, brings it very close to zero and solves the flatness problem. It also solves the horizon problem in the following way. By definition of the comoving horizon length $d_{\H}$, there are no points in interaction at time $t$ if they are separated by a spatial distance larger than $d_{\H}(t)$. Under this length, spatial regions are in causal contact and can thus interact. As a consequence, in the recent (matter and radiation dominated) history of the Universe, where $d_{\H}(t)$ is increasing, the large scale modes are the last ones to be ``freed''\footnote{The dark energy period is ignored as it concerns recent times and it acts on the modes in an inflationary-like way.}. They would have no particular reasons to keep the same properties, such as temperature, if inflation did not exist. Indeed, the fact that we observe an homogeneity in temperature at large scales in the \gls{CMB} makes us believe that these modes were previously in causal contact. Again, that contact is fulfilled if inflation last long enough because inflation is a period where $d_{\H}(t)$ is decreasing with time and thus is big enough if it last for long. For this reason inflation is also an answer to the horizon problem.

So how long inflation should last ? It has to last long enough to satisfy observational constraints. To explain that, let us consider that the inflationary period is correctly described by a de Sitter phase (which is an approximation as, because it has $H$ constant, such a phase cannot end by itself). Let us also consider that inflation starts at an initial time $t_i$ where $|\Omega_K| \sim 1$ and ends at a final time $t_f > t_i$ (where $|\Omega_K|$ is minimal). The curvature can then grow again in the decelerated phase of the Universe and we assume that it grows up to $|\Omega_K| \sim 1$ today (at $t_0$). Assuming that strict extremal conditions, we have that all the inflaton\footnote{The inflaton is the particle associated to the inflation mechanism (see after).} energy density $\rho_\phi$ is constant between $t_i$ and $t_f$ and is then entirely converted into radiation. We thus get that the ratio, between $t_0$ and $t_i$, of the curvature density -- $\rho_K \equiv - 3 K / (8 \pi G \, a^2)$ -- is (see \cite{CourseLesgourguesEPFL})\,:
\beq
\label{InflationCondition}
\left( \frac{a_i}{a_0} \right)^2 = \frac{\rho_K(a_0)}{\rho_K(a_i)} = \frac{\rho_r(a_0)}{\rho_\phi(a_i)} = \frac{\rho_r(a_0)}{\rho_\phi(a_f)} = \frac{\rho_r(a_0)}{\rho_r(a_f)} = \left( \frac{a_f}{a_0} \right)^4 ~~.
\eeq
We can introduce a quantity called the \emph{\gls{e-fold number}}\,: $N \equiv \ln a$. The equality in thus recast into\,:
\beq
N_f - N_i ~ = ~ N_0 - N_f ~~.
\eeq
In the general case, we can have $|\Omega_K| > 1$ at $t_i$ and we observe that $|\Omega_K| \ll 1$ today. The consequence of this is that the second equality sign in Eq. \rref{InflationCondition} is an inequality sign, leading to the condition\,:
\beq
N_f - N_i ~ \geq ~ N_0 - N_f ~~.
\eeq
This identity means that we need more e-folds of expansion during inflation than after to satisfy observations. We know that today $\rho_r(a_0) \simeq (10^{-4} ~ \eV)^4$. We also know that the energy scale of inflation is $< \rho_r(a_f) \simeq (10^{16} ~ \GeV)^4$ (otherwise observed CMB anisotropies would have already revealed gravitational waves) and this scale is also $> (1 ~ \TeV)^4$ from particle physics. These experimental constraints lead to the following condition on the time duration of inflation\,:
\beq
\Delta N \equiv N_f - N_i = \ln \left( \frac{\rho_r(a_f)}{\rho_r(a_0)} \right)^{1/4} ~ \in ~ [37,67] ~~.
\eeq

\subsubsection{Slow-roll inflation}
\label{AppASec212}

There are a lot of different inflation scenarios nowadays, but the simplest one to achieve the necessary behaviour is simply a single scalar field model. Indeed, if we consider an homogeneous scalar field $\phi(t)$ -- the \emph{\gls{inflaton}} -- described by a potential $V(\phi)$ and a kinetic energy $\dphi^2 / 2$, we can define some effective density and pressure as\,:
\beq
\rho_\phi = \frac12 \dphi^2 + V(\phi) ~~~~~,~~~~~ p_\phi = \frac12 \dphi^2 - V(\phi)  ~~.
\eeq
This field, non-interacting in an homogeneous Universe, obeys the \gls{EoM}\,:
\beq
\ddphi + 3 H \dphi + V'(\phi) = 0 ~~.
\eeq
We should notice the friction term $3 H \dphi$ which shows that an expansion of space decreases $\dphi$, lowering at the same time the density and pressure of the field. The equation of state for this inflationary field is, by definition\,:
\beq
-1 \leq w_\phi(t) \equiv \frac{p_\phi(t)}{\rho_\phi(t)} \leq 1 ~~.
\eeq

As we said, it is not necessary for inflation to be a really precise model. In particular, it is not necessary to give a precise expression for the potential $V(\phi)$ if it gives a long enough lasting inflationary period. A sufficient approximation is given by the slow-roll inflation description. If we assume that $\phi$ takes slowly varying values, i.e. that $\dphi^2$ always stays negligible with respect to $V(\phi)$, and that its varying speed is also always small -- i.e. that $|\ddphi| \ll |3 H \dphi|, \, |V'(\phi)|$ -- we get that $w_\phi$ is close to -1 and thus we have an inflationary field which behaves like a cosmological constant (and an expansion which is exponential).
Taking $K=0$, the Friedmann equation and the EoM give us
\beq
H^2 = \frac{8 \pi G}{3} V(\phi) ~~~~~,~~~~~ 3 H \dphi + V'(\phi) = 0 ~~.
\eeq
We can then define the so-called \emph{\gls{slow-roll parameters}}\,:
\beq
\label{SlowRollParameters}
\epsilon \equiv \frac{1}{16 \pi G} \left(\frac{V'}{V}\right)^2 = - \frac{\dot{H}}{H^2} ~~~~~,~~~~~ |\epsilon| \ll 1 ~~~~~;~~~~~ \eta \equiv \frac{1}{8 \pi G} \frac{V''}{V} = \epsilon - \frac{\ddphi}{H \dphi} ~~~~~,~~~~~ |\eta| \ll 1 ~~.
\eeq
We can also notice that the condition $(\epsilon, \eta) \ll (1,1)$ is equivalent to $(|V'/V|,|V''/V|) \ll (1,1)$ and thus a flat enough potential satisfies these slow-roll conditions. More generally, the condition to have an inflationary phase is given by $w_\phi < -1/3$, which corresponds to $\dphi^2 < V(\phi)$, or equivalently $\rho_\phi + 3 \, p_\phi < 0$. Consequently, a period of inflation, as the period of dark energy domination, violates the \gls{strong energy condition} $\rho + 3 \, p > 0$.

Inflation is a very interesting subject on its own, but it will only appear in this work through its prediction of the primordial power spectrum with especially the fact that its spectral index $n_s$ is slightly less than 1. It also implies that perturbations are Gaussian and adiabatic.

\subsection{Fluid equations and Newtonian approximation}
\label{AppASec22}

In the Universe that we observe today, the density constrast for matter, namely $\delta$, takes a very wide range of values (from very high values in neutron stars to very low ones for the voids in the Universe, taking a simple volume average). Physically, we can see this quantity in Fourier space with\,:
\beq
\delta(t,\kbf) = \delta_{\L}(t,\kbf) + \delta_{\NL}(t,\kbf) ~~,
\eeq
where $\delta_{\L}$ can be seen as the contribution to $\delta$ coming from linearized equations whereas $\delta_{\NL}$ takes the rest of contributions into account, especially the contributions of different modes to the mode $\kbf$.

To obtain the general behaviour of $\delta_{\L}(t,\kbf)$, we are now going to elaborate on the linear perturbation theory from the fluid perspective. We consider a Universe filled with an inhomogeneous fluid of matter with density $\rho(t,\rbf)$, velocity field $\vbf(t,\rbf)$ and pressure $p(t,\rbf)$. Because this fluid is made of a massive content, it reacts and contributes to a gravitational potential $\Phi(t,\rbf)$ present in this Universe, and it also has an entropy per unit mass $S(t,\rbf)$. Note that we are using here a radial coordinate $\rbf$ which corresponds to a physical coordinate (in opposition to comoving coordinates, as we will see later).

In Newtonian physics, the evolution of the 7 parameters $X \equiv \{ X_i \} = \{\rho,\vbf,p,\Phi,S\}$ describing the fluid of matter, is given by\,:
\bea
\label{NewtEq1}
& \partial_t \rho + \nabla \cdot (\rho \vbf) = 0 & ~~~~~ \mbox{(continuity eq.)} ~~, \\
\label{NewtEq2}
& \partial_t \vbf + (\vbf \cdot \nabla) \vbf = - \frac{\nabla p}{\rho} - \nabla \Phi & ~~~~~ \mbox{(Euler eq.)} ~~, \\
\label{NewtEq3}
& \nabla^2 \Phi = 4 \pi G \, \rho & ~~~~~ \mbox{(Poisson eq.)} ~~, \\
\label{NewtEq4}
&  \partial_t S + (\vbf \cdot \nabla) S = 0 & ~~~~~ \mbox{(entropy conservation)} ~.
\eea
This system of equations is actually not closed and for this reason we need to add up an extra equation given by the (general) \gls{equation of state}\,:
\beq
p = p\,(\rho, S) ~~.
\eeq
From now on, the 7 unknown functions can be found, at least in principle. We can notice that these equations are non-relativistic equations and, consequently, implicitly assume that $|\vbf| \ll c$ and $p \ll \rho c^2$. It is also assumed that there are no bifurcations in this fluid. In other words, we say that the fluid is in a single stream regime, i.e. fluid particles that are close to each other at a certain time, stay close in time (let say heuristically that the distance that separate them is always much less than the distance they traveled). This is a very good approximation for dark matter particles which interact very weakly with baryonic matter and radiation.

Splitting the 7 quantities $X_i$ in background plus first order perturbations\,:
\beq
X_i = \overline{X_i} + \delta X_i ~~,
\eeq
we can solve this system perturbatively and find, at zeroth order and assuming $\pb = 0$\,:
\bea
\dot{\rhob} + 3 H \rhob = 0 ~~,~~ \partial_t \vbbf + (\vbbf \cdot \nabla) \vbbf = - \nabla \Phib ~~,~~ \nabla^2 \Phib = 4 \pi G \, \rhob ~~,~~ \partial_t \Sb + (\vbbf \cdot \nabla) \Sb = 0 ~~.
\eea
Taking the divergence of the second relation above and using the Hubble relation at zeroth order\,:
\beq
\vbbf = H(t) \, \rbf ~~\mbox{ gives rise to }~~~ \dot{H} + H^2 = - \frac{4 \pi G}{3} \rhob ~~,
\eeq
where we have used the Poisson equation \rref{NewtEq3}. This last equation is the Friedmann equation for a pressureless matter background fluid.
In order to take into account a fluid with pressure, we need to study more closely the situation we have. Indeed, if in the Universe that we are studying there are other energy contributions interacting with our matter fluid only through gravitation, and if these contributions are homogeneous, they will only enter as an homogeneous term in the Poisson equation. They will thus enter only at the unperturbed level and will not affect orders in perturbations. On the other hand, they will modify the expansion of the Universe and could bring some relativistic pressure.
It turns out that we can consider our Newtonian system of equations to be correct at zeroth order and just assume that the evolution is given by the relativistic Friedmann equation (involving pressure).

We can thus eliminate the background and write the equations at first order in perturbations\,:
\bea
& \partial_t \delta\rho + \rhob ~ \nabla \cdot \delta\vbf + \nabla \cdot (\delta\rho ~ \vbbf) = 0  ~~, \\
& \partial_t \delta \vbf + (\delta \vbf \cdot \nabla) \vbbf + (\vbbf \cdot \nabla) \delta\vbf = - \frac{1}{\rhob} \nabla (c_s^2 \delta \rho + \tau \delta S) - \nabla \delta\Phi ~~, \\
& \nabla^2 \delta\Phi = 4 \pi G \, \delta\rho ~~, \\
&  \partial_t S + (\vbf \cdot \nabla) S = 0 ~~,
\eea
where we have used the expansion involving $c_s^2 \equiv \left(\partial \pb / \partial \rhob \right)_{\Sb}$, the \gls{speed of sound} and $\tau \equiv \left(\partial \pb / \partial \Sb \right)_{\rhob}$\,:
\beq
\delta p = c_s^2 \delta \rho + \tau \delta S ~~.
\eeq

As a final step, we can introduce the comoving coordinates $(t,\xbf)$, linked to the physical ones by\,:
\beq
\rbf = a \, \xbf ~~~~~,~~~~~ \nabla_{\rbf} = \frac{1}{a} \nabla_{\xbf} ~~~~~,~~~~~ \left.\frac{\partial}{\partial t}\right|_{\rbf} = \left.\frac{\partial}{\partial t}\right|_{\xbf} - \frac{1}{a} ( \vbf \cdot \nabla_{\xbf} ) ~~.
\eeq
Taking now the Fourier expansion of this system of equations and defining the comoving momentum $\ktbf \equiv \kbf / a$ (and $\tilde{k} = |\ktbf|$), we obtain\,:
\bea
& \partial_t \delta\rho + 3 H \delta \rho + i \ktbf \cdot \delta \vbf = 0 ~~, \\
& \partial_t \delta \vbf + H \delta\vbf = - \frac{i \ktbf}{\rhob} \nabla (c_s^2 \delta \rho + \tau \delta S) - i \ktbf \delta\Phi ~~, \\
& \tilde{k}^2 \delta\Phi = - 4 \pi G \delta\rho ~~, \\
&  \partial_t \delta S = 0 ~~,
\eea
where we used $\nabla \cdot \vbf = 3 H$ and $(\delta \vbf \cdot \nabla) \vbbf / a = \delta v \, H$.
We see that these relations do not contain pressure and we are left with 5 coupled linear first order differential equations and one algebraic equation. 

The solution of this system is then a superposition of 5 linear independent modes\,: 2 vortical modes, 2 adiabatic modes, and one entropy mode.
The \emph{\gls{entropy mode}} is a static entropy perturbation $\delta S(t,\kbf) = \delta S(\kbf)$, with other functions $\delta \rho$, $\delta \vbf$ and $\delta \Phi$ chosen appropriately, and can only appear in a multi-component fluid. As dark matter is an almost non-interacting field and that this entropy mode is not predicted by simple inflationary models, we can neglect it. Furthermore, we can see for a static universe ($H=0$ and $\rhob = {\rm cst}$) that the density fluctuation does not grow and this entropy mode thus has no impact on structure formation. The \emph{\gls{vortical modes}} are obtained by setting $\delta \rho = \delta \Phi = \delta S= 0$ and $\kbf \cdot \delta \vbf = 0$. In that case, the Euler equation is equivalent to
\beq
\partial_t \delta \vbf + H \delta \vbf = 0 ~~~~~\Rightarrow~~~~~ \delta \vbf \propto 1/a ~~.
\eeq
That explains the well-known statement that vortical modes are negligible as they decay with time as the inverse of the scale factor.
The last ones, the \emph{\gls{adiabatic modes}}, are obtained by $\delta S = 0$ and $\kbf ~ \| ~ \delta \vbf$ (so $\kbf \cdot \delta \vbf = k \delta v$).
Indeed, these conditions and the continuity equation at zeroth and first order give\,:
\beq
\label{EqAdiabModes}
\dot{\delta} + i \tilde{k} \delta v = 0 ~~\mbox{ with }~~~ \delta(t,\kbf) \equiv \frac{\delta \rho(t,\kbf)}{\rhob(t)} ~~.
\eeq

Differentiating this equation with respect to cosmic time $t$, we can eliminate $\partial_t \delta v$ from the Euler equation and then use Eq. \rref{EqAdiabModes} and the Poisson equation to get
\beq
\label{EDO2ofDelta}
\ddot{\delta} + 2 H \dot{\delta} + (c_s^2 \tilde{k}^2 - 4 \pi G \rhob) \delta = 0 ~~.
\eeq
This equation is a linear second order differential equation and it has in general two independent solutions which can be either stable or evolving in time.

\subsection{Gravitational instability and growth of structures}
\label{AppASec23}

We can notice that the stability of a mode of comoving momentum $\tilde{k}$ in Eq. \rref{EDO2ofDelta} is determined by the sign of the last term in this equation, i.e. by the sign of $c_s^2 \tilde{k}^2 - 4 \pi G \rhob$. Indeed, when this term is positive, we have two oscillatory modes and, when this term is negative, we have exponentially growing/decaying modes. As we can notice, these two different regimes happen at different wavelengths, due to the fact that this coefficient is $\tilde{k}$-dependent. Consequently, we have exponential solutions when the mode we consider has a comoving wavelength $\tilde{\lambda}$ such that\,:
\beq
\tilde{\lambda} = \frac{2 \pi}{\tilde{k}} ~~>~~ \tilde{\lambda}_{\rm J} = c_s \sqrt{\frac{\pi}{G \rhob}} ~~.
\eeq
This critical value for the wavelength is called the \emph{\gls{Jeans length}} and it says that a structure of matter larger than this length is not stable, so it either grows or decays.

Considering a sphere of density $\rho$, we understand that this structure will collapse or grow if its density is too high, in other words if its mass exceeds a certain mass, called the \emph{\gls{Jeans mass}}, given by\,:
\beq
M_{\rm J} = \frac{4 \pi}{3} \left(\frac{\lambda_{\rm J}}{2}\right)^3 \rhob = \frac{\pi}{6} \lambda_{\rm J}^3 ~ \rhob ~~.
\eeq
To give an example, we can notice that the \gls{speed of sound} of baryonic matter during the radiation era is $c_s = c / \sqrt{3} \simeq 0.6 ~ c \,$ as baryons are tightly coupled to photons. So the Jeans length is large and baryons cannot start forming structures. After the \gls{decoupling}, however, the baryonic matter is decoupled from photons and its speed of sound is very close to zero. Baryons can then start to form structures. On the other hand, dark matter is only interacting gravitationally and has a very small speed of sound even in the radiation dominated phase of the Universe. That explains why after decoupling baryons start to fall in the potential wells created by dark matter and why baryonic structures form in a faster way than it would have been expected in a Universe without dark matter.

For a very large wavelength mode, the equation \rref{EDO2ofDelta} is approximated by
\beq
\label{EDO2ofDeltaLongWVL}
\ddot{\delta} + 2 H \dot{\delta} - 4 \pi G \, \rhob \, \delta = 0 ~~,
\eeq
which has for general solution\,:
\beq
\delta(t,\kbf) = \delta_+(\kbf) D_+(t) + \delta_-(\kbf) D_-(t) ~~.
\eeq
Here, $D_+(t) \equiv D(t)$ is the \emph{\gls{linear growth function}}.
When our Universe is matter dominated, we then have $H(t) = 2 / 3t$. We can then use the Friedmann equation $4 \pi G \rhob = 3 H^2 / 2$ and look for solutions of $\delta(t,\kbf)$ of the type $\delta(t,\kbf) = \delta(\kbf) \, t^\alpha$ (with $\alpha \in \RRR$). We find easily that only two values are authorized\,: $\alpha = 2/3, -1$. Hence $D_+ \propto t^{2/3}$ and $D_- \propto t^{-1}$. Compared to the growth of perturbations going as an exponential when the Universe is static, we have structures that form only as a power law in the realistic matter era. During the radiation era, the growth of dark matter structures is even slower, and we can show under reasonable conditions that the growing mode does not grow more than by a factor 2.5 during that period. This is called the \emph{\gls{Meszaros effect}}\,: dark matter pertubations of wavelength much longer than the Jeans length are almost frozen during the radiation era.

To conclude on this subsection, we should notice that there is no analytical solution of Eq. \rref{EDO2ofDeltaLongWVL} in the case of a flat $\Lambda$CDM model. Nevertheless, a very good approximation is given by \cite{Carroll:1991mt}\,:
\beq
D(z) = \frac{1}{1+z} \frac{g(z)}{g(0)} ~~\mbox{ with }~~ g(z) \propto \, \Omega_m(z) \left[ \Omega_m^{4/7}(z) - \Omega_{\Lambda}(z) + \left(1 + \frac{\Omega_m(z)}{2} \right) \left(1 + \frac{\Omega_{\Lambda}(z)}{70}\right) \right]^{-1}
\eeq
\beq
\mbox{where \,} ~~~~~ \Omega_m(z) = \frac{\Omega_{m0} (1+z)^3}{\Omega_{m0} (1+z)^3 + \Omega_{\Lambda0}} ~~~~~,~~~~~ \Omega_\Lambda(z) = \frac{\Omega_{\Lambda0}}{\Omega_{m0} (1+z)^3 + \Omega_{\Lambda0}} ~~.
\eeq

\subsection{Reentering of the modes, the transfer function}
\label{AppASec24}

The discussion made above, leading to the expression of $D(z)$, is valid for perturbations well inside the Hubble horizon. For modes larger than this horizon, a general relativistic treatment is needed and we must deal with gauge invariance complications. Qualitatively, what happens is that modes reentering the horizon, during radiation or matter domination, have grown in amplitude. We can show that the density contrast of a mode reentering at $t_{\rm in}$ scales as
\beq
\delta(t_{\rm in},\kbf(t_{\rm in})) \propto \frac{1}{H^2(t_{\rm in}) a^2(t_{\rm in})} \propto \left\{ \begin{array}{ll} a^2(t_{\rm in}) & \mbox{for } t_{\rm in} \in \mbox{ Radiation Era} ~~, \\ a(t_{\rm in}) & \mbox{for } t_{\rm in} \in \mbox{ Matter Era} ~~. \end{array} \right.
\eeq
We thus have matter perturbations growing like $a$ both outside and inside the horizon during matter domination, and super-horizon pertubations growing like $a^2$ when sub-horizon perturbations are frozen during the radiation dominated era. We can hence use the modes reentering the horizon as references for the inside horizon modes. Thus, we can understand how these modes have been modified during the radiation domination. This is the role of the \emph{\gls{transfer function}}, which is defined as\,:
\beq
T(k) \equiv \frac{\delta(t,\kbf)}{\delta(t_{\rm in},\kbf)} \frac{D(t_{\rm in})}{D(t)} ~~.
\eeq
This function has by definition the limit $\lim_{k \rightarrow 0} T(k) = 1$ as super-horizon modes are not affected by radiation domination (they reenter during the Matter Era). We also understand that $T(k)$ has to include all physical effects changing the amplitude of perturbations during radiation domination. These effects are numerous and a precise determination of $T(k)$ is only possible by solving the full relativistic Boltzmann equation. Nevertheless, the main physical effects that we can retain are the overall suppression due to evolution, the Silk damping effect, and baryonic oscillations.

The main effect building $T(k)$ is the overall suppression from evolution. Indeed, super-horizon perturbations are frozen and passively grow as $a^2$ in radiation domination. When they reenter the horizon during matter domination, they have a larger size than if they had reentered some time before. On the other hand, a perturbation reentering during radiation domination stays almost frozen due to the Meszaros effect and only starts to grow like $D \propto a$ after matter-radiation equality ($t = t_{\rm eq}$). The consequence is a suppression of all the modes reentering at a time $t_{\rm in} < t_{\rm eq}$ by a factor $(a(t_{\rm eq}) / a(t_{\rm in}))^2$. During radiation, we also know that modes reentering have a momentum $k \propto H(t_{\rm in}) a(t_{\rm in}) \propto 1/a(t_{\rm in})$, so we have the ratio with respect to the matter-radiation equality\,: $k / k_{\rm eq} = a(t_{\rm eq})/a(t_{\rm in})$. The suppression of modes is thus going like $(k / k_{\rm eq})^{-2}$ and we have the important scaling limit\,:
\beq
\lim_{k \rightarrow \infty} T(k) \sim (k / k_{eq})^{-2} ~~.
\eeq

The Newtonian description of perturbations is valid well within the Hubble horizon. For super-horizon perturbations, we must use the full relativistic perturbation theory.

\newpage

\section{Gauge invariance and perturbation theory}
\label{AppASec3}

We will mainly use here the considerations and notations of \cite{Malik:2008im} and \cite{1992PhR...215..203M} to explain the basic notions on gauge invariance. We will nevertheless use the opposite signature for the metric.
We first mention a subtlety of language that we will avoid to address in this section. One should in principle make the distinction between ``gauge independent'' and ``gauge invariant''. Indeed, the first notion implies the second, but not the converse. To avoid any complication, we will only speak of gauge invariance, gathering hence both concepts in the same term.

\subsection{General considerations}
\label{AppASec31}

The study of the evolution of perturbations within the set up of general relativity consists in the linearization of Einstein's equations around a certain background. The problem we then have to face is that general relativity does not assume a preferred system of coordinates and there is a non-physical freedom associated to these coordinates that we call the \gls{gauge freedom}. We will describe in these sections different gauges and will give the elements describing the formulation of perturbation theory in a gauge invariant way, i.e. keeping only physically relevant quantities.

We should notice first of all that the linearized gravitational perturbation theory that we will describe here is only valid around a homogeneous and isotropic FLRW background. So most of the equations that we will present in this section will be valid at first order in perturbations only. This assumption is reasonable if we consider observations and it is enforced by the fact that in a FLRW Universe the solution of the linearized equations is also the linearization of the solution of the full (i.e. non-perturbative) equations.
Let us decompose the metric of the perturbed universe as ``background plus pertubations'' in a given system of coordinates $\{ x^\mu \}$\,:
\beq
g_{\mu\nu}(x) = g_{\mu\nu}^{(0)} + \delta g_{\mu\nu}  ~~,
\eeq
and let us choose the background to be isotropic and homogeneous so its line element is
\beq
ds_{b.g.}^2 = g_{\mu\nu}^{(0)} dx^\mu dx^\nu = - dt^2 + a^2(t) \gamma_{ij}(|\xbf|) dx^i dx^j = a(\eta)^2 (- d\eta^2 + \gamma_{ij}(|\xbf|) dx^i dx^j) ~~.
\eeq
Choosing this background geometry to be \gls{FLRW}, we take\,:
\beq
\label{FLRWMukhanov}
\gamma_{ij}(|\xbf|) = \delta_{ij} \left[ 1 + \frac{1}{4} K |\xbf|^2 \right]^{-2} ~~~\mbox{ with }~~ K = -1, 0, 1 ~~~,
\eeq
depending on whether the $\eta = \cst$ spatial hypersurfaces are closed, flat or opened\footnote{We can directly show that this definition of the FLRW metric in Eq. \rref{FLRWMukhanov} is equivalent to the definition given in Eq. \rref{FLRWlineElement} by setting $r = |\xbf| / \left( 1 + \frac{K}{4} |\xbf|^2 \right)$.}.

Using the \gls{Einstein equations}\,:
\beq
\label{EEunperturbed}
G^\mu_{~\nu} = R^\mu_{~\nu} - \frac{\delta^\mu_{~\nu}}{2} R = 8 \pi G \, T^\mu_{~\nu} ~~,
\eeq
where $R \equiv R^\mu_{~\mu}$, we get only two independent equations
\bea
\mbox{(0,0)} ~~:~~ & (a')^2 + K a^2 = - \frac{8 \pi G}{3} \, T^0_0 a^4 ~~, \\
\mbox{(i,i)} ~~:~~ & a'' + K a = - \frac{4 \pi G}{3} \, T a^3 ~~,
\eea
with $a' \equiv da / d \eta$ and $T \equiv T^\mu_{~\mu}$. In the second of these equations, we used that $T^i_{~i} = (T - T^0_0)/3$ (with no summation over i) and used the first equation. We should notice that these equations only involve time and because we choosed an isotropic background, we got that $R^i_{~j} \propto \delta^i_{~j}$. This leads to the conclusion that $T^i_{~j} \propto \delta^i_{~j}$.

The equation of matter conservation, namely $\nabla_\mu T^\mu_{~~0} \equiv T^{\mu}_{~~0;\mu} = 0$, can be written as\,:
\beq
dT^0_{~0} = - (4 T^0_{~0} - T) d\ln a ~~.
\eeq

The metric perturbations at first order can be decomposed in three independent types of perturbations\footnote{They will not be independent anymore at second order as for example we can have vector and tensor perturbations generated by spatial derivatives of scalar perturbations, this if no mechanism like inflation comes into play.}\,: scalar, vector and tensor. We will see that scalar perturbations are the only ones to grow in an expanding universe\,: vector perturbations then decay and tensor perturbations do not couple to density and pressure perturbations. It turns out that the 10 \gls{dofs} present in Einstein's equations describing a general universe are decomposed in $4$ scalar dofs, $2 \times (3-1) = 4$ vector dofs and $6 - (3+1) = 2$ tensor dofs.

The most general form of the metric scalar perturbations at first order is
\beq
\delta g_{\mu\nu}^{(s)} = - a^2(\eta) \left( \begin{array}{cc} 2 \phi & - B_{|i} \\ - B_{|i} & 2 (\psi \gamma_{ij} - E_{|ij}) \end{array} \right) ~~,
\eeq
where $|i$ denotes the simple derivative with respect to the spatial metric inside the constant-$\eta$ hypersurface, i.e. $B_{|i} = \partial_i B$.
The vector perturbations can be written as\,:
\beq
\delta g_{\mu\nu}^{(v)} = a^2(\eta) \left( \begin{array}{cc} 0 & - S_i \\ - S_i & F_{i|j} + F_{j|i} \end{array} \right) ~~,
\eeq
with $S_i^{|i} = F_i^{|i} = 0$. This divergenceless condition imposes that $S_i$ and $F_i$ have no contribution coming from the gradient of a scalar quantity.
Last, the vector perturbations contribute only to the spatial part of the metric\,:
\beq
\delta g_{\mu\nu}^{(t)} = a^2(\eta) \left( \begin{array}{cc} 0 & 0 \\ 0 & h_{ij} \end{array} \right) ~~,
\eeq
where $h_{ij}$ is a transverse contribution\,: it is trace-free ($h_i^{~i} = 0$) and divergence-free ($h_{ij}^{~~|j} = 0$), i.e. it has neither vector nor scalar contributions.

The total metric perturbation can thus be written as
\beq
\delta g_{\mu\nu} = a^2(\eta) \left( \begin{array}{cc} - 2 \phi & B_{|i} - S_i \\ B_{|i} - S_i & - 2 (\psi \gamma_{ij} - E_{|ij}) + F_{i|j} + F_{j|i} + h_{ij} \end{array} \right) ~~.
\eeq
and the most general line element of the ``background + perturbation'' metric is
\beq
ds^2 = a^2(\eta) \big( -(1+2\phi) d\eta^2 + 2(B_{|i} - S_i)d\eta dx^i + [(1- 2\psi) \gamma_{ij} - E_{|ij} + F_{i|j} + F_{j|i} + h_{ij}]dx^i dx^j \big) ~.
\eeq
This line element is often rewritten in the following way\,:
\beq
{ds}^2 = a^2(\eta) \left[ -(1+2\phi){d\eta}^2 + 2 B_i {dx^i}{d\eta} + (\gamma_{ij} + 2 C_{ij}){dx^i}{dx^j} \right]
\eeq
with the following definitions of metric perturbations\,:
\beq
\left\{
\begin{array}{l}
\delta g^{(n)}_{~00} = - 2 a^2 \phi^{(n)} \\
\delta g^{(n)}_{~0i} = a^2 B^{(n)}_{~i} \\
\delta g^{(n)}_{~ij} = 2 a^2 C^{(n)}_{~ij}
\end{array}
\right.
, ~~~~~\mbox{ with }~~~~~
\left\{
\begin{array}{l}
B_i = B_{,i} - S_i ~~~,~~ B_{[,ij]} = 0 ~ (\mbox{curl-free}) \\
C_{ij} = - \psi \delta_{ij} + E_{,ij} + F_{(i,j)} + \frac{1}{2} h_{ij}
\end{array}
\right.
~~\mbox{for n=1} ~~~,
\eeq
where $(n)$ is the order in perturbations. In the following, the equations we will derive will concern only the first order.
We recall that we have 4 scalar dofs ($\phi, \psi, B, E$), 4 vector dofs ($S_i$ and $F_i$), and 2 tensor dofs ($h_{ij}$).

\subsection{Passive and active approaches of gauge transformations}
\label{AppASec32}

There are at least two different ways to consider the meaning of gauge transformations\,: the passive and the active descriptions.
In the passive approach, one specifies two systems of coordinates and then evaluate how geometrical quantities change under the transformation of coordinates, staying at the same physical point. In the active approach, one sees how the geometrical quantities change under a mapping of coordinates (including the transformation of geometrical quantities), staying at the same coordinate point.

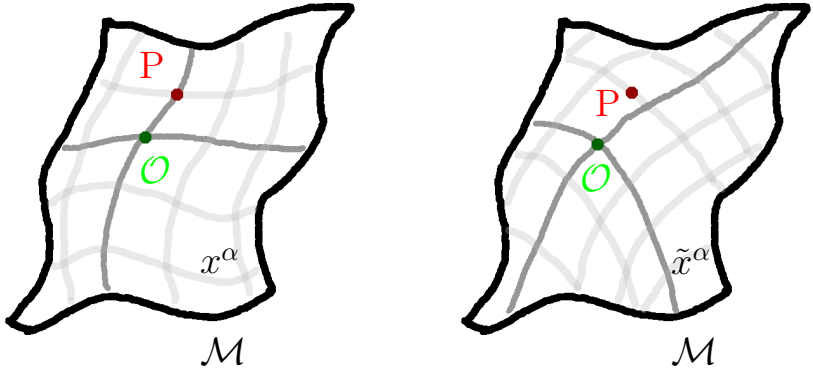
\begin{figure}[t]
\centering
\begin{tikzpicture}
\node (label) at (-3,0){
        \includegraphics[width=5cm]{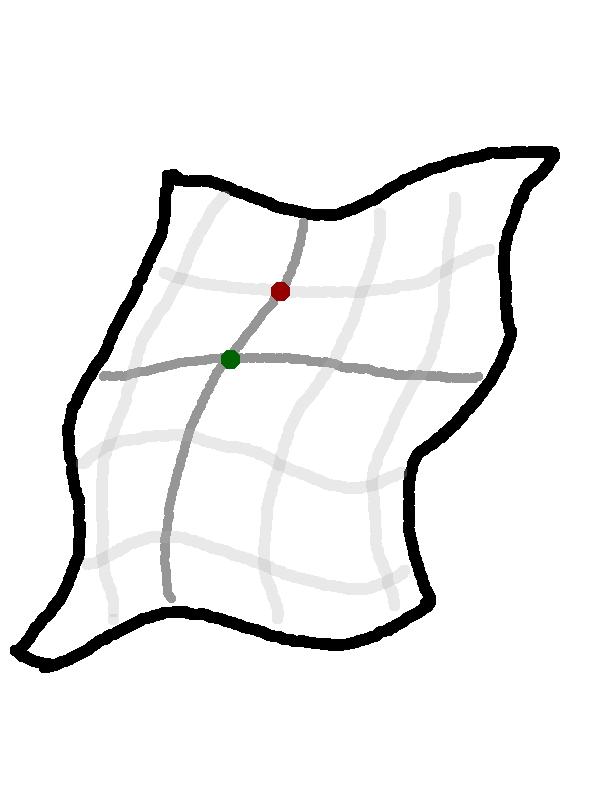}};
\node (label) at (3,0){
        \includegraphics[width=5cm]{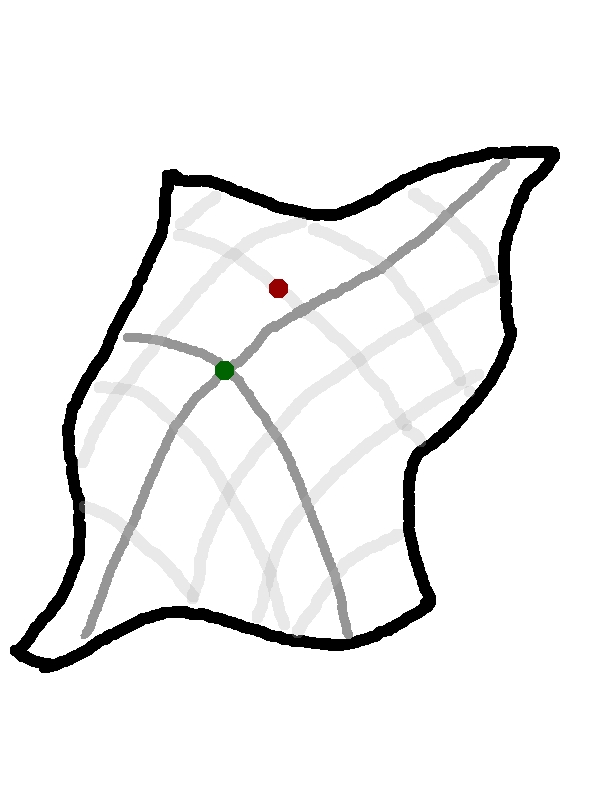}};

\node[anchor=west] at (-3,-1.3) (description) {\Large $x^\alpha$};
\node[anchor=west] at (-3,-2.5) (description) {\Large $\Mcal$};
\node[anchor=west] at (-3.8,1.3) (description) {\Large \textcolor{red}{P}};
\node[anchor=west] at (-3.8,-0.1) (description) {\Large \textcolor{green}{$\Ocal$}};

\node[anchor=west] at (3.2,-1.3) (description) {\Large $\tilde{x}^\alpha$};
\node[anchor=west] at (3.2,-2.5) (description) {\Large $\Mcal$};
\node[anchor=west] at (2.2,0.8) (description) {\Large \textcolor{red}{P}};
\node[anchor=west] at (2.0,-0.2) (description) {\Large \textcolor{green}{$\Ocal$}};

\end{tikzpicture}
\caption{Illustration of the passive approach of gauge transformations.}
\label{FigPassiveGauge}
\end{figure}

\paragraph{Passive approach\,:}
We consider a 4-dimensional manifold $\Mcal$ and a fixed point $\Prm \in \Mcal$. We can parametrize $\Mcal$ by two different systems of coordinates $x^\alpha$ and $\tilde{x}^\alpha$, as illustrated in Fig. \ref{FigPassiveGauge} . We suppose that we have a backgound function $Q^{(0)}(x^\alpha)$ which is attached to the point $\Prm$ and is not a geometrical quantity (i.e. we are not interested by the displacement of this quantity as we could be for a scalar or a vector). We thus have the perturbation defined as
\bea
\delta Q (\Prm) &=& Q(x^\alpha(\Prm)) - Q^{(0)}(x^\alpha(\Prm)) ~~\mbox{ with the $x^\alpha$-parametrization of $\Mcal$} ~~, \\
\delta \ti{Q} (\Prm) &=& \ti{Q}(\tilde{x}^\alpha(\Prm)) - Q^{(0)}(\tilde{x}^\alpha(\Prm)) ~~\mbox{ with the $\tilde{x}^\alpha$-parametrization of $\Mcal$} ~~,
\eea
where it is important to notice that $Q^{(0)}(x^\alpha(\Prm))$ has the same value as $Q^{(0)}(\tilde{x}^\alpha(\Prm))$ though we have $x^\alpha(\Prm) \neq \tilde{x}^\alpha(\Prm)$ in general. On the other hand $Q(x^\alpha(\Prm)) \neq \ti{Q}(\tilde{x}^\alpha(\Prm))$ and thus $\delta Q (\Prm) \neq \delta \ti{Q} (\Prm)$. We can thus speak about a change of variables $x^\alpha(\Prm) \rightarrow \tilde{x}^\alpha(\Prm)$ for each $\Prm \in \Mcal$ and the corresponding change $\{ \delta Q (\Prm) \rightarrow \delta \ti{Q} (\Prm), ~ \Prm \in \Mcal \}$ is called the gauge transformation associated to this change of variables.

\begin{figure}[t]
\centering
\begin{tikzpicture}
\node (label) at (0.3,0){
        \includegraphics[width=5cm]{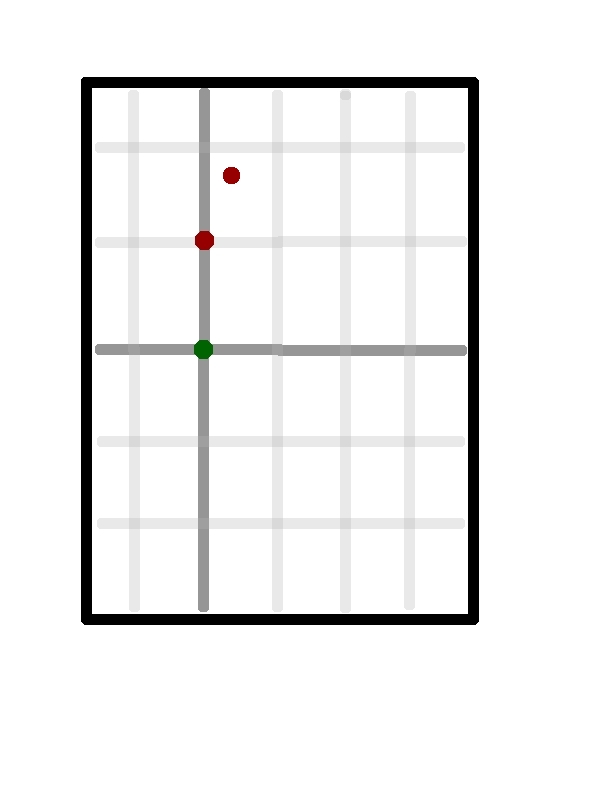}};
\node (label) at (-5,-5){
        \includegraphics[width=5cm]{./pictures/M1v1.jpg}};
\node (label) at (5,-5){
        \includegraphics[width=5cm]{./pictures/M2v1.jpg}};

\node[anchor=west] at (0.8,-1.3) (description) {\Large $X^\alpha$};
\node[anchor=west] at (0.8,-2.5) (description) {\Large $\Ncal$};
\node[anchor=west] at (-1.2,1.7) (description) {\Large \textcolor{red}{R$_1$}};
\node[anchor=west] at (-0.55,2.35) (description) {\Large \textcolor{red}{R$_2$}};
\node[anchor=west] at (-1.2,0.7) (description) {\Large \textcolor{green}{$\Ocal$}};

\node[anchor=west] at (-5,-6.3) (description) {\Large $x^\alpha$};
\node[anchor=west] at (-5,-7.5) (description) {\Large $\Mcal$};
\node[anchor=west] at (-5.8,-3.7) (description) {\Large \textcolor{red}{P}};
\node[anchor=west] at (-5.8,-5.1) (description) {\Large \textcolor{green}{$\Ocal$}};

\node[anchor=west] at (5.2,-6.3) (description) {\Large $\tilde{x}^\alpha$};
\node[anchor=west] at (5.2,-7.5) (description) {\Large $\Mcal$};
\node[anchor=west] at (4.2,-4.2) (description) {\Large \textcolor{red}{P}};
\node[anchor=west] at (4.0,-5.2) (description) {\Large \textcolor{green}{$\Ocal$}};

\node[anchor=west] at (-5,0) (description) {\Large \textbf{$\Dcal$}};
\node[anchor=west] at (-2,0) (description) {};
\node[anchor=east] at (-5,-3) (text) {};
\draw[line width = 0.1cm] (description) edge[out=180,in=90,->] (text);

\node[anchor=west] at (4.5,0) (description) {\Large \textbf{$\ti{\Dcal}$}};
\node[anchor=west] at (2,0) (description) {};
\node[anchor=east] at (5,-3) (text) {};
\draw[line width = 0.1cm] (description) edge[out=0,in=90,->] (text);

\end{tikzpicture}
\caption{Illustration of the active approach of gauge transformations.}
\label{FigActiveGauge}
\end{figure}

\paragraph{Active approach\,:}
In this approach, we consider a background manifold $\Ncal$ with a rigid system of coordinates $\{ X^\alpha \}$ on it and two diffeomorphisms $\Dcal$ and $\ti{\Dcal}$ giving rise to different systems of coordinates $\{ x^\alpha \}$ and $\{ \tilde{x}^\alpha \}$ on the manifold $\Mcal$, as illustrated in Fig. \ref{FigActiveGauge}.
Because these two diffeomorphisms are different, a point $\Prm \in \Mcal$ is the image of two different points (${\rm R}_1$ and ${\rm R}_2$) of $\Ncal$ through $\Dcal$ and $\ti{\Dcal}$. We thus have two different perturbations at the point $\Prm \in \Mcal$ simply because of the different diffeomorphisms that has been used\,:
\bea
\delta Q (\Prm) &=& Q(\Prm) - Q^{(0)}(\Dcal^{-1}(\Prm)) ~~\mbox{ with the diffeomorphism $\Dcal$} ~~, \\
\delta \ti{Q} (\Prm) &=& \ti{Q}(\Prm) - Q^{(0)}(\ti{\Dcal}^{-1}(\Prm)) ~~\mbox{ with the diffeomorphism $\tilde{\Dcal}$} ~~,
\eea
where $Q(\Prm) = \ti{Q}(\Prm)$ because we want the total perturbation at $\Prm \in \Mcal$ to be independent on the parametrization. We thus have $Q^{(0)}(\Dcal^{-1}(\Prm)) \neq Q^{(0)}(\ti{\Dcal}^{-1}(\Prm))$, which is consistent with the fact that these quantities correspond to the evaluation of $Q^{(0)}$ on two different points of the manifold $\Ncal$. To conclude that case, we see that the change $\Dcal \rightarrow \ti{\Dcal}$ induces a change in the size of the perturbations that we can call the gauge transformation\,: $\delta Q (\Prm) \rightarrow \delta \ti{Q} (\Prm)$.

\paragraph{Comparison\,:}
To explain things in a simple form, the passive approach considers a point on $\Mcal$ and move this point in two different places by two different choices of coordinates. The final perturbation $Q(x^\alpha(\Prm))$ is thus different from the perturbation in the other system of coordinates, namely $\ti{Q}(\tilde{x}^\alpha(\Prm))$ and this because they correspond to a different point on the perturbed manifold whereas the non-perturbed positions of these two points were the same. The active approach does the converse consideration. It starts with two different points on the unperturbed manifold $\Ncal$, but the diffeomorphims that we choose both lead to the same perturbed point. Consequently, the values of $Q$ in the unperturbed manifold are different, but the final value on $\Mcal$ is the same (because we are at the same physical point). The important thing is that in both cases we can trace back the change in $Q$ due to the change in coordinates and keep track of this change to avoid any non-physical value of perturbations coming from the choice of coordinates themselves.

\paragraph{Lie derivation\,:}
To be more precise on these aspects of independence of physical quantities from the choice of a gauge, let us now consider an infinitesimal change of coordinates given by\,:
\beq
\label{Transformationxxi}
x^\mu \rightarrow \tilde{x}^\mu = x^\mu + \xi^\mu ~~,
\eeq
where $\xi^\mu$ is an infinitesimal change in coordinates (considered here to be first order in perturbations).
From the explanations given above, we know that a quantity $Q$ defined on the manifold $\Mcal$ vary by an amount which is also the \gls{Lie derivative} of $Q$ in the direction of $\xi^\mu$\,:
\beq
\Delta Q = \widetilde{\delta Q} - \delta Q = \mathscr{L}_\xi Q ~~.
\eeq
These infinitesimal gauge tranformations form a group, the Lie group of infinitesimal transformations.

In general terms, one can decompose a tensorial quantity $\textbf{T}$ (scalar, vector, tensor, ...), as follows\,:
\beq
\textbf{T}(\eta,x^i) = \textbf{T}_0(\eta) + \delta\textbf{T}(\eta,x^i) ~~\mbox{ with }~~
\delta\textbf{T}(\eta,x^i) = \sum_{n=1}^{\infty} \frac{\epsilon^n}{n!}\delta\textbf{T}_n(\eta,x^i) ~~,
\eeq
where $\epsilon$ an infinitesimal parameter. In this equation, the background quantity $\textbf{T}_0(\eta)$ is gauge invariant and we need to choose a gauge to describe the perturbations.
The generator of the gauge transformation $\xi^{\mu}$ can be decomposed in perturbations as\,:
\begin{equation}
\xi^{\mu} \equiv \epsilon \xi_1^{\mu} + \frac{1}{2} \epsilon^2 \xi_2^{\mu} + \mathcal{O}(\epsilon^3) ~~.
\end{equation}
We then have the transformed tensor\,:
\begin{equation}
\widetilde{\textbf{T}} = e^{\Lscr_{\xi}} \textbf{T} ~\mbox{ with }~ e^{\Lscr_{\xi}} = 1 + \epsilon \Lscr_{\xi} + \frac{1}{2} \epsilon^2 [ \Lscr_{\xi_2} + \Lscr^2_{\xi_1} ] + \mathcal{O}(\epsilon^3) ~~,
\end{equation}
which defines the perturbed quantities\,:
\bea
\label{TransfoTensorTwithLie}
\widetilde{\textbf{T}}_0 &=& \textbf{T}_0 ~~, \nonumber \\
\delta \widetilde{\textbf{T}}_1 &=& \delta\textbf{T}_1 + \Lscr_{\xi_1} \textbf{T}_0 ~~, \nonumber \\
\delta\widetilde{\textbf{T}}_2 &=& \delta\textbf{T}_2 + 2 \Lscr_{\xi_1}\delta\textbf{T}_1 + \Lscr_{\xi_2}\textbf{T}_0  + \Lscr^2_{\xi_1}\textbf{T}_0 ~~.
\eea
In the case of a vector quantity $\textbf{T}$, with covariant components $T^\mu$, the Lie derivative gives the following transformation\,:
\beq
\Lscr_\xi T^\mu = \xi^\nu \partial_\nu T^\mu - T^\nu \partial_\nu \xi^\mu ~~.
\eeq
For a more general tensorial quantity, this last relation is generalized to the following transformation\,:
\beq
\Lscr_\xi \, T^{\mu_1 \ldots \mu_m}_{\nu_1 \ldots \nu_n} = \xi^\alpha \partial_\alpha T^{\mu_1 \ldots \mu_m}_{\nu_1 \ldots \nu_n} - \sum_{i = 1}^{m} T^{\mu_1 \ldots \alpha \ldots \mu_m}_{\nu_1 \ldots \nu_n} \partial_\alpha \xi^{\mu_i} + \sum_{j = 1}^{n} T^{\mu_1 \ldots \mu_m}_{\nu_1 \ldots \alpha \ldots \nu_n} \partial_{\nu_j} \xi^\alpha ~~.
\eeq
The Lie derivative is an important notion to study the symmetries of Riemannian spacetimes. Moreover, one can show that the derivative of the metric tensor is\,:
\beq
\Lscr_\xi \, g_{\mu\nu} = \nabla_\mu \xi_\nu + \nabla_\nu \xi_\mu ~~,
\eeq
and thus we have that $g_{\mu\nu}$ is invariant under the change of coordinates of Eq. \rref{Transformationxxi} when we have $\Lscr_\xi g_{\mu\nu} = 0$ (called \emph{Killing's equation} and where $\xi^\mu$ is then called a Killing vector).

We finally have the general mapping between the two systems of coordinates $\{ x^\mu(p) \}$ and $\{ x^\mu (q) \}$\,:
\begin{equation}
x^{\mu}(q) = e^{\xi^{\lambda}\frac{\partial}{\partial x^{\lambda}}|_p} x^{\mu}(p) = x^{\mu}(p) + \epsilon \xi^{\mu}(p) + \frac{1}{2}\epsilon^2(\xi_{1,\nu}^{\mu} \, \xi_1^{\nu} + \xi_2^{\mu})(p) ~~.
\end{equation}
and the 3-vector part of $\xi^\mu$ can be decomposed in two contributions
\beq
\xi^i = \xi^i_{tr} + \gamma^{ij} \xi_{|j} ~~~\mbox{ where }~~ \xi^i_{tr|i} = 0 ~~\mbox{ and }~~ \xi^{|i}_{~~|i} = \xi^i_{~~|i} ~~~,
\eeq 
that is to say in a transverse vector part and a scalar part.
The most general transformation which preserve the scalar nature of perturbations is thus given by the condition $\xi_{tr}^i = 0$\,:
\beq
\label{GenScalarGaugeTrans}
\eta \rightarrow \tilde{\eta} = \eta + \xi^0(\eta,\xbf) ~~~~~,~~~~~ x^i \rightarrow \tilde{x}^i = x^i + \gamma^{ij} \xi_{|j} (\eta,\xbf) ~~,
\eeq
where we recall that $\gamma^{ij}$ is the inverse background metric of spatial coordinates.

\paragraph{Gauge invariant perturbations\,:}
The infinitesimal transformation of the coordinates $x^\mu \rightarrow x^\mu + \xi^\mu$ changes the perturbation of the metric $g_{\mu\nu}$ by $\Delta g_{\mu\nu}$:
\beq
\delta g_{\mu\nu} \rightarrow \widetilde{\delta g}_{\mu\nu} = \delta g_{\mu\nu} + \Delta g_{\mu\nu} ~~.
\eeq
Let us consider an example and take the transformation of the scalar contribution $\phi$ of the component $g_{00}$. We have by definition $\tilde{g}_{\mu\nu}(\tilde{x}) = g^{(0)}_{\mu\nu}(\tilde{\eta}) + \widetilde{\delta g}_{\mu \nu}(\tilde{x})$ where
\beq
\widetilde{\delta g}_{\mu\nu}^{(s)} = - a^2(\tilde{\eta}) \left( \begin{array}{cc} 2 \tilde{\phi} & - \tilde{B}_{|i} \\ - \tilde{B}_{|i} & 2 (\tilde{\psi} \gamma_{ij} - \tilde{E}_{|ij}) \end{array} \right) ~~,
\eeq
and where $x = (\eta, \xbf)$, $\tilde{x} = (\tilde{\eta},\tilde{\xbf})$, and we have\,:
\beq
\widetilde{g}_{\mu\nu}(\tilde{x}) = \frac{\partial \tilde{x}^\alpha}{\partial x^\mu} \frac{\partial \tilde{x}^\beta}{\partial x^\nu} g_{\alpha \beta}(x) ~~~\mbox{ and }~~ \left\{ \begin{array}{l} \eta = \tilde{\eta} - \xi^0 (\tilde{\eta},\tilde{\xbf}) \\ x^i = \tilde{x}^i - \gamma^{ij} \xi_{|j}(\tilde{\eta},\tilde{\xbf}) \end{array} \right. ~~.
\eeq
In principle we should also add a tilde over $\gamma_{ij}$ and $\gamma^{ij}$ in both equations above as we must express this tensor in terms of $\tilde{x}$. Nevertheless, this tensor multiplies $\tilde{\psi}$ and $\xi_{|j}$ respectively, which are first order quantities. Consequently it is not necessary to go to that level of specification at the infinitesimal level we are considering here.
The transformation between the two metrics, with the expansion of $a(\tilde{\eta})$ as $a(\eta) ( 1 + \Hcal(\tilde{\eta}) \xi^0 )$, leads us to
\beq
\label{GTphi}
\tilde{\phi} = \phi - \partial_{\tilde{\eta}} \xi^0 - \Hcal(\tilde{\eta}) \xi^0 ~~.
\eeq
We could also have added tildes over $\xi^0$ but for simplity we kept the same notation for $\xi^0(\tilde{\eta},\tilde{\xbf})$. Also, the derivative $\partial_{\tilde{\eta}}$ can be replaced by $\partial_{\eta}$ at this order, or more simply denoted by a prime. Similarly, we can replace $\Hcal(\tilde{\eta})$ by $\Hcal(\eta)$ at this order.

Proceeding in the same way for the other scalar dofs, we get the transformations\,:
\beq
\label{GTpsiBE}
\tilde{\psi} = \psi + \Hcal \xi^0 ~~~~~,~~~~~ \tilde{B} = B + \xi^0 - \xi' ~~~~,~~~~~ \tilde{E} = E - \xi ~~.
\eeq
From these expressions, we can also build up gauge invariant combinations. Bardeen indeed showed that we can form two quantities having this property and defined as\,:
\beq
\label{GenBardeenPotentials}
\Phi = \phi + \frac{[ (B-E') a]'}{a} ~~~~~,~~~~~ \Psi = \psi - \Hcal (B - E') ~~.
\eeq
These are called the \gls{Bardeen potentials}\footnote{In fact, Bardeen potentials were $\Phi_A$ and $\Phi_H$ such that $\Phi_A = \Phi$ and $\Phi_H = - \Psi$, cf. \cite{1980PhRvD..22.1882B}.} and they have been extensively used in the literature. We should add that every combination of these variables is also gauge invariant but that $\Phi$ and $\Psi$ are the simplest gauge invariant quantities that we can write down. These two quantities are invariant under all infinitesimal scalar coordinate transformations, but we should keep in mind that they are not invariant under all finite coordinate transformations.

If we consider a 4-scalar $q(\eta,\xbf)$ on $\Mcal$, we can decompose this scalar into a background and a perturbed contribution. From homogeneity, the background contribution is only time dependent and the perturbation is of first order\,:
\beq
q(\eta,\xbf) = q^{(0)}(\eta) + \delta q(\eta,\xbf) ~~.
\eeq
As a consequence, this scalar quantity is modified under infinitesimal coordinate transformations and its variation is given by\,:
\beq
\label{Transdq}
\delta q(\eta,\xbf) ~~ \rightarrow ~~ \widetilde{\delta q}(\eta,\xbf) = \delta q(\eta,\xbf) - q^{(0)}(\eta)' \, \xi^0(\eta,\xbf) ~~,
\eeq
and thus $\delta q$ is gauge invariant only if $q^{(0)}$ is time independent. Otherwise, we can build up a gauge invariant quantity by combining $\delta q$ and  gauge perturbations, namely\,:
\beq
\delta q^{g.i.} = \delta q + (q^{(0)})' \, (B - E') ~~.
\eeq

\paragraph{Explicit formulation of gauge transformations\,:}

Let us summarize the effect of an infinitesimal transformation of the coordinates defined by the first order quantity $\xi^{\mu} = (\alpha \,,\, \beta^{|i} + \gamma^i)$. Using the notations for perturbations introduced at the end of Sec. \ref{AppASec31}, one has\,:
\begin{itemize}
\itemsep-1.5em
 \item For scalar perturbations\,:
\beq
\left\{
\begin{array}{l}
\tilde{\phi} = \phi + \mathcal{H}\alpha + \alpha' \\
\tilde{\psi} = \psi - \mathcal{H}\alpha \\
\tilde{B} = B - \alpha + \beta' \\
\tilde{E} = E + \beta
\end{array}
\right.
\eeq
\\
 \item For vector perturbations\,:
\beq
\left\{
\begin{array}{l}
\widetilde{S^{i}} = S^i - (\gamma^i)' \\
\widetilde{F^i} = F^i + \gamma^i
\end{array}
\right.
\eeq
\\
 \item For tensor perturbations\,:
\beq
\tilde{h}_{ij} = h_{ij} ~~.
\eeq
\end{itemize}
We recall that tensorial metric perturbations are not gauge invariant at second order in perturbations but they are at first.
We can also introduce the transformation of quantities which are not metric elements such as the \gls{scalar shear potential} $\sigma = E' - B$ and the spatial velocity perturbation of a fluid velocity $u^\mu = (u^0, u^i)$ which is given at first order by $u^0 = a^{-1} ( 1 - \phi )$ and $u^i = a^{-1} v^i$, with $v^i$ that can be split into scalar and vector parts\,: $v^i = \delta^{ij} \, v_{,j} + v^i_{\rm vec}$.
These two quantities change, under the previous coordinate transformations, in the following way\,:
\beq
\left\{
\begin{array}{l}
\tilde{\sigma} = \sigma + \alpha \\
\tilde{v} + \tilde{B} = v + B - \alpha \\
\tilde{v}_i + \tilde{B}_i = v_i + B_i - \alpha_{,i}
\end{array}
\right. ~~.
\eeq

\subsection{Different gauges\,: synchronous \& longitudinal}
\label{AppASec33}

We have seen that only two scalar quantities are necessary to fully describe the infinitesimal scalar coordinate transformations\,: $\xi^0$ and $\xi$. We can thus impose two conditions on the four scalar degrees of freedom (\gls{dofs}) of first order metric perturbations. Different conditions impose different gauges and different transformations of the coordinates (which involve these dofs). Choosing a gauge is thus equivalent to choosing the system of coordinates in which we want to describe perturbations. We will now present the two important gauges for this thesis\,: the synchronous and longitudinal gauges. We emphasize the point that expressions presented here, i.e. in particular the functions $\psi, \phi, B, E, \ldots$, are all first order relations that are more involved at second order (see e.g. Eq. \rref{TransfoTensorTwithLie}).

\paragraph{Synchronous gauge (SG)\,:}
It is defined by
\beq
\tilde{\phi} = 0 ~~,~~ \tilde{B} = 0 ~~~~~ \mbox{ (synchronous gauge condition) } ~~.
\eeq
We can understand from $\tilde{\phi} = 0$ that an observer at a fixed position $x_s^i$ in the synchronous gauge has a proper time which coincide with the cosmic time of the FLRW background (i.e. $d\tau = a d\eta \equiv dt$).

One can see that Eq. \rref{GTphi} and \rref{GTpsiBE} give in that case\,:
\beq
0 = \phi - \partial_{\tilde{\eta}} \xi^0 - \Hcal \xi^0 ~~,~~ \tilde{\psi} = \psi + \Hcal \xi^0 ~~,~~ 0 = B + \xi^0 - \xi' ~~,~~ \tilde{E} = E - \xi ~~,
\eeq
and so we can choose the functions $(\xi^0,\xi)$, i.e. the transformation of coordinates which set a general perturbed background (i.e. in an arbitrary system of coordinates), in such a way that these properties are satisfied and so that the system of coordinates $\{ \tilde{x}^\mu \}$ has the desired properties. The solution to be used is
\beq
\label{TransToSG}
\eta \rightarrow \eta_s = \eta + \frac{1}{a} \int a \phi \, d\eta ~~,~~ x^i \rightarrow x_s^i = x^i + \gamma^{ij} \left( \int B d\eta + \int \frac{1}{a} d\eta \int a \phi d\eta \right)_{|j} ~~.
\eeq
These coordinates $(\eta_s,x_s^i)$ are the very popular synchronous gauge coordinates\footnote{Note that the subscript ``s'' here means ``synchronous'' and not the source position as it was employed in the main body of this thesis.}.

We can see nevertheless that synchronous coordinates are not totally fixed\,: there is a residual gauge freedom under the transformation\,:
\beq
\label{ResidTransToSG}
\eta \rightarrow \eta_s = \eta + \frac{1}{a} C_1(\xbf) ~~,~~ x^i \rightarrow x_s^i = x^i + \gamma^{ij} C_{1|j}(\xbf) \int \frac{1}{a} d\eta + \gamma^{ij} C_{2|j}(\xbf) ~~,
\eeq
where $C_1(\xbf)$ and $C_2(\xbf)$ are arbitrary functions of the spatial coordinates. It was thus realized historically that this gauge can lead to ambiguities\,: the so called gauge-modes which are not physical but only due to this arbitrary freedom in the coordinates. For that reason the synchronous gauge is less used nowadays, especially for the interpretation of super-horizon perturbations.

One way out to this problem of residual gauge freedom is to impose the further condition that $\widetilde{v}_{cdm} = 0$. One then has that $C_1(\xbf) = a (v_{cdm} + B)$.

\paragraph{Longitudinal gauge\,:}
This gauge is also called conformal \tbf{Newtonian gauge (NG)} (or the \emph{orthogonal zero-shear gauge}). It is defined by
\beq
\label{DefLongitudinalGauge}
\tilde{B} = 0 ~~,~~ \tilde{E} = 0 ~~~~~ \mbox{ (longitudinal gauge condition) } ~~.
\eeq
This condition thus implies that the scalar shear potential is zero in the longitudinal gauge\,: $\sigma_\ell = E_\ell' - B_\ell = 0$.

We can see from Eq. \rref{GTpsiBE} that these conditions fix $\xi^0$ and $\xi$, as $B$ and $E$ are given in our starting system of coordinates $\{ x^\mu \}$. We then find the infinitesimal transformation of coordinates which write this metric in the longitudinal form\,:
\beq
\label{TransToLG}
\eta \rightarrow \eta_\ell = \eta - (B - E') ~~,~~ x^i \rightarrow x_\ell^i = x^i + \gamma^{ij} E_{|j}(\xbf) ~~,
\eeq

We can also see from Eq. \rref{GenBardeenPotentials} that the gauge invariant Bardeen potentials are simply equal to the variables $\phi$ and $\psi$ in the longitudinal gauge\,: $\Phi = \phi_\ell$ and $\Psi = \psi_\ell$\footnote{
We remark that the longitudinal gauge, given simply by the line element
\beq
{ds}^2 = - (1+2\Phi) {dt}^2 + (1-2\Psi) \gamma_{ij} dx^i dx^j ~~, \nonumber
\eeq
is sometimes also written, with clearly different definitions of the Bardeen potentials $\Phi$ and $\Psi$, in the following way\,:
\beq
{ds}^2 = - e^{2\Phi} {dt}^2 + e^{-2\Psi} \gamma_{ij} dx^i dx^j ~~. \nonumber
\eeq
}. It is thus very interesting to derive the EoM in the longitudinal gauge as these EoMs are the gauge invariant EoMs after we replace $(\phi_\ell,\psi_\ell)$ by $(\Phi,\Psi)$. If the stress-energy tensor has a spatial part which is diagonal, we then have $\Phi = \Psi$ which is a generalization of the Newtonian gravitational potential $\phi$.

We can do a transformation of coordinates to write the synchronous coordinates in the longitudinal gauge. For this we apply the transformation of Eq. \rref{TransToLG}
which writes a general metric in the longitudinal gauge, and we fix this general metric to actually be the synchronous metric, i.e. we set $\phi = B = 0$ (and specifying $E = E_s$, $\psi = \psi_s$). The general transformation to go from the synchronous gauge to the longitudinal one is thus\,:
\beq
\label{TransSGToLG}
\eta_s \rightarrow \eta_\ell = \eta_s + E_s' ~~,~~ x_s^i \rightarrow x_\ell^i = x_s^i + \gamma^{ij} E_{s|j}(\xbf) ~~.
\eeq
We find that the functions $\phi$ and $\psi$ in both gauges are related by the following relations\,:
\beq
\phi_\ell = - \Hcal E_s' - E_s'' ~~,~~ \psi_\ell = \psi_s + \Hcal E_s' ~~.
\eeq
Using Eq. \rref{Transdq}, one can show that the energy-density perturbations in synchronous and longitudinal gauges are linked by
\beq
\delta \epsilon_\ell = \delta \epsilon_s - \epsilon_0' E_s' ~~.
\eeq
Conversely, we can also write the longitudinal gauge in a synchronous gauge form (see \cite{1992PhR...215..203M}, p. 217).

We should note also that the instability implied by $\sigma_\ell = 0$ is known to make the use of the longitudinal gauge difficult in the super-horizon limit.

\subsection{Other interesting gauges}
\label{AppASec34}

We have presented until now the synchronous gauge and the longitudinal gauge. Other gauges are commonly used in the literature and we intend to give a short review of them here.

\paragraph{Poisson gauge\,:}

The Poisson gauge is an extension of the Newtonian gauge to the case of vector and tensor modes. One can define this gauge by the following line element (presented in Eq. \rref{PGmetricstandard}) \,:
\beq
{ds}^2 = a^2(\eta) \left[ - (1 + 2\Phi) {d\eta}^2 + 2 \omega_i {dx}^i{d\eta} + ( (1 - 2\Psi) \delta_{ij} + h_{ij}) {dx}^i {dx}^j \right]
\eeq
where $h_{~i}^i = 0$, $h_{ij}^{~~|\,i} = 0$ and $\omega_i^{~,i} = 0$.

One obtain the definition of this gauge by taking the gauge conditions of the Newtonian gauge, Eq. \rref{DefLongitudinalGauge}, and by adding to them the further conditions\,:
\begin{equation}
\widetilde{B}_{l}^i = 0 ~\mbox{ and }~ \widetilde{S}_{l}^i = 0 ~~.
\end{equation}
This defines the vectorial part of the spatial gauge transformation\,:
\begin{equation}
\gamma_{1l}^i = \left[ \int S_1^i \mathrm{d}\eta \right] + \hat{C}_1^i(\xbf)
\end{equation}
where $\hat{C}_1^i(\xbf)$ is an arbitrary 3-vector which depends on the choice of the spatial coordinates on the initial hypersurface.

\paragraph{Spatially flat gauge\,:}

It is also called the \emph{Uniform curvature gauge} or the \emph{Off-diagonal gauge} and it is defined by the gauge conditions\,:
\begin{equation}
\widetilde{\psi} = \widetilde{E} = 0 ~~~~~\mbox{ and }~~~~~ \widetilde{F}_i = 0
\end{equation}
This gauge is sometimes more practical than the longitudinal gauge, e.g. in calculations involving collapsing Pre-Big Bang scenarios.

\paragraph{Comoving orthogonal gauge\,:}

We can choose the spatial coordinates such that the 3-velocity of the fluid cancels\,:
\begin{equation}
\widetilde{v}_{i} = 0 ~~.
\end{equation}
The orthogonality of the constant-$\eta$ hypersurfaces with the 4-velocity $u^{\mu}$ imposes\,:
\begin{equation}
\widetilde{v}_i + \widetilde{B}_i = 0 ~~.
\end{equation}
One then has $\alpha_{com}$ which is fixed but $\beta_{com}$ depends on an arbitrary function $\hat{C}(x^i)$ corresponding to a translation of spatial coordinates.

\paragraph{Total matter gauge\,:}

It is also called the \emph{velocity orthogonal isotropic gauge}. One fixes the temporal part of this gauge by imposing that the \emph{total momemtum potential} cancels\,:
\begin{equation}
\widetilde{v} + \widetilde{B} = 0 ~~.
\end{equation}
The spatial part is fixed by the following\,:
\begin{equation}
\widetilde{E} = \widetilde{F}_{i} = 0 ~~.
\end{equation}

\paragraph{Uniform density gauge\,:}

One fixes the temporal part of this gauge by imposing the following conditions on the density\,:
\begin{equation}
\widetilde{\delta \rho} = 0 ~~.
\end{equation}
It remains a liberty which is the liberty of the spatial part of the gauge. One fixes it by choosing that either $\tilde{B}$, $\tilde{E}$, or $\tilde{v}$ is zero in order to fix $\beta$.

\subsection{Perturbations of Einstein's equations}
\label{AppASec35}

Now we have described the gauge issue and the proper way to deal with it through gauge invariant quantities, we come back on Einstein's equations \rref{EEunperturbed} and explain the basic elements to deal with perturbations of these equations. We attach ourself in particular to describe the properties of the stress-energy tensor as it is important for this thesis.

Let us decompose the Einstein tensor and the stress-energy tensor as
\beq
G^\mu_{~\nu} = G^{(0)\mu}_{~~~~~\nu} + \delta G^\mu_{~~\nu} ~~~~~,~~~~~ T^\mu_{~\nu} = T^{(0)\mu}_{~~~~~\nu} + \delta T^\mu_{~~\nu} ~~~,
\eeq
where the background FLRW universe implies 
\beq
G^{(0)0}_{~~~~0} = 3 \frac{\Hcal^2 + K}{a^2} ~~~~~,~~~~~ G^{(0)0}_{~~~~i} = 0 ~~~~~,~~~~~ G^{(0)i}_{~~~~j} = \frac{2 \Hcal' + \Hcal^2 + K}{a^2} \delta^i_{~j} ~~.
\eeq
and to respect the symmetry we must have
\beq
T^{(0)i}_{~~~~~0} = T^{(0)0}_{~~~~~i} = 0 ~~~~~,~~~~~ T^{(0)i}_{~~~~~j} \propto \delta^i_{~j} ~~~.
\eeq

Using that
\beq
G^{(0)\mu}_{~~~~~\nu} = 8 \pi G \, T^{(0)\mu}_{~~~~~\nu} ~~,
\eeq
we get at first order in perturbations\,:
\beq
\delta G^\mu_{~\nu} = 8 \pi G \, \delta T^\mu_{~\nu} ~~.
\eeq
Nevertheless $\delta G^{\mu}_{~\nu}$ is not gauge invariant in general and thus the RHS and LHS of these last perturbed Einstein equations are not both invariant under gauge transformations.

Under scalar perturbations given by Eq. \rref{GenScalarGaugeTrans}, $\delta G^{\mu}_{~\nu}$ transforms as\,:
\bea
\delta G^0_{~0} &\rightarrow& \widetilde{\delta G}^0_{~0} = \delta G^0_{~0} - (G^{(0)0}_{~~~~0})' \, \xi^0 ~~, \nonumber \\
\delta G^0_{~i} &\rightarrow& \widetilde{\delta G}^0_{~i} = \delta G^0_{~i} + (G^{(0)0}_{~~~~0} - \frac13 \delta G^{k}_{~k}) \xi^0_{~~|i} ~~, \nonumber \\
\delta G^i_{~j} &\rightarrow& \widetilde{\delta G}^i_{~j} = \delta G^i_{~j} - (G^{(0)i}_{~~~~j})' \, \xi^0 ~~,
\eea
where $\delta G^0_{~0}$, $\delta G^0_{~i}$ and $\delta G^i_{~j}$ are complicated expressions of $\phi$, $\psi$, $B$ and $E$. Nevertheless, it is possible to build up gauge invariant perturbations of $\delta G^{\mu}_{~\nu}$ by
\bea
\label{GaugeINvG}
\delta G^{(g.i.)0}_{~~~~~~~0} &=& \delta G^0_{~0} + (G^{(0)0}_{~~~~0})' (B - E') ~~, \nonumber \\
\delta G^{(g.i.)0}_{~~~~~~~i} &=& \delta G^0_{~i} + (G^{(0)0}_{~~~~0} - \frac13 \delta G^{k}_{~k}) (B - E')_{|i} ~~, \nonumber \\
\delta G^{(g.i.)i}_{~~~~~~~j} &=& \delta G^i_{~j} + (G^{(0)i}_{~~~~j})' ( B - E' ) ~~,
\eea
and where
\bea
\label{GpertExplicit}
\delta G^0_{~0} &=& 2 a^{-2} \left[ - 3 \Hcal (\Hcal \Phi + \Psi') + \nabla^2 \Psi + 3 K \Psi + 3 \Hcal (-\Hcal' + \Hcal^2 + K) (B - E') \right] ~~, \nonumber \\
\delta G^0_{~i} &=& 2 a^{-2} \left[ \Hcal \Phi + \Psi' + (\Hcal' - \Hcal^2 - K) (B - E')\right]_{,i} ~~, \nonumber \\
\delta G^i_{~j} &=& - 2 a^{-2} \bigg[ \left( [2\Hcal' + \Hcal^2] \Phi + \Hcal \Phi' + \Psi'' + 2 \Hcal \Psi' - K \Psi + \frac12 \nabla^2 D \right) \delta^i_j \nonumber \\
& & ~~~~~ ~~~~~ ~~~~~ + \left( \Hcal'' - \Hcal \Hcal' - \Hcal^3 - K \Hcal \right) (B-E') \delta^i_j - \frac12 \gamma^{ik} D_{|kj} \bigg] ~~, \nonumber \\
D &\equiv& \Phi - \Psi ~~.
\eea
We can define, in the same way as in Eq. \rref{GaugeINvG}, gauge invariant perturbations of the stress-energy tensor. The same relations hold in the case where $(G^{\tdev}_{~\tdev},\delta G^{\tdev}_{~\tdev})$ is replaced by $(T^{\tdev}_{~\tdev},\delta T^{\tdev}_{~\tdev})$. Doing so, we get the equality
\beq
\delta G^{(g.i.)\mu}_{~~~~~~~\nu} = 8 \pi G \, \delta T^{(g.i.)\mu}_{~~~~~~~\nu} ~~,
\eeq
where both the LHS and the RHS are now gauge invariant. We can also write down this last equation in an explicit form. In that case, we can easily see by going to the longitudinal gauge that the general gauge invariant expression $\delta G^{(g.i.)\mu}_{~~~~~~~\nu}$ is just given by Eq. \rref{GpertExplicit} where we set $B = E = 0$. The perturbation $\delta T^{(g.i.)\mu}_{~~~~~~~\nu}$ is on the other hand given by its \rref{GaugeINvG}-like form. Finally, we need to close the system by writing the EoMs of the universe content in a gauge invariant way.

\subsection{Gauge invariant EoMs}
\label{AppASec36}

The most general expression of the first order scalar perturbation of the stress-energy tensor involves 4 scalar $\delta \rho$, $\delta p$, $v$ and $\sigma$\,:
\beq
\delta T^\mu_{~\nu} = \left( \begin{array}{cc} \delta \rho & -(\rho_0 + p_0) a^{-1} v_{,i} \\ (\rho_0 + p_0) a^{-1} v_{,i} & - \delta p \delta_{ij} + \sigma_{|ij} \end{array} \right) ~~.
\eeq
If we are only dealing with a perfect fluid, $\sigma$ which corresponds to the anisotropic stress is zero and if the fluid is dissipationless we have the general expression for $T^\mu_{~\nu}$\,:
\beq
T^\mu_{~\nu} = (\rho + p) u^\mu u_\nu - p \, \delta^\mu_{~\nu} ~~.
\eeq
where in general $p = p(\rho,S)$ (the equation of state), with $S$ is the entropy per baryon ratio, and\,:
\beq
\delta p = \left(\frac{\partial p}{\partial \rho}\right)_{S} \delta \rho + \left(\frac{\partial p}{\partial S}\right)_{\rho} \delta S \equiv c_s^2 \delta \rho + \tau \delta S ~~.
\eeq

In conformal time, the scalar perturbations $\delta T^\mu_{~\nu}$ can thus be written as
\beq
\delta T^0_{~0} = \delta \rho ~~,~~ \delta T^0_{~i} = (\rho_0 + p_0) a^{-1} \delta u_i ~~,~~ \delta T^i_{~j} = - \delta p \, \delta^i_j ~~.
\eeq
To find the gauge invariant perturbations $\delta T^\mu_{~~\nu}$, we use the gauge invariant expressions of $\delta \rho$, $\delta p$ and $\delta u_i$\,:
\bea
\delta \rho^{(g.i.)} &=& \delta \rho + \rho_0' (B - E') ~~, \nonumber \\
\delta p^{(g.i.)} &=& \delta p + p_0' (B - E') ~~, \nonumber \\
\delta u_i^{(g.i.)} &=& \delta u_i + a (B - E')_{|i} ~~.
\eea
We can notice that these gauge invariant expressions correspond to the expressions of $\delta \rho$, $\delta p$ and $\delta u_i$ in the longitudinal gauge, as expected. We thus have
\beq
\delta T^{(g.i.)0}_{~~~~~~~0} = \delta \rho^{(g.i.)} ~~,~~ \delta T^{(g.i.)0}_{~~~~~~~i} = (\rho_0 + p_0) a^{-1} \delta u_i^{(g.i.)} ~~,~~ \delta T^{(g.i.)i}_{~~~~~~~j} = - \delta p^{(g.i.)} \delta^i_{~j} ~~.
\eeq

We find the gauge invariant perturbed Einstein equations to be equivalent to\,:
\bea
- 3 \Hcal (\Hcal \Phi + \Psi') + \nabla^2 \Psi + 3 K \Psi &=& 4 \pi G a^2 \delta \rho^{(g.i.)} ~~, \nonumber \\
\left( \Hcal \Phi + \Psi' \right)_{,i} &=& 4 \pi G (\rho_0 + p_0) \delta u_i^{(g.i.)} ~~, \nonumber \\
\left( [2\Hcal' + \Hcal^2] \Phi + \Hcal \Phi' + \Psi'' + 2 \Hcal \Psi' - K \Psi + \frac12 \nabla^2 D \right) \delta^i_{~j} - \frac12 \gamma^{ik} D_{|kj}&=& 4 \pi G a^2 \delta p^{(g.i.)} \delta^i_{~j} ~.~~~
\eea
From the $(i,j\neq i)$ component of $T^\mu_{~~\nu}$, one can show that we need the condition $\Phi = \Psi$ (and thus $D = 0$)\footnote{See \cite{1992PhR...215..203M}, p. 223, for more details on this statement. We should notice that the notion of the vanishing of averaging over a spatial hypersurface at first order is mentioned to justify this condition.}. We hence obtain
\bea
\nabla^2 \Phi - 3 \Hcal \Phi' - 3 (\Hcal^2 - K) \Phi &=& 4 \pi G a^2 \delta \rho^{(g.i.)} ~~, \nonumber \\
\left( a \Phi \right)_{,i}' &=& 4 \pi G (\rho_0 + p_0) \delta u_i^{(g.i.)} ~~, \nonumber \\
\Phi'' + 3 \Hcal \Phi' + ( 2\Hcal' + \Hcal^2 - K) \Phi &=& 4 \pi G a^2 \delta p^{(g.i.)} ~~~.
\eea
We notice that the first of these equations is the generalization of the Poisson equation in the case of an expanding background. Another important fact about this equation is that it is valid only in its linear regime, i.e. for $|\Phi| \ll 1$, which does not imply $|\delta \rho / \rho| \ll 1$ and so this equation can be valid even for high density contrasts (such as in clusters of galaxies or on super-horizon scales).
We directly get the general expression of the gauge invariant perturbations of the matter density and its peculiar velocity (for dust matter, the pressure is negligible and so not interesting)\,:
\bea
\frac{\delta \rho^{(g.i.)}}{\rho_0} &=& 2 \frac{\nabla^2 \Phi - 3 \Hcal \Phi' - 3 (\Hcal^2 - K) \Phi}{3 (\Hcal^2 + K)} ~~, \nonumber \\
\delta u^{(g.i.)i} &=& - \frac{( a \Phi)_{,i}'}{a^2 (\Hcal^2 - \Hcal' + K)} ~~.
\eea

\newpage

\section{The power spectrum in the linear regime}
\label{AppASec4}

\subsection{Basic facts}
\label{AppASec41}

Let us recall here the notions that we presented in Sec. \ref{Ch1Sec42} in order to illustrate and complete them. We will then, in Sec. \ref{AppASec5}, extend these notions to the non-linear spectrum.

In synchronous gauge, the Poisson equation gives a relation between the gravitational (Bardeen) potential and the matter density contrast which is given by\,:
\beq
\nabla^2 \psi(x) = 4 \pi G \rho(t) \, \delta(x) ~.
\eeq
When we go in Fourier space, assuming a flat FLRW model, this equation becomes\,:
\beq
\label{PoissonInFourierSpace}
|\psi_k|^2 = \left(\frac{3}{2}\right)^2 \frac{\Om_{m}^2(z) \Hcal^4}{k^4} |\delta_k|^2 ~~~~~ \mbox{ where } ~~~~~ \Om_m(z) = \Om_{m0} (1+z) \left( \frac{\Hcal_0}{\Hcal} \right)^2 ~~.
\eeq
The dimensionless power spectrum for the potential is defined by
\beq
\Pcal_\psi (k,\eta) \equiv \frac{k^3}{2 \pi^2} |\psi_k(\eta)|^2 ~~,
\eeq
and effectively built up from the primordial inflationary spectrum $\left(\frac{3}{5}\right)^2 \Delta_{\cal R}^2$ and the addition of a transfer function $T(k)$, to describe the sub-horizon evolution of the gravitational potential during the radiation era, and a growth factor $g(z)$ to account for its evolution during the (recent) dark energy dominated era. Consequently, one has the expression for the primordial inflationary spectrum\,:
\beq
\Delta_{\cal R}^2=A \left(\frac{k}{k_0}\right)^{n_s-1} ~~\mbox{ and }~~~ \left( A, n_s, k_0/a_0 \right) = \left( 2.45 ~10^{-9} ~,~ 0.96 ~,~ 0.002 ~{\rm Mpc}^{-1} \right) ~~,
\eeq
and the linear power spectrum is finally given by
\beq
\label{PsiPch8}
\Pcal_\psi (k) = \left(\frac{3}{5}\right)^2 \Delta_{\cal R}^2 T^2(k) \left(\frac{g(z)}{g_{\infty}}\right)^2 ~~.
\eeq

\subsection{Transfer function}
\label{AppASec42}

From the study of Eisenstein \& Hu \cite{Eisenstein:1997ik}, we can approximate the true \gls{transfer function} $T(k)$ with an effective shape in the case of density perturbations with no baryonic effects involved (such as Silk damping or acoustic oscillations). In other words, all the baryonic matter is considered as being dark matter. We then write $T(k)=T_0(k)$ with\,:
\beq
\label{EH97App}
T_0(q) = \frac{L_0}{L_0+q^2C_0(q)}  ~, ~~~~ L_0(q) = \ln(2e+1.8q) ~, 
~~~~ C_0(q) = 14.2+\frac{731}{1+62.5q} ~, ~~~~q  = \frac{k}{13.41 k_{\rm eq}} ~~,
\eeq
where $k_{\rm eq} = (\Omega_{m0} h^2/13.41) \, {\rm Mpc}^{-1}$ is the scale corresponding to matter-radiation equality and $h\equiv H_0/(100 \,{\rm km \,s}^{-1} {\rm Mpc}^{-1})$. In the following we will use $h=0.7$ and set $a_o =1$. The numerical value of $k_{\rm eq}$ is then $k_{\rm eq}\simeq 0.036 ~{\rm Mpc}^{-1}$ for $\Omega_{m0} =1$ (CDM) and $k_{\rm eq}\simeq 0.010 ~{\rm Mpc}^{-1}$ for $\Omega_{m0} =0.27$ ($\Lambda$CDM). We can easily check that the above transfer function goes to 1 for $k \ll k_{\rm eq}$,  while it falls like  $k^{-2}\log k$ for $k \gg k_{\rm eq}$.

We can complicate the description of the transfer function by including the effect of a baryonic fraction in $\Omega_{m0}$. In that case the transfer function is suppressed by the so-called \gls{Silk damping}, leading to a decrease of 10-40\%  of the transfer function in the range $k \gaq 0.01 h \,{\rm Mpc}^{-1}$.  In order to take into account this baryonic contribution we have to replace the value of $q$ used in the previous section, i.e. $q  = k/(13.41 k_{\rm eq})$, with the more accurate value given by \cite{Eisenstein:1997ik}\,:
\bea
\label{EH97eff}
q&=&\frac{k \times {\rm Mpc}}{h \Gamma} ~~, \nonumber \\
\Gamma&=&\Omega_{m0}h\left(\alpha_{\Gamma}+\frac{1-\alpha_{\Gamma}}{1+(0.43 k s)^4}\right)~, \nonumber \\
\alpha_{\Gamma}&=&1-0.328\ln(431 \Omega_{m0}h^2)\frac{\Omega_{b0}}{\Omega_{m0}}+0.38 \ln(22.3 \Omega_{m0} h^2)\left(\frac{\Omega_{b0}}{\Omega_{m0}}\right)^2~, \nonumber \\
s&=&\frac{44.5 \ln(9.83/\Omega_{m0}h^2)}{\sqrt{1+10(\Omega_{b0}h^2)^{3/4}}} \,{\rm Mpc} ~~.
\eea
Here $\Omega_{b0}$ is the baryon density parameter, $s$ is the sound horizon and $\Gamma$ is the $k$-dependent \gls{effective shape parameter}. For the numerical values, we can take $\Omega_{\Lambda 0}=0.73$, $\Omega_{m0}=0.27$, $\Omega_{b0}=0.046$, and $h=0.7$.

\begin{figure}[ht!]
\centering
\includegraphics[width=10cm]{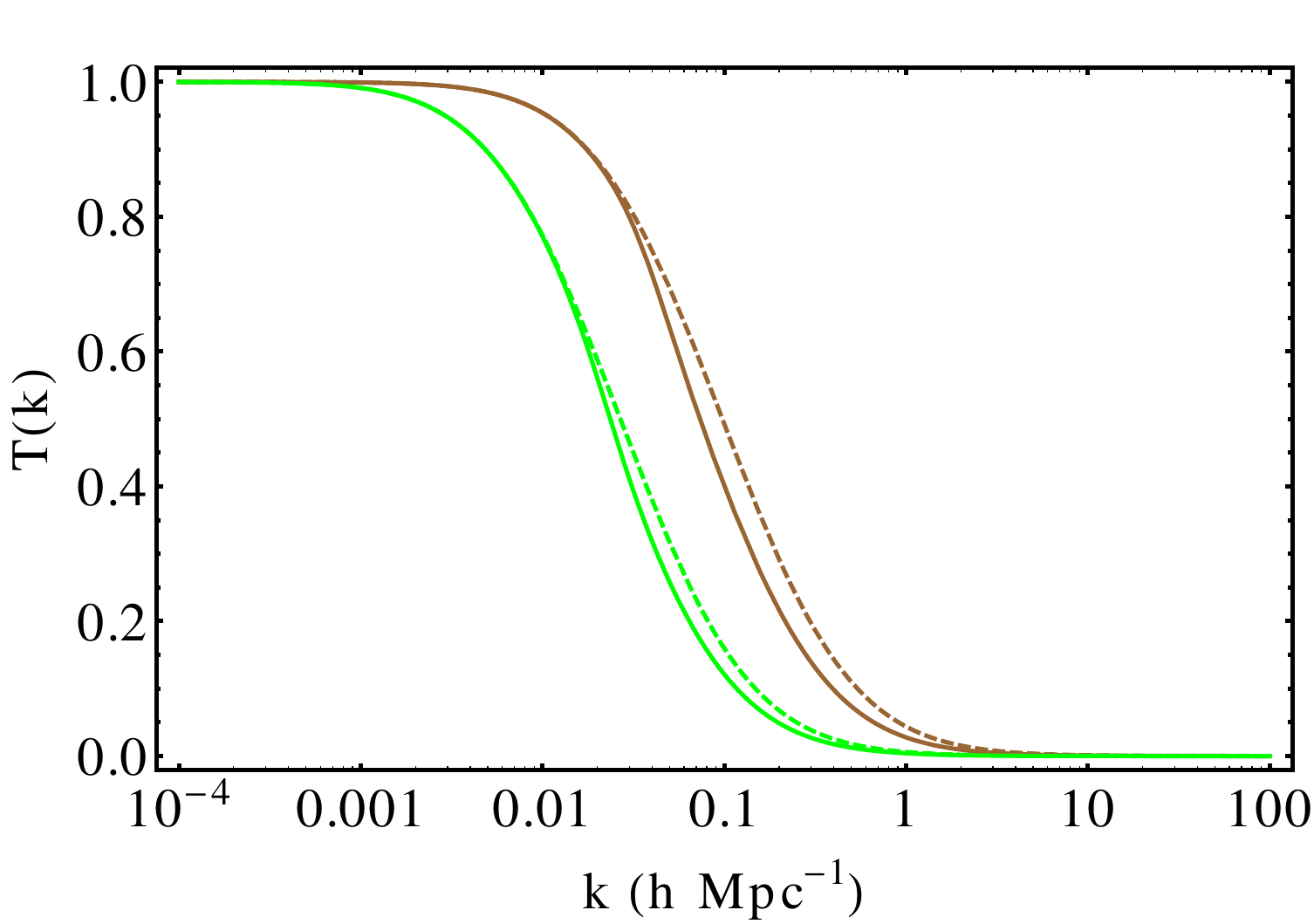}
\centering
\caption[The transfer function $T(k)$]{The transfer function used in the elaboration of the linear power spectrum $\Pcal_\psi(k)$ in different situations\,: in CDM for $\Omega_{m0} = 1$ (brown) and in $\Lambda$CDM with $(\Omega_{m0}, \Omega_{\Lambda 0}) = (0.27,0.73)$ (green). There is no $z$-dependence in $T(k)$. Plots are represented in the simple case of dark matter only (dashed), Eq. \rref{EH97App}, and in the case including a baryonic contribution (solid), Eq. \rref{EH97eff}.}
\label{FigT}
\end{figure}

We could go further in complexifying the expression of this transfer function by including the effect of baryonic oscillations. Nevertheless, to avoid unnecessary complications, and because we have not used it in our work, they are neglected here. Above considerations regarding the two different transfer functions can easily be verified in Fig. \ref{FigT}.

\subsection{The growth factor g(z)}
\label{AppASec43}

In the expression \rref{PsiPch8}, the growth factor $g(z)$ is taking into account the time dependence of $\psi_k$ due to the late dark energy phase of expansion, i.e.\,:$$\psi_k(z) = \frac{g(z)}{g(0)} \psi_k(0)$$ and its general expression in the case of $\Lambda$CDM is (see \cite{Lahav:1991wc,Carroll:1991mt})\,:
\bea
\label{gOfz}
g(z) = \frac52 \Om_m(z) \, g_{\infty} \left[ \Om_m(z)^{4/7} - \Om_\Lambda(z) + \left(1 + \frac{\Om_m(z)}{2} \right) \left(1 + \frac{1}{70} \Om_\Lambda(z) \right) \right]^{-1} \\
\label{OmOlInLCDM}
\mbox{ with }~ \Omega_m(z) = \frac{\Omega_{m0} (1+z)^3}{\Omega_{m0} (1+z)^3 + \Omega_{\Lambda0}} ~\mbox{ and }~ \Omega_\Lambda(z) = \frac{\Omega_{\Lambda0}}{\Omega_{m0} (1+z)^3 + \Omega_{\Lambda0}} ~~,
\eea
where the limit $z \rightarrow 0$ gives $g(0) = 0.7584 ~ g_{\infty}$.
In the limiting case $(\Om_{m 0} \rightarrow 1, \Om_{\Lambda 0} \rightarrow 0)$, that is to say a CDM model, we get $g(z) = g_{\infty} ~\forall~ z$. We can thus take $g_{\infty} = 1$ if we want, the important thing being that $g(z) / g(0) = 1$.

\subsection{Plot of the linear power spectrum}
\label{AppASec44}

\begin{figure}[ht!]
\centering
\includegraphics[width=10cm]{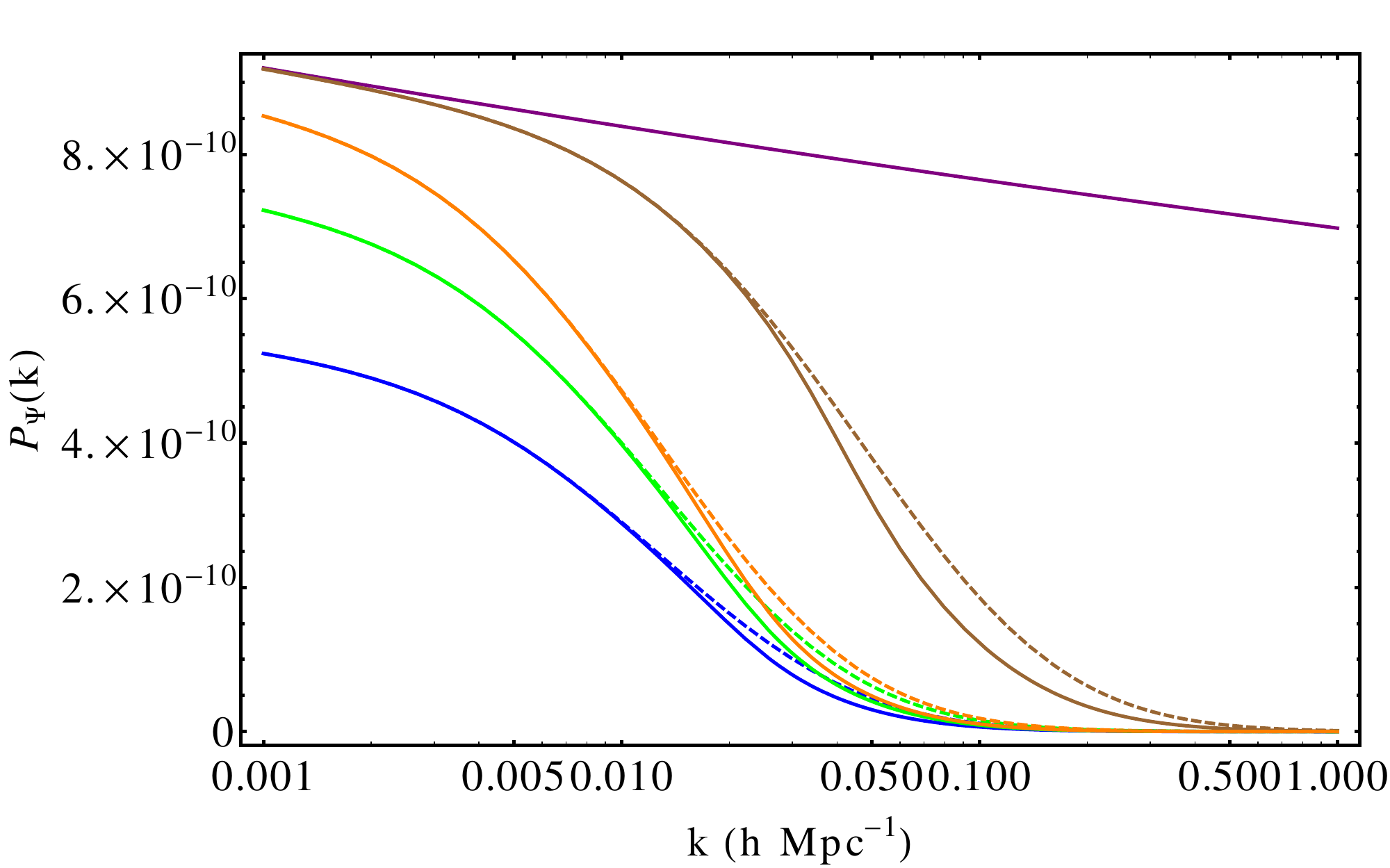}
\centering
\caption[The linear gravitational power spectrum $\Pcal_{\L}(k)$]{The linear gravitational power spectrum in different situations\,: CDM (brown) ; $\Lambda$CDM for $z=0$ (blue), $z=0.5$ (green), and $z=1.5$ (orange). It can be compared to the primordial power spectrum from inflation (purple). Dotted (solid) lines correspond to the use of the transfer function without (with) baryonic effect.}
\label{FigLPSeff}
\end{figure}

We have all the elements to draw the power spectrum for the gravitational potential, in the linear regime of perturbations, for different redshifts. This is what is done in Fig. \ref{FigLPSeff}. We remark from these plots that $\Pcal_\psi(k)$ decreases in magnitude when $z$ decreases (a modulation entirely due to the factor $g(z)$ as it can be seen in Eq. \rref{PsiPch8}). In the case of the matter density spectrum, it would be the contrary and this change in behavior is due to the $(1 + z)^2$ factor, coming from $\Omega_m^2(z)$, in Eq. \rref{PoissonInFourierSpace}.

\section{The HaloFit model}
\label{AppASec5}

In this section we will describe the HaloFit model which brings an effective description of the power spectrum in its non-linear regime. This model was first established by \cite{Smith:2002dz} and recently revisited in \cite{Takahashi:2012em} in the case of more accurate N-body simulations.

\subsection{General considerations\,: density power spectrum}
\label{AppASec51}

Let us recall here the basic considerations about the density power spectrum, already shown in Sec. \ref{Ch1Sec43}. The density spectrum is defined by the \gls{density variance}\,:
\beq
\label{DensityPS}
\sigma^2 \equiv \overline{\delta(x)\delta(x)} = \int \frac{d^3k}{(2 \pi)^3} |\delta_{\vec{k}}|^2 = \int \frac{k^2 dk}{2 \pi^2} |\delta_k|^2 = \int \Delta^2(k) ~ d\ln k ~~.
\eeq
This expression leads to the direct definition in Fourier space\,:
\beq
\label{Deltadelta}
\Delta^2(k,z) = \frac{k^3}{2 \pi^2} |\delta_k(z)|^2 = \frac{d \sigma^2}{d \ln k} ~~,
\eeq
showing that $\Delta(k,z)$ is a dimensionless quantity.

This density spectrum is related to the gravitational potential spectrum by the Poisson equation \rref{PoissonInFourierSpace} which is valid both in the linear and non-linear regimes\,:
\beq
\label{PEApp}
\Pcal_\psi^{\rm{L,NL}}(k) = \frac{9}{4} \frac{\Om_{m0}^2 \Hcal_0^4}{k^4} (1+z)^2 \Delta_{\rm{L,NL}}^2(k) ~~.
\eeq
Using the linear power spectrum given by Eq. \rref{PsiPch8}, one gets the expression for the density spectrum in the linear regime\,:
\beq
\label{OurDl}
\Delta_{\rm{L}}^2(k) = \frac{4}{9} \frac{k^4}{\Om_{m0}^2 \Hcal_0^4 (1+z)^2} \Pcal_\psi^{\rm{L}}(k) = \frac{4}{25} \frac{A \Om_{m0}^{-2} \Hcal_0^{-4}}{(1+z)^2} ~ \frac{k^{n_s+3}}{k_0^{n_s-1}} \left(\frac{g(z)}{g_0}\right)^2 T^2(k) ~~.
\eeq

We will now use the HaloFit model to generate from $\Delta_{\L}^2(k)$ the density power spectrum in its non-linear regime, i.e. $\Delta_{\NL}^2(k)$. As a last step, we will use the Poisson equation in the non-linear regime to obtain $\Pcal_\psi^{\rm{NL}}(k)$, the quantity of interest for our work.

\subsection{The NLPS from the LPS\,: the HaloFit Model}
\label{AppASec52}

The HaloFit model decomposes the matter density spectrum into a sum of two contributions:
\beq
\label{HaloFitDecom}
\Delta^2_{\NL}(k)=\Delta^2_{\Q}(k)+\Delta^2_{\H}(k) ~~,
\eeq
where roughly one settles the linear behaviour and the other deals with the non-linear regime of inhomogeneities.
Indeed, the first term in the RHS of Eq. \rref{HaloFitDecom}, $\Delta^2_{\Q}(k)$, is called the \gls{two-halo term} and is taking into account matter correlations at large distances (low $k$). The other one, $\Delta^2_{\H}(k)$, is the \gls{one-halo term} and it describes the non-linear behaviour of matter inside the halo region.
These two contributions are parametrized \cite{Smith:2002dz} by\,:
\bea
\label{DqDh}
\Delta^2_{\Q}(k) &=& \Delta^2_{\rm{L}}(k) \left[\frac{(1+\Delta_{\rm{L}}^2(k))^{\beta_n}}
{1+\alpha_n\Delta^2_{\rm{L}}(k)} \right] e^{-f(y)} ~~, \nonumber \\
\Delta^2_{\H}(k) &=& {\Delta^{2\ \prime}_{\H }(k) \over 1 + \mu_n y^{-1}+\nu_n y^{-2}} ~~\mbox{ with }~~
\Delta^{2\ \prime}_{\H}(k) = \frac{a_n ~ y^{3f_1(\Omega_{m 0})}}{1+b_ny^{f_2(\Omega_{m 0})} +\left[ c_n f_3(\Omega_{m 0}) ~ y \right]^{3-\gamma_n}} ~~,~~~~~
\eea
where $y\equiv k/k_{\sigma}$ and $f(y)=y/4+y^2/8$.
We can easily understand from its expression that $\Delta^2_{\Q}(k)$ is made to fit the linear power spectrum $\Delta^2_{\rm{L}}(k)$ when it takes small values (i.e. at large scales) and change it when it becomes of order 1 (i.e. when it reaches the weak non-linear regime). It is then rapidly negligible at $k>k_{\sigma}$ (defined later) because of its exponential suppression factor.
The other contribution $\Delta^2_{\H}(k)$ is taking into account the non-linear aspect of the power spectrum in a way which is more difficult to analyse analytically, but which is physically expected to raise the power spectrum (to enhance the strength of non-linear scales).

We can see in these expressions of Eq. \rref{DqDh} that a lot of parameters come into play. The first parameters that we can notice are functions of the cosmology in ${\Delta_{\H}^2}'(k)$, namely $\{f_i\}_{i=1,2,3}$. These $\Omega_{m 0}$-dependent functions are parametrized as follows (see \cite{Smith:2002dz})\,:
\beq
\left. 
\begin{array}{lll}
f_{1a}(\Omega_{m 0}) & = & (\Omega_{m 0})^{\;-0.0732} \\
f_{2a}(\Omega_{m 0}) & = & (\Omega_{m 0})^{\;-0.1423}\\
f_{3a}(\Omega_{m 0}) & = & (\Omega_{m 0})^{\;0.0725}\\
\end{array}
\right\} ~\mbox{ for }~ \Omega_{m 0} \le 1
~,~
\left. 
\begin{array}{lll}
f_{1b}(\Omega_{m 0}) & = & (\Omega_{m 0})^{\;-0.0307}\\
f_{2b}(\Omega_{m 0}) & = & (\Omega_{m 0})^{\;-0.0585}\\
f_{3b}(\Omega_{m 0}) & = & (\Omega_{m 0})^{\;0.0743}\\
\end{array}
\right\} ~\mbox{ for }~ \Omega_{m 0} + \Omega_{\Lambda 0} = 1 ~.
\eeq
We naturally choose this last (b) case where $\Omega_{m 0} + \Omega_{\Lambda 0} = 1$ to be consistent with the $\Lambda$CDM model\footnote{For models in which $\Omega_{\Lambda 0}$ is neither zero nor $1-\Omega_{m 0}$, \cite{Smith:2002dz} suggests to interpolate the functions $f_i$ linearly in $\Omega_{\Lambda 0}$ between the open and flat cases. This is not our case.}. This is also the case adopted by \cite{Takahashi:2012em}.

We can see that the Eq. \rref{DqDh} includes an additional parameter which is $k_\sigma$. In fact this parameter is involved in the function $e^{-f(y)}$ and it basically sets the non-linear scale. It is also present in all the other parameters that we will describe in the next subsection. It is thus a very important quantity.
For this we redefine the variance of the density perturbation presented in Eq. \rref{DensityPS}, applying to it a Gaussian filtering at a physical scale $R_{\rm G}$\,:
\beq
\label{sigmasquaredGF}
\sigma^2(R_{\rm G}) \equiv \int \Delta^2_{\rm{L}}(k,z) \, \exp(-k^2 R_{\rm G}^2) ~ d\ln k ~~,
\eeq
we then define $k_\sigma$ by normalizing this variance to 1\,:
\beq
\label{sigmasquared}
\sigma^2(R_{\rm G}= k_\sigma^{-1},z) \equiv \int \Delta^2_{\rm{L}}(k,z) \, \exp(-(k/k_\sigma)^2)\;d\ln k \equiv 1 ~~.
\eeq
In practice we compute this integral by incrementing the value of $k_\sigma$ until we reach a value for $\sigma^2$ which is close enough to unity. It is also obvious that the $\Lambda$CDM case will produce a linear spectrum $\Delta_{\L}^2(k,z)$ that depends on the redshift. As a consequence $k_\sigma$ will also be a redshift dependent quantity and so will be the $X_n$ parameters of the fit that we are now going to present.

\subsection{Two parametrizations\,: Smith \etal \& Takahashi \etal}
\label{AppASec53}

The other coefficients, namely $X_n \equiv \{a_n,b_n,c_n,\gamma_n,\alpha_n,\beta_n,\mu_n,\nu_n\}$, are found by fitting simulations and give a really accurate effective description. They are expressed as polynomials of two quantities\,: an effective spectral index $n_{\rm eff}$ and a parameter $C$ controling the curvature of the spectral index.

This \gls{effective spectral index} is given by
\beq
3+n_{\rm eff} \equiv -\left.{d \ln \sigma^2(R) 
\over d \ln R}\right|_{\sigma=1} ~~,
\eeq
and it basically describes the deviation of the linear power spectrum $\Delta_{\rm L}^2(k)$ from a power law $\sim k^{n+3}$. In our case, we start with a linear density spectrum given by Eq. \rref{OurDl} and it will clearly not be a power law.

The \gls{curvature of the spectral index} $C$ is by definition\footnote{We can show that $C=0$ when the linear matter spectrum is a power law ($\Delta_{\rm L}^2(k) \sim k^{n+3}$). In this case we get\,:
\beq
C \equiv - 2 (n+3) G(n) + 4 G(n)^2 ~~ \mbox{ with } ~~ G(n) \equiv \Gamma\left( \frac{n+5}{2} \right) \bigg/ \Gamma\left( \frac{n+3}{2} \right) = \frac{n + 3}{2} ~~,
\eeq
and that is actually zero.}
:
\bea
C &\equiv& - \left.{d^2 \ln \sigma^2(R) \over d \ln R^2}
\right|_{\sigma=1} = \left\{ - \frac{d}{d \ln R} \left[ \frac{R}{\sigma^2} \frac{d \sigma^2(R)}{d R} \right] \right\}_{R = k_\sigma^{-1}} \nonumber \\
&=& \left\{ - \frac{R}{\sigma^2} \frac{d \sigma^2}{d R} + \frac{R^2}{(\sigma^2)^2} \left(\frac{d \sigma^2}{d R}\right)^2 - \frac{R^2}{\sigma^2} \frac{d^2 \sigma^2}{d R^2} \right\}_{R = k_\sigma^{-1}} ~~.
\label{EqC}
\eea
We understand that both $n_{\rm eff}$ and $C$ will have a dependence on $z$, for the simple reason that they depend on $k_\sigma$. Their value is computed numerically after having computed the $d/d R$ derivatives analytically (these derivatives acting only in the exponential of Eq. \rref{sigmasquaredGF}) and replaced $\Delta_{\L}^2(k,z)$ by its explicit form of Eq. \rref{OurDl}.

In \cite{Smith:2002dz}, the $X_n$ coefficients are evaluated for a certain set of initial power spectra (which, we recall, are taken as power laws of $k$) and are found to be\footnote{We should note that \cite{Smith:2002dz} study $\Delta_{\rm L}^2 \sim k^{n+3}$ with a wide range of values for $n$, between $0$ and $-2$, as they are believed to bring a self-similarity behaviour. Spectra outside this range are not well tested and thus not considered. Also, the fact that we (will) obtain values of $n_{\rm eff}$ which are between $0$ and $-2$, by the use of $\Delta_L^2(k)$ given in Eq. \rref{OurDl}, is a confirmation that we are indeed authorized to use these expressions for $X_n$ in our study. From \cite{Smith:2002dz} we also learn that terms up to $n^4$ in the fit of $a_n$ are especially required in order to describe the rapid rise in amplitude of the halo term for $n<-2$.}\,:
\bea
\label{ParamHaloFitSmith}
\log_{10} a_n &=& ~~ 1.4861 + 1.8369 ~ n_{\rm eff} + 1.6762 ~ n_{\rm eff}^2 + 0.7940 ~ n_{\rm eff}^3 + 0.1670 ~ n_{\rm eff}^4 - 0.6206 ~ C ~~, \nonumber \\
\log_{10} b_n &=& ~~ 0.9463 + 0.9466 ~ n_{\rm eff} + 0.3084 ~ n_{\rm eff}^2 - 0.9400 ~ C ~~, \nonumber \\
\log_{10} c_n &=& - 0.2807 + 0.6669 ~ n_{\rm eff} + 0.3214 ~ n_{\rm eff}^2 - 0.0793 ~ C ~~, \nonumber \\
\gamma_n &=& ~~ 0.8649 + 0.2989 ~ n_{\rm eff} + 0.1631 ~ C ~~, \nonumber \\
\alpha_n &=& ~~ 1.3884 + 0.3700 ~ n_{\rm eff} - 0.1452 ~ n_{\rm eff}^2 ~~, \nonumber \\
\beta_n &=& ~~ 0.8291 + 0.9854 ~ n_{\rm eff} + 0.3401 ~ n_{\rm eff}^2 ~~, \nonumber \\
\log_{10} \mu_n &=& - 3.5442 + 0.1908 ~ n_{\rm eff} ~~, \nonumber \\
\log_{10}\nu_n &=& ~~ 0.9589 + 1.2857 ~ n_{\rm eff} ~~.
\eea
With these parameters (and others previously described), \cite{Smith:2002dz} obtain a deviation between the model and N-body simulations at scales $k<10\hompc$ which is always $< 3\%$, up to $z<3$.

In a recent paper, Takahashi \etal \cite{Takahashi:2012em} have reconsidered the halo model with a greater accuracy in simulations. This new study found the following parametrization of these parameters\,:
\bea
\label{ParamHaloFitTakahashi}
\log_{10} a_n &=& ~~ 1.5222 + 2.8553 ~ n_{\rm eff} + 2.3706 ~ n_{\rm eff}^2 + 0.9903 ~ n_{\rm eff}^3 + 0.2250 ~ n_{\rm eff}^4 - 0.6038 ~ C \nonumber \\
& & ~~ + 0.1749 ~ \Omega_{DE}(z) ( 1 + w_{DE} ) \nonumber \\
\log_{10} b_n &=& -0.5642 + 0.5864 ~ n_{\rm eff} + 0.5716 ~ n_{\rm eff}^2 - 1.5474 ~ C \nonumber \\
& & ~~ + 0.2279 ~ \Omega_{DE}(z) ( 1 + w_{DE} ) \nonumber \\
\log_{10} c_n &=& ~~ 0.3698 + 2.0404 ~ n_{\rm eff} + 0.8161 ~ n_{\rm eff}^2 + 0.5869 ~ C \nonumber \\
\gamma_n &=& ~~ 0.1971 - 0.0843 ~ n_{\rm eff} + 0.8460 ~ C \nonumber \\
\alpha_n &=& ~~ | 6.0835 + 1.3373 ~ n_{\rm eff} - 0.1959 ~ n_{\rm eff}^2 - 5.5274 ~ C | \nonumber \\
\beta_n &=& ~~ 2.0379 - 0.7354 ~ n_{\rm eff} + 0.3157 ~ n_{\rm eff}^2 + 1.2490 ~ n_{\rm eff}^3 + 0.3980 ~ n_{\rm eff}^4 - 0.1682 ~ C \nonumber \\
\mu_n &=& ~~ 0 ~~, \nonumber \\
\log_{10}\nu_n &=& ~~ 5.2105 + 3.6902 ~ n_{\rm eff} ~~,
\eea
where a new parameter, namely the dark energy density parameter $\Omega_{DE}(z)$, has been introduced. These $\Omega_{DE}(z)$-terms are redshift dependent and represent a possible deviation from a cosmological-constant nature of the acceleration. In our case (i.e. $\Lambda$CDM), it is naturally set to zero as $w_{DE} = -1$. This new parametrization is believed to be accurate up to very small scales\,: $k \leq 30 \hompc$. More precisely, \cite{Takahashi:2012em} estimates its precision to be $\sim 5 \%$ for $0 \leq z \leq 10$ and $k \leq 1 \hompc$, and to be $\sim 10 \%$ for $0 \leq z \leq 3$ and $1 \leq k \leq 10 \hompc$.

\subsection{Results on the density spectrum}
\label{AppASec54}

We will present in the following three subsections (the current one, and the next two) the results of the computation of the power spectra at the non-linear level, using the first expression for the tranfer function $T(k)$, i.e. $T_0(k)$ which does not include any baryonic effect.

\begin{figure}[ht!]
\centering
\includegraphics[width=15cm]{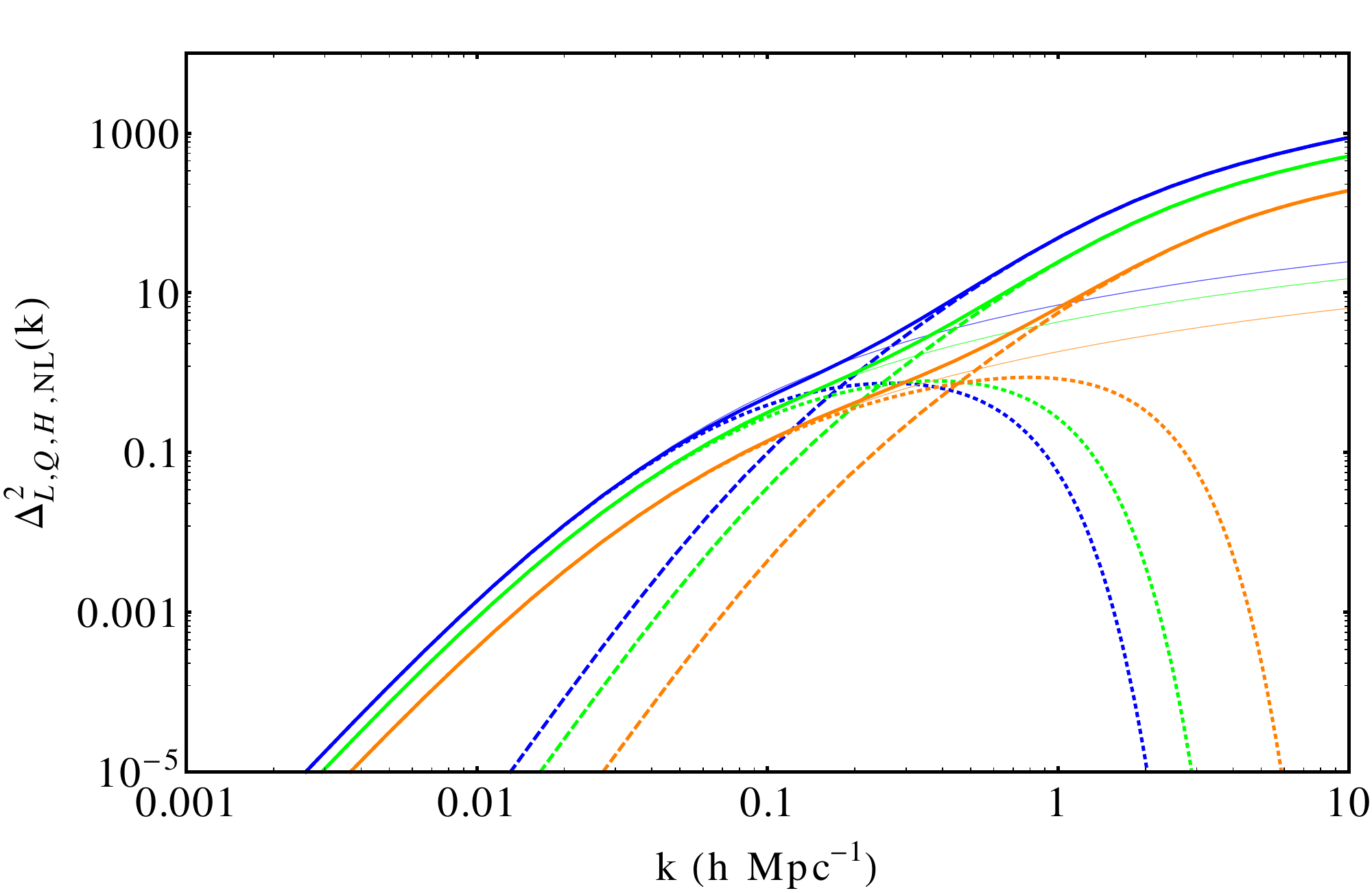}
\centering
\caption[The linear and non-linear density spectra deduced from the HaloFit model]{The linear and non-linear density spectra deduced from Smith's method is shown for $\Lambda$CDM at 3 different redshifts (z=0\,: blue ; z=0.5\,: green ; z = 1.5\,: orange). For each of them, $\Delta_{\rm{L}}^2(k)$ is in thin light color (always increasing with $k$ and increasing when $z$ decreases, as expected), $\Delta_{\rm{Q}}^2(k)$ is in dotted line (dominant at small $k$ and exponentially suppressed at $k \sim 1-10 \hompc$), $\Delta_{\rm{H}}^2(k)$ is in dashed line and contributes only to the NL regime (dominant at large k), $\Delta_{\rm{NL}}^2(k)$ is in solid line. These plots should be compared with Fig. 14 \& 15 in \cite{Smith:2002dz}.}
\label{FigNLPS}
\end{figure}

The different contributions $\{\Delta^2_{\L}(k),\Delta^2_{\H}(k),\Delta^2_{\Q}(k),\Delta^2_{\NL}(k)\}$ can be seen in Fig. \ref{FigNLPS} for the case of Smith \etal parametrization. It confirms the different properties that have been described previously. We could also obtain the same quantities from the parametrization of Takahashi \etal, but the plots would basically be the same. Nevertheless, this latter parametrization is not adapted to the CDM case. For these two reasons, we do not present the plot from Takahashi \etal parametrization. However, we will present its results on the final gravitational power spectrum. One may be astonished by the color ordering in Fig. \ref{FigNLPS} as it was the blue spectrum for gravitational potential which was the lowest one. Nevertheless, this tendency of the $\Lambda$CDM case is changed when we go to the spectrum of density contrast, and is due to the factor $(1+z)^2$ in the Poisson equation.

\begin{figure}[ht!]
\centering
\includegraphics[width=8cm]{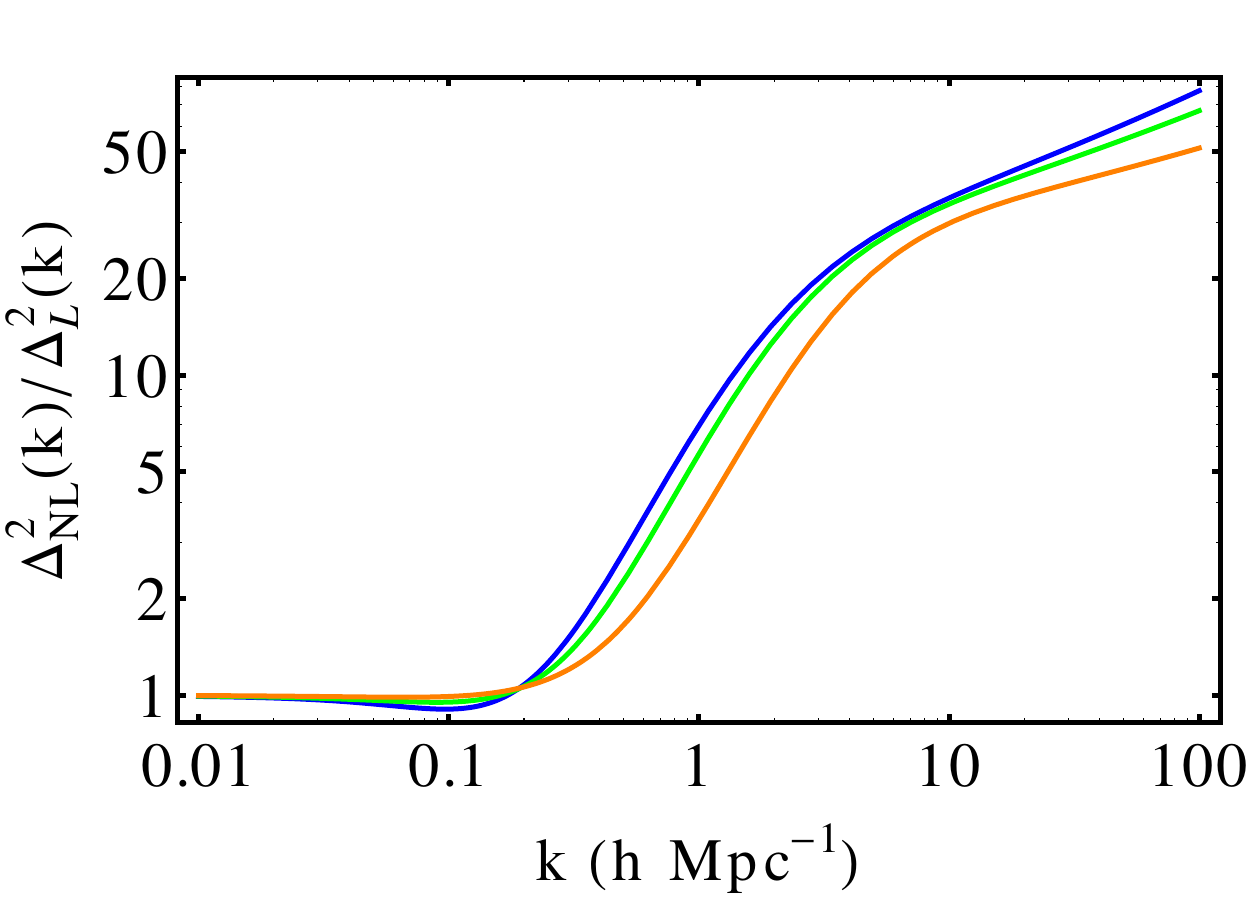}
\includegraphics[width=8cm]{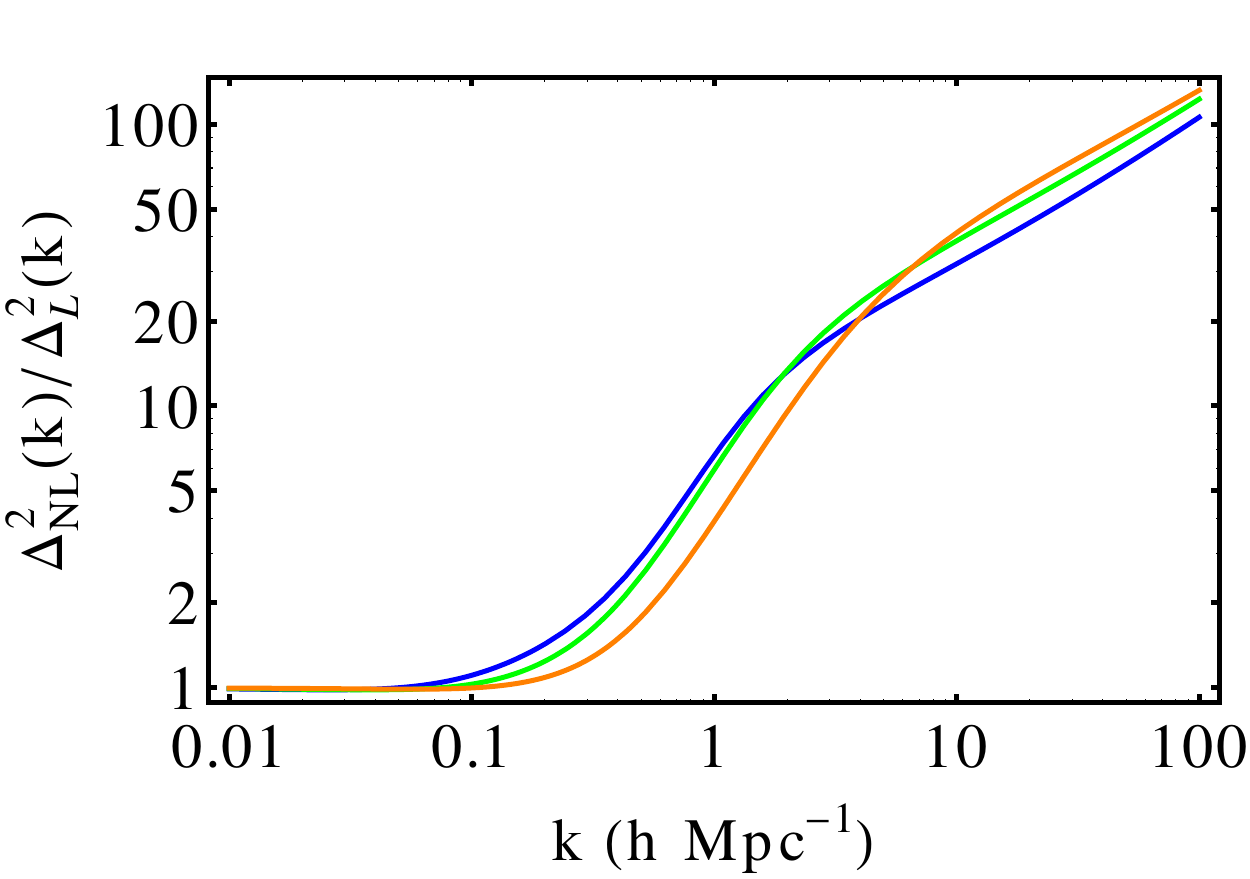}
\centering
\caption[Ratio $\Delta_{\rm{NL}}^2(k)/\Delta_{\rm{L}}^2(k)$ computed from Smith and Takahashi parametrizations]{Ratio $\Delta_{\rm{NL}}^2(k)/\Delta_{\rm{L}}^2(k)$ computed from Smith's method (left panel) and Takahashi's (right panel) for $\Lambda$CDM at 3 different redshifts (z=0\,: blue ; z=0.5\,: green ; z = 1.5\,: orange). The CDM case is not represented. These plots can be compared with Fig. 16 in \cite{Smith:2002dz}.}
\label{FigNLPSv1}
\end{figure}

We present in Fig. \ref{FigNLPSv1} the plot of the ratio $\Delta_{\rm{NL}}^2(k)/\Delta_{\rm{L}}^2(k)$ obtained from above's method. The results are given for $\Lambda$CDM and give a first element of comparison between \cite{Smith:2002dz} and \cite{Takahashi:2012em}. Takahashi's parametrization is not presented but rather similar to the one of Smith \etal.

\subsection{Final result\,: the gravitational spectrum}
\label{AppASec55}

\begin{figure}[ht!]
\centering
\includegraphics[width=16cm]{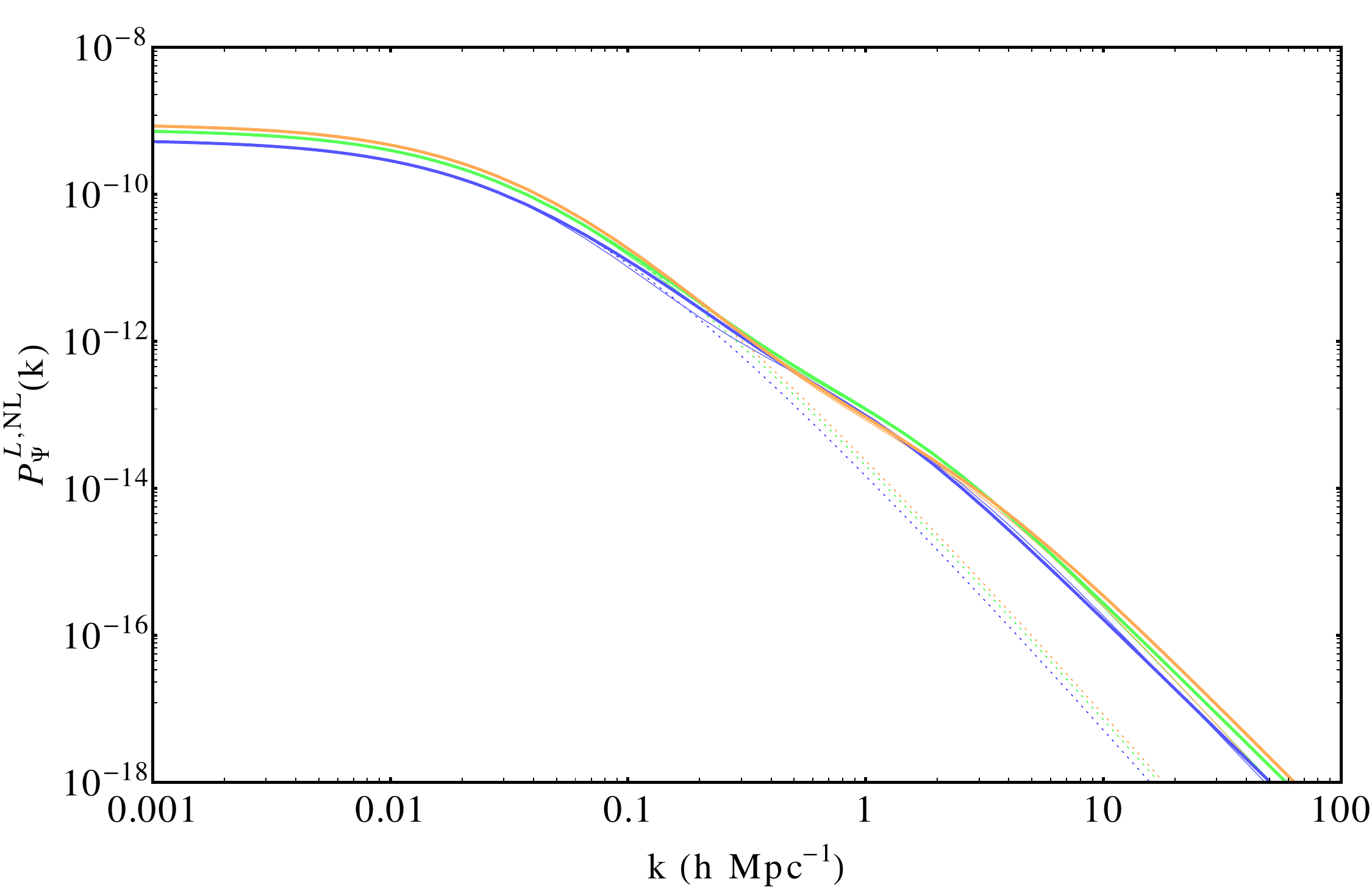}
\centering
\caption[The linear spectrum and the non-linear spectrum for the gravitational potential]{The linear spectrum and the non-linear spectrum for the gravitational potential in $\Lambda$CDM at 3 different redshifts (z=0\,: blue ; z=0.5\,: green ; z = 1.5\,: orange). For each of them, $\Pcal_{\psi}^{\rm{L}}(k)$ is in dotted line, $\Pcal_{\psi}^{\rm{NL}}(k)$ is in solid thin line (Smith) or solid thick line (Takahashi). The difference is very hard to read, and it is better to look at Fig. \ref{FigNLPSforPpsiAndFactor} in order to compare both parametrizations.}
\label{FigNLPSforPpsi}
\end{figure}

\begin{figure}[ht!]
\centering
\includegraphics[width=16cm]{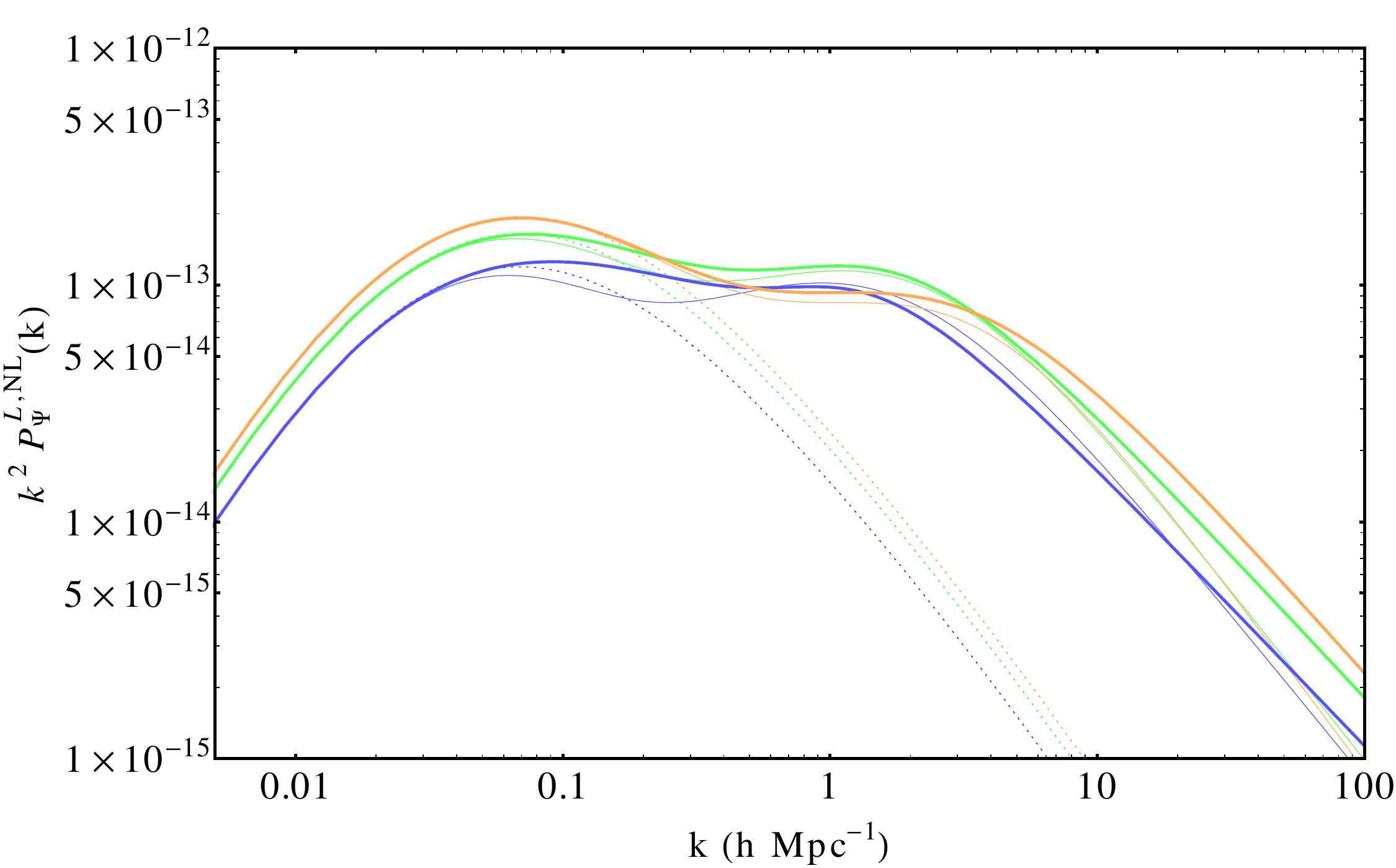}
\centering
\caption[Same as in Fig. \ref{FigNLPSforPpsi} except that spectra are multiplied by $k^2$]{Same as in Fig. \ref{FigNLPSforPpsi} except that spectra are multiplied by $k^2$ in order to have a better understanding of their non-linear behaviour.}
\label{FigNLPSforPpsiAndFactor}
\end{figure}

Using back the Poisson equation to get the power spectrum for the gravitational potential, we have\,:
\beq
\Pcal_\psi^{\rm{NL}}(k) = \frac{9}{4} \frac{\Hcal_0^4 \Om_{m0}^2}{k^4} (1+z)^2 \Delta_{\rm{NL}}^2(k) ~~.
\eeq
We should not think however that the $(1 + z)^2$ factor completely cancels out because $\Delta_{\rm{L}}^2(k)$ is proportional to $1/(1+z)^2$. Indeed, first the parameters necessary to build up $\Delta_{\rm{Q}}^2(k)$ and $\Delta_{\rm{H}}^2(k)$ depend on the model and, for the case of $\Lambda$CDM, the redshift too. But secondly, $\Delta_{\rm{Q}}^2(k)$ is a non-trivial function of $\Delta_{\rm{L}}^2(k)$ and consequently still depends on $z$. Thus the three spectra now intersect each other in the NL regime. We can indeed check that comments in Fig. \ref{FigNLPSforPpsi} and \ref{FigNLPSforPpsiAndFactor}.

\begin{figure}[ht!]
\centering
\includegraphics[width=12cm]{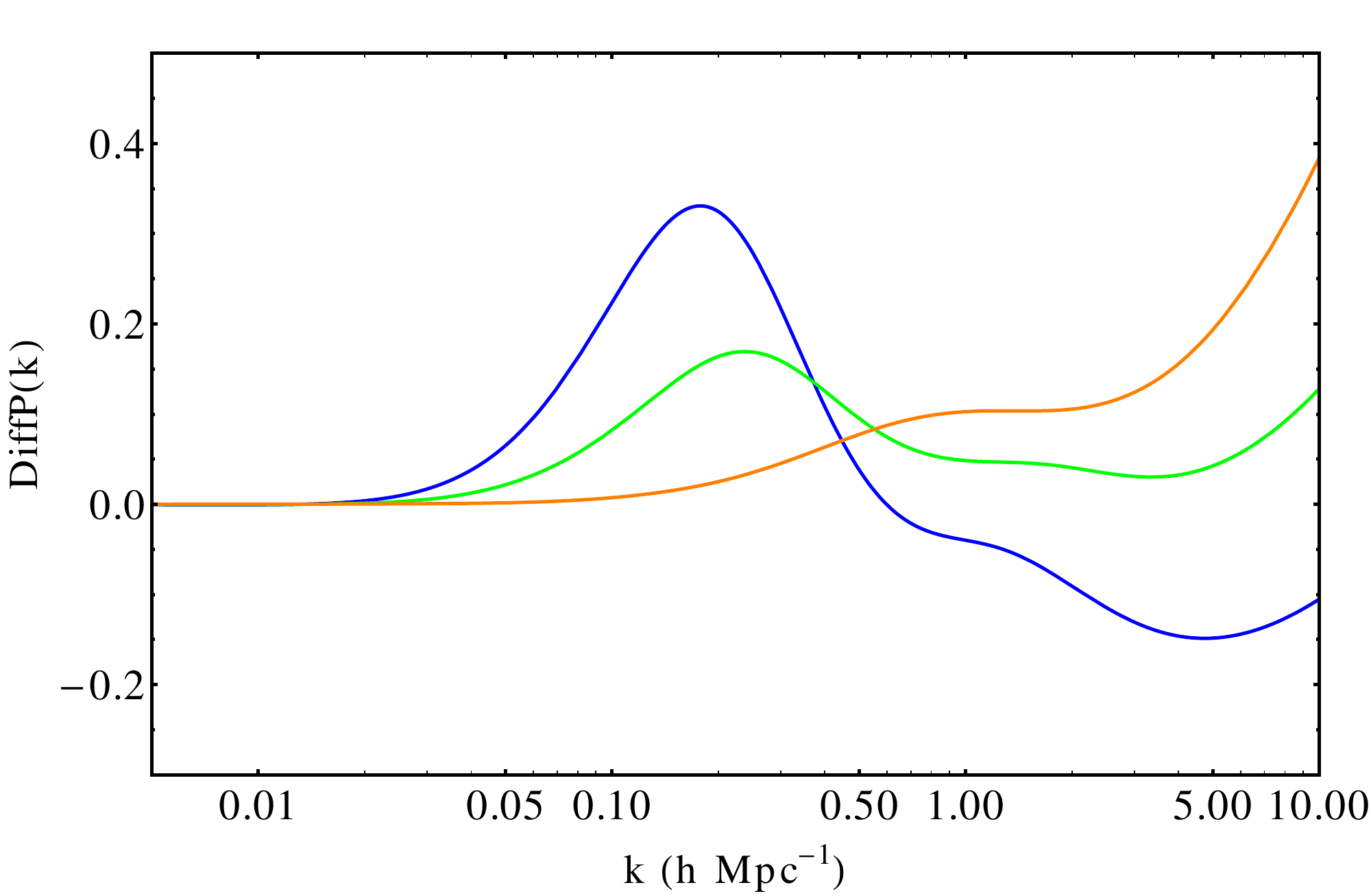}
\centering
\caption[Relative difference ${\rm DiffP}(k)$]{Relative difference for the NLPS\,: ${\rm DiffP}(k)$ ; for the 3 different redshifts in the $\Lambda$CDM model we considered. The NLPS by Smith \etal cannot be much trusted at large $k$.}
\label{DiffRelPpsiNL}
\end{figure}

As a last result, we present in Fig. \ref{DiffRelPpsiNL} the computation of $${\rm DiffP}(k) \equiv \frac{\Pcal_\psi^{\rm{NL}}(\rm{Takahashi};k) - \Pcal_\psi^{\rm{NL}}(\rm{Smith};k)}{\Pcal_\psi^{\rm{NL}}(\rm{Smith};k)} ~~,$$ for the 3 redshifts of the $\Lambda$CDM model already addressed. We can see that the two parametrizations differ only during the non-linear transition and at very small scales (large $k$). The difference, however, stays less than $40 \%$ for the redshifts under study.

\subsection{Deflection}
\label{AppASec56}

Another quantity that we can compute is the variance of density perturbations at a radius of $R_8 = 8 \, h^{-1} \Mpc$. Using a top-hat spatial window function to compute it, we have\footnote{We could also use a Gaussian window and use $\sigma_8^2 \equiv \sigma^2(R_8) = \int \Delta_{\rm L}^2(k) \exp(- k^2 R_8^2) \, d\ln k$ ~.}\,:
\beq
\sigma_8^2 \equiv \int_0^\infty \frac{d k}{k} \Delta_{\rm{L}}^2(k) \left(3\frac{j_1(k R_8)}{k R_8}\right)^2 ~\mbox{ with }~ j_1(k R_8) = \frac{\sin(k R_8) - (k R_8) \cos(k R_8)}{(k R_8)^2} ~.
\eeq
The value obtained from this expression gives a good test of the calculations we are doing. We have computed it, and the relevant physical quantities described before, in the case of $\Lambda$CDM and in the Smith \etal parametrization. These results are given in Table \ref{TableParameters}. As we can see, the $\Lambda$CDM-values for $\sigma_8^2$ have the good order of magnitude.

\begin{table}[ht!]
\caption[Parameters deduced from the identification of the non-linear spectrum of Smith \etal.]{\label{TableParameters} Parameters deduced from the identification of the non-linear spectrum of Smith \etal. We give the estimation of the parameters for $\Lambda$CDM at 3 different redshifts. The quantity $\sigma_8^2$ appears as a by-product of the estimation of other parameters. It is thus used to test the validity of the NL power spectrum.}
\vskip 0.3 cm
\centering
\begin{tabular}{|c|c|c|c|}
\hline 
 & $\Lambda CDM (z=0)$ & $\Lambda CDM (z=1.5)$ & $\Lambda CDM (z=1.5)$ \\ 
\hline 
$k_\sigma$ & 0.158 & 0.223 & 0.447 \\ 
\hline
$n_{\rm eff}$ & -1.502 & -1.648 & -1.888 \\ 
\hline 
$C$ & 0.444 & 0.387 & 0.296 \\ 
\hline 
$\sigma_8^2$ & 1.269 & 0.777 & 0.330 \\ 
\hline
\end{tabular}
\end{table}

\newpage

\section{Distance modulus for a Milne model}
\label{AppASec6}

We briefly describe here how the expression of the \gls{distance modulus} for a \gls{Milne}-type geometry can be derived, i.e.\,:
\beq
\label{MuMilne}
\mu^M = 5 \log_{10}\left[\frac{(2+z_s)z_s}{2 H_0}\right] ~~.
\eeq

In the Milne Universe, homogeneity and isotropy are assumed in their strict sense. The metric has thus the general form of a Robertson-Walker Universe (taking $c=1$) \,:
\beq
ds^2 = - dt^2 + a^2(t) \left[ \frac{dr^2}{1 - K r^2} + r^2 d\Omega^2 \right] ~~,
\eeq
and the Milne geometry consists in taking $a(t) = t$. One can then show that the associated curvature of this metric only cancels if $K = -1$, i.e. if the spatial 3-hypersurfaces have an hyperbolic geometry. In that case, we prove by the transformation (see \cite{Vishwakarma:2013fwa})\,:
\beq
t = \sqrt{\bar{t}^2 - \bar{r}^2} ~~~~~,~~~~~ r = \frac{\bar{r}}{\sqrt{\bar{t}^2 - \bar{r}^2}} ~~,
\eeq
that this metric indeed takes the flat Minkowski form $ds^2 = -d\bar{t}^2 + d\bar{r}^2 + \bar{r}^2 d^2\Omega$.

The luminosity distance is then related to the angular distance by $d_L = (1 + z)^2 d_A$ and this last distance is given by\,:
\beq
d_A(z_s) = a(t_s) \sinh\left( \int_{t_s}^{t_o} \frac{dt}{a(t)} \right) = t_s \sinh \left( \ln \left(\frac{t_o}{t_s}\right) \right) = \frac{H_0^{-1}}{1 + z_s} \sinh \left( \ln (1 + z_s) \right) ~~,
\eeq
where we have used that $H(t) = 1/t$ and $1 + z_s = a(t_o) / a(t_s)$.
Using finally that\,:
\beq
\sinh \left( \ln (1 + z_s) \right) = \frac{( 2 + z_s) z_s}{2 (1 + z_s)} ~~,
\eeq
and the expression of the distance modulus being $\mu = 5 \log_{10} d_L$, we obtain the result announced in Eq. \rref{MuMilne}.

% --------------------------------------------------------- %

%  --------------------- Chapter 9 ----------------------  %
% --------------------------------------------------------- %
%\chapter{Extra considerations on the averaging problem}
%\label{Chap9}
%\input{Chap_9_AppendixAv}
% --------------------------------------------------------- %

%  --------------------- Chapter 10 ----------------------  %
% --------------------------------------------------------- %
\chapter{R\'esum\'e (en fran\c cais)}
\label{Chap10}
\pagestyle{plain}

\label{AppC}

Nous présentons dans cette section un résumé de cette thèse en langue française. La majorité des notions de la thèse est abordée. Néanmoins, certaines discussions qui concernent moins notre travail sont significativement réduites de manière à laisser à ce chapitre une taille raisonnable. La plupart des équations importantes sont présentées et le nombre de références a été réduit de manière à ce que la lecture de ce chapitre puisse se faire sans avoir besoin de revenir souvent au document.

\section{Introduction, motivations et plan de la thèse}
\label{Ch10Sec1}

\subsection{Le modèle standard de la cosmologie}
\label{Ch10Sec11}

Les observations effectuées depuis plusieurs décennies en cosmologie ont amené à un modèle d'Univers qui semble expliquer avec une grande précision les données collectées. Dans ce modèle dit \emph{Modèle Standard} de la cosmologie, la géométrie de l'Univers est celle d'un espace homogène et isotrope, décrite de manière générale par la métrique de Friedmann-Lemaître-Robertson-Walker (FLRW)\,:
\bea
ds^2_{\rm FLRW} &=& - dt^2 + \gamma_{ij}(|\xbf|) dx^i dx^j \nonumber \\
&=& -dt^2 + a^2(t) \left[d\chi^2 + \left( \frac{1}{\sqrt{K}} \sin\left(\sqrt{K} \chi \right) \right)^2 d \Omega^2 \right] ~~, \\
\mbox{ avec } ~~~~~ d \Omega^2 &=& d\theta^2 + (\sin\theta)^2 d\phi^2 ~~. \nonumber
\eea
Dans cette métrique, exprimée en fonction du temps cosmique $t$ et des coordonnées sphériques $(\xbf) = (\chi, \theta, \phi)$, la constante $K$ détermine la courbure des sections spatiales et vaut $\{ 1, 0, -1 \}$ selon que l'Univers est fermé, plat, ou ouvert. Sous une telle forme, le facteur d'échelle $a(t)$, introduit ci-dessus, possède la dimension d'une distance\footnote{Nous avons également imposé à la vitesse de la lumière d'être égale à $c=1$ dans ces unités.}. Cette métrique se réécrit en coordonnées sphériques comme\,:
\beq
d s_{\rm FLRW}^2 = - dt^2 + a(t)^2 \left(\frac{1}{1 - K r^2} dr^2 + r^2 d \Omega^2 \right) ~~,
\eeq
et on peut redonner ici à $r$ la dimension d'une distance et considérer $a(t)$ comme une grandeur sans dimension. Le temps cosmique peut être remplacé par le temps conforme $\eta$ tel que $d\eta = dt / a(t)$.

Sous ces hypothèses d'homogénéité et d'isotropie, le tenseur énergie impulsion d'un fluide de matière s'écrit comme
\beq
T^{\mu\nu} = (\rho + p) u^{\mu} u^{\nu} + p \, g^{\mu\nu} ~~,
\eeq
avec $\rho(t)$ la densité d'énergie du fluide, $p(t)$ sa pression (relativiste), et $u^\mu$ sa 4-vitesse.
Les 10 équations d'Einstein indépendantes (dont 4 peuvent être éliminée par un choix dit de \guillemotleft \, jauge \guillemotright)\,:
\beq
G_{\mu\nu} + \Lambda g_{\mu\nu} \equiv R_{\mu\nu} - \frac12 R g_{\mu\nu} + \Lambda g_{\mu\nu} = 8 \pi G \, T_{\mu\nu} ~~,
\eeq
se réduisent alors à deux équations indépendantes que sont les équations de Friedmann\,:
\bea
\label{FrEqFried}
H^2 &=& \frac{8 \pi G}{3} \rho - \frac{K}{a^2} + \frac{\Lambda}{3} ~~, \\
\label{FrEqFried2}
\frac{\ddot{a}}{a} &=& - \frac{4 \pi G}{3} (\rho + 3p) + \frac{\Lambda}{3} ~~,
\eea
avec $H(t) \equiv \dot{a}(t)/a(t)$ le paramètre de Hubble, et l'équation de conservation\,:
\beq
\label{FrEqConserv}
\dot{\rho} + 3 H (\rho + p) = 0 ~~.
\eeq
Nous avons introduit ici un terme de constante cosmologique $\propto \Lambda$ et nous voyons que cette composante en énergie peut être écrite comme un fluide ayant les propriétés suivantes\,:
\beq
\rho_{\Lambda} = \frac{\Lambda}{8 \pi G} ~~,~~ p_{\Lambda} = - \frac{\Lambda}{8 \pi G} ~~~\mbox{ et l'équation d'état }~~ w_\Lambda \equiv \frac{p_\Lambda}{\rho_\Lambda} = -1 ~~.
\eeq
Seules deux de ces trois dernières équations sont indépendantes et elles déterminent les évolutions respectives du facteur d'échelle $a(t)$ et de la densité de matière $\rho(t)$ après que $p(t)$ eut été choisi en fonction de $\rho(t)$ par l'imposition d'une équation d'état $p(t) = w \, \rho(t)$.
L'équation \rref{FrEqFried} peut se réécrire en terme des paramètres de densité suivants\,:
\beq
\Omega_m = \frac{8 \pi G \, \rho}{3 H^2} ~~~~~,~~~~~ \Omega_\Lambda = \frac{\Lambda}{3 H^2} ~~~~~,~~~~~ \Omega_K = - \frac{K}{H^2 a^2} ~~,
\eeq
qui donnent lieu à\,:
\beq
\label{FrSumOmega}
\Omega_m + \Omega_\Lambda + \Omega_K ~ \equiv ~ \Omega_{\rm tot} + \Omega_K ~ = ~ 1 ~~.
\eeq

On notera que dans le cas d'un Univers contenant plusieurs types de matière (ou de radiation), il suffit de remplacer $\rho$ par $\sum_{i} \rho_i$ avec $\rho_i$ les densités de matière respectives des différents fluides. Si certains de ces fluides ont une pression, on doit aussi poser $p = \sum_{i} p_i$. Enfin, chacun des fluides suit une équation de conservation qui est $\dot{\rho}_i + 3 H (\rho_i + p_i) = 0$ et le système est fermé par le choix d'une équation d'état $p_i=p_i(\rho_i)$ pour chacun de ces fluides.

Le modèle standard de la cosmologie, dit modèle $\Lambda$CDM, est lui composé de différents paramètres (entre 6 et 12 selon les interprétations) qui, entre autres, définissent la composition de matière de l'Univers. Les mesures expérimentales, que nous allons brièvement décrire maintenant, donnent une composition faite de $\sim 73 \%$ d'énergie sombre (prise comme une constante cosmologique $\Lambda$), $\sim 26 \%$ de matière noire froide (CDM) dont $4.6 \%$ de matière baryonique, et des densités de radiation et de courbure négligeables aujourd'hui.

\subsection{Les Supernov\ae\ de type Ia}
\label{Ch10Sec12}

Les supernov\ae\ sont des phénomènes très rares qui se caractérisent par l'explosion d'une naine blanche ayant accrétée de la matière d'une étoile compagnon jusqu'à atteindre la masse de Chandrashekar ($\sim 1.4 \, M_{\odot}$). On observe la progression de leur intensité sur une période de l'ordre de la centaine de jours, qu'on appelle la courbe de lumière (Fig. \ref{FigLightCurve}), et on les observe sur des distances aussi grandes que la moitié de notre Univers observable. On en distingue plusieurs types dont le plus intéressant est celui des Supernov\ae\ Ia (SNe Ia) caractérisé par l'absence d'Hydrogène (et d'Hélium) et la présence de silicium dans son spectre (Fig. \ref{FigSNeSpectra}). C'est aussi celui des supernov\ae\ les plus lumineuses.

Les SNe Ia sont intéressantes du fait de l'homogénéité des conditions qui leur donnent naissance mais sont néanmoins le résultat d'un processus complexe et encore mal compris. Elles sont toutefois \guillemotleft \, standardisables \guillemotright \, (Fig. \ref{FigStretch}) et deviennent alors des étalons de distance en cosmologie. C'est en réalisant leur intérêt que l'on a pu mesurer des distances se rapportant à une dizaine de milliards d'années lumières et mettre en exergue l'existence de l'énergie sombre. Leur mesure se fait dans différentes bandes spectrales et leur analyse permet d'extraire deux quantités très utiles pour leur standardisation\,: le facteur de \emph{stretch} (étirement) $s$ et le paramètre de couleur $c$.

Comme pour tout autre objet astrophysique, la question de la magnitude est importante dans le cas des SNe Ia. On peut montrer que le module de distance, après un choix d'unité appropriée (ici le parsec \pc), est définit par
\beq
\mu(\pc) = m - M = -2.5 \log_{10} \left(\frac{\Phi(\pc)}{\Phi(10 \, \pc)}\right) ~~,
\eeq
avec $m$ la magnitude apparente de la SN Ia, $M$ sa magnitude absolue (i.e. à une distance de $10 \, \pc$) et $\Phi$ le flux lumineux de l'étoile. Ce module de distance est relié à la distance de luminosité selon\:
\beq
\mu(\Mpc) = 25 + 5 \log_{10}(d_L(\Mpc)) = 43.16 + 5 \log_{10}\left(H_0(\Mpc^{-1}) d_L(\Mpc)\right) ~~.
\eeq
C'est par le tracé de cette quantité $\mu$ en fonction du redshift (ou \guillemotleft \, décalage vers le rouge \guillemotright) des SNe Ia, soit autrement dit du diagramme de Hubble (Fig. \ref{FigHubbleDiagBin}), et l'estimation de $H_0$ que le modèle de matière CDM (pour \emph{cold dark matter}) a été mis en défaut et que le modèle $\Lambda$CDM fut introduit.
Dans ce modèle d'univers de type FLRW et plat (i.e. $K=0$), contenant de la matière (noire et baryonique) ainsi qu'une constante cosmologique, la distance de luminosité s'écrit\,:
\beq
\label{FrdLinLCDM}
d_L^{\rm \Lambda CDM}(z_s)=  \frac{1+z_s}{H_0} \int_0^{z_s} {\frac{dz'}{\left[\Om_{\Lambda0} + \Om_{m0} (1+z')^{3}\right]^{1/2}}} ~~.
\eeq
C'est par le fit (l'ajustement) des données des SNe Ia, et leur combinaison avec d'autres mesures cosmologiques (cf. Fig. \ref{FigUnion2CrossPlot}), que les paramètres cosmologiques ont été évalués à $(\Om_{\Lambda0},\Om_{m0}) = (0.73,0.27)$.

\subsection{L'Univers inhomogène}
\label{Ch10Sec13}

Les relevés d'observations aux grandes échelles, tels que SDSS et le 2dFGRS, ont révélé une structure très filamenteuse de la matière aux petites échelles et une bonne homogénéité statistique aux plus grandes (cf. Fig. \ref{Fig2dF}). Néanmoins plusieurs complications se présentent à nous lorsque l'on cherche à déduire de ces observations la vraie distribution de matière dans l'Univers. Premièrement, ce que nous observons se situe sur notre cône de lumière passé et la vraie distribution que nous voyons est celle des galaxies (qui émettent de la lumière). Le contraste de densité des galaxies est relié au contraste de matière par la relation suivante\,:
\beq
\delta_{\rm gal}(t,\xbf) = b \, \delta \, (t,\xbf) ~~~~~\mbox{ avec}~~~~~ \delta(t,\xbf) = \frac{\rho(t,\xbf) - \rho(t)}{\rho(t)} \equiv \frac{\delta\rho(t,\xbf)}{\rho(t)} ~~,
\eeq
et où $\rho(t)$ est la partie homogène de la densité de matière, $\rho(t,\xbf)$ le champ de matière complet. Le paramètre $b(t,\xbf)$ est appelé le biais et a été étudié en détail dans la littérature. Une deuxième complication, de taille, est que les galaxies ne sont en fait pas observées dans l'espace réel, mais bien dans l'espace des redshifts, i.e. par deux angles dans le ciel et le redshift $z$ qui peut être relié à la \guillemotleft \, distance \guillemotright \, sur le cône de lumière, telle que la combinaison $\eta -r$, mais ceci seulement après le choix d'un modèle d'Univers. La conséquence à cela est que les effets de vitesse (dont l'effet caractéristique nommé \emph{fingers of God}) affectent directement la mesure de la distance.

Quoi qu'il en soit, la combinaison de plusieurs méthodes, et les mesures par effet de lentilles gravitationnelles faibles (indépendantes du biais), ont permis de reconstruire le spectre de puissance de la matière $\delta(t,\xbf)$. On peut alors se placer dans l'espace de Fourier et, supposant un principe ergodique, prendre la moyenne de volume des fonctions de corrélation à 2 points et considérer que ceci correspond à la moyenne stochastique $\overline{\delta_\kbf \delta_\kbf}$\footnote{Notons que cette moyenne est habituellement dénotée par des crochets angulaires, $\lla \ldots \rra$, dans la littérature. Néanmoins, cette dernière notation sera pour nous la moyenne sur le ciel, d'où notre choix pour la moyenne stochastique.}. On obtient ensuite la valeur expérimentale du spectre de matière par sa définition\,:
\beq
\Delta^2(t,\kbf) = \frac{k^3}{2 \pi^2} |\delta_\kbf(t)|^2 ~~,
\eeq
et ses valeurs expérimentales présentée en Fig. \ref{FigExpPS}.

Toutefois, nous ne saurions utiliser directement ce spectre pour nos calculs. En effet, les inhomogénéités que nous décrirons dans le reste de ce travail seront représentées par leur champ gravitationnel et non la densité de matière. Il nous faut par conséquent utiliser l'équation de Poisson qui relie ces deux grandeurs. Dans la jauge synchone (voir Sec. \ref{AppASec33}), cette équation s'écrit comme\,:
\beq
\nabla^2 \psi(t,\xbf) = 4 \pi G \rho(t) \, \delta(t,\xbf) ~~,
\eeq
avec $\rho(t)$ la densité de fond (i.e. homogène). La relation entre $|\delta_k|^2$ et $|\psi_k|^2$ est alors donnée par\,:
\beq
|\psi_\kbf(z)|^2 = \left(\frac{3}{2}\right)^2 \frac{\Om_{m}^2(z) \Hcal^4}{k^4} |\delta_\kbf(z)|^2 = \left(\frac{3}{2}\right)^2 \frac{\Om_{m0}^2 \Hcal_0^4}{k^4} (1+z)^2 |\delta_\kbf(z)|^2 ~~,
\eeq
où nous avons fait l'hypothèse d'une matière de type poussière (i.e. sans pression) et avons utilisé la relation qui en découle\,: $\Om_m(z) = \Om_{m0} (1+z) \left( \frac{\Hcal_0}{\Hcal} \right)^2$.

D'autre part, le spectre de puissance (sans dimension) du champ gravitationnel est relié à $|\psi_\kbf|^2$ par la définition\,:
\beq
\label{FrSpectrePuissance}
\Pcal_\psi (\kbf,\eta) \equiv \frac{k^3}{2 \pi^2} |\psi_\kbf(\eta)|^2 ~,
\eeq
et où la dépendance en temps $\eta$ peut être convertie en redshift (de manière analytique dans un Univers CDM, sous forme intégrale dans le cas $\Lambda$CDM). On sait aussi relier ce spectre au spectre de puissance inflationnaire (i.e. produit par les fluctuations quantiques à la fin de l'inflation) où sa valeur est de $\left(\frac{3}{5}\right)^2 \Delta_{\cal R}^2$\,, avec \,:
\beq
\Delta_{\cal R}^2=A \left(\frac{k}{k_0}\right)^{n_s-1} ~~\mbox{ et }~~~ A=2.45 ~10^{-9} ~~~~~ , ~~~~~  n_s=0.96 ~~~~~ , ~~~~~ k_0/a_0=0.002 ~{\rm Mpc}^{-1} ~~~.
\eeq
Il est nécessaire d'ajouter à cette contribution initiale une fonction de transfert dont le rôle est de prendre en compte l'évolution des modes sub-horizon (i.e. de taille plus petite que l'horizon cosmologique) réentrant dans la période de domination par la radiation. Nous ajoutons enfin un \guillemotleft \, facteur de croissance \guillemotright \, (\emph{growth factor} dans la suite) qui a pour rôle de décrire l'évolution de ces modes depuis et entre autre durant la période récente de domination par l'énergie sombre. Nous aboutissons alors à l'expression du spectre de puissance dans son régime linéaire\,:
\beq
\label{FrPsiPIntro}
\Pcal_\psi (k) = \left(\frac{3}{5}\right)^2 \Delta_{\cal R}^2 T^2(k) \left(\frac{g(z)}{g_{\infty}}\right)^2 ~~.
\eeq

Finissons enfin cette section en décrivant brièvement la manière dont est calculée le spectre de puissance dans son régime non-linéaire. En effet, ce régime nous est nécessaire car c'est celui qui décrit la matière aux petites échelles (typiquement celles des amas de galaxies et moindres). Pour cela, il est nécessaire de repasser à une description en terme de densité de matière, la raison étant que le régime non-linéaire a été étudié au travers des simulations à N-corps où cette densité est directement mesurable. Ces simulations nous fournissent alors une paramétrisation du spectre de matière jusque dans son régime non-linéaire $\Delta_{\NL}(k)$, dit le modèle du HaloFit, en fonction de son spectre linéaire $\Delta_{\L}(k)$ et de paramètres de fit (cf. \cite{Smith:2002dz}).
En utilisant le fait que l'équation de Poisson est valable à la fois dans les régimes linéaire et non-linéaire\,:
\beq
\label{FrPE}
\Pcal_\psi^{\rm{L,NL}}(k) = \frac{9}{4} \frac{\Om_{m0}^2 \Hcal_0^4}{k^4} (1+z)^2 \Delta_{\rm{L,NL}}^2(k) ~~,
\eeq
nous obtenons directement le spectre de matière linéaire correspondant à \rref{FrPsiPIntro}\,:
\beq
\label{FrOurDl}
\Delta_{\rm{L}}^2(k) = \frac{4}{9} \frac{k^4}{\Om_{m0}^2 \Hcal_0^4 (1+z)^2} \Pcal_\psi^{\rm{L}}(k) = \frac{4}{25} \frac{A \Om_{m0}^{-2} \Hcal_0^{-4}}{(1+z)^2} ~ \frac{k^{n_s+3}}{k_0^{n_s-1}} \left(\frac{g(z)}{g_0}\right)^2 T^2(k) ~~,
\eeq
et il est alors possible d'obtenir le spectre non-linéaire $\Delta_{\NL}^2(k)$. Par une utilisation directe de \rref{FrPE} cette fois, on obtient finalement la valeur de $\Pcal_\psi^{\NL}(k)$. Cette méthode et les détails de la parametrisation du HaloFit sont expliqués en Sec. \ref{AppASec4} et \ref{AppASec5}.

\subsection{Introduction à la thèse}
\label{Ch10Sec14}

Le modèle de concordance (ou $\Lambda$CDM), basé sur l'association commune de l'énergie sombre, des matières noire et baryonique, est devenu le paradigme de référence pour la description de notre Univers dans son époque récente (cf. e.g. \cite{Nakamura:2010zzi}). Il rend compte à la fois des observations du CMB, des grandes structures et des données de SNe Ia \cite{Riess:1998cb,Perlmutter:1998np}. Néanmoins, ces dernières sont plus directement affectées par leur dépendance à un modèle théorique qui est choisi comme étant le modèle de Friedmann-Lema\^itre-Robertson-Walker (FLRW), homogène et isotrope. Il a ainsi été suggéré que la prise en compte des inhomogénéités pourrait significativement altérer ces conclusions. Par exemple, les modèles inhomogènes dans lesquels nous occupons une position privilégiée ont démontré la possibilité de simuler une énergie sombre, i.e. une accélération de l'expansion (cf. \cite{Celerier:1999hp}). Ils nécessitent néanmoins trop d'ajustements pour être réalistes. On ne peut toutefois contredire le fait que la dispersion dans les données du diagramme de Hubble est très importante et, bien que sa source principale soit probablement en grande partie due à la variabilité des SNe elles-mêmes, on ne peut s'empêcher de se demander quelle est la part de dispersion qui est causée par la présence d'inhomogénéités de matière. Au final, on peut résumer le but de la cosmologie à l'étude de deux choses\,: le contenu de l'Univers et la dynamique de ce contenu. Il apparaît ainsi très important de comprendre comment la matière, et sa structure, affectent la mesure des distances en cosmologie.
Il est aussi clair qu'un traitement plus complet de l'accélération de l'Univers est nécessaire, au moins de manière perturbative autour d'un modèle homogène, pour aboutir à des considérations précises et établir notre entière confiance dans la capacité du modèle $\Lambda$CDM à expliquer les diverses observations.

Il est de plus bien établi (cf. e.g. \cite{AppEt1933a}) que les solutions moyennées des équations d'Einstein complètes (et inhomogènes) sont, en général, différentes de celles obtenues en résolvant les équations d'Einstein moyennes (homogènes, i.e. les équations de Friedmann). En particulier, la procédure de moyenne ne commute pas avec les opérateurs différentiels agissant dans les équations d'Einstein qui sont fondamentalement de nature non-linéaire. Par conséquent, la dynamique des géométries moyennées est affectée par des contributions dites de ``backreaction'' (\guillemotleft \, rétro-action \guillemotright) engendrées par les inhomogénéités. Ces effets ont donc suscité un intérêt grandissant ces dernières années (voir \cite{Clarkson:2011zq,Buchert:2011yu,Ellis:2011hk,Buchert:2011sx}).
Pour certains auteurs \cite{Buchert:1999mc,Buchert:2002ij,Rasanen:2003fy,Rasanen:2006kp,Barausse:2005nf,Kolb:2004am,Kolb:2005da}, les inhomogénéités de matière pourraient expliquer l'accélération que nous observons aujourd'hui, rendant inutile l'introduction de l'énergie sombre et répondant en même temps au \guillemotleft \, problème de coincidence \guillemotright.
Pour d'autres \cite{Flanagan:2005dk, Hirata:2005ei,Ishibashi:2005sj,Paranjape:2008jc,Baumann:2010tm,Green:2010qy}, l'effet est complètement négligeable. La vraie réponse à ce problème pourrait se trouver à mi-chemin entre ces différentes interprétations et révéler l'importance des inhomogénéités lors de l'établissement de contraintes précises sur les paramètres de l'énergie sombre.

Si les inhomogénéités ont en effet une influence importante, cela signifie que les observations actuelles correspondent à une certaine mesure de quantités moyennes. Un accent particulier a donc été porté ces dernières années à la définition de moyennes pour correctement prendre en compte les effets des inhomogénéités. Dans la quasi-totalité de ces travaux, suivant l'exemple des articles fondateurs de T. Buchert \cite{Buchert:1999er,Buchert:2001sa,2000GReGr..32..105B,2001GReGr..33.1381B}, la géométrie homogène résultante de la procédure de moyenne a été déterminée par l'intégration sur des hypersurfaces spatiales à 3 dimensions. Néanmoins, et comme suggéré depuis longtemps \cite{Maartens1,Maartens2}, une reconstruction effective de la métrique d'espace-temps et de son évolution aux échelles cosmologiques (révélée expérimentalement par la mesure de la distance et du redshift \cite{Riess:1998cb,Perlmutter:1998np}), ne peut se faire que par la considération d'observations sur le cône de lumière passé (l'hypersurface sur laquelle se propagent tous les signaux lumineux). La procédure de moyenne doit donc être faite sur une hypersurface de genre lumière, plutôt que sur des hypersurfaces spatiales. Cette procédure de moyenne dont l'importance a été soulignée plusieurs fois dans la litérature (cf. \cite{Buchert:2011yu,Maartens1,Maartens2,Marra:2007gc,Li:2008yj,Rasanen:2008be,Rasanen:2009uw,Kolb:2009hn}), n'avait jamais été mise en oeuvre en pratique.

Nous présentons dans cette thèse une définition générale, invariante de jauge, pour la définition de quantités scalaires sur des hypersurfaces de type lumière, résolvant ainsi le problème de la moyenne sur le cône de lumière. Nous appliquons aussi cette moyenne au cas d'un observateur générique au sein d'un univers inhomogène et nous établissons les relations analogues à la règle de commutation de Buchert-Ehlers \cite{Buchert:1995fz} dans le cas de la dérivation des moyennes sur ce cône. Nous mettons aussi à profit l'invariance de jauge de ces moyennes par l'introduction d'un système de coordonnées adapté à leur simplification, que nous avons appelé coordonnées \guillemotleft \, géodésiques sur le cône de lumière \guillemotright \, (GLC). Ces coordonnées correspondent à une fixation de jauge des coordonnées dites \guillemotleft \, observationelles \guillemotright \, (cf. \cite{Maartens1,1985PhR...124..315E}) et ne sont pas totalement équivalentes. Notre définition de la moyenne se simplifie alors tout en gardant le nombre de degrés de liberté nécessaire pour son application à des métriques inhomogènes générales. Ces coordonnées GLC se révèlent aussi utiles pour le calcul de nombreuses quantités sur le cône de lumière et ont donc une application qui dépasse largement le problème de la moyenne. Entre autre, on observe que le redshift et la distance de luminosité prennent une expression très simple dans ces coordonnées.

Nous portons ensuite notre attention sur le cas d'un modèle de type FLRW et plat en présence d'inhomogénéités scalaires d'origine primordiale (inflationnaire). Ce modèle est réaliste du point de vue des observations et assez simple pour une approche analytique. Ce genre de perturbation est en général définit dans la jauge longitudinale (ou Newtonienne, \cite{Mukhanov:1990me}) et nous établissons pour cela la transformation de coordonnées entre les jauges GLC et Newtonienne. Du fait de la stochasticité des inhomogénéités considérées, les effets de ces dernières n'apparaissent qu'au second ordre en perturbation. Nous calculons ainsi, à l'aide de notre transformation, la relation de distance-redshift jusqu'au second ordre. Ce résultat très général a un intérêt en lui-même, indépendant des considérations de moyennes, et pourrait aboutir à de nombreuses applications en cosmologie de précision. De plus, ce résultat pour $d_L$ est valable quelque soit la géométrie de fond considérée (sauf en présence de caustiques où la distance est modifiée).

Associant ces deux résultat théoriques, nous étudions ensuite la moyenne de quantités scalaires $S$ développées au second ordre autour d'une géométrie de fond du type FLRW plat. Nous introduisons une moyenne d'ensemble sur les inhomogénéités et un spectre de puissance des perturbations stochastiques. La combinaison de la moyenne angulaire sur le cône de lumière et de cette moyenne stochastique amène à des contributions caractéristiques désignées comme \guillemotleft \, backreaction induite \guillemotright \, du fait de la corrélation qu'elles contiennent entre les pertubations de $S$ et la mesure d'intégration de la moyenne. L'application de cette procédure à des quantités telles que le flux et la distance de luminosité nous permet d'étudier l'effet des inhomogénéités de nature stochastiques sur la détermination des paramètres de l'énergie sombre en cosmologie de précision. Contrairement à d'autres études utilisant des moyennes spatiales (cf. \cite{Kolb:2011zz,Clarkson:2011uk}), nos résultats sont exempts aussi bien de divergences infrarouges que de divergences ultraviolettes. Toutefois, les termes de backreaction induite n'englobent pas tout les effets des inhomogénéités et un calcul complet au second ordre est établi. En commençant par un modèle simple de type CDM, nous trouvons que le flux $\Phi \sim d_L^{-2}$ n'est quasiment pas affecté par les inhomogénéités tandis que les grandeurs plus communément utilisées (comme $d_L$ ou le module de distance $\mu \sim 5 \log_{10} d_L$) recoivent des corrections plus significatives (et bien plus grandes qu'on aurait pu le penser naïvement). Ceci prouve, au moins en principe, qu'il existe une ambiguité dans la mesure des paramètres de l'énergie sombre lorsque l'on ne prend correctement en compte l'effet des inhomogénéités. La faible dépendence du flux, déjà remarquée dans le contexte différent de l'étude du binning en $z$ des données \cite{Wang:1999bz}, est très surement à rapprocher de la conservation de la quantité de lumière le long du cône passé. Les premières conclusions de notre analyse, basées sur un spectre de puissance dans son régime linéaire dans un modèle CDM \cite{Eisenstein:1997ik}, sont les suivantes. D'un côté, les inhomogénéités ne peuvent simuler une fraction conséquente d'énergie sombre, leur contribution étant à la fois trop faible et ne possède pas la dépendance en $z$ adéquate. D'un autre, ces fluctuations stochastiques apportent une dispersion relativement importante aux prédictions du modèle FLRW homogène et isotrope. Cette dispersion est indépendante de l'appareil expérimental, de la procédure d'observation ou de la dispersion dans la luminosité des sources observées. Elle pourrait, si non prise en compte, empêcher la détermination du paramètre cosmologique $\Omega_{\Lambda}(z)$ en dessous du pourcent. Une autre conclusion importante à souligner est que la moyenne sur le cône de lumière de différentes fonctions de la même observable sont affectées différemments, avec le flux qui apparaît comme la quantité la mieux conservée.

Nous avons ensuite étendu notre étude par la considération d'un modèle $\Lambda$CDM, en ajoutant l'effet des baryons, et en considérant deux paramétrisations du modèle du HaloFit \cite{Smith:2002dz,Takahashi:2012em} décrivant le spectre de densité de matière dans son régime non-linéaire (i.e. aux petites échelles). Nous avons calculé l'effet des inhomogénéités de la même manière que précédemment tout en ne considérant toutefois que les termes dominants qui ont été indentifiées dans le cas CDM. Nous trouvons que l'effet des inhomogénéités est renforcé aux grands redshifts du fait du terme dominant de lentille gravitationnelle. En dépit de cet effet, les inhomogénéités ne parviennent toujours pas à simuler une fraction importante de l'énergie sombre. La dispersion quant à elle est fortement augmentée et correspond à une possible déviation statistique de $\sim 10\%$ sur $\Omega_{\Lambda0}$. Ce résultat suggère fortement de considérer l'effet des inhomogénéités de matière dans l'interprétation des données des SNe Ia. Nous trouvons de plus que cette dispersion est en très bon accord avec les données du Union2 aux petits redshifts (où les effets de vitesse particulière dominent) mais est bien trop petite aux larges redshifts. Ceci démontre la robustesse des conclusions du modèle standard en soulignant toutefois la nécessité de considérer l'aspect non-négligeable des inhomogénéités. Enfin, nous comparons de manière précise la dispersion due à l'effet de lentille avec les tentatives récentes d'évaluation d'un signal de lentille au sein des SNe Ia \cite{Kronborg:2010uj,Jonsson:2010wx}. Nous trouvons notre prédiction en parfait accord avec ces études et pouvons donc espérer une confirmation expérimentale de cette prédiction dans les prochaines années.

\section{Moyennes sur le cône de lumière}
\label{Ch10Sec2}

\subsection{Approche historique à la Buchert}
\label{Ch10Sec21}

Le formalisme de Buchert considère l'Univers comme statistiquement homogène à partir d'une certaine échelle de distance (conformément aux observations), rempli de matière sans pression et d'une constante cosmologique (de pression négative), et cherche à dériver -- à partir des équations d'Einstein -- des équations effectives et non-perturbatives pour la cosmologie. Les observations présentes, encore peu précises et toutes de nature astrophysique (objets locaux au centre des inhomogénéités de matière), consistent en effet en des moyennes spatiales (plus justement sur le cône de lumière passé) et ne garantissent pas que les équations homogènes de Friedmann soient, stricto sensu, les bonnes pour interpréter ces observations.

Dans ce formalisme, nous considérons donc un domaine compact $\Dcal(t)$ au sein d'un espace folié par des hypersurfaces à temps $t = \cst$, dit formalisme ADM (Arnowitt-Deser-Misner). 
L'élément de distance dans ce formalisme est donné par\,:
\beq
\label{Frds2ADM}
ds^2 = - N^2 dt^2 + g_{ij} (dx^i + N^i dt) (dx^j + N^j dt) ~~,
\eeq
où $N$ est appelée la \emph{lapse function} et $N^i$ le \emph{shift vector} et $n_\mu = (-N,\vec{0}\,)$ (ou $n^\mu = (1/N,-N^i/N)$) est le vecteur normal aux hypersurfaces à $t$ constant. Nous pouvons aussi travailler (en imposant $N=1$ et $N^i=0$) dans la jauge dite synchrone\,:
\beq
\label{FrSGmetric}
ds^2_{\rm SG} = - dt^2 + g_{ij} dx^i dx^j ~~.
\eeq
Le volume $\Dcal(t)$ est choisi de manière à conserver la masse $M_\Dcal$ qu'il contient à l'instant initial $t_0$ et la moyenne d'une quantité scalaire $S$ est définie en terme d'une moyenne de volume Riemannienne (i.e. pondérée par la métrique)\,:
\beq
\lla S(t,\xbf) \rra_{\Dcal}(t) = \frac{1}{V_\Dcal} \int_\Dcal S(t,\xbf) ~ d\mu_g ~~~\mbox{ avec }~~ V_\Dcal = \int_\Dcal d\mu_g ~~~\mbox{ et }~~ d\mu_g \equiv J ~ d^3 \xbf = \sqrt{\det g_{ij}} ~ d^3\xbf ~~.
\eeq
La dérivée temporelle de cette moyenne donne alors la \emph{règle de commutation} de Buchert-Ehlers\,:
\beq
\label{FrRelNonCom}
\lla S \rra_\Dcal^{\tdev} - \langle \dot{S} \rangle_\Dcal = \lla S \Theta \rra_\Dcal - \lla S \rra_\Dcal \lla \Theta \rra_\Dcal ~~,
\eeq
qui montre que la moyenne spatiale et la dérivation temporelle ne peuvent être simplement échangées dans un univers dont le facteur d'expansion $\Theta$ dépend de l'espace.
On peut aussi définir un facteur d'échelle effectif $a_\Dcal$ et un paramètre de Hubble effectif $H_\Dcal$ en fonction du volume de ce domaine\,:
\beq
a_\Dcal(t) \equiv \left( \frac{V_\Dcal(t)}{V_\Dcal(t_i)} \right)^{1/3} ~~~\mbox{ et donc }~~ \frac{\dot{V}_\Dcal}{V_\Dcal} = 3 \frac{\dot{a}_\Dcal}{a_\Dcal} \equiv 3 H_\Dcal ~~.
\eeq
Il convient toutefois d'être prudent avec cette définition qui redonne bien $a_\Dcal(t) = a(t)$ dans le cas homogène mais qui, dans le cas non-homogène, voit $a_\Dcal(t)$ correspondre à des volumes d'intégration $\Dcal(t)$ bien différents.

\paragraph{Equations d'Einstein dans ADM\,:}
On introduit plusieurs quantités que sont le tenseur de projection sur une hypersurface à $t$ constant, qui est aussi orthogonal à la 4-vitesse du fluide (supposé statique et donc pour lequel on peut choisir $u^\mu = n^\mu$)\,:
\beq
h_{\mu\nu} \equiv g_{\mu\nu} + n_\mu n_\nu ~~~,~~ h_{\mu\nu} n^\nu = 0 ~~~\mbox{ et donc ici\,: }~~~ h_{00} = 0 ~~,~~ h_{0i} = 0 ~~,~~ h_{ij} = g_{ij} ~~,
\eeq
ainsi que le tenseur de courbure extrinsèque
\beq
K_{ij} \equiv - h^\alpha_i h^\beta_j \nabla_{\beta} ~ n_{\alpha} ~~.
\eeq

On peut alors montrer que les équations d'Einstein et l'équation de conservation se réduisent à des équations de contrainte\,:
\bea
\label{FrConstr1}
\frac12 \left( \Rcal + K^2 - K^i_{~ij} K^j_{~i} \right) &=& 8 \pi G \rho + \Lambda ~~, \\
\label{FrConstr2}
K^i_{~j\,||\,i} - K_{|j} &=& 0 ~~,
\eea
et les équations d'évolution suivantes\,:
\bea
\label{FrDyn1}
\dot{\rho} &=& K \rho ~~, \\
\label{FrDyn2}
(g_{ij})^{\tdev} &=& - 2 ~ g_{ik} K^k_{~j} ~~, \\
\label{FrDyn3}
(K^i_{~j})^{\tdev} &=& K K^i_{~j} + \Rcal^i_{~j} - (4 \pi G \rho + \Lambda) \delta^i_{~j} ~~.
\eea
La notation $||i$ désigne la dérivée covariante par rapport à la métrique spatiale tridimensionnelle $g_{ij}$ (et $|i$ est une simple dérivée $\partial_i$).
Nous avons aussi utilisé la trace $K$ de $K_{ij}$ ainsi que celle du tenseur de Ricci spatial $\Rcal \equiv \Rcal^i_{~i}$ qui représente la courbure locale à l'intérieur d'une hypersurface à $t$ constant.

On peut aussi définir le tenseur d'expansion d'un faisceau de lignes géodésiques\,:
\beq
\Theta_{\mu\nu} = \frac13 h_{\mu\nu} \Theta + \sigma_{\mu\nu} + \omega_{\mu\nu} ~~,
\eeq
qui fait intervenir respectivement le scalaire d'expansion, le tenseur symétrique de cisaillement, et le tenseur anti-symétrique de vorticité\,:
\beq
\Theta \equiv \nabla_\mu n^\mu ~~,~~ \sigma_{\mu\nu} \equiv h^\alpha_\mu h^\beta_\nu \left( \nabla_{(\alpha} n_{\beta)} - \frac13 h_{\alpha\beta} \Theta \right) ~~,~~ \omega_{\mu\nu} \equiv h^\alpha_\mu h^\beta_\nu \nabla_{[\alpha} n_{\beta]} ~~.
\eeq
Ces deux tenseurs sont sans trace ($\sigma^\mu_{~\mu} = 0 = \omega^\mu_{~\mu}$) et leurs carrés, $\sigma^2 \equiv \sigma^\mu_{~\nu} \sigma^\nu_{~\mu} / 2$ et $\omega^2 \equiv \omega^\mu_{~\nu} \omega^\nu_{~\mu} / 2$, représentent les taux de cisaillement et de rotation des lignes géodésiques.

Dans la jauge ADM, le tenseur de courbure extrinsèque est directement lié au tenseur d'expansion par $K_{ij} = - \Theta_{ij}$ et on a directement que sa trace est $K = - \Theta$. Les équations de contrainte et dynamiques se réécrivent alors comme
\bea
\label{FrNewConstr1}
\frac12 \Rcal + \frac13 \Theta^2 - \sigma^2 - \omega^2 &=& 8 \pi G \rho + \Lambda ~~, \\
\label{FrNewConstr2}
\sigma^i_{~j\,||\,i} + \omega^i_{~j\,||\,i} &=& \frac23 \Theta_{|\,j} ~~, \\
\label{FrNewDyn1}
\dot{\rho} &=& - \Theta \rho ~~, \\
\label{FrNewDyn2}
(g_{ij})^{\tdev} &=& 2 g_{ik} \sigma^k_{~j} + 2 g_{ik} \omega^k_{~j} + \frac23 \Theta g_{ij} ~~, \\
\label{FrNewDyn3}
(\sigma^i_{~j} + \omega^i_{~j})^{\tdev} &=& - \Theta (\sigma^i_{~j} + \omega^i_{~j}) - \Rcal^i_{~j} + \frac23 \delta^i_{~j} \left[ \sigma^2 + \omega^2 - \frac{\Theta^2}{3} + 8 \pi G \rho + \Lambda \right] ~~.~~~
\eea
Nous avons utilisé dans l'Eq. \rref{FrNewDyn3} la trace de l'Eq. \rref{FrDyn3}, appelée l'\emph{équation de Raychaudhuri}\,:
\beq
\label{FrRauchaudhuriEq}
\dot{\Theta} + \frac{\Theta^2}{3} + 2 \sigma^2 + 2 \omega^2 + 4 \pi G \rho - \Lambda = 0 ~~,
\eeq
et remplacé $\Rcal$ par son expression \rref{FrNewConstr1} qui est dénommée \emph{contrainte Hamitonienne}. Ces deux équations, bien que plus générales ici, gardent la même forme en gravité Newtonienne.

\paragraph{Moyennes spatiales de ces equations\,:}
On peut remarquer que le tenseur de courbure extrinsèque peut aussi s'écrire comme $K^i_{~j} = - \frac12 g^{ik} (g_{kj})^{\tdev}$. En définissant alors $J(t,\xbf) \equiv \sqrt{\det g_{ij}}$, nous avons\,:
\beq
(\ln J)^{\tdev} =  \frac12 g^{ik} (g_{ki})^{\tdev} = - K ~~~\mbox{ et donc }~~ \dot{J} = - K J = \Theta J ~~,
\eeq
et cette dernière égalité nous montre que
\beq
\lla \Theta \rra_\Dcal = \frac{1}{V_\Dcal} \int_\Dcal \frac{\dot{J}}{J} ~ J ~ d^3\xbf = \frac{1}{V_\Dcal} \left( \int_\Dcal J ~ d^3\xbf \right)^{\tdev} = \frac{\dot{V}_\Dcal}{V_\Dcal} = 3 H_\Dcal ~~.
\eeq
L'équation de conservation \rref{FrNewDyn1} s'écrit aussi simplement comme
\beq
(\rho J )^{\tdev} = 0 ~~~,~~\mbox{ de solution générale }~~ \rho(t,\xbf) = \frac{\rho(t_0,\xbf) J(t_0,\xbf)}{J(t,\xbf)} ~~.
\eeq
Ce résultat implique que la masse conservée dans $\Dcal(t)$ est constante au cours du temps\,:
\beq
M_\Dcal = \int_\Dcal \rho J ~ d^3\xbf = {\rm cst} ~~ \mbox{ et donc } ~~
\lla \rho \rra_\Dcal = \frac{1}{V_\Dcal} \int_\Dcal \rho J ~ d^3\xbf = \frac{M_\Dcal}{V_\Dcal} = \frac{M_\Dcal}{V_{\Dcal(t_0)} a_\Dcal^3} ~~.
\eeq

L'équation de Rauchaudhuri \rref{FrRauchaudhuriEq}, moyennée et en utilisant $S = \Theta$ dans l'Eq. \rref{FrRelNonCom}, vaut\,:
\beq
\lla \Theta \rra_\Dcal^{\tdev} - \frac23 \lla \Theta^2 \rra_\Dcal + \lla \Theta \rra_\Dcal^2 + 2 \lla \sigma^2 \rra_\Dcal + 2 \lla \omega^2 \rra_\Dcal + 4 \pi G \lla \rho \rra_\Dcal - \Lambda = 0 ~~.
\eeq
Elle peut s'écrire aussi\,:
\beq
3 \frac{\ddot{a}_\Dcal(t)}{a_\Dcal(t)} + 4 \pi G \frac{M_\Dcal}{V_{\Dcal_0} a_\Dcal^3} + \Lambda = \Qcal_\Dcal ~~,
\eeq
avec
\beq
\Qcal_\Dcal \equiv \frac23 \left[ \lla \Theta^2 \rra_\Dcal - \lla \Theta \rra_\Dcal^2 \right] - 2 \lla \sigma^2 \rra_\Dcal - 2 \lla \omega^2 \rra_\Dcal  ~~.
\eeq
La contrainte Hamiltonienne \rref{FrNewConstr1} moyennée s'écrit quant à elle comme
\beq
\frac12 \lla \Rcal \rra_\Dcal + \frac13 \lla \Theta^2 \rra_\Dcal - 8 \pi G \lla \rho \rra_\Dcal - \Lambda = \lla \sigma^2 \rra_\Dcal + \lla \omega^2 \rra_\Dcal ~~,
\eeq
et se révèle équivalente à
\beq
3 \left(\frac{\dot{a}_\Dcal}{a_\Dcal} \right)^2 + \frac12 \lla \Rcal \rra_\Dcal - 8 \pi G \frac{M_\Dcal}{V_{\Dcal_0} a_\Dcal^3} - \Lambda = - \frac{\Qcal_\Dcal}{2} ~~.
\eeq

On a donc trois équations qui généralisent le cas homogène et qui, lorsque l'on rajoute une courbure constante au modèle FLRW, s'écrivent\,:
\bea
3 \frac{\ddot{a}_\Dcal(t)}{a_\Dcal(t)} &=& - 4 \pi G \lla \rho \rra_\Dcal + \Qcal_\Dcal + \Lambda ~~, \\
3 H_\Dcal^2 &=& 8 \pi G \lla \rho \rra_\Dcal - \frac{3 k_{\Dcal_i}}{a_\Dcal^2} - \frac12 \left( \Wcal_\Dcal + \Qcal_\Dcal \right) + \Lambda ~~, \\
\lla \rho \rra_\Dcal^{\tdev} &=& - 3 H_\Dcal \lla \rho \rra_\Dcal ~~.
\eea
On a définit ici la déviation de courbure à la courbure constante $K_{\Dcal_i}$ selon\,:
\beq
\Wcal_\Dcal = \lla \Rcal \rra_{\Dcal} - 6 ~ \frac{K_{\Dcal_i}}{a_\Dcal^2} ~~.
\eeq

On peut vérifier que deux seulement de ces équations sont indépendantes si et seulement si nous avons la \emph{condition} nécessaire dite \emph{d'intégrabilité}\,:
\beq
\label{FrFriedBack4}
\dot{\Qcal}_\Dcal + 6 H_\Dcal \Qcal_\Dcal + \dot{\Wcal}_\Dcal + 2 H_\Dcal \Wcal_\Dcal = 0 ~~ \Longleftrightarrow ~~ a_\Dcal^{-6} \left(a_\Dcal^6 \Qcal_\Dcal \right)^{\tdev} + a_\Dcal^{-2} \left( a_\Dcal^2 \Wcal_\Dcal \right)^{\tdev} = 0 ~~.
\eeq

C'est par l'étude de ce système d'équations qu'à été proposée l'idée que les inhomogénéités puissent entrainer une accélération apparente en cosmologie et ce du fait que les vraies quantitées observées sont des quantités moyennes. On voit que cette condition d'intégrabilité impose une relation entre la moyenne sur le domaine spatial $\Dcal$ de la courbure tridimensionnelle, au sein de $\Wcal_\Dcal$, et le terme dit de rétroaction (\guillemotleft \, backreaction \guillemotright) $\Qcal_\Dcal$ qui contient les effets géométriques d'une distribution inhomogène de matière. L'étude de ces équations peut aussi être développée en fonction d'un champ scalaire effectif $\Phi_\Dcal$, dépendant du domaine étudié, et nommé le champ de \emph{morphon} (cf. Sec \ref{Ch2Sec113}).

\subsection{Moyenne sur le cône de lumière}
\label{Ch10Sec22}

La moyenne sur le cône de lumière est une généralisation de la moyenne sur une hypersurface spatiale au cas d'hypersurfaces nulles. Réexprimons brièvement cette première de manière invariante de jauge\footnote{J'écris \emph{invariante de jauge} pour traduire le terme \emph{gauge invariant}, ce que d'autres appèlent \emph{covariante de jauge} en français. J'utilise ainsi \emph{indépendante de jauge} pour traduire \emph{gauge independent}, ce que d'autres appellent \emph{invariant de jauge}, mais nous éviterons d'entrer dans ces complications.}. On considère une hypersurface spatiale déterminée par l'équation $A(x) = A_0$ où $A$ est un scalaire dans $\Mcal_4$. Le vecteur normal à cette hypersurface est de type temps et se définit comme
\beq
n_\mu =- {\pa_\mu A\over \sqrt{-\partial_\nu A \partial^\nu A} } ~~~\mbox{ avec }~~~~ n_\mu n^\mu=-1 ~~~.
\label{Frvel}
\eeq
On prend alors pour définition de la moyenne d'une quantité scalaire $S$ sur un domaine $\Dscr \subset \Mcal_4$\,:
\beq
\langle S \rangle_{\Dscr} = \frac{I(S; \Dscr)}{I(1; \Dscr)} ~~\mbox{ où }~~ I(S;\Dscr) =  \int_{\Mcal_4} d^4 x \sqrt{-g(x)} ~ W_\Dscr(x) S(x) ~~.
\eeq
$W_\Dscr$ est une fonction fenêtre qui sélectionne le domaine d'intégration. Dans le cas d'une intégration sur un domaine $\Dscr$ de l'hypersurface $A = A_0$, délimité par la condition radiale $B(x) < r_0$ avec $B > 0$ (cf. Fig. \ref{FigSpatialAveraging}), on peut (avec $\delta_D$ la distribution de Dirac et $\Theta$ celle de Heaviside) prendre la fonction\,:
\beq
\label{FrWindowSpatialAv}
W_{\Dscr}(x) = n^\mu \nabla_\mu \Theta(A(x) - A_0) \Theta(r_0 - B(x)) = \sqrt{-\partial_\mu A \partial^\mu A} ~ \delta_{\rm D}(A(x) - A_0) \Theta(r_0 - B(x)) ~~.
\eeq
Si l'on veut en revanche intégrer sur toute l'hypersurface $A(x) = A_0$, il suffit de prendre $r_0 \rightarrow \infty$.
Le coefficient $\sqrt{-\partial_\mu A \partial^\mu A}$ qui apparaît dans la fonction $W_{\Dscr}$ garantit l'invariance de jauge de cette moyenne. Il faut aussi pour cela que $B(x)$ soit un scalaire. Cette définition est de plus invariante sous les reparametrisations du type $A(x) \rightarrow A'(x)$ (à condition que les surfaces physiques soient inchangées).

On voit pourtant que transformer $A(x)$ en un scalaire de genre lumière (ou nul) ne suffit pas pour décrire la moyenne de $S$ sur une surface nulle. On a en effet $\sqrt{-\partial_\mu A \partial^\mu A} = 0$ et la moyenne précédente est non-définie. Pour définir alors cette moyenne, on introduit un scalaire $V$ définissant une hypersurface nulle et un scalaire $A$ définissant une hypersurface spatiale (analogues de $w$ et $\tau$ dans la jauge GLC).
La moyenne sur un volume d'espace-temps à 4 dimensions délimité par ces deux surfaces est alors donnée par l'utilisation de l'intégrale\,:
\beq
I(S;-;A_0,V_0)= \int_{\Mq} d^4x \sqrt{-g}~ \Theta(V_0-V) \Theta(A-A_0) ~ S(x) ~~.
\label{FrIntvol}
\eeq
Appliquée à $S = \nabla_\mu n^\mu$, on obtient par une simple intégration par parties que\,:
\small
\bea
\int_{\Mq} d^4x \sqrt{-g}~ \Theta(V_0-V)  \Theta(A-A_0)  \nabla^{\mu}  n_{\mu}
&=& -\int_{\Mq} d^4x \sqrt{-g}\, \Theta(V_0-V) \delta_{\rm D}(A-A_0)  
\sqrt{-\partial_\mu A \partial^\mu A} \nonumber \\
&+& \int_{\Mq} d^4x \sqrt{-g} \,\delta_{\rm D}(V_0-V)  \Theta(A-A_0) 
\frac{-\partial_\mu V \partial^\mu A}{\sqrt{-\partial_\nu A \partial^\nu A}} ~~.~~~~~
\label{FrGauss1}
\eea
\normalsize
Ce résultat est instructif car il contient à droite de l'égalité, d'une part une moyenne spatiale sur la base du 4-cône (tronqué) considéré (Fig. \ref{Fig3Averages}~(b)), d'autre part une moyenne sur la surface nulle du cône elle-même (Fig. \ref{Fig3Averages}~(a)). De plus, le coefficient qui apparaît dans l'intégrande de cette dernière quantité se révèle être la bonne normalisation pour définir la moyenne sur la surface nulle.

Physiquement, il n'est pas vraiment intéressant d'intégrer sur cette surface nulle car cela revient à extraire un nombre unique pour toute les observations effectuées sur le cône. La perte d'information est donc considérable. Il est plus judicieux de découper ce cône en sphères à valeurs de $A$ constantes et d'intégrer sur chacune de ces sphères. On obtient alors, en prenant $W_\Dscr \propto \delta_{\rm D}(V_0-V) \delta_{\rm D}(A-A_0)$\,:
\beq
I(S;V_0,A_0;-)=\int_{\Mq} d^4x \sqrt{-g}~ \delta_{\rm D}(V_0-V) \delta_{\rm D}(A-A_0) ~ |\partial_\mu V \partial^\mu A| S(x) ~~,
\label{Frnull2d}
\eeq  
ce qui correspond à la bonne manière -- invariante de jauge -- d'intégrer sur la 2-sphère topologique contenue dans les hypersurfaces $V(x) = V_0$ et $A(x) = A_0$ (Fig. \ref{Fig3Averages}~(c)). On peut enfin regrouper ces trois définitions dans des notations distinctes\,:
\beq
\langle S \rangle_{V_0,A_0} = \frac{I(S;V_0,A_0;-)}{I(1;V_0,A_0;-)} ~~,~~ \langle S \rangle_{V_0}^{A_0} = \frac{I(S;V_0;A_0)}{I(1;V_0;A_0)} ~~,~~ \langle S \rangle_{A_0}^{V_0} = \frac{I(S;A_0;V_0)}{I(1;A_0;V_0)} ~~,
\eeq
où l'indice du bas correspond à une distribution $\delta_D$ et celui du haut à $\Theta$.
De manière plus générale, la définition de l'intégrale utilisée peut être résumée par l'Eq. \rref{GenExprForIntI} et le tableau Tab. \ref{TabDiffNorm}.

\subsection{Généralisation des règles de Buchert-Ehlers}
\label{Ch10Sec23}

Il est possible de montrer que les définitions données dans la section précédente amènent elles aussi à des règles de commutation. En effet, la dérivation de la moyenne sur la surface du cône de lumière par rapport aux paramètres $A_0$ et $V_0$ donne\,:
\bea
\label{FrLCeq2}
\frac{\partial}{\partial A_0} \lla S \rra_{V_0}^{A_0} &=& \lla \frac{k \cdot \partial S}{k \cdot \partial A} \rra_{V_0}^{A_0} ~ + \left\{ \frac{1}{k \cdot \partial A} \left[ \nabla \cdot k - \frac{1}{2}\frac{k^{\mu}\partial_{\mu}\left((\partial A)^2\right)}{(\partial A)^2}\right] ~\Bigg|~ S\right\}_{V_0}^{A_0} ~~,
\\
\label{FrLCeq4}
\frac{\partial}{\partial V_0} \lla S \rra_{V_0}^{A_0} &=& \left\langle \frac{\partial A \cdot \partial S}{k \cdot \partial A} \right\rangle_{V_0}^{A_0} 
- \lla k \cdot \partial S \frac{(\partial A)^2}{(k \cdot \partial A)^2} \rra_{V_0}^{A_0} + \left\{ \frac{\left[\Box A-\partial^\mu A \partial_\mu \ln\left((\partial A)^2\right) \right]}{k \cdot \partial A} ~\Bigg|~ S \right\}_{V_0}^{A_0}
\nonumber \\
& &
- \left\{
\left[\nabla \cdot k \frac{(\partial A)^2}{(k \cdot \partial A)^2}+\frac{1}{2} k^\mu \partial_\mu 
\left(\frac{(\partial A)^2}{(k \cdot \partial A)^2}\right)\right] ~\Bigg|~ S \right\}_{V_0}^{A_0} ~~.
\eea
La dérivation de la moyenne sur la 2-sphère donne quant à elle\,:
\bea
\label{FrLCeq1}
\frac{\partial}{\partial A_0} \langle S \rangle_{V_0,A_0} &=& \left\langle \frac{k \cdot \partial S}{k \cdot \partial A}\right\rangle_{V_0,A_0} + ~ \left\{ \frac{\nabla \cdot k}{k \cdot \partial A} ~\Bigg|~ S \right\}_{V_0,A_0} ~~,
\\
\label{FrLCeq3}
\frac{\partial}{\partial V_0} \langle S \rangle_{V_0,A_0} &=& \left\langle \frac{\partial A \cdot \partial S}{k \cdot \partial A} \right\rangle_{V_0,A_0} - \lla k \cdot \partial S \frac{(\partial A)^2}{(k \cdot \partial A)^2} \rra_{V_0,A_0}
+ \left\{\frac{\Box A-\nabla_\mu \left(k^\mu \frac{(\partial A)^2}{k \cdot \partial A}\right)}{k \cdot \partial A} ~\Bigg|~ S \right\}_{V_0,A_0} ~~.
\eea
Nous avons ici introduit les notations simplifiées suivantes\,:
\bea
& \left\{ A ~|~ B \right\}_{V_0,A_0} ~\equiv~ \lla A ~ B \rra_{V_0,A_0} - \lla A \rra_{V_0,A_0} ~ \lla B \rra_{V_0,A_0} ~~, \\
& \left\{ A ~|~ B \right\}_{V_0}^{A_0} ~\equiv~ \lla A ~ B \rra_{V_0}^{A_0} - \lla A \rra_{V_0}^{A_0} ~ \lla B \rra_{V_0}^{A_0} ~~, \\
& k^\mu \partial_\mu S=k \cdot \partial S ~~~~~,~~~~~ 
k^\mu \partial_\mu A=k \cdot \partial A ~~~~~,~~~~~
\partial_\mu A \, \partial^\mu S=\partial A \cdot \partial S ~~,  \\
& \partial_\mu A \, \partial^\mu A=(\partial A)^2 ~~~~~,~~~~~ 
\nabla_\mu k^\mu = \nabla \cdot k ~~~~~,~~~~~ \Box=\nabla^\mu \nabla_\mu ~~.
\eea

\section{Jauge géodésique sur le cône de lumière (GLC)}
\label{Ch10Sec3}

\subsection{Définition de la jauge géodésique sur le cône de lumière}
\label{Ch10Sec31}

Les équations d'Einstein ne privilégient aucun système de coordonnées et possèdent une liberté dite de jauge. Cette liberté est similaire à celle que l'on observe en électromagnétisme lorsque l'on redéfinit le potentiel électrique ou le potentiel vecteur sans en changer les champs. On peut alors, sans changer la physique, choisir une jauge dans laquelle nos équations sont les plus simples.

Dans le cas de calculs sur le cône de lumière passé, on définit la métrique GLC (pour \emph{geodesic lightcone gauge} \cite{P1}) qui dépend de 6 fonctions arbitraires (un scalaire $\Ups$, un 2-vecteur $U^a$, et une matrice $2 \times 2$ symétrique $\gamma_{ab}$ avec $a,b = 1,2$). L'élément de distance est alors définit par
\bea
\label{Frds2GLC}
ds_{GLC}^2 = \Ups^2 dw^2-2\Ups dw d\tau+\gamma_{ab}(d\ti{\theta}^a-U^a dw)(d\ti\theta^b-U^b dw) ~~,
\eea
où la coordonnée $w$ est de type nulle (i.e. $\pa_\mu w \, \pa^\mu w=0$) et elle spécifie un cône de lumière passé au sommet duquel se trouve un observateur. Le temps $\tau$ définit le temps propre de cet observateur et est égal au temps $t$ de la jauge synchrone (quelque soit la métrique). On peut vérifier que le vecteur $\partial_{\mu} \tau$ définit son flot géodésique, soit\,:
\beq
\left( \pa^\nu \tau\right) \nabla_\nu \left( \pa_\mu \tau\right) = 0 ~~.
\eeq
Quant aux angles $\widetilde{\theta}^a$, ils paramétrisent la surface compacte topologiquement équivalente à une 2-sphère située à l'intersection des hypersurfaces $\tau = {\rm cst}$ et $w = {\rm cst}$. 
Sous forme matricielle, et en terme des coordonnées $(\tau,w,\widetilde{\theta}^a)$, la métrique et son inverse s'écrivent\,:
\beq
\label{FrGLCmetric}
g^{GLC}_{\mu\nu} =
\left(
\begin{array}{ccc}
0 & - \Ups &  \vec{0} \\
-\Ups & \Ups^2 + U^2 & -U_b \\
\vec0^{\,T}  &-U_a^T  & \gamma_{ab} \\
\end{array}
\right) ~~,~~
g_{GLC}^{\mu\nu} =
\left(
\begin{array}{ccc}
-1 & -\Ups^{-1} & -U^b/\Ups \\
-\Ups^{-1} & 0 & \vec{0} \\
-(U^a)^T/\Ups & \vec{0}^{\, T} & \gamma^{ab}
\end{array}
\right) ~,
\eeq
avec $\vec 0=(0,0)$, $U_b= (U_1, U_2)$, et où $\gamma_{ab}$ (resp. $\gamma^{ab}$) est une matrice $2 \times 2$ qui sert à descendre (resp. monter) les indices angulaires. Cette matrice $\gamma_{ab}$ paramétrise la géométrie de la 2-sphère. On peut aussi montrer la relation simple suivante\,:
\beq
\sqrt{-g} = \Ups \sqrt{|\gamma|} ~~~~~\mbox{avec}~~~~~ \gamma \equiv \det\gamma_{ab} ~~.
\eeq
Dans la limite d'un Univers FLRW, les coordonnées et les éléments de métrique se réduisent à\,:
\beq
\tau=t ~~,~~ w= r+\eta ~~,~~ \Ups = a(t) ~~,~~
U^a=0 ~~,~~ \gamma_{ab} d \tilde{\theta}^a d\tilde{\theta}^b = a^2(t) r^2 (d \theta^2 + (\sin \theta)^2 d\phi^2) ~~.
\eeq

La métrique GLC possède de fortes similitudes avec la métrique ADM. En effet, la comparaison des Eqs. \rref{Frds2GLC} et \rref{Frds2ADM} montre que l'on peut interpréter $\Ups$ comme une \emph{lapse function}, définissant la \guillemotleft \, distance \guillemotright \, entre deux cônes de lumière $w = w_0$ et $w = w_1$. Le vecteur $U^a$ est alors similaire à un \emph{shift vector} et définit le changement des coordonnées angulaires lorsque l'on passe d'une 2-sphère $\Sigma(w_0,\tau_0)$ à une autre $\Sigma(w_1,\tau_1)$. Enfin, $\gamma_{ab}$ est la métrique définissant la géométrie au sein de la 2-sphère.

Il convient aussi de noter que cette métrique, dans sa forme générale, correspond à la fixation de jauge d'un système de coordonnées dites \emph{observationelles} (\emph{observational coordinates}, \cite{1985PhR...124..315E}) pour lesquelles nous avons employé une coordonnée de type temps (i.e. $\tau$) à la place d'une coordonnée radiale (qui peut être choisie de différentes manières). D'autres différences peuvent également être relevées (cf. Sec. \ref{Ch3Sec23}).

\subsection{Applications}
\label{Ch10Sec32}

\subsubsection{Redshift et distance de luminosité}
\label{Ch10Sec321}

On comprend aisément que cette métrique GLC est parfaitement adaptée aux observations. Ceci se voit d'autant plus par la considération du redshift et de la distance de luminosité calculés dans ces coordonnées. Considérons en effet une source située sur le cône $w = w_o$ et à un temps $\tau_s$ en dénotant par ``o'' une quantité évaluée au niveau de l'observateur et par ``s'' une quantité située au niveau de la source. Les photons émis par cette source ont un moment $k_\mu = \partial_\mu w$ orthogonal à lui même ($k_\mu k^\mu = 0$) et à la 2-sphère ($\partial_\mu \ti{\theta}^a k^\mu = 0$). Le produit scalaire de $k_\mu$ avec le vecteur $n^\mu = - \partial_\mu \tau$ normal à l'hypersurface spatiale $\tau = {\rm cst}$ donne alors simplement
\beq
k_\mu n^\mu = g^{w \tau}_{GLC} = - \Ups^{-1} ~~.
\eeq
Le redshift est alors donné par le rapport de ces quantités aux positions de la source et de l'observateur,
\beq
\label{Frredshift}
(1+z_s) = \frac{(k^{\mu} u_{\mu})_s }{(k^{\mu} u_{\mu})_o} = {\Ups(w_o, \tau_o, \ti \theta^a)\over \Ups(w_o, \tau_s, \ti \theta^a)} ~~,
\eeq
et ne dépend que de $\Ups$.
La distance de luminosité $d_L$ d'une source ayant un redshift $z_s$ est elle obtenue à travers la distance angulaires $d_A$ selon l'expression\,:
\beq
d_L(z_s) = (1 + z_s)^2 d_A(z_s)
\eeq
qui découle du théorème de réciprocité d'Etherington \cite{Et1933}. On peut alors écrire, d'après l'équation de convergence de Sachs, que
\beq
\label{FrEqSachs}
{d \over d \la} \left(\ln d_A\right) = {\hat{\Theta} \over 2} \equiv {1\over 2} \nabla_\mu k^\mu ~~,
\eeq
où $\la$ est un paramètre affine allant de la source ($\la = 0$) à l'observateur ($\la = 1$) le long du chemin parcouru par le photon.
De l'égalité $k^\mu = dx^\mu/ d \la= -\da^\mu_\tau \Ups^{-1}$ nous déduisons\,:
\beq
\nabla_\mu k^\mu = - \Ups^{-1} \pa_\tau \left( \ln \sqrt{\ga}\right) = {d \over d \la} \left(\ln \sqrt{\ga}\right) ~~\mbox{avec $\ga \equiv \det \ga_{ab}$} ~~.
\eeq
L'intégration de \rref{FrEqSachs} donne alors
\beq
d_A^2(\la)= c \sqrt{\ga(\la)} ~~,
\eeq
où $c$ est une constante vis-à-vis du paramètre affine $\la$. En prenant la limite $d_A \ra 0$ (ou $\la \ra 0$) pour laquelle $\ti \theta^a \ra \theta^a$ (i.e. les angles homogènes $\theta$, $\phi$), nous obtenons que $\lim_{\la \rightarrow 0} \left[\sqrt{\ga} / d_A^2 \right] = \sin \ti \theta^1 = \sin \theta^1_o$ et donc $c=1/\sin \tilde{\theta}^1$.
Nous avons ainsi montré que la distance de luminosité pouvait s'écrire\,:
\beq
\label{FrGeneral_Eq_dA}
d_L(z_s) = (1 + z_s)^2 ~ \ga^{1/4}(z_s,w_o,\widetilde{\theta}^a) \left(\sin \tilde{\theta}^1\right)^{-1/2} ~~.
\eeq
Cette relation simple et non-perturbative donne la distance de luminosité dans un Univers quelconque et ne fait intervenir que le déterminant de la métrique sur la 2-sphère à $w = w_o$ et $\tau_s = \tau(w_o,z_s,\widetilde{\theta}^a)$.

\subsubsection{Règles de Buchert-Ehlers}
\label{Ch10Sec322}

L'utilisation de la jauge GLC pour les coordonnées $\{ x^\mu \}$ des intégrales présentes dans les règles de commutation établies en Sec. \ref{Ch10Sec23} révèle une propriété intéressante. En effet, les choix naturels $A(\tau,w,\widetilde{\theta}^a) = \tau$ et $V(\tau,w,\widetilde{\theta}^a) = w$ nous permettent de montrer que les expressions précédentes sont équivalentes à\,:
\beq
\partial_{\tau_s} \lla S \rra = \lla \partial_{\tau} S \rra + \lla S \partial_{\tau}\ln\sqrt{|\gamma|} \rra - \lla \partial_{\tau}\ln\sqrt{|\gamma|} \rra \lla S \rra ~~,
\eeq
\bea
\partial_{w_o} \lla S \rra &=& \lla \partial_{w} S + U^a \partial_a S \rra + \lla \left[ \partial_{w}\ln\sqrt{|\gamma|} \right] S \rra - \lla \partial_{w}\ln\sqrt{|\gamma|}\rra \lla S \rra \nonumber \\
& & + \lla \left[ \partial_a U^a + U^a \partial_a \ln\sqrt{|\gamma|}\right] S \rra - \lla \partial_a U^a +U^a \partial_a \ln\sqrt{|\gamma|} \rra \lla S \rra ~~.
\eea
Ces expressions sont valides à la fois pour $\lla \ldots \rra_{w_o}^{\tau_s}$ (surface nulle) et $\lla \ldots \rra = \lla \ldots \rra_{w_o,\tau_s}$ (2-sphère). C'est une des nombreuses propriétés remarquables de cette métrique GLC dont un autre exemple est le calcul du \emph{redshift drift} (cf. Sec. \ref{Ch3Sec63}).

On peut enfin montrer qu'il n'y a pas d'effet de rétroaction dans un espace homogène et isotrope. Les équations de Friedmann généralisées à un domaine $\Dscr$ se réduisent alors aux équations de Friedmann présentées en Eqs. \rref{FrEqFried} et \rref{FrEqFried2} (cf. Sec. \ref{Ch2Sec4}). Le scénario le plus proche du modèle standard que l'on puisse trouver et qui fasse intervenir un effet de rétroaction est l'Univers FLRW plat et perturbé.

\section{La distance de luminosité au second ordre en perturbations}
\label{Ch10Sec4}

\subsection{Au premier ordre}
\label{Ch10Sec41}

Nous allons calculer la distance de luminosité au second ordre dans les perturbations de métrique autour d'un Univers de type FLRW. Pour cela, nous profitons des relations simples que nous offre le choix de la métrique GLC. Néanmoins, les perturbations de métrique sont souvent exprimés dans la jauge Newtonienne (ou plus généralement de Poisson). De plus, nous connaissons le spectre de puissance des inhomogénéités dans la jauge Newtonienne, et non dans la métrique GLC.
Il nous est par conséquent nécessaire de trouver la transformation entre la métrique GLC, utile pour nos calculs, et la métrique de Poisson qui définit les inhomogénéités. Nous verrons que la nature de ces inhomogénéités, considérées comme des quantités stochastiques, impose d'évaluer cette transformation jusqu'au deuxième ordre en perturbations. On utilise donc\,:
\beq
g_{GLC}^{\alpha\beta}(x)=\frac{\partial x^\alpha}{\partial y^\mu} \frac{\partial x^\beta}{\partial y^\nu} g_{PG}^{\mu\nu}(y) ~~,
\label{FrEqBetweenGauges}
\eeq
avec $\{ x^\mu \} = (\tau, w, \ti{\theta}^a)$, $\{ y^\mu \} = (\eta, r, \theta, \phi)$ et la métrique de Poisson (\emph{Poisson Gauge})\,:
\beq
g_{PG}^{\mu \nu} = a(\eta)^{-2} {\rm diag}( -1 + 2 \tilde{\Phi}, 1 + 2 \tilde{\Psi}, (1 + 2 \tilde{\Psi}) \gamma_0^{ab} ) ~~,
\eeq
où, de manière cohérente avec les Eqs. \rref{PGmetricstandard} et \rref{PhiAndPsiInPG}, nous avons\,:
\beq
\tilde{\Phi} = \psi + \frac{1}{2} \phi^{(2)} - 2 \psi^2 ~~,~~ \tilde{\Psi} = \psi + \frac{1}{2} \psi^{(2)} + 2 \psi^2 ~~,~~ \ga_0^{ab} = {\rm diag} \left( r^{-2}, r^{-2} \sin^{-2} \theta \right) ~~.
\eeq
Les champs $\tilde{\Phi}$ et $\tilde{\Psi}$ sont appelés \emph{potentiels de Bardeen} et généralisent la notion du potentiel gravitationnel Newtonien. Nous les prenons égaux au premier ordre car nous supposons le cisaillement de la distribution de matière nul à cet ordre\footnote{Autrement, les termes de propagation du photon au premier ordre dépendent de la combinaison $\tilde{\Phi} + \tilde{\Psi}$.}.

Les composantes $(\tau,\tau)$, $(w,w)$ et $(w,\widetilde{\theta}^a)$ de $g_{GLC}^{\alpha\beta}$ nous permettent d'écrire $\{ \tau, w, \ti{\theta}^a \}$ comme un ensemble de fonctions de $(\eta, r, \theta, \phi)$ (voir Eqs. \rref{tau2order} à \rref{ExprLamb}). Pour rendre la discussion plus simple, nous ne présenterons que les résultats au premier ordre de cette transformation. On obtient donc\,:
\bea
\label{Frtauwtheta}
\tau &=& \int_{\eta_{in}}^\eta d\eta' a(\eta')\left[1 + \psi(\eta', r, \theta^a)\right] ~, \\
w &=& \eta_+ + \int_{\eta_+}^{\eta_-} dx\, \hat{\psi}(\eta_+, x, \theta^a) ~,
\label{Frtauwtheta1}\\
\ti{\theta}^a &=& \theta^a + \frac{1}{2} \int_{\eta_+}^{\eta_-} dx \, \hat{\gamma}^{ab}_0 
(\eta_+, x,\theta^a) \int_{\eta_+}^x dy\, \pa_b \hat{\psi}(\eta_+, y, \theta^a)  ~,
\label{Frtheta}
\eea
où $\hat{\psi}(\eta_+,\eta_-,\theta^a)\equiv \psi(\eta,r,\theta^a)$, $\hat \gamma^{ab}(\eta_+,\eta_-,\theta^a)\equiv \gamma^{ab}(\eta,r,\theta^a)$ et $\eta_{in}$ représente un temps assez reculé pour que l'intégrande considérée soit négligeable. Nous avons aussi introduit les variables d'ordre zero sur le cône de lumière\,:
\beq
\eta_\pm = \eta \pm r ~~\mbox{ avec\,: }~~~ \pa_\eta = \pa_+ + \pa_- ~~,~~ \pa_r = \pa_+ - \pa_- ~~,~~ \pa_\pm= {\pa \over \pa \eta_\pm}={1\over 2} \left( \pa_\eta \pm \pa_r \right) ~~.
\eeq

Les autres composantes $(\tau,w)$, $(\tau,\widetilde{\theta}^a)$ et $(\widetilde{\theta}^a,\widetilde{\theta}^b)$ donnent alors $\{ \Ups, U^a, \gamma^{ab} \}$ en terme de $(\psi,\psi^{(2)},\phi^{(2)})$ (voir Eqs. \rref{InvUps} à \rref{gammaab2}), avec au premier ordre\,:
\scriptsize
\bea
\Ups &=& a(\eta) \left[ 1 + \hat{\psi}(\eta_+,\eta_+, \theta^a) - \int_{\eta_+}^{\eta_-} dx \,\partial_+ \hat{\psi}(\eta_+, x, \theta^a)\right]
+ \int_{\eta_{in}}^\eta d\eta' a(\eta')\partial_r \psi(\eta', r, \theta^a) ~~,
\label{FrUps} \\
U^a &=& \frac{1}{2}  \hat{\gamma}^{ab}_0 \int_{\eta_+}^{\eta_-} dx\, \pa_b \hat{\psi}(\eta_+, x, \theta^a) - \frac{1}{a(\eta)}  \gamma^{ab}_0 \int_{\eta_{in}}^\eta d\eta' a(\eta') \, \pa_b \psi(\eta', r, \theta^a)
\nonumber \\
&& +\frac{1}{2} \int_{\eta_+}^{\eta_-} dx~ \partial_+ \left[\hat{\gamma}^{ab}_0(\eta_+,x,\theta^a) \int_{\eta_+}^x dy~ \pa_b\hat{\psi}(\eta_+, y, \theta^a) \right] -\frac{1}{2} \lim_{x\rightarrow \eta_+} \left[\hat{\gamma}^{ab}_0(\eta_+,x,\theta^a) \int_{\eta_+}^x dy~ \pa_b\hat{\psi}(\eta_+, y, \theta^a) \right] ~~,~~~ \\
\gamma^{ab} &=& \frac{1}{a(\eta)^2}\left\{ \left[1+2 \psi(\eta, r, \theta^a)\right]\gamma_0^{ab} 
+{1\over 2}\left[  \hat{\gamma}_0^{ac}  \int_{\eta_+}^{\eta_-} dx\, \partial_c \left(
\hat{\gamma}_0^{bd}(\eta_+, x,\theta^a) \int_{\eta_+}^x dy\, \pa_d\hat{\psi}(\eta_+, y, \theta^a)\right)
+ a \leftrightarrow b \right]\right\} ~~.
\label{Frgammaab}
\eea
\normalsize

On obtient par l'Eq. \rref{FrGeneral_Eq_dA}, après considération de certaines subtilités entre autre liées au développement perturbatif du terme $a(\eta_s) r_s$ contenu dans $\gamma \equiv \det \gamma_{ab}$ (cf. Sec. \ref{Ch4Sec21} au second ordre ou, plus simplement, en Sec. \ref{Ch5Sec1}), l'expression de la distance de luminosité au second ordre\,:
\beq
\frac{d_L(z_s, \theta^a)}{d_L^{\rm FLRW}(z_s)} = \frac{d_L(z_s, \theta^a)}{(1+z_s)a_o (\eta_o - \eta_s)} =  1 + \delta_S^{(1)}(z_s, \theta^a) + \delta_S^{(2)}(z_s, \theta^a) ~~,
\eeq
où seules les perturbations scalaires ont été considérées. Nous introduisons les grandeurs suivantes\,:
\beq
\Xi_s =  1 - \frac{1}{\Hcal_s \Delta \eta} ~~,~~ \Delta\eta = \eta_o - \eta_s^{(0)} ~~,
\eeq
\beq
Q(\eta_+, \eta_-, \theta^a) = \int_{\eta_+}^{\eta_-} dx~ \hat{\psi}(\eta_+,x,\theta^a) ~~,~~ P(\eta, r, \theta^a) = \int_{\eta_{in}}^\eta d\eta' \frac{a(\eta')}{a(\eta)} \psi(\eta',r,\theta^a) ~~.
\eeq
En définissant maintenant les quantités physiques qui suivent\,:
\beq
J  = ([\partial_+ Q]_s - [\partial_+ Q]_o) - ([\partial_r P]_s - [\partial_r P]_o) ~~,
\eeq
\beq
J_2^{(1)} = \frac12 \nabla_a \tilde{\theta}^{a (1)} = \frac12 \left(\cot \theta ~ \tilde{\theta}^{(1)} + \partial_a \tilde{\theta}^{a (1)} \right) = \int_{\eta_s^{(0)}}^{\eta_o} \frac{d \eta}{\Delta\eta} ~ \frac{\eta - \eta_s^{(0)}}{\eta_o - \eta} \Delta_2 \psi(\eta, \eta_o-\eta, \theta^a) ~~,
\eeq
la correction au premier ordre $\delta_S^{(1)}$ s'écrit comme\,:
\bea
\label{FrfinaldL}
\delta_S^{(1)}(z_s, \theta^a) &=& \Xi_s J - \frac{Q_s}{\Delta \eta} - \psi_s^{(1)} - J_2^{(1)} ~~.
\eea
Ces termes à l'ordre 1 sont bien connus dans la littérature et sont aisément interprétables. Les contributions du type $\partial_r P$ correspondent au produit scalaire $\vbf \cdot \nbf$ du vecteur vitesse (de la source ou de l'observateur) avec un vecteur unitaire $\nbf$ parallèle au chemin du photon et pointant en direction de l'observateur. La vitesse $\vbf$ s'écrit, dans la jauge Newtonienne\,:
\beq
\vbf =-\int_{\eta_{in}}^{\eta}d\eta'\frac{a(\eta')}{a(\eta)} \nabla \psi(\eta', r, \theta^a) ~~.
\eeq
Les termes du type\,:
\beq
\partial_+ Q_s-\partial_+ Q_o \sim \psi_o - \psi_s - 2 \int_{\eta_s}^{\eta_o} d \eta ~ \partial_\eta \psi(\eta,r,\theta^a) ~~,
\eeq
correspondent aux effets Sachs-Wolfe (SW) et Sachs-Wolfe intégré (ISW).
Enfin, le terme
\beq
J_2 = \int_{\eta_s^{(0)}}^{\eta_o} \frac{d \eta}{\Delta\eta} ~ \frac{\eta - \eta_s^{(0)}}{\eta_o - \eta} \Delta_2 \psi(\eta, \eta_o-\eta, \theta^a) + \Ocal(2) ~~,
\eeq
est un terme de lentille gravitationnelle.

\subsection{Au second ordre}
\label{Ch10Sec42}

On obtient la correction à la distance de luminosité au deuxième ordre dans les perturbations scalaires $\delta_S^{(2)}(z_s, \theta^a)$ de la même manière qu'au premier ordre, i.e. en développant $\ga^{1/4}(z_s,w_o,\widetilde{\theta}^a)$ et $(\sin \tilde{\theta}^1)^{-1/2}$ dans l'Eq. \rref{FrGeneral_Eq_dA}. L'expression de $d_L$ obtenue est alors du type $(1+z_s)^2 (a_s r_s) \big(1 + \Ocal(1) + \Ocal(2) \big)$ avec $a_s \equiv a(\eta_s)$ le facteur d'échelle au niveau de la source et $r_s$ sa distance radiale (cf. Eq. \rref{d_L}). Ces deux termes doivent ensuite être développés en perturbations. Le premier est obtenu en considérant la relation $1+z_s = \Ups_o / \Ups_s$ que l'on développe d'après l'expression perturbée de $\Ups$. On obtient alors une égalité du type $1+z_s = a(\eta_o) / a(\eta_s) ( 1 + \Ocal(1) + \Ocal(2) )$ et on peut développer $a(\eta_s)$ comme $a(\eta_s^{(0)}) \left( 1 + \eta_s^{(1)} \Hcal_s + \Ocal(2) \right)$. L'utilisation de l'indentité $1+z_s = a(\eta_o) / a(\eta_s^{(0)})$ impose d'annuler toutes les perturbations présentes et donne les corrections de $\eta_s = \eta_s^{(0)} + \eta_s^{(1)} + \eta_s^{(2)}$. On peut alors calculer les corrections à $r_s$ en utilisant la transformation\,: $w_s = \eta_o = \eta_s + r_s + \Ocal(1) + \Ocal(2)$. Bien sûr toutes ces opérations, dont le détail rigoureux est présenté en Sec. \ref{Ch4Sec21}, génèrent de nombreux termes au second ordre à partir du premier ordre (termes quadratiques).

Les termes obtenus et résumés en Eqs. \rref{finaldL}, \rref{X2Y2} et \rref{eps2Smoins0}, \rref{eps2oneTotwo}, sont de toutes les sortes possibles. On trouve en effet des combinaisons du premier ordre, du type\,:
\bea
& \psi_s^2 ~~,~~ (\psi_s, Q_s) \times \mbox{(Lent. Grav., ISW, Doppler)} ~~,~~ (\mbox{ISW})^2 ~~, \nonumber \\
& (\mbox{Doppler})^2 ~~,~~ \mbox{ISW} \times \mbox{Doppler} ~~,~~ (\mbox{Lent. Grav.})^2 ~~.
\eea
On y trouve aussi des termes intrinsèques à l'ordre 2\,:
\beq
\psi_s^{(2)} ~~,~~ \left(\mbox{Lent. Grav.}\right)^{(2)} = \frac12 \nabla_a \tilde{\theta}^{a (2)} ~~,~~ Q_s^{(2)} ~~,
\eeq
de nouveaux termes intégrés tels que\,:
\beq
\frac{1}{4 \Delta\eta} \int_{\eta_s^{(0)+}}^{\eta_s^{(0)-}} dx~ \left[4 \hat{\psi} ~ \partial_+ Q + \hat{\gamma}_{0}^{ab} ~ \partial_a Q ~ \partial_b Q \right] (\eta_s^{(0)+},x,\theta^a) ~~,
\eeq
et des termes de déformation angulaire\,:
\beq
(\gamma_0)_{ab} \partial_+ \tilde{\theta}^{a (1)}\partial_- \tilde{\theta}^{b (1)} ~~,~~ \partial_a \tilde{\theta}^{b (1)}\partial_b \tilde{\theta}^{a (1)} ~~,
\eeq
avec $\ti{\theta}^{a(1)}$ donné en Eq. \rref{Frtheta}.
Enfin, les perturbations de redshift engendrent un grand nombre de nouvelles contributions, e.g.\,:
\bea
& \gamma_0^{ab} \left( \partial_a P ~ \partial_b P ~,~ \partial_a Q ~ \partial_b Q ~,~ \partial_a Q ~ \partial_b P \right) ~~,~~ 
\partial_+ \int_{\eta_+}^{\eta_-} dx~  \left[ 4 \hat{\psi} ~ \partial_+ Q + \hat{\gamma}_0^{ab} \partial_a Q ~ \partial_b Q \right] ~~,~~ \\
& \int_{\eta_{in}}^\eta d\eta'~ \frac{a(\eta')}{a(\eta)} \partial_r \left[\left( \partial_r P \right)^2 + \gamma_0^{ab} \partial_a P ~ \partial_b P \right] ~~,
\eea
dont la dernière contient la partie transverse $\partial_a P$ de la vitesse particulière de la source.

Il est possible de généraliser ces calculs à des perturbations de métrique vectorielles et tensorielles. On utilise dans ce cas la jauge de Poisson au lieu de la jauge Newtonienne\,:
\beq
ds_{PG}^2 = a^2(\eta) \left( -d\eta^2 + 2 v_i d\eta d x^i + [(\gamma_0)_{ij} + \chi_{ij}] d x^i d x^j \right) ~,~ \{x^i\} = (r,\theta,\phi) ~~,
\eeq
avec $v^i$ et $\chi^{ij}$ qui vérifient\,:
\beq
\nabla_i v^i = \nabla_i \chi^{ij} = 0 ~~~~~ \mbox{(sans divergence)} ~~,~~ (\gamma_0)_{ij} \chi^{ij} = 0 ~~~~~ \mbox{(sans trace)} ~~.
\eeq
Les résultats de la transformation des coordonnées et des éléments de métrique, en supposant les perturbations vectorielles et tensorielles nulles au premier ordre (prédit par l'inflation), sont analogues et font intervenir la combinaison radiale $\frac{v^r}{2} - \frac{\chi^{rr}}{4}$ (cf. Eqs. \rref{tauVT} à \rref{thetaVT}). On obtient aussi des corrections à la distance de luminosité (cf. Eq. \rref{dLVTfinal}). Néanmoins, la moyenne de ces corrections sur la 2-sphère (Sec. \ref{Ch5Sec36}) est symétriquement nulle et il n'est pas nécessaire de les décrire plus.

\section{Moyenne stochastique sur le cône de lumière}
\label{Ch10Sec5}

\subsection{Considérations générales}
\label{Ch10Sec51}

Dans le but d'évaluer l'effet des inhomogénéités sur les observations cosmologiques de l'énergie sombre, nous allons effectuer la moyenne d'observables scalaires, une moyenne sur la 2-sphère qui se situe dans notre cône de lumière passé à un redshift $z_s$, mais aussi stochastique, de manière à s'affranchir d'une distribution particulière de matière et à la place ne considérer que le spectre de puissance de ces inhomogénéités.
On décompose le champ gravitationnel au premier ordre en modes de Fourier\,:
\beq
\psi(\eta, \xbf) = \frac{1}{(2 \pi)^{3/2}} \int d^3 k ~ \e^{i\kbf\cdot \xbf} ~ \psi_k(\eta) E(\kbf) ~~,
\label{FrPsiFourier}
\eeq
où $E$ est une variable aléatoire unitaire vérifiant\,:
\bea
\label{FrPropRandomVar}
& E^*(\kbf) = E(-\kbf) ~~~\mbox{(homogénéité statistique)} ~~, \\
& \overline{E(\kbf)} = 0 ~~,~~ \overline{E(\kbf_1) E(\kbf_2)} = \delta_{\rm D}(\kbf_1+\kbf_2) ~~~\mbox{(moyenne d'ensemble)} ~~.
\eea
Cette variable aléatoire $E$ rend les fluctuations statistiquement homogènes et isotropes.

Le spectre de puissance du champ gravitationnel est donné dans son régime linéaire par\,:
\bea
\label{FrSpecPuiss}
\Pcal_\psi (k, \eta) &\equiv& \frac{k^3}{2 \pi^2} |\psi_k(\eta)|^2 = g^2(z) T^2(k) \frac{9}{25} A \left(\frac{k}{k_0}\right)^{n_s-1} ~~.
\eea
C'est une quantité sans dimension qui dépend du spectre primordial (issu de l'inflation) $A \left( k / k_0 \right)^{n_s-1}$ dont les paramètres numériques sont connus\,:
\beq
A = 2.45 ~10^{-9} ~~,~~ n_s = 0.96 ~~~\mbox{(indice spectral)} ~~,~~ \frac{k_0}{a_0} = 0.002 ~ {\rm Mpc}^{-1} ~~.
\eeq
Il dépend aussi d'une fonction $T(k)$, la fonction de transfert, qui tient compte de l'évolution de la matière entre la fin de l'inflation et la période de matière (autrement dit, dans l'ère de radiation). Grossièrement, cette fonction décrit l'effet de réentrée progressive de ces modes et leur perte de puissance aux courtes longueurs d'onde due à la dispersion par radiation (\emph{Silk damping}) et l'oscillation (BAO) des modes de grande longueur d'onde (mais plus courte que l'horizon causal). Ce dernier phénomène n'ajoute qu'une faible modulation à l'amplitude de $T(k)$ et n'est pas considéré dans notre étude.
Enfin, le spectre de puissance fait intervenir la fonction du redshift $z$ (ou du temps $\eta$)\,:
\beq
g(z) = \frac{5g_{\infty}}{2} \Om_m(z) \Bigg[ \Om_m(z)^{4/7} - \Om_\Lambda(z) + \left[1 + \frac{\Om_m(z)}{2} \right] \left[1 + \frac{\Om_\Lambda(z)}{70} \right] \Bigg]^{-1} ~~,
\eeq
qui a déjà été décrite en Sec. \ref{Ch10Sec13}. Cette fonction fait intervenir l'expression des paramètres cosmologiques dans un Univers de type $\Lambda$CDM\,:
\beq
\Omega_m(z) = \frac{\Omega_{m0}(1+z)^3}{\Omega_{m0}(1+z)^3+\Omega_{\Lambda0}} ~~,~~ \Omega_{\Lambda}(z) = \frac{\Omega_{\Lambda0}}{\Omega_{m0}(1+z)^3+\Omega_{\Lambda0}} ~~.
\eeq

\subsection{Termes quadratiques\,: backreaction induite}
\label{Ch10Sec52}

Le but de notre étude, liée aux observations des SNe Ia, est de calculer des corrélations du type $\overline{\lla d_L(z,\nbf) \rra}$ ou e.g. $\overline{\lla d_L(z,\nbf) d_L(z,\nbf) \rra}$ et non $\overline{\lla d_L(z,\nbf) d_L(z,\nbf') \rra}$ ou encore $\overline{\lla d_L(z,\nbf) d_L(z',\nbf) \rra}$ (cf. \cite{Bonvin:2005ps} pour ces dernières). On considère donc des perturbations aux quantités présentes dans le modèle FLRW et dont l'origine est le fond stochastique des perturbations primordiales. Ces perturbations vérifient les conditions présentées en Eq. \rref{FrPropRandomVar}\,: $\overline \psi = 0$ et $\overline{\psi^2} \neq 0$. Par conséquent, la moyenne au premier ordre d'une quantité scalaire $S$ donne $\overline{S}= S^{\rm FLRW}$. Pour obtenir une correction au modèle homogène, il est donc nécessaire d'aller à l'ordre quadratique ou des ordres supérieurs (ou, encore une fois, de considérer des quantités du type $\overline{S(z, \theta^a) S(z', \theta^{\prime a})}$). Dans cette étude toutefois, nous allons considérer la moyenne stochastique, non pas directement d'une quantité scalaire $S$, mais de sa moyenne sur le cône de lumière $\lla S \rra$. Nous allons voir qu'il est alors possible d'obtenir des corrections à partir du simple premier ordre de $S$ et ce du fait du couplage, dans la moyenne, entre cette quantité scalaire et la mesure d'intégration sur le cône (amenant ainsi $\overline{\langle S \rangle} \neq S^{\rm FLRW}$). C'est cet effet généré par la combinaison des deux moyennes, sur le cône de lumière et stochastique, que nous allons ici décrire. Il convient de rappeler aussi que cette combinaison avait déjà été étudiée dans le cas d'une moyenne spatiale (cf. \cite{Li:2007ny,Clarkson:2009hr,Clarkson:2011uk}), mais jamais avec la moyenne sur le cône.

Considérons une moyenne typique sur une surface compacte $\Sg$ (topologiquement équivalente à une 2-sphère) plongée dans le cône de lumière passé $w=w_o$ et à un redshift constant $z_s$. Nous écrivons cette moyenne simplement comme\,:
\beq
\label{FrAvS}
\langle S  \rangle_\Sg =  \frac{\int_\Sg d^2 \mu \,S}{ \int_\Sg d^2 \mu} ~~,
\eeq
où $d^2 \mu$ est la mesure d'intégration adéquate obtenue par notre approche invariante de jauge (voir Eq. \rref{zaverageexact}) et $S$ est une observable scalaire (possiblement non locale). On peut séparer la contribution homogène de la perturbation au premier ordre dans ces deux quantités\,:
\bea
\label{Frexp}
d^2 \mu =  (d^2 \mu)^{(0)}(1 + \mu) ~~~~~,~~~~~  S = S^{(0)} (1 + \sigma) ~~,
\eea
et renormaliser nos intégrales de sorte que $\int (d^2 \mu)^{(0)} = 1$. On obtient alors\,:
\bea
\label{FrAvS1}
\left\langle \frac{S}{S^{(0)}}  \right\rangle =  \frac{\int (d^2 \mu)^{(0)}(1+ \mu)(1+\sigma) }{\int (d^2 \mu)^{(0)}(1+ \mu)} = 
{\langle (1+ \mu)(1+\sigma)\rangle_0  \over 1+ \langle \mu \rangle_0} ~~,
\eea
où nous avons omis le symbole $\Sg$ et avons désigné les moyennes relatives à la mesure $(d^2 \mu)^{(0)}$ par la notation $\langle \dots \rangle_0$. Ce dernier indice sera oublié dans les considérations qui font suite.

Considérons maintenant la moyenne d'ensemble (stochastique) en gardant à l'esprit que celle-ci ne se factorise pas, i.e. $\overline{AB} \ne \overline{A} ~\overline{B}$. On trouve par cette moyenne de l'Eq. \rref{FrAvS1} que\,:
\bea
\label{FrEnsAv}
\overline{\langle S/S^{(0)}  \rangle}  =  1+ \overline{(\langle \sigma\rangle+ \langle \mu \sigma\rangle)  (1+ \langle \mu \rangle)^{-1}} ~~.
\eea
Cette relation devient particulièrement intéressante lorsque l'on developpe $\mu$ et $\sigma$ en séries\,:
\bea
\mu = \sum_i \mu_i ~~~~~,~~~~~~ \sigma = \sum_i \sigma_i ~~, 
\eea
et que l'on fait l'hypothèse que $\mu_1$ et $\sigma_1$ s'annulent sous l'effet de la moyenne d'ensemble (comme toute perturbation d'origine inflationnaire). Notre résultat peut alors se développer selon\,:
\beq
\label{FrEnsAvExp}
\overline{\langle S/S^{(0)}  \rangle}  = 1 +  \overline{\langle \sigma_2 \rangle} + \rm{IBR}_2 + \overline{\langle \sigma_3 \rangle} + \rm{IBR}_3 + \dots 
\eeq
où, par l'utilisation de $\overline{\langle \mu_1\rangle  \langle \sigma_1\rangle} \neq \overline{\langle \mu_1\rangle}~  \overline{\langle \sg_1\rangle}$, nous avons\,:
\bea
\label{Fribr2}
\rm{IBR}_2 &=&  \overline{\langle \mu_1 \sigma_1\rangle} - \overline{\langle \mu_1\rangle  \langle \sigma_1\rangle} ~~, \\
\rm{IBR}_3 &=&  \overline{\langle \mu_2 \sigma_1\rangle} - \overline{\langle \mu_2\rangle  \langle \sigma_1\rangle} +
\overline{\langle \mu_1 \sigma_2\rangle} - \overline{\langle \mu_1\rangle  \langle \sigma_2\rangle} - \overline{\langle \mu_1\rangle  \langle \mu_1 \sigma_1\rangle} + \overline{\langle \mu_1\rangle \langle \mu_1\rangle \langle \sigma_1\rangle} ~~.
\eea
On observe que ce résultat de la combinaison des deux moyennes dépend, à un ordre donné, à la fois de termes provenants directement du développement de $S$ et d'autres venant de la combinaison des fluctuations de $S$ et de celles de la mesure d'intégration $d^2 \mu$. Ces derniers termes sont dit termes de \guillemotleft \, backreaction induite \guillemotright \, (\emph{Induced Backreaction}, IBR) du fait de leur capacité à engendrer des corrections au modèle homogène à partir du premier ordre de $S$. Lorsque $S$ et $d^2 \mu$ ne sont pas corrélées, ces contributions sont nulles et seules les contributions ne dépendant que de $S$ (et de la mesure homogène $d^2 \mu^{(0)}$) subsistent, i.e. $\overline{\lla \sigma_i \rra}$ avec $i \geq 2$.

Abordons finalement le calcul de la variance, i.e. de la largeur de la distribution des valeurs de $S/S^{(0)}$ autour de la valeur moyenne $\overline{\langle S/S^{(0)} \rangle}$. Cette dispersion est due, en même temps, aux fluctuations de la surface spatiotemporelle sur laquelle est faite la moyenne et à celles dues à la moyenne d'ensemble. On la définit donc en considérant les deux opérations communément\,:
\beq
\label{Frvargen}
{\rm Var}[S/S^{(0)}] \equiv \overline{\lla \left(S/S^{(0)} - \overline{\langle S/S^{(0)}\rangle} \right)^2 \rra}  = \overline{\lla (S/S^{(0)})^2 \rra}  - \left(\overline{\lla S/S^{(0)} \rra}\right)^2 ~~.
\eeq
Par l'introduction des développements de l'Eq. \rref{Frexp} et quelque calculs, cette variance devient\,:
\beq
\label{Frvargenexp}
{\rm Var}[S/S^{(0)}] =  \overline{\langle \sigma^2 (1 + \mu) \rangle(\langle 1 + \mu \rangle)^{-1}}  - \left(\overline{\langle \sigma (1 + \mu) \rangle (1+ \langle  \mu \rangle)^{-1}} \right)^2 ~~.
\eeq
On montre alors que les développements en séries de $\mu$ et $\sigma$ imposent au second terme de l'Eq. \rref{Frvargenexp} d'être au moins d'ordre quatre. Par conséquent, on trouve qu'à l'ordre dominant (ordre 2)\,:
\beq
\label{Frvargen2nd}
{\rm Var}[S/S^{(0)}] = \overline{\langle \sigma_1^2  \rangle} ~~.
\eeq
Comme pour le terme IBR$_2$, nous avons un résultat qui ne dépend que de l'ordre quadratique. De plus ici le résultat ne dépend pas des fluctuations de métrique, mais seulement du scalaire moyenné.

Notons enfin que nous aurions pu opter pour une autre définition de la variance, en considérant la dispersion stochastique de la moyenne $\langle S/S^{(0)}\rangle$. Cette définition de la variance donne l'expression suivante\,:
\beq
\label{Frvargen1}
{\rm Var'}[S/S^{(0)}] \equiv \overline{\left(\langle S/S^{(0)}\rangle- \overline{\langle S/S^{(0)}\rangle} \right)^2} = \overline{\left(\lla S/S^{(0)}\rra\right)^2}  - \left(\overline{\lla S/S^{(0)} \rra}\right)^2 ~~,
\eeq
qui après des considérations similaires à celles déjà présentées, a pour résultat\,:
\beq 
\label{Frvarp}
{\rm Var'}[S/S^{(0)}] = \overline{\langle \sigma_1 \rangle^2} ~~.
\eeq
On verra dans le cas $S \equiv d_L$ que cette dispersion est beaucoup plus petite que la première. La raison intuitive qui explique ce fait est la suivante. ${\rm Var'}[S]$ définit la variation de la moyenne spatiale $\lla S \rra$ par la considération de différentes réalisations de notre Univers inhomogène (décrit par le spectre du puissance). On comprend que pour une sphère assez grande devant la taille des inhomogénéités, et pour l'ordre en $k$ des termes considérés, cette variation doit être petite. D'un autre côté, la variance ${\rm Var}[S]$ exprime la dispersion de l'observable $S$ autour de la valeur moyenne $\overline{\lla S \rra}$, et c'est cette quantité qui fluctue de manière importante. La vraie dispersion se trouve donc dans la moyenne angulaire et non dans la moyenne stochastique (non moins nécessaire pour autant) et par conséquent nous oublierons la définition de ${\rm Var'}[S]$ pour ne garder que ${\rm Var}[S]$.

\subsection{Application à l'ordre quadratique}
\label{Ch10Sec53}

Nous décrirons brievement ici l'élaboration du calcul des termes composant IBR$_2$ dans le cas où $S$ est une observable physique. Pour des SNe Ia, cette observable est normalement le flux lumineux, $S \propto d_L^{-2}$, mais on peut également penser à moyenner la distance de luminosité $S = d_L$. En effet, le calcul de ces termes de backreaction induite permet une première évaluation de la moyenne de ces quantités, bien qu'incomplète, à partir de la transformation des coordonnées au premier ordre. Ces termes apparaissent de la manière suivante\,:
\beq
(\overline{\langle d_L^{-2} \rangle})^{-1/2} = d_L^{\rm FLRW} \left( 1 + \overline{\langle \sigma_2 \rangle} + \rm{IBR}_2 - \frac{3}{2}\overline{\langle \sigma_1^2 \rangle} \right) ~~,~~ \overline{\langle d_L \rangle} = d_L^{\rm FLRW} \left( 1 + \overline{\langle \sigma_2 \rangle} + \rm{IBR}_2 \right) ~~,
\eeq
et seul un calcul complet au second ordre (dont nous présenterons les résultats par la suite) permet la détermination de $\overline{\lla \sigma_2 \rra}$.
Toutefois, les expressions des termes de backreaction induite IBR$_2$ et de la variance au second ordre $\overline{\langle \sigma_1^2 \rangle}$ ne dépendent que du premier ordre. Ces quantités peuvent donc déjà être évaluées exactement par un calcul au premier ordre. On peut aussi remarquer qu'il n'est pas nécessaire de connaître $\mu_2$ pour le calcul au second ordre.

D'après l'Eq. \rref{FrfinaldL}, nous avons\,:
\bea
& \label{Frsigma1}
\sigma_1 = A_1 + A_2 + A_3 + A_4 + A_5 ~~, \\
& \label{FrDefA}
A_1 = -\psi_s ~~,~~  A_2 = - {Q_s}/{\Delta \eta} ~~,~~  A_3 = \Xi_s I_+ ~~,~~ A_4 = - \Xi_s I_r ~~,~~ A_5= - J_2 ~~, ~~~~~
\eea
avec $I_+$ et $I_r$ définis en Eqs. \rref{IplusIr} et \rref{Ir}.
D'autre part, puisque nous intégrons sur le ciel à un certain redshift dans notre cône de lumière passé, la mesure $d^2\mu$ est donnée par l'Eq. \rref{zaverageexact}\,:
\beq
d^2\mu=d^2 \ti{\theta} \sqrt{\gamma(w_o,\tau(z_s,w_o,\tilde{\theta}^a),\tilde{\theta}^a)} ~~.
\eeq
Il nous faut transformer l'intégration sur $d^2 \ti{\theta}$ (i.e. sur la $2$-sphère $\Sg$) en une intégration sur les coordonnées polaires usuelles $d^2 \theta= d\theta d\phi$ de la jauge Newtonienne. On vérifie aisément (en considérant le Jacobien de la transformation $\theta^a \ra \ti \theta^a$, cf. Eq. \rref{Frtheta}) que l'on a au premier ordre\,:
\beq
\int d^2 \ti{\theta} \sqrt{\gamma} = \int d\phi d\theta \sin \theta \left([a_s r_s](z_s, \theta^a)\right)^2 (1- 2 \psi_s) ~.
\eeq
La mesure non perturbée est donc donnée par $(d^2 \mu)^{(0)}= d^2 \Om/ 4 \pi$ avec $d^2\Om= d \phi d\theta \sin \theta$. L'utilisation du développement du produit $a_s r_s$ présenté en Eq. \rref{asrsOnlyOrder1} (ou \rref{asrs}) donne ensuite\,:
\beq
\label{Frmu1}
\mu_1= -2 \psi_s - 2 {Q_s}/{\Delta \eta} + 2 \Xi_s J(z_s, \theta^a) = 2\,( A_1 + A_2 + A_3 + A_4) ~~.
\eeq
Il nous est donc possible d'obtenir la contribution IBR$_2$, ainsi que la variance de $d_L/d_L^{\rm FLRW}$, en insérant les résultats de (\ref{Frsigma1}) et (\ref{Frmu1}) dans les Eqs. (\ref{Fribr2}) et (\ref{Frvargen2nd}). Les contributions à calculer sont bilinéaires dans le champ gravitationnel $\psi$ et apparaissent toutes sous la forme $\overline{\langle A_i A_j \rangle}$ ou bien $\overline{\langle A_i \rangle \langle A_j \rangle}$ et où les $A_i$ ont été définis en Eq. \rref{FrDefA}.

L'exemple le plus simple est celui de $A_1= -\psi_s$. En utilisant la notation $\overline{\lla \ldots \rra}$ pour la moyenne présentée en Eq. \rref{FrAvS}, et dénotant par $d^2 \Om/ 4 \pi$ la mesure d'intégration homogène (et normalisée), nous obtenons \,:
\bea
\label{FrTrivialExample1}
\overline {\lla \psi_s \psi_s \rra} &=& \int \frac{d^3 k ~d^3 k'}{(2 \pi)^{3}} \overline{E(\kbf) E(\kbf')} \int \frac{d^2 \Omega}{4 \pi} \left[ \psi_k(\eta_s^{(0)}) e^{i r \kbf \cdot \hat{\xbf}}\right]_{r=\eta_o - \eta_s^{(0)}} \left[ \psi_{k'}(\eta_s^{(0)}) e^{i r \kbf' \cdot \hat{\xbf}} \right]_{r=\eta_o - \eta_s^{(0)}} \nonumber \\
&=& \int \frac{d^3 k}{(2 \pi)^{3}} ~ |\psi_k(\eta_s^{(0)})|^2 \int_{-1}^{1} \frac{d(\cos\theta)}{2} \left[ e^{i k \Delta \eta \cos\theta} \right] \left[ e^{-i k \Delta \eta \cos\theta} \right] \nonumber \\
&=& \int \frac{d^3 k}{(2 \pi)^{3}}  ~ |\psi_k(\eta_s^{(0)})|^2 = \int_0^{\infty} \frac{d k}{k} ~ \Pcal_{\psi}(k,\eta_s^{(0)}) ~~,
\\
\label{FrTrivialExample1a}
\overline {\lla \psi_s \rra \lla \psi_s \rra} &=& \int \frac{d^3 k ~d^3 k'}{(2 \pi)^{3}} \overline{E(\kbf) E(\kbf')} \left[ \int \frac{d^2 \Omega}{4 \pi} \psi_k(\eta_s^{(0)}) e^{i r \kbf \cdot \hat{\xbf}} \right]_{r=\eta_o - \eta_s^{(0)}}   \left[ \int \frac{d^2 \Omega'}{4 \pi} \psi_{k'}(\eta_s^{(0)}) e^{i r \kbf' \cdot \hat{\xbf}'} \right]_{r=\eta_o - \eta_s^{(0)}} \nonumber \\
&=& \int \frac{d^3 k}{(2 \pi)^{3}} ~ |\psi_k(\eta_s^{(0)})|^2 \left[ \int_{-1}^{1} \frac{d(\cos\theta)}{2} e^{i k \Delta \eta \cos\theta} \right] \left[ \int_{-1}^{1} \frac{d(\cos\theta')}{2} e^{-i k \Delta \eta \cos\theta'} \right] \nonumber \\
&=& \int \frac{d^3 k}{(2 \pi)^{3}}  ~ |\psi_k(\eta_s^{(0)})|^2 \left( \frac{\sin(k \Delta\eta)}{k \Delta \eta} \right)^2 = \int_0^{\infty} \frac{d k}{k}  ~ \Pcal_{\psi}(k,\eta_s^{(0)}) \left( \frac{\sin(k \Delta\eta)}{k \Delta \eta} \right)^2 ~~.
\eea
Dans la seconde ligne de ces deux résultats, nous avons utilisé l'isotopie (i.e. $\psi_\kbf$ ne dépend que de $k = |\kbf|$) et avons définit les angles $\theta$ et $\theta'$ comme les angles respectifs entre $\kbf$ et $\xbf \equiv r \hat{\xbf}$ et entre $\kbf'$ et $\xbf' \equiv r \hat{\xbf}'$. Nous avons aussi introduit le spectre de puissance donné par l'Eq. \rref{FrSpectrePuissance}.

Un cas légèrement plus compliqué de terme qui combine moyenne angulaire et moyenne stochastique, avec $\Pcal_\psi$ supposé indépendant de $\eta$ (cas vérifié dans le modèle CDM), est donné par\,:
\bea
\overline{\lla \psi_s \frac{Q_s}{\Delta \eta} \rra} &=& - \frac{2}{\Delta \eta} \int \frac{d^3 k_1 d^3 k_2}{(2 \pi)^3} \psi_{k_1} \psi_{k_2} \overline{E(\kbf_1) E(\kbf_2)} \lla e^{i \kbf_1 \cdot \hat{\xbf} \Delta\eta} \int_{\eta_s}^{\eta_o} d\eta ~ e^{i \kbf_2 \cdot \hat{\xbf} (\eta_o - \eta)} \rra \nonumber \\
&=& - \frac{2}{\Delta \eta} \int_{0}^{\infty} \frac{d k}{4 \pi k} \Pcal_\psi(k) \int_{\eta_s}^{\eta_o} d\eta \lla e^{i k (\eta_s - \eta) \cos \theta } \rra \nonumber \\
&=& \int_{0}^{\infty} \frac{d k}{k} \Pcal_\psi(k) ~ \left[ - \frac{2}{k \Delta \eta} {\rm SinInt}(k \Delta \eta) \right] ~~, \\
\overline{\lla \psi_s \rra \lla \frac{Q_s}{\Delta \eta} \rra} &=& \int_{0}^{\infty} \frac{d k}{k} \Pcal_\psi(k) ~ \left[ - {\rm Sinc}(k \Delta\eta) \frac{2}{k \Delta \eta} {\rm SinInt}(k \Delta \eta) \right] ~~.
\eea
On remarque la différence dans les résultats et le fait que le calcul effectué sur le cône de lumière a une influence directe sur la nature des intégrandes finales où interviennent le sinus cardinal et le sinus intégral\,:
\beq
{\rm Sinc}(x) \equiv \frac{\sin(x)}{x} ~~~~~,~~~~~ {\rm SinInt}(x) = \int_0^x dy ~ {\rm Sinc}(y) ~~.
\eeq
On a supposé dans ces deux derniers calculs que l'on avait un Univers CDM, et ce pour avoir $\psi_k$ indépendant du temps $\eta$ et ainsi pouvoir factoriser le spectre de puissance $\Pcal_\psi(k)$.

Dans le calcul complet de IBR$_2$ ou de la variance, de nombreux termes apparaissent et ont des dépendances différentes. Dans le cas d'une géométrie de fond décrite par un modèle CDM, cette dépendance est simplifiée par le fait que les perturbations $\psi$, ou le spectre de puissance $\Pcal_\psi$, sont indépendantes du temps (i.e. $\pa_\eta \psi_k=0$). Dans ce cas, toutes les contributions peuvent être écrites de la manière suivante\,:
\bea
\overline{\langle A_iA_j \rangle} &=& \int_0^{\infty} \frac{d k}{k} ~ \Pcal_{\psi}(k) \, {\mathcal C}_{ij}(k,\eta_o,\eta_s) ~~,  
\label{Fraiaj}
\\  
\overline{\langle A_i \rangle \langle  A_j \rangle} &=& \int_0^{\infty} \frac{d k}{k} ~ \Pcal_{\psi}(k) \, {\mathcal C}_{i}(k,\eta_o,\eta_s)~ {\mathcal C}_{j} (k,\eta_o,\eta_s) ~~,
\label{Frai}
\eea
où les ${\mathcal C}_{ij}$ et ${\cal C}_i$ sont appelés coefficients spectraux. Dans ces conditions, le terme de backreaction induite $\rm{IBR}_2$, résultant du développement \rref{FrEnsAvExp} de $\overline{\langle d_L  \rangle}/ d_L^{\rm FLRW}$, peut s'écrire comme\,:
\bea
{\rm IBR}_2 &=& 2 \sum_{i=1}^4 \sum_{j=1}^5  \Big[ \overline{\langle A_iA_j \rangle} - \overline{\langle A_i \rangle \langle  A_j \rangle} \Big] \nonumber \\
&=& \int_0^{\infty} \frac{d k}{k}  ~ \Pcal_{\psi}(k) \times 2 \sum_{i=1}^4 \sum_{j=1}^5  \Big[ {\mathcal C}_{ij}(k,\eta_o,\eta_s)  - {\mathcal C}_{i}(k,\eta_o,\eta_s)~ {\mathcal C}_{j} (k,\eta_o,\eta_s) \Big] ~~.
\label{FrBR2backreaction}
\eea
La variance quant à elle, exprimée par l'Eq. \rref{Frvargen2nd}, s'écrit comme\,:
\bea
\label{FrBRVarTh}
\left({\rm Var}\left[\frac{d_L}{d_L^{\rm FLRW}}\right]\right)^{1/2} &=& \sqrt{\overline{\lla  \sigma_1^2 \rra}} = \left[ \sum_{i=1}^5 \sum_{j=1}^5 \overline{\langle A_iA_j \rangle}\right]^{1/2} \nonumber \\
&=& \left[\int_0^{\infty} \frac{d k}{k} ~ \Pcal_{\psi}(k) \sum_{i=1}^5 \sum_{j=1}^5  {\mathcal C}_{ij}(k,\eta_o,\eta_s)\right]^{1/2} ~~.
\eea
Les valeurs obtenues pour ces coefficients ${\mathcal C}_{ij}$ et ${\cal C}_i$ sont données en Table \ref{Tab1OfP2} et Table \ref{Tab2OfP2}. Cette première table contient également la limite aux petits $k$ ($k \Delta \eta \ll 1$) de ${\mathcal C}_{ij}$ et ${\mathcal C}_{i} {\mathcal C}_{j}$. Nous avons utilisé par simplicité la variable sans dimension $l= k \Da \eta$.

Certains des termes nécessitent l'utilisation du facteur d'échelle. Dans ce cas nous avons supposé le cas CDM, i.e. dominé par une matière de type poussière implicant $a(\eta) = a(\eta_o)(\eta/\eta_o)^2$, et défini\,:
\beq
f_{o,s} \equiv \int_{\eta_{in}}^{\eta_{o,s}} d\eta \frac{a(\eta)}{a(\eta_{o,s})} = \frac{\eta_{o,s}^3 - \eta_{in}^3}{3\eta_{o,s}^2} ~ \simeq \frac13 \eta_{o,s} ~~~~~ \mbox{ avec } ~~~~~ \eta_{in} \ll \eta_{o,s} ~~.
\eeq
L'utilisation du modèle CDM impose également la relation entre la durée $\Delta \eta$ et $z_s$ à l'ordre zéro\,:
\beq
\Delta \eta = \frac{2}{a_o H_0} \left[1-\frac{1}{\sqrt{1+z_s}}\right] ~~ \simeq ~~ \frac{z_s}{a_o H_0} ~~~~~ \mbox{pour} ~~~~~ z_s \ll 1 ~~.
\eeq

\subsection{Distance et flux lumineux à l'ordre 2 (complet)}
\label{Ch10Sec54}

Bien que le choix soit large, l'une des quantités les plus intéressantes à moyenner est le flux lumineux envoyé par différentes sources au redshift $z_s$ le long de notre cône passé. Il se révèle aussi être la quantité qui subit le moins l'effet des inhomogénéités. Considérons donc la définition classique du flux et sa décomposition perturbative (implicant une luminosité intrinsèque $L$ constante)\,:
\bea
\Phi = L / (4 \pi d_L^2) = \Phi_0 + \Phi_1 + \Phi_2 ~~.
\eea
On peut moyenner ce flux sur la 2-sphère à redshift constant. On trouve\,:
\beq
\langle d_L^{-2} \rangle(z_s, w_o) = (1+z_s)^{-4} \left[ \int \frac{d^2 \tilde{\theta}^a}{4 \pi} \sqrt{\gamma}(w_o, \tau_s(z_s,\tilde{\theta^a}), \tilde{\theta}^b) \right]^{-1} \equiv I_\phi(w_o, z_s)^{-1} (d_L^{\rm FLRW})^{-2} ~~,
\eeq
avec $I_\phi(w_o, z_s)^{-1}$ la correction au cas homogène et dont l'inverse vaut\,:
\bea
I_\phi(w_o, z_s) &\equiv& \frac{\Acal(w_o, z_s)}{4 \pi (d_A^{\rm FLRW})^2} = (d_A^{\rm FLRW})^{-2} \int \frac{d^2 \tilde{\theta}^a}{4 \pi} \sqrt{\gamma} (w_o, \tau_s(z_s,\tilde{\theta^a}), \tilde{\theta}^b) ~~, \\
&=& \frac{{\cal A}(w_o, z_s)}{4 \pi (a(\eta_s^{(0)}) \Delta \eta)^2} = \int \frac{d^2 \tilde{\theta}^a}{4 \pi} \sin \tilde{\theta}
\left(1+{\cal I}_1+{\cal I}_{1,1}+{\cal I}_2\right) ~~.
\eea
Nous avons dans cette dernière égalité définit les corrections qui interviennent dans la moyenne du flux (différentes de celles du flux par des termes de dérivées totales angulaires) et qui regroupent respectivement des termes d'ordre $\Ocal(1)$, $\Ocal(1)^2$ et $\Ocal(2)$ en perturbations. Ces contributions ont des expressions très compliquées qui ont été regroupées au sein des Eqs. \rref{I1}, \rref{I11} et \rref{I2} et d'une manière plus pratique par les décompositions\,:
\beq
\label{FrDecomIinTterms}
{\cal I}_{1} = \sum_{i = 1}^{3} \Tcal_i^{(1)} ~~~~~,~~~~~ {\cal I}_{1,1} = \sum_{i = 1}^{23} \Tcal_i^{(1,1)} ~~~~~,~~~~~ {\cal I}_{2} = \sum_{i = 1}^{7} \Tcal_i^{(2)} ~~~~~,
\eeq
et où le détail de ces 33 termes a été donné en Eqs. \rref{DefI1}, \rref{DefI11part2} et \rref{DefI2}.
Nous avons aussi montré les relations\,:
\beq
\label{FrIdeltarel}
\Ical_1 = 2 \bar{\delta}_S^{(1)} + (\rm{t.~d.})^{(1)} ~~~~~,~~~~~ \Ical_{1,1} + \Ical_2 = 2 \bar{\delta}_S^{(2)} + (\bar{\delta}_S^{(1)})^2 + (\rm{t.~d.})^{(2)} ~~,
\eeq
où il est intéressant de constater que la dérivée totale $(\rm{t.~d.})^{(1)}$ est égale à $2 J_2^{(1)}$ et que bien que ce terme de lentille gravitationnelle soit présent dans le flux $\Phi \propto d_L^{-2}$, il disparait totalement de sa moyenne au premier ordre. Les coefficients spectraux de chacun de ces termes présentés en Eqs. (\ref{DecomIinTterms}-\ref{DefI2}) ont été évalués en Sec. \ref{Ch5Sec34}.
Prenons la moyenne stochastique de $\langle d_L^{-2} \rangle(z_s, w_o)$, nous obtenons alors\,:
\beq
\overline{\langle d_L^{-2} \rangle}(z) = (d_L^{\rm FLRW})^{-2}\overline{(I_\Phi(z))^{-1}} \equiv (d_L^{\rm FLRW})^{-2} \left[1 + f_\Phi(z) \right] ~~,
\eeq
où
\beq
\label{Fr7a}
f_\Phi(z) \equiv \overline{\lla \Ical_1 \rra^2} - \overline{\lla \Ical_{1,1} + \Ical_2 \rra}
\eeq
est la correction au flux due aux effets des inhomogénéités par rapport au cas homogène.

On obtient la correction à la distance, sans refaire le calcul de toutes ses contributions, simplement par un développement de Taylor des quantités moyennées. En effet, la moyenne d'une fonction $F$ d'une quantité scalaire $S=S_0+S_1+S_2$ peut se décomposer comme 
\beq
\overline{\langle F(S) \rangle} = F(S_0) + F'(S_0) \overline{\langle S_1+S_2  \rangle} + F''(S_0) \overline{\langle S_1^2/2 \rangle} ~~,
\eeq
et de manière générale $\overline{\langle F(S) \rangle} \not= F(\overline{\langle S\rangle})$ du fait de l'aspect non-linéaire de la procédure de moyenne. Aussi, on a vu que $\overline{\langle S_1 \rangle}$ n'était pas toujours égal à zéro, comme dans le cas des contributions de backreaction induite.
En prenant $S= \Phi$ et en considérant $F(\Phi)= \Phi^{-1/2} \sim d_L$, on obtient\,:
\beq
\overline{\langle d_L \rangle}(z)=d_L^{\rm FLRW} \left[1+f_d(z)\right] ~~,
\eeq
avec $f_d$ relié à $f_\Phi$ selon la relation importante\,:
\beq
f_d = -{1\over 2}f_\Phi+{3\over 8}\,\overline{\langle \left(\Phi_1/\Phi_0\right)^2 \rangle} ~~.
\eeq
On peut aussi montrer que $\overline{\langle \left(\Phi_1/\Phi_0\right)^2 \rangle} = 4 \, \overline{\langle (\bar{\delta}_S^{(1)})^2 \rangle}$.
Nous verrons que ce terme se révèle capital et qu'il contient à la fois un terme Doppler et un terme de lentille gravitationnelle qui dominent par rapport aux autres effets physiques. D'une manière plus générale, on montre que\,:
\beq
f_{d^{\beta}} = - \frac{\beta}{2} f_\phi + \frac{(2 + \beta) \beta}{2} ~ \overline{\lla (\bar{\delta}_S^{(1)})^2 \rra} ~~~~~ \forall ~~~ \beta \neq 0 ~~,
\eeq
ce qui montre que toute puissance $\beta \neq -2$ dépend de ce dernier terme.

On peut aussi appliquer le développement de Taylor à la fonction $F(\Phi)= -2.5 \log_{10}\Phi + {\rm cst}$, soit le module de distance $\mu$, on obtient alors\,:
\beq
\label{Fr14}
\overline{\langle \mu \rangle} = \mu^{\rm FLRW} - 1.25 (\log_{10}e) \Big[ \, 2f_\Phi-\overline{\langle \left(\Phi_1/\Phi_0\right)^2 \rangle} \, \Big] ~~.
\eeq
Encore une fois, la déviation au modèle homogène est donnée par une fonction de $\overline{\langle (\bar{\delta}_S^{(1)})^2 \rangle}$.

La deuxième notion intéressante que l'on peut définir est la variance d'une quantité scalaire $S$. On a vu en Sec. \ref{Ch10Sec52} que deux cas sont possibles selon que l'on considère les deux moyennes (spatiale et stochastique) comme deux opérations successives ou comme une seule opération\,:
\beq
{\rm Var'}[S] \equiv \overline{\left(\langle S \rangle- \overline{\langle S \rangle} \right)^2}  = \overline{\left(\lla S \rra\right)^2}  - \left(\overline{\lla S \rra}\right)^2 ~~,~~
{\rm Var}[S] \equiv \overline{\lla \left(S - \overline{\langle S \rangle} \right)^2 \rra}  = \overline{\lla S^2 \rra}  - \left(\overline{\lla S \rra} \right)^2 ~~.
\eeq
Nous avons en fait montré, par des considérations sur les termes d'ordre $\Ocal(1)^2$ (suffisantes pour la variance), que la première de ces variances est très petite devant la seconde et moins intéressante du point de vue physique. Nous choisissons donc d'utiliser cette dernière et définissons l'écart-type (ou \guillemotleft \, dispersion \guillemotright)\,:
\begin{equation}
\sigma_S \equiv \sqrt{\overline{\lla \left(S - \overline{\langle S \rangle} \right)^2 \rra}} = \sqrt{ \overline{\lla S^2 \rra}  - \left(\overline{\lla S \rra}\right)^2} ~~.
\end{equation}

Appliquée au flux lumineux, cette définition de la dispersion donne\,:
\beq
\sigma_\Phi=\sqrt{ \overline{\langle \left(\Phi/\Phi_0\right)^2 \rangle}- \left( \overline{\langle \Phi/\Phi_0 \rangle}\right)^2}=  \sqrt{\overline{\langle \left(\Phi_1/\Phi_0\right)^2 \rangle}} ~~.
\eeq
Pour le module de distance, nous avons\,:
\beq
\label{Fr15}
\sigma_\mu = \sqrt{ \overline{\langle \mu^2 \rangle} - \left( \overline{\langle \mu \rangle}\right)^2} = 2.5 (\log_{10} e) \sqrt{\overline{\langle \left(\Phi_1/\Phi_0\right)^2 \rangle}} ~~.
\eeq
La quantité $\overline{\langle \left(\Phi_1/\Phi_0\right)^2 \rangle}$ est positive et inférieure à 1, on peut donc s'attendre à observer une dispersion significativement plus grande que la moyenne dans les cas du flux et du module de distance.

Toutes ces contributions peuvent être écrite de manière similaire à ce qui a été présenté pour l'ordre quadratique, en Eqs. \rref{Fraiaj} et \rref{Frai}, i.e. sous la forme\,:
\begin{eqnarray}
\overline{\langle X \rangle} &=& \int_0^{\infty} \frac{d k}{k} ~ {\cal P}_{\psi}(k, \eta_o) \, \Ccal_X(k,\eta_o,\eta_s^{(0)}) ~~,  \\
\overline{\langle X' \rangle \langle Y' \rangle} &=& \int_0^{\infty} \frac{d k}{k} ~ {\cal P}_{\psi}(k, \eta_o) \, \Ccal_{X'}(k,\eta_o,\eta_s^{(0)}) \, \Ccal_{Y'}(k,\eta_o,\eta_s^{(0)}) ~~.
\end{eqnarray}
Ici, $X'$ et $Y'$ ($X$) sont des fonctions générales du premier (second) ordre dépendantes de $(\eta,r,\theta^a)$ et les $\Ccal$ leurs coefficients spectraux. Cette intégration sur $k$ est faite numériquement et un cut-off doit être choisi sur des bases physiques (bien que les intégrales soient, comme nous le verrons en Sec. \ref{Ch10Sec61}, convergentes dans l'UV et l'IR). Notons maintenant qu'une complication importante émerge lorsque l'on ne considère pas le simple cas CDM. Dans celui-ci, le spectre de puissance est factorisable selon l'ordre des équations ci-dessus et les coefficients spectraux ne dépendent pas du spectre. On utilisera d'ailleurs dans la section suivante ce cas simple pour identifier les effets physiques dominants parmi les nombreux termes évalués. Dans le cas d'un Universe $\Lambda$CDM, en revanche, le spectre de puissance apparaît aussi dans les expressions des coefficients $\Ccal$ car il est directement dépendant du temps $\eta_s^{(0)}$ et donc non-factorisable. Il intervient aussi généralement au sein d'intégrales sur $\eta$ qui doivent donc être déterminées numériquement. Il faut alors en général procéder à des approximations sur le spectre lui-même.

Une deuxième complication est que nous connaissons uniquement le spectre de puissance du champ gravitationnel $\psi$, i.e. l'ordre linéaire. Pour les termes intrinsèques à l'ordre 2, tels que ceux qui mettent en jeu $\phi^{(2)}$ et $\psi^{(2)}$, nous n'avons pas de spectre associé. Pour remédier à cela, nous utilisons les équations développées par \cite{Bartolo:2005kv} pour relier le second ordre au premier. On a pour le cas CDM\,:
\bea
\phi^{(2)} &=& 2\psi^2 - 6 ~ \Ocal^{ij} \partial_i \psi \partial_j \psi + \frac{\eta^2}{14} ~ \Ocal_2^{ij} \partial_i \psi \partial_j \psi ~~, \\
\psi^{(2)} &=& -2\psi^2 + 4 ~ \Ocal^{ij} \partial_i \psi \partial_j \psi + \frac{\eta^2}{14} ~ \Ocal_2^{ij} \partial_i \psi \partial_j \psi ~~,
\eea
avec les opérateurs suivants\,:
\beq
\Ocal^{ij} = \nabla^{-2} \left( \frac{\partial^i \partial^j}{\nabla^2} - \frac{1}{3} \delta^{ij} \right) ~~,~~ \Ocal_2^{ij} = \frac{10}{3} \frac{\partial^i \partial^j}{\nabla^2} - \delta^{ij} ~~.
\eeq
Il nous est alors possible de calculer tous les termes, aux approximations du spectre près, au prix d'une augmentation du nombre de termes par la conversion de ceux à l'ordre 2 intrinsèque. La conversion est encore plus compliquée dans le cas $\Lambda$CDM, cf. \rref{PSI} et \rref{PHI}, mais encore une fois réalisable.

\section{Résultats pour le spectre non-linéaire et un Univers $\Lambda$CDM}
\label{Ch10Sec6}

Nous allons ici présenter les résultats principaux de cette thèse liés à l'estimation numérique de l'effet des inhomogénéités sur les observables cosmologiques. Pour cela, nous utiliserons une expression explicite du spectre de puissance. Nous commencerons par donner l'estimation de la taille des termes quadratiques présents dans IBR$_2$ et $\overline{\langle \sigma_1^2 \rangle}$, ceci dans le cas CDM. Nous avons déjà vu que cette dernière quantité permet une estimation complète de la variance. Nous calculerons ensuite l'effet des inhomogénéités par un calcul complet au deuxième ordre, dans un premier temps dans le cas CDM, où nous évaluerons les contributions dominantes, puis dans le cas $\Lambda$CDM. Nous traiterons d'abord le cas d'un spectre de puissance linéaire, puis celui du régime non-linéaire donné par le modèle du HaloFit.

\subsection{Termes dominants dans CDM à l'ordre quadratique, convergence UV et IR}
\label{Ch10Sec61}

\subsubsection{Fonction de transfert}
\label{Ch10Sec611}

Nous avons présenté en Sec. \ref{Ch10Sec51} l'expression du spectre de puissance dans son régime linéaire. Nous nous sommes toutefois gardé de donner une expression de la fonction de transfert $T(k)$. Nous utiliserons dans le cas présent, où nous évaluons les termes d'ordre quadratique décrits en Secs. \ref{Ch10Sec52} et \ref{Ch10Sec53}, la fonction de transfert en l'absence de contribution baryonique, dénotée par $T(k)=T_0(k)$ et simplement donnée par \cite{Eisenstein:1997ik}\,:
\beq
\label{FrEH97}
T_0(q) = \frac{L_0}{L_0+q^2C_0(q)}  ~~,~~ L_0(q) = \ln(2e+1.8q) ~~,~~ C_0(q) = 14.2+\frac{731}{1+62.5q} ~~,~~ q = \frac{k}{13.41 k_{\rm eq}} ~~.
\eeq
Ici $k_{\rm eq}\simeq 0.07 \, \Omega_{m0} ~ h^2 \Mpc^{-1}$ est le moment correspondant à l'échelle d'égalité matière-radiation, avec $h\equiv H_0/(100 \,{\rm km \,s}^{-1} {\rm Mpc}^{-1})$. Il est facile de voir que $T_0(k)$ tend vers 1 pour $k \ll k_{\rm eq}$ tandis qu'elle décroit comme $k^{-2}\log k$ pour $k \gg k_{\rm eq}$. Nous utiliserons les valeurs $h=0.7$, $a_o =1$ et $\Omega_m=1$ dans le cas de nos estimations numériques, ici dans le modèle CDM.
Dans ce cas, nous avons $k_{\rm eq}\simeq 0.036 \,{\rm Mpc}^{-1}$ et on peut définir le régime asymptotique de la fonction de transfert comme suit\,:
$T_0 \simeq 1$ pour $k \laq 10^{-3}\, {\rm Mpc}^{-1}$ et $T_0 \sim k^{-2}\log k$ pour $k \gaq 2.5\, {\rm Mpc}^{-1}$. Notons que cette deuxième échelle se trouve déjà profondément ancrée dans le régime non-linéaire. De plus amples détails sont donnés en Secs. \ref{AppASec24} et \ref{AppASec42}.

\subsubsection{Contributions dominantes, convergence IR et UV}
\label{Ch10Sec612}

Les intégrales que nous estimons sont à la fois fonction du spectre de puissance et des coefficients spectraux correspondants aux observables calculées. L'étude des termes présentés dans les Tables \ref{Tab1OfP2} et \ref{Tab2OfP2}, de leur dépendance en $k$, nous montre que les coefficients spectraux ${\mathcal C}_{ij}$ et ${\mathcal C}_{i} {\mathcal C}_{j}$ de IBR$_2$ et $\overline{\langle \sigma_1^2 \rangle}$ sont d'ordre ${\mathcal O}(1)$, mais que leur combinaison ${\mathcal C}_{ij}-{\mathcal C}_{i} {\mathcal C}_{j}$ est d'ordre ${\mathcal O}(k^2 \Delta \eta^2)$ dans la limite infrarouge (IR)\,: $k \Delta \eta \ll 1$. D'autre part le spectre de puissance est quasi constant dans cette limite ($\Pcal_\psi (k) \sim k^{n_s - 1} \simeq k^{-0.04}$). On a par conséquent que les termes IBR$_2$ et $\overline{\langle \sigma_1^2 \rangle}$ donnent une contribution négligeable dans la limite IR et on peut sans crainte poser le cut-off IR de nos intégrales comme étant $k= H_0$ (i.e. en intégrant sur tous les modes sub-horizon).
De plus, les coefficients ${\mathcal C}_{ij}$ et ${\mathcal C}_{i} {\mathcal C}_{j}$ n'augmentent pas plus rapidement que ${\mathcal O}(k^3 \Delta \eta^3)$ dans la limite ultraviolette (UV)\,: $k \Delta \eta \gg 1$. Le comportement de la fonction de transfert au carré, allant comme $k^{-4}(\log k)^2$ dans cette limite, fait donc que les intégrales que nous calculons convergent aussi dans la limite UV. On comprend donc que l'essentiel de l'effet des inhomogénéités vient du domaine des moments donné par $1/\Delta \eta \ll k \leq 2.5 \, {\rm Mpc}^{-1}$, là où la fonction de transfert ne décroit pas encore significativement. On peut en fait constater que seules deux contributions dominent sur les autres dans la Table \ref{Tab1OfP2} et que les termes ${\mathcal C}_{i} {\mathcal C}_{j}$ donnent lieu à des contributions inférieures en général à celles de ${\mathcal C}_{ij}$. Ces deux contributions dominantes sont ${\mathcal C}_{44}$ et ${\mathcal C}_{55}$. En effet, dans la limite UV, nous avons ces deux termes qui valent\,:
\bea
{\mathcal C}_{44} &\simeq & \left(1 - \frac{1}{{\mathcal H}_s\Delta \eta}\right)^2 \frac{(f_0^2 + f_s^2)}{\Delta \eta^2} \frac{k^2 \Delta \eta^2}{3} ~~,
\\
{\mathcal C}_{55} &\simeq & \frac{k^3 \Delta \eta^3}{15} {\rm SinInt}(k \Delta \eta) ~~.
\eea
On comprend aussi à partir de la dépendance temporelle de ces deux expressions que $\overline{\lla A_4 A_4 \rra}$ est la contribution la plus importante lorsque $z_s\ll 1$ tandis que $\overline{\lla A_5 A_5 \rra}$ domine pour $z_s\gg 1$ (voir Fig. \ref{Fig1and2ofP2}). Cette première correspond à un terme de vitesse particulière (effet Doppler) tandis que la seconde correspond à un terme de lentille gravitationnelle. Notons aussi que les intégrales calculées ici, bien que convergentes dans l'IR et dans l'UV, convergent plus rapidement dans ce premier cas de figure, du fait des puissances en $k$ relevées ci-dessus. La contribution IBR$_2$, dominée par $2 \overline{\lla A_4 A_4 \rra}$ (et n'ayant pas $\overline{\lla A_5 A_5 \rra}$), dépend donc très faiblement du cut-off $k_{UV}$. La contribution à la dispersion donnée par l'Eq. \rref{FrBRVarTh}, dominée par $\overline{\lla A_5 A_5 \rra}$ (mais ayant aussi $\overline{\lla A_4 A_4 \rra}$), y est plus sensible. Nous présentons en Fig. \ref{Fig1and2ofP2} l'évaluation numérique de ces deux termes $\overline{\lla A_4 A_4 \rra}$ et $\overline{\lla A_5 A_5 \rra}$ en fonction du redshift (dans un domaine correspondant à celui des SNe Ia observées actuellement) et pour différents cutoffs ($k_{UV}=0.1 \,{\rm Mpc}^{-1}$ à $k_{UV}=+\infty$). On notera finalement, pour insister, le rôle primordial de la fonction de transfert dans la détermination de la taille de l'effet des inhomogénéités. L'absence de divergence qui découle aussi de cette fonction est illustrée en Fig. \ref{FigSchemeNoDiv}.

Finalement, il est possible de calculer l'effet de l'ensemble des contributions IBR$_2$ et $\overline{\langle \sigma_1^2 \rangle}$ dans le cas CDM, i.e. en supposant qu'il n'est pas de constante cosmologique, et de comparer ce résultat de $\overline{\langle d_L \re} \pm d_L^{CDM}\sqrt{\overline{\langle \sg_1^2 \re}}$ (incluant la dispersion autour de la moyenne) avec la distance de luminosité du modèle $\Lambda$CDM. On est contraint pour cela de ne pas considérer la contribution $\overline{\lla \sigma_2 \rra}$ dans cette expression de $\overline{\langle d_L \re}$ (cf. Eq. \rref{Fribr2}), i.e. que nous prenons $\overline{\langle d_L \re}= d_L^{CDM}(1+ {\rm IBR}_2)$. La contribution manquante ne peut être calculée que par un calcul complet au second ordre, ce qui sera effectué par la suite. Nous avons établi cette comparaison du cas homogène et du cas CDM avec perturbations en traçant le module de distance $\mu(\overline{\langle d_L \rangle}) \equiv m-M = 5 \log_{10}(\overline{\langle d_L \rangle})$ auquel a été soustrait le module de distance de l'Univers de Milne dans lequel l'expansion est linéaire (cf. Appendice \ref{AppASec6})\,:
\beq
\label{Frmodulus}
\Delta(m-M) \equiv \mu(\overline{\lla d_L \rra}) - \mu^M = 5 \log_{10} \left[ \overline{\langle d_L \rangle} \right] - 5 \log_{10}\left[\frac{(2+z_s)z_s}{2 H_0}\right] ~~.
\eeq
Notons que la manière rigoureuse d'évaluer cet effet est de prendre la moyenne du module de distance, i.e. $\overline{\lla \mu \rra}$, et non $5 \log_{10} (\overline{\lla d_L \rra})$. C'est ce que nous ferons en Sec. \ref{Ch10Sec63}, en nous basant sur les Eqs. \rref{Fr14} et \rref{Fr15}.
Les résultats de ce tracé sont présentés en Fig. \ref{Fig3and4ofP2} pour des cut-offs de $k_{UV}=0.1 \,{\rm Mpc}^{-1}$ et $k_{UV}=1 \,{\rm Mpc}^{-1}$. Ils sont à comparer avec les tracés de l'Univers $\Lambda$CDM pour $\Om_\La=0.1$ et $\Om_\La= 0.73$. L'utilisation d'un cut-off plus grand aurait pour effet d'augmenter légèrement le résultat final, toutefois, le spectre de puissance adéquat à ces échelles entre dans son régime non-linéaire et ne saurait être celui utilisé jusqu'à présent.

\subsubsection{Conclusions sur IBR$_2$ et $\overline{\langle \sigma_1^2 \rangle}$}
\label{Ch10Sec613}

Il apparaît clairement sur la Fig. \ref{Fig3and4ofP2} que les corrections induites par IBR$_2$ sur la distance de luminosité dans le cas d'un modèle homogène CDM ne peuvent simuler le phénomène que nous nommons énergie sombre, ceci même lorsque l'on prend en compte la dispersion autour de la moyenne $\overline{\langle d_L \rangle}$. On voit dans ces figures que l'utilisation d'un cut-off de $1 \, \Mpc^{-1}$ impose une dispersion assez grande pour expliquer une constante cosmologique de paramètre $\Om_\La \sim 0.1$ comme une fluctuation statistique. Cependant même ce résultat, si on le veut utile pour expliquer une partie de la constante cosmologique que l'on observe aujourd'hui, doit être réinterprété dans le cadre d'un calcul au sein d'un modèle $\Lambda$CDM ayant une part d'énergie sombre. C'est ce que nous ferons en Sec. \ref{Ch10Sec63} et \ref{Ch10Sec64}.

Il y a donc différentes raisons pour effectuer ce calcul de manière complète au second ordre en perturbations. Premièrement, et comme déjà mentionné, un calcul définitif de l'effet des inhomogénéités se doit d'inclure $\overline{\langle \sg_2 \rangle}$. A priori, cette contribution peut contenir des termes similaires à ceux de la dispersion, entre autre un terme de lentille $\overline{\lla A_5 A_5 \rra}$. Dans ce cas, la moyenne $\overline{\lla d_L \rra}$ serait significativement affectée, en particulier aux grand redshifts. Nous verrons que ce terme n'apparaît pas dans $\overline{\langle \sg_2 \re}$, par une annulation quasi-miraculeuse qui est en fait liée à la conservation du flux lumineux. Ensuite, il est certain que l'effet des inhomogénéités est renforcé par la considération du spectre de puissance dans son régime non-linéaire qui autorise une intégration allant à des cut-offs plus grands que $1 \, \Mpc^{-1}$. Enfin, la détermination la plus précise possible de l'effet des inhomogénéités a une conséquence très importante pour les observations futures. En effet, cette estimation précise peut être comparée aux relevés observationels et constituer une signature des inhomogénéités qui ne saurait être confondue avec une dépendance spatiotemporelle de l'énergie sombre. Ce calcul est donc nécessaire également pour pouvoir enveler la contribution, même faible, des inhomogénéités dans les données.

Remarquons enfin que certaines combinaisons de puissances de $d_L$ ne dépendent pas de $\overline{\langle \sg_2 \rangle}$ mais seulement de $\overline{\langle \sg_1^2 \rangle}$. C'est le cas des combinaisons suivantes\,:
\beq
\overline{\lla \left(d_L/d_L^{\rm FLRW}\right)^{\alpha} \rra}  - \alpha  \overline{\lla d_L/d_L^{\rm FLRW} \rra} = 1-\alpha + \frac{\alpha (\alpha -1)}{2} \, \overline{\langle \sigma_1^2 \rangle} ~~~~~ \forall \, \alpha \in \RRR ~~.
\eeq
On peut tracer cette quantité pour un modèle inhomogène donné et différentes valeurs de $\alpha$ et comparer ces courbes à leurs équivalentes dans le cas d'un modèle $\Lambda$CDM, de manière potentiellement discriminante pour ce dernier. Néanmoins, nous ne mesurons en pratique que le flux des SNe Ia et il n'est pas exclu que les deux modèles (inhomogène et $\Lambda$CDM) donnent le même résultat pour cette observable particulière.

\subsection{Calcul complet dans CDM}
\label{Ch10Sec62}

Nous allons ici évaluer numériquement les moyennes de $\Ical_{1}$, $\Ical_{1,1}$ et $\Ical_{2}$ dans le cas du modèle CDM. Nous étudierons les termes dominants, retrouvant entre autres ceux déjà présentés pour IBR$_2$ et $\overline{\lla \sigma_1^2 \rra}$, qui nous seront utiles pour le calcul plus général (et plus complexe) dans le cas d'un modèle $\Lambda$CDM.

\subsubsection{Les contributions $\overline{\lla \Ical_1 \rra^2}$ et $\overline{\lla \Ical_{1,1} \rra}$}
\label{Ch10Sec621}

Nous calculons ici ces deux termes qui contribuent à l'expression de la correction du flux présentée en Eq. \rref{Fr7a}. On commence par $\overline{\lla \Ical_1 \rra^2}$ qui, par les Eqs. \rref{DecomIinTterms} et \rref{DefI1}, vaut\,:
\beq
\label{FrEq_with_spectralI_1^2}
\overline{\lla \Ical_1 \rra^2} = \sum_{i=1}^{3} \sum_{j=1}^{3} \overline{\lla {\Tcal^{(1)}_i} \rra \lla {\Tcal^{(1)}_j} \rra} 
= \int_0^{\infty} \frac{d k}{k} ~ {\cal P}_{\psi}(k, \eta_o) ~ \sum_{i=1}^3 \sum_{j=1}^3  \Ccal_{\Tcal^{(1)}_i}(k,\eta_o,\eta_s)~ 
\Ccal_{\Tcal^{(1)}_j}(k,\eta_o,\eta_s) ~~,
\eeq
où nous substituons dès lors $\eta_s$ à $\eta_s^{(0)}$. Encore une fois, le cas CDM impose que $\partial_\eta \psi_k=0$ et $a(\eta) = a(\eta_o)(\eta/\eta_o)^2$. On peut aussi définir (avec $\eta_{in} \ll \eta_{o,s}$)\,:
\beq
\label{FrfCDM}
f_{o,s} \equiv \int_{\eta_{in}}^{\eta_{o,s}} d\eta \frac{a(\eta)}{a(\eta_{o,s})} = \frac{\eta_{o,s}^3 - \eta_{in}^3}{3\eta_{o,s}^2} ~ \simeq \frac13 \eta_{o,s} ~~.
\eeq
Les coefficients de l'Eq. \rref{FrEq_with_spectralI_1^2} peuvent ainsi être calculés et leurs résultats sont présentés en Table \ref{Tab1}, qui est équivalent (dans ces nouvelles notations) à la Table \ref{Tab2OfP2}.

De la même manière, on estime les contributions à $\overline{\lla \Ical_{1,1} \rra}$ par\,:
\beq
\label{FrIntkI11}
\overline{\lla \Ical_{1,1} \rra} = \int_0^{\infty} \frac{d k}{k} ~ {\cal P}_{\psi}(k, \eta_o) ~ \sum_{i=1}^{23} \Ccal_{\Tcal_i^{(1,1)}} (k,\eta_o,\eta_s) ~~,
\eeq
avec les formes explicites des coefficients $\Ccal(\Tcal_i^{(1,1)})$ pour le cas CDM qui sont présentées en Sec. \ref{Ch5Sec34}.

En rassemblant les contributions de $\overline{\lla \Ical_1 \rra^2}$ et $\overline{\lla \Ical_{1,1} \rra}$ selon l'expression \rref{Fr7a}, il apparaît que les termes dominants appartiennent à $\overline{\lla \Ical_{1,1} \rra}$ et non $\overline{\lla \Ical_1 \rra^2}$. Il sont caractérisés par la dépendance en $k^2$ de leurs coefficients spectraux et se révèlent être les $\Tcal_i^{(1,1)}$ tels que $\{i=2,4,5,7,8,12,14,18,20\}$ (cf. les termes encadrés en Sec. \ref{Ch5Sec34}). En ne retenant que ces contributions, soit l'ordre dominant (\emph{Leading order}), on trouve\,:
\bea
\label{FrSumT11lead}
\left[ \sum_{i=1}^{23}\Ccal_{\Tcal_i^{(1,1)}}\right]_{\mbox{Lead}} &=& \aleph_s \frac{f_s^2 + f_o^2}{3} k^2 + \frac{2 (\Xi_s - 3)}{\Hcal_s} \frac{f_s}{3} k^2 + \Xi_s \frac{f_s^2}{3} k^2 - \Xi_s \frac{f_o^2}{3} k^2 \nonumber \\
&=& f_o^2 k^2 \left(\frac{\aleph_s}{3} - \frac{\Xi_s}{3} \right) + f_s^2 k^2 \left(\frac{\aleph_s}{3} + \frac{4 \Xi_s}{3} - 3 \right) ~~ = ~~ - \frac{k^2}{\Hcal_0^2} ~ \ti f_{1,1}(z) ~~, ~~~~~
\eea
où nous avons défini\,:
\beq
\aleph_s ~\equiv~ \Xi_s^2 - \frac{1}{\Hcal_s \Delta\eta}\left(1 - \frac{\Hcal_s'}{\Hcal_s^2}\right) ~~,
\eeq
et utilisé $\Hcal_s \simeq 2 / (3 f_s)$ (valide pour le modèle CDM). La fonction $\ti f_{1,1}(z)$ contient toute la dépendance en $z$ de ces contributions et vaut\,:
\beq
\ti f_{1,1}(z) = \frac{10-12 \sqrt{1+z}+5 z \left(2+\sqrt{1+z}\right)}{27\, (1+z) \left(-1+\sqrt{1+z}\right)^2} ~~.
\label{Frf_11}
\eeq
Elle change de signe pour une valeur $z^{\ast}=0.205$, comme présenté en Fig. \ref{Figf11f2} et visible en Fig. \ref{f1}.

\subsubsection{La contribution intrinsèque au second ordre\,: $\overline{\lla \Ical_2 \rra}$}
\label{Ch10Sec622}

Le calcul de $f_\Phi(z)$ est complet une fois la quantité $\overline{\lla \Ical_2 \rra}$ évaluée. Pour cela, il nous faut estimer les coefficients $\Ccal(\Tcal_i^{(2)})$ définis en terme des potentiels de Bardeen dans l'Eq. \rref{DefI2}. Les résultats de la Sec. \ref{Ch5Sec35}, en particulier les Eqs. (\ref{PSI}) et (\ref{PHI}) montrent que les seules contributions en $k^2$ contenues dans ces termes $\Ccal(\Tcal_i^{(2)})$ sont celles venant des termes $B_3(\eta) \nabla^{-2} \partial_i\partial^j(\partial^i \psi_o \partial_j \psi_o )$ et $B_4(\eta) \partial^i \psi_o \partial _i\psi_o$. Par conséquent, nous avons que $\phi^{(2)}\simeq \psi^{(2)}$ à l'ordre dominant. Dans le cas CDM considéré ici, nous avons $l(\eta)=1$ et $B_1(\eta)=B_2(\eta)=0$ (cf. Eqs. \rref{B1} et \rref{B2}). De plus, les modèles d'inflation usuels nous encouragent à poser $a_{nl}=1$ (absence de non-Gaussiannités). Les Eqs. \rref{PSI} et \rref{PHI} se réduisent alors à\,:
\bea
\psi^{(2)} &=&  -2\psi_o^2 - \frac{4}{3} ~ \Ocal^{ij} \partial_i \psi_o \partial_j \psi_o + B_3(\eta) ~ \Ocal_3^{ij} \partial_i \psi_o \partial_j \psi_o
+ B_4(\eta) ~ \Ocal_4^{ij} \partial_i \psi_o \partial_j \psi_o ~~,
\label{Fr412}
\\
\phi^{(2)} &=& 2\psi_o^2 + 2 ~ \Ocal^{ij} \partial_i \psi_o \partial_j \psi_o + B_3(\eta) ~ \Ocal_3^{ij} \partial_i \psi_o \partial_j \psi_o
+ B_4(\eta) ~ \Ocal_4^{ij} \partial_i \psi_o \partial_j \psi_o ~~,
\label{Fr412a}
\eea
et nous utilisons $a(\eta) = a(\eta_o) \left({\eta}/{\eta_o}\right)^2$ pour obtenir de l'Eq. \rref{B3tildeB4tilde}\,:
\beq
B_3(\eta)=\frac{20}{21}\frac{1}{ {\cal H}^2}  ~~~~,~~~~  B_4(\eta)=-\frac{2}{7}\frac{1}{{\cal H}^2} ~~.
\eeq
On peut calculer alors les termes du second ordre tels que ceux présentés en Eqs. (\ref{Final_overline_O}), (\ref{Final_overline_O3}) et (\ref{Final_overline_O4}) et voir que les termes $\Ccal(\Tcal_1^{(2)})$ et $\Ccal(\Tcal_2^{(2)})$ s'annulent. Les contributions $\Ccal(\Tcal_{3}^{(2)})$ et $\Ccal(\Tcal_{4}^{(2)})$ sont quant à elles nulles par symétrie sphérique de notre intégration. La contribution $\Tcal_{5}^{(2)}$ ne s'annule pas du fait d'une dérivée $\partial_\eta$ agissant sur les coefficients $B_A(\eta)$\,:
\beq
\Ccal(\Tcal_{5}^{(2)}) = \frac{\Xi_s}{126} (\eta_s^2 - \eta_o^2) k^2 ~~.
\eeq
Enfin les deux termes $\Ccal(\Tcal_{6}^{(2)})$ et $\Ccal(\Tcal_{7}^{(2)})$ valent\,:
\bea
\Ccal(\Tcal_{6}^{(2)}) &=& -2 - \frac{k^2 \eta_s^2}{126} + \frac{32 k}{45}
\frac{\psi_k \partial_k \psi_k^\ast}{|\psi_k|^2} + \frac{8 k^2}{45}
\frac{\psi_k \partial_k^2 \psi_k^\ast}{|\psi_k|^2} ~~, \nonumber \\
\Ccal(\Tcal_{7}^{(2)}) &=& \frac{\eta_o^3 - \eta_s^3}{189 \Delta\eta}k^2 +
\frac{16 k}{45} \frac{\psi_k \partial_k \psi_k^\ast}{|\psi_k|^2} + \frac{4
k^2}{45} \frac{\psi_k \partial_k^2 \psi_k^\ast}{|\psi_k|^2} ~~.
\eea

La somme de toute ces contributions donne enfin\,:
\beq
\label{FrGenSecOrder}
\sum_{i = 1}^{7} \Ccal(\Tcal_i^{(2)}) = - 2 + {1\over 7}\left[ \frac{\Xi_s}{2}
(f_s^2 - f_o^2) - \frac{f_s^2}{2} - \frac{f_s^3 - f_o^3}{ \Delta \eta}
\right] k^2 + \frac{16 k}{15} \frac{\psi_k \partial_k
\psi_k^\ast}{|\psi_k|^2} + \frac{4 k^2}{15} \frac{\psi_k \partial_k^2
\psi_k^\ast}{|\psi_k|^2} ~~,
\eeq
et sa contribution dominante peut s'écrire comme\,:
\beq
\left[ \sum_{i = 1}^{7} \Ccal(\Tcal_i^{(2)}) \right]_{\mbox{Lead}} = {1\over 7}\left[ \frac{\Xi_s}{2} (f_s^2 - f_o^2) - \frac{f_s^2}{2} - \frac{f_s^3 - f_o^3}{\Delta \eta} \right] k^2 = - \frac{k^2}{{\cal H}_o^2} \, \ti{f}_2(z) ~~,
\eeq
avec\,:
\beq
\ti{f}_2(z)= -\frac{1}{189} \frac{2-2 \sqrt{1+z}+z \left(9-2 \sqrt{1+z}\right)}{(1+z)(\sqrt{1+z}-1)} ~~.
\label{Frf_2}
\eeq

Le point intéressant ici est que cette contribution $\overline{\lla \Ical_2 \rra}$ est d'un ordre de grandeur inférieure aux contributions décrites pour $\overline{\lla \Ical_{1,1} \rra}$, pour des redshifts $z \sim 0.1 - 2$, et ce à cause d'une annulation partielle des termes présents. Ce fait est apparent lorsque l'on compare $\widetilde{f}_{1,1}$ et $\widetilde{f}_{2}$, comme en Fig. \ref{Figf11f2}. Ceci répond donc à la question de la taille relative du terme $\overline{\lla \sigma_2 \rra}$ par rapport au terme de backreaction induite IBR$_2$ qui avait été soulevée à partir de la transformation au premier ordre en perturbations.

\subsubsection{Résultat complet dans le modèle CDM}
\label{Ch10Sec623}

Nous avons donc trouvé la contribution totale à la correction du flux lumineux (cf. Eq. \rref{Fr7a})\,:
\beq
f_\Phi(z) = \int_0^{\infty} \frac{d k}{k} \, {\cal P}_\psi(k, z=0) \Big[f_{1,1}(k,z) +  f_{2}(k,z) \Big] ~~,
\label{Fr9}
\eeq
avec $f_{1,1}$ et $f_{2}$ des fonctions compliquées mais connues analytiquement dans le modèle CDM. Nous avons aussi remarqué que les contributions majeures à ces termes sont du type $f(k, z) \sim (k/ {\cal H}_o)^2 \ti f(z)$, permettant d'écrire avec une très bonne approximation que\,:
\beq
f_\Phi(z) \simeq \Big[\ti f_{1,1}(z) +  \ti f_{2}(z) \Big] \int_0^{\infty} \frac{d k}{k} \,\left( \frac{k}{{\cal H}_o}\right)^2{\cal P}_\psi(k, z=0) ~~, 
\label{Fr10}
\eeq
avec $\ti f_{1,1}(z)$ et $\ti f_{2}(z)$ donnés par les Eqs. (\ref{Frf_11}) et (\ref{Frf_2}). Les résultats obtenus, en prenant les bornes d'intégrations similaires à celles décrites dans les sections précédentes, sont présentés en Fig. \ref{f1}. On y voit la très bonne approximation due aux termes dominants de l'Eq. \rref{Fr10}. On voit aussi que le flux ne reçoit pas de correction importante aux grands redshifts, du fait de l'absence de terme de lentille gravitationnelle (l'origine de ceci étant liée à la conservation du flux lumineux le long du cône de lumière). Notons aussi que cette correction a une dépendance en $z$ caractéristique et n'a pas le bon signe aux larges redshifts pour simuler une composante, même petite, d'énergie sombre.

\subsection{Le cas $\Lambda$CDM avec spectre linéaire}
\label{Ch10Sec63}

Nous généraliserons ici les résultats présentés dans la section précédente au cas du modèle de concordance en cosmologie, i.e. le modèle $\Lambda$CDM, et ce en utilisant le spectre de puissance dans son régime linaire. Nous changerons néanmoins l'expression de la fonction de transfert pour la remplacer par une expression plus réaliste. La Sec. \ref{Ch6Sec32} suivante utilisera quant à elle le spectre dans son régime non-linéaire. Nous nous attacherons à ne considérer que les termes dominants d'ordre $k^2$ et $k^3$ déjà identifiés dans la section précédente. Nous considérerons aussi les cas de la distance de luminosité et du module de distance. Nous utiliserons dans la suite les valeurs numériques suivantes\,: $\Omega_{\Lambda 0}=0.73$, $\Omega_{m0}=0.27$, $\Omega_{b0}=0.046$, $h=0.7$.

\subsubsection{Corrections du flux au second ordre}
\label{Ch10Sec631}

Dans le cas d'un modèle $\Lambda$CDM, le spectre de puissance dépend du temps et peut s'écrire dans son régime linéaire comme\,:
\beq
\label{FrPsiPpar}
\Pcal_\psi (k,\eta) = \left[\frac{g(\eta)}{g(\eta_o)}\right]^2 \Pcal_\psi (k,\eta=\eta_o) ~~,
\eeq
où nous avons utilisé le \emph{growth factor} $g(\eta)$ pour décrire l'évolution temporelle des perturbations du champ gravitationnel.

Il nous est donc possible à partir de ce nouveau spectre d'évaluer les contributions dominantes du flux déjà décrites dans les sections précédentes pour le calcul de $\overline{\lla \Ical_{1,1} \rra}$. Les contributions dominantes à la correction du flux $I_\phi$, du type $k^2$, sont issues des termes rappelés en Eq. \rref{DefI11_Leading}. Notons que pour la plupart de ces contributions -- $\Tcal_{2,L}^{(1,1)}$, $\Tcal_{4,L}^{(1,1)}$, $\Tcal_{5,L}^{(1,1)}$, $\Tcal_{7,L}^{(1,1)}$, $\Tcal_{12,L}^{(1,1)}$, $\Tcal_{14,L}^{(1,1)}$ et $\Tcal_{20,L}^{(1,1)}$ -- la dépendance temporelle peut être factorisée car elle n'apparaît pas dans les facteurs exponentiels $\exp(i \, \kbf \cdot \xbf)$ (cf. e.g. Eq. \rref{TrivialExample2a}). C'est le cas des contributions qui dépendent de l'intégrale $P(\eta,r,\ti\theta^a)$.
On évalue donc aisément ces contributions en $(i)$ insérant un facteur $g(\eta_s)/g(\eta_o)$ pour chaque terme $\psi_s$ présent, et $(ii)$ en remplaçant les facteurs $f_{o,s}$ par\,:
\beq
\label{FrftildeLambdaCDM}
\ti{f}_{o,s} \equiv \int_{\eta_{in}}^{\eta_{o,s}} d\eta \frac{a(\eta)}{a(\eta_{o,s})} \frac{g(\eta)}{g(\eta_{o})} ~~.
\eeq
Pour les deux termes dominants qui restent, à savoir $\Tcal_{8,L}^{(1,1)}$ et $\Tcal_{18,L}^{(1,1)}$, nous effectuons les intégrales le long du chemin $r=\eta_o-\eta$ et la dépendance temporelle ne peut être factorisée. Il faut donc calculer les intégrales selon $\eta$ et $k$ numériquement. Toutefois, on peut approximer ces termes en remplaçant $\psi(\eta',\eta_o-\eta', \tilde{\theta}^a)$ par $\psi(\eta_s,\eta_o-\eta', \tilde{\theta}^a)$ du fait que la composante principale de ces intégrales émerge à partir de temps proches de $\eta_s$. On obtient ainsi\,:
\beq
\Ccal(\Tcal_{8,L}^{(1,1)}) = \frac{2}{3} \Xi_s \frac{\ti{f}_s}{\Hcal_s} k^2 ~~~~~,~~~~~
\Ccal(\Tcal_{18,L}^{(1,1)}) = - \frac{4}{3} \frac{\ti{f}_s}{\Hcal_s} k^2 ~~,
\eeq
identique à celui établi dans le cas CDM mais avec ${f}_s$ remplacé par $\ti{f}_s$.

Les contributions à $\overline{\lla \Ical_2 \rra}$ doivent aussi être estimées dans le cas $\Lambda$CDM. On se limite encore une fois aux opérateurs ${\cal O}_3^{ij}$ et ${\cal O}_4^{ij}$ donnés en Eqs. (\ref{PSI}), (\ref{PHI}) et \rref{Ocals}. On trouve que les contributions $\Tcal_1^{(2)}$ et $\Tcal_2^{(2)}$ donnent un effet négligeable et que les termes $\Tcal_{3}^{(2)}$ et $\Tcal_{4}^{(2)}$ s'annulent comme avant (par symétrie).
On obtient les trois contributions restantes par l'utilisation des Eqs. (\ref{Final_overline_O3}) et (\ref{Final_overline_O4})\,:
\begin{eqnarray}
\Ccal(\Tcal_{5,L}^{(2)}) &=& -\Xi_s\left[ \frac{1}{3}\left(B_3(\eta_o)-B_3(\eta_s)\right)+\left(B_4(\eta_o)-B_4(\eta_s)\right)\right] k^2 ~~,
\\ 
\Ccal(\Tcal_{6,L}^{(2)}) &=& -\left( \frac{1}{3} B_3(\eta_s) + B_4(\eta_s)\right) k^2 ~~,
\\ 
\Ccal(\Tcal_{7,L}^{(2)}) &=& \frac{2}{\Delta\eta} \int_{\eta_s}^{\eta_o} d\eta' \left( \frac{1}{3} B_3(\eta') + B_4(\eta')\right) k^2 ~~.
\end{eqnarray}
Ces contributions sont évaluées alors en utilisant les Eqs. (\ref{B3tildeB4tilde}) et (\ref{labelC}) et en changeant la variable $\eta$ au profit du redshift, avec\,:
\beq
\Hcal(z)=\frac{\Hcal_0}{1+z} \left[\Omega_{m0}(1+z)^3+\Omega_{\Lambda0}\right]^{1/2} ~~,
\eeq
\beq
\Omega_m(z)=\frac{\Omega_{m0}(1+z)^3}{\Omega_{m0}(1+z)^3+\Omega_{\Lambda0}} ~~~~~,~~~~~ \Omega_{\Lambda}(z)=\frac{\Omega_{\Lambda0}}{\Omega_{m0}(1+z)^3+\Omega_{\Lambda0}} ~~,
\eeq
On trouve, comme dans le cas CDM, que ces contributions intrinsèques au deuxième ordre sont d'un ou deux ordres de grandeur inférieures aux termes quadratiques de $\overline{\lla \Ical_{1,1} \rra}$.

Commentons enfin sur le spectre de puissance. Dans le cas $\Lambda$CDM, le spectre s'écrit d'après l'Eq. \rref{FrPsiPpar} comme\,:
\beq
\label{FrPsiPz}
\Pcal_\psi (k,z) = \left(\frac{3}{5}\right)^2 \Delta_{\cal R}^2 T^2(k) \left(\frac{g(z)}{g_{\infty}}\right)^2 ~~,
\eeq
où nous avons inclus la dépendance temporelle des perturbations gravitationnelles. Il faut aussi noter que l'échelle d'équilibre entre la période de radiation et celle de matière est réduite (ceci à cause de la diminution de $\Omega_{m0}$ lors du passage de CDM à $\Lambda$CDM). Cet effet de la réduction de puissance dans le cas $\Lambda$CDM est illustré en Fig. \ref{Fig2}. On remarque dans cette figure que la fonction de transfert $T(k)=T_0(k)$ (en trait tireté) déjà présentée ne dépend pas d'une composante baryonique de matière. Le cas où cette composante baryonique est présente est représenté en trait plein. On constate que le manque de considération de la fraction baryonique de $\Omega_{m0}$ entraine une surestimation de la puissance du spectre de près de 40\% dans le domaine $k \gaq 0.01 \, h \,{\rm Mpc}^{-1}$, et ce du fait de l'effet de \emph{Silk damping} (cf. \cite{Eisenstein:1997ik}). Nous considérons donc cet effet en remplaçant la valeur de $q$ précédemment utilisée, $q = k/(13.41 \, k_{\rm eq})$, par \cite{Eisenstein:1997ik}\,:
\bea
q&=&\frac{k \times \Mpc}{h \Gamma} ~~,\\
\Gamma&=&\Omega_{m0}h\left(\alpha_{\Gamma}+\frac{1-\alpha_{\Gamma}}{1+(0.43 k s)^4}\right) ~~,\\
\alpha_{\Gamma}&=&1-0.328\ln(431 \Omega_{m0}h^2)\frac{\Omega_{b0}}{\Omega_{m0}}+0.38 \ln(22.3 \Omega_{m0} h^2)\left(\frac{\Omega_{b0}}{\Omega_{m0}}\right)^2 ~~,\\
s&=&\frac{44.5 \ln(9.83/\Omega_{m0}h^2)}{\sqrt{1+10(\Omega_{b0}h^2)^{3/4}}} \,{\rm Mpc} ~~.
\eea
$\Omega_{b0}$ est le paramètre de densité baryonique, $s$ l'horizon du son (\emph{sound horizon}) et $\Gamma$ le paramètre de forme effectif (\emph{effective shape parameter}) qui dépend de $k$. L'effet des oscillations acoustiques de baryons (BAO) se révèle négligeable pour nos calculs. En effet, celui-ci a pour effet de superposer au spectre de puissance une oscillation qui est quasiment annulée par notre intégration sur $k$.

Les résultats de l'intégration des termes dominants contribuants à la correction du flux lumineux sont présentés en Fig. \ref{Fig3}. On voit que l'effet obtenu pour $|f_\Phi|$ est inférieur à celui qui avait été obtenu en Fig. \ref{f1}. Les raisons à cela sont la présence du \emph{growth factor} dans le spectre, le changement d'expression de la fonction de transfert et la diminution du moment $k_{\rm eq}$. Le flux, en tant que quantité minimalement affectée par les inhomogénéités, apparaît donc comme la grandeur la plus adaptée pour extraire des informations sur les paramètres cosmologiques. Ce n'est pas le cas de la distance $d_L$ ou du module de distance $\mu$ que nous allons décrire maintenant.

\subsubsection{Correction aux autres observables et dispersions}
\label{Ch10Sec632}

Considérons les autres observables que le flux et qui ont été présentées en Sec. \ref{Ch10Sec54}, à savoir la distance $d_L$ et le module $\mu$. Les corrections à ces deux grandeurs diffèrent de celle du flux $f_\Phi$ par des fonctions de la moyenne de $(\Phi_1/\Phi_0)^2$. En prenant $d_L = d_{L}^{(0)} + d_{L}^{(1)} + d_{L}^{(2)}$ et en tenant compte des explications données en Sec. \ref{Ch10Sec41} et \ref{Ch10Sec42}, nous pouvons montrer que\,:
\beq
\label{FrOBcorrection}
\frac{\Phi_1}{\Phi_0}=-2 \frac{d_{L}^{(1)}}{d_{L}^{(0)}} = - {\cal I}_1 + ({\rm t.d.})^{(1)} = 
2\left(-\Xi_s J+\frac{Q_s}{\Delta \eta}+\psi_s+J_2^{(1)}\right) ~~.
\eeq
Nous avons aussi montré que la contribution principale (\emph{\underline{L}eading contribution}) à la moyenne de ce terme, dans le cas CDM, est donnée par\,:
\beq
\label{FrAppTermFlux}
\overline{\left\langle \left(\frac{\Phi_1}{\Phi_0}\right)^2 \right\rangle}_{ L} = 4\left\{\overline{\left\langle \left(J_2^{(1)}\right)^2 \right\rangle}+
\Xi_s^2 \left[\overline{\langle  ([\partial_r P]_s)^2  \rangle}+\overline{\langle  ([\partial_r P]_o)^2  \rangle}\right]\right\} ~~.
\eeq
Nous avons déjà vu que les termes faisant intervenir $\partial_r P$ étaient des termes de vitesse particulière et qu'ils peuvent être calculés dans le modèle $\Lambda$CDM sans approximation. Leur intégrandes sont du type $k^2 {\cal P}_\psi(k, \eta_o)$. Le premier terme de ce résultat, le terme de lentille gravitationnelle (du type $k^3 {\cal P}_\psi(k, \eta_o)$), est en revanche plus difficile à évaluer. Il est en effet nécessaire de remplacer $\psi(\eta',\eta_o-\eta', \tilde{\theta}^a)$ par $\psi(\eta_s,\eta_o-\eta', \tilde{\theta}^a)$ dans ce dernier terme. On obtient au final le coefficient spectral dominant\,:
\beq
\Ccal(\left(\Phi_1/\Phi_0\right)^2_L)\simeq \frac{4}{3}\Xi_s^2 \left(\ti f_s^2 + \ti f_o^2\right)k^2 +
\frac{4}{15} \left[\frac{g(\eta_s)}{g(\eta_o)}\right]^2 \Da \eta^3 k^3 \, {\rm SinInt}(k \Da \eta) ~~.
\eeq

Du fait de l'existence de ce terme dans $\mu$ et $d_L$ et son abscence dans $\Phi$ (le terme de lentille, du moins), on peut s'attendre à avoir une correction plus grande pour ces quantités que pour le flux. C'est en effet ce que l'on observe en Fig. \ref{Fig3} où est présentée la contribution $f_d$ et où il apparaît clairement que $|f_\Phi| \ll f_d$ aux larges redshifts. Nous constatons donc encore une fois que l'effet des inhomogénéités est dominé par les vitesses particulières aux petits redshifts et dominé par les effets de lentille aux grands redshifts. Mais là ou le flux n'est pas affecté dans ce dernier cas, la distance l'est significativement.

En applicant ce même résultat au module de distance $\overline{\langle \mu \rangle} - \mu^M = \overline{\langle \mu \rangle} -  5 \log_{10} [(2+z)z/(2H_0)]$, nous estimons directement l'effet des inhomogénéités de matière sur le diagramme de Hubble, à la fois sur la moyenne (cf. Eq. \rref{Fr14}) et sur la dispersion (cf. Eq. \rref{Fr15}). Nous présentons en Fig. \ref{Fig4} le résultat de cette estimation de $\overline{\langle \mu \rangle} -\mu^M$ et sa dispersion, pour $\Om_{\La 0} = 0.73$, et le comparons aux résultats homogènes pour différentes valeurs de $\Om_{\La 0}$. Nous trouvons que la moyenne de $\mu$ n'est modifiée par l'effet des inhomogénéités, avec le spectre linéaire, qu'à la troisième décimale. Nous pouvons aussi constater sur la Fig. \ref{Fig4} que la dispersion du module de distance, aux larges redshifts, est comparable à une erreur de 2\% sur le paramètre d'énergie sombre $\Omega_{\Lambda 0}$. Nous allons voir dans la section suivante comment ces faits sont modifiés dans le cas de l'utilisation d'un spectre non-linéaire.

\subsection{Le cas $\Lambda$CDM avec spectre non-linéaire}
\label{Ch10Sec64}

\subsubsection{Spectre de puissance}
\label{Ch10Sec641}

Nous avons pu voir que le calcul de la moyenne ou de la dispersion se résument, modulo quelque approximations, au calcul d'une intégrale sur les moments $k$ des perturbations de matière. Cette intégrale sur $[0,\infty[$ doit être restreinte, pour des raisons physiques, en remplaçant $0$ par $\sim H_0$ (aucun mode ayant une longueur d'onde plus grande que l'horizon $H_0^{-1}$ de l'Univers observable) et $\infty$ par un cut-off n'exédant pas le moment où la description du spectre devient inappropriée. Pour le spectre linéaire (cf. Eq. \rref{FrSpecPuiss}) ce cut-off correspond à une échelle de l'ordre de $1 ~ \Mpc^{-1}$. Si on veut utiliser des échelles plus petites, il nous faut considérer l'aspect non-linéaire de la distribution de matière.

Le modèle effectif le plus simple pour décrire le régime non-linéaire est celui du HaloFit (voir \cite{Smith:2002dz} et plus récemment \cite{Takahashi:2012em}). Pour le comprendre, il faut définir le spectre (sans dimension) de la densité de matière\,:
\beq
\Delta^2(k) = \frac{k^3}{2 \pi^2} |\delta_k|^2 ~~,
\eeq
et utiliser l'équation de Poisson dans un Univers de type $\Lambda$CDM\,:
\beq
\nabla^2 \psi(x) = 4 \pi G \rho(t) \, \delta(x) ~~\Rightarrow~~ \Pcal_{\psi}^{\rm{L,NL}}(k) = \frac{9}{4} \frac{\Om_{m0}^2 \Hcal_0^4}{(1+z)^{-2} k^4} ~ \Delta_{\rm{L,NL}}^2(k) ~~.
\eeq
Cette dernière relation est valide dans les deux régimes. Ceci nous permet d'écrire le spectre de la densité de matière en fonction de celui du champ gravitationnel dans le régime linéaire\,:
\beq
\Delta_{\rm{L}}^2(k) = \frac{4}{25} \frac{A \Om_{m0}^{-2} \Hcal_0^{-4}}{(1+z)^2} ~ \frac{k^{n_s+3}}{k_0^{n_s-1}} \left(\frac{g(z)}{g_0}\right)^2 T^2(k) ~~.
\eeq

Le modèle du HaloFit, basé sur les résultats de simulations à N-corps, donne alors l'expression du spectre de la matière dans le régime non-linéaire. Cette expression est une paramétrisation effective, somme d'un terme représentant la corrélation entre deux halos séparés et d'un autre représentant la matière à l'intérieur d'un halo donné\,:
\beq
\Delta^2_{\NL}(k) = \Delta^2_{\Q}(k) + \Delta^2_{\H}(k) ~~,
\eeq
\bea
\mbox{avec }~~~ \Delta^2_{\Q}(k) &=& \Delta^2_{\rm{L}}(k) \left[\frac{(1+\Delta_{\rm{L}}^2(k))^{\beta_n}}
{1+\alpha_n\Delta^2_{\rm{L}}(k)} \right] e^{-f(y)} ~~,~~ y\equiv \frac{k}{k_{\sigma}} ~~,~~ f(y)=\frac{y}{4} + \frac{y^2}{8} ~~, \\
\mbox{et }~~ \Delta^2_{\H}(k) &=& {\Delta^{2\ \prime}_{\H}(k) \over 1 + \mu_n y^{-1}+\nu_n y^{-2}} ~~,~~ \Delta^{2\ \prime}_{\H}(k) = \frac{a_n ~ y^{3f_1(\Omega)}}{1+b_ny^{f_2(\Omega)} + \left[ c_n f_3(\Omega) ~ y \right]^{3-\gamma_n}} ~~,~~~~~
\eea
où $\{\alpha_n,\beta_n,\gamma_n,\mu_n,\nu_n,a_n,b_n,c_n\}$ sont des paramètres déterminés par trois grandeurs (cf. Eq. \rref{ParamHaloFitSmith}).

Le premier de ces trois paramètres est le moment $k_\sigma$ correspondant à l'échelle d'entrée dans le régime non-linéaire. Il est obtenu en normalisant à $1$ la variance de la densité de matière\,:
\beq
\sigma^2(k_\sigma^{-1}) \equiv \int \Delta^2_{\rm{L}}(k,z) ~ \exp \left(-(k/k_\sigma)^2 \right) ~ d\ln k \equiv 1 ~~.
\eeq
Les deux autres grandeurs sont l'indice spectral effectif du spectre et la courbure de cet indice\,:
\beq
n_{\rm eff} \equiv - \bigg[ 3 + \frac{d \ln \sigma^2(R)}{d \ln R} \bigg]_{R=k_\sigma^{-1}} ~~~~~,~~~~~ C \equiv - \frac{d^2 \ln \sigma^2(R)}{d \ln R^2} \Bigg|_{R=k_\sigma^{-1}} ~~.
\eeq
La première de ces grandeurs, et par conséquent les deux autres ainsi que les paramètres du modèle du HaloFit, dépendent du redshift (autrement dit du temps) et ce de manière non-triviale. On a donc une paramétrisation du spectre jusque dans le régime non-linéaire. Pour \cite{Smith:2002dz}, le spectre est physiquement valide jusque $k \sim 1 ~h~ \Mpc^{-1}$, tandis que \cite{Takahashi:2012em} donne une confiance sur le résultat allant jusque $k \sim 30 ~h~ \Mpc^{-1}$. Le spectre non-linéaire obtenu par l'implémentation de ces deux paramétrisations du modèle du HaloFit, \cite{Smith:2002dz} et \cite{Takahashi:2012em}, est illustré en Fig. \ref{Fig5}.

\subsection{Résultats numériques et comparaison avec le régime linéaire}
\label{Ch10Sec642}

Nous avons déjà mentionné la dépendance temporelle du spectre de puissance dans le cas $\Lambda$CDM. Il faut maintenant comprendre que le spectre non-linéaire dépend d'autant plus du temps (ou du redshift) que chacun des paramètres cités dans la section précédente a aussi une dépendance en temps. Il devient alors impossible de calculer l'effet des différents termes inhomogènes sans recourir à de nouvelles approximations. Pour cela, nous avons utilisé l'approximation du spectre suivante\,:
\beq
\Pcal_\psi^{\rm{NL}}(k, z) = \frac{g^2(z)}{g^2(z^*)} \Pcal_\psi^{\rm{NL}}(k, z^*) ~~,
\label{Fr76}
\eeq
où nous avons choisi $z^*=z_s/2$ pour essayer de minimiser les imprécisions. En effet, notre étude porte sur des redshifts tels que $0.015 \leq z \leq 2$ et cette approximation ne devient discutable que lorsque $z_s=2$ (i.e. $z^*=1$) et que l'on estime nos termes en $z=0.015$ et $z=2$. Dans ce cas, nous sous-estimons le spectre de $40\%$ pour $z=0.015$ et le surestimons d'environ $80\%$ pour $z=2$. Ces erreurs sont néanmoins présentes seulement pour deux étroites bandes de moment centrées sur $k=1 \, h \,{\rm Mpc}^{-1}$ et $k=2 \, h \,{\rm Mpc}^{-1}$. On estime donc la précision globale de notre evaluation à $\sim 10\%$. Il est alors possible d'estimer les intégrales des diverses quantités $d_L^{-2}$, $d_L$, $\mu$ comme précédemment, pour chaque redshift, en intégrand numériquement sur $k$. Le résultat final de nos calculs est donné en Figs. \ref{Fig6} et \ref{Fig7}, et ce en utilisant le spectre de puissance donné par le modèle du HaloFit de \cite{Takahashi:2012em} et incluant une contribution baryonique. On constate que la distance $d_L$ reçoit une correction de l'ordre de $10^{-3}$ autour de $z=2$ alors que la correction au flux est deux ordres de grandeur moindre. La comparaison de ces résultats avec les Figs. \ref{Fig3} et \ref{Fig4} montre que le spectre de puissance considéré dans son régime non-linéaire a augmenté l'effet des inhomogénéités sur les moyennes des observables. Ces effets restent néanmoins très difficiles à observer expérimentalement. On peut aussi voir sur la partie droite de la Fig. \ref{Fig7} que la considération du spectre non-linéaire ne rend l'effet sur la moyenne que difficilement identifiable visuellement.
En revanche, la dispersion théorique due aux inhomogénéités est augmentée significativement. En effet, la dispersion sur $\mu$ est de l'ordre de $10\%$ à $z = 2$, ce qui montre que les modèles homogènes ayant une valeur de $\Omega_{\Lambda 0}$ entre $0.68$ et $0.78$ se situent à moins d'une déviation standard ($1 \, \sigma$) de notre prédiction pour un modèle inhomogène avec $\Omega_{\Lambda 0}=0.73$.

\subsection{Comparaison avec la dispersion intrinsèque des SNe Ia}
\label{Ch10Sec643}

Attachons nous ici à comparer notre prédiction théorique de la dispersion $\sigma_{\mu}$ avec la dispersion intrinsèque estimée par les données des SNe Ia. Pour ne présenter que cette dispersion $\sg_\mu$, nous l'avons tracé en Fig. \ref{Fig8} (et déjà visible en Fig. \ref{Fig7}). Dans ce tracé, le trait plein représente la contribution totale de $\sg_\mu$ obtenue par l'Eq. (\ref{Fr15}) en fonction de $z$. Cette dispersion tolale est dans une très bonne approximation donnée par la somme des termes de vitesse et de lentille gravitationnelle (comme déjà évoqués dans l'étude de l'Eq. \rref{FrAppTermFlux}). Cette dispersion possède une évolution en $z$ bien caractéristique avec une valeur minimale proche de $0.016$ pour $z = 0.285$.

La variance totale $\sigma_{\mu}^{\rm obs}$ associée aux observations peut quant à elle être décomposée en diverses composantes (cf. e.g. \cite{March:2011xa,Conley:2011ku,Vishwakarma:2010nc})\,:
\beq
(\sigma_{\mu}^{\rm obs})^2 =  (\sigma_{\mu}^{\rm fit})^2 + (\sigma_{\mu}^{z})^2  + (\sigma_{\mu}^{\rm int})^2 ~~,
\eeq
où $\sigma_{\mu}^{\rm fit}$ est la dispersion liée à la méthode utilisée pour ajuster les courbes de lumière (e.g. la méthode SALT-II \cite{Guy:2007dv}) mais aussi la modélisation du processus amenant à la supernova. $\sigma_{\mu}^{z}$ représente l'incertitude sur la vitesse particulière des SNe Ia et la précision des mesures spectroscopiques. Enfin, $\sigma_{\mu}^{\rm int}$ représente toutes les autres sources d'incertitudes non prises en compte par le modèle homogène. Cette dernière quantité peut donc être redéfinie, par la connaissance de la dispersion due à l'effet de lentille gravitationnelle que nous avons estimé précédemment, de la manière suivante\,:
\beq
(\sigma_{\mu}^{\rm int})^2 = (\widehat{\sigma_{\mu}^{\rm int}})^2 + (\sigma_{\mu}^{\rm lens})^2  ~~.
\eeq
Ici $\widehat{\sigma_{\mu}^{\rm int}}$ est la dispersion intrinsèque restante des SNe Ia.
La précision actuelle des données (cf. \cite{Kowalski:2008ez,Guy:2007dv}) et du diagramme de Hubble ne permet pas de mettre en lumière une forte dépendance en $z$ de $\sigma_{\mu}^{\rm int}$\,: nous avons $\sigma^{\rm int} = 0.15 \pm 0.02$ pour les SNe Ia les plus proches et $\sigma^{\rm int} = 0.12 \pm 0.02$ pour les plus éloignées. D'un autre côté, on voit dans le résultat de nos calculs, présenté en Fig. \ref{Fig8}, que notre prédiction pour le terme de lentille est linéaire pour $z \gaq 0.3$. On trouve en effet que cette prédiction est très bien approximée par $\sigma_{\mu}^{\rm lens}(z) = 0.056 \, z$, ce qui la rend parfaitement compatible avec les observations (elle est $< 0.12$ jusque $z \sim 2$).

Il est aussi encore plus intéressant de noter que cette prédiction est en parfait accord avec les tentatives récentes de mesure de cet effet de lentille au sein des SNe Ia, par \cite{Kronborg:2010uj}\,:
\beq
\left( \sigma_{\mu}^{\rm lens} \right)_{\mbox{Kronborg}} = (0.05 \pm 0.022) \, z ~~,
\eeq
ou encore \cite{Jonsson:2010wx}\,:
\beq
\left( \sigma_{\mu}^{\rm lens} \right)_{\mbox{J\"onsson}} = (0.055^{+0.039}_{-0.041}) \, z ~~.
\eeq
Cette compatibilité de notre calcul avec ces mesures exprimentales est clairement identifiable dans la Fig. \ref{Fig8}. Les simulations de lentille (cf. \cite{Holz:2004xx}) ont aussi apporté une estimation de $0.088 \,z$, en parfait accord avec notre prédiction. Remarquons toutefois que d'autres auteurs n'ont pas relevé d'effet de lentille significatif avec la précision actuelle des données (cf. \cite{Williams:2004jr, Menard:2004sm, Karpenka:2012ys}).
Il est plus que probable que les prochaines données sur les supernovae permettront de confirmer ou d'infirmer la présence de cet effet de lentille dans la dispersion du diagramme de Hubble. Ceci constitue donc un test du modèle de concordance. Notons enfin que notre prédiction de la dispersion liée à la vitesse particulière (cf. Fig. \ref{Fig8}) est très bien approximée par la loi de puissance suivante\,:
\beq
\sg_\mu^{\rm Doppler}(z) \sim 0.00323 \, z^{-1} ~~.
\eeq
Ce résultat est en très bon accord avec ceux de \cite{Hui:2005nm}.

\subsection{Dispersion et comparaison aux données expérimentales}
\label{Ch10Sec645}

Nous avons finalement effectué la comparaison entre nos prédictions sur le module de distance et les données issues de la compilation \guillemotleft \, Union 2 \guillemotright, comme présenté en Fig. \ref{ThAndData}. Comme il est possible de le voir à l'oeil nu, la dispersion théorique que nous avons calculé est moins importante que la dispersion des données, en particulier aux grands redshifts. Il apparait donc que la composante principale de la dispersion des données aux petits redshifts est essentiellement due aux vitesses particulières et que notre prédiction en terme de perturbations inhomogènes est une bonne explication à cela. D'un autre côté, notre dispersion calculée aux grands redshifts et due essentiellement au terme de lentille gravitationnelle ne suffit absolument pas à expliquer la dispersion des données, révélant des sources de dispersion non prises en compte dans notre calcul. Il est en fait logique d'aboutir à une telle conclusion. Les SNe Ia sont des phénomènes physiques très complexes et notre calcul ne prend en compte que l'effet des inhomogénéités, celui qui nous intéresse dans cette étude. Des effets comme l'absorption de la lumière par le milieu interstellaire ou encore des effets de lentilles fortes sont possible. De plus, et il s'agit probablement d'un facteur important, les supernovae ont une dispersion intrinsèque dans leur luminosité qui est très difficile à estimer.

On peut en fait se réjouir d'un tel résultat car il montre que le modèle standard de la cosmologie est robuste aux effets des inhomogénéités présentés ici et que le décalage dans le diagramme de Hubble qui a donné lieu à la découverte de l'accélération de l'Univers n'est pas une simple fluctuation statistique. Dans le cas d'une dispersion plus petite que notre prédiction, nous serions forcés de soustraire les effets de dispersion pour chacune des sources avant de pouvoir conclure à l'existence de l'énergie sombre. Nous avons donc montré par nos calculs au second ordre que les inhomogénéités ne pouvaient pas jouer le rôle de cette énergie sombre. Toutefois, il est très important de remarquer que l'effet que nous avons obtenu est plus grand que celui qui était naïvement attendu, il dépend sensiblement de l'observable physique considérée et est susceptible d'être détecté dans les prochaines années, donnant ainsi un autre test des inhomogénéités et de meilleures contraintes sur les propriétés de l'énergie sombre.

\section{Résumé et Conclusions}
\label{Ch10Sec7}

Nous avons étudié dans cette thèse le problème général de la moyenne sur le cône de lumière et ses implications en cosmologie de manière à améliorer notre compréhension de l'Univers et de sa nature inhomogène.

En premier lieu, nous avons décrit le modèle standard de la cosmologie en donnant en particulier les équations utilisées pour interpréter les données expérimentales dans le cadre du modèle de concordance $\Lambda$CDM (les equations de Friedmann). Nous avons aussi décrit l'utilisation des supernov\ae\ de type Ia qui sont à la base de la découverte de l'énergie sombre et les considérations de base relatives aux grandes structures.

Nous avons ensuite donné la définition de la moyenne de quantités scalaires sur des hypersurfaces de type lumière. Cette définition, inspirée par la moyenne spatiale du formalisme de Buchert \cite{Buchert:1999er,Buchert:2001sa}, possède des caractéristiques communes avec leur généralisation invariante de jauge \cite{Gasperini:2009wp,Gasperini:2009mu}. En particulier, cette définition est invariante sous les transformations de coordonnées, de jauge et sous les reparamétrisations de ses hypersurfaces. On a ainsi pu écrire les relations équivalentes à la règle de commutation de Buchert-Ehlers qui est à la base des questions de backreaction. Cette moyenne sur le cône de lumière, longtemps soulignée comme importante, n'avait jamais été définie dans la littérature.

Ensuite, nous avons introduit un système de coordonnées qui simplifie grandement la description de quantités physiques sur notre cône de lumière passé. Nous les avons appelées coordonnées \guillemotleft \, géodésiques sur le cône de lumière \guillemotright \, (``Geodesic Light-Cone'' coordinates) et celles-ci se sont révélées utiles pour une gamme d'applications bien plus large que la moyenne. Elles sont aussi assez proches d'un autre système de coordonnées dites \guillemotleft \, observationelles \guillemotright \cite{1985PhR...124..315E,Maartens1} et correspondent à une fixation de jauge de ces dernières qui contiennent un degré de liberté supplémentaire. D'autres différences sont à noter, comme le fait que nos coordonnées contiennent le temps $\tau$ de l'observateur tandis que les coordonnées observationnelles utilisent une distance $y$ sur le cône passé (qui est nulle ou de type spatial). L'étude de cette jauge est fortement susceptible d'apporter des résultats pour toutes les applications liées aux signaux lumineux en cosmologie. Ces coordonnées GLC forment un système dans lequel les photons se propagent à coordoonées $w = w_o$ et $\ti \theta^a$ constantes. Cette redéfinition des angles le long de la propagation fait que les expressions du redshift $z$ et de la distance de luminosité $d_L$ sont très simples dans cette jauge (et directement exprimées en terme de ses éléments de métrique). Nous avons aussi décrit l'exemple de la transformation entre les coordonnées GLC et les métriques synchrone et LTB. Ces coordonnées ont enfin été utilisées pour simplifier l'expression générale de la moyenne sur le cône de lumière passé et les relations de commutation de Buchert-Ehlers et ont été appliquées par l'exemple de la moyenne du redshift drift.

Les effets de backreaction disparaissent dans le cas particulier d'un modèle FLRW et il nous a donc été nécessaire d'évaluer la distance de luminosité dans le cas d'un modèle perturbé. Du fait de l'aspect stochastique des inhomogénéités introduites, la moyenne de leurs effets disparaît au premier ordre lorsque l'on prend leur moyenne d'ensemble. Le calcul perturbatif de la distance a donc dû être poussé jusqu'au second ordre et a été exprimé en terme des éléments de métrique de la jauge de Poisson du fait que le spectre de puissance des inhomogénéités est connu dans cette jauge (et non dans la jauge GLC). L'expression finale de cette distance est riche de sense physique et inclut des termes de vitesse particulière, les effets SW et ISW, l'effet de lentille, les distortions de redshift et un grand nombre de combinaisons relatives à ces effets. Ce résultat est donc très intéressant pour la cosmologie et pourrait être utilisé dans un grand nombre de problèmes. Encore une fois, la métrique GLC joue un rôle capital pour l'établissement de ce résultat. Les perturbations vectorielles et tensorielles n'ont pas été oubliées, mais considérées seulement au second ordre (du fait de leur origine inflationnaire).

Nous avons ensuite abordé la question de la taille relative de l'effet des inhomogénéités sur la distance de luminosité, le flux lumineux et le module de distance. Pour cela nous avons dû combiner la moyenne sur le cône de lumière avec la moyenne d'ensemble sur les perturbations stochastiques. Dans une premier temps, nous avons effectué un calcul partiel basé sur l'ordre quadratique mais suffisant pour mettre en lumière la présence de termes dit de \guillemotleft \, backreaction induite \guillemotright. Ces termes sont intéressants car il regroupent les expressions des scalaires moyennés et de la mesure d'intégration. Ils permettent aussi l'évaluation exacte de la dispersion de ces grandeurs. Nous avons donné l'expression des coefficients spectraux relatifs à ces termes et avons ensuite considéré le cas complet au second ordre. Nous avons trouvé que la moyenne de la distance de luminosité différait de celle du flux lumineux par une contribution (de flux)\,: $\overline{\lla (\Phi_1 / \Phi_0)^2 \rra}$. Cette contribution est proportionnelle à la moyenne du carré de la correction au premier ordre de la distance. Elle contient entre autres les termes de vitesse particulière et de lentille gravitationnelle. Nous avons aussi montré que les perturbations vectorielles et tensorielles disparaissent sous l'effet de notre moyenne (et ce par symétrie). Quant aux contributions intrinsèques au second ordre, elles ont été reliées par l'utilisation de \cite{Bartolo:2005kv} à celles du premier ordre.

Faisant usage de ces résultats théoriques, nous sommes passés à l'évaluation numérique de leurs effets en choisissant une forme explicite pour la fonction de transfert du spectre de puissance linéaire (produit par le spectre primordial). Nous avons tout d'abord considéré le cas simple d'un univers CDM avec une fonction de transfert ne contenant pas de contribution baryonique. Nous avons utilisé ce cas pour l'évaluation des termes de backreaction induite et de la variance $\overline{\lla \sigma_1^2 \rra}$. Nous avons constaté, comme prévu, que l'effet dominant qui agit sur les SNe Ia à petits redshifts est celui de leur vitesse particulière (due aux inhomogénéités) et aux grands redshifts celui de lentille gravitationnelle.
Ce cas nous a aussi aidé à voir que nos intégrales sont libres de toute divergence, ultraviolette ou infrarouge. Ce résultat est à comparer avec celui des moyennes spatiales où des divergences UV existent et il est dû à la construction de notre moyenne, sur le cône de lumière et de manière invariante de jauge. Une autre conclusion a été que la moyenne de la distance de luminosité est très faiblement affectée par les inhomogénéités, sauf par l'effet de vitesse à petits redshifts. Toutefois, ce calcul était imcomplet et donc incertain.

Nous avons donc établi le résultat numérique du calcul complet au second ordre, dans le cas du flux et encore au sein d'un modèle CDM. Nous avons pu constater que le flux est très faiblement affecté aux large redshifts, sa correction relative restant de l'ordre de $\sim 10^{-5}$, ce qui fait du flux une quantité très spéciale pour laquelle les effets de lentille s'annulent lorsque l'on prend la moyenne sur le cône de lumière. Ceci est bien sûr relié à la conservation du flux le long du cône. Ce calcul a aussi montré que les inhomogénéités ne pouvaient simuler une constante cosmologique\,: la correction à la moyenne est trop faible et ne possède par le bon signe aux grand redshifts. L'effet sur la distance est quant à lui toujours positif mais possède une variation bien caractéristique du fait des deux effets de vitesse et de lentille. La correction à la distance peut atteindre la valeur de $\sim 10^{-2}$ à $z=2$. La considération du module de distance a montré que l'effet sur la moyenne reste négligeable (hormis aux petits redshifts) mais que la dispersion des inhomogénéités entraine une erreur de $\sim 1-2 \, \%$ sur $\Omega_{\Lambda0}$.

Nous avons ensuite considéré le cas réaliste d'un modèle $\Lambda$CDM et ajouté l'effet des baryons (\emph{Silk damping}) dans la fonction de transfert \cite{Eisenstein:1997ij}. Ces deux modifications ont une conséquence directe sur le spectre de puissance linéaire, en réduisant ses valeurs, et font donc décroître les corrections du flux et de la distance d'environ un ordre de grandeur. Nous avons ensuite considéré le régime non-linéaire du spectre à travers le modèle du HaloFit \cite{Smith:2002dz,Takahashi:2012em}. Ceci a eu pour conséquence de renforcer l'effet des inhomogénéités et ce en particulier sur la correction à la distance du fait de sa sensibilité aux effets de lentille (atteignant ainsi $\sim 10^{-3}$ à $z=2$).
Le module de distance a quant à lui révélé un changement négligeable sur sa moyenne mais a vu sa dispersion augmenter et correspondre à une erreur de $\sim 10\%$ sur la valeur de $\Omega_{\Lambda0}$. Ce résultat est donc très important et démontre que les inhomogénéités ne peuvent être négligées, en particulier au sein de la dispersion des quantités observables. Nous avons finalement comparé notre prédiction théorique avec la dispersion des données du Union2 et les récentes tentatives d'estimation des contributions de vitesse et de l'effet de lentilles dans la dispersion des SNe Ia. Notre résultat se révèle être en parfait accord avec les données et devrait être confirmé dans les prochaines observations de cosmologie de précision.

\bigskip

Nos conclusions, élaborées à l'aide d'un système de coordonnées dites GLC et l'introduction d'une prescription invariante de jauge pour la moyenne sur le cône de lumière passé, peuvent être résumées comme suit. Nous avons trouvé que la combinaison de la moyenne sur le cône avec une moyenne d'ensemble sur des perturbations stochastiques autour d'un modèle d'Univers FLRW plat, appliquée à différentes observables ($d_L$, $\Phi \propto d_L^{-2}$ et $\mu \sim 5 \log_{10} d_L$) ne pouvait simuler l'énergie sombre. Ce résultat peut nous donner confiance en la robustesse du modèle standard de la cosmologie. D'un autre côté, les inhomogénéités moyennées ont un impact bien plus grand que ce qui était naïvement attendu. Leur influence est principalement caractérisée par la vitesse particulière et l'effet de lentille gravitationnelle et elles affectent la distance plus que le flux (qui est la quantité minimalement affectée aux grands redshifts) par un facteur $\sim 100$ à $z = 2$. Cela montre que le flux est bel-et-bien la quantité idéale à mesurer et que les autres observables sont bien plus sensibles aux inhomogénéités. Enfin, nous avons comparé nos prédictions avec la dispersion intrinsèque des SNe Ia et montré qu'une part de cette dispersion devrait être explicables dans les futures observations capables de détecter (et de soustraire) ce signal de contamination. Par conséquent, notre travail souligne que, bien que les inhomogénéités ne puissent expliquer l'énergie sombre (du moins dans le cadre de notre calcul), nous pouvons avec confiance croire que l'absence de considération de leur effet rendra impossible une détermination des paramètres de l'énergie sombre en dessous du pourcent et pourrait même aboutir à une mauvaise interprétation des données comme, par exemple, la possible découverte d'une dépendence spatiotemporelle de cette énergie qui ne serait enfin que le reflet des inhomogénéités.

% - -------------------------------------------------------- %

% --------
\newpage
~
\newpage
% --------
\end{appendices}

% ----- Lists ----- %
\newpage
\listoffigures
\listoftables
% ----------------- %

% ----- Publications ----- %
\chapter*{PhD publications}
\addcontentsline{toc}{chapter}{PhD publications}
\label{SecPhDPubli}
\pagestyle{plain}

We present here a brief description of the different papers that have been published during the elaboration of this manuscript, in chronological order between 2011 and 2013.

\newcommand{\bfph}[1]{\textbf{\emph{#1}}}

\begin{enumerate}

\item %***** ***** ***** ***** ***** ***** ***** ***** ***** *****
%\bibitem{Gasperini:2011us}
\small
\bfph{Light-cone averaging in cosmology: Formalism and applications}, \\
\normalsize
M.~Gasperini, G.~Marozzi, F.~Nugier and G.~Veneziano, \\
JCAP {\bf 1107} (2011) 008, [arXiv:1104.1167 [astro-ph.CO]]. Reference \cite{P1}. \\
%%CITATION = ARXIV:1104.1167;%%

% \texttt{ \justify
We give in this paper the first definition of the average on the past lightcone of a geodetic observer. This average is gauge invariant and invariant under reparametrization and coordinates transformations. Generalizations of Ehlers-Buchert commutation relations are derived. The average is applied to different scalars : the luminosity distance (formally) and the redshift drift. \\
% } \\

\item %***** ***** ***** ***** ***** ***** ***** ***** ***** *****
%\bibitem{BenDayan:2012pp}
\small
\bfph{Backreaction on the luminosity-redshift relation from gauge invariant light-cone averaging}, \\
\normalsize
I.~Ben-Dayan, M.~Gasperini, G.~Marozzi, F.~Nugier and G.~Veneziano, \\
JCAP {\bf 1204} (2012) 036, [arXiv:1202.1247 [astro-ph.CO]]. Reference \cite{P2}. \\
%%CITATION = ARXIV:1202.1247;%%

This paper stands as a first attempt to apply lightcone averaging to the luminosity distance, making use of the average on a topological 2-sphere embedded on the past lightcone and a constant redshift hypersurface. The transformation between the GLC coordinates and the Poisson gauge is derived only at first order and the estimation of the backreaction effect, intrinsically a second order effect in inhomogeneities, is only partial. It nevertheless gives a first sight on the calculation features. We introduce the linear power spectrum of inflationary origin which describes inhomogeneities. It is found that the integrals computed to estimate this effect are both free from IR and UV divergences. Terms coming from Doppler effect and weak lensing of light signals are identified. It is not known at this time if the backreaction effect is big or small. \\

\item %***** ***** ***** ***** ***** ***** ***** ***** ***** *****
%\bibitem{BenDayan:2012ct}
\small
\bfph{Do stochastic inhomogeneities affect dark-energy precision measurements?}, \\
\normalsize
I.~Ben-Dayan, M.~Gasperini, G.~Marozzi, F.~Nugier and G.~Veneziano, \\
PRL \textbf{110}, 021301 (2013), [arXiv:1207.1286 [astro-ph.CO]]. Reference \cite{P3}. \\
%%CITATION = ARXIV:1207.1286;%%

In this PRL letter, we describe the results of a full computation of the luminosity distance at second order in perturbations, using the linear power spectrum to describe inhomogeneities. The effect on the average of the luminosity distance is found to be small, corrections are $\sim 10^{-5}$ for the luminosity flux, but $\sim 10^{-3}$ for the distance itself, this because of the presence of a lensing term in this latter quantity. Hence some observables in cosmology are more sensitively dependent on inhomogeneities than others, the flux being the minimalized observable. The effect of the variance is important, equivalent to a 2\% change in $\Omega_{\Lambda}$ with a linear matter power spectrum. \\

\item %***** ***** ***** ***** ***** ***** ***** ***** ***** *****
%\bibitem{BenDayan:2012wi}
\small
\bfph{The second-order luminosity-redshift relation in a generic inhomogeneous cosmology}, \\
\normalsize
I.~Ben-Dayan, G.~Marozzi, F.~Nugier and G.~Veneziano, \\
JCAP {\bf 1211} (2012) 045, [arXiv:1209.4326 [astro-ph.CO]]. Reference \cite{P4}. \\
%%CITATION = ARXIV:1209.4326;%%

The detailed derivation of the expression of the luminosity distance at second order in perturbations is given, including scalar, vector and tensor inhomogeneities. This expression is interesting on its own and have a large range of applications in cosmology, independently from the question of the backreaction of inhomogeneities. Among the effects encountered, we have normal and integrated Sachs-Wolfe effects, peculiar velocity effects (Doppler), weak lensing and almost all their possible combinations at second order (including integrated effects). \\

\item %***** ***** ***** ***** ***** ***** ***** ***** ***** *****
%\bibitem{BenDayan:2013gc}
\small
\bfph{Average and dispersion of the luminosity-redshift relation in the  concordance model}, \\
\normalsize
I.~Ben-Dayan, M.~Gasperini, G.~Marozzi, F.~Nugier and G.~Veneziano, \\
JCAP {\bf 06} (2013) 002, [arXiv:1302.0740 [astro-ph.CO]]. Reference \cite{P5}. \\
%%CITATION = ARXIV:1302.0740;%%

The full calculation is produced. The average of the luminosity distance is found to be small as indicated in our previous studies. Details of the PRL letter on the average (both over the sky and stochastically) are given.
A non-linear power spectrum elaborated from the HaloFit model is used. The effect is enhanced by small scales, but the order of magnitude stays the same. We find that the variation of $\Omega_{\Lambda}$ can go up to 10\%. We also find that this non-linear power spectrum prediction is in very good agreement with SNe Ia measurements, for the Doppler effect (at small redshifts) and the lensing dispersion (at large redshifts). Our work hence gives a clear theoretical prediction which could be compared with data in the next couple of years. \\

\end{enumerate}

\begin{center}

This thesis has lead me to give a certain number of seminars, in Europe and United States. \\
Here is a list of seminars that I gave concerning its content.

\fontsize{11}{13}\selectfont

\bigskip

\small
\begin{tabular}{c}
\tbf{Journ\'ee des doctorants du LPT ENS 2011}, \\
LPT ENS, Paris, France -- 26 October 2011 \\
\tbf{String, Cosmology \& Gravity Student Conference 2011}, \\
London Mathematical Society, London, England -- 2-4 November 2011 \\
\tbf{Backreaction: Where do we stand?}, \\
IAP, Paris, France -- 22-23 November 2011 \\
\tbf{S\'eminaire INformel des JEunes (SINJE)}, \\
LPT Orsay, Paris-Sud 11, France -- 27 Feb. 2011 \\
\tbf{Summer School on Cosmology}, \\
ICTP, Trieste, Italy -- 16-27 July 2012 \\
\tbf{Journ\'ee des doctorants du LPT ENS 2012}, \\
LPT ENS, Paris, France -- 7 November 2012 \\
\tbf{Particle and Astrophysics Seminar}, \\
ITP, University of Zurich, Switzerland -- 14 Nov. 2012 \\
\tbf{Cosmology \& Particle Phys. Sem.}, \\
Geneva Cosmology Group, Univ. Geneva -- 16 Nov. 2012 \\
\tbf{Cosmology Lunch Discussion}, \\
Dep. Astrophysical Sciences, Princeton Univ. -- 10 Dec. 2012 \\
\tbf{High Energy Theory Seminar}, \\
Center for Particle Cosmology, UPenn, USA -- 11 Dec. 2012 \\
\tbf{HEP Informal Talk}, \\
Center for Cosmology \& Particle Physics, New York Univ. -- 13 Dec. 2012 \\
\tbf{APC Colloquium}, \\
Laboratoire AstroParticule \& Cosmologie, Paris -- 5 March 2012 \\
\tbf{Video Conference Seminar}, \\
Los Alamos National Laboratory -- 5 July 2013 \\
\tbf{Student Talk at Les Houches}, \\
\'Ecole de Physique des Houches, France -- 26 July 2013
\end{tabular}
\end{center}
\normalsize

% ------------------------ %

% ----- Bibliography ----- %
\newpage
\bibliographystyle{utphys}
\bibliography{biblioClean}
% ------------------------ %

% ----- Glossary ------------------------- %
\newpage
\label{SecGloss}
\addcontentsline{toc}{chapter}{Glossary}
\printglossary[type=main,style=listgroup]
% ---------------------------------------- %

\newpage

\addtocounter{page}{-2}

\thispagestyle{empty}

% ----- Backcover ----- %

\setlength{\oddsidemargin}{-0.5cm}
\setlength{\evensidemargin}{-0.5cm}
%\setlength{\oddsidemargin}{0.2cm}
%\setlength{\evensidemargin}{-1.2cm}

% 4ième de couverture :
\pagestyle{empty}

\ifthenelse{\isodd{\thepage}}
 {\newpage
  ~
  \newpage}
 {}

\begin{tikzpicture}[remember picture,overlay]  
\node [xshift=-17.5cm,yshift=-3.0cm] at (current page.north east)
    {\includegraphics[width=5cm,height=2.5cm]{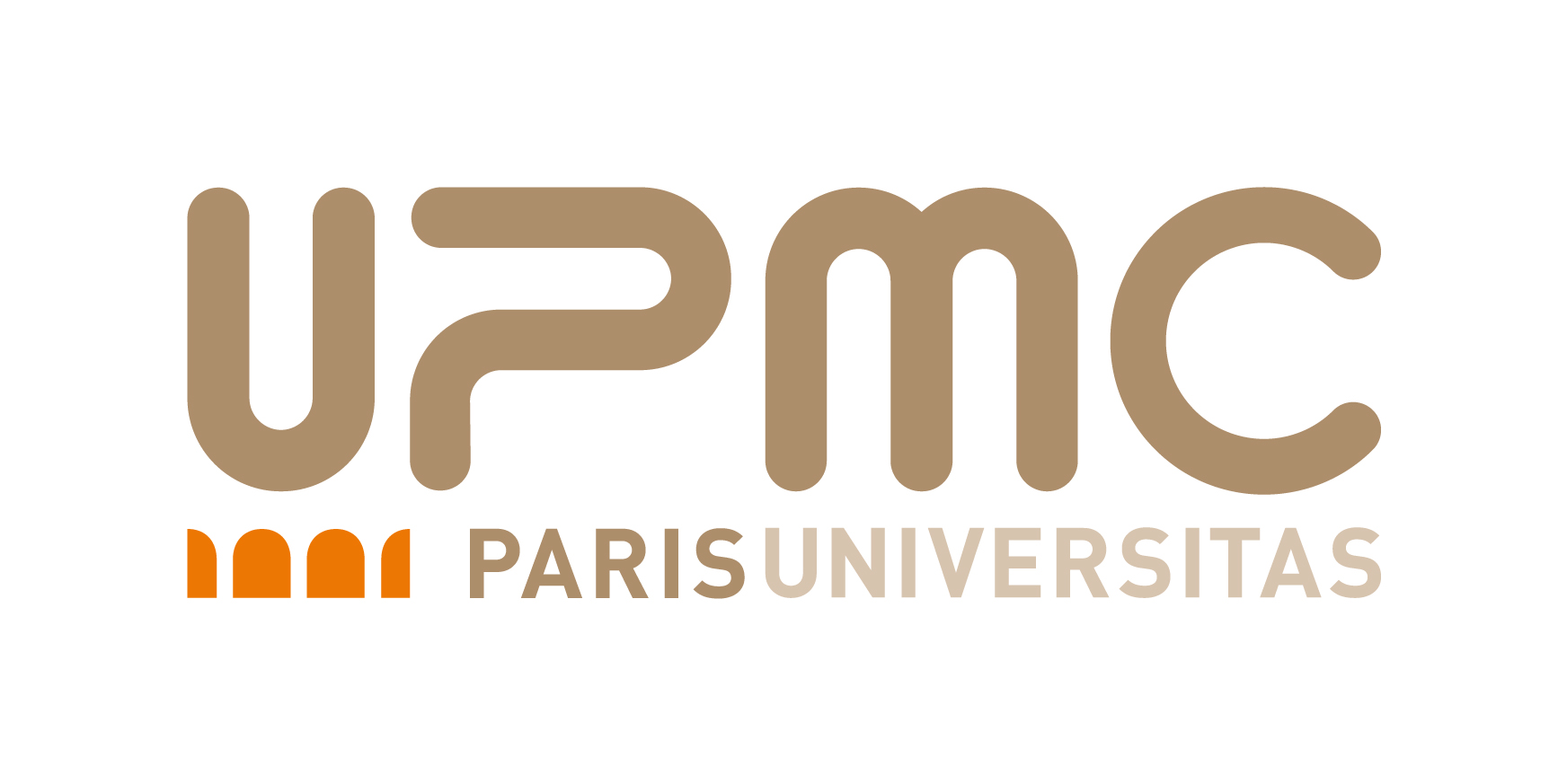}};
\node [xshift=-3.5cm,yshift=-3.0cm] at (current page.north east)
    {\includegraphics[width=2.5cm,height=2.5cm]{./pictures/Logo_ENS.jpg}};
\end{tikzpicture}

\vspace{-2cm}

\begin{center}
\Large{\textsc{Fabien Nugier}}

\smallskip

\Large{\textbf{Moyennes sur le C\^{o}ne de Lumi\`{e}re \\ et Cosmologie de Pr\'{e}cision}}

\bigskip

\large{\textbf{Lightcone Averaging \\ and Precision Cosmology}}
\end{center}

\vfill

\Large{\textbf{R\'{e}sum\'{e}\,:}}

\small

Le premier objectif de cette thèse est de répondre au manque en cosmologie de description sur le cône de lumière passé\,: hypersurface nulle où se propagent tous les signaux observés. Son deuxième est d'évaluer l'importance des inhomogéneités dans la détermination des paramètres de l'énergie sombre, de nature encore inconnue et pour laquelle l'influence des structures de matière a été proposée comme une alternative cosmologique.
Nous définissons une jauge \guillemotleft géodésique sur le cône de lumière \guillemotright ~ qui simplifie grandement l'étude de la propagation lumineuse dans l'Univers. Nous établissons ensuite une moyenne, invariante de jauge, pour les quantités scalaires sur le cône et calculons l'effet des inhomogénéités sur le flux lumineux en considérant leur spectre de puissance jusque son régime non-linéaire. Leur effet sur le flux est négligeable tandis que des observables comme la distance de luminosité, évaluée jusqu'au second ordre en perturbations, sont bien plus affectées. Nous étudions aussi le module de distance à la base de la découverte de l'énergie sombre par les supernov\ae\ (SNe) Ia. La moyenne de ce module est peu affectée par les inhomogénéités mais sa variance l'est sensiblement. Ces résultats montrent, dans leur cadre, qu'une alternative à l'énergie sombre par un effet des structures est impossible mais, en même temps, soulignent leur importance sur la dispersion des données du diagramme de Hubble.
Les effets physiques dominants sont la vitesse des SNe et l'effet de lentille faible. Nos prédictions sur la dispersion se révèlent en très bon accord avec les premières analyses de SNe visant à détecter un signal de lentilles et seront vérifiables dans les années futures.

\bigskip
\textbf{Mots-Cl\'{e}s\,:} Cosmologie, Relativité Générale, Energie Sombre, Univers Inhomog\`{e}nes, Supernov\ae, Lentilles Gravitationnelles.

\vfill

\Large{\textbf{Summary\,:}}

\small

The first objective of this thesis is to fill the lack in cosmology of a description on the past light cone\,: the null hypersurface on which observed signals propagate. Its second goal is to evaluate the importance of inhomogeneities in the determination of dark energy parameters, whose real nature is still unknown and for which the influence of matter structures has been proposed as a cosmological alternative.
We start with the definition of a ``geodesic lightcone'' gauge simplifying greatly the study of light propagation in the Universe. We then define a gauge invariant average of scalars on the past light cone and compute the effect of inhomogeneities on the luminosity flux by considering their power spectrum up to its non-linear regime. The effect on the flux is negligible whereas observables like the luminosity distance, computed at second order in perturbations, are much more affected. We also study the distance modulus at the basis of the discovery of dark energy from type Ia supernov\ae\ (SNe). The average of this modulus is only slightly affected by inhomogeneities but its variance sensitively depends on them. These results show, within their hypotheses, that an alternative to dark energy by an effect of structures is impossible but, at the same time, emphasize their importance on the dispersion of data in the Hubble diagram.
The dominating physical effects turn out to be peculiar velocity of SNe and weak lensing. Our predictions on the dispersion are in very good agreement with first analyses carried out so far to detect a signal of lensing among SNe Ia data and will be testable in the near future.

\bigskip
\textbf{Keywords\,:} Cosmology, General Relativity, Dark Energy, Inhomogeneous Universes, Supernov\ae, Gravitational Lensing.

\vfill

% --------------------- %

\end{document}